\documentclass[]{cwru} % Use `singlespaced` option to make the mainmatter text single-spaced
\pdfoutput=1
\usepackage{physics}
\usepackage[numbers,sort&compress]{natbib}

\usepackage{zref-xr}
\usepackage{graphics}
\usepackage{booktabs}
\usepackage{caption}
\usepackage{floatrow}
\usepackage{epstopdf}
\usepackage{epsfig}
\usepackage[toc,page]{appendix}
\usepackage{placeins}
\usepackage{epsf,epic}
\usepackage{marvosym }
\usepackage{amsmath}
\usepackage{tensor}
\usepackage[utf8]{inputenc}
\usepackage{pgfplots}
\pgfplotsset{compat=newest}
\usepackage{amssymb}
\usepackage{amsfonts}
\usepackage{wrapfig}
\usepackage{multirow}
\usepackage{bm}
\usepackage{dcolumn}
\usepackage[caption=false]{subfig}
\usepackage{upgreek}
\usepackage[euler]{textgreek}
\usepackage{braket}
\usepackage[]{xcolor}
\usepackage[]{hyperref}
\hypersetup{
    colorlinks=true,
    citecolor=blue,
    linkcolor=purple,
    filecolor=magenta,      
    urlcolor=blue,
    }
\usepackage{enumitem}
\usepackage{etaremune}
\newcommand{\stkout}[1]{\ifmmode\text{\sout{\ensuremath{#1}}}\else\sout{#1}\fi}
\newcommand{\bs}{\boldsymbol}
\newcommand{\pd}{\partial}
\newcommand{\pr}{^{\prime}}
\newcommand{\dpr}{^{\prime\prime}}

\newcommand{\mbk}{\mathbf{k}}
\newcommand{\mbkpr}{\mathbf{k}^{\prime}}
\newcommand{\mbkdpr}{\mathbf{k}^{\prime \prime}}

\newcommand{\veps}{\varepsilon}
\newcommand{\dprime}{\prime\prime}
\newcommand{\tprime}{\prime\prime\prime}
\title{Quantum Geometry and Nonlinear Responses in Magnetic and Topological Quantum Materials}
\author{Mandela Mehraeen}
\date{January, 2026} % Graduate Date
\doctype{dissertation} % dissertation/thesis
\degree{Doctor of Philosophy}
\department{Physics}
\defensedate{July 10, 2025}

\begin{document}

\advisor{Dr. Shulei Zhang}
\committee{Dr. Walter Lambrecht}
\committee{Dr. Harsh Mathur}
\committee{Dr. Stanisław Szarek}

% The organization of the dissertation must follow the order below:
% 
% Title page
% Committee Approval Sheet
% Copyright page (only if copyrighting)
% Dedication page (optional)
% Table of Contents
% List of Tables
% List of Figures
% Preface (optional)
% Acknowledgements (optional)
% List of Abbreviations (optional)
% Glossary (optional)
% Abstract
% --TEXT--
% Appendix
% Bibliography

\maketitle
\makeapprovalsheet

\frontmatter
\tableofcontents

\begin{listofpublications}

\noindent \cite{mehraeen2020fermion}
M. Mehraeen and S. S. Gousheh, Fermion number 1/2 of sphalerons and spectral mirror symmetry, \href{https://doi.org/10.1140/epjc/s10052-020-08451-4}{\color{blue}Eur. Phys. J. C \textbf{80}, 891 (2020)}

\noindent \cite{abbaslu2021generation}
S. Abbaslu, S. Rostam Zadeh, M. Mehraeen and S. S. Gousheh, The generation of matter–antimatter asymmetries and hypermagnetic fields by the chiral vortical effect of transient fluctuations, \href{https://doi.org/10.1140/epjc/s10052-021-09272-9}{\color{blue}Eur. Phys. J. C \textbf{81}, 500 (2021)}.

\noindent \cite{mehraeen2022spin}
M. Mehraeen and S. S.-L. Zhang, Spin anomalous-Hall unidirectional magnetoresistance, \href{https://doi.org/10.1103/PhysRevB.105.184423}{\color{blue}Phys. Rev. B \textbf{105}, 184423 (2022)}.

\noindent \cite{shim2022unidirectional}
S. Shim, M. Mehraeen, J. Sklenar, J. Oh, J. Gibbons, H. Saglam, A. Hoffmann, S. S.-L. Zhang, and N. Mason, Unidirectional Magnetoresistance in Antiferromagnet/Heavy-Metal Bilayers, \href{https://doi.org/10.1103/PhysRevX.12.021069}{\color{blue}Phys. Rev. X \textbf{12}, 021069 (2022)}.

\noindent \cite{zhang2022large}
Y. Zhang, V. Kalappattil, C. Liu, M. Mehraeen, S. S.-L. Zhang, J. Ding, U. Erugu, Z. Chen, J. Tian, K. Liu, J. Tang and M. Wu, Large magnetoelectric resistance in the topological Dirac semimetal $\alpha$-Sn, \href{https://doi.org/10.1126/sciadv.abo0052}{\color{blue}Sci. Adv. \textbf{8}, eabo0052 (2022)}.

\noindent \cite{mehraeen2023quantum}
M. Mehraeen, P. Shen and S. S.-L. Zhang, Quantum unidirectional magnetoresistance, \href{https://doi.org/10.1103/PhysRevB.108.014411}{\color{blue}Phys. Rev. B \textbf{108}, 014411 (2023)}.

\noindent \cite{mehraeen2024proximity}
M. Mehraeen and S. S.-L. Zhang, Proximity-induced nonlinear magnetoresistances on topological insulators, \href{https://doi.org/10.1103/PhysRevB.109.024421}{\color{blue}Phys. Rev. B \textbf{109}, 024421 (2024)}.

\noindent \cite{damerio2024magnetoresistive}
S. Damerio, A. Sunil, W. Janus, M. Mehraeen, S. S.-L. Zhang and C. O. Avci, Magnetoresistive detection of perpendicular switching in a magnetic insulator, \href{https://doi.org/10.1038/s42005-024-01604-x}{\color{blue} Commun. Phys., \textbf{7}, 114 (2024)}.

\noindent \cite{mehraeen2024quantum}
M. Mehraeen, Quantum kinetic theory of quadratic responses, \href{https://doi.org/10.1103/PhysRevB.110.174423}{\color{blue}Phys. Rev. B \textbf{110}, 174423 (2024)}.

\noindent \cite{shim2025spin}
S. Shim, M. Mehraeen, J. Sklenar, A. Hoffmann, S. S.-L. Zhang, and N. Mason, Spin-polarized antiferromagnetic metals, \href{https://doi.org/10.1146/annurev-conmatphys-042924-123620}{\color{blue} Annu. Rev. Condens. Matter Phys. 16 (2025)}.

\noindent \cite{mehraeen2025quantum}
M. Mehraeen, Quantum Response Theory and Momentum-Space Gravity, \href{https://doi.org/10.1103/t6nt-qzws}{\color{blue}Phys. Rev. Lett. \textbf{135}, 156302 (2025)}.

\noindent \cite{jain2025nonlinear}
A. Jain, W. J. Jankowski, M. Mehraeen, R.-J. Slager, Nonlinear Odd Viscoelastic Effect, \href{https://arxiv.org/abs/2511.22706}{\color{blue}arXiv:2511.22706 (2025)}.

\noindent \cite{jain2026topological}
A. Jain, W. J. Jankowski, M. Mehraeen, R.-J. Slager, Topological Acoustic Diode, \href{https://arxiv.org/abs/2601.20951}{\color{blue}arXiv:2601.20951 (2026)}.

\end{listofpublications}

\cleardoublepage
\phantomsection
\addcontentsline{toc}{chapter}{List of Tables}
\listoftables

\cleardoublepage
\phantomsection
\addcontentsline{toc}{chapter}{List of Figures}
\listoffigures

\begin{acknowledgments}
\noindent I would like to thank my PhD advisor, Prof. Shulei Zhang, for his guidance and support throughout my doctoral studies and for helping me broaden my horizons as a researcher. I am also grateful for all the collaborators with whom I had the privilege of interacting, which helped further enrich my PhD experience. Furthermore, I would like to thank the doctoral committee members, Prof. Walter Lambrecht, Prof. Harsh Mathur and Prof. Stanislaw Szarek for taking the time to referee my defense and for sharing their thoughts and insights. Finally, I am grateful to my family and friends--many of whom I had the pleasure of meeting in Cleveland--for their encouragement, friendship and support.
\end{acknowledgments}

% If you're using `glossaries` package

%\cleardoublepage
%\phantomsection
%\addcontentsline{toc}{chapter}{List of Acronyms}
%\printglossary[type=\acronymtype]

%\cleardoublepage
%\phantomsection
%\addcontentsline{toc}{chapter}{Glossary}
%\printglossary

\begin{abstract}
This dissertation explores various nonlinear responses that arise from the rich interplay between quantum geometry, disorder, magnetism and topology in quantum materials. In addition to presenting generalizations of quantum kinetic theory, Kubo formulas and semiclassical Boltzmann transport theory to the nonlinear response regime, we discuss several predictions of novel transport effects and physical insights that emerge from these developments.

First, we present a comprehensive generalization of quantum kinetic theory to the nonlinear response regime via a disorder-averaged density-matrix formalism. In the presence  of an electrostatic potential and random impurities, we solve the quantum Liouville equation to second order in an applied electric field and derive the carrier densities and equations of motion. In addition to the anomalous velocity arising from the Berry curvature and the Levi-Civita connection of the quantum metric tensor, a host of extrinsic velocities emerge in the equations of motion, reflecting the various possibilities for random interband walks of the carriers in this transport regime. Furthermore, several scattering and conduction channels arise, which can be classified in terms of distinct physical processes, revealing numerous unexplored mechanisms for generating nonlinear responses in disordered condensed matter systems. 

We then present a quantum response approach to momentum-space gravity in dissipative multiband systems, which dresses both the quantum geometry--through an interband Weyl transformation--and the equations of motion. In addition to clarifying the roles of the contorsion and symplectic terms, we introduce the three-state quantum geometric tensor as a necessary element in the geometric classification of nonlinear responses and discuss the significance of the emergent terms from a gravitational point of view. We also identify a dual quantum geometric drag force in momentum space that provides an entropic source term for the multiband matrix of Einstein field equations.

Next, we predict a spin anomalous-Hall unidirectional magnetoresistance in conducting bilayers composed of a ferromagnetic layer and a nonmagnetic layer, which does not rely on the spin Hall effect in the normal metal layer|in stark contrast to the well-studied unidirectional spin-Hall magnetoresistance|but, instead, arises from the spin anomalous Hall effect in the ferromagnetic layer. Physically, it is the charge-spin conversion induced by the spin anomalous Hall effect that conspires with the structural inversion asymmetry to generate a net nonequilibrium spin density in the ferromagnetic layer, which, in turn, modulates the resistance of the bilayer when the direction of the applied current or the magnetization is reversed. The dependences of the spin AH-UMR effect on materials and geometric parameters are analyzed and compared with other nonlinear magnetoresistances. In particular, we show that, in magnetic bilayers where spin anomalous Hall and spin Hall effects are comparable, the overall UMR may undergo a sign change when the thickness of either layer is varied, suggesting a scheme to quantify the spin Hall or spin anomalous Hall angle via a nonlinear transport measurement. 

Furthermore, we predict unidirectional magnetoresistance effects arising in a bilayer composed of a nonmagnetic metal and a ferromagnetic insulator, whereby both longitudinal and transverse resistances vary when the direction of the applied electric field is reversed or the magnetization of the ferromagnetic layer is rotated. In the presence of spin-orbit coupling, an electron wave incident on the interface of the bilayer undergoes a spin rotation and a momentum-dependent phase shift. Quantum interference between the incident and reflected waves furnishes the electron with an additional velocity that is even in the in-plane component of the electron's wavevector, giving rise to quadratic magnetotransport that is rooted in the wave nature of electrons. The corresponding unidirectional magnetoresistances exhibit decay lengths at the scale of the Fermi wavelength--distinctive signatures of the quantum nonlinear magnetotransport effect.

Finally, we employ quadratic-response Kubo formulas to investigate the nonlinear magnetotransport in bilayers composed of a topological insulator and a magnetic insulator, and predict both unidirectional magnetoresistance and nonlinear planar Hall effects driven by interfacial disorder and spin-orbit scattering. These effects exhibit strong dependencies on the Fermi energy relative to the strength of the exchange interaction between the spins of Dirac electrons and the interfacial magnetization. In particular, as the Fermi energy becomes comparable to the exchange energy, the nonlinear magnetotransport coefficients can be greatly amplified and their dependencies on the magnetization orientation deviate significantly from conventional sinusoidal behavior. These findings may not only deepen our understanding of the origin of nonlinear magnetotransport in magnetic topological systems but also open new pathways to probe the Fermi and exchange energies via transport measurements.
\end{abstract}

\mainmatter

\chapter{Quantum Kinetic Theory of Quadratic Responses}

\section{Introduction}

Owing to its fundamental interest and practical significance, the study of carrier transport in disordered condensed matter systems has a rich and multifaceted history, which has inspired a variety of complementary theoretical approaches, including Kubo formulas~\cite{kubo1957statistical}, Keldysh theory~\cite{keldysh2024diagram} and density-matrix methods~\cite{karplus1954hall, kohn1957quantum, luttinger1958quantum, luttinger1958theory}. Despite their seemingly different approaches to obtaining response functions, which range from evaluating dressed propagators and vertex functions via diagrammatic methods to ensemble-averaging observables weighted by carrier distribution functions, the underlying commonality of all these methods may be regarded as the attempt to quantify the evolution of the density operator in the presence of disorder and other interactions.

In light of this shared connection, and with the goal of deriving a more complete description of carrier conduction on macroscopic scales, a recurring research theme in the linear response regime has been the attempt to incorporate the results of diagrammatic methods into quantum kinetic theory~\cite{sinitsyn2005disorder, sinitsyn2007anomalous, sinitsyn2007semiclassical, culcer2017interband, sekine2017quantum, xiao2017semiclassical, xiao2019temperature, atencia2022semiclassical}. The former approach allows for a robust treatment of perturbations and is readily generalizable to many-body problems~\cite{mahan2000many, bruus2004many}. The kinetic approach, while also providing a systematic perturbative analysis~\cite{vasko2006quantum}, deals directly with carrier distribution functions and thus often allows for a more transparent transition to the physically intuitive and relatively simple semiclassical limit~\cite{xiao2010berry}.

Through a series of recent works, a proposal has been put forward of a modified semiclassical framework that is based on solving the quantum Liouville equation for the disorder-averaged density matrix~\cite{culcer2017interband, sekine2017quantum, atencia2022semiclassical}. This is a versatile approach that allows for a systematic analysis of the effects of disorder scattering on physical observables and has recently been applied to a variety of problems in response theory, including the study of resonance structures~\cite{bhalla2020resonant, bhalla2022resonant}, probing quantum geometry~\cite{cullen2021generating, bhalla2022resonant} and exploring nonlinear electric and thermal responses~\cite{bhalla2023quantum, atencia2023disorder, varshney2023quantum}. In this approach, the effect of disorder naturally arises not only in the carrier distributions--which is traditionally the case~\cite{ashcroft1976solid}--but also in the equations of motion governing the carrier dynamics. Thus, in the linear response regime, in addition to the familiar group and (intrinsic) anomalous velocities~\cite{karplus1954hall, kohn1957quantum, adams1959energy, chang1995berry, chang1996berry, sundaram1999wave}, the semiclassical equations of motion are modified by an additional disorder-dependent term dubbed the extrinsic velocity~\cite{atencia2022semiclassical}. In this picture, just as the intrinsic anomalous velocity is related to interband coherence effects mediated by an electric field, the extrinsic velocity can similarly be thought of as arising from extrinsic interband coherence effects due to the Berry connection and disorder. 

As a consequence of introducing the disorder average at the level of the quantum mechanical density matrix, the modified semiclassical equations can now be regarded as describing the averaged motion of carriers after many scattering events. The extrinsic velocity thus quantifies the average change in the carrier velocity after multiple random interband walks due to disorder scattering~\cite{atencia2022semiclassical} and is related to the side-jump velocity associated with the coordinate shift at scattering events~\cite{sinitsyn2006coordinate}. 

In addition to the equations of motion, the extrinsic velocity also modifies the linear response functions. For the specific case of the anomalous Hall effect~\cite{karplus1954hall, luttinger1958theory, smit1958spontaneous, berger1970side, nozieres1973simple}, it has been shown that including the current density associated with the extrinsic velocity results in a total conductivity that agrees with that obtained via diagrammatic methods in the noncrossing approximation in the presence of disorder~\cite{atencia2022semiclassical}. This is a notable result, as it allows one to obtain diagrammatically equivalent results by a relatively simple and physically transparent semiclassical approach. In addition, it can improve the accuracy of computational methods, which are commonly performed using the semiclassical method~\cite{wang2006abinitio, wang2007fermi, gradhand2012first, he2012berry, chen2013weyl, bianco2014how, chen2014anomalous, olsen2015valley, feng2016first, dai2017negative, martiny2019tunable, wuttke2019berry, du2020berry, he2020giant, he2021superconducting}.

Going beyond linear responses, the past decade has witnessed a surge of interest in nonlinear spin and charge transport phenomena in a variety of materials systems with differing magnetic order and band topology~\cite{du2021nonlinear, ideue2021symmetry, ortix2021nonlinear, nagaosa2024nonreciprocal, shim2024spin}. Several novel transport phenomena have emerged, including unidirectional~\cite{avci2015magnetoresistance, avci2015unidirectional, olejnik2015electrical, zhang2016theory, yasuda2016large, avci2018origins, lv2018unidirectional, duy2019giant, guillet2020observation, zelezny2021unidirectional, guillet2021large, hasegawa2021enhanced, liu2021magnonic, liu2021chirality, chang2021large, shim2022unidirectional, mehraeen2022spin, ding2022unidirectional, lou2022large, mehraeen2023quantum, cheng2023unidirectional, fan2023observation, zheng2023coexistence,  mehraeen2024proximity, zou2024nonreciprocal, zhao2024large, aoki2024evaluation, huang2024spin, kao2024unconventional} and bilinear~\cite{rikken2001electrical, he2018bilinear, dyrdal2020spin, zhang2022large, wang2022large, fu2022bilinear, golub2023electrical,  marx2024nonlinear, boboshko2024bilinear, kim2024spin} magnetoresistance effects, as well as a variety of nonlinear Hall effects~\cite{sodemann2015quantum, low2015topological, yasuda2017current, du2018band, facio2018strongly, you2018berry, zhang2018electrically, zhang2018berry, he2019nonlinear, ma2019observation, kang2019nonlinear, du2019disorder, wang2021intrinsic, li2021nonlinear, zeng2021nonlinear, wang2022observation, gao2023quantum, wang2023quantum, kaplan2023general, das2023intrinsic, ma2023anomalous, zhuang2024intrinsic, wang2024intrinsic}. From a fundamental point of view, nonlinear responses provide an opportunity to probe the nontrivial quantum geometry of Bloch states when linear responses are prohibited by symmetry~\cite{liu2024quantum}, thereby allowing for an extension of this analysis to a more diverse range of quantum materials with different lattice and space-time symmetries. Recent observations of the nonlinear Hall effect in time-reversal-symmetric conditions~\cite{ma2019observation, kang2019nonlinear} and intrinsic quadratic responses in topological antiferromagnets~\cite{gao2023quantum, wang2023quantum} with combined parity-inversion time-reversal symmetry not only yield valuable insights into the quantum geometry of these materials systems through the Berry curvature dipole or quantum metric tensor, but can also be harnessed for useful applications, such as Neel vector detection in parity-inversion time-reversal symmetric antiferromagnets~\cite{wang2021intrinsic, liu2021intrinsic}.

Furthermore, even when linear effects dominate the response of the system, the unidirectional nature of quadratic responses with respect to the electric field implies that they can rather straightforwardly be distinguished from the linear-response signal through a simple reversal in the direction of the applied electric field. In addition, as has recently been demonstrated in a variety of magnetic heterostructures~\cite{avci2015unidirectional, olejnik2015electrical, yasuda2016large, lv2018unidirectional, duy2019giant, zelezny2021unidirectional, guillet2021large, hasegawa2021enhanced, liu2021magnonic, chang2021large, shim2022unidirectional, ding2022unidirectional, lou2022large, cheng2023unidirectional, fan2023observation, zheng2023coexistence,  mehraeen2024proximity, zhao2024large, aoki2024evaluation, huang2024spin}, nonlinear transport effects such as unidirectional magnetoresistance are typically also odd in the in-plane magnetization, and are thus suitable for magnetic order detection, regardless of the presence of linear responses. Given the complementary applications and different origins and symmetry properties that nonlinear responses have with respect to their linear-response counterparts and the growing number of nonlinear transport effects being discovered, it is highly desirable to develop a unifying theoretical framework to capture these effects.

Physically, several mechanisms have been identified for generating nonlinear responses, which can generally be attributed to the combined effects of disorder scattering, band topology and the quantum geometry of Bloch states--realized through the Berry curvature and quantum metric tensors. And a number of theoretical extensions of linear-response methods have been proposed to accommodate nonlinear effects within the modern perspective of transport, including Kubo formulas~\cite{parker2019diagrammatic, du2021quantum, rostami2021gauge, kaplan2023unifying, mckay2024charge}, modified kinetic frameworks~\cite{xiao2019theory, nandy2019symmetry, bhalla2023quantum, atencia2023disorder, ba2023nonlinear, huang2023scaling}, drift-diffusion models~\cite{zhang2016theory, mehraeen2022spin} and nonequilibrium Keldysh theory~\cite{freimuth2021theory}.

\begin{figure}[hpt]
%\vspace{-.7cm}
\captionsetup[subfigure]{labelformat=empty}
    \sidesubfloat[]{\includegraphics[width=0.7\linewidth,trim={3cm -1cm 3cm -1cm}]{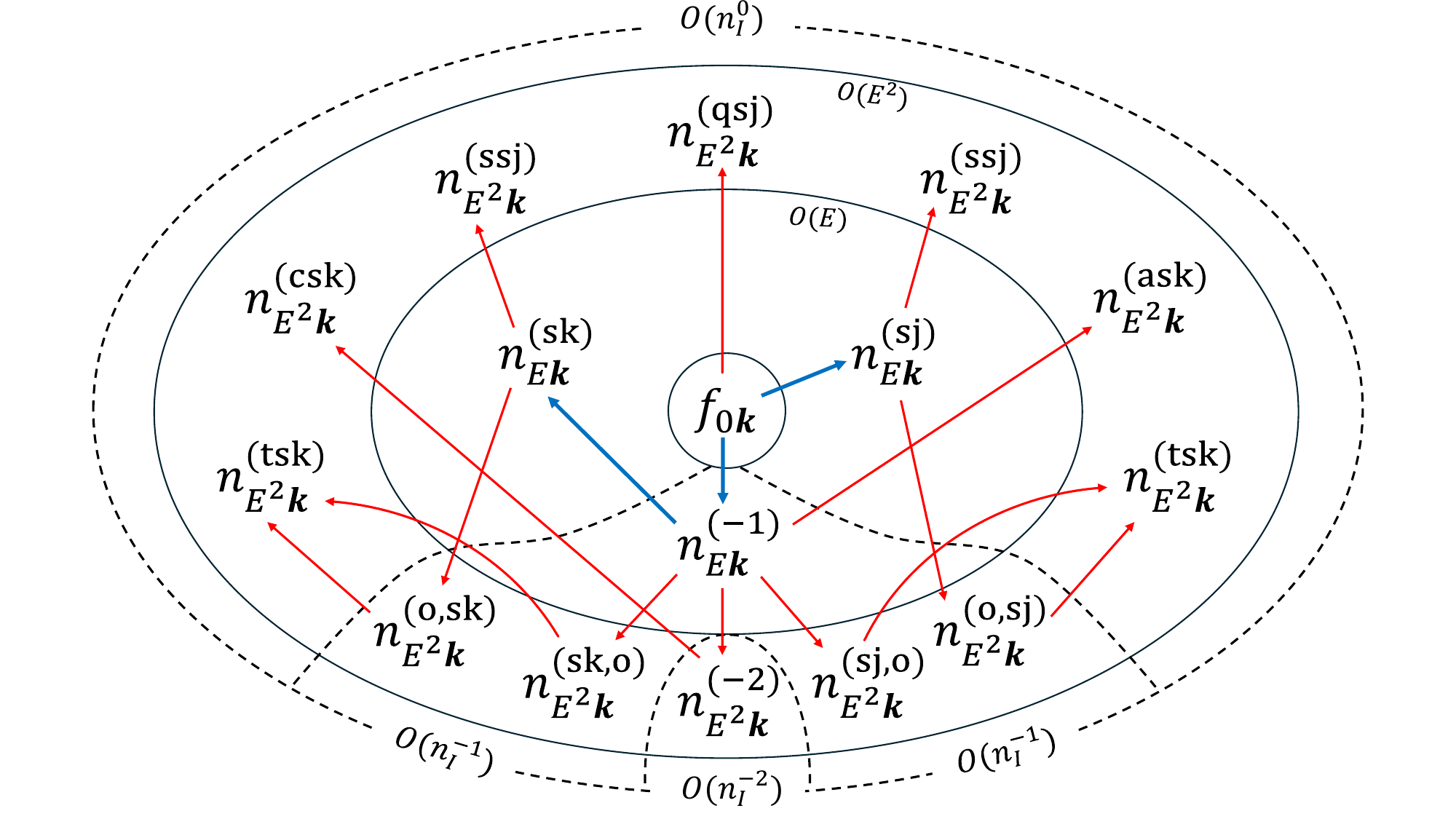}\label{fig_qkt_fig1a}}
%\quad%
%%%%%%%%%
    \caption{Schematic map of carrier densities at first and second order in the applied electric field and their physical origins, starting from the equilibrium (Fermi-Dirac) distribution $f_{0\mbk}$. The three blue central arrows represent, ordinary, side-jump and skew scattering in linear response theory, while the surrounding red arrows arise in quadratic response theory. The dashed lines separate different orders in the impurity density.}
    \label{fig_qkt_fig1}
\end{figure}

In this chapter~\cite{mehraeen2024quantum}, we present a generalization of the linear response theory proposed in recent works~\cite{culcer2017interband, sekine2017quantum, atencia2022semiclassical} to the nonlinear response regime. In the presence of the electrostatic interaction and in the weak-disorder limit, we systematically solve the quantum Liouville equation to second order in the applied electric field and to three orders in the impurity density, revealing an extensive pattern of scattering processes. As illustrated in Fig.~\ref{fig_qkt_fig1}, disorder scattering in linear response theory is represented by the triality of ordinary, side-jump and skew scattering~\cite{smit1958spontaneous, berger1970side, nagaosa2010anomalous}. In the quadratic response regime, we show that this triality underpins a relatively large number of scattering processes that contribute to the carrier density at different orders in the impurity density.

Furthermore, we uncover several field-independent and -dependent extrinsic velocities, which, in conjunction with the linear-response extrinsic velocity, account for disorder corrections to the equations of motion up to second order in the applied field. Specifically, we find that the equation of motion for the carrier position in the presence of disorder reads
\begin{equation}
\label{eom}
\begin{split}
\dot{x}_{\mu}^a
&=
v_{\mu \mbk}^a
-
\varepsilon_{\mu \nu \rho} \dot{k}_{\nu}^a \Omega_{\rho \mbk}^a
+
\dot{k}_{\nu}^a \dot{k}_{\rho}^a
\sum_{b} M_{\mbk}^{ab}
\Gamma_{\nu \rho \mu \mbk}^{ba}
+
\alpha_{\mu \mbk}^{a} 
+
\beta_{\mu \mbk}^{a}
+
\gamma_{\mu \mbk}^{a}
+
\kappa_{\mu \mbk}^{a}
+
\chi_{\mu \mbk}^{a},
\end{split}
\end{equation}
where $a$ is the band index and Greek letters represent spatial indices. Here
$\bs{v}_{\mbk}^a 
=
\bs{\pd}_{\mbk} \veps_{\mbk}^a /\hbar$ is the group velocity of carriers in the energy band $\veps_{\mbk}^a$, $\bs{\Omega}_{\mbk}^{ab}
=
\bs{\pd}_{\mbk} 
\times
\bs{\mathcal{A}}_{\mbk}^{ab}$ is the Berry curvature, with 
$\bs{\Omega}_{\mbk}^{a}
\equiv
\bs{\Omega}_{\mbk}^{aa}$ and $\bs{\mathcal{A}}_{\mbk}^{ab}
 =
 i \braket{u_{\mbk a} | \bs{\pd}_{\mbk} u_{\mbk b}}$ the Berry connection, with $\ket{u_{\mbk a}}$ the periodic part of the Bloch state. And
 $\Gamma_{\mu \nu \rho \mbk}^{a b}
=
\frac{1}{2} 
( \pd^{\mbk}_{\rho} g_{\mu \nu \mbk}^{ab}
+
\pd^{\mbk}_{\nu} g_{\mu \rho \mbk}^{ab}
-
\pd^{\mbk}_{\mu} g_{\nu \rho \mbk}^{ab})$ is a Bloch-space Levi-Civita connection component of the band-resolved quantum metric~\cite{provost1980riemannian, cheng2010quantum},
$
g_{\mu \nu \mbk}^{ab}
=
\text{Re}(\mathcal{A}_{\mu \mbk}^{ab} \mathcal{A}_{\nu \mbk}^{ba})
$.
The remaining terms in Eq.~(\ref{eom}), which we present in Sec.~\ref{sec_emergent}, are  the extrinsic velocities that emerge from the solution of the quantum Liouville equation and constitute the disorder corrections to the carrier motion.

Physically, the intrinsic terms can be understood as representing the effects of purely field-induced corrections to the Bloch states as a result of interband mixing, while the extrinsic velocities can be interpreted as encapsulating the averaged effects of random walks between energy bands due to impurity scattering, which may or may not depend on the electric field. In this sense, the approach presented here places the effects of electrostatic and disorder interactions on a more equal footing as far as the carrier dynamics is concerned and includes both their effects in the equations of motion. As we show below, the various carrier densities and velocities that emerge allow for numerous physically distinct conductivity channels to exist, reflecting the multitude of mechanisms that have partially been explored in the literature.

The remainder of the chapter is organized as follows. In Sec.~\ref{sec_qke}, we introduce the model and derive the quantum kinetic equation. In Sec.~\ref{sec_linear}, we briefly review the derivation of the linear-response densities. In Sec.~\ref{sec_quad}, we discuss in detail the derivation of the quadratic density matrix. In Sec.~\ref{application}, we apply the theory to study the nonlinear transport in a model of 2D Dirac fermions. Finally, we present concluding remarks and offer an outlook.  

%\vspace{-1cm}

\section{Quantum kinetic equation}
\label{sec_qke}

We begin our analysis with the Hamiltonian
\begin{equation}
H
=
H^0 + V(\mathbf{r}) + U(\mathbf{r}),
\end{equation}
where $H^0$ describes the unperturbed system, $V(\mathbf{r})=e \mathbf{E} \cdot \mathbf{r}$ is the electrostatic potential and $U(\mathbf{r})$ is the disorder potential, which we assume describes scalar point scatterers with a white noise distribution, $\braket{U(\mathbf{r})}=0$,  
$\braket{U(\mathbf{r}) U(\mathbf{r\pr})}=n_I U_0^2 \delta(\mathbf{r} - \mathbf{r}\pr)$, and neglect higher-order non-Gaussian disorder correlations. Here, $n_I$ is the impurity density, $U_0$ measures the strength of the impurity interaction and $\braket{\cdots}$ indicates the disorder average~\footnote{It should be stressed that the choice of scalar disorder here is merely one of convenience and the theory is readily applicable to spin-dependent disorder profiles as well.}. We work within the crystal momentum representation, $\ket{\psi_{\mathbf{k}a}}
=
e^{i \mathbf{k} \cdot \mathbf{r}}
\ket{u_{\mathbf{k}a}}$, where $\ket{\psi_{\mathbf{k}a}}$ is an eigenstate of the unperturbed system, with 
$H^0 \ket{\psi_{\mathbf{k}a}} = \veps_{\mbk}^a\ket{\psi_{\mathbf{k}a}}$. The disorder averages then read $\braket{U_{\mbk \mbkpr}^{a a\pr}} = 0$ and 
$\braket{U_{\mbk \mbkpr}^{a a\pr} 
U_{\mbkpr \mbk}^{b\pr b}} 
=
n_I U_0^2 \braket{u_{\mbk a} | u_{\mbkpr a\pr}}
\braket{u_{\mbkpr b\pr} | u_{\mbk b}}$, where
$U_{\mbk \mbkpr}^{a a\pr}
\equiv
\Braket{\psi_{\mbk a} | U | \psi_{\mbkpr a\pr}}$.

Consider the quantum Liouville equation for the single-particle density operator 
\begin{equation}
\label{qke}
\frac{\pd \rho}{\pd t} + \frac{i}{\hbar} \left[ H, \rho \right]
=
0.
\end{equation}
We express the density operator as the sum of a disorder-averaged part and a fluctuating part, 
$\rho = \braket{\rho} + \delta \rho$. Then, within the Born approximation, Eq.~(\ref{qke}) decomposes as
\begin{subequations}
\label{qke_decomp}
\begin{gather}
\label{qke_avg}
\pd_t \braket{\rho}
+
\frac{i}{\hbar} \left[ H^0, \braket{\rho} \right]
+
J \left( \braket{\rho}\right)
=
- \frac{i}{\hbar} \left[ V, \braket{\rho} \right],
\\
\label{qke_fluc}
\pd_t \delta \rho
+
\frac{i}{\hbar} \left[ H^0, \delta \rho \right]
=
- \frac{i}{\hbar} \left[ U, \braket{\rho} \right]
-
\frac{i}{\hbar} \left[ V, \delta \rho \right],
\end{gather}
\end{subequations}
where $J \left( \braket{\rho}\right) \equiv \frac{i}{\hbar} \braket{ \left[ U, \delta \rho \right]}$ is the collision integral due to disorder scattering. For the purpose of studying nonlinear responses, we consider the field expansion of the density operator,
$\braket{\rho}
=
\sum_{n=0} \braket{\rho_{E^n}}$
and 
$\delta \rho
=
\sum_{n=0} \delta \rho_{E^n}$ (with the convention 
$A_0 \equiv A_{E^0}$), which yields the solution
\begin{equation}
\label{delta_rho_0}
\delta \rho_0 (t)
=
\int_{-\infty}^{\infty} \frac{d\veps}{2 \pi i}
G^R(\veps) \left[ U, \braket{\rho_0 (t)}\right] G^A(\veps),
\end{equation}
for the field-independent part and the recursive solution
\begin{equation}
\label{delta_rho_n}
\begin{split}
\delta \rho_{E^n} (t)
=&
\int_{-\infty}^{\infty} \frac{d\veps}{2\pi i}
G^R(\veps) \mathcal{R}_{E^n}(t) G^A(\veps),
\end{split}
\end{equation}
for the higher-order fluctuations, where
$\mathcal{R}_{E^n}(t)
=
[ U, \braket{\rho_{E^n} (t)} ] + [ V, \delta \rho_{E^{n-1}} (t)]$
and
$G^{R/A}(\veps)
=
\left( \veps - H^0 \pm i \delta \right)^{-1}$ 
is the retarded (advanced) Green's function of the unperturbed system~\footnote{Here, as in Ref.~\cite{atencia2022semiclassical}, we assume the Markovian approximation, whereby time derivatives of the fluctuations can be neglected.}. To exploit the sum separability of the recursive solution, it is convenient to henceforth introduce the notation with square brackets as 
$\delta \rho_{E^n}(\braket{\rho_{0}}, \ldots, \braket{\rho_{E^n}} )
\equiv
\sum_{l=0}^n
\delta \rho_{E^n} \left[ \braket{\rho_{E^l}} \right]$
to specify individual contributions to the fluctuations.

In order to solve for the disorder-averaged part of the density operator, we first express Eq.~(\ref{qke_avg}) in the crystal momentum representation, which gives rise to the quantum kinetic equation
\begin{equation}
\label{qke_avgk}
\pd_t f_{\mbk}
+
\frac{i}{\hbar} \left[ H_{\mbk}^0, f_{\mbk} \right]
+
J_{\mbk} \left( f_{\mbk}\right)
=
\frac{e}{\hbar} \mathbf{E} \cdot \bs{\mathcal{D}}_{\mbk} f_{\mbk},
\end{equation}
where $f_{\mbk}^{ab} \equiv
\braket{\psi_{\mbk a} | \braket{\rho} | \psi_{\mbk b}}$ is the distribution function and $\bs{\mathcal{D}}_{\mbk}$ is the Berry covariant derivative, which is obtained from the the k-space representation of the position operator~\cite{blount1962formalisms}, 
$\mathbf{r}_{\mbk \mbk\pr}
=
-i \bs{\pd}_{\mbk\pr} \delta_{\mbk\pr \mbk}
+
\delta_{\mbk\pr \mbk} \bs{\mathcal{A}}_{\mbk}$, resulting in the familiar form 
\begin{equation}
\bs{\mathcal{D}}_{\mbk} \mathcal{O}_{\mbk}
=
 \bs{\pd}_{\mbk} \mathcal{O}_{\mbk}
 -
 i \left[
 \bs{\mathcal{A}}_{\mbk}, \mathcal{O}_{\mbk} \right],
\end{equation}
for a generic k-diagonal operator $ \mathcal{O}_{\mbk}$. To second order in the electric field, the momentum-space density operator reads
$f_{\mbk}
=
f_{0\mbk} + f_{E\mbk} + f_{E^2\mbk}$, with $f_{0\mbk}$ the equilibrium distribution, which--for most practical purposes--is the Fermi-Dirac distribution, and we further decompose the field-dependent parts to diagonal and off-diagonal components in band space as 
$f_{E\mbk} = n_{E\mbk} + S_{E\mbk}$ and $f_{E^2\mbk} = n_{E^2\mbk} + S_{E^2\mbk}$. Then, through a perturbative analysis in powers of the electric field and disorder parameter, we can solve for both the diagonal part $n_{\mbk}$ and the off-diagonal components $S_{\mbk}$. To achieve this, the collision integral is also decomposed using the square-bracket notation,  
$J_{\lambda \mbk} \left( f_{\eta \mbk} \right)
\equiv
\frac{i}{\hbar}
\Braket{
\left[ U , \delta \rho_{\lambda} \left[ \braket{\rho_{\eta} } \right] \right]
}_{\mbk}$, with $\lambda, \eta = 0,E,E^2$, the explicit form of which is presented in Sec.~\ref{App_collision_int}.

\section{Linear density matrix}
\label{sec_linear}

First, we revisit the quantum kinetic equation at linear order in the applied electric field. This transport regime has been extensively studied in the literature and the discussion presented in this section is not new. However, since the contributions to the linear-response densities and their physical interpretations are required for obtaining and classifying quadratic-response quantities, for completeness--as well as for notational uniformity--we briefly review here the main results. A detailed treatment of this can be found in Ref.~\cite{atencia2022semiclassical}.

As $n_{E\mbk}$ is at most linear the scattering time, we start at $O(n_I^{-1})$ in the perturbative analysis and take 
$n_{E\mbk} = n_{E\mbk}^{(-1)} + n_{E\mbk}^{(0)}$. Assuming a steady-state solution for the band-diagonal carrier density, inspection of Eq.~(\ref{qke_avgk}) reveals that the lowest-order contribution from $S_{E\mbk}$ is at $O(n_I^{0})$. Thus, the field-linear contribution to the quantum kinetic equation is decomposed as
\begin{subequations}
\label{qke_avgk_E}
\begin{gather}
\label{qke_avgk_E_a}
J_{0\mbk}^a (n_{E\mbk}^{(-1)})
=
\frac{e}{\hbar} \mathbf{E} \cdot \bs{\pd}_{\mbk} f_{0\mbk}^a,
\\
\label{qke_avgk_E_b}
\pd_t S_{E \mbk}^{(0)}
+
\frac{i}{\hbar} \left[ H^0, S_{E \mbk}^{(0)} \right]
=
D_{E \mbk}^{(0)}
+
I_{E \mbk}^{(0)},
\\
\label{qke_avgk_E_c}
J_{0\mbk} ( f_{E\mbk}^{(0)})
+
J_{E\mbk} ( f_{0\mbk})
= 0,
\end{gather}
\end{subequations}
where Eqs.~(\ref{qke_avgk_E_a}) and (\ref{qke_avgk_E_b}) are--respectively--the diagonal and off-diagonal elements of Eq.~(\ref{qke_avgk}) at $O(n_I^0)$, while Eq.~(\ref{qke_avgk_E_c}) represents the $O(n_I)$ terms. In Eq.~(\ref{qke_avgk_E_b}), the evolution of the off-diagonal density matrix $S_{E \mbk}^{(0)}$ is governed by a term originating from the covariant derivative,
$D_{E \mbk}^{(0) ab}
=
\frac{ie}{\hbar} \mathbf{E} \cdot \mathbf{\mathcal{A}}_{\mbk}^{ab}
\left( f_{0\mbk}^a - f_{0 \mbk}^b \right)$, as well as a collision-integral-type term 
$I_{E \mbk}^{(0) ab}
=
- J_{0 \mbk}^{ab} ( n_{E \mbk}^{(-1)} )$. 

Assuming a transport time $\tau_{\mbk}^a$ that takes into account both self-energy and vertex corrections in the diagrammatic language, Eq.~(\ref{qke_avgk_E_a}) is readily solved as
\begin{equation}
\label{n_Ek_-1}
n_{E \mbk a}^{(-1)}
=
\frac{e}{\hbar} 
\tau_{\mbk}^a \mathbf{E} \cdot \bs{\pd}_{\mbk}
f_{0 \mbk}^{a}.
\end{equation}
For a general anisotropic system, extracting the carrier density from the collision integral is often challenging, which translates to difficulty in obtaining a closed-form solution for the transport time. However, in the isotropic limit, $\tau_{\mbk} \rightarrow \tau_{0\mbk}$, one obtains the familiar form with the angular weighting factor arising from vertex corrections~\cite{mahan2000many}
\begin{equation}
\label{tau_0k}
\frac{1}{\tau_{0 \mbk}^a}
=
\frac{2\pi}{\hbar} 
\sum_{\mbkpr b}
\delta ( \veps_{0 k}^a - \veps_{0 k^{\prime}}^b)
\Braket{
U_{\mbk \mbkpr}^{ab} U_{\mbkpr \mbk} ^{ba} 
}
\left[ 1 - \cos (\phi - \phi\pr) \right],
\end{equation}
with $\veps_{0 k}^a$ an energy eigenvalue of the isotropic subsystem and $\phi$ the azimuthal angle in momentum space. This solution for the transport time can then be used as a perturbative basis for obtaining the full transport time in weakly anisotropic systems.  

The solution to Eq.~(\ref{qke_avgk_E_b}) is also straightforward, leading to the following form for the band-off-diagonal density matrix
\begin{equation}
\label{S_Ek_0}
S_{E \mbk}^{(0)} 
=
\hbar \int_{-\infty}^{\infty} \frac{d\veps}{2 \pi}
G^R (\veps) \left[ D_{E \mbk}^{(0)}
+
I_{E \mbk}^{(0)} \right] G^A (\veps),
\end{equation}
which we express here as 
$S_{E \mbk}^{(0)}
=
S_{E \mbk}^{(0)} \left[D_{E \mbk}^{(0)} \right]
+
S_{E \mbk}^{(0)} \left[ I_{E \mbk}^{(0)} \right]$ for brevity and relegate the full forms to Sec.~\ref{App_off-diagonal}. The off-diagonal nature of $S_{E \mbk}^{(0)} \left[ I_{E \mbk}^{(0)} \right]$ given by Eq.~(\ref{S_Ek_0I}) implies that the contribution of this term to the equilibrium collision integral consists of antisymmetric parts of the Gaussian disorder distribution, which correspond to skew scattering in the semiclassical theory. $D_{E \mbk}^{(0)}$, however, is a function of the equilibrium distribution function. Therefore, its contribution is included in the electric-field correction to $f_{0\mbk}$, which semiclassically describes a side-jump process. This then reveals the physical nature of Eq.~(\ref{qke_avgk_E_c}); introducing the decomposition 
$n_{E\mbk}^{(0)}
=
n_{E\mbk}^{(\text{sk})} + n_{E\mbk}^{(\text{sj})}$, we arrive at the two equations
\begin{subequations}
\label{linear_carrier_eq}
\begin{align}
\label{J0k_nEk_a}
&J_{0\mbk}^a \left( n_{E\mbk}^{(\text{sk})} \right)
+
J_{0\mbk}^a \left( S_{E \mbk}^{(0)} \left[ I_{E \mbk}^{(0)} \right] \right)
=
0,
\\
\label{J0k_nEk_b}
&J_{0\mbk}^a \left( n_{E\mbk}^{(\text{sj})} \right)
+
J_{E\mbk}^a \left( f_{0\mbk} \right)
+
J_{0\mbk}^a \left( S_{E \mbk}^{(0)} \left[ D_{E \mbk}^{(0)} \right] \right)
=
0,
\end{align}
\end{subequations}
which are then readily solved for the skew-scattering and side-jump carrier densities (see Sec.~\ref{App_carrier_densities} for the explicit forms of the carrier densities). This completes the derivation of the field-linear density matrix.

\section{Quadratic density matrix}
\label{sec_quad}

We now extend this framework to nonlinear responses. In the quadratic response regime, wherein one is also interested in the second-order deviation of an electron distribution from equilibrium, there are also $O(\tau^2)$ contributions to the density operator. Thus, the appropriate expansion to consider is 
$n_{E^2\mbk} = n_{E^2\mbk}^{(-2)} + n_{E^2\mbk}^{(-1)} + n_{E^2\mbk}^{(0)}$, where we retain sub-subleading disorder corrections as well. This is an essential formal inclusion, partly due to the general expectation that intrinsic effects should appear at any order in the field expansion in the absence of disorder. Similar to the linear response regime--where band-off-diagonal elements do not appear at leading order in the disorder--for steady-state solutions, one can readily verify that $S_{E^2\mbk}^{(-2)}$ will not contribute to the density operator. Thus, the field-quadratic decomposition of the quantum kinetic equation results in the five general equations
\begin{subequations}
\label{qke_avgk_E2}
\begin{gather}
\label{qke_avgk_E2_a}
J_{0\mbk}^a \left(n_{E^2\mbk}^{(-2)}\right)
=
\frac{e}{\hbar} \mathbf{E} \cdot \bs{\pd}_{\mbk} n_{E\mbk a}^{(-1)},
\\
\label{qke_avgk_E2_b}
\pd_t S_{E^2 \mbk}^{(-1)}
+
\frac{i}{\hbar} \left[ H^0, S_{E^2 \mbk}^{(-1)} \right]
=
D_{E^2 \mbk}^{(-1)}
+
I_{E^2 \mbk}^{(-1)},
\\
\label{qke_avgk_E2_c}
J_{0 \mbk}^a \left( f_{E^2\mbk}^{(-1)}\right)
+
J_{E\mbk}^a \left( f_{E\mbk}^{(-1)}\right)
=
\frac{e}{\hbar} \mathbf{E} \cdot
\left( \bs{\mathcal{D}}_{\mbk} f_{E\mbk}^{(0)} \right)^a,
\\
\label{qke_avgk_E2_d}
\begin{split}
\pd_t S_{E^2 \mbk}^{(0)}
+
\frac{i}{\hbar} \left[ H^0, S_{E^2 \mbk}^{(0)} \right]
&=
D_{E^2 \mbk}^{(0)}
+
D_{E^2 \mbk}^{\prime (0)}
+
I_{E^2 \mbk}^{(0)}
+
I_{E^2 \mbk}^{\prime (0)}
+
I_{E^2 \mbk}^{\prime \prime (0)},
\end{split}
\\
\label{qke_avgk_E2_e}
J_{0\mbk} \left( f_{E^2\mbk}^{(0)} \right)
+
J_{E\mbk} \left( f_{E\mbk}^{(0)} \right)
+
J_{E^2\mbk} \left( f_{0\mbk}\right) = 0.
\end{gather}
\end{subequations}
The solution to Eq.~(\ref{qke_avgk_E2_a}) for the nonlinear density is simply
\begin{equation}
\label{n_E2_-2}
n_{E^2 \mbk a}^{(-2)}
=
\left( \frac{e}{\hbar}\right)^2 E_{\mu} E_{\nu} \tau_{\mbk}^a \pd_{\mbk}^{\mu}
( \tau_{\mbk}^a \pd_{\mbk}^{\nu} f_{0 \mbk}^a ),
\end{equation}
which quantifies the ordinary scattering of electrons at second order in the field expansion. Below, we discuss in detail the solutions of the remaining four equations and their associated physical processes. 

\subsection{Off-diagonal elements}

Consider first the off-diagonal elements. The evolution equation for $S_{E^2 \mbk}^{(-1)}$, Eq.~(\ref{qke_avgk_E2_b}), is quite similar to that of $S_{E \mbk}^{(0)}$, given by Eq.~(\ref{qke_avgk_E_b}), and may be regarded as its quadratic-response counterpart, with the analogous higher-order driving terms
$D_{E^2 \mbk}^{(-1) ab}
=
\frac{ie}{\hbar} \mathbf{E} \cdot \mathbf{\mathcal{A}}_{\mbk}^{ab}
( n_{E \mbk a}^{(-1)} - n_{E \mbk b}^{(-1)})$ and
$I_{E^2 \mbk}^{(-1) ab}
=
- J_{0 \mbk}^{ab} ( n_{E^2 \mbk}^{(-2)} )$ and the solution
\begin{equation}
\label{S_E2k_-1}
S_{E^2 \mbk}^{(-1)} 
=
\hbar \int_{-\infty}^{\infty} \frac{d\veps}{2 \pi}
G^R (\veps) \left[ D_{E^2 \mbk}^{(-1)}
+
I_{E^2 \mbk}^{(-1)} \right] G^A (\veps),
\end{equation}

A quick comparison between Eq.~(\ref{S_Ek_0_decomp}) and Eq.~(\ref{S_E2k_-1_decomp}) reveals that $S_{E^2 \mbk}^{(-1)}$ simply corresponds to $S_{E \mbk}^{(0)}$ with the densities replaced by their higher-order counterparts in the electric field. The situation is different for the other off-diagonal component, $S_{E^2 \mbk}^{(0)}$, which includes terms that have no linear-response counterparts. As indicated in Eq.~(\ref{qke_avgk_E2_d}), its evolution is determined by five distinct driving terms, which consist of two covariant-derivative-type terms and three collision integral-type terms, given by 
\begin{subequations}
\label{S_E2k_0_terms}
\begin{align}
\label{S_E2k_0_terms_a}
D_{E^2 \mbk}^{(0) ab}
&=
\frac{ie}{\hbar} \mathbf{E} \cdot \mathbf{\mathcal{A}}_{\mbk}^{ab}
\left[ n_{E \mbk a}^{(0)} - n_{E \mbk b}^{(0)} \right],
\\
\label{S_E2k_0_terms_b}
D_{E^2 \mbk}^{\prime (0) ab}
&=
\frac{e}{\hbar} \mathbf{E} \cdot
\left[
\bs{\mathcal{D}}_{\mbk} S_{E \mbk}^{(0)}
\right]^{ab},
\\
\label{S_E2k_0_terms_c}
I_{E^2 \mbk}^{(0) ab}
&=
- J_{0 \mbk}^{ab} \left( n_{E^2 \mbk}^{(-1)} \right),
\\
\label{S_E2k_0_terms_d}
I_{E^2 \mbk}^{\prime (0) ab}
&=
- J_{E \mbk}^{ab} \left( n_{E \mbk}^{(-1)} \right),
\\
\label{S_E2k_0_terms_e}
I_{E^2 \mbk}^{\prime \prime (0) ab}
&=
- J_{0 \mbk}^{ab} \left( S_{E^2 \mbk}^{(-1)} \right),
\end{align}
\end{subequations}
with the familiar formal solution
\begin{equation}
\label{S_E2k_0}
\begin{split}
S_{E^2 \mbk}^{(0)} 
&=
\hbar \int_{-\infty}^{\infty} \frac{d\veps}{2 \pi}
G^R (\veps)
\left[ D_{E^2 \mbk}^{(0)}
+
D_{E^2 \mbk}^{\prime (0)}
+
I_{E^2 \mbk}^{(0)}
+
I_{E^2 \mbk}^{\prime (0)}
+
I_{E^2 \mbk}^{\prime \prime (0)} \right]
G^A (\veps).
\end{split}
\end{equation}
It is worth noting that the two driving terms $D_{E^2 \mbk}^{\prime (0)}$ and $I_{E^2 \mbk}^{\dprime (0)}$ are themselves functions of off-diagonal density matrix elements and can therefore be further decomposed as
\begin{subequations}
\label{S_E2k_Dpr_Idpr}
\begin{align}
S_{E^2 \mbk}^{(0)} \left[D_{E^2 \mbk}^{\prime (0)} \right]
&=
S_{E^2 \mbk}^{(0)} \left[D_{E^2 \mbk}^{\prime (0)} [ D_{E \mbk}^{(0)}]
\right]
+
S_{E^2 \mbk}^{(0)} \left[D_{E^2 \mbk}^{\prime (0)} [ I_{E \mbk}^{(0)}]
\right],
\\
S_{E^2 \mbk}^{(0)} \left[I_{E^2 \mbk}^{\dprime (0)} \right]
&=
S_{E^2 \mbk}^{(0)} \left[I_{E^2 \mbk}^{\dprime (0)} [ D_{E^2 \mbk}^{(-1)}]
\right]
+
S_{E^2 \mbk}^{(0)} \left[I_{E^2 \mbk}^{\dprime (0)} [ I_{E^2 \mbk}^{(-1)}]
\right],
\end{align}
\end{subequations}

We thus observe that, in comparison to the relatively simple picture in the linear response regime, there are a large number of terms which contribute to the off-diagonal density matrix elements in the quadratic response regime, reflecting a wealth of interconnected scattering processes that are illustrated schematically in Fig.~\ref{fig_qkt_fig1}. We next discuss the derivation of the carrier densities and the ensuing physical classification of these processes.
\vspace{-.2cm}
\subsection{Mixed scattering}

Consider Eq.~(\ref{qke_avgk_E2_c}), which must be solved for $n_{E^2 \mbk}^{(-1)}$. Based on the previously stated correspondence between $S_{E^2 \mbk}^{(-1)}$ and $S_{E \mbk}^{(0)}$, it is straightforward to conclude that this equation is essentially the field-quadratic counterpart to Eq.~(\ref{qke_avgk_E_c}) and thus describes the special (skew, side-jump) scattering of ordinary-scattered electrons. A difference, however, with the linear response is the additional covariant-derivative term on the right-hand side of Eq.~(\ref{qke_avgk_E2_c}), which includes terms that physically describe the ordinary scattering of electrons having previously undergone side jump or skew scattering at linear order in the electric field [see Eqs.~(\ref{n_Ek^sk}) and (\ref{n_Ek^sj})]. The conclusion, therefore, is that $n_{E^2 \mbk}^{(-1)}$ measures the density of electrons which undergo a mixture of consecutive special and ordinary scatterings and thus consists of the four distinct contributions
\begin{equation}
\label{n_E^2ka^-1}
n_{E^2 \mbk a}^{(-1)}
=
n_{E^2 \mbk a}^{(\text{sj,o})}
+
n_{E^2 \mbk a}^{(\text{sk,o})}
+
n_{E^2 \mbk a}^{(\text{o,sj})}
+
n_{E^2 \mbk a}^{(\text{o,sk})},
\end{equation}
where
$n_{E^2 \mbk}^{(\text{sp,o})}$ is the density of ordinary-scattered electrons which subsequently experience special scattering (sp=sj,sk), and $n_{E^2 \mbk}^{(\text{o,sp})}$ is the density of special-scattered electrons which then undergo ordinary scattering. As a result, Eq.~(\ref{qke_avgk_E2_c}) is also decomposed as
%\vspace{-.1cm}
\begin{subequations}
\begin{align}
\begin{split}
&J_{0\mbk}^a \left( n_{E^2\mbk}^{(\text{sj,o})} \right)
+
J_{0\mbk}^a \left( S_{E^2 \mbk}^{(-1)} \left[ D_{E^2 \mbk}^{(-1)} \right] \right)
+
J_{E \mbk}^a \left( n_{E \mbk}^{(-1)}\right)
=
- \frac{i e}{\hbar} \mathbf{E} \cdot
\left[ \bs{\mathcal{A}}_{\mbk} ,
S_{E \mbk}^{(0)} \right]^a,
\end{split}
\\
&J_{0\mbk}^a \left( n_{E^2\mbk}^{(\text{sk,o})} \right)
+
J_{0\mbk}^a \left( S_{E^2 \mbk}^{(-1)} ]\left[ I_{E^2 \mbk}^{(-1)} \right] \right)
=
0,
\\
&J_{0\mbk}^a \left( n_{E^2 \mbk}^{(\text{o,sj})} \right)
=
\frac{e}{\hbar} \mathbf{E} \cdot 
\bs{\pd}_{\mbk} n_{E \mbk a}^{(\text{sj})},
\\
&J_{0\mbk}^a \left( n_{E^2 \mbk}^{(\text{o,sk})} \right)
=
\frac{e}{\hbar} \mathbf{E} \cdot 
\bs{\pd}_{\mbk} n_{E \mbk a}^{(\text{sk})},
\end{align}
\end{subequations}
the solutions of which--presented in Sec.~\ref{App_carrier_densities}--yield the mixed-scattering carrier densities.

\subsection{Zeroth-order scattering}

We next turn to Eq.~(\ref{qke_avgk_E2_e}), the last of the five main equations arising from the quantum kinetic equation in the quadratic response regime, which must be solved for $n_{E^2 \mbk}^{(0)}$. Similar to the special scattering processes in the linear regime, this includes the density matrix elements which are formally of $O(n_I^0)$ in the disorder expansion, but can nevertheless arise from the concerted actions of disorder and the geometry of Bloch states. A relatively large number of collision integrals contribute to the density matrix at this order, leading to a multitude of scattering processes. Upon inspection, these are classified as five physically distinct processes, which we label as
\begin{equation}
n_{E^2 \mbk}^{(0)}
=
n_{E^2 \mbk}^{\text{(ssj)}}
+
n_{E^2 \mbk}^{\text{(qsj)}}
+
n_{E^2 \mbk}^{\text{(tsk)}}
+
n_{E^2 \mbk}^{\text{(csk)}}
+
n_{E^2 \mbk}^{\text{(ask)}},
\end{equation}
resulting in the decomposition
\begin{subequations}
\label{zeroth_collision}
\begin{align}
\label{J0k_nE^2k_c}
&J_{0\mbk}^a \left( n_{E^2 \mbk}^{(\text{ssj})} \right)
+
J_{0\mbk}^a \left( S_{E^2 \mbk}^{(0)}\left[ D_{E^2 \mbk}^{(0)} \right] \right)
+
J_{E\mbk}^a \left( n_{E \mbk}^{(0)} \right)
=
0,
\\
\label{J0k_nE^2k_d}
&J_{0\mbk}^a \left( n_{E^2 \mbk}^{(\text{csk})} \right)
+
J_{0\mbk}^a 
\left( S_{E^2 \mbk}^{(0)} \left[I_{E^2 \mbk}^{\dprime (0)} [ I_{E^2 \mbk}^{(-1)}]
\right] \right)
=
0,
\\
\label{J0k_nE^2k_e}
&J_{0\mbk}^a \left( n_{E^2 \mbk}^{(\text{tsk})} \right)
+
J_{0\mbk}^a \left( S_{E^2 \mbk}^{(0)}\left[ I_{E^2 \mbk}^{(0)} \right] \right)
=0,
\\
\label{J0k_nE^2k_a}
&J_{0\mbk}^a \left( n_{E^2\mbk}^{(\text{ask})} \right)
+
\mathcal{I}_1
= 0,
\\
\label{J0k_nE^2k_b}
&J_{0\mbk}^a \left( n_{E^2\mbk}^{(\text{qsj})} \right)
+
\mathcal{I}_2
= 0,
\end{align}
\end{subequations}
with the lengthier expressions $\mathcal{I}_1$ and $\mathcal{I}_2$ presented in Sec.~\ref{app_I_1}. Below, we elucidate this classification and elaborate on the physical significance of each term.
\vspace{-.2cm}
\subsubsection{Secondary side jump}

Consider first Eq.~(\ref{J0k_nE^2k_c}). Note that this equation is structurally analogous to Eq.~(\ref{J0k_nEk_a}) and thus describes a side-jump process, with the difference that it acts on $n_{E\mbk}^{(0)}$. It is then clear that $n_{E^2 \mbk}^{(\text{ssj})}$ measures the density the electrons that are first subjected to a side-jump or skew scattering process and then undergo an additional side jump. This secondary side-jump density is thus itself comprised of two terms
\begin{equation}
n_{E^2 \mbk a}^{\text{(ssj)}}
=
n_{E^2 \mbk a}^{\text{(sj,sj)}}
+
n_{E^2 \mbk a}^{\text{(sj,sk)}},
\end{equation}
each of which is obtained by solving the relevant contribution to Eq.~(\ref{J0k_nE^2k_c}).
\vspace{-.2cm}
\subsubsection{Tertiary skew scattering}

Given the presence of secondary side-jump scatterings, it is natural to ask whether a special scattering event could be proceeded by a skew-scattering process as well. To answer this in the affirmative, consider next Eq.~(\ref{J0k_nE^2k_e}). Based on the form of the collision integral
$J_{0\mbk}^a \left( S_{E^2 \mbk}^{(0)}\left[ I_{E^2 \mbk}^{(0)} \right] \right)$ given in Sec.~\ref{App_collision_int}, it is evident that this measures the skew scattering of electrons which have initially experienced mixed scattering. Thus, as illustrated in Fig.~\ref{fig_qkt_fig1}, overall, this describes a three-stage scattering process at least, which we refer to as tertiary skew scattering. From the fourfold decomposition of $n_{E^2\mbk}^{(-1)}$ given by Eq.~(\ref{n_E^2ka^-1}), it is clear that $n_{E^2 \mbk a}^{\text{(tsk)}}$ also decomposes as
\begin{equation}
n_{E^2 \mbk a}^{\text{(tsk)}}
=
n_{E^2 \mbk a}^{\text{(sk,sj,o)}}
+
n_{E^2 \mbk a}^{\text{(sk,o,sj)}}
+
n_{E^2 \mbk a}^{\text{(sk,sk,o)}}
+
n_{E^2 \mbk a}^{\text{(sk,o,sk)}},
\end{equation}
yielding four distinct contributions.
\vspace{-.5cm}

\subsubsection{Quadratic side jump}

In the quadratic response regime, there is yet another source of side-jump scattering. In addition to the secondary side jump, which arises from collision integrals containing corrections to $n_{E\mbk}^{(0)}$ that are linear in the electric field, one must also take into consideration field-quadratic corrections to the equilibrium distribution function itself. These quadratic side-jump contributions are obtained by solving Eq.~(\ref{J0k_nE^2k_b}) for $n_{E^2 \mbk}^{\text{(qsj)}}$, which reads
\begin{equation}
\label{n_E2k_qsj}
\begin{split}
n_{E^2\mbk a}^{(\text{qsj})}
&=
- \tau_{\mbk}^a
\left\{
J_{0\mbk}^a 
\left( S_{E^2 \mbk}^{(0)} \left[D_{E^2 \mbk}^{\prime (0)} [ D_{E \mbk}^{(0)}]
\right] \right)
+
J_{E\mbk}^a \left( S_{E \mbk}^{(0)} \left[ D_{E \mbk}^{(0)} \right] \right)
+
J_{E^2 \mbk}^a 
\left( f_{0 \mbk} \right)
\right\}.
\end{split}
\end{equation}
The explicit form of $n_{E^2 \mbk}^{\text{(qsj)}}$ is rather lengthy and can be found in Sec.~\ref{App_carrier_densities}. It is interesting to note, however, the appearance of the quantum metric in Eq.~(\ref{N_qsj_8}), which then compels one to interpret the quadratic side jump as being partly driven by the quantum metric tensor. This is in line with the findings of Ref.~\cite{gao2014field}, which relates the quantum metric to the field-induced positional shift of the carriers.
\vspace{-.55cm}
\subsubsection{Cubic skew scattering}

In addition to the tertiary skew scattering, the carrier density of which contains $\braket{UU}^2$-type terms, there is another source of skew scattering, which consists of terms that are cubic in $\braket{UU}$ and acts on the density of ordinary-scattered electrons, $n_{E^2 \mbk}^{(-2)}$, guaranteeing that the resultant density is of O($n_I^0$). The carrier density from this cubic skew scattering is obtained by solving Eq.~(\ref{J0k_nE^2k_d}), yielding
\begin{equation}
n_{E^2 \mbk a}^{(\text{csk})}
=
- \tau_{\mbk}^a
J_{0\mbk}^a 
\left( S_{E^2 \mbk}^{(0)} \left[I_{E^2 \mbk}^{\dprime (0)} [ I_{E^2 \mbk}^{(-1)}]
\right] \right),
\end{equation}
where, once again, we relegate the lengthy explicit form to Sec.~\ref{App_carrier_densities}. 

\subsubsection{Anomalous skew scattering}

Lastly, we consider Eq.~(\ref{J0k_nE^2k_a}), the solution of which is
\begin{equation}
\begin{split}
n_{E^2\mbk a}^{(\text{ask})}
=&
-\tau_{\mbk}^a
\left\{
J_{0\mbk}^a 
\left( S_{E^2 \mbk}^{(0)} \left[D_{E^2 \mbk}^{\prime (0)} [ I_{E \mbk}^{(0)}]
\right] \right)
+
J_{0\mbk}^a 
\left( S_{E^2 \mbk}^{(0)} \left[I_{E^2 \mbk}^{\dprime (0)} [ D_{E^2 \mbk}^{(-1)}]
\right] \right)
\right.
\\
&\left.+
J_{E\mbk}^a 
\left( S_{E^2 \mbk}^{(0)} \left[I_{E^2 \mbk}^{\prime (0)} \right] \right)
+
J_{E\mbk}^a 
\left( S_{E \mbk}^{(0)} \left[I_{E \mbk}^{(0)} \right] \right)
\right\}.
\end{split}
\end{equation}
This describes an anomalous skew scattering process, whereby electrons which scatter ordinarily at linear order in the electric field, $n_{E \mbk}^{(-1)}$, subsequently undergo a field-dependent scattering with a skew scattering disorder profile. As is clear from the relevant expressions in Sec.~\ref{App_carrier_densities}, the resultant carrier density is thus comprised of terms proportional to the product of the covariant derivative and the antisymmetric parts of $\braket{UU} \braket{UU}$-type terms.
\vspace{-.5cm}

\subsection{General remarks on the formalism}

Having presented the derivation of the density matrix, we note that a feature of the density-matrix formalism that becomes evident in the kinetic approach is the ability to provide a detailed mapping of the scattering channels that arise through the various processes and the interconnections between the various carrier densities. One could argue that this feature is slightly obscured in other methods, such as diagrammatic considerations, which typically do not deal directly with carrier densities. An alternative distinction scheme, which complements the viewpoint presented here, is to classify all non-ordinary scatterings as either side jump or skew scattering, where the former process would involve contributions from the conduction band only, while the latter receives contributions from both the valence and conduction bands. Despite this, as we illustrate for a specific model in Section~\ref{application}, one could argue for the merits of a more detailed distinction between the various scatterings, as processes which would otherwise fall within the same classification subtype turn out to have noticeably different scaling signatures in the response of the system.

In closing this section, it is also useful to briefly compare the formalism presented in this chapter to another recent quantum kinetic approach discussed in Ref.~\cite{bhalla2023quantum}. There, the collision integral is approximated as being proportional to the density matrix via a constant relaxation time such that scatterings are accounted for by introducing intraband and interband relaxation times for the diagonal and off-diagonal components of the quantum kinetic equation, respectively. In the present approach, we assume the diagonal elements of the collision integral are proportional to a general momentum-dependent transport time, the solution of which is obtained from extracting the carrier density from the collision integral. This results in a transport time that is physically equivalent to including self-energy and vertex corrections in its derivation~\cite{mahan2000many}. Furthermore, the solution of the off-diagonal quantum kinetic equation at different field and impurity orders emerges once the relevant carrier densities are known. Therefore, the transport time manifests in the off-diagonal density matrix as well through the carrier densities. Nevertheless, as is evident in both approaches, at higher orders beyond linear responses, intraband and interband processes are intricately connected, leading to a variety of physical phenomena in the nonlinear response regime.
\vspace{.5cm}

\section{Emergent Velocities and Modified Semiclassics}
\label{sec_emergent}

Having delved into the solution of the quantum Liouville equation, we may now proceed to evaluate the modified semiclassical equations of motion and, subsequently, the disorder-dressed quadratic current density. The former is found by adopting the proposed prescription outlined in Ref.~\cite{atencia2022semiclassical}, the essential idea of which is to reexpress the ensemble average of the velocity operator entirely in terms of the diagonal elements of the density matrix, \textit{i.e.}, the carrier densities. That is, we solve the equation
\begin{equation}
\label{rdot_v}
\sum_{\mbk a} \dot{\mathbf{r}}^a n_{\mbk a}
=
\sum_{\mbk a b} \mathbf{v}_{\mbk}^{b a} f_{\mbk}^{ab},
\end{equation}
for the time evolution of the carrier position 
$\dot{\mathbf{r}}^a$, where 
$\mathbf{v}_{\mbk} \equiv
(1/ \hbar) \bs{\mathcal{D}}_{\mbk} H_{\mbk}^0$ is the velocity operator. Using the results from the previous section, the right-hand side of Eq.~(\ref{rdot_v}) reads
\begin{equation}
\label{rdot_v_expanded}
\sum_{\mbk a b} \mathbf{v}_{\mbk}^{b a} f_{\mbk}^{ab}
=
\sum_{\mbk a}
\bs{v}_{\mbk}^a n_{\mbk a}
-
\frac{i}{\hbar} \sum_{\mbk a b}  \left(\veps_{\mbk}^a - \veps_{\mbk}^b \right)
\bs{\mathcal{A}}_{\mbk}^{b a} 
S_{\mbk}^{ab}.
\end{equation}
The first term on the right, which is already diagonal, describes the ordinary group velocity experienced by all the various carriers. Our goal then is to band-diagonalize the second term. For the linear-response term, inserting Eq.~(\ref{S_Ek_0}) leads to contributions from the anomalous and extrinsic velocities
\begin{equation}
\begin{split}
\label{S_Ek_contrib}
-\frac{i}{\hbar} \sum_{\mbk a b}
\left(\veps_{\mbk}^a - \veps_{\mbk}^b \right) \mathcal{A}_{\mbk}^{b a}
S_{E \mbk}^{(0)ab}
&=
\sum_{\mbk a}
\left[
\frac{e}{\hbar} \mathbf{E} \times \bs{\Omega}_{\mbk}^a f_{0 \mbk}^a
+
\bs{\alpha}_{\mbk a} n_{E \mbk a}^{(-1)}
\right],
\end{split}
\end{equation}
where the extrinsic velocity is defined as~\cite{atencia2022semiclassical}
\begin{equation}
\bs{\alpha}_{\mbk}^a
=
\frac{1}{\hbar} \int_{-\infty}^{\infty} \frac{d\veps}{2\pi}
\Braket{
\left[ U, G^A(\veps) \left[ U, \bs{\mathcal{A}}\pr \right] G^R(\veps)\right]
}_{\mbk}^a,
\end{equation}
with the prime indicating that only off-diagonal elements of the Berry connection are evaluated--reflecting the gauge invariance of the extrinsic velocity. Explicitly, this reads
\begin{equation}
\label{alpha_ka}
\bs{\alpha}_{\mbk}^a
=
\frac{2}{\hbar} \sum_{\mbkpr b} \text{Im}
\left(
\frac{
\Braket{U_{\mbk \mbkpr}^{ab}
\left[ U, \bs{\mathcal{A}}^{\prime} \right]_{\mbkpr \mbk}^{ba}}}
{\veps_{\mbk}^a - \veps_{\mbkpr}^b - i \delta}
\right).
\end{equation}

We now extend this analysis to the quadratic response contributions. The first term to consider is
$S_{E^2 \mbk}^{(-1)}$. Based on the previously stated analogy with its linear response counterpart $S_{E \mbk}^{(0)}$, it is straightforward to verify that
\begin{equation}
\begin{split}
-\frac{i}{\hbar} \sum_{\mbk a b}
\left(\veps_{\mbk}^a - \veps_{\mbk}^b \right) \mathcal{A}_{\mbk}^{ba}
S_{E^2 \mbk}^{(-1)ab}
&=
\sum_{\mbk a}
\left[
\frac{e}{\hbar} \mathbf{E} \times \bs{\Omega}_{\mbk}^a n_{E \mbk a}^{(-1)}
+
\bs{\alpha}_{\mbk}^a n_{E^2 \mbk a}^{(-2)}
\right].
\end{split}
\end{equation}
Thus the effect of $S_{E^2 \mbk}^{(-1)}$ is to simply describe the contributions of the anomalous and extrinsic velocities in the quadratic response regime, where we note that the anomalous contribution is that of the familar Berry curvature dipole~\cite{sodemann2015quantum}, albeit with the full disorder dependence of the transport time. Similarly, for the $S_{E^2 \mbk}^{(-1)}$ contributions from 
$D_{E^2 \mbk}^{(0)}$ and
$I_{E^2 \mbk}^{(0)}$, inserting the relevant terms of Eq.~(\ref{S_E2k_0}) into Eq.~(\ref{rdot_v_expanded}) yields   
\begin{equation}
\begin{split}
- \sum_{\mbk a b}
\mathcal{A}_{\mbk}^{b a}
\left[ D_{E^2 \mbk}^{(0)} + I_{E^2 \mbk}^{(0)} \right]^{ab}
&=
\sum_{\mbk a}
\left[
\frac{e}{\hbar} \mathbf{E} \times \bs{\Omega}_{\mbk}^a n_{E \mbk a}^{(0)}
+
\bs{\alpha}_{\mbk}^a n_{E^2 \mbk a}^{(-1)}
\right],
\end{split}
\end{equation}
showing that the higher-order carriers also experience these two velocities. 

We next turn to the contributions from Eq.~(\ref{S_E2k_0}) which have no counterparts in the linear response regime. The first term to consider is 
$D_{E^2 \mbk}^{\prime (0)}$, given by Eq.~(\ref{S_E2k_0_terms_b}). After a little algebra, it can be shown that the contribution of this term to Eq.~(\ref{rdot_v_expanded}) is also comprised of an intrinsic part and an extrinsic part as
\begin{equation}
\begin{split}
\label{eom_metric_beta}
- \sum_{\mbk a b}
\mathcal{A}_{\mu \mbk}^{b a}
D_{E^2 \mbk}^{\prime (0) ab}
&=
\sum_{\mbk a}
\left[
\frac{e^2}{\hbar^2} E^{\nu} E^{\rho}
\sum_{b}
M_{\mbk}^{ab}
\Gamma_{\nu \rho \mu \mbk}^{b a} 
f_{0\mbk}^a
+
\beta_{\mu \mbk}^{a} n_{E \mbk a}^{(-1)},
\right],
\end{split}
\end{equation}
where 
$M_{\mbk}^{ab}
\equiv
2 \hbar / (\varepsilon_{\mbk}^a - \varepsilon_{\mbk}^b)
$ with $a \neq b$. We thus see that from the off-diagonal density matrix, the geodesic term naturally emerges through this approach in the equations of motion and carrier transport, which is quite reasonable, given the Riemannian structure on the underlying manifold of quantum states~\cite{provost1980riemannian}.

The remaining term in Eq.~(\ref{eom_metric_beta}) arises from a new extrinsic velocity, $\bs{\beta}_{\mbk}$, given by
\begin{equation}
\begin{split}
\bs{\beta}_{\mbk}^a
&=
\frac{i}{\hbar}
\int_{-\infty}^{\infty} \frac{d\veps}{2 \pi}
\int_{-\infty}^{\infty} \frac{d\veps\pr}{2 \pi}
\Braket{
\left[U, G^A(\veps\pr)
\left[U, G^A(\veps)
\left[V, \bs{\mathcal{A}}\pr \right] G^R(\veps)
\right]
G^R(\veps\pr)
\right]
}_{\mbk}^a,
\end{split}
\end{equation}
or explicitly as
\begin{equation}
\bs{\beta}_{\mbk}^a
=
\frac{2e}{\hbar} \sum_{\mbkpr b} \text{Re}
\left(
\frac{
\Braket{U_{\mbk \mbkpr}^{ab}
\left[ U, \bs{\mathcal{F}} \right]_{\mbkpr \mbk}^{ba}}}
{\veps_{\mbk}^a - \veps_{\mbkpr}^b - i \delta}
\right),
\end{equation}
with 
$\bs{\mathcal{F}}_{\mbk}^{ab}
=
- \left[ \left( 
\mathbf{E} \cdot \bs{\mathcal{D}} \right) \bs{\mathcal{A}}^{\prime} \right]_{\mbk}^{ab} 
/
( \veps_{\mbk}^a - \veps_{\mbk}^b + i \delta )$, 
which is linear in the applied electric field and thus appears only in the nonlinear transport regime, in contrast to the field-independent $\bs{\alpha}_{\mbk}$, which manifests in both linear and nonlinear responses. This then suggests a general pattern, in which extrinsic velocities that are at most of $O(E^{n-1})$ contribute to the $n$-th order response, in agreement with a recent study based on a semiclassical Boltzmann transport analysis~\cite{huang2023scaling}. This is indeed the case for the remaining quadratic-response velocities, as we show below.

Moving on to Eq.~(\ref{S_E2k_0_terms_d}), the contribution from 
$I_{E^2 \mbk}^{\prime(0)}$ is also that of an extrinsic velocity, which reads
\begin{equation}
- \sum_{\mbk a b}
\mathcal{A}_{\mu \mbk}^{b a}
I_{E^2 \mbk}^{\prime(0) ab} 
=
\sum_{\mbk a}
\gamma_{\mu \mbk}^{a} n_{E \mbk a}^{(-1)},
\end{equation}
with
\begin{equation}
\begin{split}
\bs{\gamma}_{\mbk}^a
&=
\frac{i}{\hbar}
\int_{-\infty}^{\infty} \frac{d\veps}{2 \pi}
\int_{-\infty}^{\infty} \frac{d\veps\pr}{2 \pi}
\Braket{
\left[U, G^A(\veps\pr)
\left[V, G^A(\veps)
\left[ U, \bs{\mathcal{A}}\pr \right] G^R(\veps)
\right]
G^R(\veps\pr)
\right]
}_{\mbk}^a,
\end{split}
\end{equation}
or
\begin{equation}
\bs{\gamma}_{\mbk}^a
=
\frac{2e}{\hbar} \sum_{\mbkpr b} \text{Re}
\left(
\frac{
\Braket{U_{\mbk \mbkpr}^{ab}
\left[ \left( 
\mathbf{E} \cdot \bs{\mathcal{D}} \right) \bs{\mathcal{G}} \right]_{\mbkpr \mbk}^{ba}}
}
{\veps_{\mbk}^a - \veps_{\mbkpr}^b - i \delta}
\right),
\end{equation}
with 
$\bs{\mathcal{G}}_{\mbk \mbkpr}^{ab}
=
- \left[ U, \bs{\mathcal{A}}^{\prime} \right]_{\mbk \mbkpr}^{ab} 
/
( \veps_{\mbk}^a - \veps_{\mbkpr}^b + i \delta )$. As can be seen, this differs from $\bs{\beta}_{\mbk}$ by an exchange in the orders of the electrostatic and disorder potential in the formal definition.

Lastly, the contribution arising from $I_{E^2 \mbk}^{\prime \prime (0)}$ can be expressed as two separate extrinsic velocities acting on the linear and quadratic densities
\begin{equation}
- \sum_{\mbk a a\pr}
\mathcal{A}_{\mu \mbk}^{a\pr a}
I_{E^2 \mbk}^{\prime \prime (0) aa\pr}
=
\sum_{\mbk a}
\left[ \kappa_{\mu \mbk}^{a} n_{E \mbk a}^{(-1)} + \chi_{\mu \mbk}^{a} n_{E^2 \mbk a}^{(-2)} \right],
\end{equation}
with
\begin{equation}
\bs{\kappa}_{\mbk}^a
=
i e \int_{-\infty}^{\infty} \frac{d\veps}{2\pi} 
\left[
G^R(\veps)  \mathbf{E} \cdot \bs{\mathcal{A}}\pr G^A(\veps) , \bs{\alpha} \right]_{\mbk}^a,
\end{equation}
\begin{equation}
\begin{split}
\bs{\chi}_{\mbk}^a
&=
\int_{-\infty}^{\infty} \frac{d\veps}{2 \pi}
\int_{-\infty}^{\infty} \frac{d\veps\pr}{2 \pi}
\Braket{
\left[U G^A(\veps\pr) G^A(\veps)
,
\bs{\alpha} G^R(\veps) U G^R(\veps\pr)
\right] + \text{h.c.} }_{\mbk}^a,
\end{split}
\end{equation}
which explicitly read
\begin{equation}
\bs{\kappa}_{\mbk}^a
=
2e \sum_b \text{Re}
\left(
\frac{\mathbf{E} \cdot \bs{\mathcal{A}}_{\mbk}^{\prime ab} \bs{\alpha}_{\mbk}^{\prime ba} 
}
{\veps_{\mbk}^a - \veps_{\mbk}^b - i \delta }
\right),
\end{equation}
and
\begin{equation}
\begin{split}
\bs{\chi}_{\mbk}^a
&=
2 \sum_{bc} \text{Re}
\left[
\frac{
\braket{U_{\mbk \mbkpr}^{ab} U_{\mbkpr \mbk}^{ca}} \bs{\alpha}_{\mbkpr}^{bc}}
{(\veps_{\mbk}^a - \veps_{\mbkpr}^b - i \delta)
(\veps_{\mbkpr}^b - \veps_{\mbkpr}^c + i \delta)}
-
\frac{
\braket{U_{\mbk \mbkpr}^{bc} U_{\mbkpr \mbk}^{ca}} \bs{\alpha}_{\mbk}^{ab}}
{(\veps_{\mbkpr}^c - \veps_{\mbk}^a - i \delta)
(\veps_{\mbk}^a - \veps_{\mbk}^b + i \delta)}
\right].
\end{split}
\end{equation}
It is perhaps interesting to note from these terms that the off-diagonal elements of the linear-response velocity $\bs{\alpha}_{\mbk}$ also contribute to the transport, albeit as higher-order extrinsic terms.

Gathering all the above terms, and using the equation of motion for the carrier momentum, 
$\dot{\mathbf{k}}^a = -e \mathbf{E}/\hbar$,
we ultimately arrive at the equation of motion for the carrier position given by Eq.~(\ref{eom}). And at second order in the electric field, the total current density--defined as 
$\mathbf{j}
\equiv
-e \text{Tr} (\mathbf{v} \braket{\rho})$--is expressed as
\begin{equation}
\label{j_E2_tot}
\begin{split}
j_{\mu E^2}
=&
-e \sum_{\mbk a} \left[ 
v_{\mu \mbk}^a  n_{E^2 \mbk a}
+
\frac{e}{\hbar} \varepsilon_{\mu \nu \rho} E_{\nu} \Omega_{\rho \mbk}^a n_{E \mbk a}
+
\left( \frac{e}{\hbar} \right)^2 E_{\nu} E_{\rho}
\sum_{b} M_{\mbk}^{ab}
\Gamma_{\nu \rho \mu \mbk}^{b a} f_{0 \mbk}^a
\right.
\\
&\left.
+
\alpha_{\mu \mbk}^{a} \left( 
n_{E^2 \mbk a}^{(-2)} + n_{E^2 \mbk a}^{(-1)}\right)
+
\left( \beta_{\mu \mbk}^{a}
+
\gamma_{\mu \mbk}^{a}
+
\kappa_{\mu \mbk}^{a} \right) n_{E \mbk a}^{(-1)}
+
\chi_{\mu \mbk}^{a} n_{E^2 \mbk a}^{(-2)},
\right].
\end{split}
\end{equation}

Several remarks regarding these results are in order. First, as one would expect, the leading-order disorder contribution to the quadratic current is
\begin{equation}
\label{j_E2_gr}
\bs{j}_{E^2}^{(-2)}
=
-e \sum_{\mbk a} \bs{v}_{\mbk}^a  n_{E^2 \mbk a}^{(-2)},
\end{equation}
which arises from the group velocity of ordinary-scattered carriers. At next order in the impurity density, three terms contribute, which are related to the group velocity of mixed-scattered carriers, the Berry curvature dipole~\cite{sodemann2015quantum} and the leading order contribution from the (linear-response) extrinsic velocity as
\begin{equation}
\label{j_E2_mixexbc}
\bs{j}_{E^2}^{(-1)}
=
\bs{j}_{E^2}^{(\text{mix})}
+
\bs{j}_{E^2}^{(\text{ex})}
+
\bs{j}_{E^2}^{(\text{BC})},
\end{equation}
with
\begin{subequations}
\begin{align}
\label{j_E2_mix}
&\bs{j}_{E^2}^{(\text{mix})}
=
-e \sum_{\mbk a} \bs{v}_{\mbk}^a  n_{E^2 \mbk a}^{(-1)},
\\
\label{j_E2_ex}
&\bs{j}_{E^2}^{(\text{ex})}
=
-e \sum_{\mbk a}  \bs{\alpha}_{\mbk}^{a} n_{E^2 \mbk a}^{(-2)},
\\
\label{j_E2_BC}
&\bs{j}_{E^2}^{(\text{BC})}
=
- \frac{e^2}{\hbar} \sum_{\mbk a} \mathbf{E} \times \bs{\Omega}_{\mbk}^a  n_{E \mbk a}^{(-1)}.
\end{align}
\end{subequations}
The remaining terms are then of $O(n_I^0)$. One such term is the anomalous velocity experienced by $n_{E\mbk}^{(0)}$ carriers, leading to anomalous nonlinear currents arising from side-jump and skew scattering processes. Interestingly, such nonlinear responses have also been recently proposed in Refs.~\cite{ma2023anomalous, atencia2023disorder}. Another $O(n_I^0)$ term in Eq.~(\ref{j_E2_tot}) is that of the quantum metric, which has recently garnered significant interest as an intrinsic nonlinear response~\cite{wang2021intrinsic, bhalla2022resonant, gao2023quantum, wang2023quantum, das2023intrinsic, hetenyi2023fluctuations, zhuang2024intrinsic, wang2024intrinsic} and makes a natural appearance in the density-matrix formalism. The remaining $O(n_I^0)$ terms are related to the group velocity of the carriers due to zeroth-order scattering and the various extrinsic velocities experienced by the carriers, revealing the intricate interplay between disorder and quantum geometry at this order.

In order to gain better insight into the emergence of extrinsic velocities in the carrier dynamics and transport in the nonlinear response regime, it is helpful to first revisit the linear response of the system, in which the nonequilibrium Fermi-surface carriers $n_{E\mbk}^{(-1)}$ experience the extrinsic velocity $\bs{\alpha}_{\mbk}$. In the presence of disorder, the electrons undergo random scatterings off the impurities, which leads to fluctuations in their velocities as they scatter between different energy bands. The net effect of this process is an overall change in the average velocity of the electrons within each energy band, which is captured by $\bs{\alpha}_{\mbk}^a$ and contributes to the transport at the Fermi level through the nonequilibrium carriers.

Extending this picture to the quadratic responses, it is clear that Eq.~(\ref{j_E2_ex}) describes a similar effect of $\bs{\alpha}_{\mbk}$ on the field-corrected distribution $n_{E^2\mbk}^{(-2)}$. The effect at the Fermi surface can be understood through an integration by parts
\begin{equation}
\label{ell}
-e \sum_{\mbk a}  \bs{\alpha}_{\mbk}^{a} n_{E^2 \mbk a}^{(-2)}
=
\frac{e^2}{\hbar} \sum_{\mbk a} n_{E \mbk a}^{(-1)} \mathbf{E} \cdot \bs{\pd}_{\mbk}   \bs{\ell}_{\alpha \mbk}^a,
\end{equation}
where we have introduced the extrinsic transport length vector
$\bs{\ell}_{\alpha \mbk}^a
=
\tau_{\mbk}^a \bs{\alpha}_{\mbk}^a$, which measures the average displacement the carriers experience over the transport time by the random interband walks on the Fermi surface. From Eq.~(\ref{ell}), we see that this average displacement receives a correction from the electric field--absent in the linear response--which results in a nonlinear current contribution at $O(n_I^{-1})$. At next order in the impurity density, a similar argument can be applied for the contribution of $\bs{\chi}_{\mbk}$ as well as for that of $\bs{\alpha}_{\mbk}$, with the difference that in the latter case, the mixed-scattering carriers $n_{E^2 \mbk}^{(-1)}$ come into play, indicating additional intermediate side-jump and skew-scattering processes.

Furthermore, the emergence of $\bs{\beta}_{\mbk}$, $\bs{\gamma}_{\mbk}$ and $\bs{\kappa}_{\mbk}$ can be attributed to the fact that the electric field itself can also induce interband coherence effects through its coupling with the position operator in the electrostatic interaction. In the absence of disorder, this field-mediated interband mixing can be understood as giving rise to the intrinsic Berry curvature and quantum metric contributions in Eq.~(\ref{eom}). And when disorder is present, the combined effect of the two interactions can give rise to new means of interband mixing such that the resulting extrinsic velocities acquire field dependencies. This is in contrast to the linear response regime, where--within the weak-disorder limit--the electrostatic and scattering corrections to the Bloch states can be separated, resulting in only independent corrections to physical observables~\cite{xiao2017semiclassical}.

\section{Application to massive Dirac fermions}
\label{application}

\subsection{Broken time reversal symmetry}

To illustrate the leading and subleading responses that emerge from the theory, consider a minimal model of 2D tilted massive Dirac fermions
\begin{equation}
\label{H_Dirac}
H_{\mbk}^0
=
v \bs{\sigma} \cdot \mbk + \mbk \cdot \mathbf{t} + m \sigma_z,
\end{equation}
where $m$ is the mass gap energy, while $v$ and $\mathbf{t} = (t_x, t_y)$ measure the strengths of the spin-orbit interaction and tilting of the Dirac cones, respectively. This is a fairly ubiquitous model used to describe the transport-relevant band structure in a host of quantum materials and may be used to develop more detailed band structures~\cite{goerbig2008tilted, du2018band, ma2019observation}. The eigenenergies read 
$\veps_{\mbk}^a
=
\veps_{0 k}^a + \mbk \cdot \mathbf{t}$ where 
$\veps_{0 k}^a = a h_k$ is the energy of the isotropic subsystem in the absence of tilting, with 
$h_k = \sqrt{k^2v^2+m^2}$ and $a=\pm$. Note that the tilting breaks the time reversal symmetry of the energy bands as well as the Hamiltonian, as opposed to the mass, which only breaks the time reversal symmetry at the level of the Hamiltonian. The Bloch eigenstates are expressed as
\vspace{-.2cm}
\begin{equation}
%\vspace{-.5cm}
\ket{u_{\mathbf{k}a}}
=
\begin{pmatrix}
\left( \frac{1+ a}{2}\right) 
\cos \frac{\theta}{2}
+
\left( \frac{1-a}{2}\right) 
\sin \frac{\theta}{2}
\\
e^{i\phi}
\left[
\left( \frac{1+a}{2}\right) 
\sin \frac{\theta}{2}
-
\left( \frac{1-a}{2}\right) 
\cos \frac{\theta}{2}
\right]
\end{pmatrix},
\vspace{-.2cm}
\end{equation}
with $\theta = \cos^{-1}(m/h_k)$. For simplicity, we assume the weakly anisotropic case $t \ll v$, such that the eccentricity of the Fermi surface is small. We also assume the Fermi level lies in the upper band. The transport time is hence expressed as
\begin{equation}
\vspace{0cm}
\label{tau_k}
\frac{1}{\tau_{\mbk}^a}
=
\frac{2\pi}{\hbar} 
\sum_{\mbkpr b}
\delta ( \veps_{\mbk}^a - \veps_{\mbkpr}^b)
\Braket{
U_{\mbk \mbkpr}^{ab} U_{\mbkpr \mbk} ^{ba} 
}
\left( 1
- 
\frac{\mathbf{e}_E \cdot \bs{v}_{\mbkpr}^b}
{\mathbf{e}_E \cdot \bs{v}_{\mbk}^a}
\right),
\vspace{-0cm}
\end{equation}
with $\mathbf{e}_E$ the unit vector along the electric field, where we note that in the isotropic limit, Eq.~(\ref{tau_k}) reduces to Eq.~(\ref{tau_0k}). To first order in tilting, this yields
\begin{equation}
\begin{split}
\frac{1}{\tau_{\mbk}^a}
=&
\frac{1}{\tau_{0 k}}
\left( 1 
-
\frac{a h_k}{v^2}
\frac{\mathbf{e}_E \cdot \mathbf{t}}
{\mathbf{e}_E \cdot \mathbf{k}}
\right)
-
\frac{a n_I U_0^2}{32 \hbar}
\left( \frac{h_k}{v^2} \right)^2
\left[
\left(
11 + 4 \cos 2\theta + \cos 4\theta \right)
\frac{\mathbf{e}_E \cdot \mathbf{t}}
{\mathbf{e}_E \cdot \mathbf{k}}
\right.
\\
&+
\left.
\left(
21 + 12 \cos 2\theta - \cos 4\theta \right)
\mathbf{t} \cdot 
\frac{\mathbf{e}_{\mbk}}{k}
\right],
\end{split}
\end{equation}
with $\mathbf{e}_{\mbk} = \mbk/k$ and the isotropic transport time given by
\begin{equation}
\frac{1}{\tau_{0 k}}
=
\frac{n_I U_0^2}{4 \hbar} \frac{h_k}{v^2}
\left( 1 + 3 \cos^2 \theta \right).
\end{equation}
To evaluate the nonlinear conductivities, we start with the leading-order contribution in disorder, Eq.~(\ref{j_E2_gr}), which arises from ordinary scattering. Without loss of generality, let us set the electric field along the $x$ direction, so that fewer elements of the conductivity tensor need to be evaluated. Inserting Eq.~(\ref{n_E2_-2}) into Eq.~(\ref{j_E2_gr}) and using the approximation
$\pd f_{0\mbk}^a / \pd \veps_{\mbk}^a
\simeq
- \delta(\veps_{\mbk}^a - \veps_F)
$, the quadratic conductivities associated with the (longitudinal) nonlinear response and nonlinear Hall effect are obtained as
\begin{subequations}
\label{sigma_ord}
\begin{align}
\sigma_{xxx}^{(\text{o})}
&=
- \frac{e^3}{8 \pi \hbar} (n_I U_0^2)^{-2}
\frac{v^4}{h_{k_F}^2} t_x
\mathcal{S}_{xxx}^{(\text{o})}(\theta),
\\
\sigma_{yxx}^{(\text{o})}
&=
- \frac{e^3}{8 \pi \hbar} (n_I U_0^2)^{-2}
\frac{v^4}{h_{k_F}^2} t_y
\mathcal{S}_{yxx}^{(\text{o})}(\theta),
\end{align}
\end{subequations}
respectively, where the superscript (o) indicates the ordinary-scattering origin of the conductivity. Here, the functions
$\mathcal{S}_{\mu xx}(\theta)$ encapsulate the mass dependence and are presented in Sec.~\ref{app_dirac}. 

To understand the dependence of the nonlinear current associated with Eq.~(\ref{sigma_ord}) on the tilting vector, it suffices to consider the symmetries of the terms in Eq.~(\ref{j_E2_gr}), namely the quadratic carrier density $n_{E^2 \mbk a}^{(-2)}$ and the group velocity of the Dirac fermions, $\bs{v}_{\mbk}^a = (a v^2/ \hbar h_k) \mbk + \mathbf{t}/\hbar$. In the limit $t_{\mu} = 0$, the group velocity is odd under the mirror symmetry transformation $k_{\mu} \rightarrow - k_{\mu}$, while the carrier density term is even. Thus the corresponding integral in Eq.~(\ref{j_E2_gr}) vanishes, which leads to the conclusion $\sigma_{\mu xx}^{(\text{o})} \propto t_{\mu}$. For the conductivities that are discussed below, a similar symmetry analysis can be applied, from which one can infer the angular dependence on the tilting vector for each term.

\begin{figure}[hpt]
\captionsetup[subfigure]{labelformat=empty}
    \sidesubfloat[]{\includegraphics[width=0.32\linewidth,trim={0cm 0cm 0cm 0cm}]{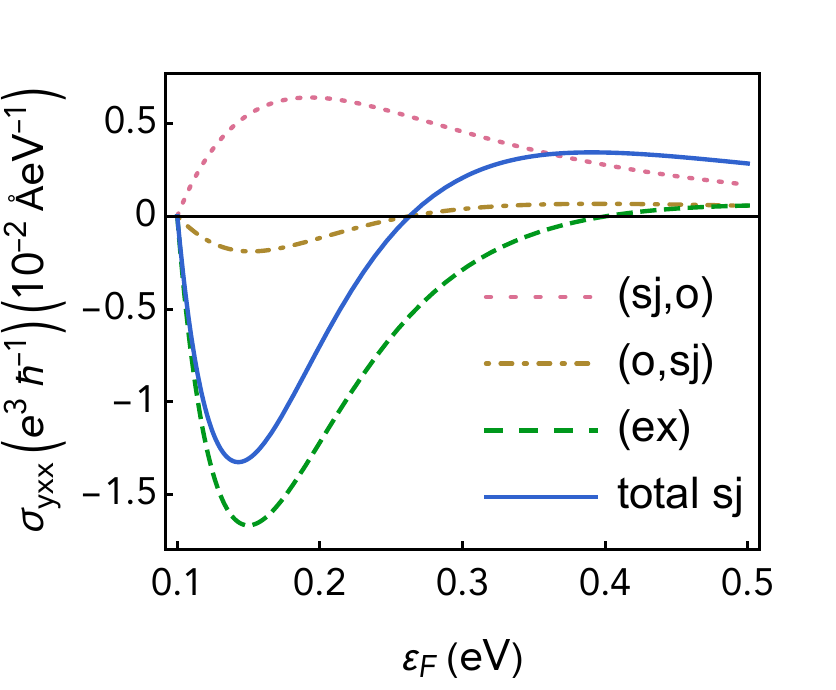}\label{fig_qkt_fig2a}}
%\quad%
    \sidesubfloat[]{\includegraphics[width=0.32\linewidth,trim={0cm 0cm 0cm 0cm}]{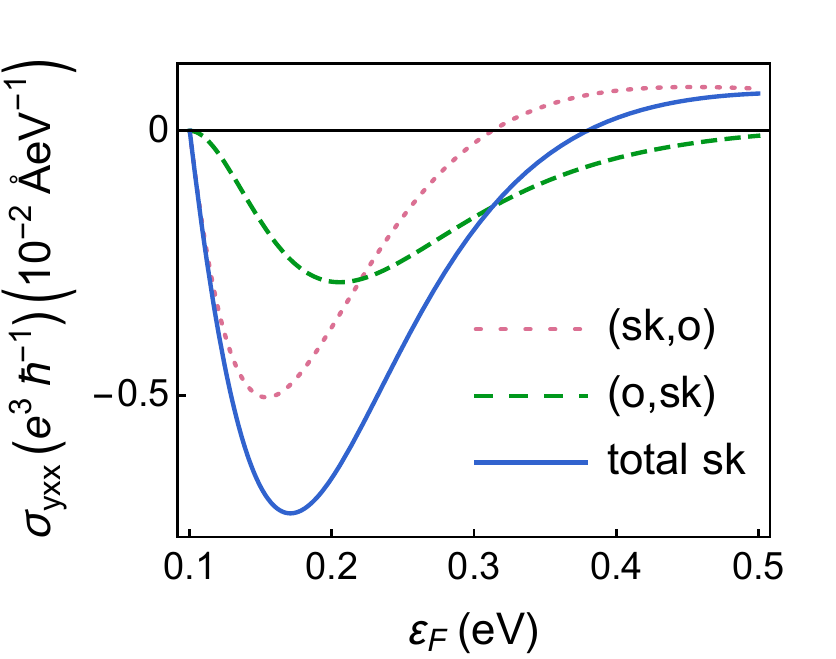}\label{fig_qkt_fig2b}}
%%%%%%%    
    \sidesubfloat[]{\includegraphics[width=0.32\linewidth,trim={0cm 0cm 0cm 0cm}]{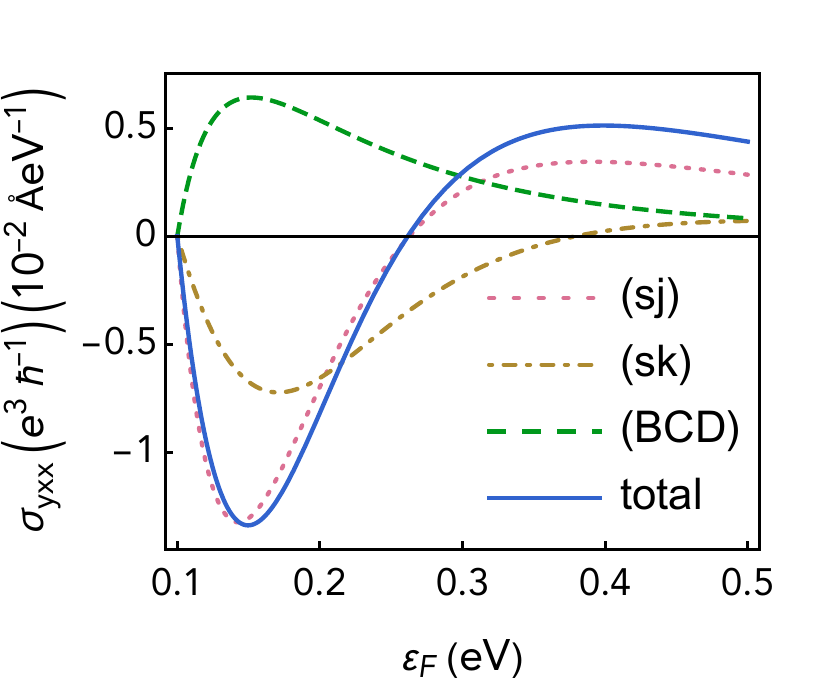}\label{fig_qkt_fig2c}}
%%%%%%%%%
    \caption{Scalings of leading-order contributions to the nonlinear Hall conductivity in the presence of time reversal symmetry as a function of the Fermi energy, starting from the bottom of the conduction band. Parameters used: $m=0.1$ eV, $t_x = 0.1$ eV  \AA, $t_y = 0$, $v=1$ eV \AA, and $n_I U_0^2=10^2$ eV$^2$ \AA$^2$.}
    \label{fig_qkt_fig2}
\end{figure}

We next look at the subleading contributions to the nonlinear current, which are those of the mixed scattering, Berry curvature dipole and extrinsic velocity. For simplicity, we do not consider the sub-subleading processes in this work, as they are higher order effects and can be neglected in the weak disorder limit. And to obtain analytical results in the weak tilting limit, we use the isotropic transport time. The current density arising from mixed scattering is given by Eq.~(\ref{j_E2_mix}), where the explicit forms of the relevant carrier densities for the tilted Dirac model are presented in full in Eqs.~(\ref{n_Ek_sj_dirac} - \ref{N_3k_sko_dirac}). Inserting these into Eq.~(\ref{j_E2_mix}), we obtain the side-jump conductivities
\begin{subequations}
\begin{align}
\begin{split}
\sigma_{xxx}^{(\text{sj,o})}
&=
- \frac{e^3}{8 \pi \hbar} (n_I U_0^2)^{-1}
\frac{v^2}{h_{k_F}^2} t_y 
\mathcal{S}_{xxx}^{(\text{sj,o})}(\theta),
\end{split}
\\
\sigma_{yxx}^{(\text{sj,o})}
&=
\frac{e^3}{8 \pi \hbar} (n_I U_0^2)^{-1}
\frac{v^2}{h_{k_F}^2} t_x
\mathcal{S}_{yxx}^{(\text{sj,o})}(\theta),
\end{align}
\end{subequations}
and
\begin{subequations}
\begin{align}
\sigma_{xxx}^{(\text{o,sj})}
&=
\frac{e^3}{8 \pi \hbar} (n_I U_0^2)^{-1}
\frac{v^2}{h_{k_F}^2} t_y
\mathcal{S}_{xxx}^{(\text{o,sj})}(\theta),
\\
\sigma_{yxx}^{(\text{o,sj})}
&=
- \frac{e^3}{2 \pi \hbar} (n_I U_0^2)^{-1}
\frac{v^2}{h_{k_F}^2} t_x
\mathcal{S}_{yxx}^{(\text{o,sj})}(\theta),
\end{align}
\end{subequations}
as well as the skew-scattering contributions
\begin{subequations}
\begin{align}
\sigma_{xxx}^{(\text{sk,o})}
&=
-\frac{e^3}{64 \pi \hbar} (n_I U_0^2)^{-1}
\frac{v^2}{h_{k_F}^2} t_y
\mathcal{S}_{xxx}^{(\text{sk,o})}(\theta),
\\
\sigma_{yxx}^{(\text{sk,o})}
&=
- \frac{e^3}{32 \pi \hbar} (n_I U_0^2)^{-1}
\frac{v^2}{h_{k_F}^2} t_x
\mathcal{S}_{yxx}^{(\text{sk,o})}(\theta),
\end{align}
\end{subequations}
and
\begin{subequations}
\begin{align}
\sigma_{xxx}^{(\text{o,sk})}
&=
\frac{e^3}{4 \pi \hbar} (n_I U_0^2)^{-1}
\frac{v^2}{h_{k_F}^2} t_y
\mathcal{S}_{xxx}^{(\text{o,sk})}(\theta),
\\
\sigma_{yxx}^{(\text{o,sk})}
&=
-\frac{e^3}{\pi \hbar} (n_I U_0^2)^{-1}
\frac{v^2}{h_{k_F}^2} t_x
\mathcal{S}_{yxx}^{(\text{o,sk})}(\theta).\end{align}
\end{subequations}
Next, we consider the leading-order extrinsic velocity contribution, Eq.~(\ref{j_E2_ex}). Evaluating the energy-conserving part of the extrinsic velocity in Eq.~(\ref{alpha_ka}) for the massive Dirac model, we obtain
\begin{equation}
\begin{split}
\label{alpha_k+}
\bs{\alpha}_{\mbk}^+
=&
\frac{n_I U_0^2}{4 \hbar v}
\left[
\sin 2\theta \mathbf{e}_{\phi} 
-
\frac{1}{2 v}
\cos \theta \left(3 + \cos 2\theta \right) \mathbf{e}_{z} \times \mathbf{t}
+
\frac{1}{4 v}
\sin \theta \sin 2\theta
\left\{
\left(
\mathbf{e}_{2\phi}  \cdot \mathbf{t} \right) \mathbf{e}_{E}
\right.
\right.
\\
&+
\left.
\left.
\left[
\mathbf{e}_{z} \cdot \left( \mathbf{e}_{2\phi} \times \mathbf{t} \right) \right]
\mathbf{e}_{z} \times \mathbf{e}_{E}
\right\}
\right],
\end{split}
\end{equation}
which generalizes the expression obtained in Ref.~\cite{atencia2022semiclassical} to the tilted case. Here, $\mathbf{e}_{\phi}$ and $\mathbf{e}_{2 \phi}$ are the azimuthal unit vectors corresponding to the angles $\phi$ and $2\phi$, respectively, and $\mathbf{e}_{z}$ is the unit vector in the $z$ direction, which is perpendicular to the system plane. Note this implies that, in the presence of tilting, the extrinsic velocity is no longer limited to the transverse direction and can, in principle, also give rise to a longitudinal current. Indeed,  inserting Eq.~(\ref{alpha_k+}) into the current expression given by Eq.~(\ref{j_E2_ex}), we arrive at the conductivities
\begin{subequations}
\begin{align}
\sigma_{xxx}^{(\text{ex})}
&=
\frac{e^3}{8 \pi \hbar} (n_I U_0^2)^{-1}
\frac{v^2}{h_{k_F}^2} t_y
\mathcal{S}_{xxx}^{(\text{ex})}(\theta),
\\
\sigma_{yxx}^{(\text{ex})}
&=
- \frac{e^3}{8 \pi \hbar} (n_I U_0^2)^{-1}
\frac{v^2}{h_{k_F}^2} t_x
\mathcal{S}_{yxx}^{(\text{ex})}(\theta).
\end{align}
\end{subequations}
The last term to consider at this order in the disorder expansion is that of the familiar Berry curvature dipole, the leading order contribution of which is due to the anomalous velocity felt by the ordinary-scattered electrons
\begin{equation}
\label{j_E2_BC}
\delta \bs{j}_{E^2}^{(\text{BC})}
=
- \frac{e^2}{\hbar} \sum_{\mbk a} \mathbf{E} \times \bs{\Omega}_{\mbk}^a  n_{E \mbk a}^{(-1)}.
\end{equation}
The out-of-plane component of the Berry curvature vector is given by 
$\Omega_{z k}^a
=
- a \cos \theta v^2 / 2 h_k^2$, which yields the nonlinear Hall conductivity
\begin{equation}
\sigma_{yxx}^{(\text{BC})}
=
\frac{e^3}{\pi \hbar} (n_I U_0^2)^{-1}
\frac{v^2}{h_{k_F}^2} t_x
\mathcal{S}_{yxx}^{(\text{BC})}(\theta).
\end{equation}
Gathering all the above terms, the total nonlinear conductivities are obtained, which are relatively complicated expressions. It is thus useful to consider two physically relevant limits. The first is the low-energy limit whereby the Fermi level lies close to the mass gap, \textit{i.e.}, $k_Fv \ll m$. In this case, we find
\begin{subequations}
\begin{align}
\sigma_{xxx}
&
\overset{k_Fv \ll m}{\simeq}
- \frac{3 e^3}{8 \pi \hbar} (n_I U_0^2)^{-2}
\frac{v^4}{m^2} t_x,
\\
\sigma_{yxx}
&
\overset{k_Fv \ll m}{\simeq}
- \frac{e^3}{8 \pi \hbar} (n_I U_0^2)^{-2}
\frac{v^4}{m^2} t_y.
\end{align}
\end{subequations}
where we assume the disorder is weak enough to neglect subleading processes. Interestingly, this results in the following ratio between the the nonlinear Hall and longitudinal conductivities
\begin{equation}
\frac{\sigma_{yxx}}{\sigma_{xxx}}
\overset{k_Fv \ll m}{=}
\frac{1}{3} \cot \phi_t,
\end{equation}
with $\phi_t= \cot^{-1}(t_x/t_y)$, in agreement with results obtained through diagrammatic~\cite{mehraeen2023quantum, mehraeen2024proximity} and Boltzmann transport~\cite{marinescu2023magnetochiral} approaches for various systems, which suggests a general relationship between the two conductivity components in the weak-disorder limit. And at the other extreme, when the Fermi level lies much higher than the gap, $k_Fv \gg m$, we find
\begin{subequations}
\begin{align}
\sigma_{xxx}
&
\overset{k_Fv \gg m}{\simeq}
\frac{11 e^3}{\pi \hbar} (n_I U_0^2)^{-2}
\frac{v^2}{k_F^2} t_x,
\\
\sigma_{yxx}
&
\overset{k_Fv \gg m}{\simeq}
\frac{e^3}{\pi \hbar} (n_I U_0^2)^{-2}
\frac{v^2}{k_F^2} t_y,
\end{align}
\end{subequations}
and the ratio
\begin{equation}
\frac{\sigma_{yxx}}{\sigma_{xxx}}
\overset{k_Fv \gg m}{=}
\frac{1}{11} \cot \phi_t.
\end{equation}
This suggests that, in addition to the individual dependencies of the conductivities on the material and disorder parameters, these differing ratios may prove useful in exploring the band topology and materials parameter dependencies of various systems via nonlinear transport measurements.
\vspace{-0cm}
\subsection{Restoring time reversal symmetry}
\vspace{-0cm}
In order to better explore the contributions of the subleading terms and their different scalings with the system parameters, it is useful to also consider the time-reversal symmetric counterpart of the model given by Eq.~(\ref{H_Dirac}), which can be obtained by extending the system to include the transformed Hamiltonian with negative tilt and mass, $(\mathbf{t}, m) \rightarrow - ( \mathbf{t} ,  m)$. Note that negating the mass corresponds to the transformation $\theta \rightarrow \pi - \theta$ in the angular functions $\mathcal{S}_{\mu xx} (\theta)$ given by Eqs.~(\ref{S_muxx^o}-\ref{S_yxx^BC}) in Sec. \ref{app_dirac}. Upon doing so, it straightforward to verify that the ordinary-scattering conductivity does not contribute to the nonlinear transport in the presence of time reversal symmetry, as
\begin{equation}
\sigma_{\mu xx}^{(\text{o})} (- \mathbf{t} , - m)
=
- \sigma_{\mu xx}^{(\text{o})} (\mathbf{t} , m).
\end{equation}
Therefore, the subleading terms in the broken time reversal system are the leading-order terms in the extended time-reversal-invariant system and will dominate the transport. And their contributions to the transport are obtained by simply doubling the broken time reversal expressions of the conductivities presented above.

In Fig.~\ref{fig_qkt_fig2}, we plot the leading-order contributions to the nonlinear Hall effect for the time-reversal-symmetric case as a function of the Fermi energy, which reveals several notable features: as shown in Fig.~\ref{fig_qkt_fig2a}, three terms comprise the conventional side-jump contribution to the nonlinear transport, including the extrinsic velocity. As can be seen, there are noticeable qualitative and quantitative differences between the contributions, such that the sum of the side-jump terms alone can give rise to a sign change of the conductivity (see the solid blue curve). A similar situation holds for the skew-scattering conductivities, plotted in Fig.~\ref{fig_qkt_fig2b}. Interestingly, while the contributions from $n_{E^2\mbk}^{(\text{sk,o})}$ and $n_{E^2\mbk}^{(\text{o,sk})}$ act constructively near the band bottom, at higher Fermi energies, they contribute with opposite signs. Finally, in Fig.~\ref{fig_qkt_fig2c}, we plot the total nonlinear Hall conductivities involving the Berry curvature dipole and disorder contributions, which again exhibit Fermi-energy-dependent sign changes from the competition between intrinsic and extrinsic effects, in relatively good agreement with previous studies of the nonlinear Hall effect based on a Boltzmann transport formalism with the coordinate shift~\cite{du2019disorder, du2021nonlinear}.

\section{Collision integrals}
\label{App_collision_int}

Inserting Eqs.~(\ref{delta_rho_0}) and (\ref{delta_rho_n}) into the general definition of the collision integral, to second order in the electric field, we obtain
\begin{subequations}
\begin{align}
\label{J_0k_fk}
J_{0\mbk} ( f_{\lambda \mbk} )
&=
\frac{1}{\hbar}
\int_{-\infty}^{\infty} \frac{d\veps}{2\pi}
\Braket{
\left[ U, G^R(\veps) \left[ U, \braket{\rho_{\lambda}} \right] G^A(\veps)
\right]
}_{\mbk},
\\
\label{J_Ek_fk}
J_{E \mbk} ( f_{\lambda \mbk} )
&=
- \frac{i}{\hbar}
\int_{-\infty}^{\infty} \frac{d\veps}{2\pi}
\int_{-\infty}^{\infty} \frac{d\veps\pr}{2\pi}
\Braket{
\left[ U, G^R(\veps)
\left[ V, G^R(\veps\pr)
\left[ U, \braket{\rho_{\lambda}} \right] G^A(\veps\pr) \right]
G^A(\veps)
\right]
}_{\mbk},
\\
\begin{split}
\label{J_E2k_fk}
J_{E^2 \mbk} ( f_{\lambda \mbk} )
&=
- \frac{1}{\hbar}
\int_{-\infty}^{\infty} \frac{d\veps}{2\pi}
\int_{-\infty}^{\infty} \frac{d\veps\pr}{2\pi}
\int_{-\infty}^{\infty} \frac{d\veps\dpr}{2\pi}
\\
&\times
\Braket{
\left[ U, G^R(\veps)
\left[ V, G^R(\veps\pr)
\left[ V, G^R(\veps\dpr)
\left[ U, \braket{\rho_{\lambda}} \right] G^A(\veps\dpr) 
\right] G^A(\veps\pr)
\right] G^A(\veps)
\right]
}_{\mbk},
\end{split}
\end{align}
\end{subequations}
where the integrals are evaluated with the relation
\begin{equation}
\int_{\infty}^{\infty}\frac{d\veps}{2 \pi i} 
G^A_{\mbk a}(\veps)
G^R_{\mbk^{\prime} b}(\veps)
=
\frac{1}{\veps_{\mbk}^a - \veps_{\mbkpr}^b + i \delta},
\end{equation}
and the useful identity
\begin{equation}
\label{dirac_derivative}
\frac{\delta (x)}{x^{n}}
=
\frac{(-1)^n}{n!} \frac{d^n}{dx^n}\delta (x).
\end{equation}
The explicit forms of the collision integrals are then derived once the off-diagonal density matrix elements are known, which are presented in the next section.

\section{Off-diagonal density matrix}
\label{App_off-diagonal}

Inserting the various driving terms into the formal solution of the off-diagonal density matrix, one finds
\begin{subequations}
\label{S_Ek_0_decomp}
\begin{gather}
\label{S_Ek_0D}
S_{E \mbk}^{(0) ab} \left[D_{E \mbk}^{(0)} \right]
=
e \mathbf{E} \cdot 
\bs{\mathcal{A}}_{\mbk}^{\prime ab}
\frac{f_{0 \mbk}^a - f_{0 \mbk}^{b}}{\veps_{\mbk}^a - \veps_{\mbk}^b},
\\
\label{S_Ek_0I}
S_{E \mbk}^{(0) ab} \left[ I_{E \mbk}^{(0)} \right]
=
i \pi
\sum_{\mbkpr c}
\frac{\Braket{U_{\mbk \mbk\pr}^{ac} U_{\mbk\pr \mbk}^{cb}}}
{\veps_{\mbk}^a - \veps_{\mbk}^b}
\left[
\left(n_{E \mbk a}^{(-1)} - n_{E \mbkpr c}^{(-1)}\right) \delta(\veps_{\mbk}^a 
-
\veps_{\mbkpr}^c)
+
\left(n_{E \mbk b}^{(-1)} - n_{E \mbkpr c}^{(-1)}\right) \delta(\veps_{\mbk}^b 
-
\veps_{\mbkpr}^c)
\right],
\end{gather}
\end{subequations}
at linear order in the electric field. And, similarly, for the analogous quadratic-response  counterparts, we obtain
\begin{subequations}
\label{S_E2k_-1_decomp}
\begin{gather}
\label{S_E2k_-1D}
S_{E^2 \mbk}^{(-1) ab} \left[D_{E^2 \mbk}^{(-1)} \right]
=
e \mathbf{E} \cdot 
\bs{\mathcal{A}}_{\mbk}^{\prime ab}
\frac{n_{E \mbk a}^{(-1)} - n_{E \mbk b}^{(-1)}}{\veps_{\mbk}^a - \veps_{\mbk}^b },
\\
\label{S_E2k_-1I}
S_{E^2 \mbk}^{(-1) ab} \left[ I_{E^2 \mbk}^{(-1)} \right]
=
i \pi
\sum_{\mbkpr c}
\frac{\Braket{U_{\mbk \mbk\pr}^{ac} U_{\mbk\pr \mbk}^{cb}}}
{\veps_{\mbk}^a - \veps_{\mbk}^b}
\left[
(n_{E^2 \mbk a}^{(-2)} - n_{E^2 \mbkpr c}^{(-2)} ) \delta(\veps_{\mbk}^a 
-
\veps_{\mbkpr}^c)
+
(n_{E^2 \mbk b}^{(-2)} - n_{E^2 \mbkpr c}^{(-2)}) \delta(\veps_{\mbk}^b 
-
\veps_{\mbkpr}^c)
\right],
\end{gather}
\end{subequations}
as well as
\begin{subequations}
\label{S_E2k_0_decomp}
\begin{gather}
\label{S_E2k_0D}
S_{E^2 \mbk}^{(0) ab} \left[D_{E^2 \mbk}^{(0)} \right]
=
e \mathbf{E} \cdot 
\bs{\mathcal{A}}_{\mbk}^{\prime ab}
\frac{n_{E \mbk a}^{(0)} - n_{E \mbk b}^{(0)}}{\veps_{\mbk}^a - \veps_{\mbk}^b },
\\
\label{S_E2k_0I}
S_{E^2 \mbk}^{(0) ab} \left[ I_{E^2 \mbk}^{(0)} \right]
=
i \pi
\sum_{\mbkpr c}
\frac{\Braket{U_{\mbk \mbk\pr}^{ac} U_{\mbk\pr \mbk}^{cb}}}
{\veps_{\mbk}^a - \veps_{\mbk}^b}
\left[
(n_{E^2 \mbk a}^{(-1)} - n_{E^2 \mbkpr c}^{(-1)} ) \delta(\veps_{\mbk}^a 
-
\veps_{\mbkpr}^c)
+
( n_{E^2 \mbk b}^{(-1)} - n_{E^2 \mbkpr c}^{(-1)} ) \delta(\veps_{\mbk}^b 
-
\veps_{\mbkpr}^c)
\right].
\end{gather}
\end{subequations}
The contribution from 
$I_{E^2 \mbk}^{\prime (0)}$ is obtained by inserting Eq.~(\ref{J_Ek_fk}) into the relevant term in Eq.~(\ref{S_E2k_0}), which yields
\begin{equation}
\label{S_E2k_0Ipr}
\begin{split}
S_{E^2 \mbk}^{(0) ab} \left[ I_{E^2 \mbk}^{\prime (0)} \right]
&=
- \frac{ e}
{\veps_{\mbk}^a - \veps_{\mbk}^b }
\mathbf{E} \cdot
\sum_{\mbkpr c}
\frac{1}
{\veps_{\mbk}^b - \veps_{\mbkpr}^c + i \delta}
\Biggl\{
\Braket{i U_{\mbk \mbk\pr}^{ac} 
(\bs{\pd}_{\mbk} + \bs{\pd}_{\mbkpr})
\left(
U_{\mbk\pr \mbk}^{cb}
\frac{n_{E \mbk b}^{(-1)} - n_{E \mbkpr c}^{(-1)}}
{\veps_{\mbk}^b - \veps_{\mbkpr}^c + i \delta}
\right)}
\\
&+
\sum_d
\left[
\Braket{U_{\mbk \mbk\pr}^{ac} U_{\mbk\pr \mbk}^{db}} \bs{\mathcal{A}}_{\mbkpr}^{cd}
\frac{n_{E \mbk b}^{(-1)} - n_{E \mbkpr d}^{(-1)}}
{\veps_{\mbk}^b - \veps_{\mbkpr}^d + i \delta}
-
\Braket{U_{\mbk \mbk\pr}^{ac} U_{\mbk\pr \mbk}^{cd}} \bs{\mathcal{A}}_{\mbk}^{db}
\frac{n_{E \mbk d}^{(-1)} - n_{E \mbkpr c}^{(-1)}}
{\veps_{\mbk}^d - \veps_{\mbkpr}^c + i \delta}
\right]
\Biggr\}
\\
&-
\frac{e}
{\veps_{\mbk}^a - \veps_{\mbk}^b}
\mathbf{E} \cdot
\sum_{\mbkpr c}
\frac{1}
{\veps_{\mbkpr}^c - \veps_{\mbk}^a + i \delta}
\Biggl\{
\Braket{i U_{\mbkpr \mbk}^{cb} 
(\bs{\pd}_{\mbk} + \bs{\pd}_{\mbkpr})
\left(
U_{\mbk \mbkpr}^{ac}
\frac{n_{E \mbk a}^{(-1)} - n_{E \mbkpr c}^{(-1)}}
{\veps_{\mbkpr}^c - \veps_{\mbk}^a + i \delta}
\right)}
\\
&+
\sum_d
\left[
\Braket{U_{\mbk \mbk\pr}^{dc} U_{\mbkpr \mbk}^{cb}} \bs{\mathcal{A}}_{\mbk}^{ad}
\frac{n_{E \mbk d}^{(-1)} - n_{E \mbkpr c}^{(-1)}}
{\veps_{\mbkpr}^c - \veps_{\mbk}^d + i \delta}
-
\Braket{U_{\mbk \mbk\pr}^{ad} U_{\mbk\pr \mbk}^{cb}} \bs{\mathcal{A}}_{\mbkpr}^{dc}
\frac{n_{E \mbk a}^{(-1)} - n_{E \mbkpr d}^{(-1)}}
{\veps_{\mbkpr}^d - \veps_{\mbk}^a + i \delta}
\right]
\Biggr\}.
\end{split}
\end{equation}
The find the remaining contributions from $D_{E^2 \mbk}^{\prime (0)}$ and $I_{E^2 \mbk}^{\dprime (0)}$, recalling Eq.~(\ref{S_E2k_Dpr_Idpr}), we arrive at the following terms
\begin{equation}
\begin{split}
S_{E^2 \mbk}^{(0) ab} \left[D_{E^2 \mbk}^{\prime (0)} [ D_{E \mbk}^{(0)}]
\right]
=&
- \frac{i e \mathbf{E} \cdot \bs{\pd}_{\mbk}
S_{E \mbk}^{(0) ab} \left[D_{E \mbk}^{(0)} \right]}
{\veps_{\mbk}^a - \veps_{\mbk}^b}
-
\frac{e^2 E^{\mu} E^{\nu}}
{\veps_{\mbk}^a - \veps_{\mbk}^b}
\Big\{
\mathcal{A}_{\mu \mbk}^{\prime ab}
(\mathcal{A}_{\nu \mbk}^{a}
-
\mathcal{A}_{\nu \mbk}^{b})
\frac{f_{0\mbk}^a - f_{0\mbk}^b}
{\veps_{\mbk}^a - \veps_{\mbk}^b}
\\
&-
\sum_{c}
\mathcal{A}_{\mu \mbk}^{\prime ac}
\mathcal{A}_{\nu \mbk}^{\prime cb}
\left(
\frac{f_{0\mbk}^a - f_{0\mbk}^c}
{\veps_{\mbk}^a - \veps_{\mbk}^c}
-
\frac{f_{0\mbk}^b - f_{0\mbk}^c}
{\veps_{\mbk}^b - \veps_{\mbk}^c}
\right)
\Bigg\},
\end{split}
\end{equation}
\begin{equation}
\begin{split}
S_{E^2 \mbk}^{(0) ab} &\left[D_{E^2 \mbk}^{\prime (0)} [ I_{E \mbk}^{(0)}]
\right]
=
- \frac{i e \mathbf{E} \cdot}
{\veps_{\mbk}^a - \veps_{\mbk}^b }
\Big\{
\bs{\pd}_{\mbk}
S_{E \mbk}^{(0) ab} \left[I_{E \mbk}^{(0)} \right]
\\
&+
\pi \sum_{\mbkpr cd}
\frac{
\bs{\mathcal{A}}_{\mbk}^{ac}
\Braket{U_{\mbk \mbk\pr}^{cd} U_{\mbk\pr \mbk}^{db}}
}
{\veps_{\mbk}^c - \veps_{\mbk}^b }
\left[
( n_{E \mbk b}^{(-1)} - n_{E \mbkpr d}^{(-1)} ) 
\delta(\veps_{\mbk}^b - \veps_{\mbkpr}^d)
+
( n_{E \mbk c}^{(-1)} - n_{E \mbkpr d}^{(-1)} )
\delta(\veps_{\mbk}^c - \veps_{\mbkpr}^d)
\right]_{c \neq b}
\\
&-
\pi \sum_{\mbkpr cd}
\frac{
\bs{\mathcal{A}}_{\mbk}^{cb}
\Braket{U_{\mbk \mbk\pr}^{ad} U_{\mbk\pr \mbk}^{dc}}
}
{\veps_{\mbk}^a - \veps_{\mbk}^c}
\left[
( n_{E \mbk a}^{(-1)} - n_{E \mbkpr d}^{(-1)} ) 
\delta(\veps_{\mbk}^a - \veps_{\mbkpr}^d)
+
( n_{E \mbk c}^{(-1)} - n_{E \mbkpr d}^{(-1)} )
\delta(\veps_{\mbk}^c - \veps_{\mbkpr}^d)
\right]_{c \neq a}
\Big\},
\end{split}
\end{equation}
\begin{equation}
\begin{split}
S_{E^2 \mbk}^{(0) ab} &\left[I_{E^2 \mbk}^{\dprime (0)} [ D_{E^2 \mbk}^{(-1)}]
\right]
=
i \pi e \mathbf{E} \cdot \sum_{\mbkpr cd}
\\
& \times \Bigg\{
\left[
\frac{
\Braket{U_{\mbk \mbk\pr}^{ac} U_{\mbkpr \mbk}^{cd}} 
\bs{\mathcal{A}}_{\mbk}^{\prime db}
}
{\veps_{\mbk}^a - \veps_{\mbk}^b }
\frac{n_{E \mbk d}^{(-1)} - n_{E \mbk b}^{(-1)}}
{\veps_{\mbk}^d - \veps_{\mbk}^b }
-
\frac{
\Braket{U_{\mbk \mbk\pr}^{ac} U_{\mbk\pr \mbk}^{db}}
\bs{\mathcal{A}}_{\mbkpr}^{\prime cd}
}
{\veps_{\mbk}^a - \veps_{\mbk}^b }
\frac{n_{E \mbkpr c}^{(-1)} - n_{E \mbkpr d}^{(-1)}}
{\veps_{\mbkpr}^c - \veps_{\mbkpr}^d }
\right]
\delta(\veps_{\mbk}^b - \veps_{\mbkpr}^c)
\\
&-
\left[
\frac{
\Braket{U_{\mbk \mbk\pr}^{ad} U_{\mbkpr \mbk}^{cb}}
\bs{\mathcal{A}}_{\mbkpr}^{\prime dc}
}
{\veps_{\mbk}^a - \veps_{\mbk}^b }
\frac{n_{E \mbkpr d}^{(-1)} - n_{E \mbkpr c}^{(-1)}}
{\veps_{\mbkpr}^d - \veps_{\mbkpr}^c }
-
\frac{
\Braket{U_{\mbk \mbk\pr}^{dc} U_{\mbk\pr \mbk}^{cb}}
\bs{\mathcal{A}}_{\mbk}^{\prime ad}
}
{\veps_{\mbk}^a - \veps_{\mbk}^b }
\frac{n_{E \mbk a}^{(-1)} - n_{E \mbk d}^{(-1)}}
{\veps_{\mbk}^a - \veps_{\mbk}^d }
\right]
\delta(\veps_{\mbk}^a - \veps_{\mbkpr}^c)
\bigg\},
\end{split}
\end{equation}

\begin{equation}
\label{S_E2_Ipr_I-1}
\begin{split}
&S_{E^2 \mbk}^{(0) ab} \left[I_{E^2 \mbk}^{\dprime (0)} [ I_{E^2 \mbk}^{(-1)}]
\right]
=
- \frac{\pi^2}
{\veps_{\mbk}^a - \veps_{\mbk}^b}
\sum_{\mbkpr \mbkdpr cde}
\delta(\veps_{\mbk}^b - \veps_{\mbkpr}^c)
\\
&\times
\bigg\{
\frac{
\Braket{U_{\mbk \mbk\pr}^{ac} U_{\mbk\pr \mbk}^{cd}}
\Braket{U_{\mbk \mbkdpr}^{de} U_{\mbkdpr \mbk}^{eb}}
}
{\veps_{\mbk}^d - \veps_{\mbk}^b }
\left[
( n_{E^2 \mbk b}^{(-2)} - n_{E^2 \mbkdpr e}^{(-2)}) 
\delta(\veps_{\mbk}^b - \veps_{\mbkdpr}^e )
+
( n_{E^2 \mbk d}^{(-2)} - n_{E^2 \mbkdpr e}^{(-2)} )
\delta(\veps_{\mbk}^d - \veps_{\mbkdpr}^e)
\right]_{d \neq b}
\\
&-
\frac{
\Braket{U_{\mbk \mbk\pr}^{ac} U_{\mbk\pr \mbk}^{db}}
\Braket{U_{\mbkpr \mbkdpr}^{ce} U_{\mbkdpr \mbkpr}^{ed}}
}
{\veps_{\mbkpr}^c - \veps_{\mbkpr}^d }
\left[
( n_{E^2 \mbkpr d}^{(-2)} - n_{E^2 \mbkdpr e}^{(-2)}) 
\delta(\veps_{\mbkpr}^d - \veps_{\mbkdpr}^e )
+
( n_{E^2 \mbkpr c}^{(-2)} - n_{E^2 \mbkdpr e}^{(-2)} )
\delta(\veps_{\mbkpr}^c - \veps_{\mbkdpr}^e)
\right]_{d \neq c}
\Big\}
\\
&-
\frac{\pi^2}
{\veps_{\mbk}^a - \veps_{\mbk}^b}
\sum_{\mbkpr \mbkdpr cde}
\delta(\veps_{\mbk}^a - \veps_{\mbkpr}^c)
\\
&\times
\bigg\{
\frac{
\Braket{U_{\mbk \mbk\pr}^{dc} U_{\mbk\pr \mbk}^{cb}}
\Braket{U_{\mbk \mbkdpr}^{ae} U_{\mbkdpr \mbk}^{ed}}
}
{\veps_{\mbk}^a - \veps_{\mbk}^d}
\left[
( n_{E^2 \mbk d}^{(-2)} - n_{E^2 \mbkdpr e}^{(-2)}) 
\delta(\veps_{\mbk}^d - \veps_{\mbkdpr}^e )
+
( n_{E^2 \mbk a}^{(-2)} - n_{E^2 \mbkdpr e}^{(-2)} )
\delta(\veps_{\mbk}^a - \veps_{\mbkdpr}^e)
\right]_{d \neq a}
\\
&-
\frac{
\Braket{U_{\mbk \mbk\pr}^{ad} U_{\mbk\pr \mbk}^{cb}}
\Braket{U_{\mbkpr \mbkdpr}^{de} U_{\mbkdpr \mbkpr}^{ec}}
}
{\veps_{\mbkpr}^d - \veps_{\mbkpr}^c}
\left[
( n_{E^2 \mbkpr c}^{(-2)} - n_{E^2 \mbkdpr e}^{(-2)}) 
\delta(\veps_{\mbkpr}^c - \veps_{\mbkdpr}^e )
+
( n_{E^2 \mbkpr d}^{(-2)} - n_{E^2 \mbkdpr e}^{(-2)} )
\delta(\veps_{\mbkpr}^d - \veps_{\mbkdpr}^e)
\right]_{d \neq c}
\Big\}.
\end{split}
\end{equation}
Thus, all the off-diagonal density matrix elements are expressed now in terms of diagonal terms. When inserted into the various collision integrals, the carrier densities are derived, the forms of which are presented in the next section.

\section{$\mathcal{I}_1$ and $\mathcal{I}_2$ in Eq.~(\ref{zeroth_collision})}
\label{app_I_1}

\begin{equation}
\mathcal{I}_1
=
J_{0\mbk}^a 
\left( S_{E^2 \mbk}^{(0)} \left[D_{E^2 \mbk}^{\prime (0)} [ I_{E \mbk}^{(0)}]
\right] \right)
+
J_{0\mbk}^a 
\left( S_{E^2 \mbk}^{(0)} \left[I_{E^2 \mbk}^{\prime (0)} \right] \right)
+
J_{E\mbk}^a 
\left( S_{E \mbk}^{(0)} \left[I_{E \mbk}^{(0)} \right] \right)
+
J_{0\mbk}^a 
\left( S_{E^2 \mbk}^{(0)} \left[I_{E^2 \mbk}^{\dprime (0)} [ D_{E^2 \mbk}^{(-1)}]
\right] \right),
\end{equation}
\begin{equation}
\mathcal{I}_2
=
J_{0\mbk}^a 
\left( S_{E^2 \mbk}^{(0)} \left[D_{E^2 \mbk}^{\prime (0)} [ D_{E \mbk}^{(0)}]
\right] \right)
+
J_{E\mbk}^a \left( S_{E \mbk}^{(0)} \left[ D_{E \mbk}^{(0)} \right] \right)
+
J_{E^2 \mbk}^a 
\left( f_{0 \mbk} \right).
\end{equation}

\section{Carrier densities}
\label{App_carrier_densities}

\subsection{Linear-response densities}

The linear-response carrier densities are obtained by solving Eq.~(\ref{linear_carrier_eq}). As outlined above, this yields~\cite{atencia2022semiclassical}
\begin{equation}
\label{n_Ek^sk}
\begin{split}
&n_{E\mbk a}^{(\text{sk})}
=
-\frac{2\pi^2}{\hbar} \tau_{\mbk}^a
\sum_{\mbkpr \mbkdpr bcd}
\delta(\veps_{\mbk}^a 
-
\veps_{\mbkpr}^b)
\\
&\times
\text{Im} \Biggl\{
\frac{
\Braket{U_{\mbk \mbk\pr}^{ab} U_{\mbkpr \mbk}^{bc}}
\Braket{U_{\mbk \mbkdpr}^{cd} U_{\mbkdpr \mbk}^{da}}} 
{\veps_{\mbk}^a - \veps_{\mbk}^c} 
\left[
\left( n_{E\mbk a}^{(-1)} - n_{E\mbkdpr d}^{(-1)}\right) \delta(\veps_{\mbk}^a 
-
\veps_{\mbkdpr}^d)
+
\left( n_{E\mbk c}^{(-1)} - n_{E\mbkdpr d}^{(-1)}\right) \delta(\veps_{\mbk}^c 
-
\veps_{\mbkdpr}^d)
\right]_{c \neq a}
\\
&+
\frac{
\Braket{U_{\mbk \mbk\pr}^{ab} U_{\mbkpr \mbk}^{ca}}
\Braket{U_{\mbkpr \mbkdpr}^{bd} U_{\mbkdpr \mbkpr}^{dc}}} 
{\veps_{\mbkpr}^b - \veps_{\mbkpr}^c}
\left[
\left( n_{E\mbkpr b}^{(-1)} - n_{E\mbkdpr d}^{(-1)}\right) \delta(\veps_{\mbkpr}^b 
-
\veps_{\mbkdpr}^d)
+
\left( n_{E\mbkpr c}^{(-1)} - n_{E\mbkdpr d}^{(-1)}\right) \delta(\veps_{\mbkpr}^c 
-
\veps_{\mbkdpr}^d)
\right]_{c \neq b} \biggr\} ,
\end{split}
\end{equation}
and
\begin{equation}
\label{n_Ek^sj}
n_{E\mbk a}^{(\text{sj})}
=
- \frac{2\pi}{\hbar} \tau_{\mbk}^a
\frac{\pd f_{0\mbk}^a}{\pd \veps_{\mbk}^a}
e \mathbf{E} \cdot \sum_{\mbkpr b}
\delta(\veps_{\mbk}^a 
-
\veps_{\mbkpr}^b)
\Bigl\{
\text{Im}
\Braket{
\left[
\left( \bs{\pd}_{\mbk} + \bs{\pd}_{\mbkpr} \right)
U_{\mbk \mbk\pr}^{ab}
\right] U_{\mbk\pr \mbk}^{ba}}
-
\Braket{
U_{\mbk \mbk\pr}^{ab} U_{\mbk\pr \mbk}^{ba}}
\left( \bs{\mathcal{A}}_{\mbk}^a
-
\bs{\mathcal{A}}_{\mbkpr}^b\right)
\Bigr\}.
\end{equation}

The various quadratic-response carrier densities are considerably more numerous. However, they are evaluated in a fairly similar manner, the explicit forms of which are presented below.

\subsection{Mixed scattering}

The contribution from $n_{E^2 \mbk a}^{\text{(sj,o)}}$ is expressed as
\begin{equation}
\label{n_E2k_sjo}
n_{E^2 \mbk a}^{\text{(sj,o)}}
=
\mathcal{N}^{(\text{sj,o})}_{1,\mbk a}
+
\mathcal{N}^{(\text{sj,o})}_{2,\mbk a},
\end{equation}
with
\begin{equation}
\begin{split}
\mathcal{N}^{(\text{sj,o})}_{1,\mbk a}=&
- \frac{2 \pi}{\hbar^2}  e E^{\mu} \tau_{\mbk}^a
\sum_{\mbkpr b}
\delta ( \veps_{\mbk}^a - \veps_{\mbkpr}^b )
\pd_{\mbkpr}^{\rho}
\left[
\frac{v_{\rho \mbkpr}^b}{\lvert \bs{v}_{\mbkpr}^b \rvert^2}
\text{Im}
\Braket{
U_{\mbk \mbk\pr}^{ab} 
\left[
\left( \pd_{\mbk}^{\mu} + \pd_{\mbkpr}^{\mu} \right) U_{\mbk\pr \mbk} ^{ba} 
\right]
}
\left(
n_{E \mbk a}^{(-1)} - n_{E \mbkpr b}^{(-1)} 
\right)
\right]
\\
&-
\frac{2 \pi}{\hbar^2}  e E^{\mu} \tau_{\mbk}^a
\sum_{\mbkpr b}
\delta ( \veps_{\mbk}^a - \veps_{\mbkpr}^b )
\pd_{\mbkpr}^{\rho}
\left[
\frac{v_{\rho \mbkpr}^b}{\lvert \bs{v}_{\mbkpr}^b \rvert^2}
\left(
\mathcal{A}_{\mu \mbk}^a - \mathcal{A}_{\mu \mbkpr}^b 
\right)
\Braket{
U_{\mbk \mbk\pr}^{ab} U_{\mbk\pr \mbk} ^{ba} 
}
\left(
n_{E \mbk a}^{(-1)} - n_{E \mbkpr b}^{(-1)} 
\right)
\right],
\end{split}
\end{equation}
\begin{equation}
\begin{split}
\mathcal{N}^{(\text{sj,o})}_{2,\mbk a}
=&
- \frac{4 \pi}{\hbar}  e E^{\mu} \tau_{\mbk}^a
\sum_{\mbkpr bc}
\delta ( \veps_{\mbk}^a - \veps_{\mbkpr}^b )
\\
& \times 
\left(
\text{Re}
\left[
\Braket{
U_{\mbk \mbk\pr}^{ab} U_{\mbk\pr \mbk} ^{bc}
}
\mathcal{A}_{\mu \mbk}^{\prime ca}
\right]
\frac{
n_{E \mbk a}^{(-1)} - n_{E \mbkpr b}^{(-1)}
}
{\veps_{\mbk}^a - \veps_{\mbk}^c}
+
\text{Re}
\left[
\Braket{
U_{\mbk \mbk\pr}^{ab} U_{\mbk\pr \mbk} ^{ca}
}
\mathcal{A}_{\mu \mbkpr}^{\prime bc}
\right]
\frac{
n_{E \mbk a}^{(-1)} - n_{E \mbkpr b}^{(-1)}
}
{\veps_{\mbk}^a - \veps_{\mbkpr}^c}
\right),
\end{split}
\end{equation}
where the velocity factors here and elsewhere arise from application of Eq.~(\ref{dirac_derivative}). The remaining mixed-scattering terms read
\begin{equation}
\label{n_E2_sko}
\begin{split}
n_{E^2 \mbk a}^{(\text{sk,o})}
=&
- \frac{2\pi^2}{\hbar} \tau_{\mbk}^a
\sum_{\mbkpr \mbkdpr bcd}
\delta(\veps_{\mbk}^a 
-
\veps_{\mbkpr}^b)
\Biggl\{
\frac{
\text{Im}
\left[
\Braket{U_{\mbk \mbk\pr}^{ab} U_{\mbkpr \mbk}^{bc}}
\Braket{U_{\mbk \mbkdpr}^{cd} U_{\mbkdpr \mbk}^{da}}
\right]
} 
{\veps_{\mbk}^a - \veps_{\mbk}^c}
\\
& \times
\left[
( n_{E^2 \mbk a}^{(-2)} - n_{E^2 \mbkdpr d}^{(-2)}) 
\delta(\veps_{\mbk}^a -\veps_{\mbkdpr}^d)
+
( n_{E^2\mbk c}^{(-2)} - n_{E^2\mbkdpr d}^{(-2)}) 
\delta(\veps_{\mbk}^c -\veps_{\mbkdpr}^d)
\right]_{c \neq a}
\\
&+ 
\frac{
\text{Im}
\left[
\Braket{U_{\mbk \mbk\pr}^{ab} U_{\mbkpr \mbk}^{ca}}
\Braket{U_{\mbkpr \mbkdpr}^{bd} U_{\mbkdpr \mbkpr}^{dc}}
\right]
} 
{\veps_{\mbk}^a - \veps_{\mbkpr}^c}
\\
& \times
\left[
( n_{E^2\mbkpr b}^{(-2)} - n_{E^2\mbkdpr d}^{(-2)})
\delta(\veps_{\mbkpr}^b -\veps_{\mbkdpr}^d)
+
( n_{E^2\mbkpr c}^{(-2)} - n_{E^2\mbkdpr d}^{(-2)}) 
\delta(\veps_{\mbkpr}^c - \veps_{\mbkdpr}^d)
\right]_{c \neq b} \biggr\} ,
\end{split}
\end{equation}

\begin{subequations}
\begin{align}
n_{E^2 \mbk a}^{(\text{o,sj})}
&=
\frac{e}{\hbar} \tau_{\mbk}^a
\mathbf{E} \cdot 
\bs{\pd}_{\mbk} n_{E \mbk a}^{(\text{sj})},
\\
n_{E^2 \mbk a}^{(\text{o,sk})}
&=
\frac{e}{\hbar} \tau_{\mbk}^a 
\mathbf{E} \cdot 
\bs{\pd}_{\mbk} n_{E \mbk a}^{(\text{sk})}.
\end{align}
\end{subequations}

\subsection{Secondary side jump}

The secondary side jump density is given by
\begin{equation}
n_{E^2 \mbk a}^{\text{(ssj)}}
=
\mathcal{N}^{(\text{ssj})}_{1,\mbk a}
+
\mathcal{N}^{(\text{ssj})}_{2,\mbk a},
\end{equation}
where

\begin{equation}
\begin{split}
\mathcal{N}^{(\text{ssj})}_{1,\mbk a}
=&
- \frac{2 \pi}{\hbar^2}  e E^{\mu} \tau_{\mbk}^a
\sum_{\mbkpr b}
\delta ( \veps_{\mbk}^a - \veps_{\mbkpr}^b )
\pd_{\mbkpr}^{\rho}
\left[
\frac{v_{\rho \mbkpr}^b}{\lvert \bs{v}_{\mbkpr}^b \rvert^2}
\text{Im}
\Braket{
U_{\mbk \mbk\pr}^{ab} 
\left[
\left( \pd_{\mbk}^{\mu} + \pd_{\mbkpr}^{\mu} \right) U_{\mbk\pr \mbk} ^{ba} 
\right]
}
\left(
n_{E \mbk a}^{(0)} - n_{E \mbkpr b}^{(0)} 
\right)
\right]
\\
&-
\frac{2 \pi}{\hbar^2}  e E^{\mu} \tau_{\mbk}^a
\sum_{\mbkpr b}
\delta ( \veps_{\mbk}^a - \veps_{\mbkpr}^b )
\pd_{\mbkpr}^{\rho}
\left[
\frac{v_{\rho \mbkpr}^b}{\lvert \bs{v}_{\mbkpr}^b \rvert^2}
\left(
\mathcal{A}_{\mu \mbk}^a - \mathcal{A}_{\mu \mbkpr}^b 
\right)
\Braket{
U_{\mbk \mbk\pr}^{ab} U_{\mbk\pr \mbk} ^{ba} 
}
\left(
n_{E \mbk a}^{(0)} - n_{E \mbkpr b}^{(0)} 
\right)
\right],
\end{split}
\end{equation}

\begin{equation}
\begin{split}
\mathcal{N}^{(\text{ssj})}_{2,\mbk a}
=&
- \frac{2 \pi}{\hbar}  e E^{\mu} \tau_{\mbk}^a
\sum_{\mbkpr bc}
\frac{
\text{Re}
\left[
\Braket{
U_{\mbk \mbk\pr}^{ab} U_{\mbk\pr \mbk} ^{bc}
}
\mathcal{A}_{\mu \mbk}^{\prime ca}
\right]
}
{\veps_{\mbk}^a - \veps_{\mbk}^c}
\\
& \times
\left[
\left(
n_{E \mbk a}^{(0)} - n_{E \mbkpr b}^{(0)}
\right)
\delta ( \veps_{\mbk}^a - \veps_{\mbkpr}^b ) 
-
\left(
n_{E \mbk c}^{(0)} - n_{E \mbkpr b}^{(0)}
\right)
\delta ( \veps_{\mbk}^c - \veps_{\mbkpr}^b ) 
\right]
\\
&-
\frac{2 \pi}{\hbar}  e E^{\mu} \tau_{\mbk}^a
\sum_{\mbkpr bc}
\frac{
\text{Re}
\left[
\Braket{
U_{\mbk \mbk\pr}^{ab} U_{\mbk\pr \mbk} ^{ca}
}
\mathcal{A}_{\mu \mbkpr}^{\prime bc}
\right]
}
{\veps_{\mbkpr}^b - \veps_{\mbkpr}^c}
\\
& \times
\left[
\left(
n_{E \mbk a}^{(0)} - n_{E \mbkpr b}^{(0)}
\right)
\delta ( \veps_{\mbk}^a - \veps_{\mbkpr}^b ) 
-
\left(
n_{E \mbk a}^{(0)} - n_{E \mbkpr c}^{(0)}
\right)
\delta ( \veps_{\mbk}^a - \veps_{\mbkpr}^c ) 
\right].
\end{split}
\end{equation}

\subsection{Tertiary skew scattering}

Similar to the previous skew scattering expressions, the tertiary skew scattering density reads
\begin{equation}
\begin{split}
n_{E^2 \mbk a}^{(\text{tsk})}
=&
- \frac{2\pi^2}{\hbar} \tau_{\mbk}^a
\sum_{\mbkpr \mbkdpr bcd}
\delta(\veps_{\mbk}^a 
-
\veps_{\mbkpr}^b)
\Biggl\{
\frac{
\text{Im}
\left[
\Braket{U_{\mbk \mbk\pr}^{ab} U_{\mbkpr \mbk}^{bc}}
\Braket{U_{\mbk \mbkdpr}^{cd} U_{\mbkdpr \mbk}^{da}}
\right]
} 
{\veps_{\mbk}^a - \veps_{\mbk}^c}
\\
&\times
\left[
( n_{E^2 \mbk a}^{(-1)} - n_{E^2 \mbkdpr d}^{(-1)}) 
\delta(\veps_{\mbk}^a -\veps_{\mbkdpr}^d)
+
( n_{E^2\mbk c}^{(-1)} - n_{E^2\mbkdpr d}^{(-1)}) 
\delta(\veps_{\mbk}^c -\veps_{\mbkdpr}^d)
\right]_{c \neq a}
\\
&+ 
\frac{
\text{Im}
\left[
\Braket{U_{\mbk \mbk\pr}^{ab} U_{\mbkpr \mbk}^{ca}}
\Braket{U_{\mbkpr \mbkdpr}^{bd} U_{\mbkdpr \mbkpr}^{dc}}
\right]
} 
{\veps_{\mbk}^a - \veps_{\mbkpr}^c}
\\
&\times
\left[
( n_{E^2\mbkpr b}^{(-1)} - n_{E^2\mbkdpr d}^{(-1)})
\delta(\veps_{\mbkpr}^b -\veps_{\mbkdpr}^d)
+
( n_{E^2\mbkpr c}^{(-1)} - n_{E^2\mbkdpr d}^{(-1)}) 
\delta(\veps_{\mbkpr}^c - \veps_{\mbkdpr}^d)
\right]_{c \neq b} \biggr\}.
\end{split}
\end{equation}

\subsection{Quadratic side jump}

The quadratic side jump density is rather lengthy, as it is comprised of the three collision integrals given by Eq.~(\ref{n_E2k_qsj}). After a number of cancellations, the remaining terms are expressed as
\begin{equation}
n_{E^2\mbk a}^{(\text{qsj})}
=
\sum_{i=1}^{8}
\mathcal{N}^{(\text{qsj})}_{i,\mbk a},
\end{equation}
where
\begin{equation}
\begin{split}
\mathcal{N}^{(\text{qsj})}_{1,\mbk a}
=&
- \frac{\pi}{4} \hbar e^2 \tau_{\mbk}^a
\frac{\pd^4 f_{0\mbk}^a}{\pd(\veps_{\mbk}^a)^4}
\sum_{\mbkpr b}
\delta ( \veps_{\mbk}^a - \veps_{\mbkpr}^b )
\Braket{U_{\mbk \mbk\pr}^{ab} U_{\mbk\pr \mbk}^{ba}}
\left[
\mathbf{E} \cdot \left(\bs{v}_{\mbk}^a + \bs{v}_{\mbkpr}^b \right)
\right]^2
\\
&-
\pi \hbar e^2  E^{\mu} E^{\nu} \tau_{\mbk}^a
\frac{\pd^3 f_{0\mbk}^a}{\pd(\veps_{\mbk}^a)^4}
\sum_{\mbkpr b}
\delta ( \veps_{\mbk}^a - \veps_{\mbkpr}^b )
\biggl\{
\pd_{\mbkpr}^{\rho} 
\left[ \frac{v_{\rho \mbkpr}^b}{\lvert \bs{v}_{\mbkpr}^b \rvert^2} 
\Braket{U_{\mbk \mbk\pr}^{ab} U_{\mbk\pr \mbk}^{ba}}
v_{\mu \mbk}^a \left(v_{\nu \mbk}^a + v_{\nu \mbkpr}^b \right)
\right]
\\
&+ 
\frac{1}{3} \left( \pd_{\mbk}^{\mu} v_{\nu \mbk}^a
+
2 \pd_{\mbkpr}^{\mu} v_{\nu \mbkpr}^b \right)
\Braket{U_{\mbk \mbk\pr}^{ab} U_{\mbk\pr \mbk}^{ba}}
+
\left(v_{\mu \mbk}^a + v_{\mu \mbkpr}^b \right)
\text{Re}
\Braket{
U_{\mbk \mbk\pr}^{ab} 
\left[
\left( \pd_{\mbk}^{\nu} + \pd_{\mbkpr}^{\nu} \right) U_{\mbk\pr \mbk} ^{ba} 
\right]
}
\biggr\},
\end{split}
\end{equation}
\begin{equation}
\begin{split}
\mathcal{N}^{(\text{qsj})}_{2,\mbk a}
=&
- \frac{\pi}{\hbar}  e^2 E^{\mu} E^{\nu}\tau_{\mbk}^a
\frac{\pd^2 f_{0\mbk}^a}{\pd(\veps_{\mbk}^a)^2}
\sum_{\mbkpr b}
\delta ( \veps_{\mbk}^a - \veps_{\mbkpr}^b )
\\
&\times
\biggl\{
\frac{1}{2} \pd_{\mbkpr}^{\rho}
\left(
\frac{v_{\rho \mbkpr}^b}{\lvert \bs{v}_{\mbkpr}^b \rvert^2}
\pd_{\mbkpr}^{\sigma}
\left[
\frac{v_{\sigma \mbkpr}^b}{\lvert \bs{v}_{\mbkpr}^b \rvert^2}
\left(
v_{\mu \mbk}^a v_{\nu \mbk}^a
-
v_{\mu \mbkpr}^b v_{\nu \mbkpr}^b
\right)
\Braket{U_{\mbk \mbk\pr}^{ab} U_{\mbk\pr \mbk}^{ba}}
\right]
\right)
\\
&+
\pd_{\mbkpr}^{\rho}
\left[
\frac{v_{\rho \mbkpr}^b}{\lvert \bs{v}_{\mbkpr}^b \rvert^2}
\left(
\pd_{\mbk}^{\mu} v_{\nu \mbk}^a
+
\pd_{\mbkpr}^{\mu} v_{\nu \mbkpr}^b
\right)
\Braket{U_{\mbk \mbk\pr}^{ab} U_{\mbk\pr \mbk}^{ba}}
\right]
\\
&+
\pd_{\mbkpr}^{\rho}
\left[
\frac{v_{\rho \mbkpr}^b}{\lvert \bs{v}_{\mbkpr}^b \rvert^2}
\left(
3 v_{\mu \mbk}^a
+
v_{\mu \mbkpr}^b
\right)
\text{Re}
\Braket{
U_{\mbk \mbk\pr}^{ab} 
\left[
\left( \pd_{\mbk}^{\nu} + \pd_{\mbkpr}^{\nu} \right) U_{\mbk\pr \mbk} ^{ba} 
\right]
}
\right]
\\
&+
\text{Re}
\Braket{
U_{\mbk \mbk\pr}^{ab} 
\left[
\left( \pd_{\mbk}^{\mu} + \pd_{\mbkpr}^{\mu} \right)
\left( \pd_{\mbk}^{\nu} + \pd_{\mbkpr}^{\nu} \right) U_{\mbk\pr \mbk} ^{ba} 
\right]
}
\biggr\},
\end{split}
\end{equation}
\begin{equation}
\begin{split}
\mathcal{N}^{(\text{qsj})}_{3,\mbk a}
=&
- \frac{\pi}{\hbar^2}  e^2 E^{\mu} E^{\nu}\tau_{\mbk}^a
\frac{\pd f_{0\mbk}^a}{\pd \veps_{\mbk}^a}
\sum_{\mbkpr b}
\delta ( \veps_{\mbk}^a - \veps_{\mbkpr}^b )
\\
&\times
\biggl\{
\pd_{\mbkpr}^{\rho}
\left(
\frac{v_{\rho \mbkpr}^b}{\lvert \bs{v}_{\mbkpr}^b \rvert^2}
\pd_{\mbkpr}^{\sigma}
\left[
\frac{v_{\sigma \mbkpr}^b}{\lvert \bs{v}_{\mbkpr}^b \rvert^2}
\left(
v_{\mu \mbk}^a - v_{\mu \mbkpr}^b 
\right)
\text{Re}
\Braket{
U_{\mbk \mbk\pr}^{ab} 
\left[
\left( \pd_{\mbk}^{\nu} + \pd_{\mbkpr}^{\nu} \right) U_{\mbk\pr \mbk} ^{ba} 
\right]
}
\right]
\right)
\\
&+
2 \pd_{\mbkpr}^{\rho}
\left(
\frac{v_{\rho \mbkpr}^b}{\lvert \bs{v}_{\mbkpr}^b \rvert^2}
\text{Re}
\Braket{
U_{\mbk \mbk\pr}^{ab} 
\left[
\left( \pd_{\mbk}^{\mu} + \pd_{\mbkpr}^{\mu} \right)
\left( \pd_{\mbk}^{\nu} + \pd_{\mbkpr}^{\nu} \right) U_{\mbk\pr \mbk} ^{ba} 
\right]
}
\right)
\biggr\},
\end{split}
\end{equation}
\begin{equation}
\begin{split}
\mathcal{N}^{(\text{qsj})}_{4,\mbk a}
=&
\frac{4 \pi}{\hbar^2}  e^2 E^{\mu} E^{\nu}\tau_{\mbk}^a
\frac{\pd f_{0\mbk}^a}{\pd \veps_{\mbk}^a}
\sum_{\mbkpr b}
\delta ( \veps_{\mbk}^a - \veps_{\mbkpr}^b )
\\
&\times
\biggl\{
\pd_{\mbkpr}^{\rho}
\left[
\frac{v_{\rho \mbkpr}^b}{\lvert \bs{v}_{\mbkpr}^b \rvert^2}
\left(
\mathcal{A}_{\mu \mbk}^a - \mathcal{A}_{\mu \mbkpr}^b 
\right)
\text{Im}
\Braket{
U_{\mbk \mbk\pr}^{ab} 
\left[
\left( \pd_{\mbk}^{\nu} + \pd_{\mbkpr}^{\nu} \right) U_{\mbk\pr \mbk} ^{ba} 
\right]
}
\right]
\\
&+
\frac{1}{2} \pd_{\mbkpr}^{\rho}
\left(
\frac{v_{\rho \mbkpr}^b}{\lvert \bs{v}_{\mbkpr}^b \rvert^2}
\left(
\mathcal{A}_{\mu \mbk}^a - \mathcal{A}_{\mu \mbkpr}^b 
\right)
\left(
\mathcal{A}_{\nu \mbk}^a - \mathcal{A}_{\nu \mbkpr}^b 
\right)
\Braket{U_{\mbk \mbk\pr}^{ab} U_{\mbk\pr \mbk}^{ba}}
\right)
\biggr\}
\\
&+
\frac{2 \pi}{\hbar}  e^2 E^{\mu} E^{\nu}\tau_{\mbk}^a
\frac{\pd^2 f_{0\mbk}^a}{\pd (\veps_{\mbk}^a)^2}
\sum_{\mbkpr b}
\delta ( \veps_{\mbk}^a - \veps_{\mbkpr}^b )
\biggl\{
\left(
\mathcal{A}_{\mu \mbk}^a - \mathcal{A}_{\mu \mbkpr}^b 
\right)
\text{Im}
\Braket{
U_{\mbk \mbk\pr}^{ab} 
\left[
\left( \pd_{\mbk}^{\nu} + \pd_{\mbkpr}^{\nu} \right) U_{\mbk\pr \mbk} ^{ba} 
\right]
}
\\
&+
\frac{1}{2}
\left(
\mathcal{A}_{\mu \mbk}^a - \mathcal{A}_{\mu \mbkpr}^b 
\right)
\left(
\mathcal{A}_{\nu \mbk}^a - \mathcal{A}_{\nu \mbkpr}^b 
\right)
\Braket{U_{\mbk \mbk\pr}^{ab} U_{\mbk\pr \mbk}^{ba}}
\biggr\},
\end{split}
\end{equation}
\begin{equation}
\begin{split}
\mathcal{N}^{(\text{qsj})}_{5,\mbk a}
=&
\pi  e^2 E^{\mu} E^{\nu} \tau_{\mbk}^a
\sum_{\mbkpr b c}
\frac{\pd^2 f_{0\mbkpr}^b}{\pd (\veps_{\mbkpr}^b)^2}
\frac{
\text{Im}
\left[
\Braket{U_{\mbk \mbk\pr}^{ab} U_{\mbk\pr \mbk}^{bc}} \mathcal{A}^{\prime ca}_{\mu \mbk}
\right]
}
{\veps_{\mbk}^a - \veps_{\mbk}^c}
\\
&\times
\left[
\left(
v_{\mu \mbk}^a +v_{\mu \mbkpr}^b 
\right)
\delta ( \veps_{\mbk}^a - \veps_{\mbkpr}^b )
-
\left(
v_{\mu \mbk}^c +v_{\mu \mbkpr}^b 
\right)
\delta ( \veps_{\mbk}^c - \veps_{\mbkpr}^b )
\right]
\\
&+
\frac{2 \pi}{\hbar}  e^2 E^{\mu} E^{\nu} \tau_{\mbk}^a
\frac{\pd f_{0\mbk}^a}{\pd \veps_{\mbk}^a}
\sum_{\mbkpr b c}
\delta ( \veps_{\mbk}^a - \veps_{\mbkpr}^b )
\frac{
\text{Im}
\Braket{
U_{\mbk \mbk\pr}^{ab} 
\left[
\left( \pd_{\mbk}^{\nu} + \pd_{\mbkpr}^{\nu} \right) \left( U_{\mbk\pr \mbk} ^{bc} 
\mathcal{A}^{\prime ca}_{\mu \mbk} \right)
\right]
} 
}
{\veps_{\mbk}^a - \veps_{\mbk}^c}
\\
&+
2 \pi  e^2 E^{\mu} E^{\nu} \tau_{\mbk}^a
\sum_{\mbkpr b c}
\delta ( \veps_{\mbk}^a - \veps_{\mbkpr}^b )
\frac{
\text{Im}
\left[
\Braket{U_{\mbk \mbk\pr}^{ab} U_{\mbk\pr \mbk}^{bc}} \mathcal{A}^{\prime ca}_{\mu \mbk}
\right]
}
{\left( \veps_{\mbk}^a - \veps_{\mbk}^c \right)^2}
v_{\nu \mbk}^c
\left(
\frac{\pd f_{0\mbk}^a}{\pd \veps_{\mbk}^a}
-
\frac{\pd f_{0\mbk}^c}{\pd \veps_{\mbk}^c}\right)
\\
&-
\frac{2 \pi}{\hbar}  e^2 E^{\mu} E^{\nu} \tau_{\mbk}^a
\sum_{\mbkpr b c}
\delta ( \veps_{\mbk}^c- \veps_{\mbkpr}^b )
\frac{\pd f_{0\mbk}^c}{\pd \veps_{\mbk}^c}
\frac{
\text{Im}
\Braket{
U_{\mbk \mbk\pr}^{ab} 
\left[
\left( \pd_{\mbk}^{\nu} + \pd_{\mbkpr}^{\nu} \right) \left( U_{\mbk\pr \mbk} ^{bc} 
\right)
\right]
} \mathcal{A}^{\prime ca}_{\mu \mbk} 
}
{\veps_{\mbk}^a - \veps_{\mbk}^c},
\end{split}
\end{equation}
\begin{equation}
\begin{split}
\mathcal{N}^{(\text{qsj})}_{6,\mbk a}
=&
\frac{2 \pi}{\hbar}  e^2 E^{\mu} E^{\nu} \tau_{\mbk}^a
\frac{\pd f_{0\mbk}^a}{\pd \veps_{\mbk}^a}
\sum_{\mbkpr b c}
\delta ( \veps_{\mbk}^a - \veps_{\mbkpr}^b )
\frac{
\text{Im}
\Braket{
U_{\mbk \mbk\pr}^{ab} 
\left[
\left( \pd_{\mbk}^{\nu} + \pd_{\mbkpr}^{\nu} \right) \left( U_{\mbk\pr \mbk} ^{ca} 
\mathcal{A}^{\prime bc}_{\mu \mbkpr} \right)
\right]
} 
}
{\veps_{\mbk}^a - \veps_{\mbkpr}^c}
\\
&+
2 \pi  e^2 E^{\mu} E^{\nu} \tau_{\mbk}^a
\frac{\pd f_{0\mbk}^a}{\pd \veps_{\mbk}^a}
\sum_{\mbkpr b c}
\delta ( \veps_{\mbk}^a - \veps_{\mbkpr}^b )
\frac{
\text{Im}
\left[
\Braket{U_{\mbk \mbk\pr}^{ab} U_{\mbk\pr \mbk}^{ca}} \mathcal{A}^{\prime bc}_{\mu \mbkpr}
\right]
}
{\left( \veps_{\mbk}^a - \veps_{\mbkpr}^c \right)^2}
\left[
v_{\nu \mbkpr}^c
-
\frac{\bs{v}_{\mbkpr}^b \cdot \bs{v}_{\mbkpr}^c}
{\lvert \bs{v}_{\mbkpr}^b \rvert^2}
\left(
v_{\nu \mbk}^a - v_{\nu \mbkpr}^b
\right)
\right]
\\
&-
2 \pi  e^2 E^{\mu} E^{\nu} \tau_{\mbk}^a
\sum_{\mbkpr b c}
\delta ( \veps_{\mbk}^a - \veps_{\mbkpr}^b )
\frac{\pd f_{0\mbkpr}^c}{\pd \veps_{\mbkpr}^c}
\frac{
\text{Im}
\left[
\Braket{U_{\mbk \mbk\pr}^{ab} U_{\mbk\pr \mbk}^{ca}} \mathcal{A}^{\prime bc}_{\mu \mbkpr}
\right]
}
{\left( \veps_{\mbk}^a - \veps_{\mbkpr}^c \right)^2}
\left[
v_{\nu \mbkpr}^c
+
\frac{\bs{v}_{\mbkpr}^b \cdot \bs{v}_{\mbkpr}^c}
{\lvert \bs{v}_{\mbkpr}^b \rvert^2}
\left(
v_{\nu \mbk}^a - v_{\nu \mbkpr}^b
\right)
\right]
\\
&+
4 \pi  e^2 E^{\mu} E^{\nu} \tau_{\mbk}^a
\sum_{\mbkpr b c}
\delta ( \veps_{\mbk}^a - \veps_{\mbkpr}^b )
\text{Im}
\left[
\Braket{U_{\mbk \mbk\pr}^{ab} U_{\mbk\pr \mbk}^{ca}} \mathcal{A}^{\prime bc}_{\mu \mbkpr}
\right]
\frac{\bs{v}_{\mbkpr}^b \cdot \bs{v}_{\mbkpr}^c}
{\lvert \bs{v}_{\mbkpr}^b \rvert^2}
\left(
v_{\nu \mbk}^a - v_{\nu \mbkpr}^b 
\right)
\frac{
f_{0 \mbk}^a - f_{0 \mbkpr}^c
}
{\left( \veps_{\mbk}^a - \veps_{\mbkpr}^c \right)^3 } 
\\
&+
\frac{2 \pi}{\hbar}  e^2 E^{\mu} E^{\nu} \tau_{\mbk}^a
\frac{\pd f_{0\mbk}^a}{\pd \veps_{\mbk}^a}
\sum_{\mbkpr b c}
\delta ( \veps_{\mbk}^a- \veps_{\mbkpr}^c )
\frac{
\text{Im}
\Braket{
U_{\mbk \mbk\pr}^{ab} 
\left[
\left( \pd_{\mbk}^{\nu} + \pd_{\mbkpr}^{\nu} \right) \left( U_{\mbk\pr \mbk} ^{ca} 
\right)
\right]
} \mathcal{A}^{\prime bc}_{\mu \mbkpr} 
}
{\veps_{\mbk}^a - \veps_{\mbkpr}^b},
\end{split}
\end{equation}
\begin{equation}
\begin{split}
\mathcal{N}^{(\text{qsj})}_{7,\mbk a}
=&
\frac{2 \pi}{\hbar}  e^2 E^{\mu} E^{\nu} \tau_{\mbk}^a
\frac{\pd f_{0\mbk}^a}{\pd \veps_{\mbk}^a}
\sum_{\mbkpr b c}
\delta ( \veps_{\mbk}^a- \veps_{\mbkpr}^b )
\left(
\mathcal{A}_{\nu \mbk}^a - \mathcal{A}_{\nu \mbkpr}^b 
\right)
\frac{
\text{Im}
\Braket{
U_{\mbk \mbk\pr}^{ab} U_{\mbk\pr \mbk} ^{bc} 
} \mathcal{A}^{\prime ca}_{\mu \mbk} 
}
{\veps_{\mbk}^a - \veps_{\mbk}^c}
\\
&-
\frac{2 \pi}{\hbar}  e^2 E^{\mu} E^{\nu} \tau_{\mbk}^a
\sum_{\mbkpr b c}
\delta ( \veps_{\mbk}^c- \veps_{\mbkpr}^b )
\frac{\pd f_{0\mbk}^c}{\pd \veps_{\mbk}^c}
\left(
\mathcal{A}_{\nu \mbk}^c - \mathcal{A}_{\nu \mbkpr}^b 
\right)
\frac{
\text{Im}
\Braket{
U_{\mbk \mbk\pr}^{ab} U_{\mbk\pr \mbk} ^{bc} 
} \mathcal{A}^{\prime ca}_{\mu \mbk} 
}
{\veps_{\mbk}^a - \veps_{\mbk}^c}
\\
&+
\frac{4 \pi}{\hbar}  e^2 E^{\mu} E^{\nu} \tau_{\mbk}^a
\frac{\pd f_{0\mbk}^a}{\pd \veps_{\mbk}^a}
\sum_{\mbkpr b c}
\delta ( \veps_{\mbk}^a- \veps_{\mbkpr}^b )
\left(
\mathcal{A}_{\nu \mbk}^a - \mathcal{A}_{\nu \mbkpr}^b 
\right)
\frac{
\text{Im}
\Braket{
U_{\mbk \mbk\pr}^{ab} U_{\mbk\pr \mbk} ^{ca} 
} \mathcal{A}^{\prime bc}_{\mu \mbkpr} 
}
{\veps_{\mbk}^a - \veps_{\mbkpr}^c},
\end{split}
\end{equation}
\begin{equation}
\label{N_qsj_8}
\begin{split}
\mathcal{N}^{(\text{qsj})}_{8,\mbk a}
=&
\frac{2 \pi}{\hbar}  e^2 E^{\mu} E^{\nu} \tau_{\mbk}^a
\sum_{\mbkpr b c}
\delta ( \veps_{\mbk}^a- \veps_{\mbkpr}^b )
\Braket{
U_{\mbk \mbk\pr}^{ab} U_{\mbk\pr \mbk} ^{ba}}
\\
&\times
\left[
\frac{g_{\mu\nu,\mbk}^{ac}}
{\veps_{\mbk}^a - \veps_{\mbk}^c}
\left(
\frac{\pd f_{0\mbk}^a}{\pd \veps_{\mbk}^a}
-
\frac{f_{0\mbk}^a - f_{0\mbk}^c}
{\veps_{\mbk}^a - \veps_{\mbk}^c}
\right)
-
\frac{g_{\mu\nu,\mbkpr}^{bc}}
{\veps_{\mbk}^a - \veps_{\mbkpr}^c}
\left(
\frac{\pd f_{0\mbk}^a}{\pd \veps_{\mbk}^a}
-
\frac{f_{0\mbk}^a - f_{0\mbkpr}^c}
{\veps_{\mbk}^a - \veps_{\mbkpr}^c}
\right)
\right].
\end{split}
\end{equation}

\subsection{Cubic skew scattering}

The density of cubic skew-scattered carriers arises from the lengthy off-diagonal element given by Eq.~(\ref{S_E2_Ipr_I-1}) and results in the expression
\begin{equation}
n_{E^2\mbk a}^{(\text{csk})}
=
\sum_{i=1}^{8}
\mathcal{N}^{(\text{csk})}_{i,\mbk a},
\end{equation}
with
\begin{equation}
\begin{split}
\mathcal{N}^{(\text{csk})}_{1,\mbk a}
=&
\frac{2\pi^3}{\hbar}
\tau_{\mbk}^a
\sum_{\mbkpr \mbkdpr \mbk_1 bcdef}
\delta(\veps_{\mbk}^a - \veps_{\mbkpr}^b )
\delta(\veps_{\mbk}^a - \veps_{\mbk_1}^d )
\frac{
\text{Re}
\left[
\Braket{U_{\mbk \mbkpr}^{ab} U_{\mbkpr \mbk}^{bc}}
\Braket{U_{\mbk \mbk_1}^{cd} U_{\mbk_1 \mbk}^{de}}
\Braket{U_{\mbk \mbkdpr}^{ef} U_{\mbkdpr \mbk}^{fa}}
\right]_{\substack{c,e \neq a}}
}
{(\veps_{\mbk}^a - \veps_{\mbk}^c)
(\veps_{\mbk}^a - \veps_{\mbk}^e)}
\\
&\times
\left[
( n_{E^2 \mbk a}^{(-2)} - n_{E^2 \mbkdpr f}^{(-2)}) 
\delta(\veps_{\mbk}^a -\veps_{\mbkdpr}^f)
+
( n_{E^2\mbk_1 d}^{(-2)} - n_{E^2\mbkdpr f}^{(-2)}) 
\delta(\veps_{\mbk_1}^d -\veps_{\mbkdpr}^f)
\right],
\end{split}
\end{equation}
\begin{equation}
\begin{split}
\mathcal{N}^{(\text{csk})}_{2,\mbk a}
=&
\frac{2\pi^3}{\hbar}
\tau_{\mbk}^a
\sum_{\mbkpr \mbkdpr \mbk_1 bcdef}
\delta(\veps_{\mbk}^a - \veps_{\mbkpr}^b )
\delta(\veps_{\mbk}^a - \veps_{\mbk_1}^d )
\frac{
\text{Re}
\left[
\Braket{U_{\mbk \mbkpr}^{ab} U_{\mbkpr \mbk}^{bc}}
\Braket{U_{\mbk \mbk_1}^{cd} U_{\mbk_1 \mbk}^{ea}}
\Braket{U_{\mbk_1 \mbkdpr}^{df} U_{\mbkdpr \mbk_1}^{fe}}
\right]_{\substack{c \neq a \\ e \neq d}}
}
{(\veps_{\mbk}^a - \veps_{\mbk}^c)
(\veps_{\mbk_1}^d - \veps_{\mbk_1}^e)}
\\
&\times
\left[
( n_{E^2 \mbk_1 e}^{(-2)} - n_{E^2 \mbkdpr f}^{(-2)}) 
\delta(\veps_{\mbk_1}^e -\veps_{\mbkdpr}^f)
+
( n_{E^2 \mbk_1 d}^{(-2)} - n_{E^2 \mbkdpr f}^{(-2)}) 
\delta(\veps_{\mbk_1}^d -\veps_{\mbkdpr}^f)
\right],
\end{split}
\end{equation}
\begin{equation}
\begin{split}
\mathcal{N}^{(\text{csk})}_{3,\mbk a}
=&
\frac{2\pi^3}{\hbar}
\tau_{\mbk}^a
\sum_{\mbkpr \mbkdpr \mbk_1 bcdef}
\delta(\veps_{\mbk}^a - \veps_{\mbkpr}^b )
\delta(\veps_{\mbk}^c - \veps_{\mbk_1}^d )
\frac{
\text{Re}
\left[
\Braket{U_{\mbk \mbkpr}^{ab} U_{\mbkpr \mbk}^{bc}}
\Braket{U_{\mbk \mbk_1}^{ce} U_{\mbk_1 \mbk}^{da}}
\Braket{U_{\mbk_1 \mbkdpr}^{ef} U_{\mbkdpr \mbk_1}^{fd}}
\right]_{\substack{c \neq a \\ e \neq d}}
}
{(\veps_{\mbk}^a - \veps_{\mbk}^c)
(\veps_{\mbk_1}^e - \veps_{\mbk_1}^d)}
\\
&\times
\left[
( n_{E^2 \mbk_1 e}^{(-2)} - n_{E^2 \mbkdpr f}^{(-2)}) 
\delta(\veps_{\mbk_1}^e -\veps_{\mbkdpr}^f)
+
( n_{E^2\mbk_1 d}^{(-2)} - n_{E^2\mbkdpr f}^{(-2)}) 
\delta(\veps_{\mbk_1}^d -\veps_{\mbkdpr}^f)
\right],
\end{split}
\end{equation}
\begin{equation}
\begin{split}
\mathcal{N}^{(\text{csk})}_{4,\mbk a}
=&
\frac{2\pi^3}{\hbar}
\tau_{\mbk}^a
\sum_{\mbkpr \mbkdpr \mbk_1 bcdef}
\delta(\veps_{\mbk}^a - \veps_{\mbkpr}^b )
\delta(\veps_{\mbk}^c - \veps_{\mbk_1}^d )
\frac{
\text{Re}
\left[
\Braket{U_{\mbk \mbkpr}^{ab} U_{\mbkpr \mbk}^{bc}}
\Braket{U_{\mbk \mbk_1}^{ed} U_{\mbk_1 \mbk}^{da}}
\Braket{U_{\mbk \mbkdpr}^{cf} U_{\mbkdpr \mbk}^{fe}}
\right]_{\substack{a,e \neq c}}
}
{(\veps_{\mbk}^a - \veps_{\mbk}^c)
(\veps_{\mbk}^e - \veps_{\mbk_1}^c)}
\\
&\times
\left[
( n_{E^2 \mbk e}^{(-2)} - n_{E^2 \mbkdpr f}^{(-2)}) 
\delta(\veps_{\mbk}^e -\veps_{\mbkdpr}^f)
+
( n_{E^2 \mbk c}^{(-2)} - n_{E^2 \mbkdpr f}^{(-2)}) 
\delta(\veps_{\mbk}^c -\veps_{\mbkdpr}^f)
\right],
\end{split}
\end{equation}
\begin{equation}
\begin{split}
\mathcal{N}^{(\text{csk})}_{5,\mbk a}
=&
\frac{2\pi^3}{\hbar}
\tau_{\mbk}^a
\sum_{\mbkpr \mbkdpr \mbk_1 bcdef}
\delta(\veps_{\mbk}^a - \veps_{\mbkpr}^b )
\delta(\veps_{\mbkpr}^c - \veps_{\mbk_1}^d )
\frac{
\text{Re}
\left[
\Braket{U_{\mbk \mbkpr}^{ab} U_{\mbkpr \mbk}^{ca}}
\Braket{U_{\mbkpr \mbk_1}^{bd} U_{\mbk_1 \mbkpr}^{de}}
\Braket{U_{\mbkpr \mbkdpr}^{ef} U_{\mbkdpr \mbkpr}^{fc}}
\right]_{\substack{b,e \neq c}}
}
{(\veps_{\mbk}^a - \veps_{\mbkpr}^c)
(\veps_{\mbkpr}^c - \veps_{\mbkpr}^e)}
\\
&\times
\left[
( n_{E^2 \mbkpr c}^{(-2)} - n_{E^2 \mbkdpr f}^{(-2)}) 
\delta(\veps_{\mbkpr}^c -\veps_{\mbkdpr}^f)
+
( n_{E^2\mbkpr e}^{(-2)} - n_{E^2\mbkdpr f}^{(-2)}) 
\delta(\veps_{\mbkpr}^e -\veps_{\mbkdpr}^f)
\right],
\end{split}
\end{equation}
\begin{equation}
\begin{split}
\mathcal{N}^{(\text{csk})}_{6,\mbk a}
=&
\frac{2\pi^3}{\hbar}
\tau_{\mbk}^a
\sum_{\mbkpr \mbkdpr \mbk_1 bcdef}
\delta(\veps_{\mbk}^a - \veps_{\mbkpr}^b )
\delta(\veps_{\mbkpr}^c - \veps_{\mbk_1}^d )
\frac{
\text{Re}
\left[
\Braket{U_{\mbk \mbkpr}^{ab} U_{\mbkpr \mbk}^{ca}}
\Braket{U_{\mbkpr \mbk_1}^{bd} U_{\mbk_1 \mbkpr}^{ec}}
\Braket{U_{\mbk_1 \mbkdpr}^{df} U_{\mbkdpr \mbk_1}^{fe}}
\right]_{\substack{c \neq b \\ e \neq d}}
}
{(\veps_{\mbk}^a - \veps_{\mbkpr}^c)
(\veps_{\mbk_1}^d - \veps_{\mbk_1}^e)}
\\
&\times
\left[
( n_{E^2 \mbk_1 e}^{(-2)} - n_{E^2 \mbkdpr f}^{(-2)}) 
\delta(\veps_{\mbk_1}^e -\veps_{\mbkdpr}^f)
+
( n_{E^2 \mbk_1 d}^{(-2)} - n_{E^2 \mbkdpr f}^{(-2)}) 
\delta(\veps_{\mbk_1}^d -\veps_{\mbkdpr}^f)
\right],
\end{split}
\end{equation}
\begin{equation}
\begin{split}
\mathcal{N}^{(\text{csk})}_{7,\mbk a}
=&
\frac{2\pi^3}{\hbar}
\tau_{\mbk}^a
\sum_{\mbkpr \mbkdpr \mbk_1 bcdef}
\delta(\veps_{\mbk}^a - \veps_{\mbkpr}^b )
\delta(\veps_{\mbkpr}^b - \veps_{\mbk_1}^d )
\frac{
\text{Re}
\left[
\Braket{U_{\mbk \mbkpr}^{ab} U_{\mbkpr \mbk}^{ca}}
\Braket{U_{\mbkpr \mbk_1}^{be} U_{\mbk_1 \mbkpr}^{dc}}
\Braket{U_{\mbk_1 \mbkdpr}^{ef} U_{\mbkdpr \mbk_1}^{fd}}
\right]_{\substack{c \neq b \\ e \neq d}}
}
{(\veps_{\mbk}^a - \veps_{\mbkpr}^c)
(\veps_{\mbk_1}^e - \veps_{\mbk_1}^d)}
\\
&\times
\left[
( n_{E^2 \mbk_1 e}^{(-2)} - n_{E^2 \mbkdpr f}^{(-2)}) 
\delta(\veps_{\mbk_1}^e -\veps_{\mbkdpr}^f)
+
( n_{E^2\mbk_1 d}^{(-2)} - n_{E^2\mbkdpr f}^{(-2)}) 
\delta(\veps_{\mbk_1}^d -\veps_{\mbkdpr}^f)
\right],
\end{split}
\end{equation}

\begin{equation}
\begin{split}
\mathcal{N}^{(\text{csk})}_{8,\mbk a}
=&
- \frac{2\pi^3}{\hbar}
\tau_{\mbk}^a
\sum_{\mbkpr \mbkdpr \mbk_1 bcdef}
\delta(\veps_{\mbk}^a - \veps_{\mbkpr}^b )
\delta(\veps_{\mbkpr}^b - \veps_{\mbk_1}^d )
\frac{
\text{Re}
\left[
\Braket{U_{\mbk \mbkpr}^{ab} U_{\mbkpr \mbk}^{ca}}
\Braket{U_{\mbkpr \mbk_1}^{ed} U_{\mbk_1 \mbkpr}^{dc}}
\Braket{U_{\mbkpr \mbkdpr}^{bf} U_{\mbkdpr \mbkpr}^{fe}}
\right]_{\substack{c,e \neq b}}
}
{(\veps_{\mbk}^a - \veps_{\mbkpr}^c )
(\veps_{\mbk}^a - \veps_{\mbkpr}^e )}
\\
&\times
\left[
( n_{E^2 \mbk e}^{(-2)} - n_{E^2 \mbkdpr f}^{(-2)}) 
\delta(\veps_{\mbk}^e -\veps_{\mbkdpr}^f)
+
( n_{E^2 \mbk c}^{(-2)} - n_{E^2 \mbkdpr f}^{(-2)}) 
\delta(\veps_{\mbk}^c -\veps_{\mbkdpr}^f)
\right].
\end{split}
\end{equation}

\subsection{Anomalous skew scattering}

The carrier density associated with anomalous skew scattering is given by
\begin{equation}
n_{E^2\mbk a}^{(\text{ask})}
=
\sum_{i=1}^{30}
\mathcal{N}^{(\text{ask})}_{i,\mbk a},
\end{equation}
\begin{equation}
\begin{split}
\mathcal{N}^{(\text{ask})}_{1,\mbk a}
&=
\frac{2 \pi^2}{\hbar^2}  e E^{\mu} \tau_{\mbk}^a
\sum_{\mbkpr \mbkdpr bcd}
\delta ( \veps_{\mbk}^a - \veps_{\mbkpr}^b )
\delta ( \veps_{\mbk}^a - \veps_{\mbkdpr}^d )
\\
&\times
\text{Re}
\Biggl\{
\frac{
\Braket{
U_{\mbk \mbk\pr}^{ab} U_{\mbk\pr \mbk} ^{bc} 
}
}
{\veps_{\mbk}^a - \veps_{\mbk}^c}
\pd_{\mbkdpr}^{\rho}
\left(
\frac{v_{\rho \mbkdpr}^d}{\lvert \bs{v}_{\mbkdpr}^d \rvert^2}
\Braket{
U_{\mbk \mbkdpr}^{cd} 
\left( \pd_{\mbk}^{\mu} + \pd_{\mbkdpr}^{\mu} \right) 
\left[
U_{\mbkdpr \mbk} ^{da}
\left(
n_{E \mbk a}^{(-1)} - n_{E \mbkdpr d}^{(-1)} 
\right) 
\right]
}
\right)
\\
&+
\frac{1}{2}
\frac{
\Braket{
U_{\mbk \mbk\pr}^{ab} U_{\mbk\pr \mbk} ^{bc} 
}
}
{\veps_{\mbk}^a - \veps_{\mbk}^c}
\pd_{\mbkdpr}^{\rho}
\left(
\frac{v_{\rho \mbkdpr}^d}{\lvert \bs{v}_{\mbkdpr}^d \rvert^2}
\pd_{\mbkdpr}^{\sigma}
\left[
\frac{v_{\sigma \mbkdpr}^d}{\lvert \bs{v}_{\mbkdpr}^d \rvert^2}
\left(
v_{\mu \mbk}^a - v_{\mu \mbkdpr}^d 
\right)
\Braket{
U_{\mbk \mbkdpr}^{cd}U_{\mbkdpr \mbk} ^{da} 
}
\left(
n_{E \mbk a}^{(-1)} - n_{E \mbkdpr d}^{(-1)} 
\right) 
\right]
\right)
\Biggr\}_{\substack{c \neq a}},
\end{split}
\end{equation}
\begin{equation}
\begin{split}
\mathcal{N}^{(\text{ask})}_{2,\mbk a}
&=
\frac{2 \pi^2}{\hbar^2}  e E^{\mu} \tau_{\mbk}^a
\sum_{\mbkpr \mbkdpr bcd}
\delta ( \veps_{\mbk}^a - \veps_{\mbkpr}^b )
\delta ( \veps_{\mbk}^a - \veps_{\mbkdpr}^d )
\\
&\times
\Biggl\{
\frac{1}{2}
\pd_{\mbkpr}^{\rho}
\left(
\frac{v_{\rho \mbkpr}^b}{\lvert \bs{v}_{\mbkpr}^b \rvert^2}
\pd_{\mbkpr}^{\sigma}
\left[
\frac{v_{\sigma \mbkpr}^b}{\lvert \bs{v}_{\mbkpr}^b \rvert^2}
\left(
v_{\mu \mbk}^a - v_{\mu \mbkpr}^b 
\right)
\frac{
\text{Re}
\left[
\Braket{
U_{\mbk \mbk\pr}^{ab} U_{\mbk\pr \mbk} ^{bc} 
}
\Braket{
U_{\mbk \mbkdpr}^{cd} U_{\mbkdpr \mbk} ^{da}
}
\right]
}
{\veps_{\mbk}^a - \veps_{\mbk}^c}
\left(
n_{E \mbk a}^{(-1)} - n_{E \mbkdpr d}^{(-1)} 
\right) 
\right]
\right)
\\
&+
\pd_{\mbkpr}^{\rho}
\left(
\frac{v_{\rho \mbkpr}^b}{\lvert \bs{v}_{\mbkpr}^b \rvert^2}
\text{Re}
\Braket{
U_{\mbk \mbk\pr}^{ab} 
\left( \pd_{\mbk}^{\mu} + \pd_{\mbkpr}^{\mu}\right)
\left[
U_{\mbk\pr \mbk} ^{bc}
\frac{
\Braket{
U_{\mbk \mbkdpr}^{cd} U_{\mbkdpr \mbk} ^{da} 
}
}
{\veps_{\mbk}^a - \veps_{\mbk}^c}
\left(
n_{E \mbk a}^{(-1)} - n_{E \mbkdpr d}^{(-1)} 
\right)
\right]  
}
\right)
\\
&+
\pd_{\mbkdpr}^{\rho}
\left(
\frac{v_{\rho \mbkdpr}^d}{\lvert \bs{v}_{\mbkdpr}^d \rvert^2}
\pd_{\mbkpr}^{\sigma}
\left[
\frac{v_{\sigma \mbkpr}^b}{\lvert \bs{v}_{\mbkpr}^b \rvert^2}
v_{\mu \mbk}^a
\frac{
\text{Re}
\left[
\Braket{
U_{\mbk \mbk\pr}^{ab} U_{\mbk\pr \mbk} ^{bc} 
}
\Braket{
U_{\mbk \mbkdpr}^{cd} U_{\mbkdpr \mbk} ^{da}
}
\right]
}
{\veps_{\mbk}^a - \veps_{\mbk}^c}
\left(
n_{E \mbk a}^{(-1)} - n_{E \mbkdpr d}^{(-1)} 
\right) 
\right]
\right)
\Biggr\}_{\substack{c \neq a}},
\end{split}
\end{equation}
\begin{equation}
\begin{split}
\mathcal{N}^{(\text{ask})}_{3,\mbk a}
=&
- \frac{2 \pi^2}{\hbar}  e E^{\mu} \tau_{\mbk}^a
\sum_{\mbkpr \mbkdpr bcd}
\delta ( \veps_{\mbk}^a - \veps_{\mbkpr}^b )
\delta ( \veps_{\mbk}^a - \veps_{\mbkdpr}^d )
\\
&\times
\text{Re}
\Biggl\{
\frac{
\Braket{
U_{\mbk \mbk\pr}^{ab} U_{\mbk\pr \mbk} ^{bc} 
}
}
{\veps_{\mbk}^a - \veps_{\mbk}^c}
\pd_{\mbk}^{\mu}
\left[
\frac{v_{\rho \mbkdpr}^d}{\lvert \bs{v}_{\mbkdpr}^d \rvert^2}
\frac{
\Braket{
U_{\mbk \mbkdpr}^{cd} U_{\mbkdpr \mbk}^{da}
}
}
{\veps_{\mbk}^a - \veps_{\mbk}^c}
\left(
n_{E \mbk a}^{(-1)} - n_{E \mbkdpr d}^{(-1)} 
\right) 
\right]
\\
&+
\frac{
\Braket{
U_{\mbk \mbk\pr}^{ab} U_{\mbk\pr \mbk} ^{bc} 
}
}
{\veps_{\mbk}^a - \veps_{\mbk}^c}
\pd_{\mbkdpr}^{\rho}
\left[
\frac{v_{\rho \mbkdpr}^d}{\lvert \bs{v}_{\mbkdpr}^d \rvert^2}
v_{\mu \mbk}^a
\frac{
\Braket{
U_{\mbk \mbkdpr}^{cd} U_{\mbkdpr \mbk}^{da}
}
}
{\veps_{\mbk}^a - \veps_{\mbk}^c}
\left(
n_{E \mbk a}^{(-1)} - n_{E \mbkdpr d}^{(-1)} 
\right) 
\right]
\Biggr\}_{\substack{c \neq a}},
\end{split}
\end{equation}
\begin{equation}
\begin{split}
\mathcal{N}^{(\text{ask})}_{4,\mbk a}
&=
- \frac{2 \pi^2}{\hbar^2}  e E^{\mu} \tau_{\mbk}^a
\sum_{\mbkpr \mbkdpr bcd}
\delta ( \veps_{\mbk}^a - \veps_{\mbkpr}^b )
\delta ( \veps_{\mbk}^c - \veps_{\mbkdpr}^d )
\\
&\times
\text{Re}
\Biggl\{
\frac{
\Braket{
U_{\mbk \mbk\pr}^{ab} U_{\mbk\pr \mbk} ^{bc} 
}
}
{\veps_{\mbk}^a - \veps_{\mbk}^c}
\pd_{\mbkdpr}^{\rho}
\left(
\frac{v_{\rho \mbkdpr}^d}{\lvert \bs{v}_{\mbkdpr}^d \rvert^2}
\Braket{
\left( \pd_{\mbk}^{\mu} + \pd_{\mbkdpr}^{\mu} \right) 
\left[
U_{\mbk \mbkdpr}^{cd} 
\left(
n_{E \mbk c}^{(-1)} - n_{E \mbkdpr d}^{(-1)} 
\right) 
\right]
U_{\mbkdpr \mbk} ^{da}
}
\right)
\\
&+
\frac{1}{2}
\frac{
\Braket{
U_{\mbk \mbk\pr}^{ab} U_{\mbk\pr \mbk} ^{bc} 
}
}
{\veps_{\mbk}^a - \veps_{\mbk}^c}
\pd_{\mbkdpr}^{\rho}
\left(
\frac{v_{\rho \mbkdpr}^d}{\lvert \bs{v}_{\mbkdpr}^d \rvert^2}
\pd_{\mbkdpr}^{\sigma}
\left[
\frac{v_{\sigma \mbkdpr}^d}{\lvert \bs{v}_{\mbkdpr}^d \rvert^2}
\left(
v_{\mu \mbk}^c - v_{\mu \mbkdpr}^d 
\right)
\Braket{
U_{\mbk \mbkdpr}^{cd}U_{\mbkdpr \mbk} ^{da} 
}
\left(
n_{E \mbk c}^{(-1)} - n_{E \mbkdpr d}^{(-1)} 
\right) 
\right]
\right)
\Biggr\}_{\substack{c \neq a}},
\end{split}
\end{equation}
\begin{equation}
\begin{split}
\mathcal{N}^{(\text{ask})}_{5,\mbk a}
&=
\frac{2 \pi^2}{\hbar^2}  e E^{\mu} \tau_{\mbk}^a
\sum_{\mbkpr \mbkdpr bcd}
\delta ( \veps_{\mbk}^a - \veps_{\mbkpr}^b )
\delta ( \veps_{\mbk}^c - \veps_{\mbkdpr}^d )
\\
&\times
\Biggl\{
\frac{1}{2}
\pd_{\mbkpr}^{\rho}
\left(
\frac{v_{\rho \mbkpr}^b}{\lvert \bs{v}_{\mbkpr}^b \rvert^2}
\pd_{\mbkpr}^{\sigma}
\left[
\frac{v_{\sigma \mbkpr}^b}{\lvert \bs{v}_{\mbkpr}^b \rvert^2}
\left(
v_{\mu \mbk}^a - v_{\mu \mbkpr}^b 
\right)
\frac{
\text{Re}
\left[
\Braket{
U_{\mbk \mbk\pr}^{ab} U_{\mbk\pr \mbk} ^{bc} 
}
\Braket{
U_{\mbk \mbkdpr}^{cd} U_{\mbkdpr \mbk} ^{da}
}
\right]
}
{\veps_{\mbk}^a - \veps_{\mbk}^c}
\left(
n_{E \mbk c}^{(-1)} - n_{E \mbkdpr d}^{(-1)} 
\right) 
\right]
\right)
\\
&+
\pd_{\mbkpr}^{\rho}
\left(
\frac{v_{\rho \mbkpr}^b}{\lvert \bs{v}_{\mbkpr}^b \rvert^2}
\text{Re}
\Braket{
U_{\mbk \mbk\pr}^{ab} 
\left( \pd_{\mbk}^{\mu} + \pd_{\mbkpr}^{\mu}\right)
\left[
U_{\mbk\pr \mbk} ^{bc}
\frac{
\Braket{
U_{\mbk \mbkdpr}^{cd} U_{\mbkdpr \mbk} ^{da} 
}
}
{\veps_{\mbk}^a - \veps_{\mbk}^c}
\left(
n_{E \mbk c}^{(-1)} - n_{E \mbkdpr d}^{(-1)} 
\right)
\right]  
}
\right)
\\
&+
\pd_{\mbkdpr}^{\rho}
\left(
\frac{v_{\rho \mbkdpr}^d}{\lvert \bs{v}_{\mbkdpr}^d \rvert^2}
\pd_{\mbkpr}^{\sigma}
\left[
\frac{v_{\sigma \mbkpr}^b}{\lvert \bs{v}_{\mbkpr}^b \rvert^2}
v_{\mu \mbk}^c
\frac{
\text{Re}
\left[
\Braket{
U_{\mbk \mbk\pr}^{ab} U_{\mbk\pr \mbk} ^{bc} 
}
\Braket{
U_{\mbk \mbkdpr}^{cd} U_{\mbkdpr \mbk} ^{da}
}
\right]
}
{\veps_{\mbk}^a - \veps_{\mbk}^c}
\left(
n_{E \mbk c}^{(-1)} - n_{E \mbkdpr d}^{(-1)} 
\right) 
\right]
\right)
\Biggr\}_{\substack{c \neq a}},
\end{split}
\end{equation}
\begin{equation}
\begin{split}
\mathcal{N}^{(\text{ask})}_{6,\mbk a}
=&
- \frac{2 \pi^2}{\hbar}  e E^{\mu} \tau_{\mbk}^a
\sum_{\mbkpr \mbkdpr bcd}
\delta ( \veps_{\mbk}^a - \veps_{\mbkpr}^b )
\delta ( \veps_{\mbk}^c - \veps_{\mbkdpr}^d )
\\
&\times
\text{Re}
\Biggl\{
\frac{
\Braket{
U_{\mbk \mbk\pr}^{ab} U_{\mbk\pr \mbk} ^{bc} 
}
}
{\veps_{\mbk}^a - \veps_{\mbk}^c}
\pd_{\mbk}^{\mu}
\left[
\frac{v_{\rho \mbkdpr}^d}{\lvert \bs{v}_{\mbkdpr}^d \rvert^2}
\frac{
\Braket{
U_{\mbk \mbkdpr}^{cd} U_{\mbkdpr \mbk}^{da}
}
}
{\veps_{\mbk}^a - \veps_{\mbk}^c}
\left(
n_{E \mbk c}^{(-1)} - n_{E \mbkdpr d}^{(-1)} 
\right) 
\right]
\\
&+
\frac{
\Braket{
U_{\mbk \mbk\pr}^{ab} U_{\mbk\pr \mbk} ^{bc} 
}
}
{\veps_{\mbk}^a - \veps_{\mbk}^c}
\pd_{\mbkdpr}^{\rho}
\left[
\frac{v_{\rho \mbkdpr}^d}{\lvert \bs{v}_{\mbkdpr}^d \rvert^2}
v_{\mu \mbk}^c
\frac{
\Braket{
U_{\mbk \mbkdpr}^{cd} U_{\mbkdpr \mbk}^{da}
}
}
{\veps_{\mbk}^a - \veps_{\mbk}^c}
\left(
n_{E \mbk c}^{(-1)} - n_{E \mbkdpr d}^{(-1)} 
\right) 
\right]
\Biggr\}_{\substack{c \neq a}},
\end{split}
\end{equation}
\begin{equation}
\begin{split}
\mathcal{N}^{(\text{ask})}_{7,\mbk a}
&=
- \frac{2 \pi^2}{\hbar^2}  e E^{\mu} \tau_{\mbk}^a
\sum_{\mbkpr \mbkdpr bcd}
\delta ( \veps_{\mbk}^a - \veps_{\mbkpr}^b )
\delta ( \veps_{\mbk}^a - \veps_{\mbkdpr}^d )
\\
&\times
\text{Re}
\Biggl\{
\frac{
\Braket{
U_{\mbk \mbk\pr}^{ab} U_{\mbk\pr \mbk} ^{ca} 
}
}
{\veps_{\mbk}^a - \veps_{\mbkpr}^c}
\pd_{\mbkdpr}^{\rho}
\left(
\frac{v_{\rho \mbkdpr}^d}{\lvert \bs{v}_{\mbkdpr}^d \rvert^2}
\Braket{
U_{\mbkdpr \mbkpr}^{dc}
\left( \pd_{\mbkpr}^{\mu} + \pd_{\mbkdpr}^{\mu} \right) 
\left[
U_{\mbkpr \mbkdpr} ^{bd}
\left(
n_{E \mbkpr b}^{(-1)} - n_{E \mbkdpr d}^{(-1)} 
\right) 
\right]
}
\right)
\\
&+
\frac{1}{2}
\frac{
\Braket{
U_{\mbk \mbk\pr}^{ab} U_{\mbk\pr \mbk} ^{ca} 
}
}
{\veps_{\mbk}^a - \veps_{\mbkpr}^c}
\pd_{\mbkdpr}^{\rho}
\left(
\frac{v_{\rho \mbkdpr}^d}{\lvert \bs{v}_{\mbkdpr}^d \rvert^2}
\pd_{\mbkdpr}^{\sigma}
\left[
\frac{v_{\sigma \mbkdpr}^d}{\lvert \bs{v}_{\mbkdpr}^d \rvert^2}
\left(
v_{\mu \mbkpr}^b - v_{\mu \mbkdpr}^d 
\right)
\Braket{
U_{\mbkpr \mbkdpr}^{bd}U_{\mbkdpr \mbkpr} ^{dc} 
}
\left(
n_{E \mbkpr b}^{(-1)} - n_{E \mbkdpr d}^{(-1)} 
\right) 
\right]
\right)
\Biggr\}_{\substack{c \neq b}},
\end{split}
\end{equation}
\begin{equation}
\begin{split}
\mathcal{N}^{(\text{ask})}_{8,\mbk a}
&=
\frac{2 \pi^2}{\hbar^2}  e E^{\mu} \tau_{\mbk}^a
\sum_{\mbkpr \mbkdpr bcd}
\delta ( \veps_{\mbk}^a - \veps_{\mbkpr}^b )
\delta ( \veps_{\mbk}^a - \veps_{\mbkdpr}^d )
\\
&\times
\Biggl\{
\frac{1}{2}
\pd_{\mbkpr}^{\rho}
\left(
\frac{v_{\rho \mbkpr}^b}{\lvert \bs{v}_{\mbkpr}^b \rvert^2}
\pd_{\mbkpr}^{\sigma}
\left[
\frac{v_{\sigma \mbkpr}^b}{\lvert \bs{v}_{\mbkpr}^b \rvert^2}
\left(
v_{\mu \mbk}^a - v_{\mu \mbkpr}^b 
\right)
\right.
\right.
\\
&\left.
\left.
\times
\frac{
\text{Re}
\left[
\Braket{
U_{\mbk \mbk\pr}^{ab} U_{\mbk\pr \mbk} ^{ca} 
}
\Braket{
U_{\mbkpr \mbkdpr}^{bd} U_{\mbkdpr \mbkpr} ^{dc}
}
\right]
}
{\veps_{\mbkpr}^b - \veps_{\mbkpr}^c}
\left(
n_{E \mbkpr b}^{(-1)} - n_{E \mbkdpr d}^{(-1)} 
\right) 
\right]
\right)
\\
&+
\pd_{\mbkpr}^{\rho}
\left(
\frac{v_{\rho \mbkpr}^b}{\lvert \bs{v}_{\mbkpr}^b \rvert^2}
\pd_{\mbkdpr}^{\sigma}
\left[
\frac{v_{\sigma \mbkdpr}^d}{\lvert \bs{v}_{\mbkdpr}^d \rvert^2}
v_{\mu \mbk}^a
\frac{
\text{Re}
\left[
\Braket{
U_{\mbk \mbk\pr}^{ab} U_{\mbk\pr \mbk} ^{ca} 
}
\Braket{
U_{\mbkpr \mbkdpr}^{bd} U_{\mbkdpr \mbkpr} ^{dc}
}
\right]
}
{\veps_{\mbkpr}^b - \veps_{\mbkpr}^c}
\left(
n_{E \mbkpr b}^{(-1)} - n_{E \mbkdpr d}^{(-1)} 
\right) 
\right]
\right)
\\
&+
\frac{1}{2}
\pd_{\mbkdpr}^{\rho}
\left(
\frac{v_{\rho \mbkdpr}^d}{\lvert \bs{v}_{\mbkdpr}^d \rvert^2}
\pd_{\mbkdpr}^{\sigma}
\left[
\frac{v_{\sigma \mbkdpr}^d}{\lvert \bs{v}_{\mbkdpr}^d \rvert^2}
\left(
v_{\mu \mbk}^a + v_{\mu \mbkpr}^b 
\right)
\right.
\right.
\\
&\left.
\left.
\times
\frac{
\text{Re}
\left[
\Braket{
U_{\mbk \mbk\pr}^{ab} U_{\mbk\pr \mbk} ^{ca} 
}
\Braket{
U_{\mbkpr \mbkdpr}^{bd} U_{\mbkdpr \mbkpr} ^{dc}
}
\right]
}
{\veps_{\mbkpr}^b - \veps_{\mbkpr}^c}
\left(
n_{E \mbkpr b}^{(-1)} - n_{E \mbkdpr d}^{(-1)} 
\right) 
\right]
\right)
\Biggr\}_{\substack{c \neq b}},
\end{split}
\end{equation}
\begin{equation}
\begin{split}
\mathcal{N}^{(\text{ask})}_{9,\mbk a}
=&
\frac{2 \pi^2}{\hbar^2}  e E^{\mu} \tau_{\mbk}^a
\sum_{\mbkpr \mbkdpr bcd}
\delta ( \veps_{\mbk}^a - \veps_{\mbkpr}^b )
\delta ( \veps_{\mbk}^a - \veps_{\mbkdpr}^d )
\\
&\times
\Biggl\{
\pd_{\mbkpr}^{\rho}
\left(
\frac{v_{\rho \mbkpr}^b}{\lvert \bs{v}_{\mbkpr}^b \rvert^2}
\text{Re}
\Braket{
U_{\mbk \mbk\pr}^{ab} 
\left( \pd_{\mbk}^{\mu} + \pd_{\mbkpr}^{\mu}\right)
\left[
U_{\mbk\pr \mbk} ^{ca}
\frac{
\Braket{
U_{\mbkpr \mbkdpr}^{bd} U_{\mbkdpr \mbkpr} ^{dc} 
}
}
{\veps_{\mbkpr}^b - \veps_{\mbkpr}^c}
\left(
n_{E \mbkpr b}^{(-1)} - n_{E \mbkdpr d}^{(-1)} 
\right)
\right]  
}
\right)
\\
&+
\pd_{\mbkdpr}^{\rho}
\left(
\frac{v_{\rho \mbkdpr}^d}{\lvert \bs{v}_{\mbkdpr}^d \rvert^2}
\text{Re}
\Braket{
U_{\mbk \mbk\pr}^{ab} 
\left( \pd_{\mbk}^{\mu} + \pd_{\mbkpr}^{\mu}\right)
\left[
U_{\mbk\pr \mbk} ^{ca}
\frac{
\Braket{
U_{\mbkpr \mbkdpr}^{bd} U_{\mbkdpr \mbkpr} ^{dc} 
}
}
{\veps_{\mbkpr}^b - \veps_{\mbkpr}^c}
\left(
n_{E \mbkpr b}^{(-1)} - n_{E \mbkdpr d}^{(-1)} 
\right)
\right]  
}
\right)
\Biggr\}_{\substack{c \neq b}},
\end{split}
\end{equation}
\begin{equation}
\begin{split}
\mathcal{N}^{(\text{ask})}_{10,\mbk a}
=&
\frac{2 \pi^2}{\hbar}  e E^{\mu} \tau_{\mbk}^a
\sum_{\mbkpr \mbkdpr bcd}
\delta ( \veps_{\mbk}^a - \veps_{\mbkpr}^b )
\delta ( \veps_{\mbk}^a - \veps_{\mbkdpr}^d )
\\
&\times
\text{Re}
\Biggl\{
\frac{
\Braket{
U_{\mbk \mbk\pr}^{ab} U_{\mbk\pr \mbk} ^{ca} 
}
}
{\veps_{\mbk}^a - \veps_{\mbkpr}^c}
\pd_{\mbkpr}^{\mu}
\left[
\frac{
\Braket{
U_{\mbkpr \mbkdpr}^{bd} U_{\mbkdpr \mbkpr}^{dc}
}
}
{\veps_{\mbkpr}^b - \veps_{\mbkpr}^c}
\left(
n_{E \mbkpr b}^{(-1)} - n_{E \mbkdpr d}^{(-1)} 
\right) 
\right]
\\
&+
\frac{
\Braket{
U_{\mbk \mbk\pr}^{ab} U_{\mbk\pr \mbk} ^{ca} 
}
}
{\veps_{\mbk}^a - \veps_{\mbkpr}^c}
\pd_{\mbkdpr}^{\rho}
\left[
\frac{v_{\rho \mbkdpr}^d}{\lvert \bs{v}_{\mbkdpr}^d \rvert^2}
v_{\mu \mbkpr}^b
\frac{
\Braket{
U_{\mbkpr \mbkdpr}^{bd} U_{\mbkdpr \mbkpr}^{dc}
}
}
{\veps_{\mbkpr}^b - \veps_{\mbkpr}^c}
\left(
n_{E \mbkpr b}^{(-1)} - n_{E \mbkdpr d}^{(-1)} 
\right) 
\right]
\Biggr\}_{\substack{c \neq b}},
\end{split}
\end{equation}
\begin{equation}
\begin{split}
\mathcal{N}^{(\text{ask})}_{11,\mbk a}
&=
\frac{2 \pi^2}{\hbar^2}  e E^{\mu} \tau_{\mbk}^a
\sum_{\mbkpr \mbkdpr bcd}
\delta ( \veps_{\mbk}^a - \veps_{\mbkpr}^b )
\delta ( \veps_{\mbkpr}^c - \veps_{\mbkdpr}^d )
\\
&\times
\text{Re}
\Biggl\{
\frac{
\Braket{
U_{\mbk \mbk\pr}^{ab} U_{\mbk\pr \mbk} ^{ca} 
}
}
{\veps_{\mbk}^a - \veps_{\mbkpr}^c}
\pd_{\mbkdpr}^{\rho}
\left(
\frac{v_{\rho \mbkdpr}^d}{\lvert \bs{v}_{\mbkdpr}^d \rvert^2}
\Braket{
U_{\mbkpr \mbkdpr}^{bd}
\left( \pd_{\mbkpr}^{\mu} + \pd_{\mbkdpr}^{\mu} \right) 
\left[
U_{\mbkdpr \mbkpr} ^{dc}
\left(
n_{E \mbkpr c}^{(-1)} - n_{E \mbkdpr d}^{(-1)} 
\right) 
\right]
}
\right)
\\
&+
\frac{1}{2}
\frac{
\Braket{
U_{\mbk \mbk\pr}^{ab} U_{\mbk\pr \mbk} ^{ca} 
}
}
{\veps_{\mbk}^a - \veps_{\mbkpr}^c}
\pd_{\mbkdpr}^{\rho}
\left(
\frac{v_{\rho \mbkdpr}^d}{\lvert \bs{v}_{\mbkdpr}^d \rvert^2}
\pd_{\mbkdpr}^{\sigma}
\left[
\frac{v_{\sigma \mbkdpr}^d}{\lvert \bs{v}_{\mbkdpr}^d \rvert^2}
\left(
v_{\mu \mbkpr}^c - v_{\mu \mbkdpr}^d 
\right)
\Braket{
U_{\mbkpr \mbkdpr}^{bd}U_{\mbkdpr \mbkpr} ^{dc} 
}
\left(
n_{E \mbkpr c}^{(-1)} - n_{E \mbkdpr d}^{(-1)} 
\right) 
\right]
\right)
\Biggr\}_{\substack{c \neq b}},
\end{split}
\end{equation}
\begin{equation}
\begin{split}
\mathcal{N}^{(\text{ask})}_{12,\mbk a}
=&
\frac{2 \pi^2}{\hbar^2}  e E^{\mu} \tau_{\mbk}^a
\sum_{\mbkpr \mbkdpr bcd}
\delta ( \veps_{\mbk}^a - \veps_{\mbkpr}^b )
\delta ( \veps_{\mbkpr}^c - \veps_{\mbkdpr}^d )
\\
&\times
\Biggl\{
\frac{1}{2}
\pd_{\mbkpr}^{\rho}
\left(
\frac{v_{\rho \mbkpr}^b}{\lvert \bs{v}_{\mbkpr}^b \rvert^2}
\pd_{\mbkpr}^{\sigma}
\left[
\frac{v_{\sigma \mbkpr}^b}{\lvert \bs{v}_{\mbkpr}^b \rvert^2}
\left(
v_{\mu \mbk}^a - v_{\mu \mbkpr}^b 
\right)
\right.
\right.
\\
&\left.
\left.
\times
\frac{
\text{Re}
\left[
\Braket{
U_{\mbk \mbk\pr}^{ab} U_{\mbk\pr \mbk} ^{ca} 
}
\Braket{
U_{\mbkpr \mbkdpr}^{bd} U_{\mbkdpr \mbkpr} ^{dc}
}
\right]
}
{\veps_{\mbkpr}^b - \veps_{\mbkpr}^c}
\left(
n_{E \mbkpr c}^{(-1)} - n_{E \mbkdpr d}^{(-1)} 
\right) 
\right]
\right)
\\
&+
\pd_{\mbkpr}^{\rho}
\left(
\frac{v_{\rho \mbkpr}^b}{\lvert \bs{v}_{\mbkpr}^b \rvert^2}
\pd_{\mbkdpr}^{\sigma}
\left[
\frac{v_{\sigma \mbkdpr}^d}{\lvert \bs{v}_{\mbkdpr}^d \rvert^2}
\left(
v_{\mu \mbk}^a - v_{\mu \mbkpr}^b + v_{\mu \mbkpr}^c
\right)
\right.
\right.
\\
&\left.
\left.
\times
\frac{
\text{Re}
\left[
\Braket{
U_{\mbk \mbk\pr}^{ab} U_{\mbk\pr \mbk} ^{ca} 
}
\Braket{
U_{\mbkpr \mbkdpr}^{bd} U_{\mbkdpr \mbkpr} ^{dc}
}
\right]
}
{\veps_{\mbkpr}^b - \veps_{\mbkpr}^c}
\left(
n_{E \mbkpr c}^{(-1)} - n_{E \mbkdpr d}^{(-1)} 
\right) 
\right]
\right)
\\
&+
\frac{1}{2}
\pd_{\mbkdpr}^{\rho}
\left(
\frac{v_{\rho \mbkdpr}^d}{\lvert \bs{v}_{\mbkdpr}^d \rvert^2}
\pd_{\mbkdpr}^{\sigma}
\left[
\frac{v_{\sigma \mbkdpr}^d}{\lvert \bs{v}_{\mbkdpr}^d \rvert^2}
\left(
v_{\mu \mbk}^a - v_{\mu \mbkpr}^b + 2 v_{\mu \mbkpr}^c 
\right)
\right.
\right.
\\
&\left.
\left.
\times
\frac{
\text{Re}
\left[
\Braket{
U_{\mbk \mbk\pr}^{ab} U_{\mbk\pr \mbk} ^{ca} 
}
\Braket{
U_{\mbkpr \mbkdpr}^{bd} U_{\mbkdpr \mbkpr} ^{dc}
}
\right]
}
{\veps_{\mbkpr}^b - \veps_{\mbkpr}^c}
\left(
n_{E \mbkpr c}^{(-1)} - n_{E \mbkdpr d}^{(-1)} 
\right) 
\right]
\right)
\Biggr\}_{\substack{c \neq b}},
\end{split}
\end{equation}
\begin{equation}
\begin{split}
\mathcal{N}^{(\text{ask})}_{13,\mbk a}
=&
\frac{2 \pi^2}{\hbar^2}  e E^{\mu} \tau_{\mbk}^a
\sum_{\mbkpr \mbkdpr bcd}
\delta ( \veps_{\mbk}^a - \veps_{\mbkpr}^b )
\delta ( \veps_{\mbkpr}^c - \veps_{\mbkdpr}^d )
\\
&\times
\Biggl\{
\pd_{\mbkpr}^{\rho}
\left(
\frac{v_{\rho \mbkpr}^b}{\lvert \bs{v}_{\mbkpr}^b \rvert^2}
\text{Re}
\Braket{
U_{\mbk \mbk\pr}^{ab} 
\left( \pd_{\mbk}^{\mu} + \pd_{\mbkpr}^{\mu}\right)
\left[
U_{\mbk\pr \mbk} ^{ca}
\frac{
\Braket{
U_{\mbkpr \mbkdpr}^{bd} U_{\mbkdpr \mbkpr} ^{dc} 
}
}
{\veps_{\mbkpr}^b - \veps_{\mbkpr}^c}
\left(
n_{E \mbkpr c}^{(-1)} - n_{E \mbkdpr d}^{(-1)} 
\right)
\right]  
}
\right)
\\
&+
\pd_{\mbkdpr}^{\rho}
\left(
\frac{v_{\rho \mbkdpr}^d}{\lvert \bs{v}_{\mbkdpr}^d \rvert^2}
\text{Re}
\Braket{
U_{\mbk \mbk\pr}^{ab} 
\left( \pd_{\mbk}^{\mu} + \pd_{\mbkpr}^{\mu}\right)
\left[
U_{\mbk\pr \mbk} ^{ca}
\frac{
\Braket{
U_{\mbkpr \mbkdpr}^{bd} U_{\mbkdpr \mbkpr} ^{dc} 
}
}
{\veps_{\mbkpr}^b - \veps_{\mbkpr}^c}
\left(
n_{E \mbkpr c}^{(-1)} - n_{E \mbkdpr d}^{(-1)} 
\right)
\right]  
}
\right)
\Biggr\}_{\substack{c \neq b}},
\end{split}
\end{equation}
\begin{equation}
\begin{split}
\mathcal{N}^{(\text{ask})}_{14,\mbk a}
=&
\frac{2 \pi^2}{\hbar}  e E^{\mu} \tau_{\mbk}^a
\sum_{\mbkpr \mbkdpr bcd}
\delta ( \veps_{\mbk}^a - \veps_{\mbkpr}^b )
\delta ( \veps_{\mbkpr}^c - \veps_{\mbkdpr}^d )
\\
&\times
\text{Re}
\Biggl\{
\frac{
\Braket{
U_{\mbk \mbk\pr}^{ab} U_{\mbk\pr \mbk} ^{ca} 
}
}
{\veps_{\mbk}^a - \veps_{\mbkpr}^c}
\pd_{\mbkpr}^{\mu}
\left[
\frac{
\Braket{
U_{\mbkpr \mbkdpr}^{bd} U_{\mbkdpr \mbkpr}^{dc}
}
}
{\veps_{\mbkpr}^b - \veps_{\mbkpr}^c}
\left(
n_{E \mbkpr c}^{(-1)} - n_{E \mbkdpr d}^{(-1)} 
\right) 
\right]
\\
&+
\frac{
\Braket{
U_{\mbk \mbk\pr}^{ab} U_{\mbk\pr \mbk} ^{ca} 
}
}
{\veps_{\mbk}^a - \veps_{\mbkpr}^c}
\pd_{\mbkdpr}^{\rho}
\left[
\frac{v_{\rho \mbkdpr}^d}{\lvert \bs{v}_{\mbkdpr}^d \rvert^2}
v_{\mu \mbkpr}^c
\frac{
\Braket{
U_{\mbkpr \mbkdpr}^{bd} U_{\mbkdpr \mbkpr}^{dc}
}
}
{\veps_{\mbkpr}^b - \veps_{\mbkpr}^c}
\left(
n_{E \mbkpr c}^{(-1)} - n_{E \mbkdpr d}^{(-1)} 
\right) 
\right]
\Biggr\}_{\substack{c \neq b}},
\end{split}
\end{equation}
\begin{equation}
\begin{split}
\mathcal{N}^{(\text{ask})}_{15,\mbk a}
=&
- \frac{2 \pi^2}{\hbar^2}  e E^{\mu} \tau_{\mbk}^a
\sum_{\mbkpr \mbkdpr bcd}
\delta ( \veps_{\mbk}^a - \veps_{\mbkpr}^b )
\delta ( \veps_{\mbk}^a - \veps_{\mbkdpr}^d )
\\
&\times
\text{Im}
\Biggl\{
\frac{
\Braket{
U_{\mbk \mbk\pr}^{ab} U_{\mbk\pr \mbk} ^{bc} 
}
}
{\veps_{\mbk}^a - \veps_{\mbk}^c}
\pd_{\mbkdpr}^{\rho}
\left[
\frac{v_{\rho \mbkdpr}^d}{\lvert \bs{v}_{\mbkdpr}^d \rvert^2}
\Braket{
U_{\mbk \mbkdpr}^{cd} U_{\mbkdpr \mbk}^{da}}
\left( \mathcal{A}_{\mu \mbk}^a 
- 
\mathcal{A}_{\mu \mbkdpr}^d \right) 
\left(
n_{E \mbk a}^{(-1)} - n_{E \mbkdpr d}^{(-1)} 
\right) 
\right]
\Biggr\}_{\substack{c \neq a}}
\\
&-
\frac{2 \pi^2}{\hbar^2}  e E^{\mu} \tau_{\mbk}^a
\sum_{\mbkpr \mbkdpr bcd}
\delta ( \veps_{\mbk}^a - \veps_{\mbkpr}^b )
\delta ( \veps_{\mbk}^c - \veps_{\mbkdpr}^d )
\\
&\times
\text{Im}
\Biggl\{
\frac{
\Braket{
U_{\mbk \mbk\pr}^{ab} U_{\mbk\pr \mbk} ^{bc} 
}
}
{\veps_{\mbk}^a - \veps_{\mbk}^c}
\pd_{\mbkdpr}^{\rho}
\left[
\frac{v_{\rho \mbkdpr}^d}{\lvert \bs{v}_{\mbkdpr}^d \rvert^2}
\Braket{
U_{\mbk \mbkdpr}^{cd} U_{\mbkdpr \mbk}^{da}}
\left( \mathcal{A}_{\mu \mbk}^c 
- 
\mathcal{A}_{\mu \mbkdpr}^d \right) 
\left(
n_{E \mbk c}^{(-1)} - n_{E \mbkdpr d}^{(-1)} 
\right) 
\right]
\Biggr\}_{\substack{c \neq a}},
\end{split}
\end{equation}
\begin{equation}
\begin{split}
\mathcal{N}^{(\text{ask})}_{16,\mbk a}
=&
- \frac{2 \pi^2}{\hbar^2}  e E^{\mu} \tau_{\mbk}^a
\sum_{\mbkpr \mbkdpr bcd}
\delta ( \veps_{\mbk}^a - \veps_{\mbkpr}^b )
\delta ( \veps_{\mbk}^a - \veps_{\mbkdpr}^d )
\\
&\times
\pd_{\mbkpr}^{\rho}
\Biggl\{
\frac{v_{\rho \mbkpr}^b}{\lvert \bs{v}_{\mbkpr}^b \rvert^2}
\frac{
\text{Im}
\left[
\Braket{
U_{\mbk \mbk\pr}^{ab} U_{\mbk\pr \mbk} ^{bc} 
}
\Braket{
U_{\mbk \mbkdpr}^{cd} U_{\mbkdpr \mbk} ^{da} 
}
\right]
}
{\veps_{\mbk}^a - \veps_{\mbk}^c}
\left( \mathcal{A}_{\mu \mbk}^a 
- 
\mathcal{A}_{\mu \mbkpr}^b \right) 
\left(
n_{E \mbk a}^{(-1)} - n_{E \mbkdpr d}^{(-1)} 
\right) 
\Biggr\}_{\substack{c \neq a}}
\\
&-
\frac{2 \pi^2}{\hbar^2}  e E^{\mu} \tau_{\mbk}^a
\sum_{\mbkpr \mbkdpr bcd}
\delta ( \veps_{\mbk}^a - \veps_{\mbkpr}^b )
\delta ( \veps_{\mbk}^c - \veps_{\mbkdpr}^d )
\\
&\times
\pd_{\mbkpr}^{\rho}
\Biggl\{
\frac{v_{\rho \mbkpr}^b}{\lvert \bs{v}_{\mbkpr}^b \rvert^2}
\frac{
\text{Im}
\left[
\Braket{
U_{\mbk \mbk\pr}^{ab} U_{\mbk\pr \mbk} ^{bc} 
}
\Braket{
U_{\mbk \mbkdpr}^{cd} U_{\mbkdpr \mbk} ^{da} 
}
\right]
}
{\veps_{\mbk}^a - \veps_{\mbk}^c}
\left( \mathcal{A}_{\mu \mbk}^a 
- 
\mathcal{A}_{\mu \mbkpr}^b \right) 
\left(
n_{E \mbk c}^{(-1)} - n_{E \mbkdpr d}^{(-1)} 
\right) 
\Biggr\}_{\substack{c \neq a}},
\end{split}
\end{equation}
\begin{equation}
\begin{split}
\mathcal{N}^{(\text{ask})}_{17,\mbk a}
=&
\frac{2 \pi^2}{\hbar}  e E^{\mu} \tau_{\mbk}^a
\sum_{\mbkpr \mbkdpr bcd}
\delta ( \veps_{\mbk}^a - \veps_{\mbkpr}^b )
\delta ( \veps_{\mbk}^a - \veps_{\mbkdpr}^d )
\\
&\times
\frac{
\text{Im}
\left[
\Braket{
U_{\mbk \mbk\pr}^{ab} U_{\mbk\pr \mbk} ^{bc} 
}
\Braket{
U_{\mbk \mbkdpr}^{cd} U_{\mbkdpr \mbk} ^{da} 
}
\right]_{\substack{c \neq a}}
}
{(\veps_{\mbk}^a - \veps_{\mbk}^c)^2}
\left( \mathcal{A}_{\mu \mbk}^a 
- 
\mathcal{A}_{\mu \mbk}^c \right) 
\left(
n_{E \mbk a}^{(-1)} - n_{E \mbkdpr d}^{(-1)} 
\right)
\\
&+
\frac{2 \pi^2}{\hbar}  e E^{\mu} \tau_{\mbk}^a
\sum_{\mbkpr \mbkdpr bcd}
\delta ( \veps_{\mbk}^a - \veps_{\mbkpr}^b )
\delta ( \veps_{\mbk}^c - \veps_{\mbkdpr}^d )
\\
&\times
\frac{
\text{Im}
\left[
\Braket{
U_{\mbk \mbk\pr}^{ab} U_{\mbk\pr \mbk} ^{bc} 
}
\Braket{
U_{\mbk \mbkdpr}^{cd} U_{\mbkdpr \mbk} ^{da} 
}
\right]_{\substack{c \neq a}}
}
{(\veps_{\mbk}^a - \veps_{\mbk}^c)^2}
\left( \mathcal{A}_{\mu \mbk}^a 
- 
\mathcal{A}_{\mu \mbk}^c \right) 
\left(
n_{E \mbk c}^{(-1)} - n_{E \mbkdpr d}^{(-1)} 
\right),
\end{split}
\end{equation}
\begin{equation}
\begin{split}
\mathcal{N}^{(\text{ask})}_{18,\mbk a}
=&
- \frac{2 \pi^2}{\hbar^2}  e E^{\mu} \tau_{\mbk}^a
\sum_{\mbkpr \mbkdpr bcd}
\delta ( \veps_{\mbk}^a - \veps_{\mbkpr}^b )
\delta ( \veps_{\mbk}^a - \veps_{\mbkdpr}^d )
\\
&\times
\text{Im}
\Biggl\{
\frac{
\Braket{
U_{\mbk \mbk\pr}^{ab} U_{\mbk\pr \mbk} ^{ca} 
}
}
{\veps_{\mbk}^a - \veps_{\mbkpr}^c}
\pd_{\mbkdpr}^{\rho}
\left[
\frac{v_{\rho \mbkdpr}^d}{\lvert \bs{v}_{\mbkdpr}^d \rvert^2}
\Braket{
U_{\mbkpr \mbkdpr}^{bd} U_{\mbkdpr \mbkpr}^{dc}}
\left( \mathcal{A}_{\mu \mbkpr}^b 
- 
\mathcal{A}_{\mu \mbkdpr}^d \right) 
\left(
n_{E \mbkpr b}^{(-1)} - n_{E \mbkdpr d}^{(-1)} 
\right) 
\right]
\Biggr\}_{\substack{c \neq b}}
\\
&-
\frac{2 \pi^2}{\hbar^2}  e E^{\mu} \tau_{\mbk}^a
\sum_{\mbkpr \mbkdpr bcd}
\delta ( \veps_{\mbk}^a - \veps_{\mbkpr}^b )
\delta ( \veps_{\mbkpr}^c - \veps_{\mbkdpr}^d )
\\
&\times
\text{Im}
\Biggl\{
\frac{
\Braket{
U_{\mbk \mbk\pr}^{ab} U_{\mbk\pr \mbk} ^{ca} 
}
}
{\veps_{\mbk}^a - \veps_{\mbkpr}^c}
\pd_{\mbkdpr}^{\rho}
\left[
\frac{v_{\rho \mbkdpr}^d}{\lvert \bs{v}_{\mbkdpr}^d \rvert^2}
\Braket{
U_{\mbkpr \mbkdpr}^{bd} U_{\mbkdpr \mbkpr}^{dc}}
\left( \mathcal{A}_{\mu \mbkpr}^c 
- 
\mathcal{A}_{\mu \mbkdpr}^d \right) 
\left(
n_{E \mbkpr c}^{(-1)} - n_{E \mbkdpr d}^{(-1)} 
\right) 
\right]
\Biggr\}_{\substack{c \neq b}},
\end{split}
\end{equation}
\begin{equation}
\begin{split}
\mathcal{N}^{(\text{ask})}_{19,\mbk a}
=&
- \frac{2 \pi^2}{\hbar^2}  e E^{\mu} \tau_{\mbk}^a
\sum_{\mbkpr \mbkdpr bcd}
\delta ( \veps_{\mbk}^a - \veps_{\mbkpr}^b )
\delta ( \veps_{\mbk}^a - \veps_{\mbkdpr}^d )
\\
&\times
\pd_{\mbkpr}^{\rho}
\Biggl\{
\frac{v_{\rho \mbkpr}^b}{\lvert \bs{v}_{\mbkpr}^b \rvert^2}
\frac{
\text{Im}
\left[
\Braket{
U_{\mbk \mbk\pr}^{ab} U_{\mbk\pr \mbk} ^{ca} 
}
\Braket{
U_{\mbkpr \mbkdpr}^{bd} U_{\mbkdpr \mbkpr} ^{dc} 
}
\right]
}
{\veps_{\mbkpr}^b - \veps_{\mbkpr}^c}
\left( \mathcal{A}_{\mu \mbk}^a 
- 
\mathcal{A}_{\mu \mbkpr}^b \right) 
\left(
n_{E \mbkpr b}^{(-1)} - n_{E \mbkdpr d}^{(-1)} 
\right) 
\Biggr\}_{\substack{c \neq b}}
\\
&-
\frac{2 \pi^2}{\hbar^2}  e E^{\mu} \tau_{\mbk}^a
\sum_{\mbkpr \mbkdpr bcd}
\delta ( \veps_{\mbk}^a - \veps_{\mbkpr}^b )
\delta ( \veps_{\mbk}^c - \veps_{\mbkdpr}^d )
\\
&\times
\pd_{\mbkdpr}^{\rho}
\Biggl\{
\frac{v_{\rho \mbkdpr}^d}{\lvert \bs{v}_{\mbkdpr}^d \rvert^2}
\frac{
\text{Im}
\left[
\Braket{
U_{\mbk \mbk\pr}^{ab} U_{\mbk\pr \mbk} ^{ca} 
}
\Braket{
U_{\mbkpr \mbkdpr}^{bd} U_{\mbkdpr \mbkpr} ^{dc} 
}
\right]
}
{\veps_{\mbkpr}^b - \veps_{\mbkpr}^c}
\left( \mathcal{A}_{\mu \mbk}^a 
- 
\mathcal{A}_{\mu \mbkpr}^b \right) 
\left(
n_{E \mbkpr b}^{(-1)} - n_{E \mbkdpr d}^{(-1)} 
\right) 
\Biggr\}_{\substack{c \neq b}},
\end{split}
\end{equation}
\begin{equation}
\begin{split}
\mathcal{N}^{(\text{ask})}_{20,\mbk a}
=&
- \frac{2 \pi^2}{\hbar^2}  e E^{\mu} \tau_{\mbk}^a
\sum_{\mbkpr \mbkdpr bcd}
\delta ( \veps_{\mbk}^a - \veps_{\mbkpr}^b )
\delta ( \veps_{\mbkpr}^c - \veps_{\mbkdpr}^d )
\\
&\times
\pd_{\mbkpr}^{\rho}
\Biggl\{
\frac{v_{\rho \mbkpr}^b}{\lvert \bs{v}_{\mbkpr}^b \rvert^2}
\frac{
\text{Im}
\left[
\Braket{
U_{\mbk \mbk\pr}^{ab} U_{\mbk\pr \mbk} ^{ca} 
}
\Braket{
U_{\mbkpr \mbkdpr}^{bd} U_{\mbkdpr \mbkpr} ^{dc} 
}
\right]
}
{\veps_{\mbkpr}^b - \veps_{\mbkpr}^c}
\left( \mathcal{A}_{\mu \mbk}^a 
- 
\mathcal{A}_{\mu \mbkpr}^b \right) 
\left(
n_{E \mbkpr c}^{(-1)} - n_{E \mbkdpr d}^{(-1)} 
\right) 
\Biggr\}_{\substack{c \neq b}}
\\
&-
\frac{2 \pi^2}{\hbar^2}  e E^{\mu} \tau_{\mbk}^a
\sum_{\mbkpr \mbkdpr bcd}
\delta ( \veps_{\mbk}^a - \veps_{\mbkpr}^b )
\delta ( \veps_{\mbkpr}^c - \veps_{\mbkdpr}^d )
\\
&\times
\pd_{\mbkdpr}^{\rho}
\Biggl\{
\frac{v_{\rho \mbkdpr}^d}{\lvert \bs{v}_{\mbkdpr}^d \rvert^2}
\frac{
\text{Im}
\left[
\Braket{
U_{\mbk \mbk\pr}^{ab} U_{\mbk\pr \mbk} ^{ca} 
}
\Braket{
U_{\mbkpr \mbkdpr}^{bd} U_{\mbkdpr \mbkpr} ^{dc} 
}
\right]
}
{\veps_{\mbkpr}^b - \veps_{\mbkpr}^c}
\left( \mathcal{A}_{\mu \mbk}^a 
- 
\mathcal{A}_{\mu \mbkpr}^b \right) 
\left(
n_{E \mbkpr c}^{(-1)} - n_{E \mbkdpr d}^{(-1)} 
\right) 
\Biggr\}_{\substack{c \neq b}},
\end{split}
\end{equation}
\begin{equation}
\begin{split}
\mathcal{N}^{(\text{ask})}_{21,\mbk a}
=&
\frac{2 \pi^2}{\hbar}  e E^{\mu} \tau_{\mbk}^a
\sum_{\mbkpr \mbkdpr bcd}
\delta ( \veps_{\mbk}^a - \veps_{\mbkpr}^b )
\delta ( \veps_{\mbk}^a - \veps_{\mbkdpr}^d )
\\
&\times
\frac{
\text{Im}
\left[
\Braket{
U_{\mbk \mbk\pr}^{ab} U_{\mbk\pr \mbk} ^{ca} 
}
\Braket{
U_{\mbkpr \mbkdpr}^{bd} U_{\mbkdpr \mbkpr} ^{dc} 
}
\right]_{\substack{c \neq b}}
}
{(\veps_{\mbk}^a - \veps_{\mbk}^c)^2}
\left( \mathcal{A}_{\mu \mbkpr}^b 
- 
\mathcal{A}_{\mu \mbkpr}^c \right) 
\left(
n_{E \mbkpr b}^{(-1)} - n_{E \mbkdpr d}^{(-1)} 
\right)
\\
&+
\frac{2 \pi^2}{\hbar}  e E^{\mu} \tau_{\mbk}^a
\sum_{\mbkpr \mbkdpr bcd}
\delta ( \veps_{\mbk}^a - \veps_{\mbkpr}^b )
\delta ( \veps_{\mbkpr}^c - \veps_{\mbkdpr}^d )
\\
&\times
\frac{
\text{Im}
\left[
\Braket{
U_{\mbk \mbk\pr}^{ab} U_{\mbk\pr \mbk} ^{ca} 
}
\Braket{
U_{\mbkpr \mbkdpr}^{bd} U_{\mbkdpr \mbkpr} ^{dc} 
}
\right]_{\substack{c \neq b}}
}
{(\veps_{\mbk}^a - \veps_{\mbkpr}^c)^2}
\left( \mathcal{A}_{\mu \mbkpr}^b 
- 
\mathcal{A}_{\mu \mbkpr}^c \right) 
\left(
n_{E \mbkpr c}^{(-1)} - n_{E \mbkdpr d}^{(-1)} 
\right),
\end{split}
\end{equation}
\begin{equation}
\begin{split}
\mathcal{N}^{(\text{ask})}_{22,\mbk a}
=&
- \frac{2 \pi^2}{\hbar}  e E^{\mu} \tau_{\mbk}^a
\sum_{\mbkpr \mbkdpr bcde}
\delta ( \veps_{\mbk}^a - \veps_{\mbkpr}^b )
\frac{
\text{Im}
\left[
\Braket{
U_{\mbk \mbk\pr}^{ab} U_{\mbk\pr \mbk} ^{bc} 
}
\Braket{
U_{\mbk \mbkdpr}^{cd} U_{\mbkdpr \mbk} ^{ea} 
}
\mathcal{A}_{\mu \mbkdpr}^{\prime de}
\right]_{\substack{c \neq a}}
}
{\veps_{\mbk}^a - \veps_{\mbk}^c}
\\
&\times
\left[
\frac{
n_{E \mbk a}^{(-1)} - n_{E \mbkdpr d}^{(-1)}
}
{\veps_{\mbk}^a - \veps_{\mbkdpr}^e}
\delta ( \veps_{\mbk}^a - \veps_{\mbkdpr}^d ) 
+
\frac{
n_{E \mbk c}^{(-1)} - n_{E \mbkdpr e}^{(-1)}
}
{\veps_{\mbk}^c - \veps_{\mbkdpr}^d}
\delta ( \veps_{\mbk}^c - \veps_{\mbkdpr}^e )
\right.
\\
&\left. 
+
\frac{
n_{E \mbk a}^{(-1)} - n_{E \mbkdpr e}^{(-1)}
}
{\veps_{\mbk}^a - \veps_{\mbkdpr}^d}
\delta ( \veps_{\mbk}^a - \veps_{\mbkdpr}^e )
+
\frac{
n_{E \mbk c}^{(-1)} - n_{E \mbkdpr d}^{(-1)}
}
{\veps_{\mbk}^c - \veps_{\mbkdpr}^e}
\delta ( \veps_{\mbk}^c - \veps_{\mbkdpr}^d )
\right],
\end{split}
\end{equation}
\begin{equation}
\begin{split}
\mathcal{N}^{(\text{ask})}_{23,\mbk a}
=&
- \frac{2 \pi^2}{\hbar}  e E^{\mu} \tau_{\mbk}^a
\sum_{\mbkpr \mbkdpr bcde}
\delta ( \veps_{\mbk}^a - \veps_{\mbkpr}^b )
\frac{
\text{Im}
\left[
\Braket{
U_{\mbk \mbk\pr}^{ab} U_{\mbk\pr \mbk} ^{bc} 
}
\Braket{
U_{\mbk \mbkdpr}^{cd} U_{\mbkdpr \mbk} ^{de} 
}
\mathcal{A}_{\mu \mbk}^{\prime ea}
\right]_{\substack{c \neq a}}
}
{(\veps_{\mbk}^a - \veps_{\mbk}^c)
(\veps_{\mbk}^a - \veps_{\mbk}^e)}
\\
&\times
\left[
\left(
n_{E \mbk a}^{(-1)} - n_{E \mbkdpr d}^{(-1)}
\right)
\delta ( \veps_{\mbk}^a - \veps_{\mbkdpr}^d ) 
\right.
-
\left.
\left(
n_{E \mbk e}^{(-1)} - n_{E \mbkdpr d}^{(-1)}
\right)
\delta ( \veps_{\mbk}^e - \veps_{\mbkdpr}^d ) 
\right]
\\
&+
\frac{2 \pi^2}{\hbar}  e E^{\mu} \tau_{\mbk}^a
\sum_{\mbkpr \mbkdpr bcde}
\left[
\delta ( \veps_{\mbk}^a - \veps_{\mbkpr}^b )
-
\delta ( \veps_{\mbk}^e - \veps_{\mbkpr}^b )
\right]
\frac{
\text{Im}
\left[
\Braket{
U_{\mbk \mbk\pr}^{ab} U_{\mbk\pr \mbk} ^{bc} 
}
\Braket{
U_{\mbk \mbkdpr}^{cd} U_{\mbkdpr \mbk} ^{de} 
}
\mathcal{A}_{\mu \mbk}^{\prime ea}
\right]_{\substack{e \neq c}}
}
{(\veps_{\mbk}^a - \veps_{\mbk}^e)
(\veps_{\mbk}^c - \veps_{\mbk}^e)}
\\
&
\times
\left[
\left(
n_{E \mbk c}^{(-1)} - n_{E \mbkdpr d}^{(-1)}
\right)
\delta ( \veps_{\mbk}^c - \veps_{\mbkdpr}^d ) 
\right.
+
\left.
\left(
n_{E \mbk e}^{(-1)} - n_{E \mbkdpr d}^{(-1)}
\right)
\delta ( \veps_{\mbk}^e - \veps_{\mbkdpr}^d ) 
\right],
\end{split}
\end{equation}
\begin{equation}
\begin{split}
\mathcal{N}^{(\text{ask})}_{24,\mbk a}
=&
- \frac{2 \pi^2}{\hbar}  e E^{\mu} \tau_{\mbk}^a
\sum_{\mbkpr \mbkdpr bcde}
\delta ( \veps_{\mbk}^a - \veps_{\mbkpr}^b )
\frac{
\text{Im}
\left[
\Braket{
U_{\mbk \mbk\pr}^{ab} U_{\mbk\pr \mbk} ^{bc} 
}
\Braket{
U_{\mbk \mbkdpr}^{ed} U_{\mbkdpr \mbk} ^{da} 
}
\mathcal{A}_{\mu \mbk}^{\prime ce}
\right]_{\substack{c \neq a}}
}
{(\veps_{\mbk}^a - \veps_{\mbk}^c)
(\veps_{\mbk}^c - \veps_{\mbk}^e)}
\\
&
\times
\left[
\left(
n_{E \mbk c}^{(-1)} - n_{E \mbkdpr d}^{(-1)}
\right)
\delta ( \veps_{\mbk}^c - \veps_{\mbkdpr}^d ) 
\right.
-
\left.
\left(
n_{E \mbk e}^{(-1)} - n_{E \mbkdpr d}^{(-1)}
\right)
\delta ( \veps_{\mbk}^e - \veps_{\mbkdpr}^d ) 
\right]
\\
&-
\frac{2 \pi^2}{\hbar}  e E^{\mu} \tau_{\mbk}^a
\sum_{\mbkpr \mbkdpr bcde}
\delta ( \veps_{\mbk}^a - \veps_{\mbkpr}^b )
\frac{
\text{Im}
\left[
\Braket{
U_{\mbk \mbk\pr}^{ab} U_{\mbk\pr \mbk} ^{bc} 
}
\Braket{
U_{\mbk \mbkdpr}^{ed} U_{\mbkdpr \mbk} ^{da} 
}
\mathcal{A}_{\mu \mbk}^{\prime ce}
\right]_{\substack{c,e \neq a}}
}
{(\veps_{\mbk}^a - \veps_{\mbk}^c)
(\veps_{\mbk}^a - \veps_{\mbk}^e)}
\\
&\times
\left[
\left(
n_{E \mbk a}^{(-1)} - n_{E \mbkdpr d}^{(-1)}
\right)
\delta ( \veps_{\mbk}^a - \veps_{\mbkdpr}^d ) 
\right.
+
\left.
\left(
n_{E \mbk e}^{(-1)} - n_{E \mbkdpr d}^{(-1)}
\right)
\delta ( \veps_{\mbk}^e - \veps_{\mbkdpr}^d ) 
\right],
\end{split}
\end{equation}
\begin{equation}
\begin{split}
\mathcal{N}^{(\text{ask})}_{25,\mbk a}
=&
- \frac{2 \pi^2}{\hbar}  e E^{\mu} \tau_{\mbk}^a
\sum_{\mbkpr \mbkdpr bcde}
\delta ( \veps_{\mbk}^a - \veps_{\mbkpr}^b )
\frac{
\text{Im}
\left[
\Braket{
U_{\mbk \mbk\pr}^{ab} U_{\mbk\pr \mbk} ^{ca} 
}
\Braket{
U_{\mbkpr \mbkdpr}^{bd} U_{\mbkdpr \mbkpr} ^{ec} 
}
\mathcal{A}_{\mu \mbkdpr}^{\prime de}
\right]_{\substack{c \neq b}}
}
{\veps_{\mbk}^a - \veps_{\mbkpr}^c}
\\
&\times
\left[
\frac{
n_{E \mbkpr b}^{(-1)} - n_{E \mbkdpr d}^{(-1)}
}
{\veps_{\mbk}^a - \veps_{\mbkdpr}^e}
\delta ( \veps_{\mbk}^a - \veps_{\mbkdpr}^d ) 
+
\frac{
n_{E \mbkpr c}^{(-1)} - n_{E \mbkdpr e}^{(-1)}
}
{\veps_{\mbkpr}^c - \veps_{\mbkdpr}^d}
\delta ( \veps_{\mbkpr}^c - \veps_{\mbkdpr}^e ) 
\right.
\\
&+
\left.
\frac{
n_{E \mbkpr b}^{(-1)} - n_{E \mbkdpr e}^{(-1)}
}
{\veps_{\mbk}^a - \veps_{\mbkdpr}^d}
\delta ( \veps_{\mbk}^a - \veps_{\mbkdpr}^e )
+
\frac{
n_{E \mbkpr c}^{(-1)} - n_{E \mbkdpr d}^{(-1)}
}
{\veps_{\mbkpr}^c - \veps_{\mbkdpr}^e}
\delta ( \veps_{\mbkpr}^c - \veps_{\mbkdpr}^d )
\right],
\end{split}
\end{equation}
\begin{equation}
\begin{split}
\mathcal{N}^{(\text{ask})}_{26,\mbk a}
=&
- \frac{2 \pi^2}{\hbar}  e E^{\mu} \tau_{\mbk}^a
\sum_{\mbkpr \mbkdpr bcde}
\delta ( \veps_{\mbk}^a - \veps_{\mbkpr}^b )
\frac{
\text{Im}
\left[
\Braket{
U_{\mbk \mbk\pr}^{ab} U_{\mbk\pr \mbk} ^{ca} 
}
\Braket{
U_{\mbkpr \mbkdpr}^{bd} U_{\mbkdpr \mbkpr} ^{de} 
}
\mathcal{A}_{\mu \mbkpr}^{\prime ec}
\right]_{\substack{c \neq b}}
}
{(\veps_{\mbk}^a - \veps_{\mbkpr}^c)
(\veps_{\mbkpr}^c - \veps_{\mbkpr}^e)}
\\
&\times
\left[
\left(
n_{E \mbkpr c}^{(-1)} - n_{E \mbkdpr d}^{(-1)}
\right)
\delta ( \veps_{\mbkpr}^c - \veps_{\mbkdpr}^d ) 
\right.
-
\left.
\left(
n_{E \mbkpr e}^{(-1)} - n_{E \mbkdpr d}^{(-1)}
\right)
\delta ( \veps_{\mbkpr}^e - \veps_{\mbkdpr}^d ) 
\right]
\\
&-
\frac{2 \pi^2}{\hbar}  e E^{\mu} \tau_{\mbk}^a
\sum_{\mbkpr \mbkdpr bcde}
\delta ( \veps_{\mbk}^a - \veps_{\mbkpr}^b )
\frac{
\text{Im}
\left[
\Braket{
U_{\mbk \mbk\pr}^{ab} U_{\mbk\pr \mbk} ^{ca} 
}
\Braket{
U_{\mbkpr \mbkdpr}^{bd} U_{\mbkdpr \mbkpr} ^{de} 
}
\mathcal{A}_{\mu \mbkpr}^{\prime ec}
\right]_{\substack{c,e \neq b}}
}
{(\veps_{\mbk}^a - \veps_{\mbkpr}^c)
(\veps_{\mbk}^a - \veps_{\mbkpr}^e)}
\\
&\times
\left[
\left(
n_{E \mbkpr b}^{(-1)} - n_{E \mbkdpr d}^{(-1)}
\right)
\delta ( \veps_{\mbk}^a - \veps_{\mbkdpr}^d ) 
\right.
+
\left.
\left(
n_{E \mbkpr e}^{(-1)} - n_{E \mbkdpr d}^{(-1)}
\right)
\delta ( \veps_{\mbkpr}^e - \veps_{\mbkdpr}^d ) 
\right],
\end{split}
\end{equation}
\begin{equation}
\begin{split}
\mathcal{N}^{(\text{ask})}_{27,\mbk a}
=&
- \frac{2 \pi^2}{\hbar}  e E^{\mu} \tau_{\mbk}^a
\sum_{\mbkpr \mbkdpr bcde}
\delta ( \veps_{\mbk}^a - \veps_{\mbkpr}^b )
\frac{
\text{Im}
\left[
\Braket{
U_{\mbk \mbk\pr}^{ab} U_{\mbk\pr \mbk} ^{ca} 
}
\Braket{
U_{\mbkpr \mbkdpr}^{ed} U_{\mbkdpr \mbkpr} ^{dc} 
}
\mathcal{A}_{\mu \mbkpr}^{\prime be}
\right]_{\substack{c \neq b}}
}
{(\veps_{\mbk}^a - \veps_{\mbkpr}^c)
(\veps_{\mbk}^a - \veps_{\mbkpr}^e)}
\\
&\times
\left[
\left(
n_{E \mbkpr b}^{(-1)} - n_{E \mbkdpr d}^{(-1)}
\right)
\delta ( \veps_{\mbkpr}^b - \veps_{\mbkdpr}^d ) 
\right.
-
\left.
\left(
n_{E \mbkpr e}^{(-1)} - n_{E \mbkdpr d}^{(-1)}
\right)
\delta ( \veps_{\mbkpr}^e - \veps_{\mbkdpr}^d ) 
\right]
\\
&-
\frac{2 \pi^2}{\hbar}  e E^{\mu} \tau_{\mbk}^a
\sum_{\mbkpr \mbkdpr bcde}
\delta ( \veps_{\mbk}^a - \veps_{\mbkpr}^b )
\frac{
\text{Im}
\left[
\Braket{
U_{\mbk \mbk\pr}^{ab} U_{\mbk\pr \mbk} ^{ca} 
}
\Braket{
U_{\mbkpr \mbkdpr}^{ed} U_{\mbkdpr \mbkpr} ^{dc} 
}
\mathcal{A}_{\mu \mbkpr}^{\prime be}
\right]_{\substack{e \neq c}}
}
{(\veps_{\mbk}^a - \veps_{\mbkpr}^e)
(\veps_{\mbkpr}^e - \veps_{\mbkpr}^c)}
\\
&\times
\left[
\left(
n_{E \mbkpr c}^{(-1)} - n_{E \mbkdpr d}^{(-1)}
\right)
\delta ( \veps_{\mbkpr}^c - \veps_{\mbkdpr}^d ) 
\right.
+
\left.
\left(
n_{E \mbkpr e}^{(-1)} - n_{E \mbkdpr d}^{(-1)}
\right)
\delta ( \veps_{\mbkpr}^e - \veps_{\mbkdpr}^d ) 
\right],
\end{split}
\end{equation}
\begin{equation}
\begin{split}
\mathcal{N}^{(\text{ask})}_{28,\mbk a}
=&
- \frac{2 \pi^2}{\hbar}  e E^{\mu} \tau_{\mbk}^a
\sum_{\mbkpr \mbkdpr bcde}
\delta ( \veps_{\mbk}^a - \veps_{\mbkpr}^e )
\frac{
\text{Im}
\left[
\Braket{
U_{\mbk \mbk\pr}^{ab} U_{\mbk\pr \mbk} ^{ca} 
}
\Braket{
U_{\mbkpr \mbkdpr}^{ed} U_{\mbkdpr \mbkpr} ^{dc} 
}
\mathcal{A}_{\mu \mbkpr}^{\prime be}
\right]_{\substack{e \neq c}}
}
{(\veps_{\mbk}^a - \veps_{\mbkpr}^b)
(\veps_{\mbk}^a - \veps_{\mbkpr}^c)}
\\
&\times
\left[
\left(
n_{E \mbkpr c}^{(-1)} - n_{E \mbkdpr d}^{(-1)}
\right)
\delta ( \veps_{\mbkpr}^c - \veps_{\mbkdpr}^d ) 
\right.
+
\left.
\left(
n_{E \mbkpr e}^{(-1)} - n_{E \mbkdpr d}^{(-1)}
\right)
\delta ( \veps_{\mbkpr}^e - \veps_{\mbkdpr}^d ) 
\right]
\\
&-
\frac{2 \pi^2}{\hbar}  e E^{\mu} \tau_{\mbk}^a
\sum_{\mbkpr \mbkdpr bcde}
\delta ( \veps_{\mbk}^a - \veps_{\mbkpr}^b )
\frac{
\text{Im}
\left[
\Braket{
U_{\mbk \mbk\pr}^{ab} U_{\mbk\pr \mbk} ^{ca} 
}
\Braket{
U_{\mbkpr \mbkdpr}^{ed} U_{\mbkdpr \mbkpr} ^{dc} 
}
\mathcal{A}_{\mu \mbkpr}^{\prime be}
\right]_{\substack{c \neq b,e}}
}
{(\veps_{\mbk}^a - \veps_{\mbkpr}^c)
(\veps_{\mbkpr}^c - \veps_{\mbkpr}^e)}
\\
&\times
\left[
\left(
n_{E \mbkpr c}^{(-1)} - n_{E \mbkdpr d}^{(-1)}
\right)
\delta ( \veps_{\mbkpr}^c - \veps_{\mbkdpr}^d ) 
\right.
+
\left.
\left(
n_{E \mbkpr e}^{(-1)} - n_{E \mbkdpr d}^{(-1)}
\right)
\delta ( \veps_{\mbkpr}^e - \veps_{\mbkdpr}^d ) 
\right],
\end{split}
\end{equation}
\begin{equation}
\begin{split}
\mathcal{N}^{(\text{ask})}_{29,\mbk a}
=&
- \frac{2 \pi^2}{\hbar}  e E^{\mu} \tau_{\mbk}^a
\sum_{\mbkpr \mbkdpr bcde}
\left[
\delta ( \veps_{\mbk}^a - \veps_{\mbkpr}^b )
-
\delta ( \veps_{\mbk}^a - \veps_{\mbkpr}^c )
\right]
\frac{
\text{Im}
\left[
\Braket{
U_{\mbk \mbk\pr}^{ab} U_{\mbk\pr \mbk} ^{cd} 
}
\Braket{
U_{\mbk \mbkdpr}^{de} U_{\mbkdpr \mbk} ^{ea} 
}
\mathcal{A}_{\mu \mbkpr}^{\prime bc}
\right]_{\substack{d \neq a}}
}
{(\veps_{\mbk}^a - \veps_{\mbk}^d)(\veps_{\mbkpr}^b - \veps_{\mbkpr}^c)}
\\
&
\times
\left[
\left(
n_{E \mbk a}^{(-1)} - n_{E \mbkdpr e}^{(-1)}
\right)
\delta ( \veps_{\mbk}^a - \veps_{\mbkdpr}^e ) 
\right.
+
\left.
\left(
n_{E \mbk d}^{(-1)} - n_{E \mbkdpr e}^{(-1)}
\right)
\delta ( \veps_{\mbk}^d - \veps_{\mbkdpr}^e ) 
\right],
\end{split}
\end{equation}
\begin{equation}
\begin{split}
\mathcal{N}^{(\text{ask})}_{30,\mbk a}
=&
- \frac{2 \pi^2}{\hbar}  e E^{\mu} \tau_{\mbk}^a
\sum_{\mbkpr \mbkdpr bcde}
\left[
\delta ( \veps_{\mbk}^a - \veps_{\mbkpr}^b )
-
\delta ( \veps_{\mbkpr}^b - \veps_{\mbk}^d )
\right]
\frac{
\text{Im}
\left[
\Braket{
U_{\mbk \mbk\pr}^{ab} U_{\mbk\pr \mbk} ^{cd} 
}
\Braket{
U_{\mbkpr \mbkdpr}^{be} U_{\mbkdpr \mbkpr} ^{ec} 
}
\mathcal{A}_{\mu \mbk}^{\prime da}
\right]_{\substack{c \neq b}}
}
{(\veps_{\mbk}^a - \veps_{\mbk}^d)(\veps_{\mbkpr}^b - \veps_{\mbkpr}^c)}
\\
&
\times
\left[
\left(
n_{E \mbkpr b}^{(-1)} - n_{E \mbkdpr e}^{(-1)}
\right)
\delta ( \veps_{\mbkpr}^b - \veps_{\mbkdpr}^e ) 
\right.
+
\left.
\left(
n_{E \mbkpr c}^{(-1)} - n_{E \mbkdpr e}^{(-1)}
\right)
\delta ( \veps_{\mbkpr}^c - \veps_{\mbkdpr}^e ) 
\right].
\end{split}
\end{equation}

\section{Carrier densities and conductivities of Dirac fermions}
\label{app_dirac}

\subsection{Carrier densities}
For the model of Dirac fermions given by Eq.~(\ref{H_Dirac}), to first order in tilting, the linear-response carrier densities associated with special scattering are obtained as
\begin{equation}
\label{n_Ek_sj_dirac}
n_{E \mbk}^{(\text{sj}) +}
=
\frac{e}{2 \hbar} \frac{n_I U_0^2}{v} E_x
\tau_{\mbk}^+ 
\frac{\pd f_{0 \mbk}^+}{\pd \veps_{\mbk}^+}
\cos \theta
\left[ \sin \theta \sin \phi
+
\frac{t_x}{4 v} \sin^2 \theta \sin 2\phi
-
\frac{t_y}{4 v} (3 + \cos 2\theta + \sin^2 \theta \cos 2\phi)
\right],
\end{equation}
\begin{equation}
\begin{split}
n_{E \mbk}^{(\text{sk}) +}
&=
\frac{3 e}{64 \hbar^2}
( n_I U_0^2 )^2
\frac{h_k}{v^3} E_x
(\tau_{\mbk}^+)^2 
\frac{\pd f_{0 \mbk}^+}{\pd \veps_{\mbk}^+}
\sin \theta \sin 2\theta
\biggl\{
\sin \theta \sin \phi
-
\frac{t_x}{6v} (1 + 5 \cos^2 \theta) \sin 2\phi
\\
&\hspace{.4\linewidth} -
\frac{t_y}{v} \left[1 + \cos^2 \theta
-
\frac{1}{6} (1 + 5 \cos^2 \theta) \cos 2\phi \right]
\biggr\},
\end{split}
\end{equation}
\begin{equation}
\begin{split}
n_{E \mbk}^{(\text{sk}) -}
&=
- \frac{3 e}{64 \hbar^2}
( n_I U_0^2 )^2
\frac{h_k}{v^3} E_x
\tau_{\mbk}^+ \tau_{\mbk}^-
\frac{\pd f_{0 \mbk}^+}{\pd \veps_{\mbk}^+}
\sin \theta \sin 2\theta
\biggl\{
\sin \theta \sin \phi
-
\frac{5 t_x}{6v} (1 + \cos^2 \theta) \sin 2\phi
\\
&\hspace{.4\linewidth} +
\frac{t_y}{v} \left[\frac{1}{3} - \cos^2 \theta
+
\frac{5}{6} (1 + \cos^2 \theta) \cos 2\phi \right]
\biggr\},
\end{split}
\end{equation}
where we note that $n_{E \mbk}^{(\text{sj}) -}=0$ when the Fermi level lies in the upper band. Note also that in the limit of vanishing tilting, these reduce to the forms obtained in Ref.~\cite{atencia2022semiclassical}.

Moving on to the mixed scattering densities in the quadratic responses, recalling Eq.~(\ref{n_E2k_sjo}), we find
\begin{equation}
\begin{split}
\mathcal{N}^{(\text{sj,o})}_{1,\mbk +}
=&
\frac{e^2}{4 \hbar} \frac{n_I U_0^2}{v h_k}
E_x^2 
\left(\tau_{\mbk}^+\right)^2 v_{x \mbk}^+
\frac{\pd f_{0 \mbk}^+}{\pd \veps_{\mbk}^+} \cos \theta
\left[
\sin \phi \sin \theta
+
\frac{t_x}{v} \sin 2\phi \sin^2 \theta
\right.
\\
&-
\left.
\frac{t_y}{v} \left( \cos 2 \theta + \sin^2 \theta \cos 2\phi \right)
\right],
\end{split}
\end{equation}
and
\begin{equation}
\mathcal{N}^{(\text{sj,o})}_{2,\mbk +}
=
- \frac{e^2}{4 \hbar} \frac{v^2}{h_k^2}
E_x^2 
\tau_{\mbk}^+
\frac{\pd f_{0 \mbk}^+}{\pd \veps_{\mbk}^+}
\frac{
\sin 2\theta \sin^2 \theta}
{5 + 3 \cos 2\theta}
\left[
\frac{t_x}{v} 
\left(3 \sin \phi + \sin 3\phi \right)
+
\frac{t_y}{v} 
\left(3 \cos \phi - \cos 3 \phi \right)
\right].
\end{equation}
The skew scattering contribution given by Eq.~(\ref{n_E2_sko}) may also be decomposed as
\begin{equation}
n_{E^2 \mbk +}^{(\text{sk,o})}
=
\mathcal{N}_{1, \mbk +}^{(\text{sk,o})}
+
\mathcal{N}_{2, \mbk +}^{(\text{sk,o})},
\end{equation}
where
\begin{equation}
\begin{split}
\mathcal{N}_{1, \mbk +}^{(\text{sk,o})}
=&
\frac{e^2}{128 \hbar} E_x^2 \tau_{\mbk}^+
\frac{\pd f_{0 \mbk}^+}{\pd \veps_{\mbk}^+}
\frac{v^2}{h_k^2}
\frac{ \sin 2\theta }
{ ( 5 + 3 \cos 2\theta)^4 }
\bigg\{
\frac{3 t_x}{v} \sin \phi (3487
+
6504 \cos 2\theta 
+
2188 \cos 4\theta
\\
&
+
88 \cos 6\theta
+
21 \cos 8\theta
)
+
\frac{2 t_y}{v} \cos \phi (9463
+
9912 \cos 2\theta 
+
1100 \cos 4\theta
+
8 \cos 6\theta
\\
&-
3 \cos 8\theta
)
\biggr\},
\end{split}
\end{equation}

\begin{equation}
\begin{split}
\mathcal{N}_{2, \mbk +}^{(\text{sk,o})}
=&
\frac{3 e^2}{8 \hbar} E_x^2 \tau_{\mbk}^+
\frac{\pd^2 f_{0 \mbk}^+}{\pd (\veps_{\mbk}^+)^2}
\frac{v^2}{h_k}
\frac{ \sin^2\theta \sin 2\theta }
{ ( 5 + 3 \cos 2\theta)^3 }
\bigg\{
\frac{t_x}{v} \sin \phi (197
+
148 \cos 2\theta 
+
7 \cos 4\theta
)
\\
&+
\frac{t_y}{v} \cos \phi (29
+
4 \cos 2\theta 
-
\cos 4\theta
)
\biggr\},
\end{split}
\end{equation}
and
\begin{equation}
n_{E^2 \mbk -}^{(\text{sk,o})}
=
\mathcal{N}_{1, \mbk -}^{(\text{sk,o})}
+
\mathcal{N}_{2, \mbk -}^{(\text{sk,o})}
+
\mathcal{N}_{3, \mbk -}^{(\text{sk,o})},
\end{equation}
with
\begin{equation}
\begin{split}
\mathcal{N}_{1, \mbk -}^{(\text{sk,o})}
=&
- \frac{e^2}{128 \hbar} E_x^2 \tau_{\mbk}^-
\frac{\pd f_{0 \mbk}^+}{\pd \veps_{\mbk}^+}
\frac{v^2}{h_k^2}
\frac{ \sin 2\theta }
{ ( 5 + 3 \cos 2\theta)^4 }
\bigg\{
\frac{t_x}{v} \sin \phi (45677
+
68896 \cos 2\theta 
+
20628 \cos 4\theta
\\
&-
96 \cos 6\theta
+
63 \cos 8\theta
)
-
\frac{t_y}{v} \cos \phi (16523
+
32240 \cos 2\theta 
+
13100 \cos 4\theta
-
432 \cos 6\theta
\\
&+
9 \cos 8\theta
)
\biggr\},
\end{split}
\end{equation}

\begin{equation}
\begin{split}
\mathcal{N}_{2, \mbk -}^{(\text{sk,o})}
=&
- \frac{e^2}{16 \hbar} E_x^2 \tau_{\mbk}^-
\frac{\pd^2 f_{0 \mbk}^+}{\pd (\veps_{\mbk}^+)^2}
\frac{v^2}{h_k}
\frac{ \sin^2\theta \sin 2\theta }
{ ( 5 + 3 \cos 2\theta)^3 }
\bigg\{
\frac{t_x}{v} \sin \phi (1819
+
1516 \cos 2\theta 
+
57 \cos 4\theta
)
\\
&-
\frac{t_y}{v} \cos \phi (477
+
596 \cos 2\theta 
+
15 \cos 4\theta
)
\biggr\},
\end{split}
\end{equation}

\begin{equation}
\label{N_3k_sko_dirac}
\begin{split}
\mathcal{N}_{3, \mbk -}^{(\text{sk,o})}
=&
\frac{e^2}{16 \hbar^2} (n_I U_0^2)^2 E_x^2 \tau_{\mbk}^+ \tau_{\mbk}^-
\left[
\pd_{\mbk}^x (\tau_{\mbk}^+ v_{x \mbk}^+)
\frac{\pd f_{0 \mbk}^+}{\pd \veps_{\mbk}^+}
+
\hbar \tau_{\mbk}^+ (v_{x \mbk}^+)^2
\frac{\pd^2 f_{0 \mbk}^+}{\pd (\veps_{\mbk}^+)^2}
\right]
\frac{h_k}{v^4} \sin 2\theta
\\
&\times \left(
\frac{t_x}{v} \sin \phi
-
\frac{t_y}{v} \cos \phi
\right).
\end{split}
\end{equation}
And the remaining mixed scattering terms are readily evaluated by taking the derivatives of the linear response densities.

\begin{table}
\caption{Transport coefficients of the lonigitudinal and transverse nonlinear conductivities.}

\begin{tabular}{l c c c c c}
{} & $n=0$ & $n=1$ & $n=2$ & $n=3$ & \multicolumn{1}{c}{$n=4$}
\\
\hline
$a_{2 n}^{(\text{o})}$ & 6207 & 6328 & -788 & 520 & 21
\\
$b_{2 n}^{(\text{o})}$ & 2665 & 2168 & -684 & -56 & 3
\\
$a_{2 n}^{(\text{sj,o})}$ & 109 & 139 & 15 & 0 & 0
\\
$b_{2 n}^{(\text{sj,o})}$ & 229 & 84 & 7 & 0 & 0
\\
$a_{2 n+1}^{(\text{o,sj})}$ & 9823 & 5301 & 1215 & 45 & 0 
\\
$b_{2 n}^{(\text{o,sj})}$ & 105 & 148 & 3 & 0 & 0
\\
$a_{2 n+1}^{(\text{sk,o})}$ & 3662 & 7856 & -1624 & -1717 & 15
\\
$b_{2 n+1}^{(\text{sk,o})}$ & 114852 & 59382 & 13258 & 987 & -63 
\\
$a_{2 n}^{(\text{o,sk})}$ & 7422 & 10159 & 2770 & 129 & 0 
\\
$b_{2 n}^{(\text{o,sk})}$ & 389 & 428 & 15 & 0 & 0
\\
$a_{2 n+1}^{(\text{ex})}$ & 6583 & 3057 & 555 & 45 & 0 
\\
$b_{2 n+1}^{(\text{ex})}$ & 15833 & 7375 & 1269 & 99 & 0 
\\
\end{tabular}
\label{tab1}
\end{table}

\subsection{Conductivities}

Here, we present the explicit forms of the functions appearing in the nonlinear conductivities in Section~(\ref{application}). The first functions read
\vspace{-.2cm}
\begin{subequations}
\label{S_muxx^o}
\begin{align}
\mathcal{S}_{xxx}^{(\text{o})}(\theta)
&=
\sum_{n=0}^{4}
\frac{a_{2 n}^{(\text{o})} \cos 2 n\theta}
{( 5 + 3 \cos 2 \theta )^4},
\\
\mathcal{S}_{yxx}^{(\text{o})}(\theta)
&=
\sum_{n=0}^{4}
\frac{b_{2 n}^{(\text{o})} \cos 2 n\theta}
{( 5 + 3 \cos 2 \theta )^4},
\end{align}
\end{subequations}
%\vspace{-.2cm}
where $a_{2 n}^{(\text{o})}$ and $b_{2 n}^{(\text{o})}$ are numerical factors that emerge from the analytical integrations, the values of which are presented in Table~\ref{tab1}. The remaining functions read
\vspace{-.35cm}
\begin{subequations}
\begin{align}
\mathcal{S}_{xxx}^{(\text{sj,o})}(\theta)
&=
\sum_{n=0}^{2}
\frac{a_{2 n}^{(\text{sj,o})} \cos 2 n\theta \sin \theta \sin 2\theta} 
{( 5 + 3 \cos 2 \theta )^3},
\\
\mathcal{S}_{yxx}^{(\text{sj,o})}(\theta)
&=
\sum_{n=0}^{2}
\frac{b_{2 n}^{(\text{sj,o})} \cos 2 n\theta \sin \theta \sin 2\theta} 
{( 5 + 3 \cos 2 \theta )^3},
\end{align}
\end{subequations}
\begin{subequations}
\begin{align}
\mathcal{S}_{xxx}^{(\text{o,sj})}(\theta)
&=
\sum_{n=0}^{3}
\frac{a_{2n+1}^{(\text{o,sj})} \cos (2n+1) \theta} 
{( 5 + 3 \cos 2 \theta )^4},
\\
\mathcal{S}_{yxx}^{(\text{o,sj})}(\theta)
&=
\sum_{n=0}^{2}
\frac{b_{2n}^{(\text{o,sj})} \cos 2n \theta \sin \theta \sin 2\theta} 
{( 5 + 3 \cos 2 \theta )^4},
\end{align}
\end{subequations}
\begin{subequations}
\begin{align}
\mathcal{S}_{xxx}^{(\text{sk,o})}(\theta)
&=
\sum_{n=0}^{4}
\frac{a_{2n+1}^{(\text{sk,o})} \cos (2n+1)\theta \sin^2 \theta} 
{( 5 + 3 \cos 2 \theta )^5},
\\
\mathcal{S}_{yxx}^{(\text{sk,o})}(\theta)
&=
\sum_{n=0}^{4}
\frac{b_{2n+1}^{(\text{sk,o})} \cos (2n+1)\theta \sin^2 \theta} 
{( 5 + 3 \cos 2 \theta )^5},
\end{align}
\end{subequations}
\begin{subequations}
\begin{align}
\mathcal{S}_{xxx}^{(\text{o,sk})}(\theta)
&=
\sum_{n=0}^{3}
\frac{a_{2n}^{(\text{o,sk})} \cos 2n\theta \sin \theta \sin 2 \theta} 
{( 5 + 3 \cos 2 \theta )^5},
\\
\mathcal{S}_{yxx}^{(\text{o,sk})}(\theta)
&=
\sum_{n=0}^{2}
\frac{b_{2n}^{(\text{o,sk})} \cos 2n \theta \sin^3 \theta \sin 2 \theta} 
{( 5 + 3 \cos 2 \theta )^5},
\end{align}
\end{subequations}
\begin{subequations}
\begin{align}
\mathcal{S}_{xxx}^{(\text{ex})}(\theta)
&=
\sum_{n=0}^{3}
\frac{a_{2n+1}^{(\text{ex})} \cos (2n+1)\theta \sin^2 \theta} 
{( 5 + 3 \cos 2 \theta )^4},
\\
\mathcal{S}_{yxx}^{(\text{ex})}(\theta)
&=
\sum_{n=0}^{3}
\frac{b_{2n+1}^{(\text{ex})} \cos (2n+1)\theta \sin^2 \theta} 
{( 5 + 3 \cos 2 \theta )^4},
\end{align}
\end{subequations}
\begin{equation}
\label{S_yxx^BC}
\mathcal{S}_{yxx}^{(\text{BC})}(\theta)
=
\frac{ (7 + 3 \cos 2\theta) \sin \theta \sin 2\theta} 
{( 5 + 3 \cos 2 \theta )^2}.
\end{equation}

\section{Conclusion and Outlook}

In this chapter, we have presented a nonlinear response theory based on a recently-developed density-matrix formalism in the linear response regime~\cite{culcer2017interband, sekine2017quantum, atencia2022semiclassical}, whereby the density matrix is decomposed into disorder-averaged and fluctuating parts. By treating the diagonal and off-diagonal elements of the density matrix in band space on an equal footing, the full contribution of the disorder interaction can be taken into account in a systematic manner. And through the perturbative solution of the quantum Liouville equation in the presence of electrostatic and weak disorder potentials, we have shown how the combined actions of the electric field and random impurities give rise to a multitude of scattering channels in the quadratic response regime, resulting in a classification of carrier densities in terms of distinct physical processes.

Moreover, in addition to the previously known linear-response extrinsic velocity, several extrinsic velocities have been identified, which only appear in the nonlinear response of the system, revealing the rich structure of interband coherence effects in this transport regime. As an application of this theory, we have studied the quadratic response in a prototypical model of tilted Dirac fermions, and have derived the leading and subleading disorder contributions to the nonlinear conductivities.

This suggests several follow-up studies. One possible direction would be to extend the analysis to nonlinear spin currents, which would allow for a detailed classification of spin-orbit torques in the nonlinear response regime. Another direction would be to apply the theory to more complicated disorder profiles, which have recently been shown to play a role in generating nonlinear responses~\cite{dyrdal2020spin, mehraeen2024proximity, boboshko2024bilinear}. Yet another possible direction would be to explore the zeroth-order contributions in specific materials--as has recently been initiated through other approaches as well~\cite{huang2023scaling, gong2024nonlinear}--in order to develop a better understanding of the contributions of the quadratic-response extrinsic velocities and their interplay with quantum geometry in this transport regime.

In closing, we note some physically relevant limits that can be relaxed in future studies based on the present approach. First, we have focussed our analysis on DC responses. Generalizing this to the AC regime, which is particularly significant in the study of nonlinear optical responses, will allow for the continued exploration of the resonance structures of nonlinear responses. This can have interesting signatures in quantum materials~\cite{bhalla2020resonant} and can be utilized as useful probes of quantum geometry~\cite{bhalla2022resonant}. In addition, in this initial study, we have limited our analysis to low temperatures. Relaxing this enables further analysis of finite-temperature effects via the quantum kinetic approach, including the effects of thermal magnons and nonlinear thermoelectric responses~\cite{varshney2023quantum}. Finally, while it is customary to include the effect of a magnetic field in semiclassical transport studies of spin-orbit-coupled materials via the Zeeman interaction--similar to the exchange coupling in magnetic systems, a thorough analysis would also require the inclusion of the magnetic field in the derivation of the equations of motion. This leads to additional contributions, such as the coupling to the Berry curvature and modifying the density of states, which can then manifest in the nonlinear response of the system~\cite{gao2014field}.

\chapter{Quantum Response Theory and Momentum-Space Gravity}

\section{Introduction}

The space of physical states of a quantum mechanical system--complex projective Hilbert space--has the structure of a Kahler manifold~\cite{nakahara2018geometry}. This has inspired various geometric and topological interpretations of physical phenomena in quantum systems over the years~\cite{ashtekar1999geometrical, xiao2010berry}. And, in several areas of exploration, this quantum state geometry has recently garnered immense interest~\cite{torma2023essay, liu2024quantum, yu2024quantum, chen2024quantum, shim2025spin, jiang2025revealing, verma2025quantum}, with the identification of the role of the quantum metric~\cite{provost1980riemannian}--the real part of the quantum geometric tensor (QGT)--in an increasing number of physical effects~\cite{neupert2013measuring, claassen2015position, ahn2020low, watanabe2021chiral, wang2021intrinsic, liu2021intrinsic, bhalla2022resonant, gao2023quantum, wang2023quantum, hetenyi2023fluctuations, das2023intrinsic, komissarov2024quantum, onishi2024fundamental, fang2024quantum, kang2024measurements, jankowski2025optical}.

While the dynamical significance of the Berry curvature has been known for decades~\cite{karplus1954hall, kohn1957quantum, adams1959energy, chang1995berry, chang1996berry, sundaram1999wave}, it is perhaps the emergence of the quantum-metric Levi-Civita connection~\cite{gao2014field} that truly reveals the geometric underpinning of carrier dynamics and highlights the geodesic nature of the motion of Bloch electrons. Combining this with the Einstein field equations (EFE) arising from the Riemannian structure of quantum state manifolds has led to the intriguing recent proposal of momentum-space gravity~\cite{smith2022momentum} in condensed matter systems.

In this framework, the quantum metric in momentum space is viewed as the dual of the classical spacetime metric and the intrinsic dynamics is described by a dual Lorentz force. Furthermore, for mixed states, the momentum-space EFE acquire a source term that depends on the von Neumann entropy, which reveals intriguing similarities with efforts in understanding the fundamental nature of gravity and spacetime~\cite{bekenstein1973black, hawking1975particle, ruppeiner1979thermodynamics, jacobson1995thermodynamics, padmanabhan2010thermodynamical, verlinde2011origin, carroll2016what, bianconi2025gravity} and suggests a materials setting for exploring the connections between gravity and thermodynamics. Given the semiclassical nature of this approach based on wavepacket dynamics, a natural question that arises is the form this theory of gravity takes in momentum space from the vantage point of quantum response theory.

\begin{figure}[t]
%\vspace{-.7cm}
\captionsetup[subfigure]{labelformat=empty}
    \sidesubfloat[]{\includegraphics[width=0.5\linewidth,trim={1cm 0cm 1cm 1cm}]{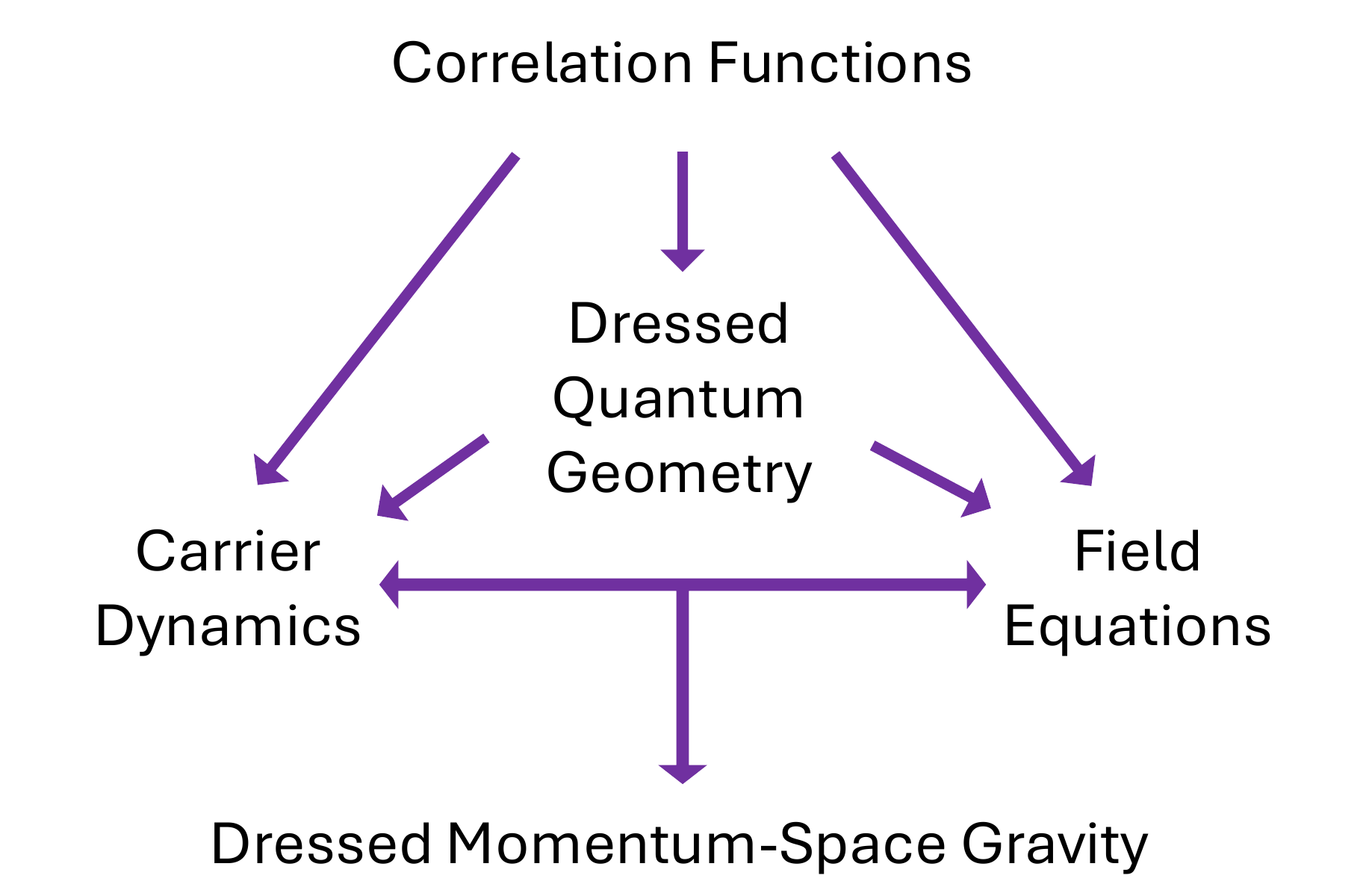}\label{fig_qmsg_fig1a}}
%\quad%
%%%%%%%%%
    \caption{Schematic summary of this chapter. Correlation functions corresponding to conductivity Kubo formulas are applied to simultaneously dress with dissipation the quantum geometry, carrier dynamics and momentum-space field equations. The latter two yield dressed versions of the two central equations of general relativity, resulting in dressed gravity in momentum space.}
    \label{fig_qmsg_fig1}
\end{figure}

Here~\cite{mehraeen2025quantum}, we address this question through a diagrammatic approach, with the general idea underlying this chapter presented schematically in Fig.~\ref{fig_qmsg_fig1}. We first apply Kubo formulas to propose a diagrammatic generalization of carrier dynamics to multiband systems in the presence of finite dissipation.  To this end, we build on recent developments~\cite{ahn2022riemannian, bouhon2023quantum, mitscherling2024gauge, avdoshkin2024multi, jankowski2024quantized, jankowski2025enhancing} in two-state~\footnote{Here, we refer to an $n$-state quantum geometric object as one that carries $n$ band indices. For example, $g_{\mu\nu}^a$ and $g_{\mu\nu}^{ab}$ are the single- and two-state quantum metric tensors, respectively.} quantum geometry to propose the \textit{three-state} QGT and show that it appears already at the quadratic-response level. We also discuss the appearance of the quantum geometric \textit{contorsion} tensor~\cite{nakahara2018geometry} and analogous symplectic terms when the full Berry covariant derivative is used to derive the Hermitian connection components. 

Physically, much intuitive insight has been gained from viewing the Berry curvature as an intrinsic gauge field strength tensor in momentum space~\cite{xiao2010berry}. Following this, and the standard argument in general relativity~\cite{misner1973gravitation, wald2010general}, we argue that once the choice of connection is made to be Levi-Civita, the remaining terms from the Hermitian connection and three-state QGT in the carrier position's equation of motion can then be interpreted as intrinsic matter fields arising from the multistate quantum geometry.

A noteworthy feature of the present approach is the appearance of dressed quantum geometric quantities in the equations of motion and EFE. While we discuss the gravitational significance of this later on, here we mention that there is increasing ongoing effort to define quantum geometry in interacting and disordered systems ~\cite{souza2000polarization, Michishita2022dissipation, chen2022measurement, kashihara2023quantum, zhou2024sloqvist, romeral2025scaling, sukhachov2025effect}. The approach which naturally emerges in the present derivation is closest to that of Ref.~\cite{Michishita2022dissipation} and, with the gravitational context in mind, is essentially a dissipation-induced multiband Weyl transformation of the quantum geometry. We apply this result to show that the dissipative correction to the EFE can be related to the entropic cost of dressing the metric through a dual momentum-space \textit{drag force} induced by the dissipative scattering. This, in turn, can be thought of as providing a diagrammatic generalization of the corresponding semiclassical arguments to dissipative multiband systems. We conclude by presenting an outlook on possible insights that may be obtained from these results.

\section{Multistate quantum geometry}

In the presence of an applied electric field, the electric dipole interaction $e \mathbf{E} \cdot \mathbf{r}$ is included to the system Hamiltonian, where $\mathbf{r}$ is the position operator. In the k-space representation, this takes the form~\cite{blount1962formalisms}
$\mathbf{r}_{\mbk \mbk\pr}
=
-i \bs{\pd}_{\mbk\pr} \delta_{\mbk\pr \mbk}
+
\delta_{\mbk\pr \mbk} \bs{\mathcal{A}}_{\mbk}$, where $\mathcal{A}_{\mu \mbk}^{ab}
 =
 i \braket{u_{\mbk}^a | \pd_{\mu}^{\mbk} u_{\mbk}^b}$ is the Berry connection and $\ket{u_{\mbk}^a}$ the periodic part of the Bloch state~\footnote{Regarding notation, we use Greek letters for real- or momentum-space indices--with Einstein summation implied--and Latin letters for band indices. Furthermore, $\mathcal{O}_{(\mu\nu)} \equiv (\mathcal{O}_{\mu\nu} +  \mathcal{O}_{\nu\mu})/2$ and $\mathcal{O}_{[\mu\nu]} \equiv (\mathcal{O}_{\mu\nu} -  \mathcal{O}_{\nu\mu})/2$.}. We henceforth work locally in momentum space and thus drop the momentum parameter from the notation. The commutator of the position operator and local momentum-space operators naturally defines a Berry covariant derivative
 $\bs{\mathcal{D}} \mathcal{O}
 \equiv
 -i [\mathbf{r},\mathcal{O}]$, which takes the form
\begin{equation}
(\mathcal{D}_{\mu} \mathcal{O})^{ab}
=
 \pd_{\mu} \mathcal{O}^{ab}
 -
 i \left[
 \mathcal{A}_{\mu}, \mathcal{O} \right]^{ab}.
\end{equation}
It should be stressed that this includes the full Berry connection, which is decomposed into diagonal and off-diagonal elements in band space as
$\mathcal{A}_{\mu}^{ab}
=
a_{\mu}^{a} \delta^{ab}
+
\mathcal{A}_{\mu}^{\pr ab}$.

The transition dipole matrix element $\mathcal{A}_{\mu}^{\pr ab}$ can be viewed as a component of the complex-valued vielbein matrix $e_{\mu}^{ab}
=
\mathcal{A}_{\mu}^{\pr ab} \ket{u^a} \bra{u^b}$ in band space~\cite{ahn2022riemannian}, thereby providing a geometric rationale for using off-diagonal Berry connection components as fundamental elements in the construction of multistate quantum geometric quantities. And the Hilbert-Schmidt inner product 
$\braket{A,B} = \text{Tr}(A^{\dagger} B)$
induces complex Riemannian structure on the manifold of quantum states and allows for the definition of quantum geometric invariants. The simplest such two-state object is the well-known QGT, which is obtained from the inner product of tangent basis vectors,
$Q_{\mu\nu}^{ab}
\equiv
\braket{e_{\nu}^{ab}, e_{\mu}^{ab}}
=
\mathcal{A}_{\mu}^{\pr ab} \mathcal{A}_{\nu}^{\pr ba}$, and which obeys the projector calculus identity
\begin{equation}
\label{eq_proj2}
i^2 \bra{u^a} P^a \pd_{\mu} P^b \pd_{\nu} P^a
\ket{u^a}
=
Q_{\mu\nu}^{ab}
-
\delta^{ab} \sum_c Q_{\mu\nu}^{ac},
\end{equation}
with the projection operator given by
$P^a = \ket{u^a}\bra{u^a}$. Note that for $b \neq a$, Eq.~(\ref{eq_proj2}) may be regarded as an alternative definition of the QGT in the projector calculus language. Indeed, this approach can be used to identify other elements of two-state quantum geometry~\cite{avdoshkin2024multi, mitscherling2024gauge}, with the two-state QGT being the primary member. The real and imaginary parts of this tensor define the two-state quantum metric and Berry curvature tensors,
$g_{\mu \nu}^{ab}
=
\text{Re} (Q_{\mu\nu}^{ab})$ and
$\Omega_{\mu \nu}^{ab}
=
-2 \text{Im} (Q_{\mu\nu}^{ab})$. Furthermore, we note that the two-state QGT is basically the band resolution of single-state quantum geometry, and yields the familiar Fubini-Study metric and (single-state) Berry curvature tensors once one sums over intermediate states, as is done in the second term on the right in Eq.~(\ref{eq_proj2}).

Moving beyond the QGT, the position operator acting on the tangent basis vectors induces a Hermitian connection~\cite{ahn2022riemannian}, the components of which are given by
$C_{\mu \nu \rho}^{ba}
=
\mathcal{A}_{\mu}^{\pr ab}
(\mathcal{D}_{\nu} \mathcal{A}_\rho^{\pr})^{ba}$. Using the standard definition of the (two-state) torsion tensor,
$\mathcal{T}_{\mu \nu \rho}^{ba}
=
2 C_{\mu [\nu \rho]}^{ba}$, we note that the real part of the Hermitian connection is expressed as
\begin{equation}
\text{Re} \left( C_{\mu \nu \rho}^{ba}\right)
=
\Gamma_{\mu \nu \rho}^{ba}
+
\text{Re} \left( K_{\mu \nu \rho}^{ba} \right),
\end{equation}
where
$\Gamma_{\nu \rho \mu}^{ba}
=
\frac{1}{2} 
( \pd_{\mu} g_{\nu \rho}^{ba}
+
\pd_{\rho} g_{\mu \nu}^{ba}
-
\pd_{\nu} g_{\mu \rho}^{ba})$
is a Levi-Civita connection component of the band-resolved quantum metric and the second term, given by
\begin{equation}
K_{\mu \nu \rho}^{ba}
=
\frac{1}{2}
\left(
\mathcal{T}_{\mu \nu \rho}^{ba}
-
\mathcal{T}_{\nu \mu \rho}^{ba}
-
\mathcal{T}_{\rho \mu \nu}^{ba}
\right),
\end{equation}
is identified as the quantum geometric contorsion tensor~\cite{nakahara2018geometry}, which contains the torsionful part of the metric connection and arises once one uses the full Berry covariant derivative to define the Hermitian connection, as opposed to using only the diagonal Berry connection components. We note that the contorsion tensor inherits the antisymmetric property
$K_{\mu \nu \rho}^{ba}
=
K_{[\mu \nu] \rho}^{ba}$ from the torsion tensor.

Motivated by a geometric characterization of nonlinear responses, we now present an extension of this framework beyond the two-state formalism by introducing the three-state QGT 
\begin{equation}
Q_{\mu \nu \rho}^{abc}
=
\mathcal{A}_{\mu}^{\pr ab} \mathcal{A}_{\nu}^{\pr bc} \mathcal{A}_{\rho}^{\pr ca},
\end{equation}
which cyclically connects three distinct states, and represents a natural generalization of its two-state counterpart. Formally, this can be defined by utilizing the non-Abelian QGT~\cite{provost1980riemannian, ma2010abelian, bouhon2023quantum, jankowski2025enhancing} in operator form, composed of basis vector derivatives, as
$\mathcal{Q}_{\rho \nu}^c
=\ket{\pd_{\rho} u^c} \bra{\pd_{\nu} u^c}$, which allows for the definition
\begin{equation}
Q_{\mu \nu \rho}^{abc}
\equiv
\Braket{\mathcal{Q}_{\rho \nu}^c, e_{\mu}^{ab}},
\end{equation}
with
$\mathcal{(Q}_{\nu \rho}^c)^{ba}
=
\mathcal{A}_{\nu}^{\pr bc} \mathcal{A}_{\rho}^{\pr ca}$. This naturally generalizes the two-state projector identity given by Eq.~(\ref{eq_proj2}) to the three-state identity
\begin{equation}
\label{eq_proj3}
\begin{split}
i^3 \bra{u^a} P^a \pd_{\mu} P^b \pd_{\nu} P^c \pd_{\rho} P^a
\ket{u^a}
=
Q_{\mu\nu\rho}^{abc}
-
\delta^{ab} \sum_d Q_{\mu\nu\rho}^{adc}
+
\delta^{ab} \sum_d Q_{\mu\nu\rho}^{acd}
-
\delta^{bc} \sum_d Q_{\mu\nu\rho}^{abd},
\end{split}
\end{equation}
lending further support to the proposition that the three-state QGT defined above is indeed a central object in higher-state quantum geometry.

We point out that the real and imaginary parts of the three-state QGT are decomposed into linear combinations of the non-Abelian quantum metric and Berry curvature tensors as
\begin{subequations}
\begin{align}
\text{Re} \left( Q_{\mu\nu\rho}^{abc} \right)
&=
\mathcal{A}_{\mu}^{\pr (ab)}
(\textbf{g}_{\nu \rho}^c)^{ba}
-
\frac{i}{2}
\mathcal{A}_{\mu}^{\pr [ab]}
(\bs{\Omega}_{\nu \rho}^c)^{ba},
\\
\text{Im} \left( Q_{\mu\nu\rho}^{abc} \right)
&=
-i \mathcal{A}_{\mu}^{\pr [ab]}
(\textbf{g}_{\nu \rho}^c)^{ba}
-
\frac{1}{2}
\mathcal{A}_{\mu}^{\pr (ab)}
(\bs{\Omega}_{\nu \rho}^c)^{ba},
\end{align}
\end{subequations}
thereby exhibiting a more complex structure compared to the two-state counterpart. The symmetric and antisymmetric parts,
$S_{\mu\nu\rho}^{abc}
=
Q_{\mu(\nu\rho)}^{abc}$
and
$A_{\mu\nu\rho}^{abc}
=
Q_{\mu[\nu\rho]}^{abc}$, are also expressible in terms of these quantities. Specifically, we note that the antisymmetric term is the band resolution of the torsion tensor, i.e.,
$\sum_c A_{\mu\nu\rho}^{abc}
=
-i \mathcal{T}_{\mu\nu\rho}^{ba}/2$, while 
$S_{\mu\nu\rho}^{abc}$ is its symmetric counterpart. As is shown in the next section, the three-state QGT already makes an appearance at the quadratic-response level in dissipative multiband systems and is required for a complete geometric characterization of the response functions.

\section{Dressed carrier dynamics from diagrammatics}

Motivated by recent developments via density-matrix methods in generalizing carrier dynamics~\cite{atencia2022semiclassical,  mehraeen2024quantum}, here, we obtain the equation of motion for the carrier position by taking the route of Kubo formulas. Using recently developed diagrammatic approaches within the Matsubara formalism~\cite{parker2019diagrammatic, Michishita2021effects}, the linear and quadratic ac  conductivities are obtained. In order to obtain dissipative carrier dynamics, we note that the dc current density, 
$j_{\mu}
= 
\sigma_{\mu\nu}E^{\nu}
+
\sigma_{\mu\nu\rho}E^{\nu} E^{\rho}$, is related to the equation of motion of the carrier position and the carrier density $n$ as
$j_{\mu}
=
-e \text{Tr}(n \dot{x}_{\mu})$~\cite{atencia2022semiclassical, mehraeen2024quantum}. Therefore, upon evaluating  the correlation functions, and using the equation of motion of the carrier momentum,
$\dot{k}^{\mu} = -e E^{\mu}/\hbar$,
we arrive at the equation of motion for the carrier position, which is expressed as
\begin{equation}
\label{eom}
\begin{split}
\dot{x}_{\mu}^a
&=
v_{\mu}^a
+
\dot{k}^{\nu} \sum_b 
(
\mathcal{Z}_{\Omega}^{ab} \tilde{\Omega}_{\mu \nu}^{ab}
-
\mathcal{Z}_{g}^{ab} \tilde{g}_{\mu \nu}^{ba}
+
\dot{k}^{\rho} m^{ab} \mathcal{Z}_{\Gamma}^{ab}
\tilde{\Gamma}_{\nu \rho \mu}^{ba})
\\
&+
\dot{k}^{\nu} \dot{k}^{\rho}
\sum_{b} 
( \mathcal{Z}_{m}^{ab} 
\tilde{g}_{\nu \rho}^{ba} \pd_{\mu}
+
\mathcal{M}_{\mu \nu \rho}^{ba}
+
\sum_c \mathcal{N}_{\mu \nu \rho}^{abc} ) m^{ab},
\end{split}
\end{equation}
where $v_{\mu}^a = \pd_{\mu} \veps^a / \hbar$ is the group velocity and the $\mathcal{Z}$'s are various renormalization factors that depend on the dissipation parameter. Eq.~(\ref{eom}) is one of the central results of this chapter and represents a diagrammatic generalization of carrier dynamics to dissipative multiband systems. As can be seen, in addition to dressing the known terms related to the Berry curvature and Levi-Civita connection, which are interpreted as the (multiband) momentum-space magnetic field and geodesic contributions, respectively, several additional contributions are identified. The remainder of this section is devoted to a discussion of these terms and their physical significance, particularly within the context of the gravitational interpretation.

The matrix $m^{ab} = \hbar/ \veps^{ab}$ (with
$\veps^{ab} \equiv \veps^a - \veps^b$) couples to the Christoffel symbol in the equation of motion and may thus be regarded as the local multiband generalization of the effective mass term in Ref.~\cite{smith2022momentum}. Physically, this implies that the effective gravitational force the electron in band $a$ experiences from the Levi-Civita connection between bands $b$ and $a$ is inversely proportional to the local energy difference between the two bands. Interestingly, this force can be attractive or repulsive depending on the sign of $\veps^{ab}$. And since the effective mass is a local quantity, its variation also contributes to the equation of motion, which explains the 
$\pd_{\mu} m^{ab}$ term in the equation of motion.

It is worth elaborating further on the appearance of dressed quantum geometric quantities in Eq.~(\ref{eom}) as opposed to bare ones. With the exception of the Levi-Civita connection, these are defined as
$\tilde{\mathcal{O}}^{ab}
\equiv
\lambda^{ab} \mathcal{O}^{ab}$,
where 
$\lambda^{ab}= [1+ (\eta^{ab})^2]^{-1}$, 
and $\eta^{ab}=\gamma/\veps^{ab}$ is a dimensionless measure of the dissipation strength, with $\gamma$ the self-energy~\footnote{We also note that this choice of dressing has the favorable property of preserving the $a \leftrightarrow b$ symmetries of the quantum metric and Berry curvature in band space.}. We stress that dressing the quantum geometry is an essential inclusion for a gravitational interpretation to hold, as, in general relativity, the matter distribution and the (spacetime) curvature are related and influence each other~\cite{misner1973gravitation}.

In our case, the introduction of scatterers to the system has the effect of screening the quantum geometry through the interband scattering function $\lambda^{ab}$, which acts as a dissipation-induced multiband Weyl transformation on the various geometric terms in the equation of motion. Physically, this accounts for the spectral broadening of the electron propagator from self-energy corrections due to scattering, and results in an effective rescaling of the two-state QGT in the linear response function. And the remaining dissipative corrections that arise in the quadratic responses can then be naturally absorbed in the renormalization functions, implying that the local ratios of the momentum-space forces an electron in a given band experiences from the various geometric terms are also affected by scattering, reflecting the nontrivial modification of the dynamics by dissipation.

The additional quantum geometric objects that appear in Eq.~(\ref{eom}) are given by
\begin{equation}
\mathcal{M}_{\mu \nu \rho}^{ba}
=
\mathcal{Z}_{K}^{ab}
\text{Re} (\tilde{K}_{\mu \nu \rho}^{ba})
+
\mathcal{Z}_{C}^{ab}
\text{Im}
\left(
\tilde{C}_{\mu \nu \rho}^{ba}
-
\tilde{C}_{\nu \mu \rho}^{ba}
-
\tilde{C}_{\rho \nu \mu}^{ba}
\right),
\end{equation}
which contains the contribution from the rest of the dressed Hermitian connection--including the contorsion--and
\begin{equation}
\label{N_munurho}
\begin{split}
\mathcal{N}_{\mu \nu \rho}^{abc}
&=
\mathcal{Z}_1^{abc} \text{Re} (\tilde{S}_{\mu\nu\rho}^{abc})
+
\mathcal{Z}_2^{abc} \text{Im} (\tilde{S}_{\mu\nu\rho}^{abc})
+
\mathcal{Z}_3^{abc} \text{Re} (\tilde{S}_{\nu\mu\rho}^{abc})
\\
&+
\mathcal{Z}_4^{abc} \text{Im} (\tilde{S}_{\nu\mu\rho}^{abc})
+
\mathcal{Z}_5^{abc} \text{Re} (\tilde{A}_{\nu\mu\rho}^{abc})
+
\mathcal{Z}_6^{abc} \text{Im} (\tilde{A}_{\nu\mu\rho}^{abc}),
\end{split}
\end{equation}
which encapsulates the contributions from the dressed three-state QGT~\footnote{Note that we define the dressed three-state QGT in terms of the first two band indices as $\tilde{Q}_{\mu\nu\rho}^{abc} = Q_{\mu\nu\rho}^{abc}/1+(\eta^{ab})^2$}. Together, $\mathcal{M}$ and  $\mathcal{N}$ can be intuitively viewed as describing intrinsic momentum-space matter fields~\cite{misner1973gravitation, wald2010general} that help capture the quantum geometry of Bloch electrons, including the symplectic and non-Abelian contributions, at second order in the carrier momentum.

Finally, consider now the term proportional to $\tilde{g}_{\mu\nu}^{ba}$ in Eq.~(\ref{eom}), which includes a dissipative linear-response addition to the equation of motion
\begin{equation}
\label{eq_drag}
\dot{x}_{\mu}^a
\supset
-2 \dot{k}^{\nu} \sum_b \eta^{ab} 
\tilde{g}_{\mu\nu}^{ba}
\end{equation}
that is also not captured in the semiclassical or density-matrix approaches. Applying the dual-space transformation $\mathbf{x} \leftrightarrow \mathbf{k}$ as discussed in Ref.~\cite{smith2022momentum} to this term reveals that, if the geodesic term is to be regarded as the momentum-space gravitational force, then the contribution from Eq.~(\ref{eq_drag}) can be thought of as arising from a dual drag force in momentum space induced by the scattering. In the next section, we discuss this more within the context of the EFE.

\section{Dissipative Einstein field equations}

We now discuss the effect of dissipation on the multiband momentum-space EFE and how it sources the field equations within the present formalism. In the absence of dissipation, the quantum metric is Fubini-Study, which is the canonical Riemannian metric of complex projective Hilbert space. This is an Einstein metric, with the Ricci tensor being proportional to the metric. Therefore, the bare metric satisfies the sourceless EFE, corresponding to the vacuum solution of the gravitational theory. Consider next the dissipative case. For convenience, we initially drop band indices from the notation, such that the indices $ab$ are implied for the various quantities that appear (i.e.,
$ g_{\mu\nu} \rightarrow g_{\mu\nu}^{ab}, R_{\mu\nu} \rightarrow R_{\mu\nu}^{ab}$, etc.). The inverse of the dressed metric is identified as
$\tilde{g}^{\mu\nu}= \lambda^{-1} g^{\mu\nu}$, such that
$\tilde{g}^{\mu \rho} \tilde{g}_{\rho \nu}
=
g^{\mu \rho} g_{\rho \nu}
=
\delta^{\mu}_{\nu}$, i.e., the dressed (bare) metric raises and lowers indices in the presence (absence) of dissipation. And the screened Levi-Civita connection is defined in the usual sense~\cite{wald2010general},
$\tilde{\nabla} \tilde{g} = 0$, which implies that the dressed Christoffel symbol is that of the dressed metric
\begin{equation}
\tilde{\Gamma}_{\mu\nu\rho}
=
\lambda \Gamma_{\mu\nu\rho}
+
\frac{1}{2} \left(
g_{\mu\nu}\nabla_{\rho} \lambda
+
g_{\mu\rho}\nabla_{\nu} \lambda
-
g_{\nu\rho}\nabla_{\mu} \lambda
\right).
\end{equation}
This can then be used to derive the dressed Riemann tensor
$\tensor{\tilde{R}}{^{\mu}_\nu_\rho_\sigma}$
and its contractions, namely the Ricci tensor
$\tilde{R}_{\mu\nu}
=
\tensor{\tilde{R}}{^{\rho}_\mu_\rho_\nu}$ and Ricci scalar
$\tilde{R}= \tensor{\tilde{R}}{^{\mu}_\mu}$.
Combining terms, we arrive at the dissipative EFE
\begin{equation}
\label{EFE}
\tilde{R}_{\mu\nu}
-\frac{1}{2} \tilde{R} \tilde{g}_{\mu\nu}
+
\Lambda \tilde{g}_{\mu\nu}
=
T_{\mu\nu},
\end{equation}
where $\Lambda$ and $T_{\mu\nu}$, which emerge entirely as a result of dissipative scattering, are identified as the local momentum-space cosmological constant and stress tensor, respectively, and are given, up to $O(\eta^2)$, by
\begin{subequations}
\begin{align}
\Lambda
&=
\nabla^2 \eta^2,
\\
T_{\mu\nu}
&=
\frac{n-2}{2}
\left( \nabla_{\mu} \nabla_{\nu} - g_{\mu\nu} \nabla^2 \right)\eta^2,
\end{align}
\end{subequations}
with $n$ the dimension of the system. This assignment is rather arbitrary as far as the EFE is concerned, and $\Lambda$ may well be absorbed into the stress tensor~\cite{misner1973gravitation}. What matters here instead is that the dressed Einstein tensor
$\tilde{G}_{\mu\nu}
=
\tilde{R}_{\mu\nu}
-
\tilde{R} \tilde{g}_{\mu\nu}/2$ does not vanish, which physically implies that dissipation is providing a source term for the momentum-space EFE.

To gain further insight into this result, it is helpful to recall that the modification of the EFE emerged from the screening of the metric. Following the semiclassical results, one could ask whether there is an entropic explanation for this. To answer this in a consistent manner within the present response-theory formalism, we note that, to second order in the electric field, the rate of local entropy production by microscopic scattering processes is determined by the symmetric part of the linear conductivity tensor~\cite{ziman2001electrons}
\begin{equation}
\dot{S}
=
\beta \mathbf{j} \cdot \mathbf{E}
=
\beta \sigma_{(\mu\nu)} E^{\mu} E^{\nu},
\end{equation}
which is related to Joule heating and has a quantum geometric contribution, $\dot{S}_{\text{g}}$, identified as giving rise to the momentum-space drag force, Eq.~(\ref{eq_drag}). This corresponds to the local conductivity contribution
$
- 2 e^2 / \hbar \sum_{ab}
f^a \eta^{ab} \tilde{g}_{\mu\nu}^{ab}
$
upon restoring band indices. Introducing the band resolution of the entropy rate as
$\dot{S} = \sum_{ab} \dot{S}^{ab}$, to $O(\eta^2)$ we find
\begin{equation}
\dot{S}_{\text{g}}^{ab}
=
- 2 \hbar \beta \dot{k}^{\mu} \dot{k}^{\nu} f^a g_{\mu\nu}^{ab}\eta^{ab}.
\end{equation}
This demonstrates that the source term in the multiband EFE  can be reexpressed as a function of the local entropy rate associated with dissipative interband processes, thereby providing a quantum-response generalization of the semiclassical arguments to dissipative multiband systems. This demonstrates that the source term in the multiband EFE  can be reexpressed as a function of the local entropy rate associated with dissipative interband processes, thereby providing a quantum-response generalization of the arguments in Ref.~\cite{smith2022momentum} to dissipative multiband systems. Furthermore, in this sense, one can think of momentum-space gravity as essentially being a manifestation of local entropy changes in momentum-space, which is reminiscent of similar insights on gravity in spacetime. In this light, one could also think of the quantum geometric drag force related to Eq.~(\ref{eq_drag}) as a counterpart of the entropic force discussed in Ref.~\cite{verlinde2011origin}, with the difference that the latter--inspired by holographic arguments--is typically not associated with a fundamental field. In our case, this role is played by the quantum metric itself.

\section{Geometric origin of dissipative responses and dynamics}
In this section, after elaborating on the basic diagrammatic framework for evaluating conductivities in the presence of finite dissipation, we take the dc limit and derive the equation of motion that arises from the response functions. In doing so, we also show how both linear and nonlinear responses can be fully reexpressed in the geometric language, which, to our knowledge, has not been demonstrated before. Indeed, while there are several geometric presentations of \textit{specific} quadratic responses in the literature, a complete geometric classification of quadratic response theory that takes into account all Feynman diagrams simultaneously is yet to be done. In the derivations of the nonlinear responses presented below, several known elements of two-state quantum geometry make an appearance, including the quantum metric, Berry curvature and quantum connection. Furthermore, while two-state quantum geometry is certainly a necessary element, we find that it is not sufficient, as there are terms in the quadratic response functions which do not take closed two-state forms. This should perhaps not come as a surprise, as it would seem reasonable to question the complete classification of the three-point functions of quadratic response theory (involving traces over three distinct states) via two-state objects. As it turns out, we find that it is precisely the three-state QGT that is the missing necessary ingredient to complete the geometric classification.

\subsection{Response functions and frequency summations}

\begin{figure}[hpt]
%\vspace{-.7cm}
\captionsetup[subfigure]{labelformat=empty}
    \sidesubfloat[]{\includegraphics[width=0.51\linewidth,trim={1cm 1cm 1cm .5cm}]{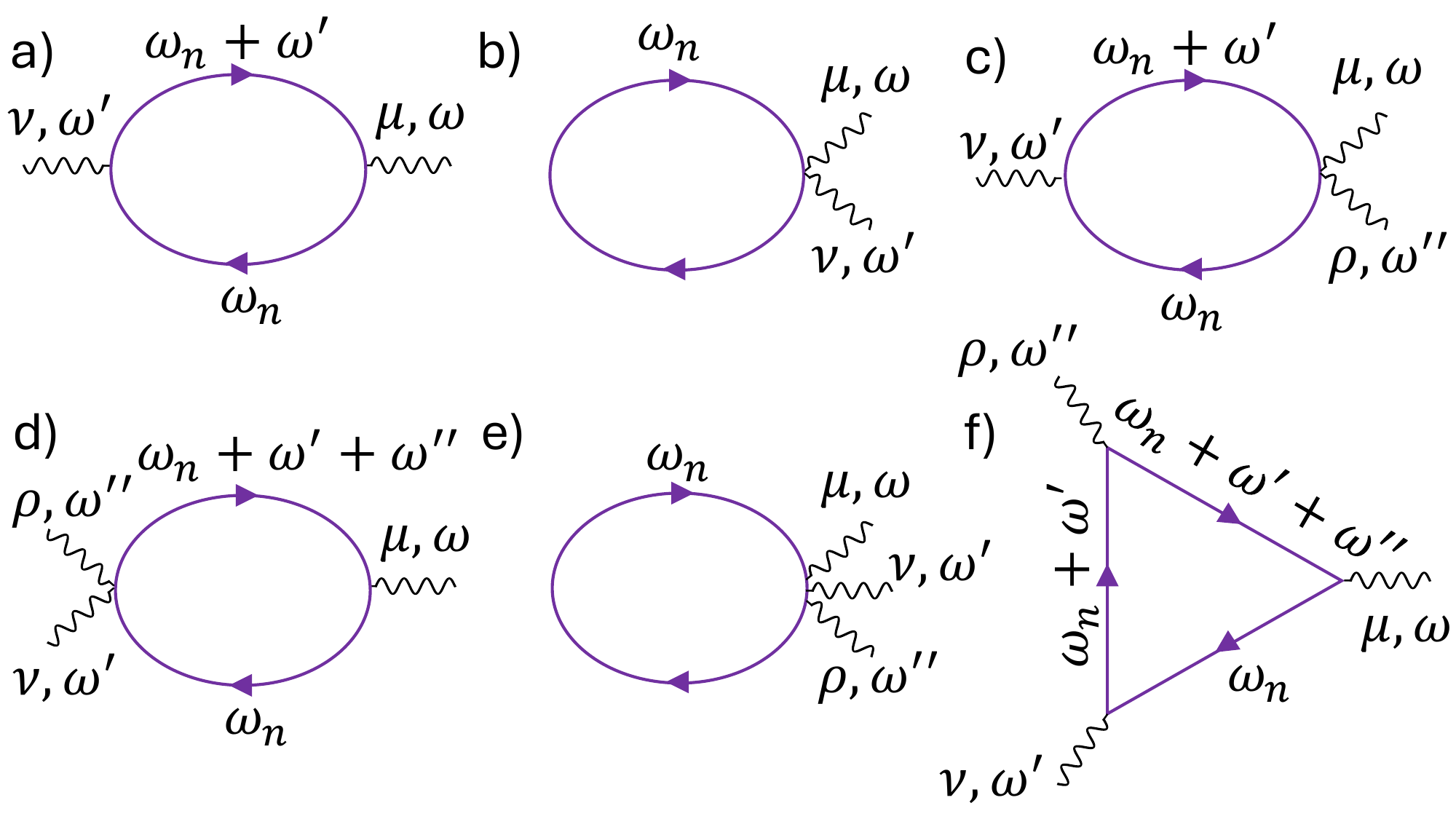}\label{fig_qmsg_fig1a}}
%\quad%
%%%%%%%%%
    \caption{Linear [(a) and (b)] and quadratic [(c)-(f)] conductivity diagrams. Loops and legs are electron and photon propagators, respectively, and vertices imply velocity operator insertions, with the index $\mu$ reserved for the output. The Feynman rules resulting in Eqs.~(\ref{eq_sigma1}) and (\ref{eq_sigma2}) consist of the following factors and procedures: 1) A factor of $1/k!$ for every set of $k$ connected photons. 2) Factors of $e/\hbar$ for the output photon and $ie/\chi$ for each input photon, with $\chi=\omega^{\pr},\omega^{\dpr}$. 3) Trace over momentum and band indices of the loop. 4) Matsubara frequency summation $1/\beta \sum_n$.}
    \label{figS1}
\end{figure}

Using a diagrammatic approach within the Matsubara formalism~\cite{parker2019diagrammatic}, and applying the Feynman rules presented in Fig.~{\ref{figS1}}, the linear and quadratic ac  conductivities are obtained as
\begin{equation}
\label{eq_sigma1}
\begin{split}
\sigma_{\mu \nu} (\omega; \omega^{\pr})
=&
\frac{ie^2}{\hbar \omega^{\pr}}\frac{1}{\beta}\sum_n
\text{Tr} \left[
h_{\mu} \mathcal{G}(\omega_n) 
h_{\nu} \mathcal{G}(\omega_n + \omega^{\pr})
+
h_{\mu \nu} \mathcal{G}(\omega_n)
\right],
\end{split}
\end{equation}
and
\begin{equation}
\label{eq_sigma2}
\begin{split}
\sigma_{\mu \nu \rho}
(\omega; \omega^{\pr}, \omega^{\dpr})
=&
\frac{e}{2 \hbar}
\frac{(ie)^2}{\omega^{\pr} \omega^{\dpr}}\frac{1}{\beta}\sum_n
\text{Tr} \left[
h_{\mu \rho} \mathcal{G}(\omega_n) 
h_{\nu} \mathcal{G}(\omega_n + \omega^{\pr})
+
\frac{1}{2} h_{\mu} \mathcal{G}(\omega_n) 
h_{\nu \rho} \mathcal{G}( \omega_n + \omega^{\pr} +\omega^{\dpr})
\right.
\\
&+
\left.
\frac{1}{2} h_{\mu \nu \rho} \mathcal{G}(\omega_n)
+
h_{\mu} \mathcal{G}(\omega_n) 
h_{\nu} \mathcal{G}(\omega_n + \omega^{\pr})
h_{\rho} \mathcal{G}(\omega_n + \omega^{\pr}
+ \omega^{\dpr})
\right.
\\
&+
\left.
\{(\nu, \omega^{\pr}) \leftrightarrow
(\rho, \omega^{\dpr}) \}
\right],
\end{split}
\end{equation}
where $\omega$ is the frequency of the output photon and $\omega^{\pr}, \omega^{\dpr}$ are input frequencies. The Matsubara Green's function is given by
$\mathcal{G}(\omega_n)
=
(i\omega_n - H - \Sigma)^{-1}$, where $H$ is the system Hamiltonian, with
$H \ket{\psi^a} = \veps^a \ket{\psi^a}$. And for the self-energy, we make the phenomenelogical approximation $\Sigma = -i\gamma/2$, with $\gamma$ measuring the strength of dissipation induced by scattering events. Finally, velocity operators are obtained by applying successive Berry covariant derivatives,
$h_{\mu_1 \ldots \mu_n}
=
\mathcal{D}_{\mu_1 \ldots \mu_n} H$.

%\vspace{-1cm}

To evaluate Eqs.~(\ref{eq_sigma1}) and (\ref{eq_sigma2}), the effect of the dissipation parameter $\gamma$ is included through a phenomenological shift of the frequency factors in the complex plane,
$\omega \rightarrow \omega + i\gamma$~\cite{parker2019diagrammatic}. And the Matsubara Green's function is expressed in the spectral representation as
\begin{equation}
\mathcal{G}^a (\omega_n)
=
\int_{-\infty}^{\infty} \frac{d\veps}{2\pi}
\frac{A^a(\veps)}{i\omega_n - \veps},
\end{equation}
where the denominator is the bare Green's function,
$\mathcal{G}_0^a (\omega_n)= (i \omega_n - \veps^a)^{-1}$ and
$A^a (\veps)= -2 \text{Im} G_R^a (\veps)$
is the spectral function~\cite{mahan2000many}, with $G_R$ the retarded Green's function. The frequency summations that are relevant to the evaluation of the conductivities are given by~\cite{mahan2000many, Michishita2021effects}
\begin{subequations}
\begin{align}
\frac{1}{\beta} \sum_n
\mathcal{G}_0^a (\omega_n)
&=
f^a,
\\
\frac{1}{\beta} \sum_n
\mathcal{G}_0^a (\omega_n)
\mathcal{G}_0^b (\omega_n+ \omega^{\pr})
&=
\frac{f^a - f^b}{\omega^{\pr} + \veps^{ab}},
\\
\frac{1}{\beta} \sum_n
\mathcal{G}_0^a (\omega_n)
\mathcal{G}_0^b (\omega_n+ \omega^{\pr})
\mathcal{G}_0^c (\omega_n+ \omega^{\pr}
+ \omega^{\dpr})
&=
\frac{
(\omega^{\dpr} - \veps^{cb})(f^a - f^b)
+
(\omega^{\pr} - \veps^{ba})(f^c - f^b)
}
{(\omega^{\pr} - \veps^{ba})
(\omega^{\dpr} - \veps^{cb})
(\omega^{\pr} + \omega^{\dpr} - \veps^{ca})},
\end{align}
\end{subequations}
with the shorthand $f^a \equiv f(\veps^a)$ for the Fermi distribution. In addition, an approximation is needed to obtain analytical results in the presence of dissipation. Taking
$\gamma \beta \ll1$
results in the identity~\cite{Michishita2021effects}
\begin{equation}
\int_{-\infty}^{\infty}
\frac{d\veps}{2 \pi}
A^a(\veps) 
F \left( \veps, \{\omega_n\} \right) f(\veps)
\simeq
F \left( \veps^a \pm \frac{i \gamma}{2}, 
\{\omega_n\} \right) f (\veps^a \pm \frac{i \gamma}{2} ),
\end{equation}
where $F$ is a general function that includes velocity operator components and the sign is chosen $\pm$ when $F$ is analytic in the upper/lower complex plane of the energy integrand. We note that this approximation is not as restricting as may seem, as values as large as $\gamma \beta=0.5$ have been shown to yield good agreement with numerical results as far as the conductivity is concerned~\cite{Michishita2021effects}. We apply these steps to Eqs.~(\ref{eq_sigma1}) and (\ref{eq_sigma2}) and take the dc limit
$\omega, \omega^{\pr}, \omega^{\dpr} \rightarrow 0$, yielding the frequency summations
\begin{subequations}
\label{freq_sum_dress}
\begin{align}
&\frac{1}{\beta} \sum_n
\mathcal{G}^a (\omega_n)
\mathcal{G}^b (\omega_n+ \omega^{\pr})
\rightarrow
\frac{ f (\veps^a + \frac{i \gamma}{2})
- f (\veps^b - \frac{i \gamma}{2})}
{\veps^{ab} + i \gamma}
\simeq
\frac{f^a - f^b}{\veps^{ab} + i\gamma}
+
i \frac{\gamma}{2}
\left(
\frac{\pd f^a}{\pd \veps^a} + \frac{\pd f^b}{\pd \veps^b}
\right)
\frac{1}{\veps^{ab} + i\gamma},
\\
&\frac{1}{\beta} \sum_n
\mathcal{G}^a (\omega_n)
\mathcal{G}^b (\omega_n+ \omega^{\pr})
\mathcal{G}^c (\omega_n+ \omega^{\pr}
+ \omega^{\dpr})
\rightarrow
\frac{ f (\veps^a + \frac{i \gamma}{2})
- f (\veps^b - \frac{i \gamma}{2})}
{(\veps^{ba} - i \gamma) (\veps^{ca} - i \gamma)}
+
\frac{ f (\veps^c - \frac{i \gamma}{2})
- f (\veps^b + \frac{i \gamma}{2})}
{(\veps^{cb} - i \gamma) (\veps^{ca} - i \gamma)}
\\
&\simeq
\frac{f^a - f^b}
{(\veps^{ba} - i\gamma) (\veps^{ca} - i\gamma)}
+
\frac{f^c - f^b}
{(\veps^{cb} - i\gamma) (\veps^{ca} - i\gamma)}
+
i \frac{\gamma}{2}
\left(
\frac{\pd f^a}{\pd \veps^a} + \frac{\pd f^b}{\pd \veps^b}
\right)
\frac{1}{(\veps^{ba} - i\gamma) (\veps^{ca} - i\gamma)}
\\
&\;\;\;\;-
i \frac{\gamma}{2}
\left(
\frac{\pd f^c}{\pd \veps^c} + \frac{\pd f^b}{\pd \veps^b}
\right)
\frac{1}{(\veps^{cb} - i\gamma) (\veps^{ca} - i\gamma)}.
\end{align}
\end{subequations}

\subsection{DC conductivities and dressed equation of motion}

Using the explanations of the previous section, the dc response functions can be evaluated. From Eq.~(\ref{freq_sum_dress}), it is clear that both the distribution function and its derivatives appear in the responses. As is shown below, for the purpose of deriving the equation of motion at nonlinear order, it is necessary to include the latter terms in order to obtain the correct signs of the carrier densities in the nonlinear response functions. 

The linear dc conductivity reads
\begin{equation}
\label{sig_munu_expanded}
\begin{split}
\sigma_{\mu\nu}
&=
\frac{e^2}{\hbar \gamma} \sum_{\mbk a}
\left(
f^a h_{\mu\nu}^a
+
\sum_b  \frac{f^a - f^b}{\veps^{ab} + i \gamma}
h_{\nu}^{ab} h_{\mu}^{ba}
\right)
\\
&=
\frac{e^2}{\gamma} \sum_{\mbk a}
f^a \pd_{\nu} v_{\mu}^a
+
\frac{i e^2}{\hbar} \sum_{\mbk ab}
(f^a - f^b) \frac{Q_{\mu\nu}^{ab}}{1 - i \eta^{ab}}
\\
&=
- \frac{e^2}{\gamma} \sum_{\mbk a}
\pd_{\nu} f^a v_{\mu}^a
+
\frac{e^2}{\hbar} \sum_{\mbk ab}
f^a \tilde{\Omega}_{\mu\nu}^{ab}
-
\frac{2 e^2}{\hbar} \sum_{\mbk ab}
f^a \eta^{ab} \tilde{g}_{\mu\nu}^{ab},
\end{split}
\end{equation}
yielding the linear current density
\begin{equation}
\frac{j_{\mu}^1}{-e}
=
\sum_{\mbk a}
n_1^a v_{\mu}^a
+
\dot{k}^{\nu} \sum_{\mbk ab}
f^a \tilde{\Omega}_{\mu\nu}^{ab}
-
2 \dot{k}^{\nu} \sum_{\mbk ab}
f^a \eta^{ab} \tilde{g}_{\mu\nu}^{ab},
\end{equation}
with the linear carrier density given by $n_1^a = - \tau \dot{k}^{\nu} \pd_{\nu} f^a$ and $\tau = \hbar/ \gamma$ the relaxation time. Note that Eq.~(\ref{sig_munu_expanded}) reveals the natural choice of dressing for the quantum geometric quantities in the sense that the Weyl-transformed Berry curvature and quantum metric represent the effective geometry the carriers experience in the presence of dissipation within the relaxation time approximation. This is then consistently applied in the evaluation of the nonlinear responses as well, the calculation of which is considerably more detailed. However, the contributions can be systematically analyzed and classified as follows. Taking into account the four terms in Eq.~(\ref{eq_sigma2}), we decompose the nonlinear conductivity tensor as
\begin{subequations}
\begin{align}
\sigma_{\mu\nu\rho}
&=
\sigma_{\mu\nu\rho}^\text{A}
+
\sigma_{\mu\nu\rho}^\text{B},
\\
\sigma_{\mu\nu\rho}^\text{A}
&=
\sigma_{\mu\nu\rho}^\text{I,A}
+
\sigma_{\mu\nu\rho}^\text{II,A}
+
\sigma_{\mu\nu\rho}^\text{III,A}
+
\sigma_{\mu\nu\rho}^\text{IV,A},
\\
\sigma_{\mu\nu\rho}^\text{B}
&=
\sigma_{\mu\nu\rho}^\text{I,B}
+
\sigma_{\mu\nu\rho}^\text{II,B}
+
\sigma_{\mu\nu\rho}^\text{III,B}
+
\sigma_{\mu\nu\rho}^\text{IV,B},
\end{align}
\end{subequations}
where the labels ``A" and ``B" specify Fermi distribution terms and Fermi distribution derivative terms, respectively, as per Eq.~(\ref{freq_sum_dress}). The former terms yield the contributions
\begin{equation}
\begin{split}
\sigma_{\mu\nu\rho}^\text{I,A}
&=
\frac{e^3}{2 \hbar \gamma^2}
\sum_{\mbk a}
f^a h_{\mu \nu \rho}^a
+ (\nu \leftrightarrow \rho)
=
\frac{e^3}{2 \hbar \gamma^2} \sum_{\mbk a} f^a
\left[
\hbar \pd_{\nu} \pd_{\rho} v_{\mu}^a
-
2 \sum_b \pd_{\mu} (g_{\mu\nu}^{ab} \veps^{ab})
\right]
\\
&-
\frac{i e^3}{2 \hbar \gamma^2} \sum_{\mbk ab} 
(f^a - f^b) \mathcal{A}_{\mu}^{\pr ab}
\left[
-i \pd_{\nu} (\mathcal{A}_{\rho}^{\pr ba} \veps^{ab})
-
i \mathcal{A}_{\nu}^{\pr ba} \pd_{\rho} \veps^{ab}
-
\sum_c (\mathcal{A}_{\nu}^{bc} \mathcal{A}_{\rho}^{ca} \veps^{ac}
-
\mathcal{A}_{\nu}^{ca} \mathcal{A}_{\rho}^{bc} \veps^{cb})
\right],
\end{split}
\end{equation}
%%%%%
\begin{equation}
\begin{split}
\sigma_{\mu\nu\rho}^\text{II,A}
&=
\frac{e^3}{\hbar \gamma^2} \sum_{\mbk ab}
\frac{f^a - f^b}{\veps^{ab} - i \gamma}
h_{\mu\rho}^{ab} h_{\nu}^{ba}
+ (\nu \leftrightarrow \rho)
\\
&=
\frac{e^3}{\hbar \gamma^2} \sum_{\mbk ab} 
\frac{f^a - f^b}{\veps^{ab} - i \gamma}
\mathcal{A}_{\nu}^{\pr ba} \veps^{ab}
\left[
\pd_{\mu} (\mathcal{A}_{\rho}^{\pr ab} \veps^{ab})
+
\mathcal{A}_{\mu}^{\pr ab} \pd_{\rho} \veps^{ab}
+
i \sum_c (\mathcal{A}_{\mu}^{ac} \mathcal{A}_{\rho}^{cb} \veps^{bc}
-
\mathcal{A}_{\rho}^{ac} \mathcal{A}_{\mu}^{cb} \veps^{ca})
\right],
\end{split}
\end{equation}
%%%%%
\begin{equation}
\begin{split}
\sigma_{\mu\nu\rho}^\text{III,A}
&=
\frac{e^3}{2 \hbar \gamma^2} \sum_{\mbk ab}
\frac{f^a - f^b}{\veps^{ab} + i \gamma}
h_{\nu\rho}^{ab} h_{\mu}^{ba}
+ (\nu \leftrightarrow \rho)
\\
&=
\frac{e^3}{2 \hbar \gamma^2} \sum_{\mbk ab} 
\frac{f^a - f^b}{\veps^{ab} + i \gamma}
\mathcal{A}_{\mu}^{\pr ba} \veps^{ab}
\left[
\pd_{\nu} (\mathcal{A}_{\rho}^{\pr ab} \veps^{ab})
+
\mathcal{A}_{\nu}^{\pr ab} \pd_{\rho} \veps^{ab}
+
i \sum_c (\mathcal{A}_{\nu}^{ac} \mathcal{A}_{\rho}^{cb} \veps^{bc}
-
\mathcal{A}_{\rho}^{ac} \mathcal{A}_{\nu}^{cb} \veps^{ca})
\right],
\end{split}
\end{equation}
%%%%%
\begin{equation}
\begin{split}
\sigma_{\mu\nu\rho}^\text{IV,A}
=&
\frac{e^3}{\hbar \gamma^2} \sum_{\mbk abc}
\left[
\frac{f^a - f^c}
{(\veps^{ac} - i \gamma) (\veps^{ab} - i \gamma)}
+
\frac{f^b - f^c}
{(\veps^{cb} - i \gamma) (\veps^{ab} - i \gamma)}
\right]
h_{\mu}^{ab} h_{\nu}^{bc} h_{\rho}^{ca}
+ (\nu \leftrightarrow \rho)
\\
=&
\frac{e^3}{\hbar \gamma^2} \sum_{\mbk abc}
\left[
\frac{f^a - f^c}
{(\veps^{ac} - i \gamma) (\veps^{ab} - i \gamma)}
+
\frac{f^b - f^c}
{(\veps^{cb} - i \gamma) (\veps^{ab} - i \gamma)}
\right]
\\
&\times
\left[
Q_{\nu\rho}^{ac} \delta^{ab} (\veps^{ac})^2 \pd_{\mu} \veps^a
+
Q_{\mu\rho}^{ab} \delta^{bc} (\veps^{ab})^2 \pd_{\nu} \veps^b
+
Q_{\mu\nu}^{ab} \delta^{ac} (\veps^{ab})^2 \pd_{\rho} \veps^a
-
i Q_{\mu\nu\rho}^{abc} \veps^{ab} \veps^{bc} \veps^{ca}
\right].
\end{split}
\end{equation}
Combining the four terms and after a little algebra, we arrive at the expression
\begin{equation}
\label{pre_sig2_A}
\begin{split}
\sigma_{\mu\nu\rho}^\text{A}
=&
\frac{e^3}{2 \gamma^2}
\sum_{\mbk a} f^a \pd_{\nu} \pd_{\rho} v_{\mu}^a
-
\frac{e^3}{2 \hbar \gamma} \sum_{\mbk a b} f^a \tilde{\Omega}_{\mu\nu}^{ab} \frac{\pd_{\rho} \veps^{ab}}{\veps^{ab}}
\frac{1- 3 (\eta^{ab})^2}{1 + (\eta^{ab})^2}
-
\frac{e^3}{\hbar} \sum_{\mbk a b} f^a \tilde{g}_{\mu\nu}^{ab} \pd_{\rho} \left( \frac{1}{\veps^{ab}} \right)
\frac{3- (\eta^{ab})^2}{1 + (\eta^{ab})^2}
\\
&+
\frac{e^3}{\hbar} \sum_{\mbk a b} f^a
\tilde{g}_{\nu \rho}^{ab} \pd_{\mu}
\left( \frac{1}{\veps^{ab}} \right)
\frac{1- (\eta^{ab})^2}{1 + (\eta^{ab})^2}
-
\frac{e^3}{\hbar \gamma^2}
\sum_{\mbk a b} f^a
\tilde{g}_{\nu \rho}^{ab} \pd_{\mu} \veps^{ab}
-
\frac{2 e^3}{\hbar} \sum_{\mbk a b} f^a
\frac{1}{\veps^{ab}}
\tilde{\Gamma}_{\nu\rho\mu}^{ab}
+
\sum_{i=1}^4 \mathcal{F}_i
\end{split}
\end{equation}
where
\begin{equation}
\begin{split}
\mathcal{F}_1
=&
- \frac{e^3}{\hbar \gamma} \sum_{\mbk a b} f^a
\text{Im} \bigg\{
\frac{\mathcal{A}_{\mu}^{\pr ab} 
(\mathcal{D}_{\nu} \left[ \mathcal{A}_{\rho}^{\pr} , H \right])^{ba}}
{\veps^{ab} - i \gamma} \bigg\}
\\
=&
-\frac{e^3}{\hbar \gamma} \sum_{\mbk a b} f^a
\frac{\veps^{ab}}{(\veps^{ab})^2 + \gamma^2}
\left[
- \frac{1}{2} \Omega_{\mu\rho}^{ab} \pd_{\nu} \veps^{ab}
+
\text{Im} (C_{\mu\nu\rho}^{ba}) \veps^{ab}
-
\sum_c \text{Re} (S_{\mu\nu\rho}^{abc}) (\veps^{bc} - \veps^{ca})
\right]
\\
&-
\frac{e^3}{\hbar} \sum_{\mbk a b} f^a
\frac{1}{(\veps^{ab})^2 + \gamma^2}
\left[
g_{\mu\rho}^{ab} \pd_{\nu} \veps^{ab}
+
\text{Re} (C_{\mu\nu\rho}^{ba}) \veps^{ab}
+
\sum_c \text{Im} (S_{\mu\nu\rho}^{abc}) (\veps^{bc} - \veps^{ca})
\right],
\end{split}
\end{equation}
%%%%%%%%
\begin{equation}
\begin{split}
\mathcal{F}_2
=&
\frac{2 e^3}{\hbar \gamma} \sum_{\mbk a b} f^a
\frac{\veps^{ab}}{(\veps^{ab})^2 + \gamma^2}
\text{Im} \{
\mathcal{A}_{\nu}^{\pr ab} 
(\mathcal{D}_{\mu} \left[ \mathcal{A}_{\rho}^{\pr} , H \right])^{ba}\}
\\
=&
\frac{2 e^3}{\hbar \gamma} \sum_{\mbk a b} f^a
\frac{\veps^{ab}}{(\veps^{ab})^2 + \gamma^2}
\left[
\text{Im} (C_{\nu\mu\rho}^{ba}) \veps^{ab}
-
\sum_c \text{Re} (S_{\nu\mu\rho}^{abc}) (\veps^{bc} - \veps^{ca})
-
\frac{1}{2} \text{Im} (\mathcal{T}_{\nu\mu\rho}^{ba}) \veps^{ab}
\right],
\end{split}
\end{equation}
%%%%%%%%
\begin{equation}
\begin{split}
\mathcal{F}_3
=&
\frac{2 e^3}{\hbar \gamma^2} \sum_{\mbk a b} f^a
\frac{
\text{Im} \{ \mathcal{A}_{\nu}^{\pr ab} 
\left[ \mathcal{A}_{\mu}^{\pr} , \left[ \mathcal{A}_{\rho}^{\pr} , H \right] \right]^{ba} \}}
{1 + (\eta^{ab})^2}
\\
=&
\frac{2 e^3}{\hbar \gamma^2} \sum_{\mbk a b} f^a
\left[
\text{Re} (\tilde{\mathcal{T}}_{\nu\mu\rho}^{ba}) \veps^{ab}
+
\sum_c \text{Im} (\tilde{Q}_{\nu\mu\rho}^{abc}) \veps^{bc}
-
\sum_c \text{Im} (\tilde{Q}_{\nu\rho\mu}^{abc}) \veps^{ca}
\right],
\end{split}
\end{equation}
%%%%%%%%
\begin{equation}
\begin{split}
\mathcal{F}_{4}
=&
\frac{2 e^3}{\hbar \gamma^2} \sum_{\mbk a} f^a
\text{Im} \big\{
\left( 
\left[
\left[\alpha_{\mu} (\gamma) , H \right]
,
\left[ \mathcal{A}_{\nu}^{\pr} , H \right]
\right] 
\left[ \alpha_{\rho} (-\gamma) , H \right]
\right)_{\mbk}^{a}
\big\}
\\
=&
\frac{2 e^3}{\hbar \gamma^2} \sum_{\mbk abc} f^a
\bigg\{
\text{Im} (Q_{\rho\nu\mu}^{abc})
\frac{ \veps^{ac} \veps^{ab} - \gamma^2}{ [ (\veps^{ac})^2 + \gamma^2 ]
[ (\veps^{ab})^2 + \gamma^2 ] }
-
\text{Re} (Q_{\rho\nu\mu}^{abc})
\frac{ \gamma ( \veps^{ac} + \veps^{ab} ) }{ [ (\veps^{ac})^2 + \gamma^2 ]
[ (\veps^{ab})^2 + \gamma^2 ] }
\\
&-
\text{Im} (Q_{\rho\mu\nu}^{abc})
\frac{ \veps^{cb} \veps^{ab} - \gamma^2}{ [ (\veps^{ab})^2 + \gamma^2 ]
[ (\veps^{bc})^2 + \gamma^2 ] }
+
\text{Re} (Q_{\rho\mu\nu}^{abc})
\frac{ \gamma ( \veps^{cb} + \veps^{ab} ) }{ [ (\veps^{ab})^2 + \gamma^2 ]
[ (\veps^{bc})^2 + \gamma^2 ] }
\bigg\}
\veps^{ab} \veps^{bc} \veps^{ca},
\end{split}
\end{equation}
with $ [ \alpha_{\mu} (\gamma) ]^{ab} \equiv \mathcal{A}_{\mu}^{\pr ab}/( \veps^{ab} - i \gamma)$. Reinserting these back into Eq.~(\ref{pre_sig2_A}) and collecting similar geometric terms, the conductivity reads
\begin{equation}
\label{sig2_A_supp}
\begin{split}
\sigma_{\mu\nu\rho}^{A}
=&
\frac{e^3}{2 \gamma^2}
\sum_{\mbk a} \pd_{\nu} \pd_{\rho} f^a  v_{\mu}^a
-
\frac{e^3}{\hbar^2} \sum_{\mbk ab} f^a
\tilde{\Omega}_{\mu\nu}^{ab} \pd_{\rho} m^{ab}
\frac{2 \eta^{ab}}{1 + (\eta^{ab})^2}
-
\frac{e^3}{\hbar^2} \sum_{\mbk ab} f^a
\tilde{g}_{\mu\nu}^{ab} \pd_{\rho} m^{ab}
\frac{1- (\eta^{ab})^2}{1 + (\eta^{ab})^2}
\\
&+
\frac{e^3}{\hbar^2} \sum_{\mbk ab} \pd_{\rho} f^a
\tilde{g}_{\mu\nu}^{ab}  m^{ab}
-
\frac{e^3}{\hbar^2} \sum_{\mbk ab} f^a
m^{ab} \tilde{\Gamma}_{\nu\rho\mu}^{ab}
+
\frac{e^3}{\hbar^2} \sum_{\mbk ab} f^a
\tilde{g}_{\nu\rho}^{ab} \pd_{\mu} m^{ab}
\frac{1}{(\eta^{ab})^2} 
\frac{1+ 2 (\eta^{ab})^2}{1 + (\eta^{ab})^2}
\\
&-
\frac{e^3}{\hbar^2} \sum_{\mbk ab} f^a m^{ab}
\text{Re} (\tilde{K}_{\mu\nu\rho}^{ba})
-
\frac{e^3}{\hbar^2} \sum_{\mbk ab} f^a m^{ab}
\frac{1}{\eta^{ab}}
\text{Im}
\left(
\tilde{C}_{\mu \nu \rho}^{ba}
-
\tilde{C}_{\nu \mu \rho}^{ba}
-
\tilde{C}_{\rho \nu \mu}^{ba}
\right)
\\
&+
\frac{e^3}{\hbar^2} \sum_{\mbk abc} f^a m^{ab}
\left[
\text{Re} (\tilde{S}_{\mu\nu\rho}^{abc})
- \eta^{ab} \text{Im} (\tilde{S}_{\mu\nu\rho}^{abc})
\right]
\left( \frac{1}{\eta^{bc}} - \frac{1}{\eta^{ca}} \right)
\\
&+
\frac{2 e^3}{\hbar^2} \sum_{\mbk abc} f^a m^{ab}
\text{Re} (\tilde{S}_{\nu\mu\rho}^{abc})
\frac{1}{\eta^{ab}}
\left[
\frac{\eta^{bc}}{\eta^{ca}}
\frac{ 1 + \eta^{ab} \eta^{bc} }{1 + (\eta^{bc})^2}
-
\frac{\eta^{ca}}{\eta^{bc}}
\frac{ 1 + \eta^{ab} \eta^{ca} }{1 + (\eta^{ca})^2}
\right]
\\
&+
\frac{2 e^3}{\hbar^2} \sum_{\mbk abc} f^a m^{ab}
\text{Im} (\tilde{S}_{\nu\mu\rho}^{abc})
\frac{1}{\eta^{ab}}
\left[
\frac{\eta^{bc}}{\eta^{ca}}
\frac{ \eta^{ab} - \eta^{bc} }{1 + (\eta^{bc})^2}
-
\frac{\eta^{ca}}{\eta^{bc}}
\frac{ \eta^{ab} - \eta^{ca} }{1 + (\eta^{ca})^2}
\right]
\\
&-
\frac{2 e^3}{\hbar^2} \sum_{\mbk abc} f^a m^{ab}
\text{Re} (\tilde{A}_{\nu\mu\rho}^{abc})
\frac{1}{\eta^{ab}}
\left[
\frac{1}{\eta^{ca}}
\frac{ \eta^{ab} - \eta^{bc} }{1 + (\eta^{bc})^2}
+
\frac{1}{\eta^{bc}}
\frac{ \eta^{ab} - \eta^{ca} }{1 + (\eta^{ca})^2}
\right]
\\
&+
\frac{2 e^3}{\hbar^2} \sum_{\mbk abc} f^a m^{ab}
\text{Im} (\tilde{A}_{\nu\mu\rho}^{abc})
\frac{1}{\eta^{ab}}
\left[
\frac{\eta^{bc}}{\eta^{ca}}
\frac{ \eta^{ab} - \eta^{bc} }{1 + (\eta^{bc})^2}
+
\frac{\eta^{ca}}{\eta^{bc}}
\frac{ \eta^{ab} - \eta^{ca} }{1 + (\eta^{ca})^2}
\right],
\end{split}
\end{equation}
which reveals the various contributions from the dressed two-state quantum geometric objects. The remaining terms, which cannot be expressed within the two-state formalism, are precisely captured by the symmetric and antisymmetric parts of the three-state QGT. This highlights an interesting--and perhaps useful--parallel with the appearance of the two-state QGT in dissipative linear response theory. As can be seen from Eq.~(\ref{sig_munu_expanded}), the two-state QGT makes an appearance in the response function, which cannot be summed out in general. For example, focussing on the the Berry curvature term responsible for the anomalous Hall effect, 
$\dot{k}^{\nu} \sum_{\mbk ab}
f^a \tilde{\Omega}_{\mu\nu}^{ab}
$,
we note that in the clean limit, one can sum out the intermediate states of the two-state Berry curvature, leaving behind the single-band Berry curvature. That is, in this limit, one could remove all indications that this term is essentially an (intrinsic) interband coherence effect and express it as a single-state object. However, this is no longer the case for the dressed Berry curvature and one should more appropriately consider its band resolution, as a result of the additional interband mixing arising from the dissipation. In this sense, the appearance of the three-state QGT can be regarded as the quadratic-response counterpart to the above argument.

We next repeat the above steps for the distribution derivative terms. This yields the conductivity contribution
\begin{equation}
\begin{split}
\sigma_{\mu\nu\rho}^\text{B}
=&
- \frac{3 e^3}{2 \gamma^2}
\sum_{\mbk a} f^a \pd_{\nu} \pd_{\rho} v_{\mu}^a
-
\frac{e^3}{4 \hbar \gamma} \sum_{\mbk a b} 
\frac{\pd f^a}{\pd \veps^a}
\tilde{\Omega}_{\mu\nu}^{ab} \pd_{\rho} \veps^{ab}
\frac{1+ 5 (\eta^{ab})^2}{1 + (\eta^{ab})^2}
+
\frac{e^3}{\hbar \gamma} \sum_{\mbk a b} 
\pd_{\rho} f^a \tilde{\Omega}_{\mu\nu}^{ab}
\\
&+
\frac{e^3}{2\hbar} \sum_{\mbk a b}
\frac{\pd f^a}{\pd \veps^a}
\pd_{\mu} \tilde{g}_{\nu\rho}^{ab}
+
\frac{e^3}{\hbar \gamma^2} \sum_{\mbk a b}
\frac{\pd f^a}{\pd \veps^a}
\tilde{g}_{\nu \rho}^{ab} \pd_{\mu} \veps^{ab} \veps^{ab}
\frac{1+ 2 (\eta^{ab})^2}{1 + (\eta^{ab})^2}
\\
&-
\frac{e^3}{\hbar \gamma^2} \sum_{\mbk a b}
\pd_{\rho}f^a \tilde{g}_{\mu\nu}^{ab} \veps^{ab}
\left[ 1+ 3 (\eta^{ab})^2 \right]
-
\frac{e^3}{2 \hbar} \sum_{\mbk a b}
\frac{\pd f^a}{\pd \veps^a} \tilde{g}_{\mu\nu}^{ab}
\frac{\pd_{\rho} \veps^{ab}}{\veps^{ab}} 
\frac{1 - 3 (\eta^{ab})^2}{1 + (\eta^{ab})^2}
+
\sum_{i=1}^4 \mathcal{K}_i,
\end{split}
\end{equation}
where
\begin{equation}
\begin{split}
\mathcal{K}_1
=&
\frac{e^3}{2 \hbar \gamma} \sum_{\mbk a b}
\frac{\pd f^a}{\pd \veps^a} \veps^{ab}
\text{Im} \bigg\{
\frac{\mathcal{A}_{\mu}^{\pr ab} 
(\mathcal{D}_{\nu} \left[ \mathcal{A}_{\rho}^{\pr} , H \right])^{ba}}
{\veps^{ab} - i \gamma} \bigg\}
\\
=&
\frac{e^3}{ 2 \hbar \gamma} \sum_{\mbk a b}
\frac{\pd f^a}{\pd \veps^a}
\frac{1}{ 1 + (\eta^{ab})^2 }
\left[
- \frac{1}{2} \Omega_{\mu\rho}^{ab} \pd_{\nu} \veps^{ab}
+
\text{Im} (C_{\mu\nu\rho}^{ba}) \veps^{ab}
-
\sum_c \text{Re} (S_{\mu\nu\rho}^{abc}) (\veps^{bc} - \veps^{ca})
\right]
\\
&+
\frac{e^3}{ 2 \hbar} \sum_{\mbk a b}
\frac{\pd f^a}{\pd \veps^a}
\frac{ \veps^{ab} }{(\veps^{ab})^2 + \gamma^2}
\left[
g_{\mu\rho}^{ab} \pd_{\nu} \veps^{ab}
+
\text{Re} (C_{\mu\nu\rho}^{ba}) \veps^{ab}
+
\sum_c \text{Im} (S_{\mu\nu\rho}^{abc}) (\veps^{bc} - \veps^{ca})
\right],
\end{split}
\end{equation}
%%%%%%%%
\begin{equation}
\begin{split}
\mathcal{K}_2
=&
- \frac{e^3}{\hbar \gamma} \sum_{\mbk a b}
\frac{\pd f^a}{\pd \veps^a}
\frac{\text{Im} \{
\mathcal{A}_{\nu}^{\pr ab} 
(\mathcal{D}_{\mu} \left[ \mathcal{A}_{\rho}^{\pr} , H \right])^{ba}\}}{1 + (\eta^{ab})^2}
\\
=&
- \frac{e^3}{\hbar \gamma} \sum_{\mbk a b}
\frac{\pd f^a}{\pd \veps^a}
\left[
\text{Im} (\tilde{C}_{\nu\mu\rho}^{ba}) \veps^{ab}
-
\sum_c \text{Re} (\tilde{S}_{\nu\mu\rho}^{abc}) (\veps^{bc} - \veps^{ca})
-
\frac{1}{2}
\text{Im} (\tilde{\mathcal{T}}_{\nu\mu\rho}^{ba}) \veps^{ab}
\right],
\end{split}
\end{equation}
%%%%%%%%
\begin{equation}
\begin{split}
\mathcal{K}_3
=&
\frac{ e^3 }{ \hbar } \sum_{\mbk a b}
\frac{\pd f^a}{\pd \veps^a}
\frac{\veps^{ab}}{(\veps^{ab})^2 + \gamma^2}
\text{Im} \{ \mathcal{A}_{\nu}^{\pr ab} 
\left[ \mathcal{A}_{\mu}^{\pr} , \left[ \mathcal{A}_{\rho}^{\pr} , H \right] \right]^{ba} \}
\\
=&
\frac{e^3}{\hbar} \sum_{\mbk a b}
\frac{\pd f^a}{\pd \veps^a}
\left[
\text{Re} (\tilde{\mathcal{T}}_{\nu\mu\rho}^{ba})
+
\sum_c \text{Im} (\tilde{Q}_{\nu\mu\rho}^{abc})
\frac{\veps^{bc}}{\veps^{ab}}
-
\sum_c \text{Im} (\tilde{Q}_{\nu\rho\mu}^{abc})
\frac{\veps^{ca}}{\veps^{ab}}
\right],
\end{split}
\end{equation}
%%%%%%%%
\begin{equation}
\begin{split}
\mathcal{K}_{4}
=&
- \frac{e^3}{\hbar \gamma} \sum_{\mbk a}
\frac{\pd f^a}{\pd \veps^a}
\text{Re} \big\{
\left( 
\left[
\left[\alpha_{\mu} (\gamma) , H \right]
,
\left[ \mathcal{A}_{\nu}^{\pr} , H \right]
\right] 
\left[ \alpha_{\rho} (-\gamma) , H \right]
\right)_{\mbk}^{a}
\big\}
\\
=&
\frac{e^3}{\hbar \gamma} \sum_{\mbk abc}
\frac{\pd f^a}{\pd \veps^a}
\bigg\{
\text{Re} (Q_{\rho\nu\mu}^{abc})
\frac{ \veps^{ac} \veps^{ab} - \gamma^2}{ [ (\veps^{ac})^2 + \gamma^2 ]
[ (\veps^{ab})^2 + \gamma^2 ] }
+
\text{Im} (Q_{\rho\nu\mu}^{abc})
\frac{ \gamma ( \veps^{ac} + \veps^{ab} ) }{ [ (\veps^{ac})^2 + \gamma^2 ]
[ (\veps^{ab})^2 + \gamma^2 ] }
\\
&-
\text{Re} (Q_{\rho\mu\nu}^{abc})
\frac{ \veps^{cb} \veps^{ab} - \gamma^2}{ [ (\veps^{ab})^2 + \gamma^2 ]
[ (\veps^{bc})^2 + \gamma^2 ] }
-
\text{Im} (Q_{\rho\mu\nu}^{abc})
\frac{ \gamma ( \veps^{cb} + \veps^{ab} ) }{ [ (\veps^{ab})^2 + \gamma^2 ]
[ (\veps^{bc})^2 + \gamma^2 ] }
\bigg\}
\veps^{ab} \veps^{bc} \veps^{ca},
\end{split}
\end{equation}
resulting in the contribution
\begin{equation}
\label{sig2_B}
\begin{split}
\sigma_{\mu\nu\rho}^{B}
=&
-\frac{3 e^3}{2 \gamma^2}
\sum_{\mbk a} \pd_{\nu} \pd_{\rho} f^a  v_{\mu}^a
+
\frac{e^3}{\hbar \gamma} \sum_{\mbk ab} \pd_{\rho} f^a
\tilde{\Omega}_{\mu\nu}^{ab}
-
\frac{e^3}{2 \hbar \gamma} \sum_{\mbk ab}
\frac{\pd f^a}{\pd \veps^a}
\tilde{\Omega}_{\mu\nu}^{ab} \pd_{\rho} \veps^{ab}
\frac{1 + 3 (\eta^{ab})^2}{1 + (\eta^{ab})^2}
\\
&+
\frac{e^3}{2 \hbar} \sum_{\mbk ab}
\frac{\pd f^a}{\pd \veps^a}
\pd_{\rho} \tilde{g}_{\mu\nu}^{ab}
-
\frac{3 e^3}{\hbar \gamma} \sum_{\mbk ab}
\pd_{\rho} f^a \tilde{g}_{\mu\nu}^{ab} \eta^{ab}
-
\frac{e^3}{\hbar \gamma^2} \sum_{\mbk ab}
\pd_{\rho} f^a \tilde{g}_{\mu\nu}^{ab} \veps^{ab}
\\
&-
\frac{e^3}{\hbar^2} \sum_{\mbk ab}
\frac{\pd f^a}{\pd \veps^a}
\tilde{g}_{\mu\nu}^{ab} \veps^{ab} \pd_{\rho} m^{ab}
\frac{(\eta^{ab})^2}{1 + (\eta^{ab})^2}
+
\frac{e^3}{2 \hbar} \sum_{\mbk ab}
\frac{\pd f^a}{\pd \veps^a}
\tilde{\Gamma}_{\nu\rho\mu}^{ab}
\\
&+
\frac{e^3}{2 \hbar^2} \sum_{\mbk ab}
\frac{\pd f^a}{\pd \veps^a}
\tilde{g}_{\nu\rho}^{ab} \veps^{ab} \pd_{\mu} m^{ab}
\left[ 
\frac{(\eta^{ab})^2}{1 + (\eta^{ab})^2}
-
\frac{2}{(\eta^{ab})^2} 
\frac{1+ 2 (\eta^{ab})^2}{1 + (\eta^{ab})^2}
\right]
\\
&+
\frac{e^3}{2 \hbar} \sum_{\mbk ab}
\frac{\pd f^a}{\pd \veps^a}
\text{Re} (\tilde{K}_{\mu\nu\rho}^{ba})
+
\frac{e^3}{2 \hbar} \sum_{\mbk ab}
\frac{\pd f^a}{\pd \veps^a}
\frac{1}{\eta^{ab}}
\text{Im}
\left(
\tilde{C}_{\mu \nu \rho}^{ba}
-
\tilde{C}_{\nu \mu \rho}^{ba}
-
\tilde{C}_{\rho \nu \mu}^{ba}
\right)
\\
&-
\frac{e^3}{2 \hbar} \sum_{\mbk abc}
\frac{\pd f^a}{\pd \veps^a}
\left[
\text{Re} (\tilde{S}_{\mu\nu\rho}^{abc})
- \eta^{ab} \text{Im} (\tilde{S}_{\mu\nu\rho}^{abc})
\right]
\left( \frac{1}{\eta^{bc}} - \frac{1}{\eta^{ca}} \right)
\\
&+
\frac{e^3}{\hbar} \sum_{\mbk abc}
\frac{\pd f^a}{\pd \veps^a}
\text{Re} (\tilde{S}_{\nu\mu\rho}^{abc})
\left[
\frac{\eta^{bc}}{\eta^{ca}}
\frac{ \eta^{ab} - \eta^{bc} }{1 + (\eta^{bc})^2}
-
\frac{\eta^{ca}}{\eta^{bc}}
\frac{ \eta^{ab} - \eta^{ca} }{1 + (\eta^{ca})^2}
\right]
\\
&-
\frac{e^3}{\hbar} \sum_{\mbk abc}
\frac{\pd f^a}{\pd \veps^a}
\text{Im} (\tilde{S}_{\nu\mu\rho}^{abc})
\left[
\frac{\eta^{bc}}{\eta^{ca}}
\frac{ 1 + \eta^{ab} \eta^{bc} }{1 + (\eta^{bc})^2}
-
\frac{\eta^{ca}}{\eta^{bc}}
\frac{ 1 + \eta^{ab} \eta^{ca} }{1 + (\eta^{ca})^2}
\right]
\\
&+
\frac{e^3}{\hbar} \sum_{\mbk abc}
\frac{\pd f^a}{\pd \veps^a}
\text{Re} (\tilde{A}_{\nu\mu\rho}^{abc})
\left[
\frac{1}{\eta^{ca}}
\frac{ 1 + \eta^{ab} \eta^{bc} }{1 + (\eta^{bc})^2}
+
\frac{1}{\eta^{bc}}
\frac{ 1 + \eta^{ab} \eta^{ca} }{1 + (\eta^{ca})^2}
\right]
\\
&-
\frac{e^3}{\hbar} \sum_{\mbk abc}
\frac{\pd f^a}{\pd \veps^a}
\text{Im} (\tilde{A}_{\nu\mu\rho}^{abc})
\left[
\frac{\eta^{bc}}{\eta^{ca}}
\frac{ 1 + \eta^{ab} \eta^{bc} }{1 + (\eta^{bc})^2}
+
\frac{\eta^{ca}}{\eta^{bc}}
\frac{ 1 + \eta^{ab} \eta^{ca} }{1 + (\eta^{ca})^2}
\right].
\end{split}
\end{equation}

Adding all the various conductivity contributions, the total current density can be obtained. To do so, we note that the distribution derivative terms in Eq.~(\ref{sig2_B}) can be reexpressed with the identity
\begin{equation}
\sum_{\mbk} \frac{\pd f^a}{\pd \veps^a} \mathcal{O}
=
- \sum_{\mbk} f^a \pd^{\sigma}
\left[
\frac{1}{\hbar} \frac{v_{\sigma}^a}{(\bs{v}^a)^2} \mathcal{O}
\right]
=
- \sum_{\mbk} f^a
\bigg\{ \pd^{\sigma} , 
\frac{1}{\hbar} \frac{v_{\sigma}^a}{(\bs{v}^a)^2} \bigg\}\mathcal{O},
\end{equation}
where $\mathcal{O}$ is an arbitrary momentum- and band-dependent function and the curly brackets in the last term denote the anticommutator.

Collecting terms from the linear and nonlinear conductivities, we ultimately arrive at the total current density
\begin{equation}
\label{j_mu_supp}
\begin{split}
\frac{j_{\mu}}{-e}
=&
\sum_{\mbk a} (n_1^a + n_2^a) v_{\mu}^a
+
\dot{k}^{\nu} \sum_{\mbk ab} (f^a + n_1^a)
\mathcal{Z}_{\Omega}^{ab} \tilde{\Omega}_{\mu \nu}^{ab}
-
\dot{k}^{\nu} \sum_{\mbk ab} (f^a + n_1^a)
\mathcal{Z}_{g}^{ab} \tilde{g}_{\mu \nu}^{ab}
\\
&+
\dot{k}^{\nu} \dot{k}^{\rho}
\sum_{\mbk ab} f^a m^{ab}
\mathcal{Z}_{\Gamma}^{ab}
\tilde{\Gamma}_{\nu \rho \mu}^{ba}
+
\dot{k}^{\nu} \dot{k}^{\rho}
\sum_{\mbk ab} f^a m^{ab}
\mathcal{Z}_{\Gamma}^{ab}
\tilde{\Gamma}_{\nu \rho \mu}^{ba}
+
\dot{k}^{\nu} \dot{k}^{\rho} \sum_{\mbk ab} f^a 
\mathcal{Z}_{m}^{ab} 
\tilde{g}_{\nu \rho}^{ab} \pd_{\mu} m^{ab}
\\
&+
\dot{k}^{\nu} \dot{k}^{\rho}
\sum_{\mbk ab} f^a m^{ab}
\left[
\mathcal{Z}_{K}^{ab}
\text{Re} (\tilde{\mathcal{K}}_{\mu\nu\rho}^{ba})
+
\mathcal{Z}_{C}^{ab}
\text{Im}
\left(
\tilde{C}_{\mu \nu \rho}^{ba}
-
\tilde{C}_{\nu \mu \rho}^{ba}
-
\tilde{C}_{\rho \nu \mu}^{ba}
\right)
\right]
\\
&+
\dot{k}^{\nu} \dot{k}^{\rho}
\sum_{\mbk abc} f^a m^{ab}
\left[
\mathcal{Z}_{1}^{abc}
\text{Re} (\tilde{S}_{\mu\nu\rho}^{abc})
+
\mathcal{Z}_{2}^{abc}
\text{Im} (\tilde{S}_{\mu\nu\rho}^{abc})
+
\mathcal{Z}_{3}^{abc}
\text{Re} (\tilde{S}_{\nu\mu\rho}^{abc})
+
\mathcal{Z}_{4}^{abc}
\text{Im} (\tilde{S}_{\nu\mu\rho}^{abc})
\right]
\\
&+
\dot{k}^{\nu} \dot{k}^{\rho}
\sum_{\mbk abc} f^a m^{ab}
\left[
\mathcal{Z}_{5}^{abc}
\text{Re} (\tilde{A}_{\nu\mu\rho}^{abc})
+
\mathcal{Z}_{6}^{abc}
\text{Im} (\tilde{A}_{\nu\mu\rho}^{abc})
\right],
\end{split}
\end{equation}
where $n_2^a = - \tau \dot{k}^{\nu} \pd_{\nu} n_1^a$ yields the nonlinear Drude weight and the various renormalization functions that appear are given by
\begin{equation}
\mathcal{Z}_{\Omega}^{ab}
=
1 + \dot{k}^{\rho} \pd_{\rho} m^{ab}
\frac{2 \eta^{ab}}{1 + (\eta^{ab})^2}
+
\frac{1}{2} \bigg\{ \pd^{\sigma} , 
\frac{v_{\sigma}^a}{(\bs{v}^a)^2}
\dot{k}^{\rho} \pd_{\rho} \ln (m^{ab})
\frac{1}{\eta^{ab}} 
\frac{1 + 3 (\eta^{ab})^2}{1 + (\eta^{ab})^2}
\bigg\},
\end{equation}
%%%%%%%%
\begin{equation}
\begin{split}
\mathcal{Z}_{g}^{ab}
=&
2 \eta^{ab} - \dot{k}^{\rho} \pd_{\rho} m^{ab}
\frac{1 -  (\eta^{ab})^2 }{ 1 + (\eta^{ab})^2}
-
\frac{1}{2} \bigg\{ 
\pd^{\sigma} , 
\frac{v_{\sigma}^a}{(\bs{v}^a)^2}
\bigg\}
\dot{k}^{\rho} \pd_{\rho}
+
\frac{1}{(\eta^{ab})^2} 
\left[
m^{ab} , \dot{k}^{\rho} \pd_{\rho}
\right]
\\
&+
\bigg\{ \pd^{\sigma} , 
\frac{v_{\sigma}^a}{(\bs{v}^a)^2}
\dot{k}^{\rho} \pd_{\rho} \ln (m^{ab})
\frac{(\eta^{ab})^2}{1 + (\eta^{ab})^2}
\bigg\},
\end{split}
\end{equation}
%%%%%%%%
\begin{equation}
\mathcal{Z}_{m}^{ab}
=
- \frac{1}{(\eta^{ab})^2}
\frac{1+ 2 (\eta^{ab})^2}{1 + (\eta^{ab})^2}
+
\frac{1}{2} \bigg\{ 
\pd^{\sigma} ,
\frac{1}{m^{ab}}
\frac{v_{\sigma}^a}{(\bs{v}^a)^2}
\left[
\frac{(\eta^{ab})^2}{1 + (\eta^{ab})^2}
-
\frac{2}{(\eta^{ab})^2}
\frac{ 1 + 2 (\eta^{ab})^2}{1 + (\eta^{ab})^2}
\right]
\bigg\},
\end{equation}
%%%%%%%%
\begin{equation}
\mathcal{Z}_{\Gamma}^{ab}
=
\mathcal{Z}_{K}^{ab}
=
1
+
\frac{1}{2 m^{ab}} \bigg\{ 
\pd^{\sigma} , 
\frac{v_{\sigma}^a}{(\bs{v}^a)^2}
\bigg\}
\end{equation}
%%%%%%%
\begin{equation}
\mathcal{Z}_{C}^{ab}
=
\frac{1}{\eta^{ab}}
+
\frac{1}{2 m^{ab}} \bigg\{ 
\pd^{\sigma} , 
\frac{v_{\sigma}^a}{(\bs{v}^a)^2}
\frac{1}{\eta^{ab}}
\bigg\}
\end{equation}
%%%%%%%
\begin{equation}
\mathcal{Z}_{1}^{abc}
=
\frac{1}{\eta^{ca}}
-
\frac{1}{\eta^{bc}}
+
\frac{1}{2 m^{ab}} \bigg\{ 
\pd^{\sigma} , 
\frac{v_{\sigma}^a}{(\bs{v}^a)^2}
\left(
\frac{1}{\eta^{ca}}
-
\frac{1}{\eta^{bc}}
\right)
\bigg\},
\end{equation}
%%%%%%%
\begin{equation}
\mathcal{Z}_{2}^{abc}
=
\frac{\eta^{ab}}{\eta^{bc}}
-
\frac{\eta^{ab}}{\eta^{ca}}
+
\frac{1}{2 m^{ab}} \bigg\{ 
\pd^{\sigma} , 
\frac{v_{\sigma}^a}{(\bs{v}^a)^2}
\left(
\frac{\eta^{ab}}{\eta^{bc}}
-
\frac{\eta^{ab}}{\eta^{ca}}
\right)
\bigg\},
\end{equation}
%%%%%%%%
\begin{equation}
\begin{split}
\mathcal{Z}_{3}^{abc}
=&
\frac{2}{\eta^{ab}}
\left[
\frac{\eta^{ca}}{\eta^{bc}}
\frac{1+\eta^{ab}\eta^{ca}}{1+(\eta^{ca})^2}
-
\frac{\eta^{bc}}{\eta^{ca}}
\frac{1+\eta^{ab}\eta^{bc}}{1+(\eta^{bc})^2}
\right]
\\
&-
\frac{1}{m^{ab}} \bigg\{ 
\pd^{\sigma} , 
\frac{v_{\sigma}^a}{(\bs{v}^a)^2}
\left[
\frac{\eta^{ca}}{\eta^{bc}}
\frac{\eta^{ab} - \eta^{ca}}{1+(\eta^{ca})^2}
-
\frac{\eta^{bc}}{\eta^{ca}}
\frac{\eta^{ab} - \eta^{bc}}{1+(\eta^{bc})^2}
\right]
\bigg\},
\end{split}
\end{equation}
%%%%%%%%%
\begin{equation}
\begin{split}
\mathcal{Z}_{4}^{abc}
=&
\frac{2}{\eta^{ab}}
\left[
\frac{\eta^{ca}}{\eta^{bc}}
\frac{\eta^{ab} - \eta^{ca}}{1+(\eta^{ca})^2}
-
\frac{\eta^{bc}}{\eta^{ca}}
\frac{\eta^{ab} - \eta^{bc}}{1+(\eta^{bc})^2}
\right]
\\
&+
\frac{1}{m^{ab}} \bigg\{ 
\pd^{\sigma} , 
\frac{v_{\sigma}^a}{(\bs{v}^a)^2}
\left[
\frac{\eta^{ca}}{\eta^{bc}}
\frac{1+\eta^{ab}\eta^{ca}}{1+(\eta^{ca})^2}
-
\frac{\eta^{bc}}{\eta^{ca}}
\frac{1+\eta^{ab}\eta^{bc}}{1+(\eta^{bc})^2}
\right]
\bigg\},
\end{split}
\end{equation}
%%%%%%%%
\begin{equation}
\begin{split}
\mathcal{Z}_{5}^{abc}
=&
\frac{2}{\eta^{ab}}
\left[
\frac{1}{\eta^{ca}}
\frac{\eta^{ab} - \eta^{bc}}{1+(\eta^{bc})^2}
+
\frac{1}{\eta^{bc}}
\frac{\eta^{ab} - \eta^{ca}}{1+(\eta^{ca})^2}
\right]
\\
&+
\frac{1}{m^{ab}} \bigg\{ 
\pd^{\sigma} , 
\frac{v_{\sigma}^a}{(\bs{v}^a)^2}
\left[
\frac{1}{\eta^{ca}}
\frac{1 + \eta^{ab}\eta^{bc}}{1+(\eta^{bc})^2}
+
\frac{1}{\eta^{bc}}
\frac{1 + \eta^{ab}\eta^{ca}}{1+(\eta^{ca})^2}
\right]
\bigg\},
\end{split}
\end{equation}
%%%%%%%%
\begin{equation}
\begin{split}
\mathcal{Z}_{6}^{abc}
=&
- \frac{2}{\eta^{ab}}
\left[
\frac{\eta^{bc}}{\eta^{ca}}
\frac{\eta^{ab} - \eta^{bc}}{1+(\eta^{bc})^2}
+
\frac{\eta^{ca}}{\eta^{bc}}
\frac{\eta^{ab} - \eta^{ca}}{1+(\eta^{ca})^2}
\right]
\\
&+
\frac{1}{m^{ab}} \bigg\{ 
\pd^{\sigma} , 
\frac{v_{\sigma}^a}{(\bs{v}^a)^2}
\left[
\frac{\eta^{bc}}{\eta^{ca}}
\frac{1 + \eta^{ab}\eta^{bc}}{1+(\eta^{bc})^2}
+
\frac{\eta^{ca}}{\eta^{bc}}
\frac{1 + \eta^{ab}\eta^{ca}}{1+(\eta^{ca})^2}
\right]
\bigg\}.
\end{split}
\end{equation}
From these, we observe that the distribution derivative terms in the response functions not only affect the carrier densities, but can also be thought of as inducing gradient corrections to the various geometric structures that appear in the equation of motion through the renormalization functions. Comparing Eq.~(\ref{j_mu_supp}) with the generic relation for the current density, $\sum_{\mbk a} n^a \dot{x}_{\mu}^a$, where $n^a = f^a + n_1^a + n_2^a$, the dressed equation of motion of the carrier position is obtained.

\section{Derivation of Dissipative field equations}

To derive the dissipative field equations, Eq.~(17) in the main text, it is helpful to recall the expression for the Levi-Civita connection of an arbitrary rank $(p,q)$ tensor $\mathcal{O}$~\cite{carroll2019spacetime}
\begin{equation}
\begin{split}
\nabla_{\rho} \tensor{\mathcal{O}}{^{\mu_1}^{\cdots}^{\mu_p}_{\nu_1}_{\cdots}_{\nu_q}}
&=
\pd_{\rho} \tensor{\mathcal{O}}{^{\mu_1}^{\cdots}^{\mu_p}_{\nu_1}_{\cdots}_{\nu_q}}
+
\Gamma_{\rho \sigma}^{\mu_1}
\tensor{\mathcal{O}}{^{\sigma}^{\cdots}^{\mu_p}_{\nu_1}_{\cdots}_{\nu_q}}
+
\cdots
+
\Gamma_{\rho \sigma}^{\mu_p}
\tensor{\mathcal{O}}{^{\mu_1}^{\cdots}^{\sigma}_{\nu_1}_{\cdots}_{\nu_q}}
\\
&-
\Gamma_{\rho \nu_1}^{\sigma}
\tensor{\mathcal{O}}{^{\mu_1}^{\cdots}^{\mu_p}_{\sigma}_{\cdots}_{\nu_q}}
-
\cdots
-
\Gamma_{\rho \nu_q}^{\sigma}
\tensor{\mathcal{O}}{^{\mu_1}^{\cdots}^{\mu_p}_{\nu_1}_{\cdots}_{\sigma}},
\end{split}
\end{equation}
as well as the Riemann tensor in terms of connection components
\begin{equation}
\tensor{R}{^{\mu}_\nu_\rho_\sigma}
=
\pd_{\rho} \Gamma^{\mu}_{\nu \sigma}
-
\pd_{\sigma} \Gamma^{\mu}_{\nu \rho}
+
\Gamma^{\lambda}_{\nu \sigma}
\Gamma^{\mu}_{\lambda \rho}
-
\Gamma^{\lambda}_{\nu \rho}
\Gamma^{\mu}_{\lambda \sigma}.
\end{equation}
Following the procedure to obtain dressed quantities as discussed in the main text, we perform the necessary contractions to obtain the dressed Ricci tensor and scalar as
\begin{equation}
\tilde{R}_{\mu\nu}
=
R_{\mu\nu} - \frac{1}{2} g_{\mu\nu} \nabla^2 \lambda
-
\frac{n-2}{2} \nabla_{\mu} \nabla_{\nu}\lambda,
\end{equation}
and
\begin{equation}
\tilde{R}
=
\lambda^{-1}R
-
(n+1) \nabla^2 \lambda.
\end{equation}
Combining terms results in the dissipative contributions to the Einstein tensor, $\tilde{G}_{\mu\nu} - G_{\mu\nu}$, from which the field equations are obtained.

\section{Conclusion and Outlook}

In this chapter, we have presented an extension of semiclassical momentum-space gravity to dissipative multiband systems by taking quantum response theory as the starting point. Within a diagrammatic approach that includes a phenomenological dissipation parameter, we have studied the simultaneous and interconnected dressing of the quantum geometry, carrier dynamics and momentum-space EFE, resulting in a dressed theory of momentum-space gravity. On the technical side, this work introduces and provides two equivalent definitions of the three-state QGT as the simplest quantum geometric quantity beyond the two-state picture, which paves the way for future studies of higher-state quantum geometry. In addition, we have clarified the role of the quantum geometric contorsion tensor and its relation to the full Berry covariant derivative.

A general viewpoint on quantum response theory that has become more prominent recently is to view it as a probe of the rich Riemannian geometry of quantum state manifolds. The results presented here suggest that this may be harnessed to study a variety of theories of gravitation in quantum materials via optical, magnetic or thermal means once one identifies the corresponding gravitational effect in the multiband system. In addition, this approach shines an arguably more positive light on the role of dissipation, which is ubiquitous in any realistic system, but often regarded as an obstacle in probing intrinsic materials properties. Its interpretation as a generator of gravity in momentum space suggests a path to explore the connections between thermodynamics and gravity--which have been and continue to be of great interest in fundamental physics~\cite{bekenstein1973black, hawking1975particle, ruppeiner1979thermodynamics, jacobson1995thermodynamics, padmanabhan2010thermodynamical, verlinde2011origin, carroll2016what, bianconi2025gravity}--via responses in materials systems.

The analysis of the present chapter utilized standard Kubo formulas for closed quantum systems within the relaxation-time approximation, which suggests several generalizations for future works. One possibly interesting direction to explore from the gravitational point of view is the significance of special disorder scattering (i.e., side jump and skew scattering)--or nonscalar scatterings in the quantum geometry and carrier dynamics~\cite{mehraeen2024quantum, huang2025scaling, liu2024effect, gong2024nonlinear}. The former is particularly significant, given that special scattering processes in the nonlinear response regime often have relatively complicated and arguably unintuitive classifications. Therefore, finding their gravitational correspondences may be helpful in revealing a more intuitive picture of their nature. Furthermore, by considering system-bath correlations, one could generalize the present analysis to explore the interplay between quantum geometry and thermodynamics via momentum-space gravity in open quantum systems~\cite{konopik2019quantum}.
 
Finally, we anticipate that the results presented here may also be of general interest to studies in quantum science, information and technology. A notable direction in this regard is in the active field of quantum information geometry, given its common underlying geometric framework~\cite{facchi2010classical}. In addition, given recent developments in simulating dissipative nonlinear responses via generalizations of quantum phase estimation frameworks~\cite{loaiza2024nonlinear}, it would be quite interesting to consider the possibility of leveraging this approach to simulate theories of gravity via nonlinear responses on quantum computers.

\chapter{Spin Anomalous-Hall Unidirectional Magnetoresistance}

\section{Introduction}

Originating from the interplay between magnetism and relativistic spin-orbit interaction, the anomalous Hall (AH) effect in solid-state systems with broken time-reversal symmetry has been of enduring interest for more than a century~\cite{Hall1881AHE,Nagaosa10RMP}. One class of materials that has received particular attention in studies of this effect are conducting ferromagnets~\cite{Nagaosa10RMP}, such as ferromagnetic metals. 

The AH effect in ferromagnetic metals has several salient properties. Due to the coupling of spin and orbital degrees of freedom, the effect not only generates a transverse charge current|which is perpendicular to both the magnetization and the applied electric field, but also gives rise to a transverse spin current. And, in ferromagnetic metals with strong exchange interaction, conduction-electron spins are well aligned with the local magnetization, making the coupled spin and charge currents 
controllable by varying the direction of the magnetization. Furthermore, the mobility of conduction electrons in a ferromagnetic metal is, in general, spin-dependent, enabling mutual conversion between spin and charge currents mediated by the AH effect. 

These properties have been shown to spawn unconventional magnetoresistances in the linear response regime. For instance, both a bulk anisotropic magnetoresistance and planar Hall resistance may result from two consecutive transverse scatterings of spin-polarized conduction electrons, due to the AH effect~\cite{sZhang14JAP_AMR}. In geometrically confined systems|such as ferromagnetic-metal thin films or layered structures, the anomalous-Hall induced anisotropic magnetoresistances may acquire distinctive angular dependences, owing to the modulation of the bulk spin and charge currents caused by interfacial spin accumulation and the resulting diffusive spin current~\cite{zhang14Thesis,Tomo15PRAppl_AHE-SOT, xrWang16EPL_AMR,Tomo16PRB_AH-AMR,yhWu18NC_AHMR,yzWu20NJP_AHMR}. 

In the nonlinear response regime|where the Onsager reciprocal relations no longer hold, the role of the AH effect is yet to be explored. In this chapter, we unveil a unidirectional magnetoresistance (UMR) driven by the spin AH effect in conducting ferromagnet$\mid$nonmagnet bilayer systems, whereby the resistance can be altered by reversing the direction of either the magnetization or the applied electric field. Hereafter, we shall refer to this nonlinear magnetoresistance as spin anomalous-Hall unidirectional magnetoresistance (AH-UMR). 

\begin{figure}[hpt]
\captionsetup[subfigure]{labelformat=empty}
    \sidesubfloat[]{\includegraphics[width=0.3\linewidth,trim={0cm 0cm 0cm 0cm}]{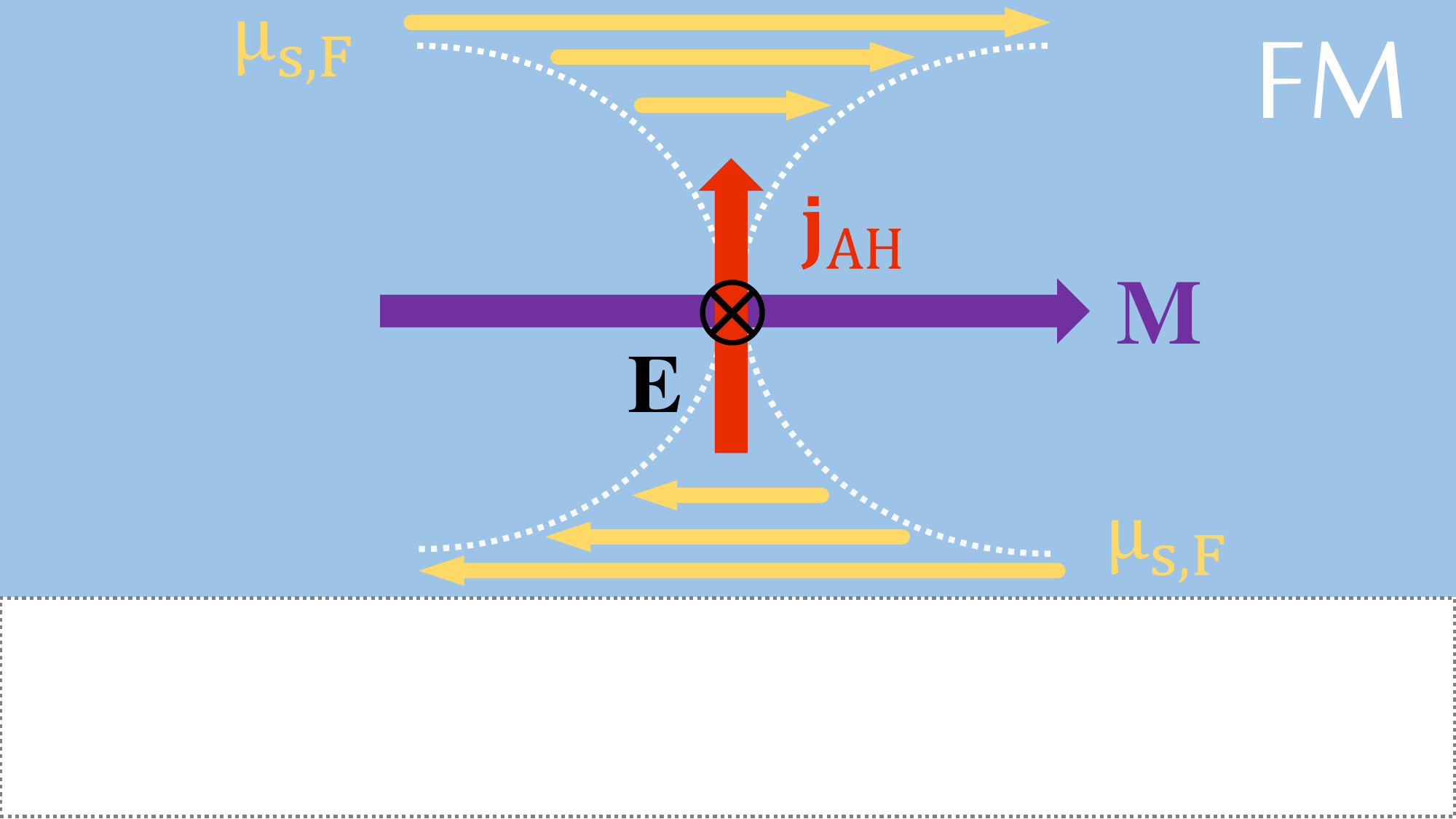}\label{fig_sahumr_fig1a}}
%\quad%
    \sidesubfloat[]{\includegraphics[width=0.3\linewidth,trim={0cm 0cm 0cm 0cm}]{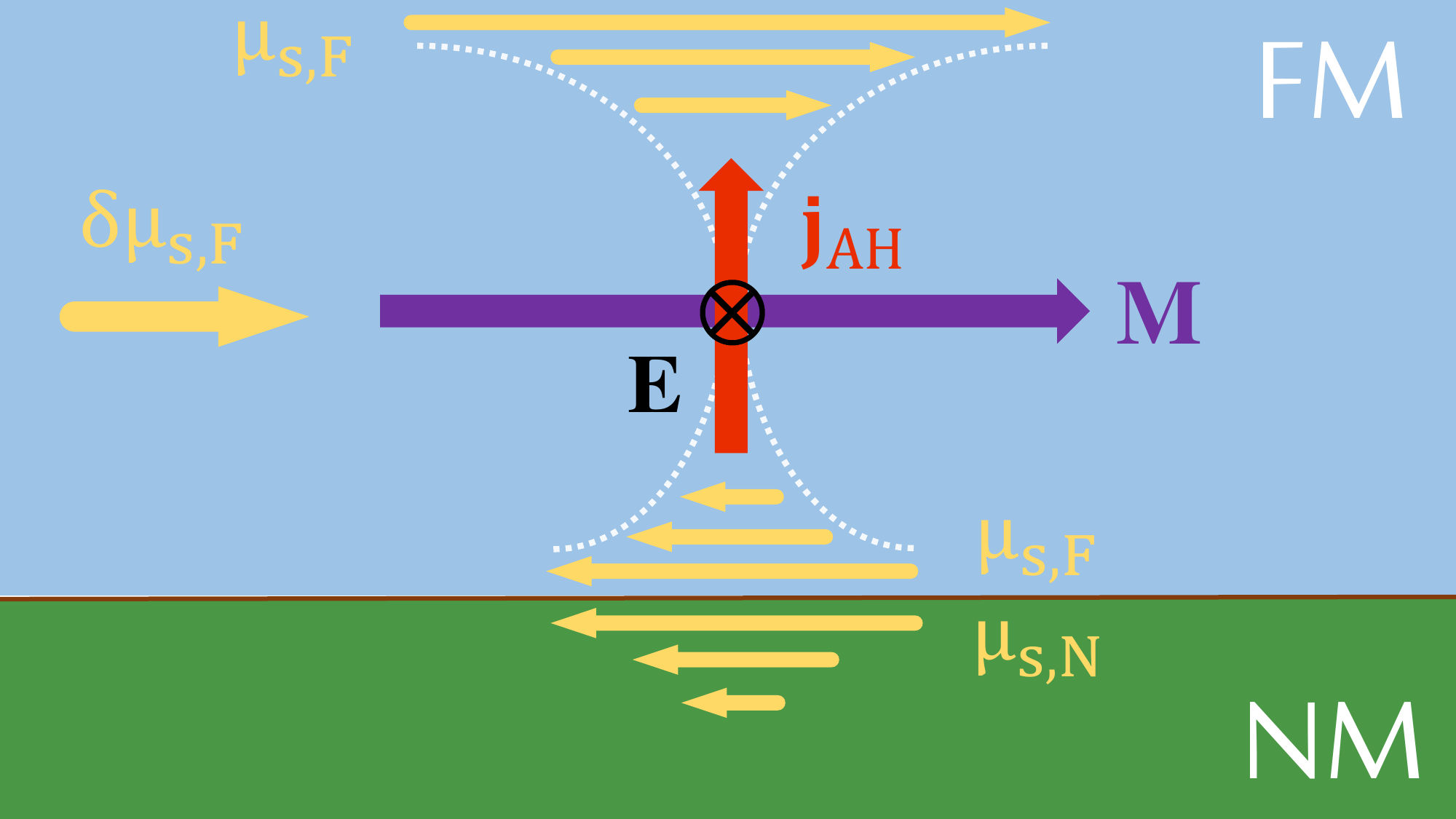}\label{fig_sahumr_fig1b}}
%%%%%%%    
    \sidesubfloat[]{\includegraphics[width=0.3\linewidth,trim={0cm 0cm 0cm 0cm}]{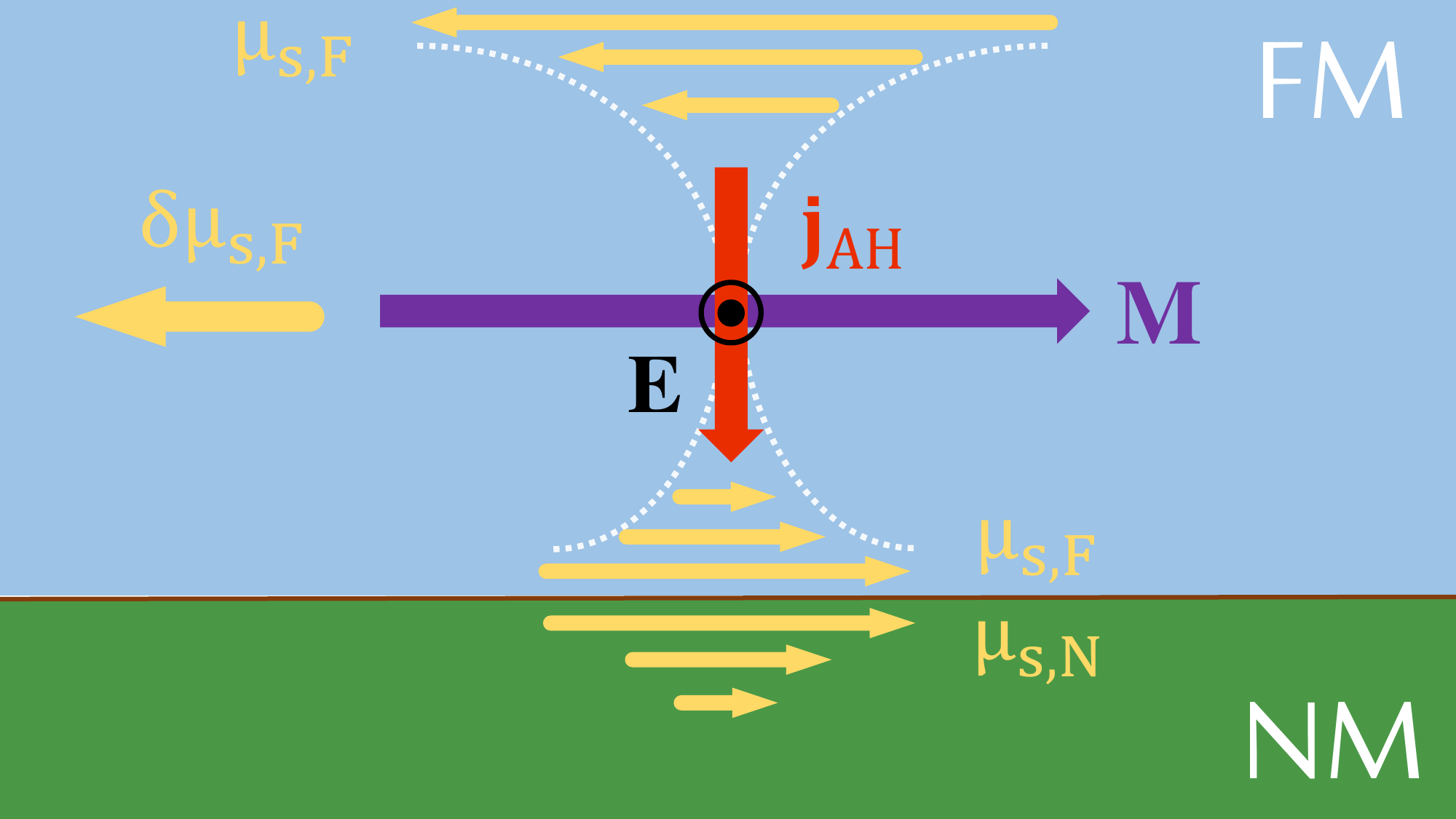}\label{fig_sahumr_fig1c}}
%%%%%%%%%
    \caption{Schematics of the spin AH-UMR effect. (a) In a single ferromagnetic-metal (FM) layer, the spin accumulation $\mu_{s,F}$ (indicated by the small yellow arrows) has an antisymmetric distribution about the center line of the layer, with no net nonequilibrium spin density (spatially-averaged $\mu_{s,F}$) induced. So the overall spin AH-UMR is also zero. (b) The presence of a neighboring nonmagnetic-metal (NM) layer induces a net nonequilibrium spin density $\delta \mu_{s,F}$ (indicated by the large yellow arrow) in the FM layer, giving rise to a finite spin AH-UMR. (c) Reversing the electric field direction flips the direction of the AH current $\mathbf{j}_{\text{AH}}$ and thus the sign of $\delta \mu_{s,F}$, thereby changing the sign of the spin AH-UMR.}
    \label{fig_sahumr_fig1}
\end{figure}

The underlying physics of the spin AH-UMR can be understood as follows. In a single ferromagnetic-metal layer, the spin current induced by the spin AH effect creates spin accumulations of opposite orientations at the top and bottom surfaces, but there is no net nonequilibrium spin density due to inversion symmetry, as shown schematically in Fig.~\ref{fig_sahumr_fig1}\textcolor{purple}{a}. This, however, is no longer the case when a nonmagnetic-metal layer is attached to the ferromagnetic-metal layer, as the spin accumulation at the interface would ``leak" into the nonmagnetic layer, resulting in a net nonequilibrium spin density in the ferromagnetic layer, which conspires with the spin asymmetry in electron mobility to produce the spin AH-UMR effect, as depicted in Figs.~\ref{fig_sahumr_fig1}\textcolor{purple}{b} and~\ref{fig_sahumr_fig1}\textcolor{purple}{c}.  

There is a key difference between the spin AH-UMR and other types of UMR effects previously studied in various magnetic systems~\cite{avci2015natphys,Jungwirth15PRB_UMR-DMS, shulei2016prb,Ferguson16APL_UMR-STO,yasuda2016prl, avci2018prl,lv2018natcomm,pnHai19JAP_UMR_TI-DMS, guillet2021prb,dWu21PRL_magnon-USMR}: for the spin AH-UMR, the nonequilibrium spin density in the presence of an external electric field emanates from the spin AH effect in the ferromagnetic layer itself, whereas for other UMRs|such as the unidirectional spin Hall magnetoresistance (USMR)~\cite{avci2015natphys} and the unidirectional Rashba magnetoresistance~\cite{guillet2021prb}|the nonequilibrium spin density is engendered 
by the spin Hall (SH) effect in the nonmagnetic layer~\cite{avci2015natphys,Jungwirth15PRB_UMR-DMS, shulei2016prb,Ferguson16APL_UMR-STO,avci2018prl,dWu21PRL_magnon-USMR} or by the Rashba-Edelstein effect due to the spin-momentum locked surface states~\cite{yasuda2016prl,lv2018natcomm,pnHai19JAP_UMR_TI-DMS,guillet2021prb,mandela21_QUMR}.

In what follows~\cite{mehraeen2022spin}, we will first examine the coupled nonlinear transport of spin and charge induced by the spin AH effect in a ferromagnetic metal that is adjacent to a nonmagnetic-metal layer with negligible SH effect. Based on a generalized drift-diffusion model, an analytical expression of the UMR coefficient|a proper characterization of the nonlinear transport phenomenon|will be derived, which reveals the dependences of the spin AH-UMR effect on specific materials and geometric parameters of the bilayer. We will then generalize our results to bilayer structures comprised of a ferromagnetic metal and a heavy metal, wherein both the spin AH-UMR and USMR are present. And we predict, for particular choices of materials combinations, that the total UMR would exhibit a sign reversal when the thickness of either the ferromagnetic-metal or the heavy-metal layer is varied, which suggests a scheme to quantify the SH and spin AH angles experimentally through a UMR measurement. We will conclude with some materials considerations on both direct and indirect detections of the spin AH-UMR as well as the enhancement of the effect.

\section{Spin-dependent drift-diffusion model}

Consider a thin-film ferromagnetic layer located at $0<z<d_F$ placed on top of a nonmagnetic layer at $-d_N<z<0$, as shown in Figs.~\ref{fig_sahumr_fig1}\textcolor{purple}{b,c}. Applying an in-plane electric field $\mathbf{E}$, the coupled drift-diffusion equations for charge and spin densities|taking into account the AH and the SH effects \cite{amin2019intrinsic} in the ferromagnetic layer|can be written as \cite{McGuire&Potter75IEEE_AMR,sZhang14JAP_AMR,kim2020generalized}
\begin{subequations}
\label{coupled}
\begin{align}
j_i
&=
\sigma 
\left(\mathcal{E}_i
+
p_{\sigma} \mathcal{E}^s_{ij}m_j \right)
-
\theta_{\text{SH}}^I \epsilon_{ijk} \mathcal{J}_{jk}
+
\theta_{\text{SH}}^A \epsilon_{ijk} m_j m_l\mathcal{J}_{kl}\,,
\\
\mathcal{J}_{ij}
&=
\sigma \left(\mathcal{E}^s_{ij} + p_{\sigma}\mathcal{E}_i m_j\right)
+
\theta_{\text{SH}}^I \epsilon_{ijk}j_k
+
\theta_{\text{SH}}^A \epsilon_{ilk} m_j m_l j_k\,,
\end{align}
\end{subequations}
where $j_i(z)$ and $\mathcal{J}_{ij}(z)$ are the local charge and spin current densities \footnote{By assuming spatial homogeneity in the $xy$ plane, all local quantities will have only a $z$ dependence.}, respectively, with the index $i$ indicating the direction of momentum flow and $j$ the spin polarization direction. $\theta_{\text{SH}}^I$ ($\theta_{\text{SH}}^A$) characterizes the strength of the isotropic (anisotropic) SH effect in the ferromagnet and $\mathbf{m}$ is a unit vector denoting the direction of the magnetization. In Eq.\,(\ref{coupled}), we have defined an effective local electric field felt by the electrons (in units with $e=1$) as $\bs{\mathcal{E}}(z) \equiv \mathbf{E} + \bs{\nabla}\mu_c(z)$, as well as an effective local `spin electric field', $\mathcal{E}^s_{ij}(z) \equiv \pd_i \mu_{s,j}(z)$. Here, $\mu_c$ and $\bs{\mu}_s$ are the charge and spin chemical potentials, respectively, whose gradients determine effective fields generated by local variations in the carrier densities \footnote{Note that, owing to the SH effect in the ferromagnet, the spin accumulation therein is, in general, in an arbitrary direction. This constitutes a generalization of the formalism presented in Ref.\,\cite{shulei2016prb}}. Furthermore, $\sigma(z)$ is the local conductivity and $p_{\sigma}$ the spin asymmetry in the linear conductivity in the ferromagnet.

To leading order in the Hall angles, the components of the spin current density polarized transverse to the magnetization|as well as the relevant boundary conditions|are decoupled from the charge current density~\cite{kim2020generalized}, the details of which are presented in Sec.~\ref{appendix_a}. Thus, Eq.~(\ref{coupled}) reduces to
\begin{subequations}
\label{coupled_vectorial}
\begin{align}
\mathbf{j}
&=
\sigma \mathbf{E} + \sigma_0 \left(\bs{\nabla}\mu_c + p_{\sigma} \bs{\nabla} \mu_s + p_{\sigma}\theta_{\text{SAH}} \mathbf{m} \times \mathbf{E}\right),
\\
\bs{\mathcal{J}}
&=
p_{\sigma}\sigma \mathbf{E} + \sigma_0 \left(p_{\sigma} \bs{\nabla}\mu_c + \bs{\nabla} \mu_s + \theta_{\text{SAH}} \mathbf{m} \times \mathbf{E}\right),
\end{align}
\end{subequations}
where $\sigma_0$ is the equilibrium conductivity, $\theta_{\text{SAH}} (\equiv \theta_{\text{SH}}^I + \theta_{\text{SH}}^A$) is the spin AH angle with $p_{\sigma} \theta_{\text{SAH}}$ being its charge counterpart, $\mathcal{J}_i \equiv \mathcal{J}_{ij}m_j$ and $\mu_s \equiv \bs{\mu}_s \cdot \mathbf{m}$ are the longitudinal components of the spin current density and spin chemical potential, respectively.

The local conductivity in Eq.\,(\ref{coupled_vectorial}) may be expressed as the sum of the conductivities of each spin channel, given by
\begin{equation}
\label{sigma_z}
\sigma^{\alpha}(z)=\nu^{\alpha}\left[n_0^{\alpha} + n^{\alpha}(z)\right]\,,
\end{equation}
where $\alpha=+(-)$ denotes the spin moment parallel (antiparallel) to the local magnetization, $n_0^{\alpha}$ and $\nu^{\alpha}$ are the equilibrium electron density and mobility of spin-$\alpha$ electrons, respectively, which in combination give rise to the longitudinal conductivity in the linear response regime ( \textit{i.e.}, $\sigma_0=\sum_{\alpha} \sigma_0^{\alpha}~\text{with}~ \sigma_0^{\alpha}=\nu^{\alpha}n_0^{\alpha}$) as well as the spin asymmetry in the conductivity by $p_{\sigma}\equiv (\sigma_0^+ - \sigma_0^-)/(\sigma_0^+ + \sigma_0^-)$. And $n^{\alpha}(z)$ are the current-induced nonequilibrium carrier densities that are responsible for the UMR. Local charge neutrality is assumed, \textit{i.e.}, $\sum_{\alpha}n^{\alpha}(z)=0$, for the metallic system.

The charge and spin chemical potentials may also be defined as $\mu_c \equiv (\mu^+ + \mu^-)/2$ and $\mu_s \equiv (\mu^+ - \mu^-)/2$, where $\mu^{\alpha}(z)$ is the spin-dependent chemical potential parallel to the magnetization, which is related to the nonequilibrium electron density through
\begin{equation}
\label{mu(z)}
\mu^{\alpha}(z)=\left[N^{\alpha}(\epsilon_F)\right]^{-1} n^{\alpha}(z) - \phi(z)\,,
\end{equation}
with $N^{\alpha}(\epsilon_F)$ the density of states of spin-$\alpha$ electrons at the Fermi level and $\phi(z)$ the spin-independent part of the chemical potential.

In the presence of spin-flip scattering, the charge- and spin-current densities satisfy, respectively, the following continuity 
equations~\cite{valet1993prb}
\begin{equation}
\label{cont}
\bs{\nabla} \cdot \mathbf{j}=
0~~\text{and}~~\bs{\nabla} \cdot \bs{\mathcal{J}}=
\frac{2}{\tau_{sf}} n_s\,,
\end{equation}
where $\tau_{sf}$ is the spin-flip relaxation time and $n_s(z)=\left(1-p_N^2\right)N\left(\epsilon_F\right)\mu_s(z)$ is the local spin density at the Fermi level, with $N(\epsilon_F)=\sum_{\alpha} N^{\alpha}(\epsilon_F)$ the total density of states and $p_N \equiv (N^{+}-N^{-})/(N^{+} +N^{-})$.

Inserting Eq.\,(\ref{coupled_vectorial}) into Eq.\,(\ref{cont}), we obtain a set of differential equations for the charge and spin chemical potentials
\begin{subequations}
\label{decomposed_cont_fm}
\begin{align}
\label{sym_cont_fm}
&\frac{d^2}{dz^2}\mu_{c,F}(z) + p_{\sigma}\frac{d^2}{dz^2}\mu_{s,F}(z)=0\,,
\\
\label{antisym_cont_fm}
&\frac{d^2}{dz^2}\mu_{s,F}(z) - \frac{\mu_{s,F}(z)}{\lambda_F^2}=0\,,
\end{align}
\end{subequations}
where $\lambda_F=\sqrt{\sigma_{0,F}(1-p_{\sigma}^2)\tau_{sf}/2N_F(\epsilon_F)(1-p_N^2)}$ is the spin diffusion length of the ferromagnetic metal.

In a nonmagnetic metal, where $p_{\sigma},p_N=0$, the charge and spin chemical potentials satisfy
\begin{subequations}
\label{decomposed_cont_nm}
\begin{align}
\label{sym_cont_nm}
&\frac{d^2}{dz^2}\mu_{c,N}(z) =0\,,
\\
\label{antisym_cont_nm}
&\frac{d^2}{dz^2}\mu_{s,N}(z) - \frac{\mu_{s,N}(z)}{\lambda_N^2}=0\,,
\end{align}
\end{subequations}
where $\lambda_N=\sqrt{\tau_{sf}\sigma_{0,N}/2N_N(\epsilon_F)}$ is the spin diffusion length of the normal metal with $N_N(\epsilon_F)$ the density of states of electrons at the Fermi level.

At the interface of the bilayer ($z=0$), we assume both the current density and chemical potential for each conduction channel are continuous (see Sec.~\ref{appendix_b} for a more detailed discussion on the boundary conditions), \textit{i.e.}, $\mu^{\alpha}_N(0^-)=\mu^{\alpha}_F(0^+)$ and $j_z^{\alpha}(0^-)=j_z^{\alpha}(0^+)$. And open boundary conditions are imposed at the two outer surfaces, \textit{i.e.}, $j_z^{\alpha}(-d_N)=j_z^{\alpha}(d_F)=0$.

\begin{figure}[hpt]
\captionsetup[subfigure]{labelformat=empty}
    \sidesubfloat[]{\includegraphics[width=0.4\linewidth,trim={0cm 0cm 0cm 0cm}]{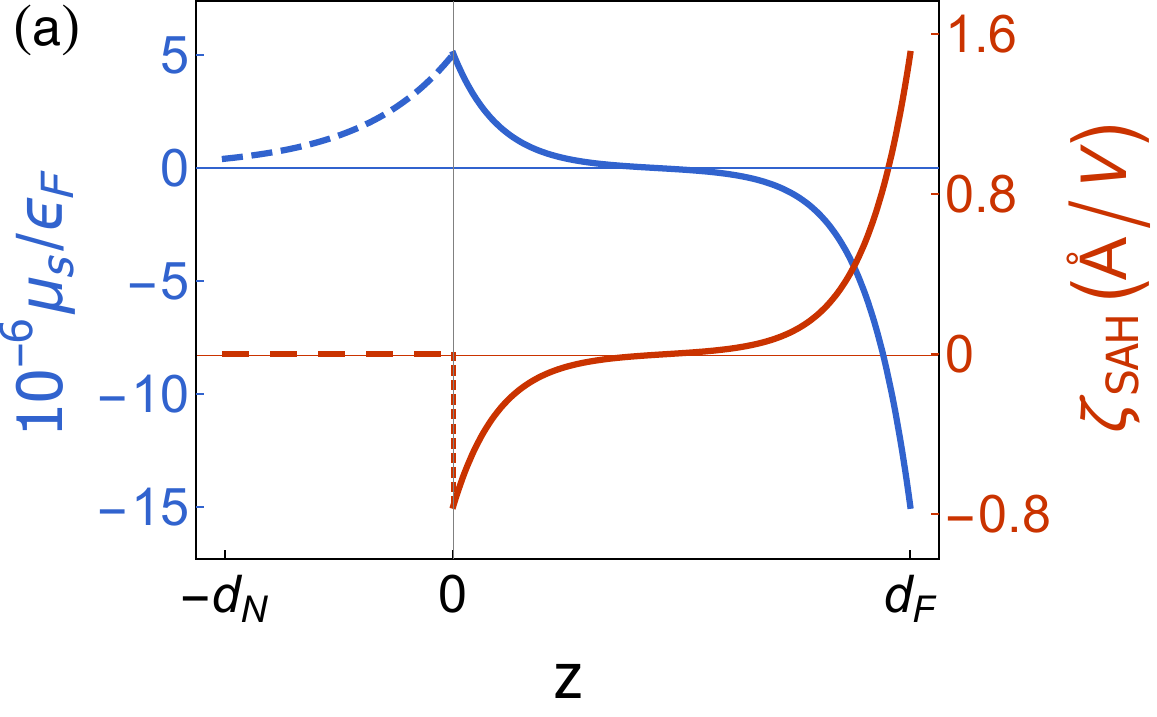}\label{fig_sahumr_fig2a}}
%\quad%
    \sidesubfloat[]{\includegraphics[width=0.4\linewidth,trim={0cm 0cm 0cm 0cm}]{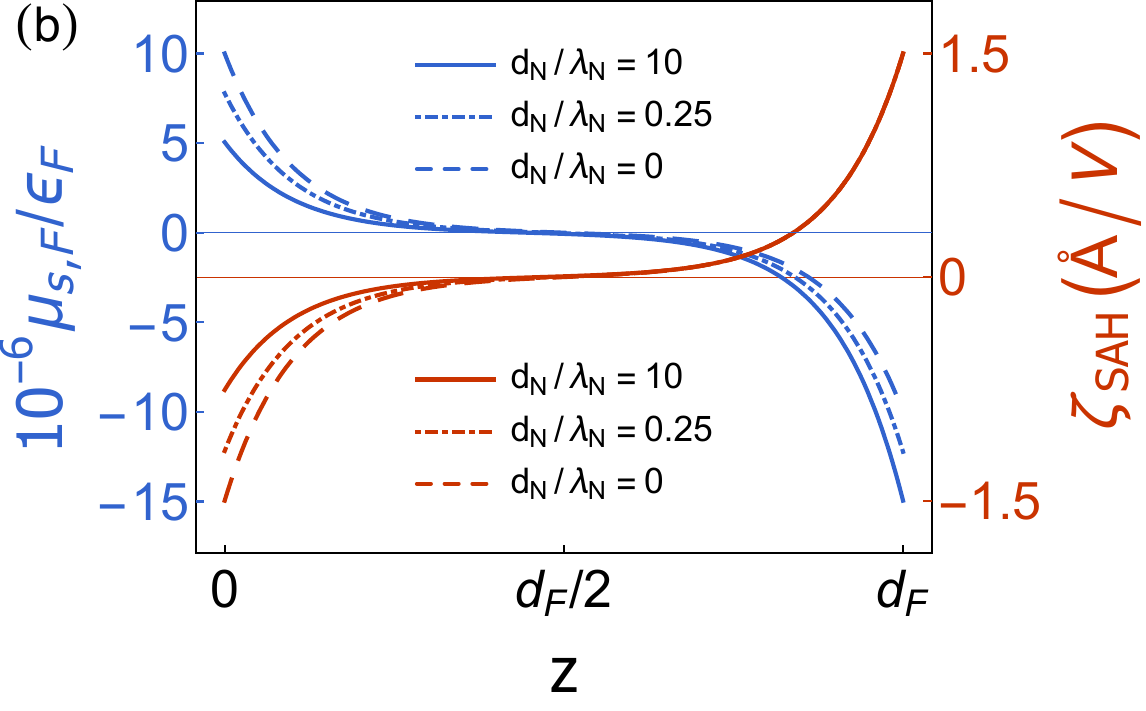}\label{fig_sahumr_fig2b}}
%%%%%%%%%
    \caption{Spatial dependences of the spin accumulation and spin AH-UMR coefficient in the bilayer system. (a) z-dependences of the spin accumulation $\mu_s$ and spin AH-UMR coefficient $\zeta_{\text{SAH}}$ in the FM and NM layers. Note the absence of the spin AH-UMR in the NM. (b) Plots of the spin accumulation and spin AH-UMR coefficient in the FM for different thicknesses of the NM layer. Parameters used: $\lambda_F=10$~nm, $\lambda_N=20$~nm, $\theta_{\text{SAH}}=0.05$, $\epsilon_F=5$~eV, $\sigma_{0,F}=\sigma_{0,N}=0.033$ (\textmu\textOmega\;cm)$^{-1}$, $p_{\sigma}=0.7$, and $p_N=0.2$.}
    \label{fig_sahumr_fig2}
\end{figure}

\section{Spin accumulation and nonlinear transport}

Without loss of generality, let us fix the electric field $\mathbf{E}$ in the $x$ direction, $\mathbf{E}=E_x \mathbf{x}$, and set the magnetization to point in the 
$y$ direction|in which case the 
magnitude of the longitudinal UMR would reach the maximum. The total charge current density, given by $\sum_{\alpha}j^{\alpha}(z)$, can be divided into two parts: $\mathbf{j}=\mathbf{j}^{(1)}+\mathbf{j}^{(2)}$, with a linear component $\mathbf{j}^{(1)}$ that is proportional to $E_x$ and a nonlinear one that is quadratic in $E_x$. The latter can be expressed as  
\begin{equation}\label{eq:j^(2)}
    j_x^{(2)}(z)=(\nu^{+}-\nu^{-})\mu_s(z)\bar{\mathcal{H}}(N^{+},N^{-})E_x
\end{equation}
where $\bar{\mathcal{H}}(N^{+},N^{-})$ is the harmonic mean of the density of states. And the spin accumulation $\mu_s$ is linear in $\theta_{\text{SAH}}E_x$, resulting in $j_x^{(2)}\propto E_x^2$. Note that $j_x^{(2)}$ only emerges in the ferromagnetic layer wherein $\nu^{+}\neq \nu^{-}$. 

In order to properly quantify the nonlinear charge current, we introduce a UMR coefficient $\zeta(z)$ as
\begin{equation}\label{eq:zeta1}
\zeta(z)
\equiv
\frac{\sigma _{xx}(z,E_{x})-\sigma
_{xx}(z,-E_{x})}{\sigma_{0}E_{x}}\;,
\end{equation} 
where $\sigma_{ij}=j_i/E_j$ is the conductivity tensor, and $\sigma_{0}$ is the linear longitudinal conductivity. The UMR coefficient $\zeta$ is so defined that its magnitude is independent of the electric field. The dimension of $\zeta$ is length per Volt, the inverse of which sets the scale of the electric field for which the nonlinear longitudinal conductivity|given by $j_{x}^{(2)}/E_x$|becomes comparable to its linear counterpart. 

The spatial distribution of the spin AH-UMR coefficient $\zeta_{\text{SAH}}$ is displayed in Fig.~\ref{fig_sahumr_fig2}\textcolor{purple}{a}, along with that of the spin chemical potential $\mu_s$ (or the spin accumulation) in the bilayer structure. There is a clear correlation between the two quantities: $\zeta_{\text{SAH}}$ goes to zero wherever $\mu_s$ vanishes. Furthermore, the spin AH-UMR completely comes from the ferromagnetic layer, in which the electron mobility is spin-dependent. Within the nonmagnetic layer, for which $\nu^{+} = \nu^{-}$, $\zeta_{\text{SAH}}$ vanishes everywhere despite the remnant $\mu_s$ near the interface. These observations are in full agreement with Eq.\,(\ref{eq:j^(2)}).

Although the nonmagnetic layer in question neither plays an active role as a spin polarizer nor accommodates any nonlinear charge transport, it is still indispensable to the generation of a net spin AH-UMR in its neighboring ferromagnetic layer. In the absence of the nonmagnetic layer, the spin accumulation $\mu_s$ and thus the local spin AH-UMR coefficient $\zeta_{\text{SAH}}$ have antisymmetric distributions about the center line of the ferromagnetic layer, as shown by the dashed lines in Fig.~\ref{fig_sahumr_fig2}\textcolor{purple}{b}. In this case, the total (spatially-averaged) spin AH-UMR is zero, as a result of the lack of a net nonequilibrium spin density. 

From a symmetry perspective, a net current-induced spin density is allowed, when and only when a system lacks inversion symmetry. For the present case, the nonmagnetic layer introduces structural inversion asymmetry, and makes a net nonequilibrium spin density achievable in the ferromagnetic layer next to it. Physically, it ``absorbs" spin accumulation at the interface from the ferromagnetic layer, leaving a net nonequilibrium spin density in the latter, as illustrated by the dash-dotted and solid lines in Fig.~\ref{fig_sahumr_fig2}\textcolor{purple}{b}.

\section{Spatially-averaged spin AH-UMR}

By taking the spatial average of the overall UMR coefficient over the thickness of the bilayer, $\bar{\zeta}\equiv \int_{-d_N}^{d_F} dz \;\zeta(z)/(d_N+d_F)$, we find that, up to $\mathcal{O}(\theta_{\text{SAH}})$, the spatially-averaged spin AH-UMR coefficient reads
\begin{equation}
\label{SAH-UMR}
\bar{\zeta}_{\text{SAH}}
=p_F \theta _{\text{SAH}}\left(\frac{\lambda_F}{\epsilon_F}\right)\mathcal{G}\left(\frac{d_{F}}{\lambda_{F}},\frac{d_{N}}{\lambda_{N}};\frac{\sigma_{0,F}}{\sigma_{0,N}},\frac{\lambda_{F}}{\lambda_{N}}\right)\,,
\end{equation}
where $p_F(=p_{\sigma} - p_N)$ characterizes the overall spin asymmetry of electron mobility for the ferromagnetic layer, and the thickness dependence of the spin AH-UMR is encapsulated in the dimensionless $\mathcal{G}$ function as
\begin{equation}
\label{eq:G}
\mathcal{G}\left(s,t;u,v \right)=\frac{3\left(\frac{uv}{(uv)\cdot s + t}\right)\tanh(s)\tanh(\frac{s}{2})}
{1+ \left(1-p_{\sigma}^2\right)\left(\frac{u}{v}\right) \tanh(s) \coth (t)}.
\end{equation}
For simplicity, we have adopted the free-electron model whereby  $N_F^{\alpha}=3n_{0,F}^{\alpha}/2\epsilon_F$ with $\epsilon_F$ the Fermi energy of conduction electrons in the ferromagnet. Equations\,(\ref{SAH-UMR}) and (\ref{eq:G}) are the main results of this chapter. 

Several remarks regarding the spin AH-UMR are in order. 

1) The spin AH-UMR coefficient, to leading order, is proportional to the spin AH angle $\theta_{\text{SAH}}$, in contrast to the USMR effect, which is proportional to the SH angle of the heavy-metal layer~\cite{shulei2016prb}.

2) The spin AH-UMR coefficient is also linear in $p_F$, as is the USMR~\cite{shulei2016prb}. This is not surprising, as the conversion of a net nonequilibrium spin density to a (nonlinear) charge current relies entirely on the spin asymmetry in electron scatterings. 

3) The ratio $\frac{\lambda_F}{\epsilon_F}$ has the same dimension as the spin AH-UMR coefficient (with $e=1$). In fact, the prefactor of the averaged UMR coefficient|the thickness independent part in Eq.\,(\ref{SAH-UMR})|sets the scale of the maximum spin AH-UMR that one can obtain for a given ferromagnet. For a typical transition metal with $p_F=0.3$, $\theta_{\text{SAH}}=0.02$, $\lambda_F=100$~nm, and $\epsilon_F=5$~eV, the upper bound of the spin AH-UMR coefficient, $\bar{\zeta}_{\text{SAH}}$, is of the order of 1\text{~\AA/V}.  

\begin{figure}[tph]
    \includegraphics[width=.5\linewidth]{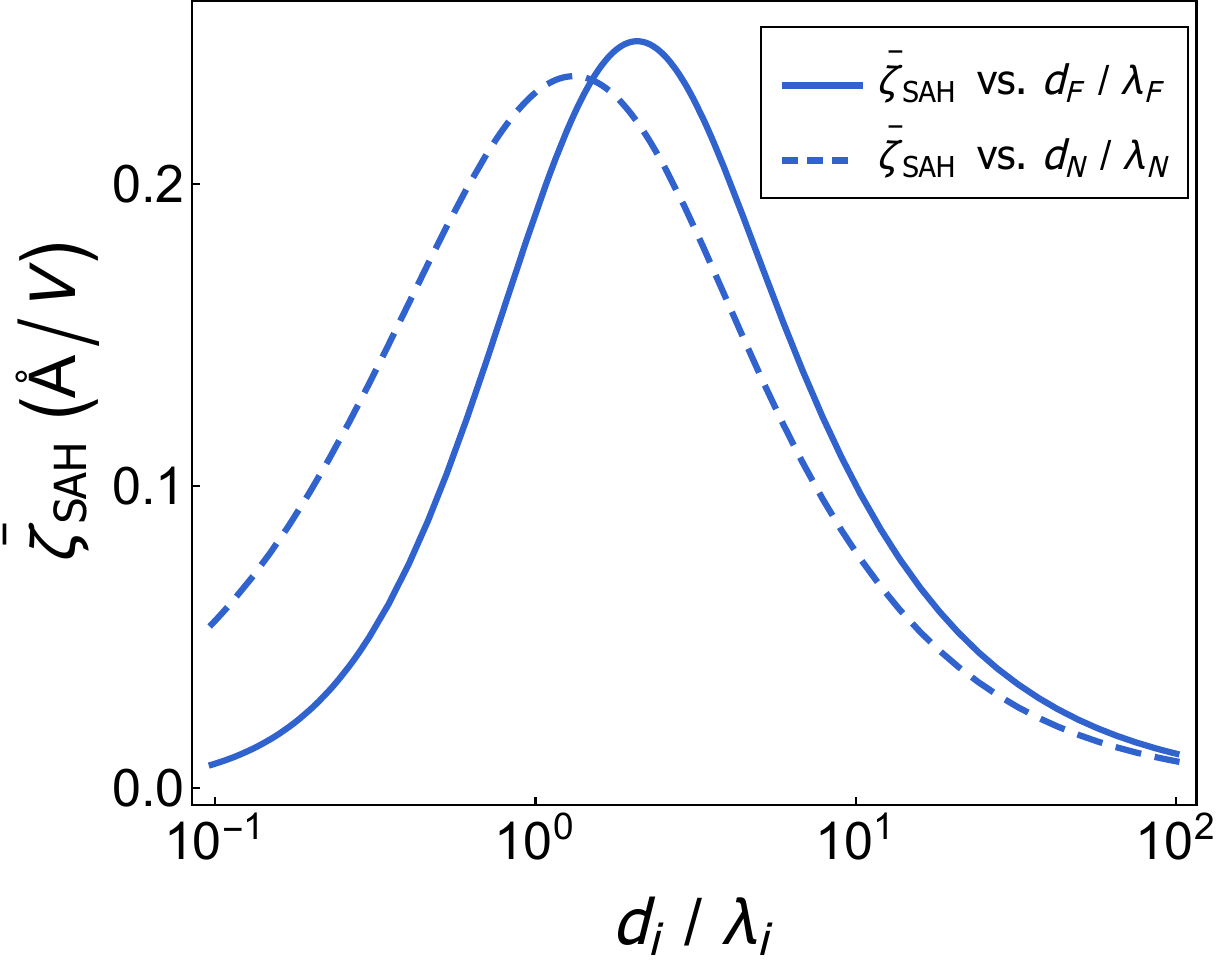}
    \caption{Thickness dependence of the spin AH-UMR coefficient. Dependence of $\bar{\zeta}_{\text{SAH}}$ on the thickness ($i=F,N$) of the FM layer for $d_N=15$ nm (solid) and on the thickness of the NM layer for $d_F=15$ nm (dashed). See the caption of Fig.~\ref{fig_sahumr_fig2} for the list of materials parameters used.}
    \label{fig_sahumr_fig3}%
\end{figure}

4) Information about how other geometric and materials parameters of a magnetic bilayer would shape the spin AH-UMR is all encoded in the dimensionless $\mathcal{G}$ function given by Eq.\,(\ref{eq:G}). The first two variables, $d_F/\lambda_F$ and $d_N/\lambda_N$, indicate that the dependences of the spin AH-UMR on the thicknesses of the ferromagnetic and nonmagnetic layers must scale with their respective spin diffusion lengths, as plotted in Fig.~\ref{fig_sahumr_fig3}. 

5) Another remarkable property of the spin AH-UMR|revealed by the $\mathcal{G}$ function|is that it increases monotonically with the ratio $\frac{\lambda_F}{\lambda_N}$, which would be useful for guiding the search for magnetic bilayers with a sizable spin AH-UMR effect.

\section{UMR sign reversal and SH/spin-AH angle quantification}

In a magnetic bilayer consisting of a ferromagnetic metal and a heavy metal, both spin AH and SH effects, in principle, may contribute to the total UMR measured in the bilayer. And their contributions turn out to be additive, \textit{i.e.}, $\bar{\zeta}=\bar{\zeta}_{\text{SAH}}+\bar{\zeta}_{\text{SH}}$ (see Sec.~\ref{appendix_b} for the full expression of $\bar{\zeta}$). The ratio of the two UMR contributions due to spin-dependent scattering|provided electron-magnon scattering is suppressed by applying a magnetic field or lowering the temperature~\cite{avci2018prl}|takes a rather neat form 
\begin{equation}
\frac{\bar{\zeta}_{\text{SAH}}}{\bar{\zeta}_{\text{SH}}}
=
\frac{\theta_{\text{SAH}} \lambda_F \tanh\left(\frac{d_F}{2\lambda_F}\right)}{\theta_{\text{SH}} \lambda_N \tanh \left(\frac{d_N}{2\lambda_N}\right)}\,.
\end{equation}
It is worthy to note that this ratio depends on only a few parameters, namely the SH/spin-AH angle, the spin diffusion length, and the thickness of each layer. 

\begin{figure}[tph]
    \includegraphics[width=.5\linewidth]{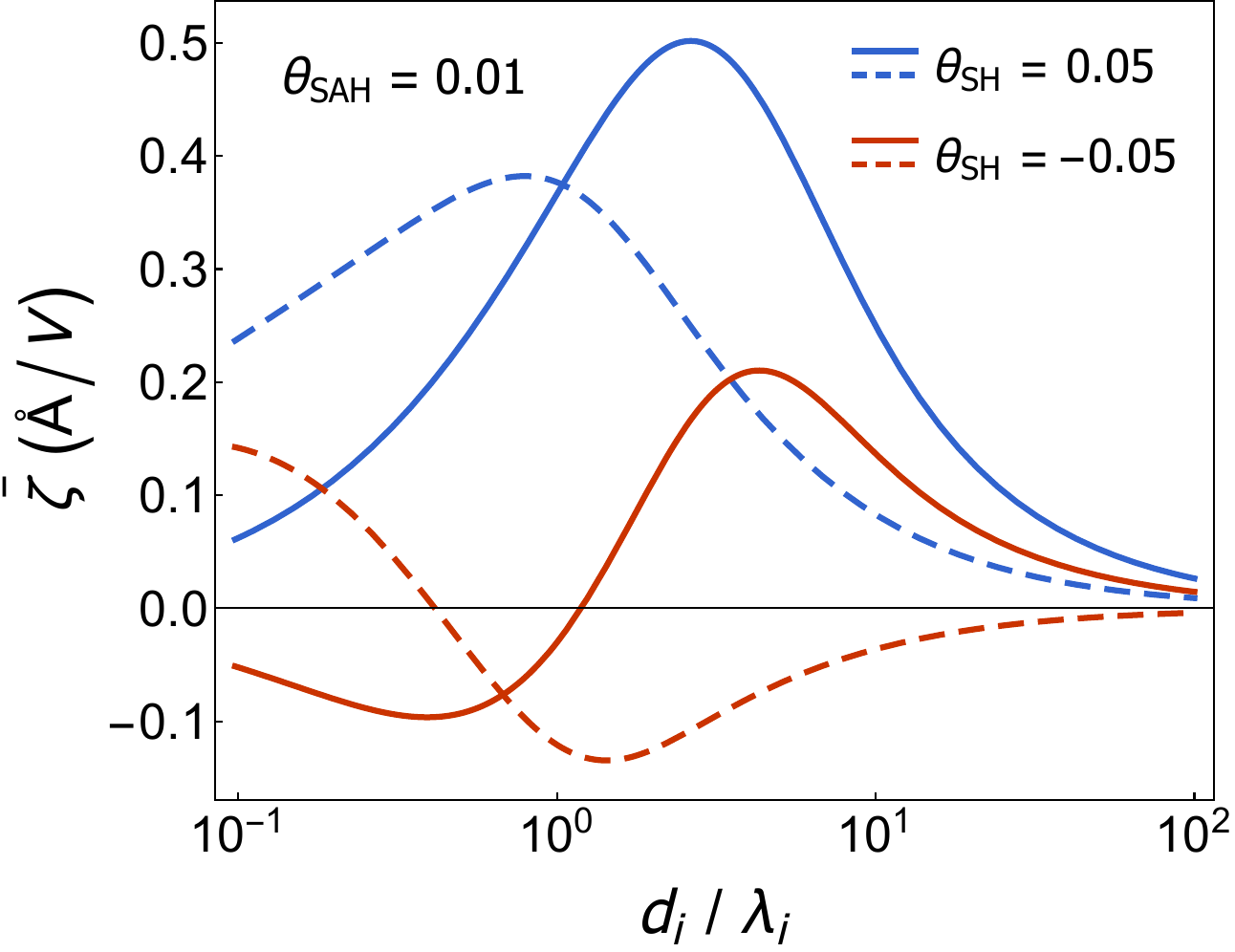}
    \caption{Thickness dependence of the total UMR coefficient. Dependence of $\bar{\zeta}$ on the thickness ($i=F,N$) of the FM layer for $d_N=5$ nm (solid) and on the thickness of the NM layer for $d_F=10$ nm (dashed). Other parameters used: $\lambda_F=20$ nm, $\lambda_N=5$ nm,  $\epsilon_F=5$ eV, $\sigma_{0,F}=\sigma_{0,N}=0.033$ (\textmu\textOmega\;cm)$^{-1}$, $p_{\sigma}=0.7$, and $p_N=0.2$.}
    \label{fig_sahumr_fig4}%
\end{figure}

The simple relation between $\bar{\zeta}_{\text{SAH}}$ and $\bar{\zeta}_{\text{SH}}$ nonetheless has a remarkable physical consequence: the total UMR coefficient of such a magnetic bilayer may exhibit qualitatively different thickness dependences, depending on the relative signs of the spin AH and SH angles. When $\theta_{\text{SAH}}$ and $\theta_{\text{SH}}$ have the same sign, the associated contributions simply add up (see the blue curves in Fig.~\ref{fig_sahumr_fig4}). It becomes more intriguing when $\theta_{\text{SAH}}$ and $\theta_{\text{SH}}$ are of opposite signs. In this case, the total UMR inevitably undergoes a sign reversal as the thickness of either layer is varied (see the red curves in Fig.\,\ref{fig_sahumr_fig4}). 

At the sign reversal point, the ratio of the Hall angles fulfills the following condition:
\begin{equation}
\frac{\theta_{\text{SH}}}{\theta_{\text{SAH}}}=-
\frac{\lambda_F \tanh\left(\frac{d_F}{2\lambda_F}\right)}{ \lambda_N \tanh \left(\frac{d_N}{2\lambda_N}\right)}\,,
\end{equation}
which can be used to experimentally quantify the SH (spin AH) angle of the heavy-metal (ferromagnetic-metal) layer, provided the Hall angle and spin diffusion length of the other layer|which serves as a reference layer|are known.

\section{Proposal for detecting spin AH-UMR}

As compared to the USMR, there are more options of materials systems for probing the spin AH-UMR effect. For the USMR, the nonmagnetic heavy-metal layer plays a central role in creating nonequilibrium spin density|via the SH effect|in the adjacent ferromagnetic layer. But this is not the case for the spin AH-UMR. For the spin AH-UMR, it is the ferromagnetic layer that serves as the spin polarizer and, hence, the choice of its neighboring layer is not necessarily limited to nonmagnetic materials with a strong SH effect. 

For instance, the spin AH-UMR effect, in principle, can also be hosted in bilayers comprised of a ferromagnetic metal and a normal metal, such as Cu, Al, or Ag. Given the weak SH effect in the normal-metal layer, the spin AH-UMR is expected to dominate over the USMR in these systems, making the detection of the former more straightforward. There is, perhaps, also a downside to these metallic bilayers|the spin AH-UMR therein is likely to be much smaller than that in magnetic bilayers with heavy-metals, as normal metals with a weak SH effect are oftentimes also poor ``spin sinks" with long spin diffusion lengths~\cite{Bass07JP-CM_spin-diffusion}, which would diminish the spin AH-UMR effect (especially when the ratio $\frac{\lambda_F}{\lambda_N}$ is small~\footnote{For instance, a UMR was not detected in Cu$|$Co bilayers~\cite{Gambardella15APL_MR-HL} (or, at least, was demonstrated to be much smaller than that in W$|$Co bilayers), which is not surprising, as the $\frac{\lambda_F}{\lambda_N}$ ratio for the former is about two orders of magnitude smaller than that for the latter.}, as was discussed in a previous section). 

The shortcoming of normal metals with long spin diffusion lengths may be compensated for by choosing a ferromagnetic layer with low carrier density. To see this, let us insert Eq.\,(\ref{eq:j^(2)}) together with $\sigma_0=\sum_{\alpha} n_0^{\alpha} \nu^{\alpha}$ into Eq.\,(\ref{eq:zeta1}), which yields 
\begin{equation}
\label{eq:zeta2}
    \zeta_{\text{SAH}} \propto \left(\frac{\nu^{+}-\nu^{-}}{\nu^{+}+\nu^{-}}\right) \frac{\theta_{\text{SAH}}}{n_0}\,.
\end{equation}
The above relation conveys a valuable piece of information: the lower the equilibrium carrier density of the ferromagnet, the larger the UMR coefficient. Thus, a more sizable spin AH-UMR is expected to arise in bilayers consisting of a normal-metal and a ferromagnetic semiconductor [\textit{e.g.}, (Ga,Mn)As]~\cite{Jungwirth15PRB_UMR-DMS} whose carrier density is usually two to three orders of magnitude smaller than that of a ferromagnetic metal. 

In magnetic bilayers comprised of nonmagnetic materials with strong spin-orbit coupling, the coexistence of the spin AH-UMR and USMR poses a challenge to differentiate the two. But, interestingly, they may also conspire to bring about a sign reversal of the overall UMR when the thickness of either layer is varied, a distinct transport signature that would not appear when either effect stands alone. This can be experimentally verified by contrasting the thickness dependences of the total UMR in two ferromagnetic-metal$|$heavy-metal bilayers, either with  different heavy-metal layers whose SH angles are of opposite signs (such Pt and $\beta$-Ta~\cite{liu2012spin}) or with different ferromagnetic-metal layers whose spin AH angles are of opposite signs (such as Fe and Gd~\cite{xxZhang10EPL_AH_Fe-Gd}). We anticipate that the results of such comparative measurements will resemble what are shown in Fig.~\ref{fig_sahumr_fig4}, with one bearing a sign change and the other not. On a related note, a sign change of the UMR was recently observed in single-crystalline Fe$|$Pt bilayers as the thickness of the Fe layer was increased~\cite{yWu21PRB_UMR-sign-revs}, implying possible competition between the spin AH-UMR and the USMR.

Other nonlinear effects that may intertwine with the spin AH-UMR are the anomalous Nernst~\cite{weiler2012local} and spin Seebeck~\cite{kikkawa2013longitudinal} effects: the Joule heating may induce a vertical temperature gradient across the ferromagnetic layer, which would in turn give rise to a nonlinear current in the direction of $\mathbf{m}\times \bs{\nabla}T$. However, one can separate the UMR contribution from the thermal contribution based on their different dependences on the direction of the applied electric field: the former is proportional to $\hat{\mathbf{z}}\cdot(\mathbf{m}\times \mathbf{E})$, whereas the latter is independent of the relative orientation between  $\mathbf{E}$ and $\mathbf{m}$, as the temperature gradient only relies on the magnitude of electric field (\textit{i.e.}, $\bs{\nabla}T \propto \lvert \mathbf{E} \rvert^2$).

\section{Decoupling of Transverse Spin Chemical Potentials}
\label{appendix_a}

Recall Eq.\,(\ref{coupled})
\begin{subequations}
\label{SM_coupled}
\begin{align}
j_i
&=
\sigma 
\left(\mathcal{E}_i
+
p_{\sigma} \mathcal{E}^s_{ij}m_j \right)
-
\theta_{\text{SH}}^I \epsilon_{ijk} \mathcal{J}_{jk}
+
\theta_{\text{SH}}^A \epsilon_{ijk} m_j m_l\mathcal{J}_{kl}\,,
\\
\mathcal{J}_{ij}
&=
\sigma \left(\mathcal{E}^s_{ij} + p_{\sigma}\mathcal{E}_i m_j\right)
+
\theta_{\text{SH}}^I \epsilon_{ijk}j_k
+
\theta_{\text{SH}}^A \epsilon_{ilk} m_j m_l j_k\,.
\end{align}
\end{subequations}
Here, $\mathcal{J}_{ij}$ is the spin current density tensor, in which the index $i$ indicates the direction of electron momentum flow and $j$ the spin polarization direction. Let us introduce the following decomposition of the spin current density
\begin{equation}
\mathcal{J}_{ij}
=
\mathcal{J}_i m_j + \mathcal{J}_i^{\perp_1} p_j + \mathcal{J}_i^{\perp_2} q_j\,,
\end{equation}
where $\mathbf{m}$, $\mathbf{p}$ and $\mathbf{q}$ are three mutually orthogonal unit vectors. $\mathcal{J}_i \equiv \mathcal{J}_{ij}m_j$ is the spin current density vector with the spins polarized along the magnetization, while $\mathcal{J}_i^{\perp_1} \equiv \mathcal{J}_{ij}p_j$ and $\mathcal{J}_i^{\perp_2} \equiv \mathcal{J}_{ij}q_j$ are two spin current density vectors with spins polarized perpendicular to the magnetization and to each other. Using the definitions of the effective local electric and spin electric fields, $\bs{\mathcal{E}}\equiv \mathbf{E} + \bs{\nabla} \mu_c$ and $\mathcal{E}^s_{ij}\equiv \pd_i \mu_{s,j}$, to leading order in the Hall angles, the charge and spin current density vectors read
\begin{subequations}
\label{SM_coupled_vectorial}
\begin{align}
&\mathbf{j}
=
\sigma \mathbf{E} + \sigma_0 \left(\bs{\nabla}\mu_c + p_{\sigma} \bs{\nabla} \mu_s + p_{\sigma}\theta_{\text{SAH}} \mathbf{m} \times \mathbf{E}\right),
\\
&\bs{\mathcal{J}}
=
p_{\sigma}\sigma \mathbf{E} + \sigma_0 \left(p_{\sigma} \bs{\nabla}\mu_c + \bs{\nabla} \mu_s + \theta_{\text{SAH}} \mathbf{m} \times \mathbf{E}\right),
\\
&\bs{\mathcal{J}}^{\perp_1}
=
\sigma_0 \left(\bs{\nabla} \mu_s^{\perp_1} + \theta_{\text{SH}}^I  \mathbf{p} \times \mathbf{E}\right),
\\
&\bs{\mathcal{J}}^{\perp_2}
=
\sigma_0 \left(\bs{\nabla} \mu_s^{\perp_2} + \theta_{\text{SH}}^I  \mathbf{q} \times \mathbf{E}\right),
\end{align}
\end{subequations}
where $\theta_{\text{SAH}} = \theta_{\text{SH}}^I + \theta_{\text{SH}}^A$ is the spin AH angle, $\mu_s^{\perp_1} \equiv \bs{\mu}_s \cdot \mathbf{p}$ and $\mu_s^{\perp_2} \equiv \bs{\mu}_s \cdot \mathbf{q}$.

In the presence of spin-flip scattering, and assuming local charge neutrality, the charge and spin current densities obey the following continuity equations
\begin{subequations}
\label{SM_cont}
\begin{align}
\bs{\nabla} \cdot \mathbf{j}
&=
0\,,
\\
\pd_i \mathcal{J}_{ij}
&=
\frac{2}{\tau_{sf}} n_{s,j}\,,
\end{align}
\end{subequations}
where $\tau_{sf}$ is the spin-flip relaxation time and $\mathbf{n}_s(z)=\left(1-p_N^2\right)N\left(\epsilon_F\right)\bs{\mu}_s(z)$ is the local spin density vector. Inserting Eq.\,(\ref{SM_coupled_vectorial}) into Eq.\,(\ref{SM_cont}), we obtain the following set of differential equations for the charge and spin chemical potentials
\begin{subequations}
\label{decomposed_cont_fm}
\begin{align}
\label{sym_cont_fm}
&\frac{d^2}{dz^2}\mu_{c}(z) + p_{\sigma}\frac{d^2}{dz^2}\mu_{s}(z)=0\,,
\\
\label{antisym_cont_fm}
&\frac{d^2}{dz^2}\mu_{s}(z) - \frac{\mu_{s}(z)}{\lambda^2}=0\,,
\\
\label{antisym_cont_perp1}
&\frac{d^2}{dz^2}\mu_{s}^{\perp_1}(z) - \frac{\mu_{s}^{\perp_1}(z)}{\lambda_{\perp}^2}=0\,,
\\
\label{antisym_cont_perp2}
&\frac{d^2}{dz^2}\mu_{s}^{\perp_2}(z) - \frac{\mu_{s}^{\perp_2}(z)}{\lambda_{\perp}^2}=0\,,
\end{align}
\end{subequations}
where $\lambda=\sqrt{\sigma_{0}(1-p_{\sigma}^2)\tau_{sf}/2N(\epsilon_F)(1-p_N^2)}$ and $\lambda_{\perp}=\sqrt{\sigma_{0}\tau_{sf}/2N(\epsilon_F)(1-p_N^2)}$ are the spin diffusion lengths parallel and perpendicular to the magnetization, respectively. We thus find that the transverse spin chemical potentials are decoupled from $\mu_c$ and $\mu_s$ to leading order in the Hall angles, as are their boundary conditions \cite{kim2020generalized}. Therefore, for the purpose of UMR analysis,  we may disregard transverse spin chemical potentials.

\section{Contribution of Boundary Resistance}
\label{appendix_b}

The presence of a boundary resistance at the bilayer interface modifies the boundary conditions on the charge and spin chemical potentials as \cite{valet1993prb}
\begin{subequations}
\label{new_densities}
\begin{align}
\label{n(z)}
\mu_{c,F}\left(0^+\right) - \mu_{c,N}\left(0^-\right)
&=
- \gamma r_b \mathcal{J}_{z}\left(0\right),
\\
\mu_{s,F}\left(0^+\right) - \mu_{s,N}\left(0^-\right)
&=
r_b \mathcal{J}_{z}\left(0\right),
\end{align}
\end{subequations}
where $\gamma$ is the interfacial spin asymmetry coefficient and $r_b$ is the boundary resistance for a unit surface of the interface.  

Resolving the continuity equations with the modified boundary conditions, the total UMR coefficient $\zeta$ of the system may be calculated. This is comprised of the SH- and spin AH-UMR coefficients,  $\zeta = \zeta_{\text{SAH}} + \zeta_{\text{SH}}$. Taking the spatial average defined as $\bar{\zeta}\equiv \int_{-d_N}^{d_F} dz \;\zeta(z)/(d_N+d_F)$, to first order in the SH and spin AH angles, we obtain
\begin{equation}
\label{UMR}
\bar{\zeta}
=
\frac{3 \left(p_{\sigma} - p_N\right)}{\epsilon_F} \left(\frac{\sigma _{0,F} \lambda_F}{\sigma _{0,F}d_F + \sigma _{0,N} d_N}\right)
\frac{\tanh \left(\frac{d_F}{\lambda_F}\right)\left[\theta_{\text{SAH}} \lambda_F \tanh \left(\frac{d_F}{2\lambda_F}\right) + \theta_{\text{SH}} \lambda_N \tanh\left(\frac{d_N}{2\lambda_N}\right)\right]}
{1+ \left(1-p_{\sigma}^2\right) \tanh \left(\frac{d_F}{\lambda_F}\right) \left[\left(\frac{\sigma_{0,F} \lambda_N}{\sigma_{0,N} \lambda_F}\right) \coth \left(\frac{d_N}{\lambda_N}\right) + \frac{r_b \sigma _{0,F}}{\lambda_F}\right]}\,,
\end{equation}
where we note the relation between the spin AH- and SH-UMR coefficients
\begin{equation}
\frac{\bar{\zeta}_{\text{SAH}}}{\bar{\zeta}_{\text{SH}}}
=
\frac{\theta_{\text{SAH}} \lambda_F \tanh\left(\frac{d_F}{2\lambda_F}\right)}{\theta_{\text{SH}} \lambda_N \tanh \left(\frac{d_N}{2\lambda_N}\right)}\,.
\end{equation}
Taking into account the interfacial resistance, we have verified that, for a typical boundary resistance of $r_b \sim 1 \text{f}\Omega\;\text{m}$ \cite{valet1993prb,Bass07JP-CM_spin-diffusion}, the effect on the transport coefficients is negligible. Thus, we may safely neglect the interfacial resistance in the present chapter.

\chapter{Quantum Unidirectional Magnetoresistance}

\section{Introduction}

As fingerprints of electron waves, quantum interference effects in electron transport have been of fundamental interest \cite{Anderson58pr_localization,AbrahamsPRL79_WL-2D,BERGMANN84Phys-rep_WL-thin-film,Webb85PRL_AB-oscillation_NM-ring,Bachtold99_AB-osc_Carbon-NT,Peng10NM_AB-TI,Spivak10RMP_QM-transport_2DEF}. And from the perspective of applications, studies of coherent quantum transport of electrons--and more generally information carriers--may also underpin the development
of future quantum devices, including quantum computers \cite{Whiticar20NC_Majorana-AB}. However, interference-based magnetoresistances have thus far been limited to the \textit{linear} response regime \cite{AbrahamsPRL79_WL-2D,BERGMANN84Phys-rep_WL-thin-film,altshuler1980prb,hikami1980}, where they remain \textit{invariant} under the reversal of the applied magnetic field.

Recently, there has been increasing interest in an emergent unidirectional
magnetoresistance (UMR) effect observed in various bilayer systems composed of a nonmagnetic layer and a ferromagnetic layer ~\cite{avci2015natphys,Olejnik15PRB_UMR-semicond,yasuda2016prl,lv2018unidirectional,guillet2021prb,liu2021magnonic,mehraeen2022spin}. This novel nonlinear magnetotransport effect features a variation in  magnetoresistance when the direction of the applied
electric field is reversed, at variance with common linear magnetoresistances, which are electric-field independent. 

While the list of systems that can host such UMRs
keeps growing, they consistently possess three essential ingredients: strong
spin-orbit coupling (SOC), structural inversion asymmetry, and broken time reversal symmetry. In
terms of microscopic mechanisms, lying at the heart of UMR effects are the
spin-momentum coupling and spin-asymmetry in electron scattering~\cite%
{shulei2016prb,Langenfeld16APL_UMR-FMR,yasuda2016prl,avci2018prl,guillet2021prb}. Quantum interference, however,
has not been demonstrated to play any role in nonlinear magnetotransport.

\begin{figure}[tph]
    \includegraphics[width=.7\linewidth]{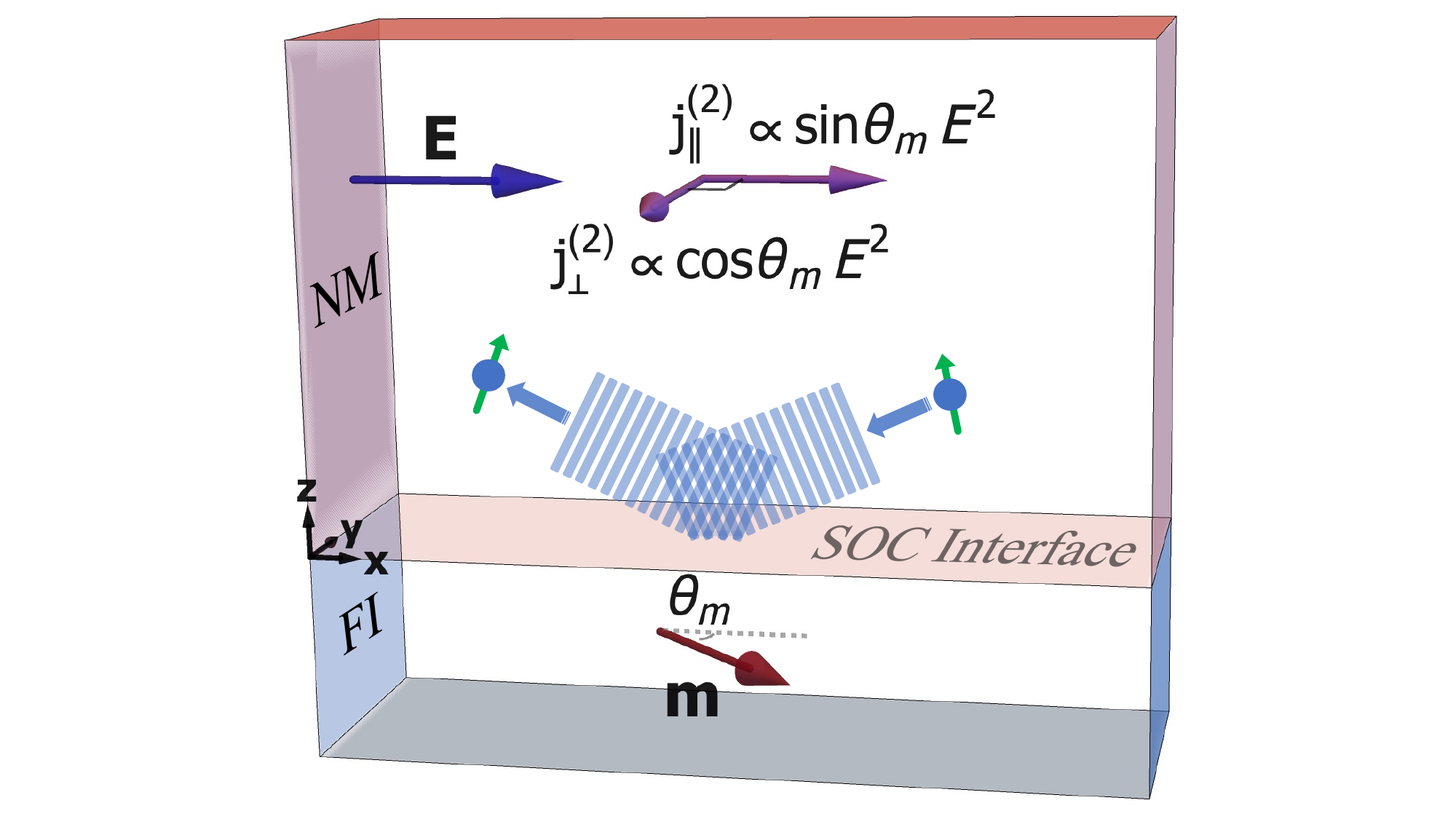}
    \caption{Schematic of the quantum unidirectional
magnetoresistance (QUMR) effect in a NM$|$FI bilayer. An electron in the NM layer scatters off the interface with spin-orbit coupling (SOC), resulting in an interference between the incident and reflected waves, which in turn gives rise to a nonlinear current. Both longitudinal and transverse components of the nonlinear current, $j_{\parallel}^{(2)}$ and $j_{\perp}^{(2)}$, are sensitive to the orientation of the in-plane magnetization $\mathbf{m}$ relative to the electric field $\mathbf{E}$, with different dependences on $\theta_m$--the angle between $\mathbf{m}$ and $\mathbf{E}$.}
    \label{fig_qumr_fig1}%
\end{figure}

In this chapter~\cite{mehraeen2023quantum}, we unveil the role of quantum interference in generating both
\textit{longitudinal} and \textit{transverse} UMRs in bilayer structures
consisting of a nonmagnetic metal (NM) and a ferromagnetic insulator (FI), in the presence of
SOC at the interface. Microscopically, the scattering of an
electron wave at such an interface hinges on the wavevector and spin
orientations of the incident electron, as shown schematically in Fig.~\ref{fig_qumr_fig1}.
Due to the interference between the incident and reflected waves, the
electron acquires an additional velocity that is \textit{even}
in the in-plane component of the wavevector (as we will show explicitly),
which in turn gives rise to a nonlinear current density of the following
form:
\begin{equation}
\mathbf{j}^{(2)}=\sigma _{\parallel }^{(2)}\mathbf{z}\cdot \left( \mathbf{E}%
\times \mathbf{m}\right) \;\mathbf{E}+\sigma _{\bot }^{(2)}\mathbf{E}\cdot
\mathbf{m}\;\mathbf{z}\times \mathbf{E},  \label{vectorform}
\end{equation}%
where $\mathbf{E}$ is the applied electric field, $\mathbf{m}$ and $\mathbf{%
z~}$are two unit vectors denoting, respectively, the directions of
magnetization of the ferromagnetic layer and the\textbf{\ }normal to the
bilayer interface. $\sigma _{\parallel }^{(2)}$ and $\sigma _{\perp }^{(2)}$, both being $\mathbf{E}$-independent, characterize, respectively, the strengths of
the longitudinal and transverse contributions to the nonlinear transport, with
the superscripts indicating that $\mathbf{j}^{(2)}$ is of the second order
in the electric field. 

Given the quantum mechanical origin of the phenomena, we shall refer to the corresponding UMRs as \textit{quantum unidirectional
magnetoresistances} (QUMRs), in order to distinguish them from their semiclassical counterparts that have heretofore been reported.

\section{The model}

For simplicity, we shall restrict ourselves to the
nonlinear magnetotransport in a NM$|$FI bilayer, whereby electrons incident on the interface are completely reflected back into the
NM layer and hence the charge transport in the ferromagnetic layer can be disregarded. Furthermore, we assume the presence of an exchange interaction coupling the electron spin and the interfacial magnetization of the FI and a Rashba-type SOC arising from the structural inversion asymmetry at the interface. Consider a setup described by Fig.~\ref{fig_qumr_fig1}, with a three-dimensional (3D) electron gas contained in a semi-infinite NM layer occupying the $z>0$ half-space and a FI layer occupying the other half.

The 3D electron gas can be described by the Hamiltonian \cite{Stiles13PRB_SOT,tokatly2015prb}
\begin{equation}
\label{Ham}
\hat{H}=\frac{\hat{\mathbf{p}}^{2}}{2m}
+
\frac{\alpha _{R}}{\hbar
}\delta (z)\hat{\boldsymbol{\sigma }}\cdot (\hat{\mathbf{p}}\times \mathbf{z}%
)-J_{ex}\delta (z)\hat{\boldsymbol{\sigma }}\cdot \mathbf{m}+V_{b}\Theta
(-z), 
\end{equation}%
where $\alpha _{R}$ and $J_{ex}$ are the coefficients of the interfacial Rashba SOC and exchange interaction, respectively, and $V_{b}$ is the height of the energy barrier,
which is greater than the Fermi energy of electrons in the NM layer.

The general scattering state may be written as
\begin{equation}
\label{gen_scat}
\boldsymbol{\psi }_{\text{scat.}}=\left\{
\begin{array}{c}
e^{i\mathbf{q}\cdot \boldsymbol{\rho }}\left( e^{-ik_{z}z}+e^{ik_{z}z}\hat{R}_{\mathbf{q}}%
\right) \boldsymbol{\chi },\;\;z>0\;\; \\
e^{i\mathbf{q}\cdot \boldsymbol{\rho }}e^{\kappa_z z}\;\hat{T}_{\mathbf{q}}\boldsymbol{\chi
},\;\;z<0\;\;%
\end{array},\right.
\end{equation}%
where $\mathbf{q}\left[ =\left( k_{x},k_{y}\right) \right] $ and $%
\boldsymbol{\rho }\left[ =\left( x,y\right) \right] $ are the wave- and
position-vector in the $x$-$y$ plane, wherein the system is translationally
invariant and allows the propagation of plane waves, $k_{z}$ is the $z$%
-component of the wavevector of the propagating wave in the NM layer, and $%
\kappa_z^{-1}\left[ =\left( 2m V_{b}/\hbar ^{2}-k_{z}^{2}\right)
^{-1/2}\right] $ characterizes the decay length of the evanescent wave in
the FI layer. $\hat{R}_{\mathbf{q}}$ and $\hat{T}_{\mathbf{q}}$ are $2\times 2$ matrices in spin
space, which describe, respectively, the spin-dependent reflection and
transmission amplitudes, and the spinors $\boldsymbol{\chi }$ are taken to
be the eigenstates satisfying $\hat{\boldsymbol{\sigma }}\cdot \mathbf{m}\;\chi ^{\pm }=\pm
\chi ^{\pm }$.

By imposing the standard boundary conditions at the Rashba interface, namely the continuity of the wavefunction and the discontinuity of its spatial derivative along the $z$ direction--as detailed in Sec.~\ref{appendixA}--we find 
\begin{equation}
\label{eq:R_q}
\hat{R}_{\mathbf{q}}
=
e^{i \left( \varphi_{\mathbf{q}} + \vartheta _{\mathbf{q}}
\hat{\boldsymbol{\sigma}}\cdot \mathbf{n}_{\mathbf{q}} \right)},
\end{equation}
and $\hat{T}_{\mathbf{q}}=1+\hat{R}_{\mathbf{q}}$, where $\varphi _{\mathbf{q}}=\arcsin \left( 2k_{z}\kappa_z /\varkappa_{\mathbf{q}}\right)$, $\vartheta _{\mathbf{q}}=\arcsin \left( 2k_{z}Q_{\mathbf{q}}/\varkappa_{\mathbf{q}}\right) $, $\varkappa_{\mathbf{q}}^{2}=\left[ Q_{%
\mathbf{q}}^{2}-(\kappa_z^{2}+k_{z}^{2})\right] ^{2}+\left( 2k_{z}Q_{\mathbf{q%
}}\right) ^{2}$, and $\mathbf{n}_{\mathbf{q}} \mathbf{=\mathbf{Q}_{%
\mathbf{q}}/}Q_{\mathbf{q}}$. Here, $\mathbf{Q}_{\mathbf{q}}\equiv \eta _{R}%
\mathbf{q}\times \mathbf{z}-\xi _{ex}\mathbf{m}$, with $\eta _{R}\equiv
2m \alpha _{R}/\hbar ^{2}$ a dimensionless constant characterizing the strength of the interfacial Rashba SOC and $\xi _{ex}\equiv 2m J_{ex}/\hbar ^{2}$ the rescaled exchange interaction. Note
that $\hat{R}_{\mathbf{q}}$ is unitary, as enforced by the conservation of probability
flux.

All key information about the reflected wave, such
as the phase shift, in particular, is encapsulated in Eq.~(\ref{eq:R_q}), from which the physical meaning of $\hat{R}_{\mathbf{q}}$ is evident: The angle $\varphi_{\mathbf{q}}$ describes a largely spin-independent phase shift, in conjunction with a spin rotation by an angle $\vartheta_{\mathbf{q}}$ about a rotation axis taken in the direction of $\mathbf{n}_{\mathbf{q}}$. Note that both $\vartheta _{\mathbf{q}}$ and $\mathbf{n}_{\mathbf{q}}$ depend, through the vector field $\mathbf{Q}_{\mathbf{q}}$, on the wavevector $\mathbf{q}$ and the magnetization $\mathbf{m}$, as the spin rotation is brought about by the interfacial exchange interaction and Rashba SOC.

\begin{figure}[tph]
\sidesubfloat[]{\includegraphics[width=0.5\linewidth,trim={1.5cm 2cm 1cm 1cm}]{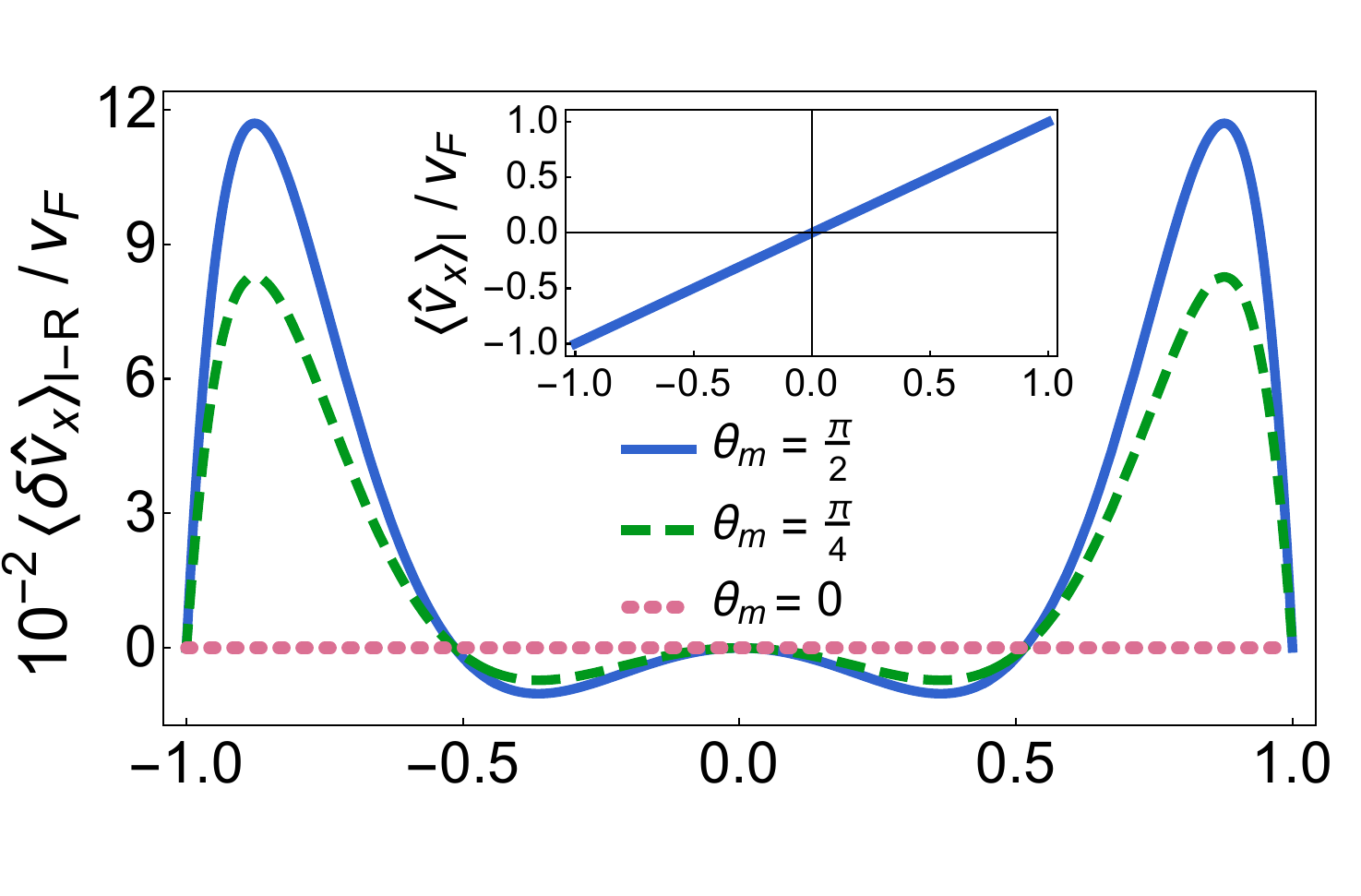}\label{fig_qumr_fig2_a}}\quad%
    \\
\sidesubfloat[]{\includegraphics[width=0.5\linewidth,trim={2.5cm 0.5cm 0.5cm 1.5cm}]{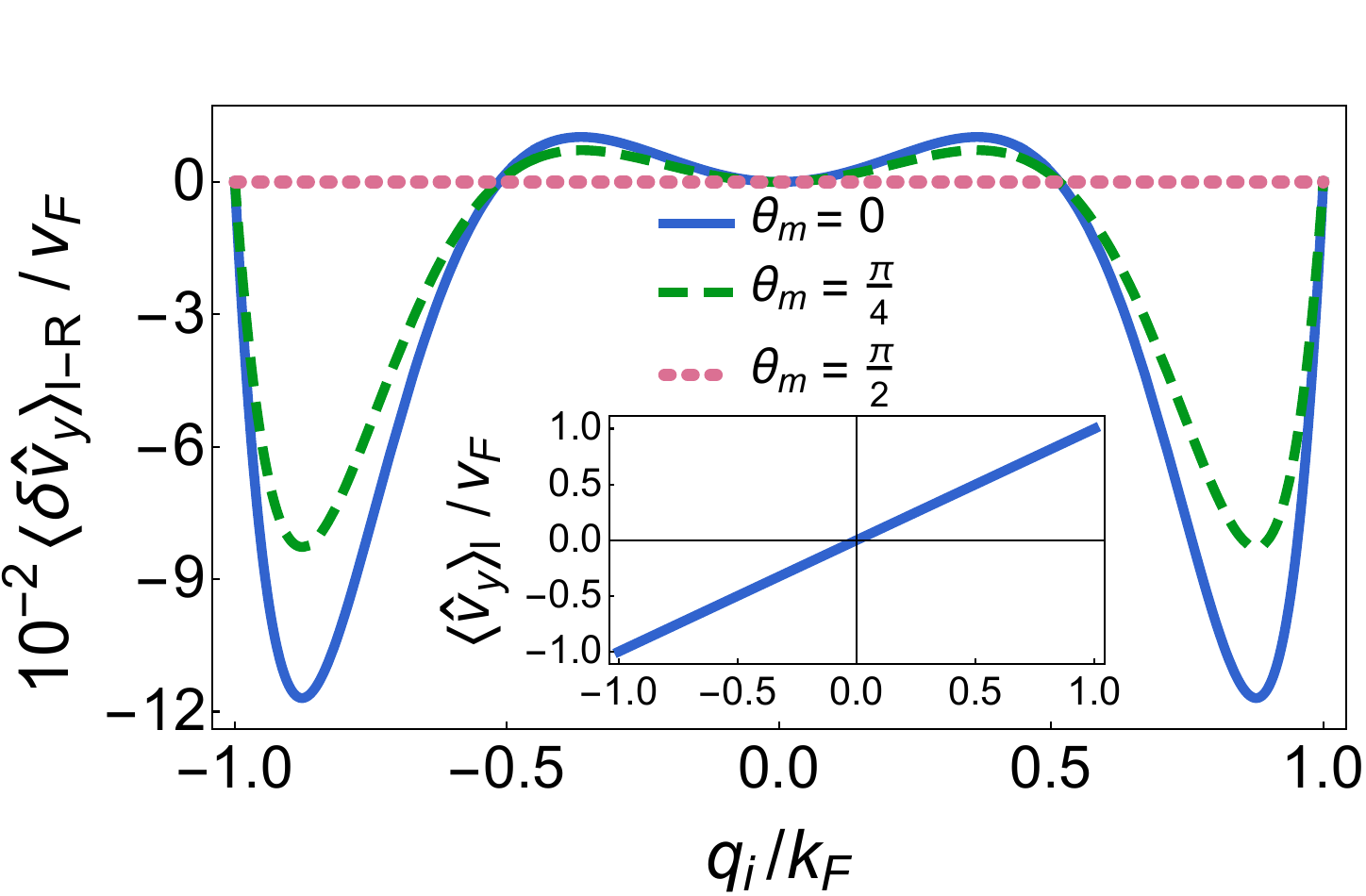}\label{fig_qumr_fig2_b}}%
    \caption{The even part of the spatial average of $\langle \hat{\boldsymbol{v}}_{\parallel}\left( \mathbf{q}%
\right) \rangle _{I\text{-}R}$, given by $\langle \protect\delta \hat{%
\boldsymbol{v}}_{\parallel}\rangle _{I\text{-}R}=\frac{1}{2}\left[ \langle \hat{\boldsymbol{%
v}}_{\parallel}(\mathbf{q})\rangle _{I\text{-}R}+\langle \hat{\boldsymbol{v}}_{\parallel}(-\mathbf{q}%
)\rangle _{I\text{-}R}\right] $, as a function of in-plane momentum for three different values of $\theta_m$. (a) Plots
of $\langle \protect\delta \hat{v}_{x}\rangle _{I\text{-}R}$ (main) and $%
\langle \hat{v}_{x}\rangle _{I}$ (inset) as functions of $q_i = q_{x}$ at $q_{y}=0$. (b) Plots of $\langle \protect\delta \hat{v}_{y}\rangle _{I\text{-}R}$
(main) and $\langle \hat{v}_{y}\rangle _{I}$ (inset) as functions of $q_i=q_{y}$ at $q_{x}=0$.}
    \label{fig_qumr_fig2}
\end{figure}

\section{Interference velocity}

To provide a heuristic picture of the QUMR effect, let us first consider the spin-averaged velocity relevant to the charge transport in the NM layer, which is given by the expectation value of the velocity operator with
respect to the spinor part of the scattering state for $z>0$, \textit{i.e.},
\begin{equation}  
\label{Eq:v}
\langle \hat{\boldsymbol{v}}\rangle \equiv \sum_{\sigma }\text{Re}\left[ \left(
\psi _{\text{scat.}}^{\sigma }\right) ^{\dag }\frac{\hat{\mathbf{p}}}{%
m}\left( \psi _{\text{scat.}}^{\sigma }\right) \right] =\langle
\hat{\boldsymbol{v}}\rangle _{I}+\langle \hat{\boldsymbol{v}}\rangle
_{R}+\langle \hat{\boldsymbol{v}}\rangle _{I\text{-}R},
\end{equation}%
where the sum runs over the spin index. This velocity can be decomposed into
three parts, as shown in Eq.~(\ref{Eq:v}). The first two terms, $\langle \hat{%
\boldsymbol{v}}\rangle _{I}$ and $\langle \hat{\boldsymbol{v}}\rangle _{R}$,
are associated with the incident and reflected waves, respectively, and more
intriguingly, there exists another term, $\langle \hat{\boldsymbol{v}}%
\rangle _{I\text{-}R}$, originating from the interference between the
incident and the reflected waves. The in-plane component of this \textit{%
interference velocity} can be expressed as
\begin{equation}  \label{Eq:v_I-R}
\langle \hat{\boldsymbol{v}}_{\parallel }\rangle _{I\text{-}R}=\frac{4\hbar
\mathbf{q}}{m} \cos \left(\vartheta _{\mathbf{q}}%
\right) \cos \left( 2k_{z}z+\varphi _{\mathbf{q}}\right),
\end{equation}
where $\hat{\boldsymbol{v}}_{\parallel }=\left( \hat{v}_{x},\hat{v}%
_{y}\right) $.

There are two crucial differences between the interference velocity $\langle
\hat{\boldsymbol{v}}\rangle _{I\text{-}R}$ and the other counterparts, as
displayed in Fig.~\ref{fig_qumr_fig2}. While $\langle \hat{\boldsymbol{v}}%
\rangle _{I}$ and $\langle \hat{\boldsymbol{v}}\rangle _{R}$ share the same
in-plane component $\frac{\hbar \mathbf{q}}{m}$ (which is
clearly odd in $\mathbf{q}$), the interference velocity $\langle \hat{%
\boldsymbol{v}}\rangle _{I\text{-}R}$ includes a constituent that is \textit{%
even} in $\mathbf{q}$~\footnote{The interference velocity also has a component that is odd in $\mathbf{q}$, which contributes to an anisotropic magnetoresistance--in linear response to the applied electric field--with an angular dependence similar to the spin-Hall magnetoresistance~\cite{Saitoh13PRL_SH-MR,xJiang14prb_Rashba-AMR,slzhang15PRB_AMR}}--which we define as $\langle \delta \hat{\boldsymbol{%
v}}_{\parallel }\rangle _{I\text{-}R}\equiv \frac{1}{2}\left[ \langle \hat{%
\boldsymbol{v}}_{\parallel }\left( \mathbf{q}\right) \rangle _{I\text{-}%
R}+\langle \hat{\boldsymbol{v}}_{\parallel }\left( -\mathbf{q}\right)
\rangle _{I\text{-}R}\right] $--due to the combined action of interference
and spin-dependent scattering at the interface, as long as the magnetization
$\mathbf{m}$ is \textit{not }perpendicular to the $x$-$y$ plane.

Furthermore, $\langle \hat{\boldsymbol{v}}\rangle _{I}$ and $\langle \hat{%
\boldsymbol{v}}\rangle _{R}$ are inert to the rotation of the magnetization $%
\mathbf{m}$, but $\langle \hat{\boldsymbol{v}}_{\parallel }\rangle _{I\text{-%
}R}$ is \textit{not}. In fact, the latter is exquisitely sensitive to the
variation in the angle formed by $\mathbf{m}$ and $\mathbf{q}$, with $%
\langle \delta \hat{v}_{x}\rangle _{I\text{-}R}$ and $\langle \delta \hat{v}%
_{y}\rangle _{I\text{-}R}$ exhibiting different angular dependences. It is
these distinctive features of the interference velocity that give rise to the QUMR, as we will evaluate below.

\section{Disorder scattering}

To take into account interfacial disorder--which leads to diffusive scattering of incident electrons, we add a random impurity potential that is localized at the interface, i.e.,  $\hat{H} \rightarrow \hat{H} + V^{\text{imp}}(\mathbf{r})$ with $V^{\text{imp}}(\mathbf{r})=V^{\text{imp}}(\bs{\rho})\delta(z)$, leading to the Dyson equation
\begin{equation}
\label{G_tilde}
\hat{\tilde{G}} \left(\mathbf{r},\mathbf{r}^{\prime}\right)
=
\hat{G}\left(\mathbf{r},\mathbf{r}^{\prime}\right)
+
\int d\mathbf{r}_1
\hat{G} \left(\mathbf{r},\mathbf{r}_1\right) 
V^{\text{imp}}\left({\mathbf{r}}_1\right)
\hat{\tilde{G}} \left(\mathbf{r}_1,\mathbf{r}^{\prime}\right),
\end{equation}
for the disordered propagator $\hat{\tilde{G}}$. And we assume the impurity potential has the white noise distribution $\braket{V^{\text{imp}} \left(\bs{\rho}\right)}
= 0$ and $\braket{V^{\text{imp}} \left(\bs{\rho}\right) V^{\text{imp}} \left(\bs{\rho}^{\prime}\right)} = (\hbar^2/2m)^2 \eta_{\gamma} \delta\left(\bs{\rho} - \bs{\rho}^{\prime}\right)$, where the dimensionless parameter $\eta_{\gamma}$ characterizes the strength of the impurity interaction and $\braket{\cdots}$ here denotes the configurational average over impurity positions.

Taking the disorder average of Eq.~(\ref{G_tilde}) for the retarded propagator and introducing the decomposition 
$\braket{\hat{\tilde{G}}^R (\mathbf{r},\mathbf{r}^{\prime};\epsilon)}
=
\int_{\mathbf{q}} e^{i \mathbf{q} \cdot (\bs{\rho} - \bs{\rho}^{\prime} )} \hat{\tilde{g}}^R_{\mathbf{q}} (z,z^{\prime};\epsilon)$ with $\int_{\mathbf{q}} \equiv \int \frac{d^2\mathbf{q}}{\left( 2\pi \right)^2}$, the solution of the dressed propagator at the interface ($z,z^{\prime}=0$) reads $\hat{\tilde{g}}^R
=
[(\hat{g}^R)^{-1} - \hat{\tilde{\Sigma}}^{\text{imp}}]^{-1}$, where $\hat{g}^R$ is the bare propagator corresponding to Eq.~(\ref{Ham}) and $\hat{\tilde{\Sigma}}^{\text{imp}} \left(\epsilon\right) = (\hbar^2/2m)^2 \eta_{\gamma} \int_{\mathbf{q}}
\hat{g}^R_{\mathbf{q}} \left(0,0;\epsilon\right)$ is the configurationally averaged interfacial self-energy. Once $\hat{\tilde{g}}^R$ is known at the interface, the dressed reflection matrix is obtained, which, using the decomposition $\hat{\tilde{\Sigma}}^{\text{imp}} \left(\epsilon\right) = \hbar^2 \tilde{\xi}_{\mu}\left(\epsilon\right) \hat{\sigma}^{\mu} /2m$ with $\mu = 0,x,y,z$, and following the steps provided in Sec.~\ref{appendixA}, may be expressed as
\begin{equation}
\label{R_tilde}
\hat{\tilde{R}}_{\mathbf{q}}
=
\frac{e^{i \tilde{\varphi}_{\mathbf{q}}}}{\tilde{\varkappa}_{\mathbf{q}}}
\left(
\nu_{\mathbf{q}}
e^{i \tilde{\vartheta} _{\mathbf{q}}
\hat{\boldsymbol{\sigma}}\cdot
\tilde{\mathbf{n}}_{\mathbf{q},\text{R}}}
+
i \nu^{\prime}_{\mathbf{q}}
e^{i \tilde{\vartheta}^{\prime} _{\mathbf{q}}
\hat{\boldsymbol{\sigma}}\cdot 
\tilde{\mathbf{n}}_{\mathbf{q},\text{I}}}
\right),
\end{equation}
where 
$\tilde{\varphi}_{\mathbf{q}}
=
\arcsin[2 (\tilde{\kappa}_z \tilde{k}_z 
+
\tilde{\mathbf{Q}}_{\mathbf{q},\text{R}}
\cdot
\tilde{\mathbf{Q}}_{\mathbf{q},\text{I}}) / \tilde{\varkappa}_{\mathbf{q}}]$, 
$\tilde{\vartheta}_{\mathbf{q}}
=
\arcsin(2 k_z \tilde{Q}_{\mathbf{q},\text{R}}/\nu_{\mathbf{q}})$ and
$\tilde{\vartheta}_{\mathbf{q}}^{\prime} 
= 
\arcsin(2 k_z \tilde{Q}_{\mathbf{q},\text{I}}/\nu_{\mathbf{q}}^{\prime})$ with
$\tilde{\varkappa}_{\mathbf{q}}^2
=
(\tilde{Q}_{\mathbf{q},\text{R}}^2 
-
\tilde{Q}_{\mathbf{q},\text{I}}^2
-
\tilde{\kappa}_z^2 + \tilde{k}_z^2)^2
+
4(\tilde{\mathbf{Q}}_{\mathbf{q},\text{R}} 
\cdot
\tilde{\mathbf{Q}}_{\mathbf{q},\text{I}}
+
\tilde{\kappa}_z \tilde{k}_z)^2$, 
$\nu_{\mathbf{q}}^2 
=
[\tilde{Q}_{\mathbf{q},\text{R}}^2 
-
\tilde{Q}_{\mathbf{q},\text{I}}^2
-
\tilde{\kappa}_z^2 - \tilde{k}_z^2
-
2 \tilde{k}_z \text{Im}(\tilde{\xi}_{0})]^2
+
(2 k_z \tilde{Q}_{\mathbf{q},\text{R}})^2$ and 
$\nu_{\mathbf{q}}^{\prime 2} 
=
4[ \tilde{\mathbf{Q}}_{\mathbf{q},\text{R}} 
\cdot
\tilde{\mathbf{Q}}_{\mathbf{q},\text{I}}
-
\tilde{\kappa}_z \text{Im}(\tilde{\xi}_{0}
)]^2
+
(2 k_z \tilde{Q}_{\mathbf{q},\text{I}})^2$. Here, we have introduced the disordered quantities 
$\tilde{k}_z \equiv k_z - \text{Im}(\tilde{\xi}_0)$, 
$\tilde{\kappa}_z \equiv \kappa_z +\text{Re}(\tilde{\xi}_0)$ and $\tilde{\mathbf{Q}}_{\mathbf{q}} \equiv \mathbf{Q}_{\mathbf{q}} + \tilde{\bs{\xi}}$ with 
$\tilde{\mathbf{Q}}_{\mathbf{q}, \text{R}}
=
\text{Re}(\tilde{\mathbf{Q}}_{\mathbf{q}})$, 
$\tilde{\mathbf{Q}}_{\mathbf{q}, \text{I}}
=
\text{Im}(\tilde{\mathbf{Q}}_{\mathbf{q}})$, 
$\tilde{\mathbf{n}}_{\mathbf{q},\text{R}}
=
\tilde{\mathbf{Q}}_{\mathbf{q},\text{R}}/
\tilde{Q}_{\mathbf{q},\text{R}}$ and 
$\tilde{\mathbf{n}}_{\mathbf{q},\text{I}}
=
\tilde{\mathbf{Q}}_{\mathbf{q},\text{I}}/
\tilde{Q}_{\mathbf{q},\text{I}}$. Note that in the limit of vanishing disorder, $\nu_{\mathbf{q}}, \tilde{\varkappa}_{\mathbf{q}} \rightarrow \varkappa_{\mathbf{q}}$, while $\nu_{\mathbf{q}}^{\prime} \rightarrow 0$ (hence $\hat{\tilde{R}}_{\mathbf{q}} \rightarrow \hat{R}_{\mathbf{q}}$). We thus see that the overall effect of the interfacial disorder is a modulation of both the amplitude and phase of the reflection matrix.

Along with the surface disorder, bulk impurities are also present, whose contribution to the propagator may be included through an additional local self-energy $\Sigma^B(\mathbf{r})$. This results in a local scattering time  $\tau(\mathbf{r})= - \hbar/2 \text{Im} [\Sigma^B(\mathbf{r})]$ and a reduction of the propagator as \cite{camblong1994theory}
\begin{equation}
\hat{\tilde{G}} \left(\mathbf{r},\mathbf{r}^{\prime}\right)
\rightarrow
\hat{\tilde{G}} \left(\mathbf{r},\mathbf{r}^{\prime}\right) \text{exp} \left[- \int \limits_{\Gamma\left[\mathbf{r}, \mathbf{r}^{\prime}\right]} \frac{ds^{\dprime}}{2 l\left(\mathbf{r}^{\dprime}\right)} \right],
\end{equation}
which arises as a result of the damping of the electron wavefunction as it propagates along the straight path $\Gamma\left[\mathbf{r}, \mathbf{r}^{\prime}\right]$ connecting the points $\mathbf{r}$ and $\mathbf{r}^{\prime}$, and accounts for the average scattering encountered by the electron. Here, $l(\mathbf{r})=v_F \tau(\mathbf{r})$ is the local scattering length of the conduction electrons. In the NM layer, the loss of momentum associated with the damping of the wavefunction corresponds to the replacement $k_z \rightarrow k_z \sqrt{1 + i k_F/k_z^2 l_n}$ at the Fermi level, where $l_n$ is the mean free path.

\section{QUMR}

\begin{figure}[tph]
    \sidesubfloat[]{\includegraphics[width=0.6\linewidth,trim={1.5cm 0.7cm 0.5cm 0}]{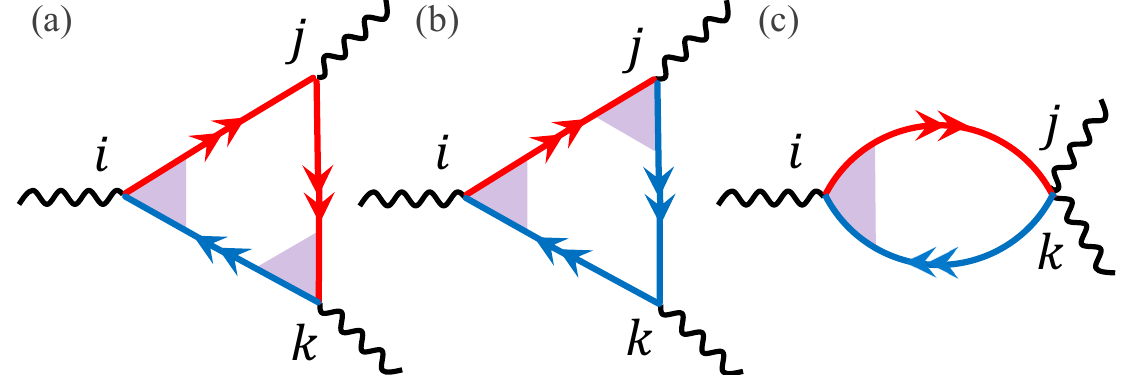}}
	\caption{Diagrammatic structure of the quadratic response. Together, (a-c) and their $j \leftrightarrow k$ counterparts comprise the dressed conductivity $\tilde{\sigma}_{ijk}$. The red (blue) double-arrowed lines represent dressed retarded (advanced) propagators, while the purple-shaded areas are dressed vertices.}
    \label{fig_qumr_fig3}
\end{figure}

To capture the nonlinear transport, we evaluate the relevant quadratic Kubo formulas. The nonlinear conductivity is then composed of two parts, $\tilde{\sigma}_{ijk}= \tilde{\sigma}_{ijk}^{(a)} + \tilde{\sigma}_{ijk}^{(b)}$, which diagrammatically correspond to dressed triangle and three-photon bubble diagrams \cite{parker2019diagrammatic, du2021quantum, rostami2021gauge}, as shown in Fig.~\ref{fig_qumr_fig3}. In general, the nonlinear conductivities will be modified by both self-energy and vertex corrections. However, as explained in Sec.~\ref{appendixB}, the contributions of the latter to the total conductivities turn out to be negligible in the weak disorder limit. Thus, to the leading order in the interfacial disorder, the conductivities may be expressed as
\begin{subequations}
\label{eq:sigma_ijk}
\begin{align}
\tilde{\sigma}_{ijk}^{(a)}
&=
\frac{2 e^3 \hbar^5}{\pi m^{3}} \int \limits_{S^I} q_i q_j q_k
\text{Im} \left[ \text{Tr} \left(
\pd_{\epsilon_F} \hat{\tilde{g}}^{R}_{\mathbf{q}, z z^{\prime}} 
\hat{\tilde{g}}^{R}_{\mathbf{q}, z^{\prime} z^{\dprime}} 
\hat{\tilde{g}}^{A}_{\mathbf{q}, z^{\dprime} z}
\right) \right],
\\
\tilde{\sigma}_{ijk}^{(b)}
&=
\frac{e^3 \hbar^3}{\pi m^2} \int \limits_{S^{II}} q_i \delta_{jk}
\text{Im} \left[ \text{Tr} \left(
\pd_{\epsilon_F} \hat{\tilde{g}}^{R}_{\mathbf{q}, z z^{\prime}} 
\hat{\tilde{g}}^A_{\mathbf{q}, z^{\prime} z}
\right) \right],
\end{align}
\end{subequations}
where $\int_{S^I} \equiv \int_{\mathbf{q}} 
\int_0^{\infty} dz^{\prime} \int_0^{\infty} dz^{\dprime}$, $\int_{S^{II}} \equiv \int_{\mathbf{q}} 
\int_0^{\infty} dz^{\prime}$, $\hat{\tilde{g}}^R_{\mathbf{q}, z z^{\prime}} \equiv \hat{\tilde{g}}^R_{\mathbf{q}} (z,z^{\prime};\epsilon_F)$ and $\hat{\tilde{g}}^A_{\mathbf{q}, z z^{\prime}} = (\hat{\tilde{g}}^R_{\mathbf{q}, z^{\prime} z})^{\dagger}$ is the advanced propagator.

Without loss of generality, let us set the electric field $\mathbf{E}$ in the $\mathbf{x}$ direction, so that we are only concerned with two tensor elements:  $\sigma _{xxx}$ and $\sigma _{yxx}$, governing the
nonlinear charge transport in the longitudinal and transverse directions, respectively. One can show that $\sigma _{xxx}$ must vanish when $\mathbf{m}$ is parallel to the $x$ axis, whereas $\sigma _{yxx}$ has to be zero when $\mathbf{m}$ is parallel to the $y$ axis (which is perpendicular to $\mathbf{E}$). These results can be understood intuitively by the following symmetry analysis. When $\mathbf{m}$ is parallel to the $x$ axis, the system is invariant under the mirror reflection in the $yz$ plane (\textit{i.e.}, $\mathcal{M}_{x}:x\rightarrow -x$) --and so are $\sigma _{ijk}$-- whereby the nonlinear current density and electric field follow the change $\{ E_{x};j^{(2)}_{x},j^{(2)}_{y}\} \rightarrow $ $\{-E_{x};-j^{(2)}_{x},j^{(2)}_{y}\} $. As a result, the quadratic response relation
must satisfy $j^{(2)}_{x}=\sigma _{xxx}E_{x}E_{x}=-j^{(2)}_{x}$, so $\sigma
_{xxx}( \mathbf{m=\pm x}) =0$. Similarly, when $\mathbf{m}$ is parallel to the $y$ axis, the mirror reflection in the $xz$ plane (\textit{i.e.}, $\mathcal{M}_{y}:y\rightarrow -y$) leads to $j^{(2)}_{y}=$ $\sigma_{yxx}E_{x}E_{x}=-j^{(2)}_{y}$, \textit{i.e.}, $\sigma _{yxx}\left(\mathbf{m=\pm y}\right) =0$.

To characterize the nonlinear transport, we introduce the UMR coefficients $\tilde{\zeta}_{\parallel }^{(2)} = \tilde{\zeta}_x^{(2)}$ and $\tilde{\zeta} _{\perp}^{(2)} = \tilde{\zeta}_y^{(2)}$, where \cite{mehraeen2022spin}
\begin{equation}
\label{Eq:zeta}
\tilde{\zeta}_i^{(2)}
\equiv
\frac{\tilde{\sigma}_{ix}(E_{x}) - \tilde{\sigma}_{ix}(-E_{x})}{\sigma_D E_x}
\simeq
-\frac{2 \tilde{\sigma} _{ixx}}{\sigma _{D}}.
\end{equation}
Here $\tilde{\sigma}_{ij}=j_{i}/E_{j}$ denotes the linear conductivity tensor, and $\sigma _{D}$ is the Drude conductivity of the NM layer. In this definition, the UMR coefficient is solely a property of the system and is independent of the strength of the applied current. Since it has the dimensions of inverse electric field, physically, one can say that it sets the scale of the electric field for which the magnitudes of the UMRs become comparable to linear-response effects \footnote{Note, however, that--owing to their directional nature--UMRs are typically observed at much weaker electric fields, as one need simply reverse the direction of the applied current to detect them.}.

\begin{figure}[tph]
    \sidesubfloat[]{\includegraphics[width=0.5\linewidth,trim={1.5cm 0.7cm 0.5cm 0}]{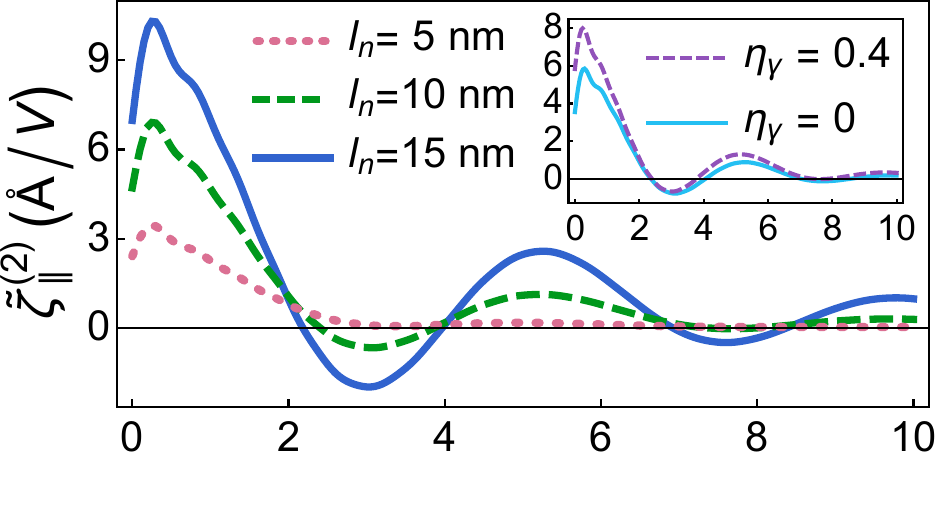}\label{fig_qumr_fig4a}}
    \\
    \sidesubfloat[]{\includegraphics[width=0.5\linewidth,trim={1.5cm 0.5cm 0.5cm 0}]{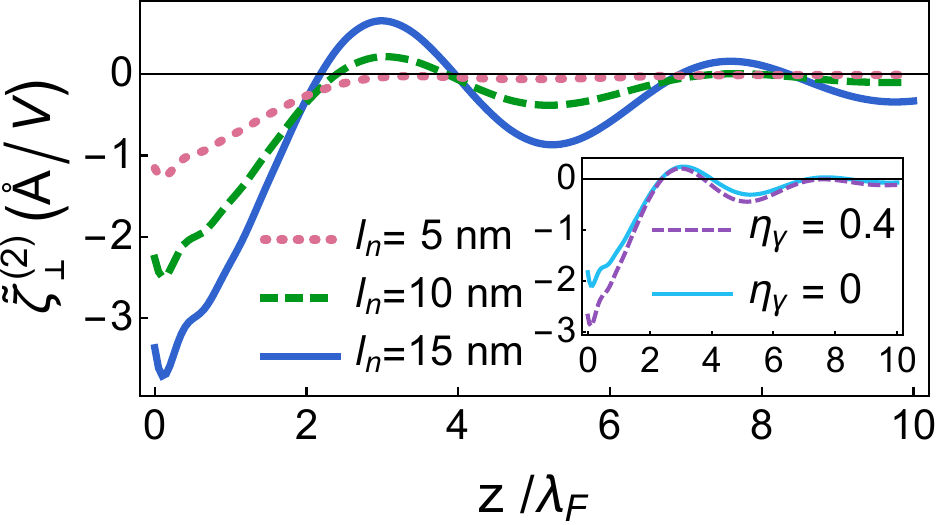}\label{fig_qumr_fig4b}}%
    \caption{Plots of the z dependences of the a) longitudinal (at $\mathbf{m}=\mathbf{y}$) and b) transverse (at $\mathbf{m}=\mathbf{x}$) disorder-averaged UMR coefficients for various values of the mean free path, with $\eta_{\gamma}= 0.2$ (main) and for different impurity strengths, with $l_n=10$ nm (inset). Other parameters used: $\eta_R = 0.2$ \cite{mihai2010current,park2013orbital,tokatly2015current,grytsyuk2016k}, $\eta_{ex} = 0.2$ \cite{kajiwara2010transmission}, $m=1.8 \times 10^{-30}$ kg, $\epsilon_F=5$ eV and $V_b=8$ eV.}
    \label{fig_qumr_fig4}
\end{figure}

Plots of the $z$ dependencies of the UMR coefficients for various values of the bulk and interfacial disorder parameters $l_n$ and $\eta_{\gamma}$ are presented in Fig.~\ref{fig_qumr_fig4}, which reveal that the UMR coefficients scale linearly with the mean free path. This is not surprising, as in the semiclassical picture, $\sigma_{ijk} \propto l_n^2$, while $\sigma_D \propto l_n$. Furthermore, as shown in the insets of Fig.~\ref{fig_qumr_fig4}, introducing the interfacial disorder leads to a slight enhancement of the UMR coefficients, which is due to the momentum relaxation of electrons caused by the interfacial disorder that effectively modulates the momentum-dependent spin-orbit scattering [see Eq.~(\ref{R_tilde})]. However, this trend is expected to be reversed when the interfacial disorder is increased further, as the vertex corrections, which constitute the diffuse scattering and contribute negatively to the magnitudes of the UMR coefficients (see Sec.~\ref{appendixB} for details), will play an increasingly important role in the quantum transport.

To shed light on the physical origin of the UMR effect, we first note from Fig.~\ref{fig_qumr_fig4} that spatial variations in $\tilde{\zeta}_{\parallel }^{(2)}$ and $\tilde{\zeta}_{\perp }^{(2)}$--comprised of an oscillatory exponential decay--occur over a length scale given by the Fermi wavelength $\lambda _{F}=2\pi/k_F$, with $k_{F}=\sqrt{2m \epsilon _{F}/\hbar ^{2}}$ the Fermi wavevector in the NM layer. This reflects the quantum nature of the nonlinear transport effect, as semiclassical UMRs typically scale with the spin diffusion length. Furthermore, as shown in Sec.~\ref{appendixB}, in the limiting case of $l_n\gg\lambda_F$, Eqs.~(\ref{eq:sigma_ijk}) may be reexpressed entirely in terms of the dressed interference velocity $\braket{\hat{\tilde{\bs{v}}}({\mathbf{q}},z)}_{I-R}$ as
\begin{equation}
\label{sigma_ijk_ballistic}
\tilde{\sigma}_{ijk} ( z ; \mathbf{m})
=
\frac{2 e^3 m^{2}}{\pi \hbar^4 k_F^2} l_n^2
\int \limits_{\mathbf{q}}
\mathcal{F}_{jk}(\mathbf{q})
\Braket{\hat{\tilde{v}}_{i}(\mathbf{q},z)}_{I-R},
\end{equation}
where
\begin{equation}
\mathcal{F}_{jk}(\mathbf{q})
=
\frac{1}{k_{z,F}} 
\left[
q_j q_k
\left(\frac{2}{k_{z,F}^2} + \frac{1}{k_F^2} - \frac{\hbar^2}{m}\pd_{\epsilon_F}\right)
+
\delta_{jk}
\right],
\end{equation}
thereby confirming the quantum-interference origin of the nonlinear magnetoresistances. In Eq.~(\ref{sigma_ijk_ballistic}), we have explicitly noted the dependence of $\tilde{\sigma}_{ijk}$ on the spatial coordinate $z$ and the magnetization $\mathbf{m}$, a property inherited from the interference velocity. Note also that the nonlinear conductivity tensor scales quadratically with the bulk disorder parameter, $\tilde{\sigma}_{ijk} \propto l_n^2$, in agreement with semiclassical expectations.

Incidentally, Eq.~(\ref{sigma_ijk_ballistic}) also reveals a simple relation between the longitudinal and transverse QUMRs. A comparison of the plots in Fig.~\ref{fig_qumr_fig4} suggests that the transverse and longitudinal UMR coefficients are related by a factor of -1/3. To confirm this, we note that the point of difference between $\tilde{\sigma}_{xxx}$ and $\tilde{\sigma}_{yxx}$ in Eq.~(\ref{sigma_ijk_ballistic}) is in the azimuthal integrals in momentum space. Performing these, we find that 
$\tilde{\sigma}_{xxx}
\propto
m_y \int_0^{2\pi} d \phi_{\mathbf{q}} \cos^4 \phi_{\mathbf{q}}
=
(3\pi/4) m_y$, while 
$\tilde{\sigma}_{yxx}
\propto
-m_x \int_0^{2\pi} d \phi_{\mathbf{q}} \cos^2 \phi_{\mathbf{q}} \sin^2 \phi_{\mathbf{q}}
=
-(\pi/4) m_x$, with $\phi_{\mathbf{q}}$ the azimuthal angle in momentum space. Thus, for an arbitrary orientation of the magnetization, this gives the simple-- yet general--relation
\begin{equation}
\label{ratio}
\frac{\tilde{\zeta}_{\perp}^{(2)}}
{\tilde{\zeta}_{\parallel}^{(2)}}
=
-\frac{1}{3} \cot \theta_{\mathbf{m}},
\end{equation}
with $\theta_{\mathbf{m}}$ the angle between the projected magnetization onto the layer plane and the applied electric field (see Fig.~\ref{fig_qumr_fig1}). And comparison of the transverse UMR coefficient when $\mathbf{m}=\mathbf{x}$ and the longitudinal UMR coefficient when $\mathbf{m} = \mathbf{y}$ readily yields the ratio -1/3 observed in Fig.~\ref{fig_qumr_fig4}. This simple relation reflects the common physical origin of the longitudinal and transverse QUMRs, and may serve as an additional transport signature of the quantum nonlinear transport effect.

\section{Materials considerations}

The most promising materials systems to
observe the QUMR effect are probably bilayers consisting of a heavy metal
and a FI, such as Au$|$YIG (Yttrium iron garnet) and $\beta $-Ta$|$YIG.
These systems possess a magnetic interface with sizable Rashba SOC and exchange interaction--two essential ingredients for generating the QUMR.

In principle, the QUMR may also arise in metallic bilayers comprising a
heavy metal and a ferromagnetic metal, but it would be accompanied by other nonlinear effects. One main competitor is the spin-Hall UMR~\cite{avci2015natphys,shulei2016prb,avci2018prl}, arising from the spin accumulation built in the
ferromagnetic-metal layer due to spin-current injection driven by the spin
Hall effect~\cite{dyakonov1971,hirsch1999prl,shufeng2000prl} in the heavy-metal layer. For a typical metallic bilayer,
such as Pt$|$Co, the corresponding UMR coefficient is about $\zeta
^{(2)}\sim 1$ \AA/V, about the same order of magnitude as the predicted
QUMR in a typical NM$|$FI bilayer. Other nonlinear effects that may
intertwine with the QUMR include the anomalous Nernst effect~\cite{syHuang11PRL_ANE} and spin Seebeck effect~\cite{uchida2008observation, uchida2010observation, qu2013prl, adachi2013theory}, arising from a vertical temperature gradient $\boldsymbol{\nabla }T\left(
\propto \mathbf{E}^{2}\right) $ across the ferromagnetic layer induced by Joule heating. Extra care would thus be needed to separate these contributions to the nonlinear resistance from the QUMR in metallic magnetic bilayers~\cite{avci2014interplay}.

\section{Scattering Amplitude Matrices}
\label{appendixA}

In this section, we derive the scattering matrix through two equivalent methods. The first approach involves imposing appropriate boundary conditions directly on the wavefunction and its spatial derivative. The main advantage of this approach is that the scattering amplitudes may be obtained in a rather straightforward manner. The second method requires the single-particle electron propagator, which, in turn, allows for a relatively transparent generalization to the case with disorder. This generalization is the focus of the last part of this section.

\subsection{Reflection Matrix from Wavefunction}

Consider a 3D electron gas in the scattering potential \cite{Stiles13PRB_SOT,tokatly2015prb}
\begin{equation}
\hat{V}_{\text{scat.}}(z)=\frac{\alpha_{R}}{\hbar} \delta(z) \hat{\boldsymbol{\sigma}}
\cdot(\hat{\mathbf{p}} \times \mathbf{z})-J_{e x} \delta(z) \hat{\boldsymbol{%
\sigma}} \cdot \mathbf{m} +V_{b} \Theta(-z)\,,
\end{equation}
where $\alpha _{R}$ and $J_{ex}$ are the coefficients of the interfacial
Rashba spin-orbit coupling (SOC) and exchange interaction, respectively, and
$V_{b}$ is the height of the energy barrier, which is greater than the Fermi
energy of electrons in the NM layer.

The general scattering state, given by Eq.~(\ref{gen_scat}), reads
\begin{equation}
\boldsymbol{\psi }_{\text{scat.}}=\left\{
\begin{array}{c}
e^{i\mathbf{q}\cdot \rho }\left( e^{-ik_{z}z}+e^{ik_{z}z}\hat{R}_{\mathbf{q}}\right)
\boldsymbol{\chi }\,,\;\;z>0\;\;~(\text{NM Layer}) \\
~~~~~~e^{i\mathbf{q}\cdot \rho }e^{\kappa_z z}\;\hat{T}_{\mathbf{q}} \boldsymbol{\chi }%
\,,\;\;~~~~~~~~~~z<0\;\;~(\text{FI Layer})%
\end{array}%
,\right.
\end{equation}%
\noindent where $\mathbf{q}\left[ =\left( k_{x},k_{y}\right) \right] $ and $%
\boldsymbol{\rho }\left[ =\left( x,y\right) \right] $ are the wave- and
position-vectors in the $x$-$y$ plane, wherein the system is translationally
invariant and allows propagation of plane waves, $k_{z}$ is the $z$%
-component of the wavevector of the propagating wave in the NM layer, and $%
\kappa_z^{-1}\left[ =\left( 2m V_{b}/\hbar ^{2}-k_{z}^{2}\right)
^{-1/2}\right] $ characterizes the decay length of the evanescent wave in
the FI layer. $\hat{R}_{\mathbf{q}}$ and $\hat{T}_{\mathbf{q}}$ are $2\times 2$ matrices in spin
space, which describe, respectively, the spin-dependent reflection and
transmission amplitudes, and the spinors $\boldsymbol{\chi }$ can be taken
as an arbitrary superposition of the eigenspinors that satisfy $\hat{%
\boldsymbol{\sigma }}\cdot \mathbf{m}\;\chi ^{\pm }=\pm \chi ^{\pm }$.

\begin{figure}[tph]
    \sidesubfloat[]{\includegraphics[width=0.25\linewidth,trim={1.5cm 0 0cm 0}]{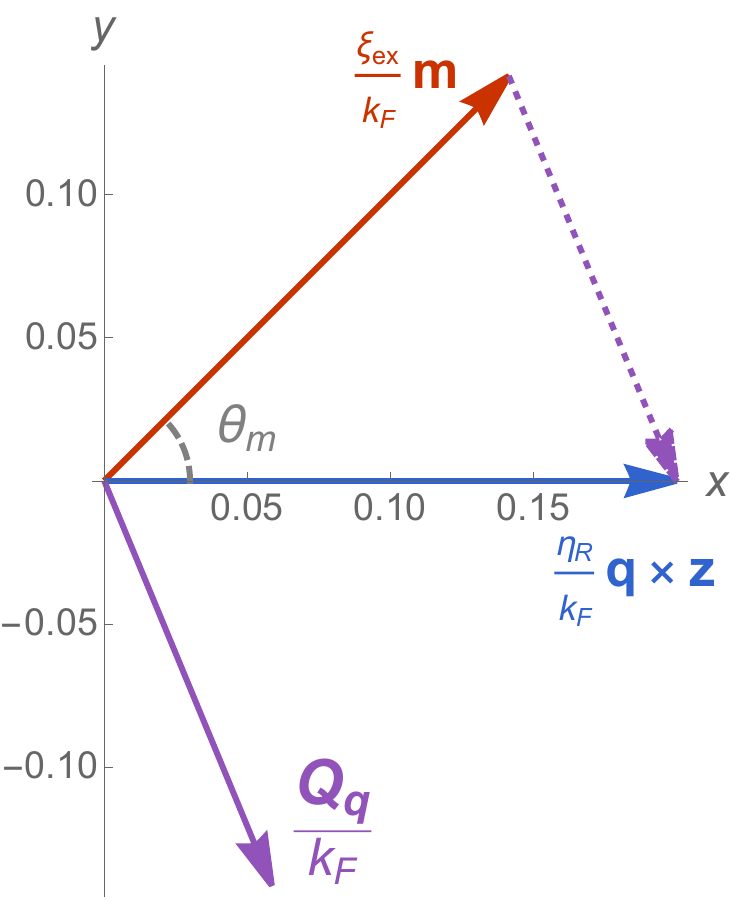}\label{fig_qumr_figS1a}}\quad%
    \sidesubfloat[]{\includegraphics[width=0.25\linewidth,trim={1.5cm 1cm 0.5cm 0}]{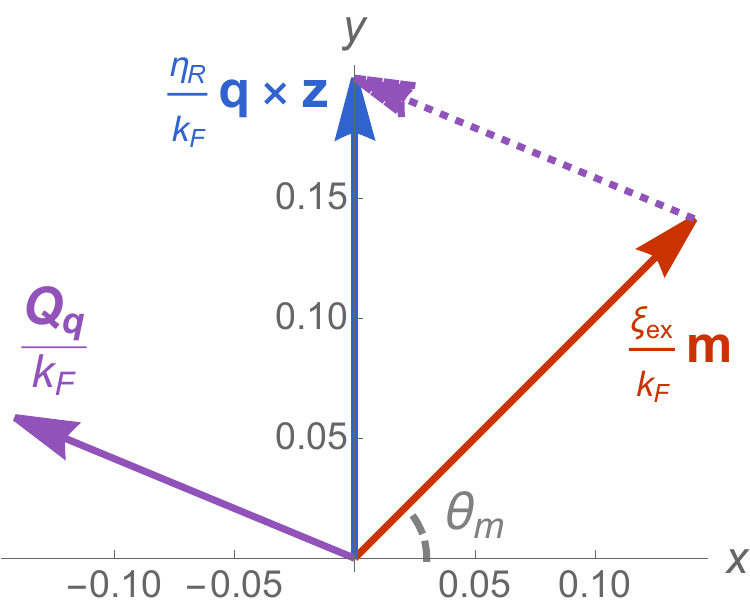}\label{fig_qumr_figS1b}}%
    \caption{Plots of the dimensionless vectors $\eta_R \mathbf{q}\times \mathbf{z}/k_F$, $\xi_{ex}\mathbf{m}/k_F$ and $\mathbf{Q_q}/k_F=(\eta_R \mathbf{q}\times \mathbf{z} - \xi_{ex}\mathbf{m})/k_F$ for the unit vectors (a) $\mathbf{q}\times \mathbf{z}/k_F=\mathbf{x}$ and (b) $\mathbf{q}\times \mathbf{z}/k_F=\mathbf{y}$. Parameters used: $\eta_R=\xi_{\text{ex}}/k_F=0.2$ and $\theta_m=\cos^{-1}(\mathbf{m}\cdot \mathbf{x})=\pi/4$.}
    \label{fig_qumr_figS1}
\end{figure}

The wavefunction obeys the following boundary conditions at the Rashba
interface
\begin{subequations}
\begin{gather}
\boldsymbol{\psi }_{\text{scat.}}\left( \rho ,0^{+}\right) =\boldsymbol{\psi
}_{\text{scat.}}\left( \rho ,0^{-}\right),
\label{BC1}
\\
\frac{d}{dz}\boldsymbol{\psi }_{\text{scat.}}\left( \rho ,0^{+}\right) -%
\frac{d}{dz}\boldsymbol{\psi }_{\text{scat.}}\left( \rho ,0^{-}\right) =\hat{%
\boldsymbol{\sigma }}\cdot \mathbf{Q_{q}}\boldsymbol{\psi }_{\text{scat.}%
}(\rho ,0),
\label{BC2}
\end{gather}
\end{subequations}
where $\mathbf{Q_{q}}=\eta _{R}\mathbf{q}\times \mathbf{z}-\xi _{ex}\mathbf{m%
}$ --as shown in Fig.~\ref{fig_qumr_figS1}-- with $\eta _{R}\equiv 2m \alpha _{R}/\hbar ^{2}$ a
dimensionless constant characterizing the strength of the interfacial Rashba
SOC and $\xi _{ex}\equiv 2m J_{ex}/\hbar ^{2}$ the rescaled
exchange interaction. Eliminating $\boldsymbol{\chi }$ and $e^{i\mathbf{q}%
\cdot \rho }$ from Eqs.~(\ref{BC1}) and~(\ref{BC2}), we obtain a set of
equations for the scattering amplitude matrices:
\begin{subequations}
\begin{gather}
1+\hat{R}_{\mathbf{q}}=\hat{T}_{\mathbf{q}},
\\
-ik_{z}\left( 1-\hat{R}_{\mathbf{q}}\right) -\kappa_z \hat{T}_{\mathbf{q}}=\hat{\boldsymbol{\sigma }}%
\cdot \mathbf{Q_{q}}\hat{T}_{\mathbf{q}}.
\end{gather}
\end{subequations}

Solving the set of equations, one can obtain the general expression of the
reflection amplitude matrix as follows
\begin{equation}
\hat{R}_{\mathbf{q}}=\frac{\mathbf{Q}_{\mathbf{q}}^{2}-(\kappa_z ^{2}+k_{z}^{2})+2ik_{z}%
\hat{\boldsymbol{\sigma }}\cdot \mathbf{Q}_{\mathbf{q}}}{(\kappa_z
-ik_{z})^{2}-\mathbf{Q}_{\mathbf{q}}^{2}}. \label{rmatrix}
\end{equation}
\noindent Introducing the change of variables $\varphi _{\mathbf{q}}=\arcsin \left( 2k_{z}\kappa_z /\varkappa_{\mathbf{q}}\right)$ and $\vartheta _{\mathbf{q}}=\arcsin \left( 2k_{z}Q_{\mathbf{q}}/\varkappa_{\mathbf{q}}\right)$ with $\varkappa_{\mathbf{q}}^{2}=\left[ Q_{%
\mathbf{q}}^{2}-(\kappa_z ^{2}+k_{z}^{2})\right] ^{2}+\left( 2k_{z}Q_{\mathbf{q%
}}\right) ^{2}$, and using the unit vector $\mathbf{n}_{\mathbf{q}} \mathbf{=\mathbf{Q}_{%
\mathbf{q}}/}Q_{\mathbf{q}}$, the reflection matrix is recast in the form
\begin{equation}
\label{SEq:R_q}
\hat{R}_{\mathbf{q}}
=
e^{i \left( \varphi_{\mathbf{q}} + \vartheta _{\mathbf{q}}
\hat{\boldsymbol{\sigma}}\cdot \mathbf{n}_{\mathbf{q}} \right)},
\end{equation}
which is Eq.~(\ref{eq:R_q}).

\subsection{Reflection Matrix from Propagator}

\subsubsection{Propagator without interfacial effects}

Let us rewrite the Hamiltonian given by Eq.~(\ref{Ham}) as
\begin{subequations}
\label{Seq:H_hat}
\begin{gather}
\hat{H}
=
H^0 + \hat{V}^{\text{int}}_{\mathbf{q}} \delta(z),
\\
H^0
=
\frac{\hbar^2 k^2}{2m} + V_b \Theta(-z),
\\
\hat{V}^{\text{int}}_{\mathbf{q}}
=
\alpha_R \,\hat{\bs{\sigma}} \cdot \mathbf{q} \times \mathbf{z} - J_{ex}\hat{\bs{\sigma}} \cdot \mathbf{m},
\end{gather}
\end{subequations}
where $\mathbf{q}$ is the in-plane momentum and $\hat{V}^{\text{int}}_{\mathbf{q}}$ is the interaction potential at the interface and consists of the interfacial Rashba SOC and exchange interactions. For the unperturbed system, $H^0$, the general scattering state for electrons with momentum $\mathbf{k}=(\mathbf{q},k_1)$ incident on the interface from the NM reads
\begin{equation}
\label{SEq:psi_k}
\psi_{\mathbf{k}}\left(\mathbf{r}\right)
=
\phi_{k_1/\kappa}(z) e^{i \mathbf{q} \cdot \bs{\rho}},
\end{equation}
where
\begin{equation}
\phi_{k_1/\kappa}(z)
=
\begin{cases}
      e^{-i k_1 z} + a_r \left( k_1, \kappa \right) e^{i k_1 z}, & z>0\; \; \text{(NM)}  \\
      a_t \left( k_1, \kappa \right) e^{\kappa z}, & z<0 \; \; \text{(FI)}  
    \end{cases}.
\end{equation}
Here, $\kappa$ quantifies the decay rate of the wavefunction in the FI and $\mathbf{r}=(\bs{\rho},z)$ is the position vector, with $a_r$ and $a_t$ the reflection and transmission amplitudes, respectively. These may be obtained from the boundary conditions on the wave function
\begin{subequations}
\begin{gather}
\phi_{k_1}\left(0^+\right)
=
\phi_{\kappa}\left(0^-\right),
\\
\frac{d}{dz}\phi_{k_1}\left(0^+\right)
=
\frac{d}{dz}\phi_{\kappa}\left(0^-\right),
\end{gather}
\end{subequations}
which imply 
\begin{equation}
a_r \left( k_1, \kappa \right)
=
\frac{ik_1 + \kappa}{ik_1 - \kappa},
\end{equation}
and $a_t=1+a_r$. The retarded unperturbed Green's function is
\begin{equation}
G^{0,R} \left(\epsilon\right)
=
\left(\epsilon - H^0 + i\delta \right)^{-1},
\end{equation}
which gives rise to the real-space propagator
\begin{equation}
\label{SEq:G_0,R}
G^{0,R} \left(\mathbf{r},\mathbf{r}^{\prime};\epsilon\right)
\equiv
\Braket{\mathbf{r} | G^{0,R} \left(\epsilon\right) | \mathbf{r}^{\prime}}
=
\sum_{\mathbf{k}} 
\frac{\psi_{\mathbf{k}}\left(\mathbf{r}\right) \psi_{\mathbf{k}}^*\left(\mathbf{r}^{\prime}\right)}
{\epsilon - \epsilon_{\mathbf{k}} + i\delta},
\end{equation}
where $\epsilon_{\mathbf{k}}= \hbar^2 k^2 /2 m + V_b \Theta(-z)$. Inserting Eq.~(\ref{SEq:psi_k}) into Eq.~(\ref{SEq:G_0,R}), we obtain the decomposition
\begin{equation}
\label{gR_z}
G^{0,R} \left(\mathbf{r},\mathbf{r}^{\prime};\epsilon\right)
=
\sum_{\mathbf{q}} e^{i\mathbf{q} \cdot(\bs{\rho} - \bs{\rho}^{\prime})} g^{0,R}_{\mathbf{q}}\left(z,z^{\prime};\epsilon\right),
\end{equation}
which, for correlations in the NM layer, leads to the solution
\begin{equation}
\label{gR_0_nm}
g^{0,R}_{\mathbf{q}}\left(z>0,z^{\prime}>0;\epsilon\right)
=
\sum_{k_1} 
\frac{\phi_{k_1}\left(z\right) \phi_{k_1}^*\left(z^{\prime}\right)}
{\epsilon - \epsilon_{\mathbf{k}} + i\delta}
=
\frac{m}{i \hbar^2 k_z(\epsilon)} \left[e^{i k_z(\epsilon) |z-z^{\prime}|} + a_r \left( k_z, \kappa_z \right) e^{i k_z(\epsilon)\left(z+z^{\prime}\right)}\right].
\end{equation}
Here, $k_1= \pm k_z(\epsilon)$ are the poles of the propagator at energy $\epsilon$, with $k_z(\epsilon)=\sqrt{[k(\epsilon)]^2 - q^2}$, $k(\epsilon)=\sqrt{2m |\epsilon| /\hbar^2}$, $\kappa_z(\epsilon)=\sqrt{k_b^2 - [k_z(\epsilon)]^2}$ and $k_b=\sqrt{2m V_b/\hbar^2}$.

Before proceeding to include interfacial effects, we also add the contribution of the bulk scatterers to the propagator. This can be achieved by modifying the Hamiltonian as $H^0 \rightarrow H^0 + \Sigma ^B(\mathbf{r})$, where $\Sigma^B(\mathbf{r})$ is the local self-energy due to bulk impurity scattering, resulting in a local scattering time  $\tau(\mathbf{r})= - \hbar/2 \text{Im} [\Sigma^B(\mathbf{r})]$ and a reduction of the propagator as \cite{camblong1994theory}
\begin{equation}
G^{0,R} \left(\mathbf{r},\mathbf{r}^{\prime};\epsilon\right)
\rightarrow
G^{0,R} \left(\mathbf{r},\mathbf{r}^{\prime};\epsilon\right) \text{exp} \left[- \int \limits_{\Gamma\left[\mathbf{r}, \mathbf{r}^{\prime}\right]} \frac{ds^{\dprime}}{2 l\left(\mathbf{r}^{\dprime}\right)} \right],
\end{equation}
which arises as a result of the damping of the electron wavefunction as it propagates along the straight path $\Gamma\left[\mathbf{r}, \mathbf{r}^{\prime}\right]$ connecting the points $\mathbf{r}$ and $\mathbf{r}^{\prime}$, and accounts for the average scattering encountered by the electron. Here, $l(\mathbf{r})=v_F \tau(\mathbf{r})$ is the local scattering length of the conduction electrons. In the NM layer, the loss of momentum associated with the damping of the wavefunction corresponds to the replacement $k_z \rightarrow k_z \sqrt{1 + i k_F/k_z^2 l_n} \simeq k_z + i \frac{k_F}{2 k_z l_n}$ at the Fermi level, where $l_n$ is the mean free path of the conduction electrons and $k_F = \sqrt{2m \epsilon_F /\hbar^2}$ is the Fermi wavenumber. Thus, the unperturbed propagator now reads
\begin{equation}
\label{gR_0_Ln}
g^{0,R}_{\mathbf{q}}\left(z,z^{\prime};\epsilon\right)
=
\frac{m}{i \hbar^2 k_z(\epsilon)} \left[e^{i k_z(\epsilon) |z-z^{\prime}|} e^{- k(\epsilon) |z-z^{\prime}|/2 k_z(\epsilon) l_n} 
+
a_r\left(k_z, \kappa_z\right) e^{i k_z(\epsilon)\left(z+z^{\prime}\right)}e^{- k(\epsilon) \left(z+z^{\prime}\right)/2 k_z(\epsilon) l_n}\right].
\end{equation}

\subsubsection{Including interfacial potential}

In the presence of spin-dependent interactions at the interface, the propagator is now a $2\times2$ matrix in spin space, which can be found using the Dyson equation
\begin{equation}
\label{GR_soc}
\hat{G}^R \left(\mathbf{r},\mathbf{r}^{\prime};\epsilon\right)
=
\hat{G}^{0,R} \left(\mathbf{r},\mathbf{r}^{\prime};\epsilon\right)
+
\int d\mathbf{r}_1 \hat{G}^{0,R} \left(\mathbf{r},\mathbf{r}_1;\epsilon\right) \hat{V}^{\text{int}}_{\mathbf{q}} \delta(z_1) \hat{G}^R \left(\mathbf{r}_1,\mathbf{r}^{\prime};\epsilon\right)\,,
\end{equation}
where
\begin{equation}
\hat{G}^{0,R} \left(\mathbf{r},\mathbf{r}^{\prime};\epsilon\right)
=
G^{0,R} \left(\mathbf{r},\mathbf{r}^{\prime};\epsilon\right) \hat{\sigma}_0\,.
\end{equation}
Inserting Eq.~(\ref{gR_z}) into Eq.~(\ref{GR_soc}), and introducing the analogous decomposition
\begin{equation}
\label{gR_hat_z}
\hat{G}^R \left(\mathbf{r},\mathbf{r}^{\prime};\epsilon\right)
=
\sum_{\mathbf{q}} e^{i\mathbf{q} \cdot(\bs{\rho} - \bs{\rho}^{\prime})} \hat{g}^R_{\mathbf{q}} \left(z,z^{\prime};\epsilon\right),
\end{equation}
the Fourier-transformed Dyson equation reads
\begin{equation}
\hat{g}^R_{\mathbf{q}}\left(z,z^{\prime};\epsilon\right)
=
\hat{g}^{0,R}_{\mathbf{q}}\left(z,z^{\prime};\epsilon\right)
+
\hat{g}^{0,R}_{\mathbf{q}}\left(z,0;\epsilon\right) \hat{V}^{\text{int}}_{\mathbf{q}}
\hat{g}^R_{\mathbf{q}}\left(0,z^{\prime};\epsilon\right),
\end{equation}
so that for interfacial correlations ($z,z^{\prime}=0$), the decomposed propagator is obtained as
\begin{equation}
\label{gR_hat_0}
\hat{g}^R_{\mathbf{q}}
=
\left(\left[\hat{g}^{0,R}_{\mathbf{q}}\right]^{-1} - \hat{V}_{\mathbf{q}}\right)^{-1}
=
\frac{2m}{\hbar^2} \left[\left(i k_z - \kappa_z \right) \hat{\sigma}_0 - \hat{\bs{\sigma}} \cdot \mathbf{Q}_{\mathbf{q}}\right]^{-1}.
\end{equation}
On the other hand, spin-dependent scattering at the interface results in spin-dependent reflection of the electron wavefunction, so that Eq.~(\ref{gR_0_Ln}) may be generalized to
\begin{equation}
\label{gR_hat_nm}
\hat{g}^R_{\mathbf{q}} \left(z,z^{\prime};\epsilon\right)
=
\frac{m}{i \hbar^2 k_z(\epsilon)} \left[e^{i k_z(\epsilon) |z-z^{\prime}|}e^{- k(\epsilon) |z-z^{\prime}|/2 k_z(\epsilon) l_n}
+
\hat{R}_{\mathbf{q}} e^{i k_z(\epsilon)\left(z+z^{\prime}\right)}e^{- k(\epsilon) \left(z+z^{\prime}\right)/2 k_z(\epsilon) l_n}\right],
\end{equation}
where $\hat{R}_{\mathbf{q}}$ is the reflection matrix. Equating Eq.~(\ref{gR_hat_nm}) at $z,z^{\prime}=0$ to Eq.~(\ref{gR_hat_0}), we arrive at the general form of the reflection matrix
\begin{equation}
\label{Seq:R_hat1}
\hat{R}_{\mathbf{q}}
=
\frac{\mathbf{Q}_{\mathbf{q}}^2 -\left(k_z^2 + \kappa_z^2 \right) + 2ik_z \hat{\bs{\sigma}} \cdot \mathbf{Q}_{\mathbf{q}}}
{\left(ik_z - \kappa_z \right)^2 - \mathbf{Q}_{\mathbf{q}}^2},
\end{equation}
or, equivalently,
\begin{equation}
\label{Seq:R_hat2}
\hat{R}_{\mathbf{q}}
=
e^{i \left( \varphi_{\mathbf{q}} + \vartheta _{\mathbf{q}}
\hat{\boldsymbol{\sigma}}\cdot \mathbf{n}_{\mathbf{q}} \right)},
\end{equation}
in agreement with Eq.~(\ref{SEq:R_q}), which was derived directly from the wave function.

\subsection{Including Interfacial Disorder}

Having derived the reflection matrix from the Green's function in the absence of interfacial disorder, we now proceed to calculate the disordered propagator by taking into account interfacial impurities, which, in general, lead to both specular and diffuse scattering off the bilayer interface. This calculation is done in two equivalent ways: once from the transition matrix, and then from a self-energy calculation. From the latter approach, we then derive the general form of the disordered reflection matrix. 

\subsubsection{Dressed propagator from transition matrix}

Recall the Dyson equation given by Eq.~(\ref{G_tilde}) 
\begin{equation}
\label{GR_imp}
\hat{\tilde{G}}^R \left(\mathbf{r},\mathbf{r}^{\prime};\epsilon\right)
=
\hat{G}^R\left(\mathbf{r},\mathbf{r}^{\prime};\epsilon\right)
+
\int d\mathbf{r}_1 \hat{G}^R \left(\mathbf{r},\mathbf{r}_1;\epsilon\right) \hat{V}^{\text{imp}}\left({\mathbf{r}}_1\right)\hat{\tilde{G}}^R \left(\mathbf{r}_1,\mathbf{r}^{\prime};\epsilon\right),
\end{equation}
where $\hat{\tilde{G}}^R \left(\mathbf{r},\mathbf{r}^{\prime};\epsilon\right)$ is the dressed propagator in the presence of both disorder and interfacial interaction, and $\hat{V}^{\text{imp}}(\mathbf{r})=\hat{V}^{\text{imp}} (\bs{\rho})\delta(z)$ is the interfacial impurity potential, for which we assume the white noise distribution $\Braket{V^{\text{imp}} \left(\bs{\rho}\right)}
= 0$ and $\Braket{V^{\text{imp}} \left(\bs{\rho}\right) V^{\text{imp}} \left(\bs{\rho}^{\prime}\right)} = \gamma^2 \delta\left(\bs{\rho} - \bs{\rho}^{\prime}\right)$ \cite{stewart2003interfacial}, where the parameter $\gamma$, or its dimensionless counterpart $\eta_{\gamma} = (2m \gamma/\hbar^2)^2$, characterizes the strength of the impurity interaction and $\braket{\cdots}$ here denotes the configurational average over impurity positions.

Disorder at the interface destroys the in-plane periodicity of the system so that the propagator is no longer diagonal in momentum space and is instead decomposed as
\begin{equation}
\label{gR_imp_z}
\hat{\tilde{G}}^R \left(\mathbf{r},\mathbf{r}^{\prime};\epsilon\right)
=
\sum_{\mathbf{q},\mathbf{q}^{\prime}} e^{i \left(\mathbf{q} \cdot\bs{\rho} - \mathbf{q} ^{\prime} \cdot \bs{\rho}^{\prime} \right)}
\hat{\tilde{g}}^R_{\mathbf{q}\mathbf{q}^{\prime}} \left(z,z^{\prime};\epsilon\right).
\end{equation}
Inserting Eq.~(\ref{gR_imp_z}) into Eq.~(\ref{GR_imp}), the Dyson equation is decomposed as
\begin{equation}
\label{gR_imp}
\hat{\tilde{g}}^R_{\mathbf{q}\mathbf{q}^{\prime}} \left(z , z^{\prime};\epsilon\right)
=
\hat{g}^R_{\mathbf{q}} \left(z,z^{\prime};\epsilon\right) \delta_{\mathbf{q}\mathbf{q}^{\prime}}
+
\hat{g}^R_{\mathbf{q}} \left(z,0;\epsilon\right) \hat{\mathcal{T}}_{\mathbf{q}\mathbf{q}^{\prime}}\left(\epsilon\right)
\hat{g}^R_{\mathbf{q}^{\prime}} \left(0,z^{\prime};\epsilon\right),
\end{equation}
where the transition matrix $\hat{\mathcal{T}}_{\mathbf{q}\mathbf{q}^{\prime}}$ obeys the integral equation
\begin{equation}
\label{T_qq}
\hat{\mathcal{T}}_{\mathbf{q}\mathbf{q}^{\prime}}\left(\epsilon\right) 
=
\hat{V}^{\text{imp}}_{\mathbf{q}\mathbf{q}^{\prime}}
+
\sum_{\mathbf{q}_1} \hat{V}^{\text{imp}}_{\mathbf{q}\mathbf{q}_1} \hat{g}^R_{\mathbf{q}_1} \left(\epsilon\right) \hat{\mathcal{T}}_{\mathbf{q}_1\mathbf{q}^{\prime}}\left(\epsilon\right).
\end{equation}
Here, $\hat{g}^R_{\mathbf{q}} \left(\epsilon\right) \equiv \hat{g}^R_{\mathbf{q}} \left(0,0;\epsilon\right)$ and the Fourier transform of the impurity potential is given by
\begin{equation}
\hat{V}^{\text{imp}}_{\mathbf{q}\mathbf{q}^{\prime}}
=
\int d\bs{\rho} \, e^{-i \bs{\rho} \cdot \left(\mathbf{q} - \mathbf{q}^{\prime}\right)} \hat{V}^{\text{imp}}\left(\bs{\rho}\right).
\end{equation}
For completeness, we also note the white noise distribution relations in momentum space
\begin{subequations}
\begin{align}
\Braket{\hat{V}^{\text{imp}}_{\mathbf{q} \mathbf{q}^{\prime}}}
&=
0,
\\
\Braket{\hat{V}^{\text{imp}}_{\mathbf{q} \mathbf{q}^{\prime}} \hat{V}^{\text{imp}}_{\mathbf{q}^{\dprime} \mathbf{q}^{\tprime}}}
&=
\gamma^2 \,\delta_{\mathbf{q} + \mathbf{q}^{\dprime}, \mathbf{q}^{\prime} + \mathbf{q}^{\tprime}}.
\end{align}
\end{subequations}
Upon ensemble averaging, translation invariance is restored and the disorder-averaged propagator becomes diagonal in momentum space,
\begin{equation}
\label{gR_imp_z_avg}
\Braket{\hat{\tilde{G}}^R \left(\mathbf{r},\mathbf{r}^{\prime};\epsilon\right)}
=
\sum_{\mathbf{q}} e^{i \mathbf{q} \cdot \left(\bs{\rho} - \bs{\rho}^{\prime} \right)} \hat{\tilde{g}}^R_{\mathbf{q}} \left(z,z^{\prime};\epsilon\right)\,,
\end{equation}
where we have used the definition $\braket{\hat{\tilde{g}}^R_{\mathbf{q} \mathbf{q}^{\prime}} \left(z,z^{\prime};\epsilon\right)} \equiv \hat{\tilde{g}}^R_{\mathbf{q}} \left(z,z^{\prime};\epsilon\right) \delta_{\mathbf{q}\mathbf{q}^{\prime}}$. Taking the configurational average of Eq.~(\ref{gR_imp}), the dressed propagator reads
\begin{equation}
\hat{\tilde{g}}^R_{\mathbf{q}} \left(z,z^{\prime};\epsilon\right)
=
\hat{g}^R_{\mathbf{q}} \left(z,z^{\prime};\epsilon\right)
+
\hat{g}^R_{\mathbf{q}} \left(z,0;\epsilon\right) \hat{\mathcal{T}}\left( \epsilon \right) \hat{g}^R_{\mathbf{q}} \left(0,z^{\prime};\epsilon\right),
\end{equation}
with the configurational average of the transition matrix defined as $\braket{\hat{\mathcal{T}}_{\mathbf{q}\mathbf{q}^{\prime}}} \equiv \hat{\mathcal{T}} \delta_{\mathbf{q}\mathbf{q}^{\prime}}$, which is found by taking the ensemble average of Eq.~(\ref{T_qq}) and using the Born approximation, leading to the solution
\begin{equation}
\label{T_s}
\hat{\mathcal{T}} \left( \epsilon \right)
=
\gamma^2 \sum_{\mathbf{q}} 
\hat{g}^R_{\mathbf{q}} \left(\epsilon\right).
\end{equation}
Thus, the dressed propagator is obtained. Below, we arrive at the same result from a self-energy calculation, which is slightly more convenient for deriving the dressed reflection matrix.

\subsubsection{Dressed propagator and reflection matrix from self-energy}

Consider the configurational average of Eq.~(\ref{GR_imp}),
\begin{equation}
\begin{split}
\label{GR_imp_2nd}
\Braket{\hat{\tilde{G}}^R \left(\mathbf{r},\mathbf{r}^{\prime};\epsilon\right)}
=&
\hat{G}^R\left(\mathbf{r},\mathbf{r}^{\prime};\epsilon\right)
\\
&+
\int d\mathbf{r}_1 \int d\mathbf{r}_2 \,
\hat{G}^R \left(\mathbf{r},\mathbf{r}_1;\epsilon\right) \Braket{\hat{V}^{\text{imp}} \left({\mathbf{r}}_1\right) 
\hat{G}^R \left(\mathbf{r}_1,\mathbf{r}_2;\epsilon\right) \hat{V}^{\text{imp}} \left({\mathbf{r}}_2\right)\hat{\tilde{G}}^R \left(\mathbf{r}_2,\mathbf{r}^{\prime};\epsilon\right)}.
\end{split}
\end{equation}
Using Eq.~(\ref{gR_imp_z_avg}) and applying the expansion \cite{brataas1994semiclassical}
\begin{equation}
\Braket{\hat{V}^{\text{imp}}\left({\mathbf{r}}_1\right)\hat{V}^{\text{imp}}\left({\mathbf{r}}_2\right)\hat{\tilde{G}}_R \left(\mathbf{r}_2,\mathbf{r}^{\prime};\epsilon\right)}
=
\Braket{\hat{V}^{\text{imp}}\left({\mathbf{r}}_1\right)\hat{V}^{\text{imp}}\left({\mathbf{r}}_2\right)}
\Braket{\hat{\tilde{G}}_R \left(\mathbf{r}_2,\mathbf{r}^{\prime};\epsilon\right)}
+
\text{vertex corrections}
,
\end{equation}
Eq.~(\ref{GR_imp_2nd}) is decomposed as
\begin{equation}
\label{gR_imp_sig}
\hat{\tilde{g}}^R_{\mathbf{q}} \left(z,z^{\prime};\epsilon\right)
=
\hat{g}^R_{\mathbf{q}} \left(z,z^{\prime};\epsilon\right)
+
\hat{g}^R_{\mathbf{q}} \left(z,0;\epsilon\right) \hat{\tilde{\Sigma}}^{\text{imp}} \left(\epsilon\right) 
\hat{\tilde{g}}^R_{\mathbf{q}} \left(0,z^{\prime};\epsilon\right),
\end{equation}
where $\hat{\tilde{\Sigma}}^{\text{imp}} \left(\epsilon\right)$ is the configurationally averaged irreducible self-energy and is given, within the Born approximation, by
\begin{equation}
\label{Sigma_tilde}
\hat{\tilde{\Sigma}}^{\text{imp}} \left(\epsilon\right)
=
\gamma^2 \sum_{\mathbf{q}}
\hat{g}^R_{\mathbf{q}} \left(\epsilon\right).
\end{equation}
Note that this is equivalent to the diagonal elements of the disorder-averaged transition matrix, Eq.~(\ref{T_s}), as one would expect \cite{bruus2004many}.

Solving Eq.~(\ref{gR_imp_sig}) for the disordered propagator at $z,z^{\prime}=0$, we find
\begin{equation}
\label{gR_imp_00}
\hat{\tilde{g}}^R
=
\left[\left(\hat{g}^R \right)^{-1} - \hat{\tilde{\Sigma}}^{\text{imp}} \right]^{-1}
=
\left[\left(\hat{g}^{0,R}_{\mathbf{q}} \right)^{-1} -\left(\hat{V}^{\text{int}}_{\mathbf{q}}
+
\hat{\tilde{\Sigma}}^{\text{imp}} \right) \right]^{-1}.
\end{equation}
Thus, in analogy with Eq.~(\ref{gR_hat_nm}), the dressed propagator may be expressed as
\begin{equation}
\label{gR_hat_imp}
\hat{\tilde{g}}^R_{\mathbf{q}}\left(z,z^{\prime};\epsilon\right)
=
\frac{m}{i \hbar^2 k_z(\epsilon)} \left[e^{i k_z(\epsilon) |z-z^{\prime}|}e^{- k(\epsilon) |z-z^{\prime}|/2 k_z(\epsilon) l_n}
+
\hat{\tilde{R}}_{\mathbf{q}} e^{i k_z(\epsilon)\left(z+z^{\prime}\right)}e^{- k(\epsilon) \left(z+z^{\prime}\right)/2 k_z(\epsilon) l_n}\right],
\end{equation}
where, using the decomposition $\hat{\tilde{\Sigma}} \left(\epsilon\right)
=
\tilde{\Sigma}_{\mu}^{\text{imp}} \left(\epsilon\right) \hat{\sigma}^{\mu}$ with $\mu=0,x,y,z$ and introducing the change of variables $\tilde{\Sigma}_{\mu}^{\text{imp}} \left(\epsilon\right) \equiv \hbar^2 \tilde{\xi}_{\mu}\left(\epsilon\right) /2m$, the dressed reflection matrix reads
\begin{equation}
\begin{split}
\label{R_hat_tilde}
\hat{\tilde{R}}_{\mathbf{q}}
&=
\frac{\tilde{\mathbf{Q}}_{\mathbf{q}}^2 -\left(\tilde{\kappa}_z^2 + \tilde{k}_z^2 \right)
\left( 1 + \frac{2i \text{Im} \,\tilde{\xi}_0}
{\tilde{\kappa}_z + i \tilde{k}_z} \right)
+ 
2i k_z \hat{\bs{\sigma}} \cdot \tilde{\mathbf{Q}}_{\mathbf{q}}}
{\left(\tilde{\kappa}_z - i\tilde{k}_z \right)^2 - \tilde{\mathbf{Q}}_{\mathbf{q}}^2},
\end{split}
\end{equation}
with $\tilde{k}_z\left(\epsilon\right) \equiv k_z\left(\epsilon\right) - \text{Im} \;\tilde{\xi}_0\left(\epsilon\right)$, $\tilde{\kappa}_z \left(\epsilon\right) \equiv \kappa_z \left(\epsilon\right) +
\text{Re} \;\tilde{\xi}_0\left(\epsilon\right)$ and $\tilde{\mathbf{Q}}_{\mathbf{q}} \equiv \mathbf{Q}_{\mathbf{q}} + \tilde{\bs{\xi}}\left(\epsilon\right)$. 

Introducing the change of variables $\tilde{\varphi}_{\mathbf{q}}
=
\arcsin[2 (\tilde{\kappa_z} \tilde{k}_z 
+
\tilde{\mathbf{Q}}_{\mathbf{q},\text{R}}
\cdot
\tilde{\mathbf{Q}}_{\mathbf{q},\text{I}}) / \tilde{\varkappa}_{\mathbf{q}}]$, 
$\tilde{\vartheta}_{\mathbf{q}}
=
\arcsin(2 k_z \tilde{Q}_{\mathbf{q},\text{R}}/\nu_{\mathbf{q}})$ and
$\tilde{\vartheta}_{\mathbf{q}}^{\prime} 
= 
\arcsin(2 k_z \tilde{Q}_{\mathbf{q},\text{I}}/\nu_{\mathbf{q}}^{\prime})$ with $\tilde{\mathbf{Q}}_{\mathbf{q}, \text{R}}
=
\text{Re}(\tilde{\mathbf{Q}}_{\mathbf{q}})$, 
$\tilde{\mathbf{Q}}_{\mathbf{q}, \text{I}}
=
\text{Im}(\tilde{\mathbf{Q}}_{\mathbf{q}})$ and
\begin{subequations}
\begin{align}
\tilde{\varkappa}_{\mathbf{q}}^2
&=
(\tilde{Q}_{\mathbf{q},\text{R}}^2 
-
\tilde{Q}_{\mathbf{q},\text{I}}^2
-
\tilde{\kappa}_z^2 + \tilde{k}_z^2)^2
+
4(\tilde{\mathbf{Q}}_{\mathbf{q},\text{R}} 
\cdot
\tilde{\mathbf{Q}}_{\mathbf{q},\text{I}}
+
\tilde{\kappa}_z \tilde{k}_z)^2,
\\
\nu_{\mathbf{q}}^2 
&=
[\tilde{Q}_{\mathbf{q},\text{R}}^2 
-
\tilde{Q}_{\mathbf{q},\text{I}}^2
-
\tilde{\kappa}_z^2 - \tilde{k}_z^2
-
2 \tilde{k}_z \text{Im}(\tilde{\xi}_{0})]^2
+
(2 k_z \tilde{Q}_{\mathbf{q},\text{R}})^2,
\\
\nu_{\mathbf{q}}^{\prime 2} 
&=
4[ \tilde{\mathbf{Q}}_{\mathbf{q},\text{R}} 
\cdot
\tilde{\mathbf{Q}}_{\mathbf{q},\text{I}}
-
\tilde{\kappa}_z \text{Im}(\tilde{\xi}_{0}
)]^2
+
(2 k_z \tilde{Q}_{\mathbf{q},\text{I}})^2,
\end{align}
\end{subequations}
the dressed reflection matrix is reexpressed as
\begin{equation}
\label{Seq:R_tilde}
\hat{\tilde{R}}_{\mathbf{q}}
=
\frac{e^{i \tilde{\varphi}_{\mathbf{q}}}}{\tilde{\varkappa}_{\mathbf{q}}}
\left(
\nu_{\mathbf{q}}
e^{i \tilde{\vartheta} _{\mathbf{q}}
\hat{\boldsymbol{\sigma}}\cdot
\tilde{\mathbf{n}}_{\mathbf{q},\text{R}}}
+
i \nu^{\prime}_{\mathbf{q}}
e^{i \tilde{\vartheta}^{\prime} _{\mathbf{q}}
\hat{\boldsymbol{\sigma}}\cdot 
\tilde{\mathbf{n}}_{\mathbf{q},\text{I}}}
\right),
\end{equation}
which is just Eq.~(\ref{R_tilde}). Here, we have introduced the unit vectors $\tilde{\mathbf{n}}_{\mathbf{q},\text{R}}
=
\tilde{\mathbf{Q}}_{\mathbf{q},\text{R}}/
\tilde{Q}_{\mathbf{q},\text{R}}$ and 
$\tilde{\mathbf{n}}_{\mathbf{q},\text{I}}
=
\tilde{\mathbf{Q}}_{\mathbf{q},\text{I}}/
\tilde{Q}_{\mathbf{q},\text{I}}$, corresponding to the real and imaginary parts of the vector $\tilde{\mathbf{Q}}_{\mathbf{q}}$, respectively. As a useful consistency check, it is worth noting that in the limit of vanishing disorder, $\nu_{\mathbf{q}}, \tilde{\varkappa}_{\mathbf{q}} \rightarrow \varkappa_{\mathbf{q}}$, while $\nu_{\mathbf{q}}^{\prime} \rightarrow 0$, hence $\hat{\tilde{R}}_{\mathbf{q}} \rightarrow \hat{R}_{\mathbf{q}}$ and Eq.~(\ref{Seq:R_hat2}) is reobtained.

Before moving on, we present a brief derivation of the disorder averaged irreducible self-energy, which  is useful when doing explicit calculations of physical quantities. Inserting Eq.~(\ref{gR_hat_nm}) into Eq.~(\ref{Sigma_tilde}) and using the previously introduced change of variables, we find
\begin{equation}
\hat{\tilde{\xi}} \left(\epsilon\right)
=
\frac{1}{8 \pi^2 i} \eta_{\gamma} \int d^2\mathbf{q}\, \frac{1}{k_z} \left(1 + \hat{R}_{\mathbf{q}} \right),
\end{equation}
with $\eta_{\gamma}$ the dimensionless strength of the interfacial impurity interaction. Keeping terms up to first order in the Rashba SOC and exchange constants in the reflection matrix $\hat{R}_{\mathbf{q}}$, given by Eq.~(\ref{Seq:R_hat1}), and integrating up to the cutoff $q_{\text{max}}=k$, we obtain the result
\begin{subequations}
\label{xi_mu}
\begin{gather}
\tilde{\xi}_0
\simeq
-\frac{1}{6} \eta_{\gamma} k
\left[ \frac{1- \left(1 - \Delta^2\right)^{\frac{3}{2}}}{\Delta} + i \Delta^2 \right],
\\
\tilde{\bs{\xi}}
\simeq
-\frac{1}{8} \eta_{\gamma} \eta_{ex} k
\left[ 2 \Delta^2 \left( 1- \Delta^2 \right)
+ i \left( \sin^{-1}\left(\Delta\right) - \Delta \left(1- 2 \Delta^2 \right) \sqrt{1- \Delta^2} \right)\right] \mathbf{m},
\end{gather}
\end{subequations}
where $\eta_{ex} \equiv \xi_{ex}/k$ is the dimensionless exchange constant and $\Delta= k/k_b= \sqrt{|\epsilon|/V_b}$ is a dimensionless quantity that measures the ratio of the electron's wavelength $k=\sqrt{2m |\epsilon| /\hbar^2}$ to the wavelength associated with the potential barrier, $k_b=\sqrt{2m V_b/\hbar^2}$. Note that at the Fermi level, we have $\eta_{ex} = \xi_{ex}/k_F$ and $\Delta=k_F/k_b$.

\section{Quadratic Response Theory}
\label{appendixB}

As magnetotransport effects that respond quadratically to an applied electric field, unidirectional magnetoresistances (UMRs) and nonlinear Hall effects cannot be captured within the formalism of linear response theory. In the language of diagrammatic response theory, this implies that, instead of the well known two-photon bubble diagrams that are widely used to calculate linear responses \cite{mahan2000many, bruus2004many}, a formal quantum calculation of UMRs and nonlinear Hall effects mandates the use of response diagrams that have \textit{three} external photon legs, namely triangle diagrams, three-photon bubble diagrams and three-photon-vertex diagrams \cite{parker2019diagrammatic,du2021quantum, rostami2021gauge}.

In this section, we present the quadratic conductivity tensors--both with and without interfacial disorder--in mixed real and momentum space, which is required for the study of quantum interference of electrons at the interface of a bilayer system. Then, through a physically relevant approximation, we derive an analytical expression for the conductivity tensor and show that it may be expressed entirely in terms of the interference velocity--even in the presence of bulk and interfacial disorder. This confirms the role of quantum interference in generating the longitudinal and transverse QUMRs and suggests the robustness of these nonlinear magnetoresistances against disorder effects.

\subsection{Conductivities without Interfacial Disorder}

\begin{figure}[tph]
    \includegraphics[width=.6\linewidth]{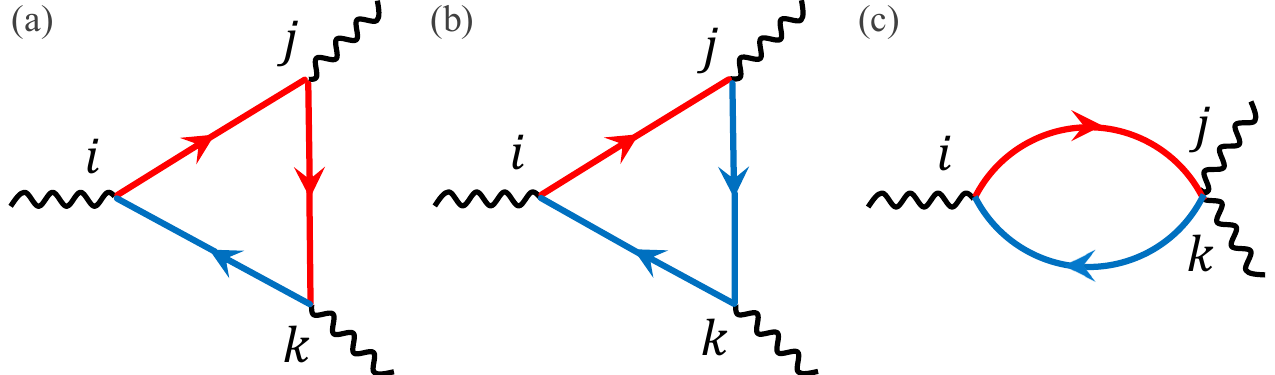}
    \caption{Diagrammatic structure of the (bare) quadratic responses in the absence of interfacial disorder. Together, figures (a)-(c), along with their $j \leftrightarrow k$ counterparts, comprise the undressed conductivity tensor $\sigma_{ijk}$. Here, red (blue) arrowed lines represent retarded (advanced) propagators, while black wavy lines are external photon legs.}
    \label{fig_qumr_figS2}%
\end{figure}

Let us first consider the simpler case where interfacial disorder is absent. In general, both the Fermi sea and the Fermi surface will contribute to the nonlinear transport. However, in the weak disorder limit, we may neglect the Fermi sea contribution and focus on the Fermi surface contribution \cite{mahan2000many}. For a Hamiltonian such as that given by Eqs.~(\ref{Seq:H_hat}), which is at most quadratic in momentum, as shown in Fig.~\ref{fig_qumr_figS2}, the local quadratic conductivity thus consists of terms arising from triangle diagrams, $\sigma_{ijk}^{(a)} (\mathbf{r})$, as well as from three-photon bubble diagrams, $ \sigma_{ijk}^{(b)} (\mathbf{r})$, so that the total conductivity is
\begin{equation}
\sigma_{ijk} (\mathbf{r})
=
\sigma_{ijk}^{(a)} (\mathbf{r}) + \sigma_{ijk}^{(b)} (\mathbf{r}),
\end{equation}

\noindent where the real-space representations of the individual conductivities read \cite{du2021quantum}
\begin{subequations}
\label{Seq:sigma_ijk_r}
\begin{align}
\begin{split}
\sigma_{ijk}^{(a)}(\mathbf{r})
=&
-\frac{e^3 \hbar^2}{\pi} \int_{-\infty}^{\infty} d\epsilon \,\pd_{\epsilon} f
\int d\mathbf{r}_1 \cdots \int d\mathbf{r}_5
\\
&\times \text{Im} \left\{\text{Tr} \left[
\braket{\mathbf{r}|\hat{v}_i |\mathbf{r}_1}
\pd_{\epsilon} \hat{G}^R\left(\mathbf{r}_1, \mathbf{r}_2; \epsilon \right) 
\braket{\mathbf{r}_2| \hat{v}_j |\mathbf{r}_3}
\hat{G}^R\left(\mathbf{r}_3, \mathbf{r}_4; \epsilon \right)
\braket{\mathbf{r}_4| \hat{v}_k |\mathbf{r}_5}
\hat{G}^A\left(\mathbf{r}_5, \mathbf{r}; \epsilon \right)
\right]\right\}
\\
&+
\left(j \leftrightarrow k \right),
\end{split}
\\
\sigma_{ijk}^{(b)}(\mathbf{r})
=
&-\frac{e^3 \hbar^2}{2 \pi} 
\int_{-\infty}^{\infty} d\epsilon \,\pd_{\epsilon} f
\int d\mathbf{r}_1 \cdots \int d\mathbf{r}_3
\\
&\times
\text{Im} \left\{\text{Tr} 
\left[
\braket{\mathbf{r}|\hat{v}_i |\mathbf{r}_1}
\pd_{\epsilon}
\hat{G}^R\left(\mathbf{r}_1, \mathbf{r}_2; \epsilon \right)
\braket{\mathbf{r}_2|\hat{v}_{jk} |\mathbf{r}_3}
\hat{G}^A\left(\mathbf{r}_3, \mathbf{r}; \epsilon \right)
\right]\right\}
+
\left(j \leftrightarrow k \right),
\end{align}
\end{subequations}
with $\pd_{\epsilon} \equiv \pd/\pd \epsilon$ and the advanced propagator related to the retarded one as
\begin{equation}
\hat{G}^A \left(\mathbf{r}^{\prime}, \mathbf{r}; \epsilon \right)
=
\left[\hat{G}^R \left(\mathbf{r}, \mathbf{r}^{\prime}; \epsilon \right)\right]^{\dagger}.
\end{equation}
In Eqs.~(\ref{Seq:sigma_ijk_r}), the replacement $j \leftrightarrow k$ ensures the intrinsic permutation symmetry of the quadratic response function. And the first- and second-order velocity operators are given by
\begin{subequations}
\begin{align}
\hat{v}^i
&=
\frac{1}{\hbar} \partial_{\mathbf{k}}^i \hat{H}_0,
\\
\hat{v}^{ij}
&=
\frac{1}{\hbar^2} \partial_{\mathbf{k}}^i \partial_{\mathbf{k}}^j \hat{H}_0,
\end{align}
\end{subequations}
with $\partial_{\mathbf{k}}^i \equiv \partial/\partial k_i$. For in-plane momenta, this leads to the real-space representations
\begin{subequations}
\begin{align}
\braket{\mathbf{r}|\hat{v}^i |\mathbf{r}_1}
&=
\frac{\hbar}{m}
\sum_{\mathbf{q}} q^i e^{i \mathbf{q} \cdot \left(\bs{\rho} - \bs{\rho}_1 \right)} \delta \left(z-z_1 \right),
\\
\braket{\mathbf{r}|\hat{v}^{ij} |\mathbf{r}_1}
&=
\frac{1}{m}
\delta^{ij} \delta \left(\mathbf{r} - \mathbf{r}_1 \right).
\end{align}
\end{subequations}
Applying Eq.~(\ref{gR_hat_z}), the conductivities now read
\begin{subequations}
\label{qumr_sigma_ijk}
\begin{align}
\begin{split}
\sigma_{ijk}^{(a)}(z)
=&
-\frac{2 e^3 \hbar^5}{\pi m^{3}}
\int \frac{d^2\mathbf{q}}{\left( 2\pi \right)^2} \int_{-\infty}^{\infty} d\epsilon \,\pd_{\epsilon} f
\int_0^{\infty} dz^{\prime} \int_0^{\infty} dz^{\dprime} \,q_i q_j q_k
\\
&\times
\text{Im} \left\{\text{Tr} \left[
\pd_{\epsilon} \hat{g}^R_{\mathbf{q}}\left(z, z^{\prime}; \epsilon \right)
\hat{g}^R_{\mathbf{q}}\left(z^{\prime}, z^{\dprime}; \epsilon \right)
\hat{g}^A_{\mathbf{q}}\left(z^{\dprime}, z ;\epsilon \right)\right]\right\},
\end{split}
\\
\sigma_{ijk}^{(b)}(z)
=
&-\frac{e^3 \hbar^3}{\pi m^2} \int \frac{d^2\mathbf{q}}{\left( 2\pi \right)^2} \int_{-\infty}^{\infty} d\epsilon \,\pd_{\epsilon} f
\int_0^{\infty} dz^{\prime} \,q_i \delta_{jk}
\,\text{Im} \left\{\text{Tr} \left[
\pd_\epsilon \hat{g}^R_{\mathbf{q}} \left(z, z^{\prime}; \epsilon \right)
\hat{g}^A_{\mathbf{q}} \left(z^{\prime}, z; \epsilon \right)\right]\right\}.
\end{align}
\end{subequations}
Inserting Eq.~(\ref{gR_hat_nm}) into Eqs.~(\ref{qumr_sigma_ijk}), the analytical form of the nonlinear conductivities in terms of the reflection matrix may be obtained, which turns out to be rather cumbersome. In the limit $l_n\gg \lambda_F$, with $\lambda_F=2\pi/k_F$ the Fermi wavelength in the NM layer, the highest-order terms in the mean free path will dominate the transport and the conductivities may be approximated as
\begin{subequations}
\label{Seq:sigma_ijk_ab_ballistic}
\begin{align}
\sigma_{ijk}^{(a)}(z)
=
&-\frac{2 e^3 m}{\pi \hbar^3} l_n^2 \int \frac{d^2\mathbf{q}}{\left( 2\pi \right)^2} \int_{-\infty}^{\infty} d\epsilon \,\pd_{\epsilon} f \,q_i q_j q_k
\left(
\frac{2k^2 + k_z^2}{k^4 k_z^3}
-
\frac{\hbar^2}{m}
\frac{1}{k^2 k_z} \pd_{\epsilon}
\right)
\text{Re} \left(e^{2i k_z z} \text{Tr} \hat{R}_{\mathbf{q}}\right),
\\
\sigma_{ijk}^{(b)}(z)
=
&-\frac{2 e^3 m}{\pi \hbar^3} l_n^2 
\int \frac{d^2\mathbf{q}}{\left( 2\pi \right)^2} 
\int_{-\infty}^{\infty} d\epsilon \,\pd_{\epsilon} f \,
q_i \delta_{jk} \frac{1}{k^2 k_z}
\text{Re} \left(e^{2i k_z z} \text{Tr} \hat{R}_{\mathbf{q}}\right).
\end{align}
\end{subequations}
Using the approximation $\pd_{\epsilon}f \simeq - \delta(\epsilon- \epsilon_F)$ and the expression for the interference velocity at the Fermi level
\begin{equation}
\Braket{\hat{\bs{v}} (\mathbf{q},z)}_{I-R}
=
\frac{2 \hbar \mathbf{q}}{m}
\text{Re} \left(e^{2i k_{z,F} z} \text{Tr} \hat{R}_{\mathbf{q}}\right),
\end{equation}
with $k_{z,F}=\sqrt{k_{F}^{2}-q^{2}}$, Eqs.~(\ref{Seq:sigma_ijk_ab_ballistic}) are recast in the form
\begin{equation}
\label{Seq:sigma_ijk_ballistic}
\sigma_{ijk}\left( z ; \mathbf{m} \right)
=
\frac{2 e^3 m^{2}}{\pi \hbar^4 k_F^2} l_n^2
\int \frac{d^2\mathbf{q}}{\left( 2\pi \right)^2}
\frac{1}{k_{z,F}} 
\left[
q_j q_k
\left(\frac{2}{k_{z,F}^2} + \frac{1}{k_F^2} - \frac{\hbar^2}{m}\pd_{\epsilon_F}\right)
+
\delta_{jk}
\right]
\Braket{\hat{v}_{i}(\mathbf{q},z)}_{I-R}.
\end{equation}
In this form, we readily see the central role played by the interference velocity $\braket{\hat{\bs{v}}({\mathbf{q}},z)}_{I-R}$ in generating the nonlinear response. Next, we study the effect of the interfacial disorder on the nonlinear response.  

\subsection{Including Interfacial Disorder}

\begin{figure}[tph]
    \includegraphics[width=.9\linewidth]{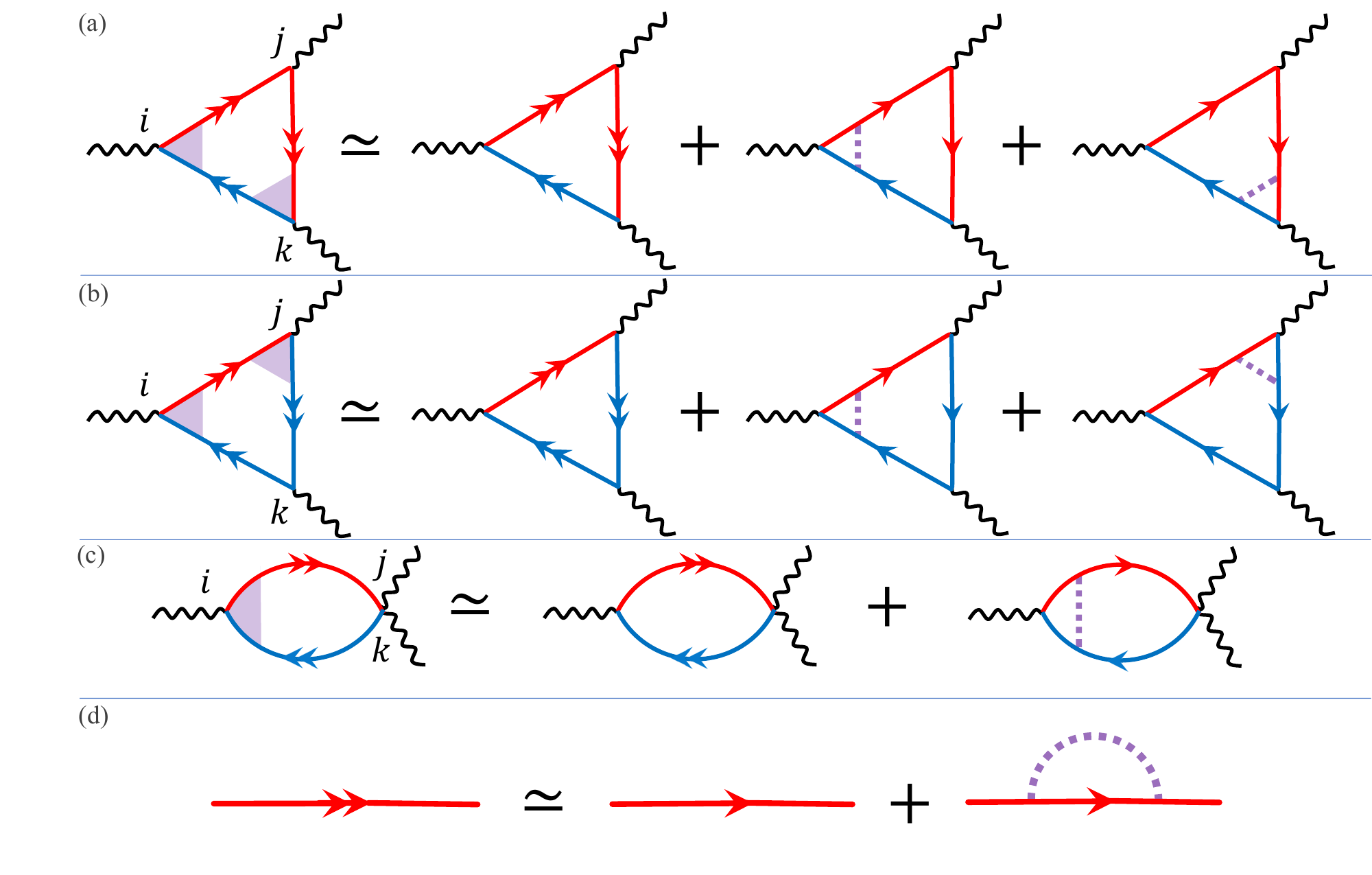}
    \caption{(a-c) Leading-order contributions to the dressed conductivity tensor, where double-arrowed lines represent dressed propagators. The purple shaded areas correspond to dressed vertices, while the leading-order disorder vertex corrections are represented by dotted purple lines. (d) The dressed propagator within the Born approximation.}
    \label{fig_qumr_figS3}%
\end{figure}

We now consider the contributions of impurities at the bilayer interface. In the presence of interfacial disorder, the propagator is no longer diagonal in momentum space and the conductivities generalize to
\begin{subequations}
\label{sigma_ijk_disorder}
\begin{align}
\begin{split}
\tilde{\sigma}_{ijk}^{(a)}(z)
=&
-\frac{e^3 \hbar^5}{\pi m^{3}} 
\left[\int \frac{d^2\mathbf{q}}{\left( 2\pi \right)^2}
\cdots
\int \frac{d^2\mathbf{q}^{\tprime}}{\left( 2\pi \right)^2}
\right]
\int_{-\infty}^{\infty} d\epsilon \,\pd_{\epsilon} f
\int_0^{\infty} dz^{\prime} \int_0^{\infty} dz^{\dprime} \,q_i q^{\prime}_j q^{\dprime}_k
\\
&\times \text{Im} \left\{\text{Tr} 
\left[
\pd_{\epsilon} 
\hat{\tilde{g}}^R_{\mathbf{q} \mathbf{q}^{\prime}}\left(z, z^{\prime}; \epsilon \right) 
\hat{\tilde{g}}^R_{\mathbf{q}^{\prime} \mathbf{q}^{\dprime}} \left(z^{\prime}, z^{\dprime}; \epsilon \right) 
\hat{\tilde{g}}^A_{ \mathbf{q}^{\dprime} \mathbf{q}^{\tprime}} \left(z^{\dprime}, z; \epsilon \right)\right]\right\}
+
\left( j \leftrightarrow k\right),
\end{split}
\\
\begin{split}
\tilde{\sigma}_{ijk}^{(b)}(z)
&=
-\frac{e^3 \hbar^3}{2 \pi m^2} 
\left[\int \frac{d^2\mathbf{q}}{\left( 2\pi \right)^2}
\cdots
\int \frac{d^2\mathbf{q}^{\dprime}}{\left( 2\pi \right)^2}
\right]
\int_{-\infty}^{\infty} d\epsilon \,\pd_{\epsilon} f
\int_0^{\infty} dz^{\prime} \,q_i \delta_{jk}
\\
&\times
\text{Im} \left\{\text{Tr} 
\left[
\pd_{\epsilon} 
\hat{\tilde{g}}^R_{\mathbf{q} \mathbf{q}^{\prime}} \left(z, z^{\prime}; \epsilon \right) 
\hat{\tilde{g}}^A_{\mathbf{q}^{\prime} \mathbf{q}^{\dprime}} \left(z^{\prime}, z; \epsilon \right) \right]\right\}
+
\left( j \leftrightarrow k\right).
\end{split}
\end{align}
\end{subequations}
In order to calculate the specular and diffuse contributions of Eqs.~(\ref{sigma_ijk_disorder}), one needs to calculate configurational averages of products of two and three propagators, which we symbolically express in the condensed notation $\braket{\hat{\tilde{G}}^2}$ and $\braket{\hat{\tilde{G}}^3}$. Let us also reexpress Eq.~(\ref{GR_imp}) in the condensed form
\begin{equation}
\hat{\tilde{G}}
=
\hat{G} + \hat{G} \hat{V}^{\text{imp}} \hat{\tilde{G}}.
\end{equation}
We then have
\begin{subequations}
\label{condensed}
\begin{align}
\Braket{\hat{\tilde{G}}^2}
&=
\Braket{\left(\hat{G} + \hat{G} \hat{V}^{\text{imp}} \hat{\tilde{G}}\right)^2}
\simeq
\Braket{\hat{\tilde{G}}}^2
+
\Braket{\left(\hat{G} \hat{V}^{\text{imp}} \hat{G} \right)^2},
\\
\begin{split}
\Braket{\hat{\tilde{G}}^3}
&=
\Braket{\left(\hat{G} + \hat{G} \hat{V}^{\text{imp}} \hat{\tilde{G}}\right)^3}
\simeq
\Braket{\hat{\tilde{G}}}^3
+
\Braket{\left(\hat{G} \hat{V}^{\text{imp}} \hat{G}\right)^2} \hat{G}
+
\Braket{\left(\hat{G} \hat{V}^{\text{imp}} \hat{G} \right) \hat{G} \left(\hat{G} \hat{V}^{\text{imp}} \hat{G} \right)}
\\
&\hspace{.28\linewidth}
+
\hat{G} \Braket{\left(\hat{G} \hat{V}^{\text{imp}} \hat{G} \right)^2},
\end{split}
\end{align}
\end{subequations}
where we have retained only the leading order contributions to the vertex corrections. In Eqs.~(\ref{condensed}), terms containing separately averaged propagators, $\braket{\hat{\tilde{G}}}^2$ and $\braket{\hat{\tilde{G}}}^3$, correspond to momentum-preserving scatterings. Hence, they constitute the specular contribution to the conductivity tensor. The vertex corrections, on the other hand, contain momentum-mixing terms and contribute to the diffuse scattering. A summary of these contributions are presented in Fig.~\ref{fig_qumr_figS3}, in which we diagrammatically highlight the approximations to the renormalized propagators and velocity vertex functions. 

As we neglect mixing terms--which appear at higher order in the interfacial disorder parameter--the dressed conductivity tensor $\tilde{\sigma}_{ijk}$ may be expressed as 
\begin{equation}
\tilde{\sigma}_{ijk}
=
\tilde{\sigma}_{ijk}^{\text{spec}}
+
\tilde{\sigma}_{ijk}^{\text{diff}},
\end{equation}
where, here and henceforth, we omit the ensemble-averaging brackets $\braket{\cdots}$ for simplicity.
Below, we present explicit calculations of the specular and diffuse contributions to the nonlinear response tensor.

\subsubsection{Specular contribution}

In the presence of specular scattering alone, as shown in Eqs.~(\ref{condensed}), the dressed conductivity is obtained in a straightforward manner by simply replacing the bare propagators in Eqs.~(\ref{sigma_ijk}) with their dressed counterparts as
\begin{subequations}
\label{Seq:sigma_ijk_spec}
\begin{align}
\begin{split}
\tilde{\sigma}_{ijk}^{(a),\text{spec}}(z)
&=
-\frac{2 e^3 \hbar^5}{\pi m^{3}} \int \frac{d^2\mathbf{q}}{\left( 2\pi \right)^2} \int_{-\infty}^{\infty} d\epsilon \,\pd_{\epsilon} f
\int_0^{\infty} dz^{\prime} \int_0^{\infty} dz^{\dprime} \,q_i q_j q_k
\\
&\times
\text{Im} \left\{\text{Tr} \left[
\pd_{\epsilon} \hat{\tilde{g}}^R_{\mathbf{q}} \left(z, z^{\dprime}; \epsilon \right)
\hat{\tilde{g}}^R_{\mathbf{q}} \left(z^{\dprime}, z^{\prime}; \epsilon \right) 
\hat{\tilde{g}}^A_{\mathbf{q}} \left(z^{\prime}, z; \epsilon \right)\right]\right\},
\end{split}
\\
\tilde{\sigma}_{ijk}^{(b),\text{spec}}(z)
&=
-\frac{e^3 \hbar^3}{\pi m^2}
\int \frac{d^2\mathbf{q}}{\left( 2\pi \right)^2}
\int_{-\infty}^{\infty} d\epsilon \,\pd_{\epsilon} f
\int_0^{\infty} dz^{\prime} \,q_i \delta_{jk}
\,\text{Im} \left\{\text{Tr} \left[
\pd_\epsilon \hat{\tilde{g}}^R_{\mathbf{q}} \left(z, z^{\prime}; \epsilon \right)
\hat{\tilde{g}}^A_{\mathbf{q}} \left(z^{\prime}, z; \epsilon \right)\right]\right\}.
\end{align}
\end{subequations}
Note that with the approximation $\pd_{\epsilon}f \simeq - \delta(\epsilon- \epsilon_F)$, Eqs.~(\ref{Seq:sigma_ijk_spec}) reduce to Eqs.~(\ref{eq:sigma_ijk}).

Similarly, the analytical approximation is obtained by the simple replacement of the reflection matrix $\hat{R}_{\mathbf{q}}$ in Eqs.~(\ref{Seq:sigma_ijk_ab_ballistic}) with its disordered counterpart $\hat{\tilde{R}}_{\mathbf{q}}$, given by Eq.~(\ref{R_hat_tilde})--or equivalently, Eq.~(\ref{Seq:R_tilde}). We thus arrive at the generalized result given by Eq.~(\ref{sigma_ijk_ballistic})
\begin{equation}
\label{sigma_ijk_ballistic_SM}
\tilde{\sigma}_{ijk}^{\text{spec}} \left( z ; \mathbf{m} \right)
=
\frac{2 e^3 m^{2}}{\pi \hbar^4 k_F^2} l_n^2
\int \frac{d^2\mathbf{q}}{\left( 2\pi \right)^2}
\frac{1}{k_{z,F}} 
\left[
q_j q_k
\left(\frac{2}{k_{z,F}^2} + \frac{1}{k_F^2} - \frac{\hbar^2}{m}\pd_{\epsilon_F}\right)
+
\delta_{jk}
\right]
\Braket{\hat{\tilde{v}}_{i}(\mathbf{q},z)}_{I-R},
\end{equation}
where the dressed interference velocity is given by
\begin{equation}
\Braket{\hat{\tilde{\bs{v}}} (\mathbf{q},z)}_{I-R}
=
\frac{2 \hbar \mathbf{q}}{m}
\text{Re} \left(e^{2i k_{z,F} z} \text{Tr} \hat{\tilde{R}}_{\mathbf{q}}\right).
\end{equation}

\subsubsection{Diffuse correction}

We now present the diffuse corrections to the conductivity tensor, which are all the diagrams in Fig.~\ref{fig_qumr_figS3} (along with their $j \leftrightarrow k$ counterparts) that include vertex corrections. Following Eqs.~(\ref{condensed}), we conclude that the diffuse corrections corresponding to dressed triangle diagrams, $\tilde{\sigma}_{ijk}^{(a),\text{diff}}$, and dressed three-photon bubble diagrams,  $\tilde{\sigma}_{ijk}^{(b),\text{diff}}$, can be expressed as
\begin{subequations}
\label{Seq:sigma_ijk_diff}
\begin{align}
\begin{split}
\tilde{\sigma}_{ijk}^{(a),\text{diff}}(z)
=&
-\frac{e^3 \hbar^5}{\pi m^{3}}
\int_{-\infty}^{\infty} d\epsilon \,\pd_{\epsilon} f\,
\text{Im}
\left\{ \text{Tr}
\left[ \hat{\mathcal{P}}_{ijk}\left( z;\epsilon\right) \right] \right\}
+
\left( j \leftrightarrow k \right),
\end{split}
\\
\tilde{\sigma}_{ijk}^{(b),\text{diff}}(z)
=
&-\frac{e^3 \hbar^3}{2 \pi m^2} 
\int_{-\infty}^{\infty} d\epsilon \,\pd_{\epsilon} f\,
\text{Im}
\left\{ \text{Tr}
\left[ \hat{\mathcal{S}}_{ijk}\left( z;\epsilon\right) \right] \right\}
+
\left( j \leftrightarrow k \right),
\end{align}
\end{subequations}
where
\begin{equation}
\hat{\mathcal{P}}_{ijk}\left( z;\epsilon\right)
=
\hat{\mathcal{P}}_{ijk}^{(1)}\left( z;\epsilon\right)
+
\hat{\mathcal{P}}_{ijk}^{(2)}\left( z;\epsilon\right)
+
\hat{\mathcal{P}}_{ijk}^{(3)}\left( z;\epsilon\right),
\end{equation}
with
\begin{subequations}
\label{F_tilde_I}
\begin{align}
\begin{split}
\hat{\mathcal{P}}_{ijk}^{(1)}\left( z;\epsilon\right)
=
&\left(\frac{\hbar^2}{2m}\right)^2 \eta_{\gamma}
\int \frac{d^2\mathbf{q}}{\left( 2\pi \right)^2}
\int \frac{d^2\mathbf{q}^{\prime}}{\left( 2\pi \right)^2}
\int_0^{\infty} dz^{\prime}
\int_0^{\infty} dz^{\dprime}\, q^{\prime}_i q_j q_k
\\
&\times 
\pd_{\epsilon}\left[
\hat{g}^R_{\mathbf{q}^{\prime}} \left(z,0;\epsilon\right) 
\hat{g}^R \left(0,z^{\prime},\mathbf{q};\epsilon\right)
\right]
\hat{g}^R_{\mathbf{q}} \left(z^{\prime},z^{\dprime};\epsilon\right)
\hat{g}^A_{\mathbf{q}} \left(z^{\dprime},0;\epsilon \right)
\hat{g}^A_{\mathbf{q}^{\prime}} \left(0,z;\epsilon\right),
\end{split}
\\
\begin{split}
\hat{\mathcal{P}}_{ijk}^{(2)}\left( z;\epsilon\right)
=
&\left(\frac{\hbar^2}{2m}\right)^2 \eta_{\gamma}
\int \frac{d^2\mathbf{q}}{\left( 2\pi \right)^2}
\int \frac{d^2\mathbf{q}^{\prime}}{\left( 2\pi \right)^2}
\int_0^{\infty} dz^{\prime}
\int_0^{\infty} dz^{\dprime}\, q_i q^{\prime}_j q_k
\\
&\times 
\pd_{\epsilon}\left[
\hat{g}^R_{\mathbf{q}} \left(z,0;\epsilon\right) 
\hat{g}^R_{\mathbf{q}^{\prime}} \left(0,z^{\prime};\epsilon\right)
\right]
\hat{g}^R_{\mathbf{q}^{\prime}}\left(z^{\prime},0;\epsilon\right)
\hat{g}^R_{\mathbf{q}} \left(0,z^{\dprime};\epsilon\right)
\hat{g}^A_{\mathbf{q}} \left(z^{\dprime},z;\epsilon\right),
\end{split}
\\
\begin{split}
\hat{\mathcal{P}}_{ijk}^{(3)}\left( z;\epsilon\right)
=
&\left(\frac{\hbar^2}{2m}\right)^2 \eta_{\gamma}
\int \frac{d^2\mathbf{q}}{\left( 2\pi \right)^2}
\int \frac{d^2\mathbf{q}^{\prime}}{\left( 2\pi \right)^2}
\int_0^{\infty} dz^{\prime}
\int_0^{\infty} dz^{\dprime}\, q_i q_j q^{\prime}_k
\\
&\times 
\pd_{\epsilon}\left[
\hat{g}^R_{\mathbf{q}} \left(z,z^{\prime};\epsilon\right) 
\right]
\hat{g}^R_{\mathbf{q}} \left(z^{\prime},0;\epsilon\right)
\hat{g}^R_{\mathbf{q}^{\prime}} \left(0, z^{\dprime};\epsilon\right)
\hat{g}^A_{\mathbf{q}^{\prime}} \left(z^{\dprime},0;\epsilon \right)
\hat{g}^A_{\mathbf{q}} \left(0,z;\epsilon\right),
\end{split}
\end{align}
\end{subequations}
and
\begin{equation}
\label{F_tilde_II}
\begin{split}
\hat{\mathcal{S}}_{ijk}\left( z;\epsilon\right)
=
&\left(\frac{\hbar^2}{2m}\right)^2 \eta_{\gamma}
\int \frac{d^2\mathbf{q}}{\left( 2\pi \right)^2}
\int \frac{d^2\mathbf{q}^{\prime}}{\left( 2\pi \right)^2}
\int_0^{\infty} dz^{\prime}\, q_i \delta_{jk}
\\
&\times
\pd_{\epsilon}\left[
\hat{g}^R_{\mathbf{q}} \left(z,0;\epsilon\right) 
\hat{g}^R_{\mathbf{q}^{\prime}} \left(0,z^{\prime};\epsilon\right)
\right]
\hat{g}^A_{\mathbf{q}^{\prime}} \left(z^{\prime},0;\epsilon\right)
\hat{g}^A_{\mathbf{q}} \left(0,z;\epsilon\right).
\end{split}
\end{equation}
While an analytical approximation to Eqs.~(\ref{Seq:sigma_ijk_diff}) is rather intractable, a numerical calculation reveals that the diffuse correction to the UMR coefficients tends to counteract the specular corrections arising from the interfacial disorder. However, the diffuse contributions turn out to be at least two orders of magnitude smaller than the overall UMR coefficient strengths. Thus, they will not affect the main physical results and may safely be neglected in the present chapter.

\section{Conclusion}

In conclusion, we have explored the quantum transport of electrons in the nonlinear response regime, and predicted a QUMR effect in NM$|$FI bilayers, which originates from the interference between electron waves approaching and reflecting off a magnetic interface with Rashba SOC. 

Several appealing properties of the QUMR have been identified, enabling both electric and magnetic control of the nonlinear magnetotransport effect. Significantly, the emergence of the QUMR effect entails variations in both longitudinal and transverse resistances whenever the direction of the electric field is reversed. Moreover, the QUMR is also sensitive to the orientation of magnetization of the FI layer, as the phase difference between the scattering waves can be tuned by the magnetization through its coupling with the spin angular momenta of conduction electrons in the NM layer.

As a final remark, we have restricted ourselves to zero temperature in the initial study of the QUMR effect. To examine finite temperature effects, one needs to consider the influence of fluctuating
scatterers on the phase coherence of the quantum magnetotransport in the nonlinear response regime, which warrants future theoretical and experimental research. And from an applications perspective, we envision this work will open new avenues for developing future quantum spintronic devices.

\chapter{Proximity-Induced Nonlinear Magnetoresistances on Topological Insulators}

\section{Introduction}

Topological-insulator (TI)-based magnetic heterostructures are appealing systems for exploring the interplay between magnetism and band topology. These hybrid systems are characterized by the coexistence of strong spin-orbit coupling (SOC), sizable magnetic exchange interaction, and Dirac surface states with spin-momentum locking. Moreover, it has been demonstrated that the Fermi level of the TI layer in these systems can be finely tuned with respect to the Dirac point~\cite{Okada16PRB_Fermi-level_TI, Kondou16NP_Fermi-level-sc_TI, sun2019large, Su21ACS_Fermi-level_TI}. These properties are remarkable in their own rights, and a blend of them makes these systems even more intriguing. A multitude of linear-response transport phenomena have attracted considerable attention, including quantum anomalous Hall~\cite{qi2006topological, yu2010quantized, chang2013experimental, checkelsky2014trajectory, liu2016quantum, chang2023colloquium}, topological Hall~\cite{wu2020ferrimagnetic, li2021topological, zhang2021giant}, spin-transfer torque~\cite{Mellnik2014,Han17PRL_SOT-TI,li2019magnetization}, and various novel magnetoresistance effects~\cite{Moghaddam20PRL_SOT-AMR-proximity-TI,chiba2017magnetic,sklenar2021proximity}, which may potentially lead to applications in many areas, ranging from classical information storage and processing~\cite{Wu21NC_MRAM-TI} to quantum computation~\cite{Fu08PRL_proximity-SC_TI_majorana,Lian18PNAS_Topo-quant-comput_MZM}. 

Going beyond linear responses, TI-based magnetic heterostructures also allow magnetotransport that violates Onsager’s reciprocity in principle, owing to the lack of both time-reversal and inversion symmetries. Corrections to linear magnetoconductivities have been observed in a few TI-based magnetic heterostructures~\cite{yasuda2016large, yasuda2017current, lv2018unidirectional, duy2019giant, wang2022observation, lv2022large}–ensuing the discovery of unidirectional magnetoresistance (UMR) effects in metallic and semiconducting magnetic bilayers~\cite{avci2015unidirectional, olejnik2015electrical}. Such nonlinear magnetoconductivities are odd under the reversal of either the direction of the applied electric field $\mathbf{E}$ or that of the magnetization (whose direction will be denoted by a unit vector $\mathbf{m}$ hereafter); i.e., 
$\sigma^{(2)}(-\mathbf{m},\mathbf{E})
=
\sigma^{(2)}(\mathbf{m},-\mathbf{E})
=
-\sigma^{(2)}(\mathbf{m},\mathbf{E})$, which are distinctly different from their linear-response counterparts and hence hold fascinating prospects for adding new functionalities in future spintronic devices.

To date, studies of nonlinear transport in magnetic layered structures have mainly been focused on controlling the corresponding magnetotransport coefficients by varying the magnetization vector with an external magnetic field. And, more specifically, the reported dependencies of the nonlinear current on the magnetization direction can be cast into the simple form~\cite{olejnik2015electrical, avci2015unidirectional, zhang2016theory, guillet2021large, zelezny2021unidirectional, zhou2021sign, liu2021magnonic, nguyen2021unidirectional, hasegawa2021enhanced, mehraeen2022spin, shim2022unidirectional, lv2022large, lou2022large, ding2022unidirectional, fan2022observation, cheng2023unidirectional, wang2023controllable, zheng2023coexistence, mehraeen2023quantum}
\begin{equation}
\label{j_1^2}
\mathbf{j}_{1}^{(2)}
=
\sigma_{\parallel,1}^{(2)} \mathbf{m} \cdot (\mathbf{z} \times \mathbf{E}) \mathbf{E}
+
\sigma_{\perp,1}^{(2)} (\mathbf{m} \cdot \mathbf{E}) \mathbf{z} \times \mathbf{E},
\end{equation}
where the unit vector $\mathbf{z}$ denotes the interface normal and $\sigma_{\parallel,1}^{(2)}$ ($\sigma_{\perp,1}^{(2)}$) is a transport coefficient that characterizes the strength of the nonlinear current longitudinal (transverse) to the applied electric field, which is denoted by $\mathbf{j}_{\parallel,1}^{(2)}$ ($\mathbf{j}_{\perp,1}^{(2)}$). Here, the superscript 2 and the subscript 1 indicate that the nonlinear current is of second order in the applied electric field and first order in the magnetization. The tunability of these nonlinear transport coefficients upon the shift of the Fermi level, however, has remained unexplored. 

\begin{figure}[hpt]
\captionsetup[subfigure]{labelformat=empty}
    \sidesubfloat[]{\includegraphics[width=0.32\linewidth,trim={1.7cm 1cm 0.5cm 0.7cm}]{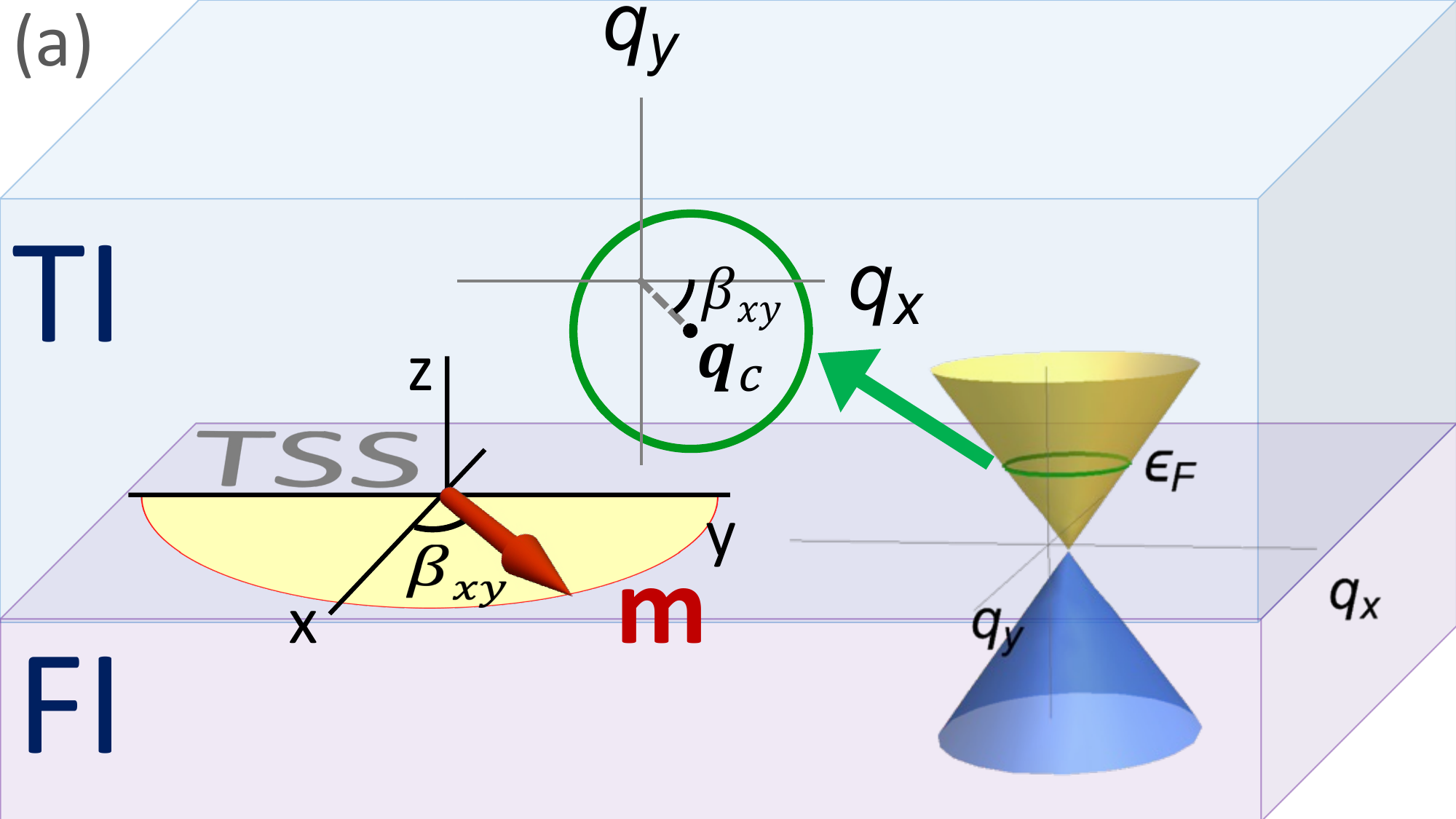}\label{fig_proximity_fig1a}}
%\quad%
    \sidesubfloat[]{\includegraphics[width=0.32\linewidth,trim={1.8cm 1cm 0.3cm 0.7cm}]{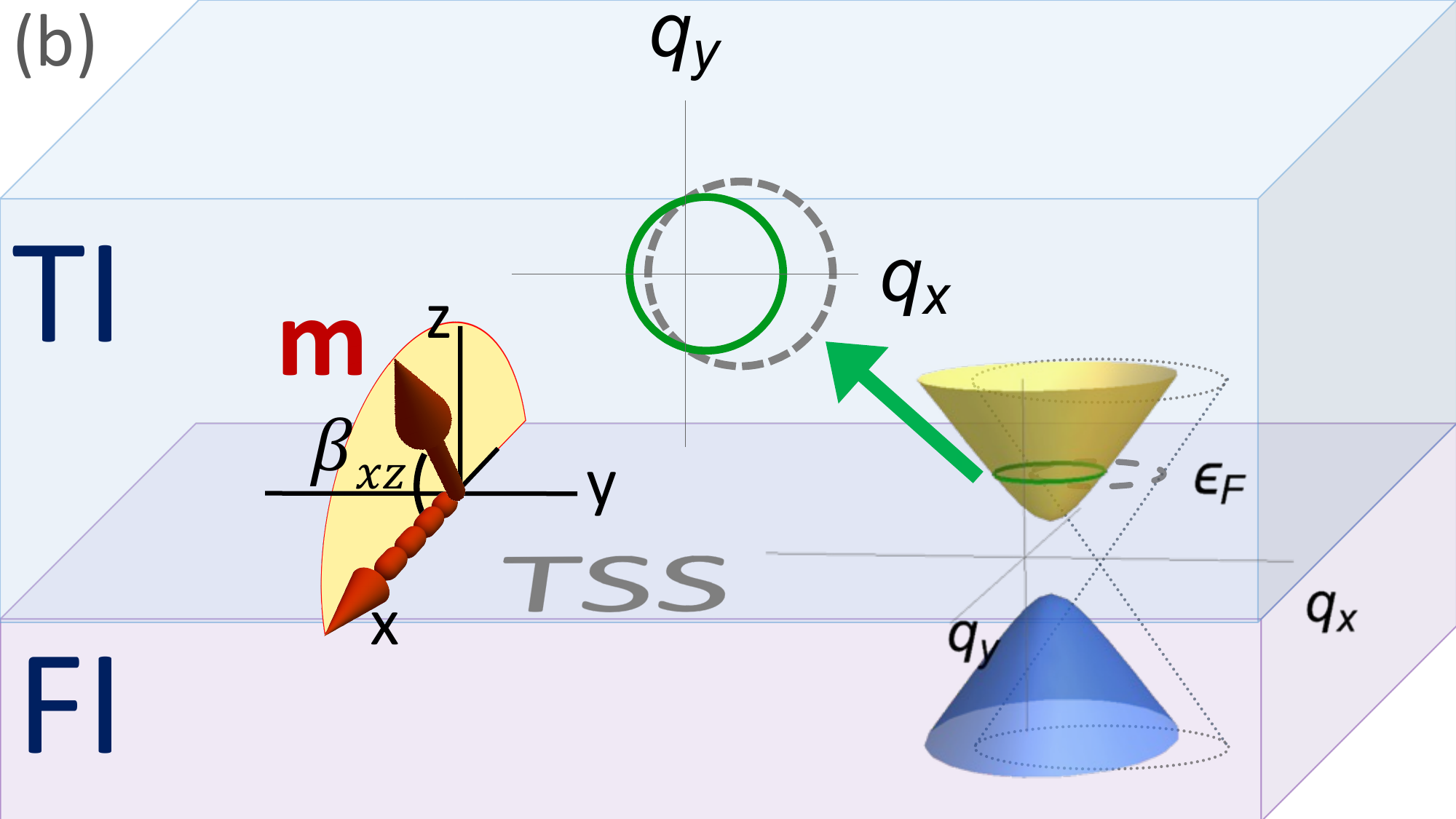}\label{fig_proximity_fig1b}}
%%%%%%%    
    \sidesubfloat[]{\includegraphics[width=0.32\linewidth,trim={1.8cm 1cm 0.5cm 0.7cm}]{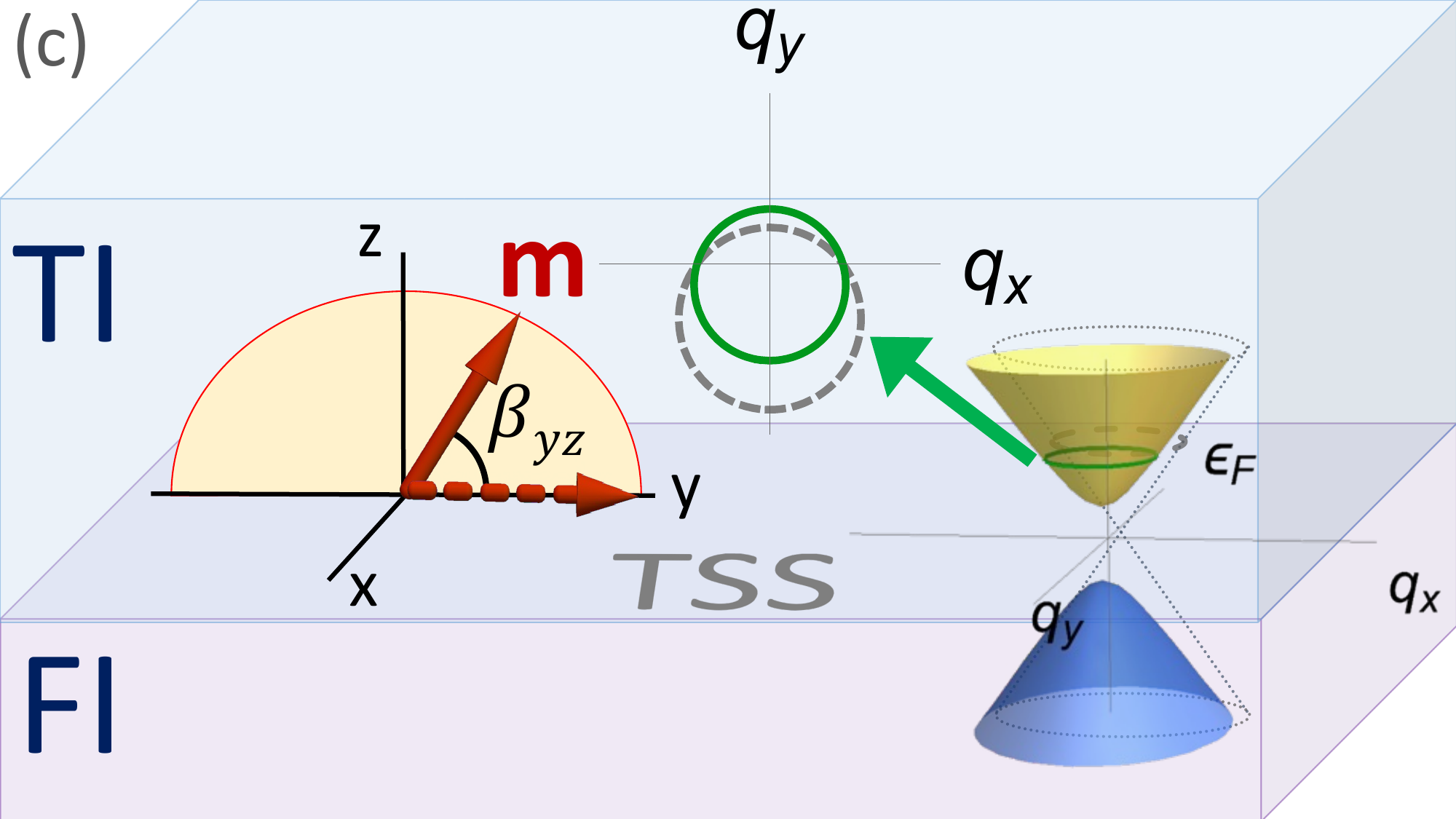}\label{fig_proximity_fig1c}}
%%%%%%%%%
    \caption{Schematics of the TI/FI bilayer system and the topological surface state (TSS). (a) In-plane rotation of the magnetization, which leads to a rigid shift of the Dirac cone and thus the Fermi contour. (b) and (c) Out-of-plane rotations of the magnetization, which gap the system. In this case, in addition to a shift, the radius of the Fermi contour is decreased as the magnetization is rotated out of the bilayer plane. Here, $\beta_{ij}$ ($i,j=x,y,z$) is the angle between the $i$ axis and the magnetization as it sweeps the $ij$ plane.}
    \label{fig_proximity_fig1}
\end{figure}

In this chapter~\cite{mehraeen2024proximity}, we theoretically investigate nonlinear magnetotransport in magnetic bilayers consisting of a TI and a ferromagnetic-insulator (FI) layer, as shown schematically in Fig.~\ref{fig_proximity_fig1}. A formal evaluation of Kubo formulas in the quadratic-response regime is performed to explore the magnetoconductivities of interest. Within this quantum approach, we find a UMR effect and a nonlinear planar Hall effect (NPHE) that are driven purely by extrinsic spin-orbit scattering at the interface. More interestingly, perhaps, unconventional dependencies of the nonlinear currents on the magnetization orientation emerge, which differ from those given by Eq.~(\ref{j_1^2}) and feature terms that are of higher order in the magnetization. Following a symmetry analysis of the nonlinear response function, the details of which are presented in Sec.~\ref{appendixA}, the general nonlinear current may be expressed as
\begin{subequations}
\label{j^2_cubic}
\begin{align}
\mathbf{j}_{\parallel}^{(2)}
&\hspace{-.05cm}
=
\hspace{-.03cm}
\mathbf{j}_{\parallel,1}^{(2)}
\left[ 1
\hspace{-.05cm}
+
\hspace{-.05cm}
\iota_{\parallel} (\mathbf{m} \cdot \mathbf{e})^2
\hspace{-.05cm}
+
\hspace{-.05cm}
\kappa_{\parallel} (\mathbf{m} \cdot \mathbf{z} \times \mathbf{e})^2
\hspace{-.05cm}
+
\hspace{-.05cm}
\lambda_{\parallel} (\mathbf{m} \cdot \mathbf{z})^2
\right],
\\
\mathbf{j}_{\perp}^{(2)}
&\hspace{-.05cm}
=
\hspace{-.03cm}
\mathbf{j}_{\perp,1}^{(2)}
\left[ 1
\hspace{-.05cm}
+
\hspace{-.05cm}
\iota_{\perp} (\mathbf{m} \cdot \mathbf{e})^2
\hspace{-.05cm}
+
\hspace{-.05cm}
\kappa_{\perp} (\mathbf{m} \cdot \mathbf{z} \times \mathbf{e})^2
\hspace{-.05cm}
+
\hspace{-.05cm}
\lambda_{\perp} (\mathbf{m} \cdot \mathbf{z})^2
\right],
\end{align}
\end{subequations}
where $\mathbf{e}$ is the unit vector along the electric field. Here, $\iota_{\parallel,\perp}$, $\kappa_{\parallel,\perp}$ and $\lambda_{\parallel,\perp}$ are dimensionless quantities that characterize the strengths of the contributions cubic in the magnetization and are, in general, functions of disorder and the exchange energy. Intriguingly, when the Fermi and exchange energies become comparable, the nonlinear magnetoresistances may be considerably amplified and the contributions of the cubic terms become relatively large, leading to strong deviations in the angular dependencies. These new features may then be used to obtain insights about the position of the Fermi level or the strength of the interfacial exchange interaction via transport measurements, as a simple alternative to optical schemes such as ARPES.

\section{Disorder scattering in TI surface states}

Let us commence with a minimal model for the surface states on a TI adjacent to a FI layer, which may be expressed as
\begin{subequations}
\label{Ham_main}
\begin{gather}
\label{H_main}
\hat{H}_{\mathbf{q}\mathbf{q}^{\prime}}
=
\hat{H}^0_\mathbf{q} \delta_{\mathbf{q}\mathbf{q}^{\prime}}
+
\hat{V}_{\mathbf{q}\mathbf{q}^{\prime}},
\\
\label{H0_main}
\hat{H}^0_\mathbf{q}
=
\hat{\bs{\sigma}} \cdot \mathbf{h}_{\mathbf{q}}
+
\Delta_{ex} \hat{\sigma}_z m_z,
\end{gather}
\end{subequations}
where $\mathbf{h}_{\mathbf{q}}
=
\hbar v_F \mathbf{q} \times \mathbf{z}
-
\Delta_{ex} \mathbf{z} \times \left( \mathbf{z} \times \mathbf{m}\right)$, with $\mathbf{q}$ the in-plane momentum, $v_F$ the Fermi velocity, $\Delta_{ex}$ the proximity-induced exchange energy and $\mathbf{m} = (m_x, m_y, m_z)$ the unit magnetization. For the impurity potential $\hat{V}_{\mathbf{q}\mathbf{q}^{\prime}}$, we assume it consists of contributions from scalar point scatterers, $\hat{U}_{\mathbf{q}\mathbf{q}^{\prime}}$, as well as from the SOC of the random structural defects \footnote{It should be stressed that the SOC disorder is a required element, as without it, the in-plane magnetization--a key UMR ingredient--can be gauged out of the problem \cite{chiba2017magnetic, dyrdal2020spin}.}, $\hat{W}_{\mathbf{q}\mathbf{q}^{\prime}}$, as \cite{sherman2003minimum, golub2004spin, strom2010edge, kimme2016backscattering, dyrdal2020spin}
\begin{subequations}
\begin{align}
\hat{V}_{\mathbf{q}\mathbf{q}^{\prime}}
&=
\hat{U}_{\mathbf{q}\mathbf{q}^{\prime}}
+
\hat{W}_{\mathbf{q}\mathbf{q}^{\prime}},
\\
\hat{U}_{\mathbf{q}\mathbf{q}^{\prime}}
&=
U_{\mathbf{q}\mathbf{q}^{\prime}}^0 \hat{\sigma}_0,
\\
\hat{W}_{\mathbf{q}\mathbf{q}^{\prime}}
&=
\frac{1}{2} W_{\mathbf{q}\mathbf{q}^{\prime}}^0
\hat{\bs{\sigma}} \cdot \left(\mathbf{q}+\mathbf{q}^{\prime}\right) \times \mathbf{z}.
\end{align}
\end{subequations}
And we assume the white noise distribution for the disorder potentials, $\braket{U_{\mathbf{q}\mathbf{q}^{\prime}}^0}=0$, $\braket{U_{\mathbf{q}\mathbf{q}^{\prime}}^0 U_{\mathbf{q}^{\prime}\mathbf{q}}^0}= n_I U_0^2$, $\braket{W_{\mathbf{q}\mathbf{q}^{\prime}}^0}=0$, and $\braket{W_{\mathbf{q}\mathbf{q}^{\prime}}^0 W_{\mathbf{q}^{\prime}\mathbf{q}}^0}= n_{\alpha} W_0^2$, where $\braket{\cdots}$ denotes the impurity average and $n_I$ and $n_{\alpha}$ are the densities of the scalar and SOC scatterers, while $U_0$ and $W_0$ measure the strengths of the disorder interactions.

\section{Scattering time and quadratic response}

\begin{figure}[tph]
\vspace{.3cm}
{\includegraphics[width=0.4\linewidth,trim={1.5cm 0.5cm 0.5cm 0}]{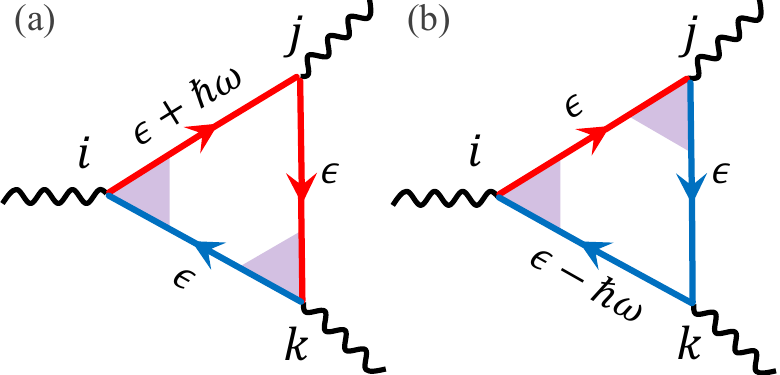}}
	\caption{Diagrammatic structure of the quadratic response. Panels (a) and (b), along with their $j \leftrightarrow k$ counterparts, are the four triangle diagrams that give rise to the UMR and NPHE. The red (blue) arrowed lines represent retarded (advanced) dressed electron Green's functions, while purple shaded areas indicate dressed vertices. The vertices are labeled by the spatial index of their external photons, while propagators are labeled by their energies.}
    \label{fig_proximity_fig2}
\end{figure}

In this section, the nonlinear magnetotransport coefficients in the system under consideration are examined by evaluating quadratic Kubo formulas, which--diagrammatically--correspond to \textit{triangle} diagrams of response theory \cite{kubo1957statistical, parker2019diagrammatic, du2021quantum, rostami2021gauge}, as shown in Fig.~\ref{fig_proximity_fig2}. This is an essential diagrammatic approach, as UMRs and nonlinear Hall effects cannot be captured by the two-photon bubble diagrams of linear response theory.

The self-energy in the Born approximation is given by
\begin{equation}
\label{selfenergy}
\Sigma^{\text{R/A}}_{\mathbf{q}\sigma}(\epsilon)
=
\sum_{\mathbf{q}^{\prime}\sigma^{\prime}}
\Braket{V_{\mathbf{q}\mathbf{q}^{\prime}}^{\sigma \sigma^{\prime}} V_{\mathbf{q}^{\prime}\mathbf{q}}^{{\sigma^{\prime} \sigma}}}
G^{0,\text{R/A}}_{\mathbf{q}^{\prime}\sigma^{\prime}}(\epsilon),
\end{equation}
where
$\sum_{\mathbf{q}} \equiv \int d^2 \mathbf{q}/(2\pi)^2$ and $G^{0,\text{R/A}}_{\mathbf{q}\sigma}(\epsilon)
=
(\epsilon - \epsilon_{\mathbf{q}\sigma} \pm i \delta)^{-1}$ 
is the retarded/advanced Green's function of the unperturbed system $\hat{H}_{\mathbf{q}}^0$, with the eigenvalue $\epsilon_{\mathbf{q}\sigma}=\sigma \sqrt{h_{\mathbf{q}}^2 + \Delta_{ex}^2 m_z^2}$ for the band $\sigma$ ($=\pm$). The scattering time is given by $\tau_{\mathbf{q}\sigma}(\epsilon)= \hbar/2 \Gamma_{\mathbf{q}\sigma} (\epsilon)$, with the scattering rate defined as $\Gamma_{\mathbf{q}\sigma} \equiv - \text{Im} \Sigma^{\text{R}}_{\mathbf{q}\sigma}$. Upon introducing the change of variables $\bs{\eta}_{\mathbf{h}}\equiv \mathbf{h}_{\mathbf{q}}/\epsilon$ and $\eta_{ex} \equiv \Delta_{ex}/\epsilon$, as detailed in Sec.~\ref{appendixB}, we find
\begin{equation}
\label{tau_qsigma}
\tau_{\mathbf{q}\sigma}
=
\frac{\hbar}{2 \Gamma_I}
\left[1+ \sigma \eta_{\text{ex}} m_z \cos \theta_{\mathbf{h}}
+
\frac{\Gamma_{\alpha}}{\Gamma_I} 
\mathcal{F}_{\sigma}
\left( \bs{\eta}_{\mathbf{h}}, \eta_{ex} ; \mathbf{m} \right)\right]^{-1},
\end{equation}
where $\Gamma_I \equiv n_I U_0^2 \epsilon / (2\hbar v_F)^2$ and  $\Gamma_{\alpha} \equiv n_{\alpha} W_0^2 \epsilon^3 /(2\hbar v_F)^4$ are the scalar and SOC disorder self-energy coefficients, respectively, and $\theta_{\mathbf{h}}
=
\cos^{-1}(\eta_{ex} m_z/\sqrt{\eta_{\mathbf{h}}^2 + \eta_{ex}^2 m_z^2})$. The dimensionless band-dependent function $\mathcal{F}_{\sigma}$--whose explicit form is presented in Sec.~\ref{appendixB}--is a rather complicated function of the Fermi velocity, exchange energy and orientation of the magnetization. As we discuss below, this nontrivial angular dependence of the scattering time on the magnetization direction plays an important role in explaining the unconventional angular dependencies of the nonlinear magnetoresistances.

The quadratic conductivity tensor is obtained by evaluating the four triangle diagrams shown in Fig.~\ref{fig_proximity_fig2}. Together, their contributions to the nonlinear dc conductivity may be succinctly expressed as \cite{du2021quantum}
\begin{equation}
\label{sigma_ijk}
\sigma^{ijk}
=
\frac{e^3 \hbar^2}{\pi}
\text{Im} \sum_{\mathbf{q}\sigma} 
\pd_{\omega}\left[ 
\mathcal{V}^i_{\mathbf{q}\sigma} \left(\epsilon_F, \epsilon_F + \hbar \omega \right)
G^R_{\mathbf{q}\sigma} \left( \epsilon_F + \hbar \omega \right)
\right]_{\omega=0}
v^j_{\mathbf{q}\sigma}
G^R_{\mathbf{q}\sigma} \left( \epsilon_F \right)
\mathcal{V}^{kF}_{\mathbf{q}\sigma}
G^{A}_{\mathbf{q}\sigma} \left( \epsilon_F \right)
+
\left( j \leftrightarrow k\right).
\end{equation}
Here $\epsilon_F$ is the Fermi energy and $G^{\text{R/A}}_{\mathbf{q}\sigma}(\epsilon)=(\epsilon - \epsilon_{\mathbf{q}\sigma} \pm i \Gamma_{\mathbf{q}\sigma})^{-1}$ is the disorder-dressed electron Green's function. The (bare) velocity operator is defined as $\hat{\mathbf{v}}_{\mathbf{q}}=\bs{\pd}_{\mathbf{q}}\hat{H}_{\mathbf{q}}^0/\hbar$, with $\pd_{\mathbf{q}}^i \equiv \pd/\pd q_i$, which--in the chiral basis of Bloch eigenstates--leads to the diagonal terms $\mathbf{v}_{\mathbf{q}\sigma}=\pd_{\mathbf{q}}\epsilon_{\mathbf{q}\sigma}/\hbar$. And $\bs{\mathcal{V}}_{\mathbf{q}\sigma} \left(\epsilon, \epsilon^{\prime}\right)$, which is presented in Sec.~\ref{appendixC}, is the disorder-averaged velocity vertex function, where $\epsilon$ and $\epsilon^{\prime}$ are, respectively, the energies of the incoming and outgoing propagators to the vertex in question and $\bs{\mathcal{V}}_{\mathbf{q}\sigma}^F \equiv \bs{\mathcal{V}}_{\mathbf{q}\sigma} \left(\epsilon_F, \epsilon_F \right)$.

\section{Angular dependencies}

Without loss of generality, let us set the electric field along the $x$ direction, $\mathbf{e}= \mathbf{x}$. Then it suffices to calculate the $\sigma_{xxx}$ and $\sigma_{yxx}$ elements of the conductivity tensor. To characterize the nonlinear transport, we introduce the longitudinal and transverse UMR coefficients $\zeta_{\parallel}^{(2)} = \zeta_x^{(2)}$ and $\zeta_{\perp}^{(2)} = \zeta_y^{(2)}$, where
\begin{equation}
\label{Eq:zeta}
\zeta_i^{(2)}
\equiv
\frac{\sigma_{ix}(E_{x}) - \sigma_{ix}(-E_{x})}{\sigma_D E_x}
\simeq
-\frac{2 \sigma_{ixx}}{\sigma _{D}},
\end{equation}
to leading order in the electric field. Here $\sigma_{ij}=j_{i}/E_{j}$ denotes the linear conductivity tensor and $\sigma_D= e^2/ [4 \pi \hbar (\eta_I+\eta_{\alpha})]$ is the Drude conductivity, with $\eta_{I} \equiv \Gamma_I/ \epsilon_F$ and $\eta_{\alpha} \equiv \Gamma_{\alpha}/ \epsilon_F$ the dimensionless disorder coefficients.

\begin{figure}[hpt]
    \sidesubfloat[]{\includegraphics[width=0.25\linewidth,trim={2cm 0cm 0.3cm 1cm}]{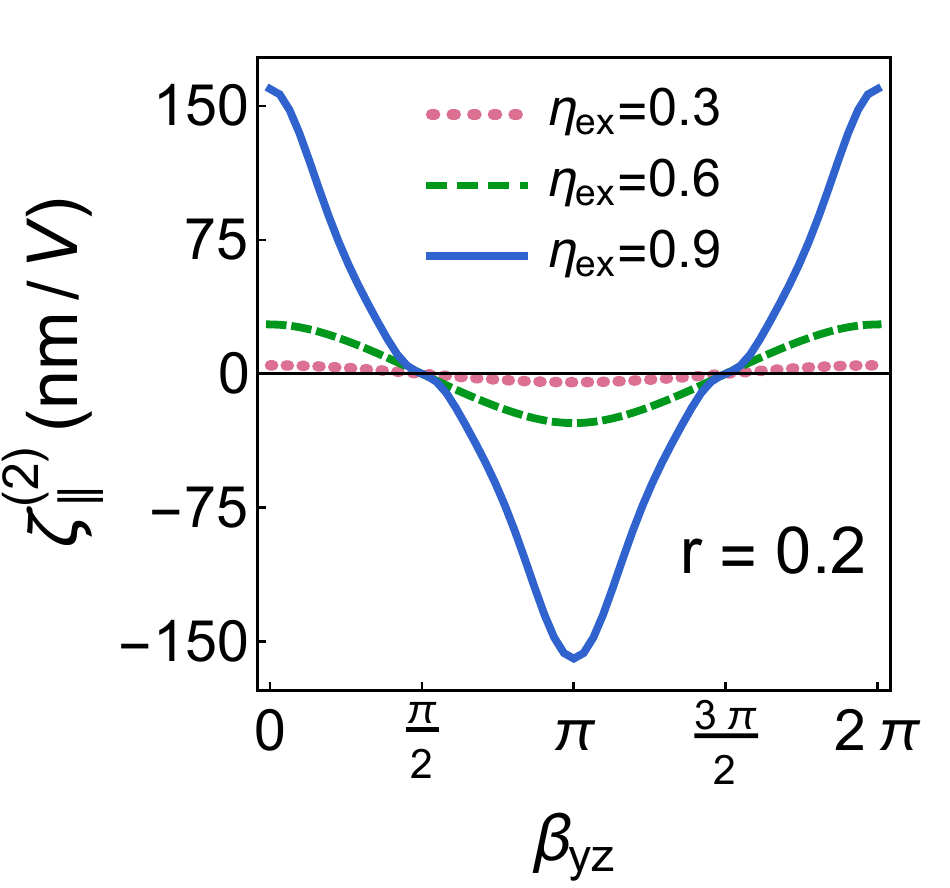}\label{fig_proximity_fig3a}}
%\quad%
    \sidesubfloat[]{\includegraphics[width=0.25\linewidth,trim={2cm 0cm 0.6cm 1cm}]{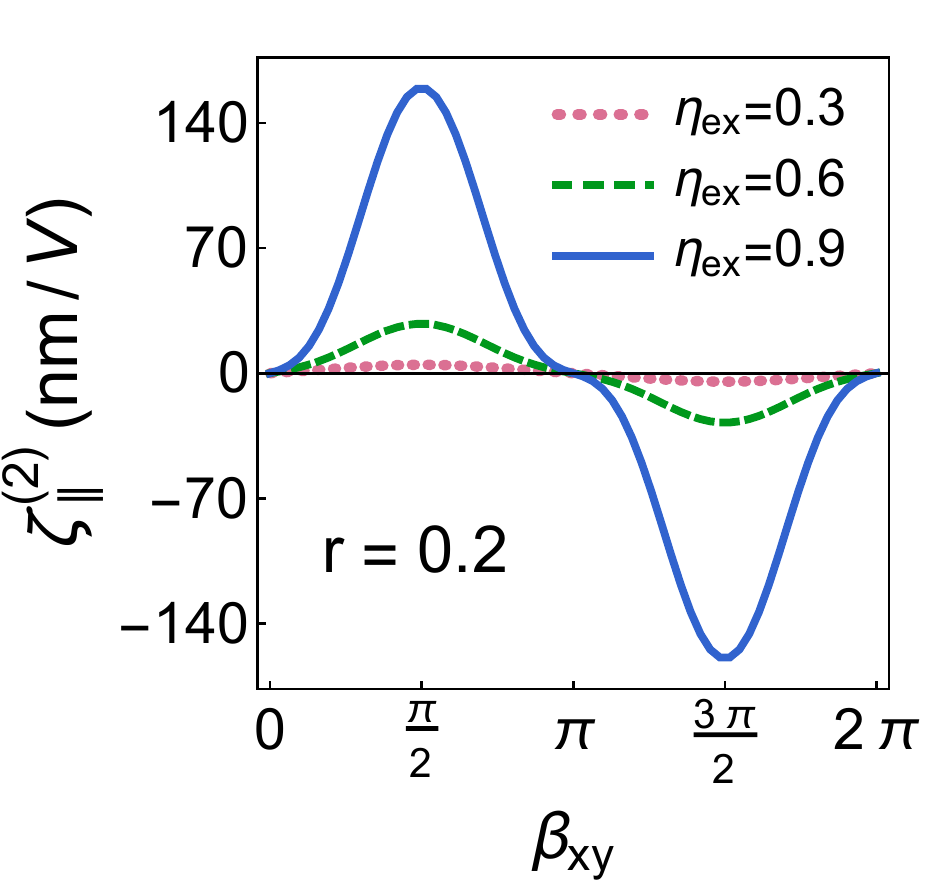}\label{fig_proximity_fig3b}}
%%%%%%%%%
\\
    \sidesubfloat[]{\includegraphics[width=0.25\linewidth,trim={1.4cm 0.9cm 0.2cm .5cm}]{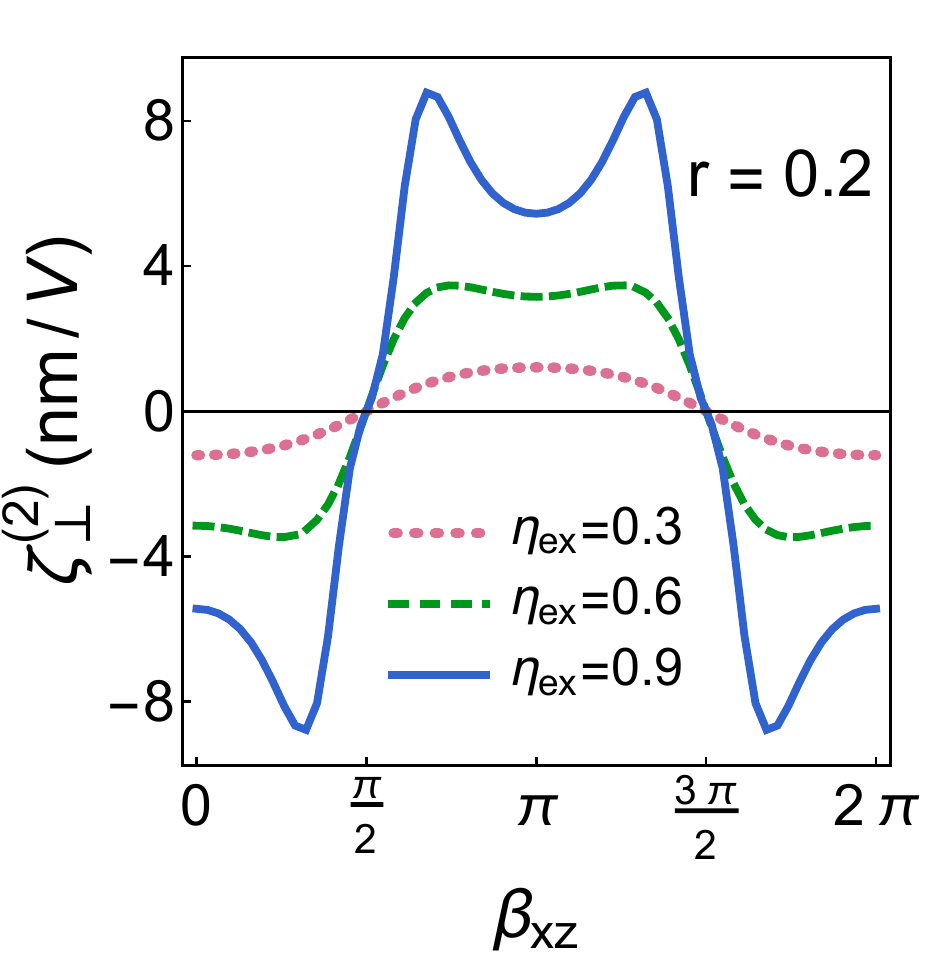}\label{fig_proximity_fig3c}}
%%%%%%%    
    \sidesubfloat[]{\includegraphics[width=0.25\linewidth,trim={1.4cm 0.9cm 0.7cm .5cm}]{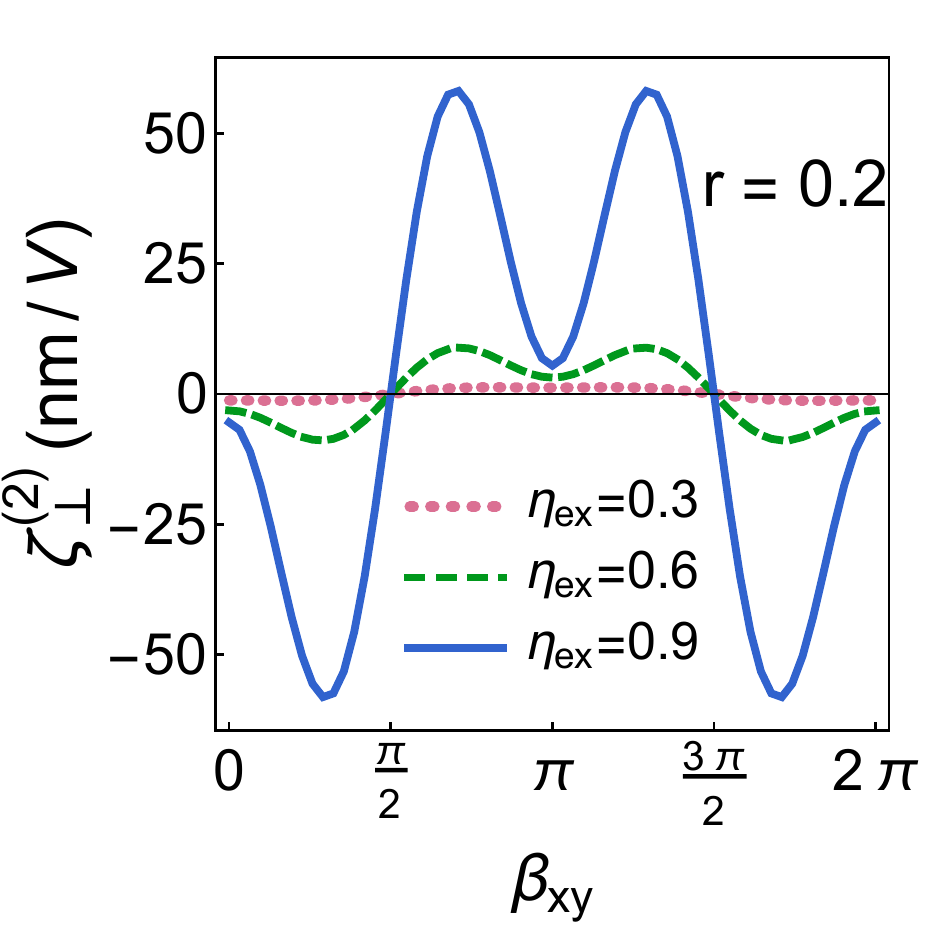}\label{fig_proximity_fig3d}}
%%%%%%%%%
    \caption{Angular dependencies of the UMR coefficients as the magnetization direction is varied in the $xy$, $yz$, and $xz$ planes (the $xz$ scan for $\zeta_{\parallel}^{(2)}$ and $yz$ scan for $\zeta_{\perp}^{(2)}$ vanish due to symmetry constraints and hence are not shown here). Parameters used: $\epsilon_F=0.5$ eV, $v_F= 5\times10^{14}$ nm/s \cite{qi2011topological} and $\eta_I = 0.01$.}
    \label{fig_proximity_fig3}
\end{figure}

Plots of the UMR coefficients for various angular sweeps of the magnetization are presented in Fig.~\ref{fig_proximity_fig3}. As shown by the blue curves in Figs.~\ref{fig_proximity_fig3a} and \ref{fig_proximity_fig3b}, we see that as the Fermi level approaches the exchange energy, \textit{i.e.}, when $\eta_{ex}$ is closer to 1, the strength of the longitudinal nonlinear magnetoresistance $\zeta_{\parallel}^{(2)}$ is significantly amplified and can be as large as 1-2 orders of magnitude stronger than when $\eta_{ex}$ is smaller (see the dashed green and dotted pink curves). Furthermore, qualitatively, the angular dependencies of the longitudinal UMR coefficient increasingly deviate from the conventional sinusoidal behavior as $\eta_{ex}$ is increased.

Quantitatively, the situation is similar for the NPHE; as shown by the blue curves in Figs~\ref{fig_proximity_fig3c} and \ref{fig_proximity_fig3d}, the transverse UMR coefficient $\zeta_{\perp}^{(2)}$ is larger when the Fermi level is closer to the exchange energy. Qualitatively, however, as $\eta_{ex}$ is increased, the NPHE no longer reaches its maximal absolute value at the expected angles $\beta_{xz}, \beta_{xy}=0,\pi$ (see the pink curves), which is when the magnetization is entirely along the $x$ axis; instead, there is an emergent bifurcation of the peaks and troughs such that the maximal absolute values of the NPHE are obtained when the magnetization is only partially along the $x$ axis \footnote{It is worth noting that these angular dependence trends of the NPHE for both out-of-plane and in-plane sweeps of the magnetization resemble the behavior predicted for the planar Hall effect in the linear response regime \cite{chiba2017magnetic}.}. 

Further analysis reveals that the angular profiles of both the UMR and NPHE are also sensitive to the ratio of the scalar to SOC disorder present in the system, $r \equiv \eta_I / \eta_{\alpha}$ (see Sec.~\ref{appendixD} for details). To capture the angular dependencies of the nonlinear responses, combining Eqs.~(\ref{j^2_cubic}) and (\ref{Eq:zeta}), the corresponding UMR coefficients with the magnetization direction being varied in the three orthogonal planes are expressed as
\begin{subequations}
\label{fits_out}
\begin{align}
\left.\zeta_{\parallel}^{(2)}
\right|_{\substack{m_x=0}}
&\simeq
f_{\parallel}(\eta_{ex}, r) \cos \beta_{yz}
+
g_{\parallel}(\eta_{ex}, r) \cos^3 \beta_{yz},
\\
\left. \zeta_{\perp}^{(2)}
\right|_{\substack{m_y=0}}
&\simeq
f_{\perp}(\eta_{ex}, r) \cos \beta_{xz}
+
g_{\perp}(\eta_{ex}, r) \cos^3 \beta_{xz},
\end{align}
\end{subequations}
for the out-of-plane sweeps of the magnetization and
\begin{subequations}
\label{fits_out}
\begin{align}
\left.\zeta_{\parallel}^{(2)}
\right|_{\substack{m_z=0}}
&\simeq
h_{\parallel}(\eta_{ex}, r) \sin \beta_{xy}
+
k_{\parallel}(\eta_{ex}, r) \sin^3 \beta_{xy},
\\
\left. \zeta_{\perp}^{(2)}
\right|_{\substack{m_z=0}}
&\simeq
h_{\perp}(\eta_{ex}, r) \cos \beta_{xy}
+
k_{\perp}(\eta_{ex}, r) \cos^3 \beta_{xy},
\end{align}
\end{subequations}
for the in-plane sweep, where the functions on the right-hand side--which, in general, depend on $\eta_{ex}$ and $r$--are given by
\begin{subequations}
\begin{align}
f_{\parallel}(\eta_{ex}, r)
&=
-\frac{2 \sigma_{\parallel,1}^{(2)}}{\sigma_D}
(1 + \lambda_{\parallel}),
\\
g_{\parallel}(\eta_{ex}, r)
&=
-\frac{2 \sigma_{\parallel,1}^{(2)}}{\sigma_D}
(\kappa_{\parallel} - \lambda_{\parallel}),
\\
h_{\parallel}(\eta_{ex}, r)
&=
-\frac{2 \sigma_{\parallel,1}^{(2)}}{\sigma_D}
(1 + \iota_{\parallel}),
\\
k_{\parallel}(\eta_{ex}, r)
&=
-\frac{2 \sigma_{\parallel,1}^{(2)}}{\sigma_D}
(\kappa_{\parallel} - \iota_{\parallel}),
\end{align}
\end{subequations}
and
\begin{subequations}
\begin{align}
f_{\perp}(\eta_{ex}, r)
&=
-\frac{2 \sigma_{\perp,1}^{(2)}}{\sigma_D}
(1 + \lambda_{\perp}),
\\
g_{\perp}(\eta_{ex}, r)
&=
-\frac{2 \sigma_{\perp,1}^{(2)}}{\sigma_D}
(\iota_{\perp} - \lambda_{\perp}),
\\
h_{\perp}(\eta_{ex}, r)
&=
-\frac{2 \sigma_{\perp,1}^{(2)}}{\sigma_D}
(1 + \kappa_{\perp}),
\\
k_{\perp}(\eta_{ex}, r)
&=
-\frac{2 \sigma_{\perp,1}^{(2)}}{\sigma_D}
(\iota_{\perp} - \kappa_{\perp}).
\end{align}
\end{subequations}
To quantify the deviations of the UMR coefficients from sinusoidal behavior for various sweeps of the magnetization, we introduce the dimensionless ratios $a_{\parallel} \equiv g_{\parallel}/f_{\parallel}$, $b_{\parallel} \equiv k_{\parallel}/h_{\parallel}$, $a_{\perp} \equiv g_{\perp}/f_{\perp}$ and $b_{\perp} \equiv k_{\perp}/h_{\perp}$. The values of these ratios for $\eta_{ex}=0.3$ and $\eta_{ex}=0.6$ are presented in Table~\ref{tab1}, from which it is evident that the cubic terms play an increasingly important--and even dominant--role as the exchange and Fermi energies become comparable. It is worth mentioning that for even higher values of $\eta_{ex}$, as indicated by the blue curves in Fig.~\ref{fig_proximity_fig3}, the UMR coefficients become increasingly nonlinear in the magnetization, such that higher-order terms in the magnetization must also be taken into account.

\begin{table}
\caption{Transport coefficients for the cubic contributions to the UMR and NPHE at $r=0.2$ and for different values of $\eta_{ex}$. For the rest of the parameters used, see the caption of Fig.~\ref{fig_proximity_fig3}.}

\begin{tabular}{l c c c c}
$\eta_{\text{ex}}$ & $a_{\parallel}$ & $b_{\parallel}$ & $a_{\perp}$ & \multicolumn{1}{c}{$b_{\perp}$}
\\
\hline
0.3 & 0.03 & 0.30 & -0.20 & -0.47
\\
0.6 & 0.09 & 1.92 & -0.59 & -0.85
\\
\end{tabular}
\label{tab1}
\end{table}

\section{Symmetry analysis of nonlinear responses}
\label{appendixA}

In this section, we derive the general form of the nonlinear current, $j^{(2)}_i = \sigma_{ijk} E^j E^k$, to third order in the magnetization for the TI/FI bilayer system described by Eqs.~(\ref{Ham_main}). To this end, consider the magnetization expansion of the quadratic conductivity tensor
\begin{equation}
\sigma^{ijk}
=
\sigma_1^{ijkl}m_l
+
\sigma_2^{ijklm}m_l m_m
+
\sigma_3^{ijklmn}m_l m_m m_n.
\end{equation}
Without loss of generality, let us set the electric field in the $\mathbf{x}$ direction so that we need only evaluate $\sigma^{xxx}$ for the longitudinal current. Note that in the absence of the exchange interaction, the system is even under the mirror reflection transformation in the $xz$ ($yz$) plane, $\mathcal{M}_{xz}$ ($\mathcal{M}_{yz}$), and thus so are $\sigma_1$, $\sigma_2$ and $\sigma_3$. Using the fact that the current and magnetization are polar and axial vectors, respectively, it is then straightforward to verify that, upon imposing $\mathcal{M}_{xz}$ and $\mathcal{M}_{yz}$, $\sigma_1^{xxxx}=\sigma_1^{xxxz}=0$, while, in general, $\sigma_1^{xxxy} \neq 0$. Similarly, for the quadratic contribution, one can verify that only terms $\propto m_x m_z$ are allowed by reflection symmetry. However, given the particle-hole symmetry and the quadratic dependence of the disorder averages on $m_z$ [see Eqs.~(\ref{vvavg})], the terms linear in $m_z$ do not contribute to the nonlinear response and $\sigma_2$, therefore, plays no role in the nonlinear transport.

Moving on to the cubic contribution, one can verify that, in addition to $\sigma_3^{xxxxyz}$, terms containing even powers of $m_y$ do not contribute to $j_x^{(2)}$. Thus, of the 10 independent components of $\sigma_3^{xxxijk}$, only 3 survive the symmetry requirements, namely $\sigma_3^{xxxyxx}$, $\sigma_3^{xxxyyy}$ and $\sigma_3^{xxxyzz}$. By an analogous argument for the transverse current, one can show that only the  independent components $\sigma_1^{yxxx}$, $\sigma_3^{yxxxxx}$, $\sigma_3^{yxxxyy}$ and $\sigma_3^{yxxxzz}$ may be nonvanishing. Therefore, the nonlinear current up to third order in the magnetization may be expressed as
\begin{subequations}
\begin{align}
j_x^{(2)}
&=
j_{x,1}^{(2)} 
\left(
1 + \iota_{\parallel} m_x^2 + \kappa_{\parallel} m_y^2 + \lambda_{\parallel} m_z^2
\right),
\\
j_y^{(2)}
&=
j_{y,1}^{(2)} 
\left(
1 + \iota_{\perp} m_x^2 + \kappa_{\perp} m_y^2 + \lambda_{\perp} m_z^2
\right),
\end{align}
\end{subequations}
where
$j_{x,1}^{(2)}
=
\sigma_1^{xxxy} m_y E_x^2$, 
$j_{y,1}^{(2)}
=
\sigma_1^{yxxx} m_x E_x^2$ and the dimensionless functions measuring the cubic contributions read
\begin{subequations}
\begin{align}
\iota_{\parallel}
&=
\frac{3 \sigma_3^{xxxyxx}}{\sigma_1^{xxxy}},
\\
\kappa_{\parallel}
&=
\frac{\sigma_3^{xxxyyy}}{\sigma_1^{xxxy}},
\\
\lambda_{\parallel}
&=
\frac{3 \sigma_3^{xxxyzz}}{\sigma_1^{xxxy}},
\end{align}
\end{subequations}
and
\begin{subequations}
\begin{align}
\iota_{\perp}
&=
\frac{\sigma_3^{yxxxxx}}{\sigma_1^{yxxx}},
\\
\kappa_{\perp}
&=
\frac{3 \sigma_3^{yxxxyy}}{\sigma_1^{yxxx}},
\\
\lambda_{\perp}
&=
\frac{3 \sigma_3^{yxxxzz}}{\sigma_1^{yxxx}}.
\end{align}
\end{subequations}
Relaxing the orientation of the electric field to point in an arbitrary direction in the $xy$ plane, we arrive at Eqs.~(\ref{j^2_cubic}), where  the functions $\iota$, $\kappa$ and $\lambda$ are now understood to be evaluated with the relevant tensor components in the basis spanned by $\mathbf{e}$, $\mathbf{z} \times \mathbf{e}$ and $\mathbf{z}$. 

\section{Self-energy and scattering time}
\label{appendixB}

The unperturbed Hamiltonian, Eq.~(\ref{H0_main}), has eigenstates given by
\begin{equation}
\ket{u_{\mathbf{q}\sigma}}
=
\begin{pmatrix}
\left( \frac{1+\sigma}{2}\right) 
\cos \frac{\theta_{\mathbf{h}}}{2}
+
\left( \frac{1-\sigma}{2}\right) 
\sin \frac{\theta_{\mathbf{h}}}{2}
\\
e^{i\phi_\mathbf{h}}
\left[
\left( \frac{1+\sigma}{2}\right) 
\sin \frac{\theta_{\mathbf{h}}}{2}
-
\left( \frac{1-\sigma}{2}\right) 
\cos \frac{\theta_{\mathbf{h}}}{2}
\right]
\end{pmatrix},
\end{equation}
where $\phi_\mathbf{h}$ is the azimuthal angle in the plane spanned by $\mathbf{h}_{\mathbf{q}}$, with $\cos \phi_\mathbf{h} = h_{x\mathbf{q}}/h_{\mathbf{q}}$ and $\sin \phi_\mathbf{h} = h_{y\mathbf{q}}/h_{\mathbf{q}}$, while $\theta_{\mathbf{h}}$ is the polar angle, with $\sin \theta_\mathbf{h} = h_{\mathbf{q}}/\sqrt{h_{\mathbf{q}}^2 + \Delta_{ex}^2 m_z^2}$. The velocity operator is defined as $\hat{\mathbf{v}}_{\mathbf{q}}=\bs{\pd}_{\mathbf{q}}\hat{H}_{\mathbf{q}}^0/\hbar$, with $\pd_{\mathbf{q}}^i \equiv \pd/\pd q_i$, which in the chiral basis of Bloch eigenstates, leads to the diagonal terms $\mathbf{v}_{\mathbf{q}\sigma}=\pd_{\mathbf{q}}\epsilon_{\mathbf{q}\sigma}/\hbar$, or
\begin{subequations}
\label{S_v_qs}
\begin{align}
v_{\mathbf{q}\sigma}^x
&=
- \sigma v_F \sin \theta_\mathbf{h} \sin \phi_{\mathbf{h}},
\\
v_{\mathbf{q}\sigma}^y
&=
\sigma v_F \sin \theta_\mathbf{h} \cos \phi_{\mathbf{h}}.
\end{align}
\end{subequations}
In the chiral Bloch basis the disorder averages read
\begin{subequations}
\label{vvavg}
\begin{align}
\Braket{V_{\mathbf{q}\mathbf{q}^{\prime}}^{\sigma \sigma^{\prime}} V_{\mathbf{q}^{\prime}\mathbf{q}}^{{\sigma^{\prime} \sigma}}}
&=
\Braket{U_{\mathbf{q}\mathbf{q}^{\prime}}^{\sigma \sigma^{\prime}} U_{\mathbf{q}^{\prime}\mathbf{q}}^{{\sigma^{\prime} \sigma}}}
+
\Braket{W_{\mathbf{q}\mathbf{q}^{\prime}}^{\sigma \sigma^{\prime}} W_{\mathbf{q}^{\prime}\mathbf{q}}^{{\sigma^{\prime} \sigma}}},
\\
\Braket{U_{\mathbf{q}\mathbf{q}^{\prime}}^{\sigma \sigma^{\prime}} U_{\mathbf{q}^{\prime}\mathbf{q}}^{{\sigma^{\prime} \sigma}}}
&=
\frac{1}{2} n_I U_0^2
\left[ 1 
+
\sigma \sigma^{\prime} 
\cos \theta_{\mathbf{h}} \cos \theta_{\mathbf{h}^{\prime}}
+
\sigma \sigma^{\prime} 
\cos \left(\phi_{\mathbf{h}} - \phi_{\mathbf{h}^{\prime}}\right)
\sin \theta_{\mathbf{h}} \sin \theta_{\mathbf{h}^{\prime}}\right],
\\
\begin{split}
\Braket{W_{\mathbf{q}\mathbf{q}^{\prime}}^{\sigma \sigma^{\prime}} W_{\mathbf{q}^{\prime}\mathbf{q}}^{{\sigma^{\prime} \sigma}}}
&=
\frac{1}{8} n_{\alpha} W_0^2
\left\{
\left[ 1 
-
\sigma \sigma^{\prime} 
\cos \theta_{\mathbf{h}} \cos \theta_{\mathbf{h}^{\prime}}\right]
\left[ \left(q_x + q_x^{\prime}\right)^2
+ \left(q_y + q_y^{\prime}\right)^2 \right]
\right.
\\
&\left. 
-
\sigma \sigma^{\prime} 
\cos \left(\phi_{\mathbf{h}} + \phi_{\mathbf{h}^{\prime}}\right)
\sin \theta_{\mathbf{h}} \sin \theta_{\mathbf{h}^{\prime}}
\left[ \left(q_x + q_x^{\prime}\right)^2
- \left(q_y + q_y^{\prime}\right)^2 \right]
\right.
\\
&\left. 
-
2 \sigma \sigma^{\prime} 
\sin \left(\phi_{\mathbf{h}} + \phi_{\mathbf{h}^{\prime}}\right)
\sin \theta_{\mathbf{h}} \sin \theta_{\mathbf{h}^{\prime}}
\left(q_x + q_x^{\prime}\right)
\left(q_y + q_y^{\prime}\right)
\right\},
\end{split}
\end{align}
\end{subequations}
where $V_{\mathbf{q}\mathbf{q}^{\prime}}^{\sigma \sigma^{\prime}} 
\equiv
\braket{u_{\mathbf{q}\sigma}| \hat{V}_{\mathbf{q}\mathbf{q}^{\prime}} |u_{\mathbf{q}^{\prime}\sigma^{\prime}}}$.
Inserting Eqs. (\ref{vvavg}) into Eq.~(\ref{selfenergy}), we obtain
\begin{equation}
\Gamma_{\mathbf{q}\sigma}
=
\Gamma_I
\left[1+ \sigma \eta_{\text{ex}} m_z \cos \theta_{\mathbf{h}}
+
\frac{\Gamma_{\alpha}}{\Gamma_I} 
\mathcal{F}_{\sigma}
\left( \bs{\eta}_{\mathbf{h}}, \eta_{ex} ; \mathbf{m}\right)\right],
\end{equation}
where the dimensionless function $\mathcal{F}_{\sigma}$ reads
\begin{equation}
\begin{split}
\mathcal{F}_{\sigma}
\left( \bs{\eta}_{\mathbf{h}}, \eta_{ex} ; \mathbf{m} \right)
&=
\left(1-\sigma \eta _{ex} m_z \cos \theta_{\mathbf{h}} \right)
\left[1 + \eta_{\mathbf{h}}^2 + \eta _{ex}^2 \left(4 - 5 m_z^2\right)
-
4 \eta_{ex} \bs{\eta}_{\mathbf{h}} \cdot \mathbf{m} \right]
\\
&+
2 \sigma \sin \theta_{\mathbf{h}}
\left(1- \eta_{ex}^2 m_z^2 \right)
\left( \eta _{\mathbf{h}} - 2 \eta_{ex} \frac{\bs{\eta}_{\mathbf{h}} \cdot \mathbf{m} }{\eta_{\mathbf{h}}}\right),
\end{split}
\end{equation}
in which we assume the Fermi level lies in the upper band ($\epsilon>0$). From this, we arrive at the momentum scattering time, Eq.~(\ref{tau_qsigma})

\section{Vertex correction and nonlinear conductivity}
\label{appendixC}

\begin{figure}[tph]
    \sidesubfloat[]{\includegraphics[width=0.25\linewidth,trim={1.5cm .3cm .5cm 1cm}]{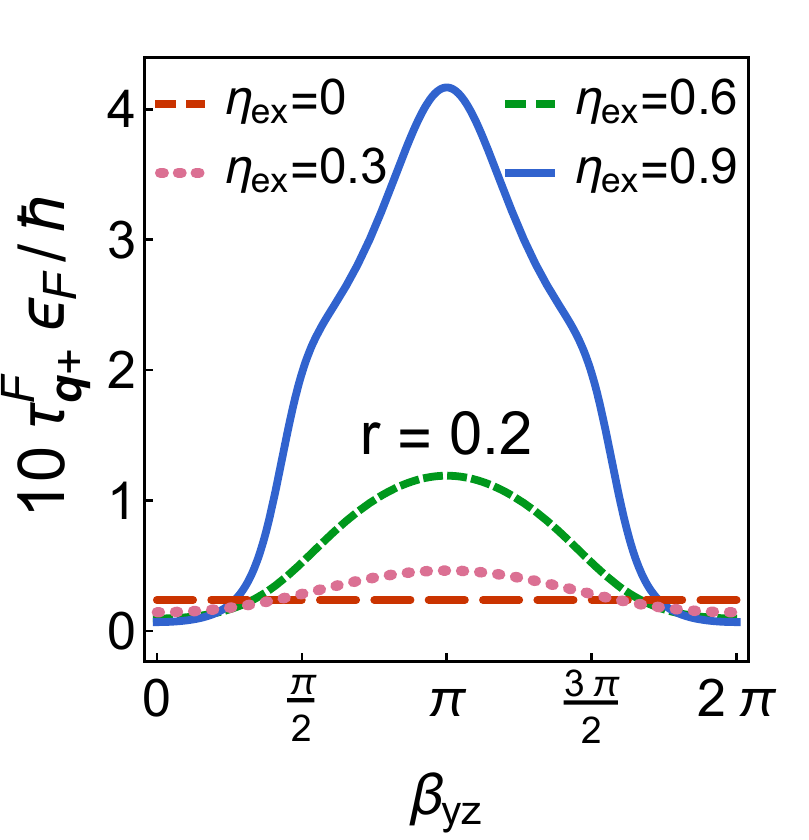}\label{fig_proximity_fig4a}}
%\quad%
    \sidesubfloat[]{\includegraphics[width=0.25\linewidth,trim={1.1cm .2cm .5cm .8cm}]{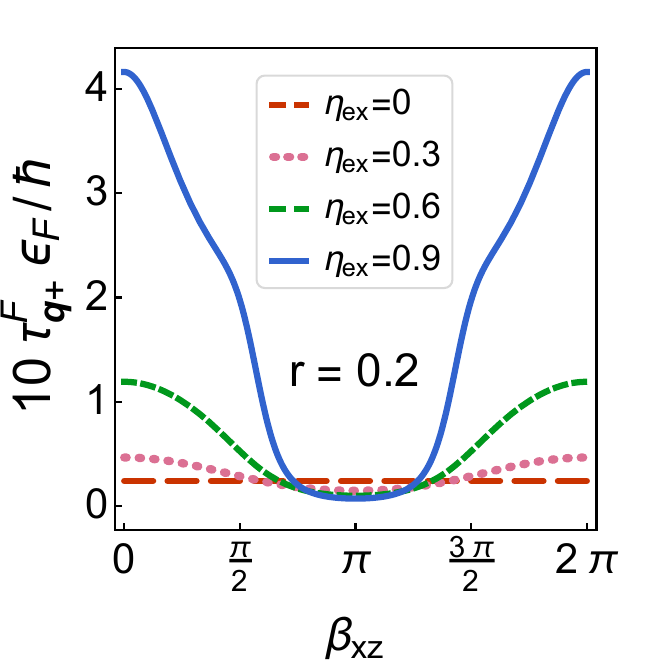}\label{fig_proximity_fig4b}}
    \\
    \sidesubfloat[]{\includegraphics[width=0.25\linewidth,trim={1.9cm .9cm .8cm .8cm}]{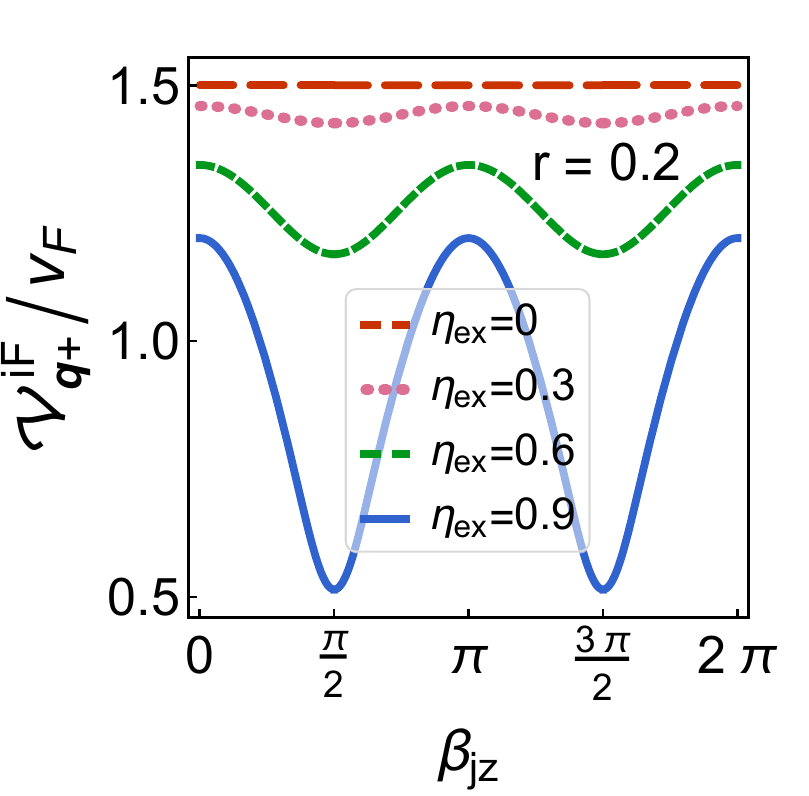}\label{fig_proximity_fig4c}}
%\quad%
    \sidesubfloat[]{\includegraphics[width=0.25\linewidth,trim={1.3cm .9cm .7cm .7cm}]{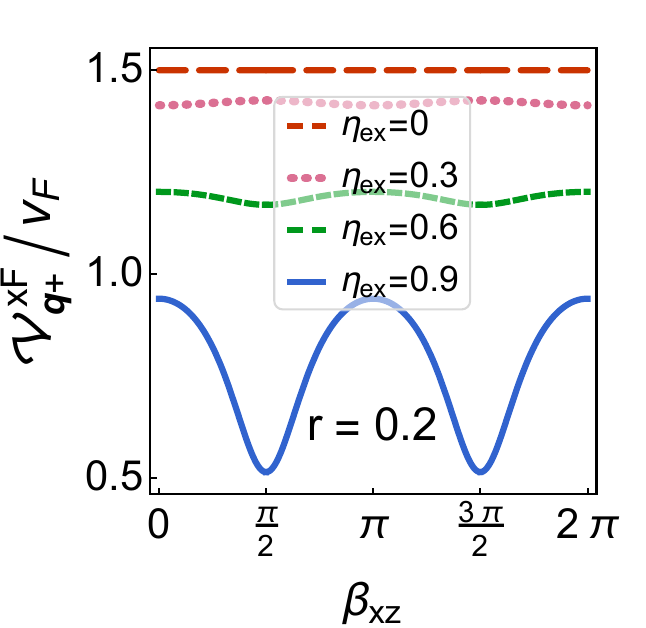}\label{fig_proximity_fig4d}}    
%%%%%%%%%
    \caption{Angular dependencies at the Fermi level.  (a) scattering time (with $q_y=0$) in the $yz$ plane, (b) $i$-th component of the velocity vertex function (with $q_j=0$) in the $jz$ plane, where $i/j = x/y$ or $y/x$. (c) scattering time (with $q_x=0$) in the $xz$ plane and (d) longitudinal velocity vertex function (with $q_y=0$) in the $xz$ plane. Parameters used: $\epsilon_F=0.5$ eV, $v_F= 5\times10^{14}$ nm/s \cite{qi2011topological} and $\eta_I = 0.01$.}
    \label{fig_proximity_figS1}
\end{figure}

The disorder-averaged velocity vertex function may be found self-consistently from the general vertex equation
\begin{equation}
\bs{\mathcal{V}}_{\mathbf{q}\sigma} \left(\epsilon, \epsilon^{\prime}\right)
=
\mathbf{v}_{\mathbf{q}\sigma}
+
\sum_{\mathbf{q}^{\prime}\sigma^{\prime}}
\Braket{V_{\mathbf{q}\mathbf{q}^{\prime}}^{\sigma \sigma^{\prime}} V_{\mathbf{q}^{\prime}\mathbf{q}}^{{\sigma^{\prime} \sigma}}}
G^{\text{A/R}}_{\mathbf{q}^{\prime}\sigma^{\prime}}\left(\epsilon\right)
\bs{\mathcal{V}}_{\mathbf{q}^{\prime}\sigma^{\prime}} \left(\epsilon, \epsilon^{\prime}\right)
G^{\text{R/A}}_{\mathbf{q}^{\prime}\sigma^{\prime}}\left(\epsilon^{\prime}\right),
\end{equation}
where $\epsilon$ and $\epsilon^{\prime}$ are, respectively, the energies of the incoming and outgoing Green's functions to the vertex in question. The near-dc behavior may be captured by setting $\epsilon^{\prime}=\epsilon+ \hbar \omega$. Then, using the identity
\begin{equation}
G^{\text{A/R}}_{\mathbf{q}\sigma}
\left( \epsilon \right)
G^{\text{R/A}}_{\mathbf{q}\sigma}
\left( \epsilon + \hbar \omega \right)
\simeq
\frac{2\pi}{\hbar} \frac{\tau_{\mathbf{q}\sigma}}{1 \mp i \omega \tau_{\mathbf{q}\sigma}}
\delta \left( \epsilon_{\mathbf{q}\sigma} - \epsilon \right),
\end{equation}
the vertex function may be approximated as
\begin{equation}
\label{V_approx}
\bs{\mathcal{V}}_{\mathbf{q}\sigma} \left(\epsilon, \epsilon + \hbar \omega\right)
\simeq
\mathbf{v}_{\mathbf{q}\sigma}
+
\frac{2\pi}{\hbar}
\sum_{\mathbf{q}^{\prime}\sigma^{\prime}}
\Braket{V_{\mathbf{q}\mathbf{q}^{\prime}}^{\sigma \sigma^{\prime}} V_{\mathbf{q}^{\prime}\mathbf{q}}^{{\sigma^{\prime} \sigma}}}
\frac{\tau_{\mathbf{q}^{\prime}\sigma^{\prime}}}{1 \mp i \omega \tau_{\mathbf{q}^{\prime}\sigma^{\prime}}}
\mathbf{v}_{\mathbf{q}^{\prime}\sigma^{\prime}} 
\delta \left( \epsilon_{\mathbf{q}^{\prime}\sigma^{\prime}} - \epsilon \right).
\end{equation}
To obtain an approximate form for the conductivity tensor, we insert Eq.~(\ref{V_approx}) into Eq.~(\ref{sigma_ijk}) and use the identity
\begin{equation}
\label{iden_cond}
\left[G^{\text{R}}_{\mathbf{q}\sigma}
\left( \epsilon \right)\right]^{m+1}
\left[G^{\text{A}}_{\mathbf{q}\sigma}
\left( \epsilon \right)\right]^{n+1}
\simeq
2\pi i^{n-m}\frac{\left(m+n\right)!}{m!n!} \left(\frac{\tau_{\mathbf{q}\sigma}}{\hbar}\right)^{m+n+1}
\delta \left( \epsilon_{\mathbf{q}\sigma} - \epsilon \right),
\end{equation}
to arrive at the following form for the in-plane conductivity components
\begin{equation}
\label{S_sigma_ijj}
\begin{split}
\sigma^{ijj}
=&
- \frac{4 e^3}{\hbar}
\text{Re} \sum_{\mathbf{q}}
\left[
\pd_{\omega}
\mathcal{V}^i_{\mathbf{q} +} \left(\epsilon_F, \epsilon_F + \hbar \omega \right) \Bigr|_{\substack{\omega=0}}
+
i \mathcal{V}_{\mathbf{q} +}^{i F} \tau_{\mathbf{q} +}^F
\left(
1 + i \frac{\pd \Gamma_{\mathbf{q} +}^F}{\pd \epsilon_F}
\right)
\right]
\\
&\times
v_{\mathbf{q} +}^j \mathcal{V}_{\mathbf{q} +}^{j F} \left( \tau_{\mathbf{q} +}^F \right)^2 \delta \left( \epsilon_{\mathbf{q} +} - \epsilon_F \right),
\end{split}
\end{equation}
where $\Gamma_{\mathbf{q} \sigma}^F$ and $\tau_{\mathbf{q} \sigma}^F$ are the self-energy and scattering time at the Fermi level. In this form, the quadratic conductivity tensor is expressed entirely in terms of the velocity vertex function and scattering time at the Fermi level. Plots of the angular dependencies of these two quantities for various sweeps of the magnetization are presented in Fig.~\ref{fig_proximity_figS1}.

\section{Angular dependencies with weak SOC disorder}
\label{appendixD}

\begin{figure}[hpt]
    \sidesubfloat[]{\includegraphics[width=0.25\linewidth,trim={.7cm 0cm .7cm 0.5cm}]{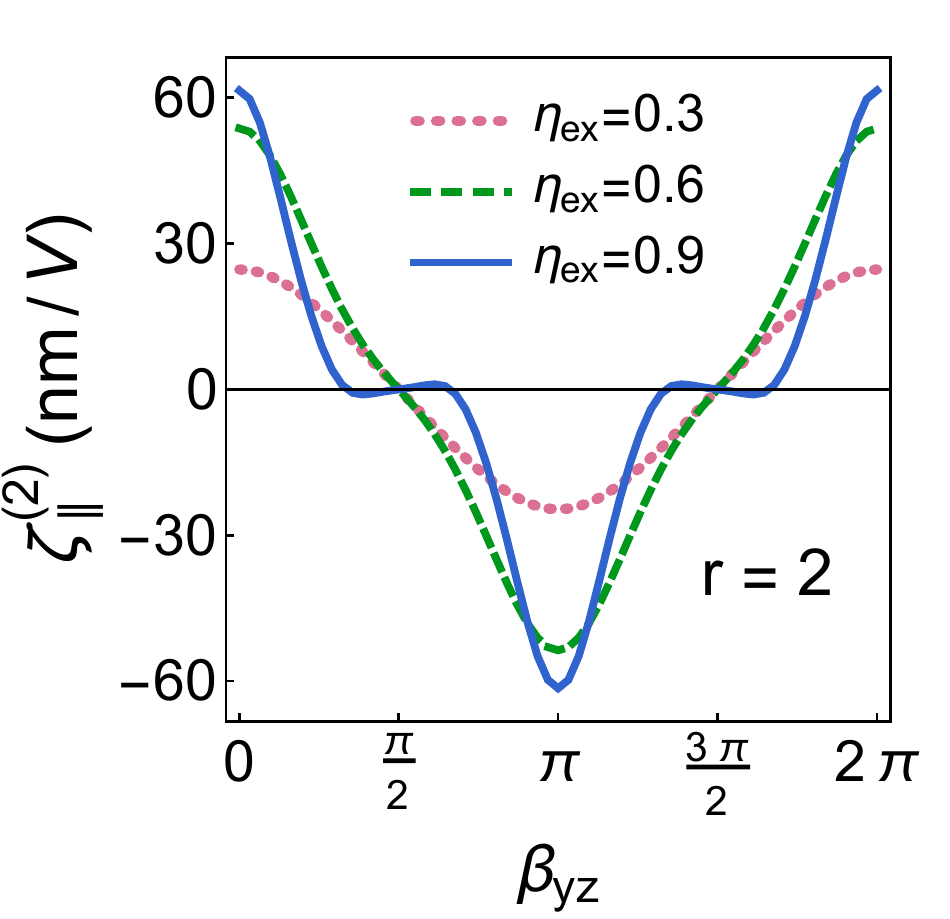}\label{fig_proximity_figS1a}}
%\quad%
    \sidesubfloat[]{\includegraphics[width=0.25\linewidth,trim={.7cm 0cm .7cm .5cm}]{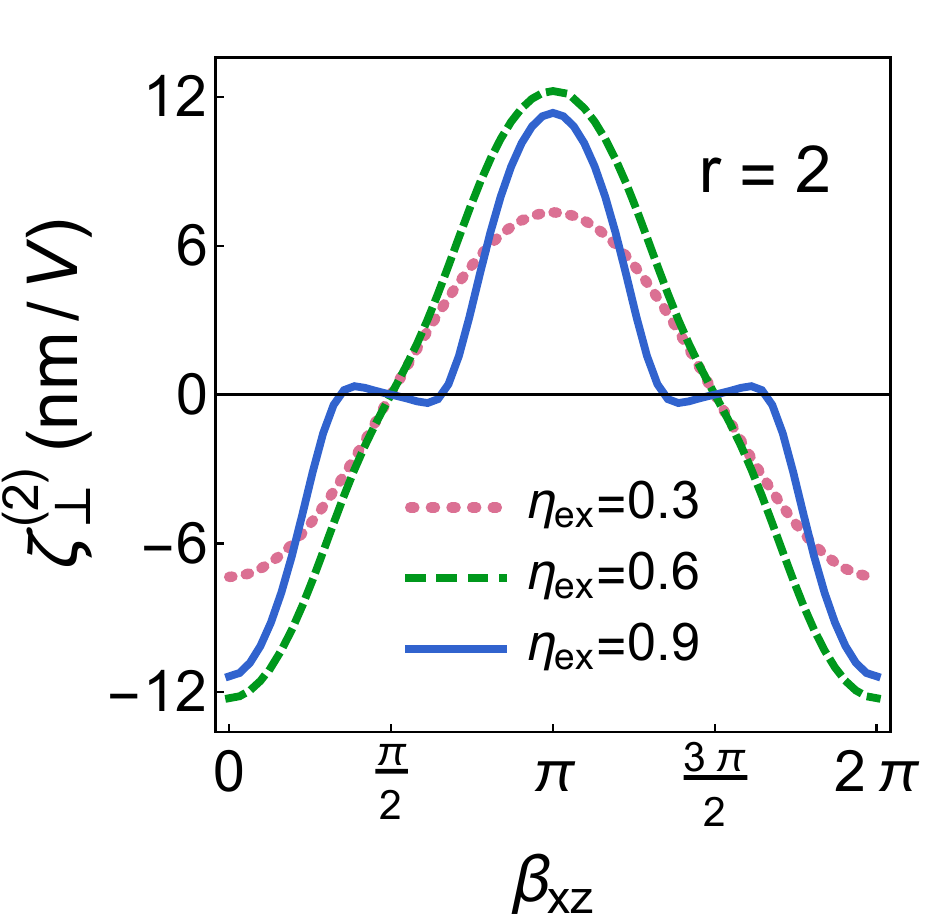}\label{fig_proximity_figS1b}}
%%%%%%%    
\\
    \sidesubfloat[]{\includegraphics[width=0.25\linewidth,trim={.7cm .7cm .7cm .5cm}]{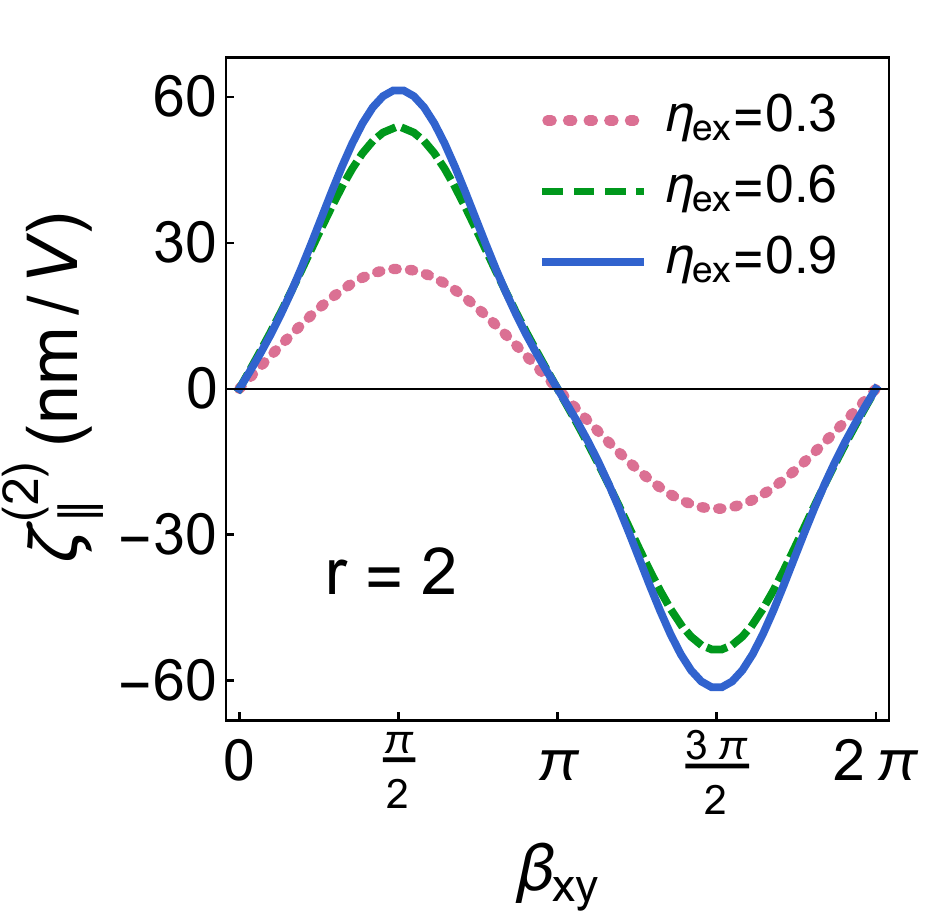}\label{fig_proximity_figS1c}}
%%%%%%%%%
    \sidesubfloat[]{\includegraphics[width=0.25\linewidth,trim={.7cm .7cm .7cm .5cm}]{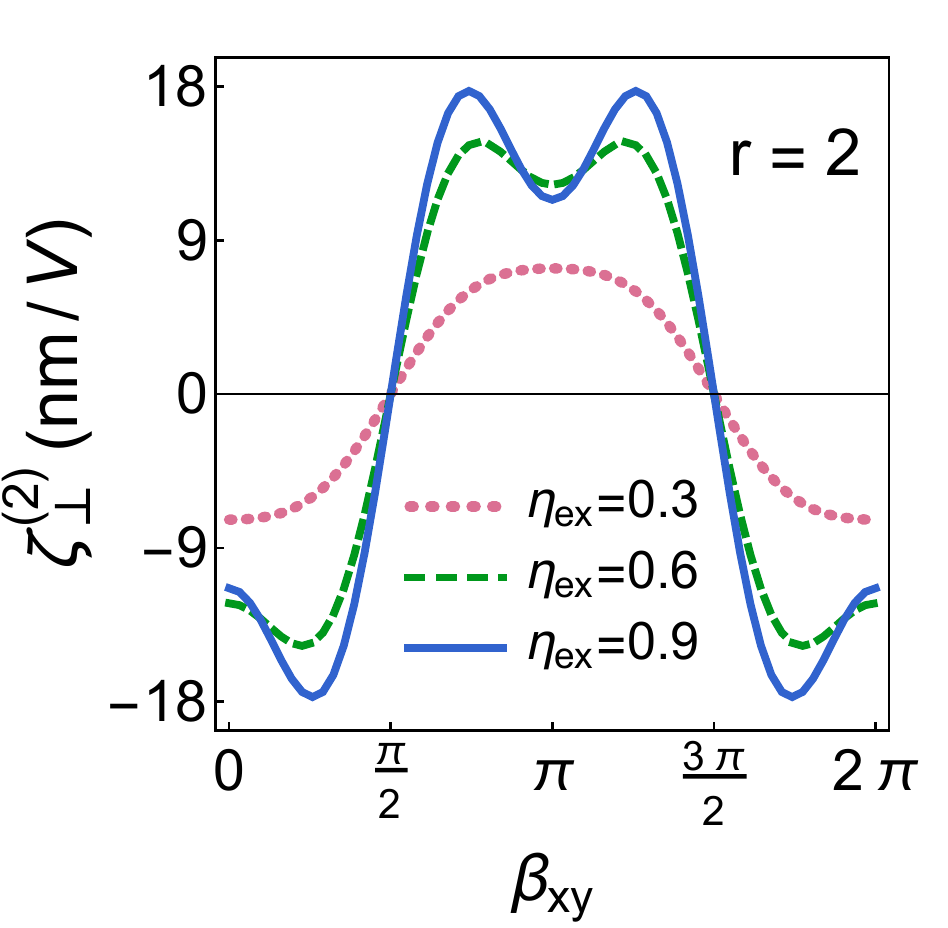}\label{fig_proximity_figS1d}}
%\quad%
 %%%%%%%%%
    \caption{Angular dependencies of the UMR coefficients for various angular sweeps of the magnetization. Here, $\beta_{ij}$ ($i,j=x,y,z$) is the angle between the $i$ axis and the magnetization as it sweeps the $ij$ plane. Parameters used: $\epsilon_F=0.5$ eV, $v_F= 5\times10^{14}$ nm/s \cite{qi2011topological} and $\eta_I = 0.01$.}
        \label{fig_proximity_figS2}
\end{figure}

Here, we analyze the angular dependencies of the UMR coefficients in the limit where the SOC disorder is weaker than the scalar disorder. We thus set $r=2$ and plot the UMR coefficients, which are presented in Fig.~\ref{fig_proximity_figS2}. The first notable feature is that compared to the plots in Fig.~\ref{fig_proximity_fig3}--wherein $r=0.2$--the amplitudes of the UMR and NPHE are generally weaker here for various in-plane and out-of-plane magnetization sweeps. This is not surprising, as the effects are generated in the presence of SOC disorder, which is also weaker here.

Another interesting feature is the qualitative behavior of the UMR coefficients when the Fermi level approaches the exchange energy. For out-of-plane sweeps of the magnetization, as shown in Figs.~\ref{fig_proximity_figS1a} and \ref{fig_proximity_figS1b}, for lower values of $\eta_{ex}$ (see the dotted pink and dashed green curves), the UMR (NPHE) displays typical sinusoidal behavior, with 2 symmetry-imposed sign changes at $\beta_{yz}$($\beta_{xz}$) $= \pi/2, 3\pi/2$. However, at $\eta_{ex}=0.9$ (see the solid blue curves), 4 additional sign change points appear, which are connected by quasiplateaus of suppressed conductivity.

In order to gain further insight into the emergence of quasiplateaus and additional sign change points in the nonlinear magnetoresistances, it is convenient to obtain an approximate analytical expression for the conductivity tensor. As the irregularities in the angular dependencies emerge when the magnetization is almost out of plane, we may treat the in-plane components of the unit magnetization vector as perturbative parameters in the quasiplateau-forming angular region. Inserting Eqs.~(\ref{S_v_qs}) and (\ref{vvavg}) into Eq.~(\ref{V_approx}), to first order in $m_x$ and $m_y$, the dressed velocity vertex function may be expressed as
\begin{equation}
\label{vertex_soln}
\begin{split}
\bs{\mathcal{V}}_{\mathbf{q}\sigma} \left(\epsilon, \epsilon + \hbar \omega\right)
&=
\mathcal{I}_{\sigma}\left( \bs{\eta}_{\mathbf{h}}, \eta_{ex}, m_z , \Gamma_{\alpha}, \Gamma_{I}; \omega \right)
\mathbf{v}_{\mathbf{q}\sigma}
\\
&+
v_F \mathbf{z} \times
\left[ \mathcal{J}_{\sigma}\left( \bs{\eta}_{\mathbf{h}}, \eta_{ex}, m_z , \Gamma_{\alpha}, \Gamma_{I}; \omega \right)
\bs{\eta}_{\mathbf{h}}
+
\eta_{ex} \mathcal{K}_{\sigma}\left( \bs{\eta}_{\mathbf{h}}, \eta_{ex}, m_z , \Gamma_{\alpha}, \Gamma_{I}; \omega \right)
\mathbf{m}
\right],
\end{split}
\end{equation}
where the dimensionless functions $\mathcal{I}_{\sigma}$, $\mathcal{J}_{\sigma}$ and $\mathcal{K}_{\sigma}$ are given by

\begin{subequations}
\begin{align}
\begin{split}
\mathcal{I}_{\sigma}\left( \bs{\eta}_{\mathbf{h}}, \eta_{ex}, m_z , \Gamma_{\alpha}, \Gamma_{I}; \omega \right)
&=
\frac{\Gamma _{\alpha} \left(1-\eta _{\text{ex}}^2 m_z^2 \right) \left[ 9 \left( 1 - \eta _{\text{ex}}^2 m_z^2 \right) + \eta _{\mathbf{h}}^2 \right]+\Gamma _I \left( 3 + \eta _{\text{ex}}^2 m_z^2 \right) \mp i \hbar \omega}
{8 \Gamma _{\alpha} \left(1 - \eta _{\text{ex}}^2 m_z^2 \right)^2 + 2 \Gamma _I \left( 1+ \eta _{\text{ex}}^2 m_z^2 \right) \mp i \hbar \omega},
\end{split}
\\%%%%%
\begin{split}
\mathcal{J}_{\sigma}\left( \bs{\eta}_{\mathbf{h}}, \eta_{ex}, m_z , \Gamma_{\alpha}, \Gamma_{I}; \omega \right)
&=
\frac{2 \Gamma _{\alpha} 
\left( 1 - \eta _{\text{ex}}^2 m_z^2 \right) 
\left( 1-\sigma  \eta _{\text{ex}} m_z \cos \theta_{\mathbf{h}} \right)}
{8 \Gamma _{\alpha } \left(1-\eta _{\text{ex}}^2 m_z^2 \right)^2 + 2 \Gamma _I \left( 1 + \eta _{\text{ex}}^2 m_z^2 \right) \mp i \hbar \omega},
\end{split}
\\%%%%%
\begin{split}
\mathcal{K}_{\sigma}\left( \bs{\eta}_{\mathbf{h}}, \eta_{ex}, m_z , \Gamma_{\alpha}, \Gamma_{I}; \omega \right)
&=
- \frac{4 \Gamma _{\alpha}
\left(1-\eta _{\text{ex}}^2 m_z^2 \right)}
{\left[ 8 \Gamma _{\alpha} 
\left( 1-\eta _{\text{ex}}^2 m_z^2 \right)^2 + 2 \Gamma _I \left(1 + \eta _{\text{ex}}^2 m_z^2 \right) \mp i \hbar \omega \right]^2}
\\
&\hspace{0.025\linewidth}\times
\Big\{
4 \Gamma _{\alpha}
\left( 1-\eta_{\text{ex}}^2 m_z^2 \right) 
\left( 1 - \eta _{\text{ex}}^2 m_z^2 - \eta_{\mathbf{h}}^2 \right)
\left( 1 - \sigma \eta _{\text{ex}} m_z \cos \theta_{\mathbf{h}} \right)
\\
&\hspace{0.025\linewidth}
- 2 \Gamma_I
\left[ 1- 3 \eta _{\text{ex}}^2 m_z^2 + \sigma \eta _{\text{ex}} m_z \left(3-\eta _{\text{ex}}^2 m_z^2 \right) 
\cos \theta_{\mathbf{h}}
- \sigma \eta_{\mathbf{h}} \left( 1 + \eta _{\text{ex}}^2 m_z^2 \right) \sin \theta_{\mathbf{h}}
\right]
\\
&\hspace{0.025\linewidth}
\mp i \hbar \omega \left[
1 - \sigma \eta _{\text{ex}} m_z \cos \theta_{\mathbf{h}} + \sigma  \eta_{\mathbf{h}} \sin \theta_{\mathbf{h}}
\right]
\Big\}.
\end{split}
\end{align}
\end{subequations}

Inserting this solution into Eq.~(\ref{S_sigma_ijj}), the longitudinal and transverse conductivities read
\begin{subequations}
\label{sigma_anal}
\begin{align}
\sigma_{xxx}
&=
-\frac{9 e^3 v_F}{32 \pi \epsilon_F^2} m_y
\,\mathcal{C} \left( \eta_{\alpha}, \eta_{I}, \eta_{ex}, m_z \right),
\\%%%%%%
\sigma_{yxx}
&=
\frac{3 e^3 v_F}{32 \pi \epsilon_F^2} m_x
\,\mathcal{C} \left( \eta_{\alpha}, \eta_{I}, \eta_{ex}, m_z \right),
\end{align}
\end{subequations}
where $\eta_{ex}$ is understood to be calculated at the Fermi level and the dimensionless function $\mathcal{C}$ is given by
\begin{equation}
\label{C_fun}
\begin{split}
\mathcal{C} \left( \eta_{\alpha}, \eta_{I}, \eta_{ex}, m_z \right)
&=
\eta _{\alpha} \eta_{ex}
\left( 1 - \eta _{\text{ex}}^2 m_z^2 \right)^2
\left[12 \eta_{\alpha}
\left( 1 - \eta _{\text{ex}}^2 m_z^2 \right)^2
+ \eta_I
\left( 1 - 3 \eta _{\text{ex}}^2 m_z^2 \right) \right] 
\\
&\times
\left[ 12 \eta_{\alpha}
\left( 1 - \eta _{\text{ex}}^2 m_z^2 \right)^2
+ \eta _I
\left( 3 + \eta _{\text{ex}}^2 m_z^2 \right) \right]^2
\\
&\times
\left[ 4 \eta_{\alpha}
\left( 1 - \eta _{\text{ex}}^2 m_z^2 \right)^2
+ \eta_I
\left( 1 + \eta _{\text{ex}}^2 m_z^2 \right) \right]^{-6},
\end{split}
\end{equation}

leading to the UMR coefficients
\begin{subequations}
\label{zeta_anal}
\begin{align}
\zeta _{\parallel }^{(2)}
&=
\frac{9 e^3 v_F}{16 \pi \sigma_D \epsilon_F^2} m_y
\,\mathcal{C} \left( \eta_{\alpha}, \eta_{I}, \eta_{ex}, m_z \right),
\\%%%%%%%%%
\zeta _{\perp}^{(2)}
&=
-\frac{3 e^3 v_F}{16 \pi \sigma_D \epsilon_F^2} m_x
\,\mathcal{C} \left( \eta_{\alpha}, \eta_{I}, \eta_{ex}, m_z \right).
\end{align}
\end{subequations}
Note that in this approximation, the ratio between the longitudinal and transverse UMR coefficients takes the simple form
\begin{equation}
\frac{\zeta _{\perp}^{(2)}}{\zeta _{\parallel}^{(2)}}
=
- \frac{1}{3} \cot \phi_{\mathbf{m}},
\end{equation}
where $\phi_{\mathbf{m}}$ is the angle between the electric field and magnetization, further confirming the common physical origin of the two nonlinear magnetoresistances.

Figs~\ref{fig_proximity_figS1a} and \ref{fig_proximity_figS1b} reveal that when an out-of-plane angular sweep of the magnetization is performed, there will be a total of 6 angles where the longitudinal or transverse UMR coefficients vanish. Two of these are the conventional angles where the in-plane magnetization vanishes ($\beta_{yz}= \pi/2, 3\pi/2$ for the UMR and $\beta_{xz}= \pi/2, 3\pi/2$ for the NPHE). The other four must than correspond to solutions of $\mathcal{C}=0$, which turn out to be $\beta_{yz}= \pi/2 \pm \theta_p/2, 3\pi/2 \pm \theta_p/2$, and similarly for $\beta_{xz}$, where $\theta_p$ is the angular arc over which the conductivities display a plateau-like profile and is given by $\theta_p = 2 \cos^{-1}[f(r)/\eta_{ex}]$, where $r=\eta_I/ \eta_{\alpha}$ and the dimensionless function $f(r)$ is
\begin{equation}
\label{theta_p}
f(r)
=
\sqrt{1 + \frac{r}{8}
\left(1- \sqrt{1+ \frac{32}{3r}}\right)}.
\end{equation}

\begin{figure}[tph]
{\includegraphics[width=0.5\linewidth,trim={0cm 0 0cm -1cm}]{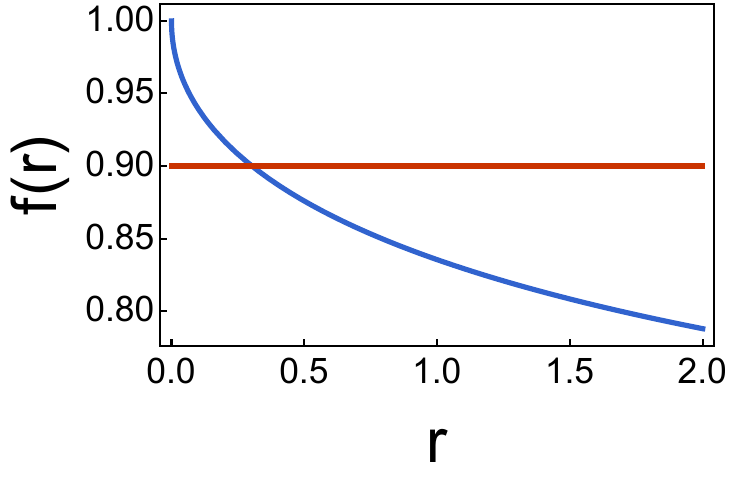}}%
    \caption{Plot of the monotonically decreasing function $f(r)$ (blue), with the red horizontal line indicating the value of $\eta_{ex}=0.9$ for comparison. The unconventional angular dependencies of the conductivities emerge when $\eta_{ex}>f(r)$.}
    \label{fig_proximity_figS3}
\end{figure}

For the parameters used in obtaining Figs~\ref{fig_proximity_figS1a} and \ref{fig_proximity_figS1b}, $\eta_{ex}=0.9$ and $r=2$, we find that $\theta_p \approx \pi/3$, which is quite sizable. From Eq. (\ref{theta_p}), we may deduce the requirement the system imposes for the additional sign changes of the nonlinear conductivity--and therefore the emergence of the quasiplateaus--to occur. From the condition that $\theta_p>0$, we conclude that the system must satisfy $\eta_{ex}>f(r)$. This requirement--or its lack thereof--is met by all the curves in Figs.~\ref{fig_proximity_fig3a} and \ref{fig_proximity_fig3c} as well as the ones presented in Fig.~\ref{fig_proximity_figS2}, further confirming the validity of the analytical approximation. In addition, as displayed in Fig.~\ref{fig_proximity_figS3}, $f(r)$ is a monotonically decreasing function. From this fact, and the requirement that $\eta_{ex}>f(r)$, we conclude that in order for the system to allow for additional sign changes of the nonlinear magnetoresistances and the subsequent emergence of plateau-like profiles, the Fermi level must be sufficiently close to the exchange energy and the density of the scalar scatterers must be sufficiently higher than that of the SOC impurities.

\section{Discussion and Conclusion}

In order to understand the physical origin of the unconventional angular dependencies of the UMR and NPHE, we note that this occurs for larger values of $\eta_{ex}$. In systems where the Fermi energy is fixed, it often suffices to consider terms only to first order in the magnetization, which typically leads to sinusoidal angular dependencies. In TIs, however, the tunability of the Fermi level implies that higher-order terms in the magnetization are expected to play an important role when $\eta_{ex} \sim 1$, thereby necessitating the cubic magnetization terms in Eqs.~(\ref{j^2_cubic}).

This is in contrast to nonlinear magnetotransport phenomena driven by hexagonal warping or particle-hole asymmetry \cite{he2018bilinear, he2019nonlinear}. In the case of warping, the six-fold symmetric deformation of the Fermi contour arises from the addition of a cubic-in-momentum term to the Dirac Hamiltonian, while particle-hole asymmetry adds a $k^2$ term to the Hamiltonian. Thus, both effects are dominant in the limit of high Fermi energy, i.e., when $\eta_{ex} \ll 1$, thereby enabling a simple way to distinguish their contributions from the ones presented in this chapter.

Another nonlinear transport effect which is stronger in the low Fermi energy limit is that generated by current-induced spin polarization. However, the contribution of this effect to the nonlinear transport can be distinguished from the mechanism predicted here by noting that current-induced spin polarization can only produce a longitudinal quadratic response with no quadratic planar Hall counterpart. This may readily be understood by noting that the emergence of a NPHE requires the magnetization to be parallel to the applied current, which, in turn, is perpendicular to the spin polarization. As a result, the magnetization cannot influence the spin polarization and no quadratic Hall response is generated~\cite{yasuda2016large, langenfeld2016exchange, sterk2019magnon, dyrdal2020spin}.

In addition, in terms of magnitude, the UMR coefficient strengths that we predict here are on the order of $10-100$ nm/V and are thus 1-2 orders of magnitude larger than the bilinear magnetoresistance effect predicted in nonmagnetic TIs \cite{dyrdal2020spin}. This can be attributed to the fact the the proximity-induced exchange interaction is much stronger than the Zeeman interaction for typical experimental magnetic field values, and suggests that TI/FI bilayers are generally a better platform for obtaining sizable nonlinear magnetotransport effects.

Yet another quadratic transport effect that can arise in TI/FI bilayers is the intrinsic nonlinear Hall effect generated by the Berry curvature dipole~\cite{sodemann2015quantum}. However, this relies on an out-of-plane magnetization, in contrast to the NPHE predicted in this chapter, which only arises when then magnetization has an in-plane component. Furthermore, the absence of a longitudinal UMR generated by the Berry curvature dipole provides an additional means to distinguish it from the effects predicted here.

A final point worth mentioning is that the minimal model we consider here for the Dirac surface states is quite general and is not limited to TI/FI bilayer systems. Thus, it is natural to expect that the nonlinear transport effects predicted in this chapter should, in principle, also arise in generally magnetized TI sysytems, including intrinsic magnetic TIs \cite{zhang2019topological, wang2020dynamical}, magnetically-doped TIs \cite{liu2009magnetic, abanin2011ordering, nomura2011surface} as well as in single TI layers in the presence of an applied magnetic field.

In conclusion, based on a formal evaluation of nonlinear Kubo formulas in the low-termperature limit, we have predicted a UMR and NPHE in a bilayer comprised of a TI and a FI which require no modification of the Dirac Hamiltonian--but instead--arise solely from extrinsic disorder scattering at the interface. Several key features and unique transport signatures have been identified, which enable the electric and magnetic tuning of the nonlinear magnetoresistance effects. We expect that this work will stimulate further theoretical and experimental studies of quantum transport in the nonlinear response regime, paving the way for potential future quantum spintronic applications.

%\begin{appendices}
%\include{appgaas}
%\include{appzgn}
%\end{appendices}
%\backmatter
%\appendix

% For your references
%TODO, add a sample Bibliography entry to ensure that this works
\bibliographystyle{nat_commune}
\bibliography{thesis.bib}

%merlin.mbs 2010-03-15 4.21a (PWD, AO, DPC)
\begin{thebibliography}{297}%
\makeatletter
\providecommand \@ifxundefined [1]{%
 \@ifx{#1\undefined}
}%
\providecommand \@ifnum [1]{%
 \ifnum #1\expandafter \@firstoftwo
 \else \expandafter \@secondoftwo
 \fi
}%
\providecommand \@ifx [1]{%
 \ifx #1\expandafter \@firstoftwo
 \else \expandafter \@secondoftwo
 \fi
}%
\providecommand \natexlab [1]{#1}%
\providecommand \enquote  [1]{``#1''}%
\providecommand \bibnamefont  [1]{#1}%
\providecommand \bibfnamefont [1]{#1}%
\providecommand \citenamefont [1]{#1}%
\providecommand \href@noop [0]{\@secondoftwo}%
\providecommand \href [0]{\begingroup \@sanitize@url \@href}%
\providecommand \@href[1]{\@@startlink{#1}\@@href}%
\providecommand \@@href[1]{\endgroup#1\@@endlink}%
\providecommand \@sanitize@url [0]{\catcode `\\12\catcode `\$12\catcode
  `\&12\catcode `\#12\catcode `\^12\catcode `\_12\catcode `\%12\relax}%
\providecommand \@@startlink[1]{}%
\providecommand \@@endlink[0]{}%
\providecommand \url  [0]{\begingroup\@sanitize@url \@url }%
\providecommand \@url [1]{\endgroup\@href {#1}{\urlprefix }}%
\providecommand \urlprefix  [0]{URL }%
\providecommand \Eprint [0]{\href }%
\@ifxundefined \urlstyle {%
  \providecommand \doi  [0]{\begingroup \@sanitize@url \@doi}%
  \providecommand \@doi [1]{\endgroup \@@startlink {\doibase
  #1}doi:\discretionary {}{}{}#1\@@endlink }%
}{%
  \providecommand \doi  [0]{doi:\discretionary{}{}{}\begingroup
  \urlstyle{rm}\Url }%
}%
\providecommand \doibase [0]{http://dx.doi.org/}%
\providecommand \Doi [0]{\begingroup \@sanitize@url \@Doi }%
\providecommand \@Doi  [1]{\endgroup\@@startlink{\doibase#1}\@@Doi}%
\providecommand \@@Doi [1]{#1\@@endlink}%
\providecommand \selectlanguage [0]{\@gobble}%
\providecommand \bibinfo  [0]{\@secondoftwo}%
\providecommand \bibfield  [0]{\@secondoftwo}%
\providecommand \translation [1]{[#1]}%
\providecommand \BibitemOpen [0]{}%
\providecommand \bibitemStop [0]{}%
\providecommand \bibitemNoStop [0]{.\EOS\space}%
\providecommand \EOS [0]{\spacefactor3000\relax}%
\providecommand \BibitemShut  [1]{\csname bibitem#1\endcsname}%
%</preamble>
\bibitem [{\citenamefont {Mehraeen}\ and\ \citenamefont
  {Gousheh}(2020)}]{mehraeen2020fermion}%
  \BibitemOpen
  \bibfield  {author} {\bibinfo {author} {\bibfnamefont {M.}~\bibnamefont
  {Mehraeen}}\ and\ \bibinfo {author} {\bibfnamefont {S.}~\bibnamefont
  {Gousheh}},\ }\bibfield  {title} {\enquote {\bibinfo {title} {{Fermion number
  1/2 of sphalerons and spectral mirror symmetry}},}\ }\href
  {https://doi.org/10.1140/epjc/s10052-020-08451-4} {\bibfield  {journal}
  {\bibinfo  {journal} {\emph {Eur. Phys. J. C}},\ }\textbf {\bibinfo {volume}
  {80}},\ \bibinfo {pages} {891}\  (\bibinfo {year} {2020})}\BibitemShut
  {NoStop}%
\bibitem [{\citenamefont {Abbaslu}\ \emph {et~al.}(2021)\citenamefont
  {Abbaslu}, \citenamefont {Rostam~Zadeh}, \citenamefont {Mehraeen},\ and\
  \citenamefont {Gousheh}}]{abbaslu2021generation}%
  \BibitemOpen
  \bibfield  {author} {\bibinfo {author} {\bibfnamefont {S.}~\bibnamefont
  {Abbaslu}}, \bibinfo {author} {\bibfnamefont {S.}~\bibnamefont
  {Rostam~Zadeh}}, \bibinfo {author} {\bibfnamefont {M.}~\bibnamefont
  {Mehraeen}}, \ and\ \bibinfo {author} {\bibfnamefont {S.}~\bibnamefont
  {Gousheh}},\ }\bibfield  {title} {\enquote {\bibinfo {title} {{The generation
  of matter--antimatter asymmetries and hypermagnetic fields by the chiral
  vortical effect of transient fluctuations}},}\ }\href
  {https://doi.org/10.1140/epjc/s10052-021-09272-9} {\bibfield  {journal}
  {\bibinfo  {journal} {\emph {Eur. Phys. J. C}},\ }\textbf {\bibinfo {volume}
  {81}},\ \bibinfo {pages} {1--14}\  (\bibinfo {year} {2021})}\BibitemShut
  {NoStop}%
\bibitem [{\citenamefont {Mehraeen}\ and\ \citenamefont
  {Zhang}(2022)}]{mehraeen2022spin}%
  \BibitemOpen
  \bibfield  {author} {\bibinfo {author} {\bibfnamefont {M.}~\bibnamefont
  {Mehraeen}}\ and\ \bibinfo {author} {\bibfnamefont {S.~S.-L.}\ \bibnamefont
  {Zhang}},\ }\bibfield  {title} {\enquote {\bibinfo {title} {Spin
  anomalous-\text{H}all unidirectional magnetoresistance},}\ }\href
  {https://link.aps.org/doi/10.1103/PhysRevB.105.184423} {\bibfield  {journal}
  {\bibinfo  {journal} {\emph {Phys. Rev. B}},\ }\textbf {\bibinfo {volume}
  {105}},\ \bibinfo {pages} {184423}\  (\bibinfo {year} {2022})}\BibitemShut
  {NoStop}%
\bibitem [{\citenamefont {Shim}\ \emph {et~al.}(2022)\citenamefont {Shim},
  \citenamefont {Mehraeen}, \citenamefont {Sklenar}, \citenamefont {Oh},
  \citenamefont {Gibbons}, \citenamefont {Saglam}, \citenamefont {Hoffmann},
  \citenamefont {Zhang},\ and\ \citenamefont {Mason}}]{shim2022unidirectional}%
  \BibitemOpen
  \bibfield  {author} {\bibinfo {author} {\bibfnamefont {S.}~\bibnamefont
  {Shim}}, \bibinfo {author} {\bibfnamefont {M.}~\bibnamefont {Mehraeen}},
  \bibinfo {author} {\bibfnamefont {J.}~\bibnamefont {Sklenar}}, \bibinfo
  {author} {\bibfnamefont {J.}~\bibnamefont {Oh}}, \bibinfo {author}
  {\bibfnamefont {J.}~\bibnamefont {Gibbons}}, \bibinfo {author} {\bibfnamefont
  {H.}~\bibnamefont {Saglam}}, \bibinfo {author} {\bibfnamefont
  {A.}~\bibnamefont {Hoffmann}}, \bibinfo {author} {\bibfnamefont {S.~S.-L.}\
  \bibnamefont {Zhang}}, \ and\ \bibinfo {author} {\bibfnamefont
  {N.}~\bibnamefont {Mason}},\ }\bibfield  {title} {\enquote {\bibinfo {title}
  {Unidirectional magnetoresistance in antiferromagnet/heavy-metal bilayers},}\
  }\href {https://link.aps.org/doi/10.1103/PhysRevX.12.021069} {\bibfield
  {journal} {\bibinfo  {journal} {\emph {Phys. Rev. X}},\ }\textbf {\bibinfo
  {volume} {12}},\ \bibinfo {pages} {021069}\  (\bibinfo {year}
  {2022})}\BibitemShut {NoStop}%
\bibitem [{\citenamefont {Zhang}\ \emph {et~al.}(2022)\citenamefont {Zhang},
  \citenamefont {Kalappattil}, \citenamefont {Liu}, \citenamefont {Mehraeen},
  \citenamefont {Zhang}, \citenamefont {Ding}, \citenamefont {Erugu},
  \citenamefont {Chen}, \citenamefont {Tian}, \citenamefont {Liu} \emph
  {et~al.}}]{zhang2022large}%
  \BibitemOpen
  \bibfield  {author} {\bibinfo {author} {\bibfnamefont {Y.}~\bibnamefont
  {Zhang}}, \bibinfo {author} {\bibfnamefont {V.}~\bibnamefont {Kalappattil}},
  \bibinfo {author} {\bibfnamefont {C.}~\bibnamefont {Liu}}, \bibinfo {author}
  {\bibfnamefont {M.}~\bibnamefont {Mehraeen}}, \bibinfo {author}
  {\bibfnamefont {S.~S.-L.}\ \bibnamefont {Zhang}}, \bibinfo {author}
  {\bibfnamefont {J.}~\bibnamefont {Ding}}, \bibinfo {author} {\bibfnamefont
  {U.}~\bibnamefont {Erugu}}, \bibinfo {author} {\bibfnamefont
  {Z.}~\bibnamefont {Chen}}, \bibinfo {author} {\bibfnamefont {J.}~\bibnamefont
  {Tian}}, \bibinfo {author} {\bibfnamefont {K.}~\bibnamefont {Liu}},  \emph
  {et~al.},\ }\bibfield  {title} {\enquote {\bibinfo {title} {Large
  magnetoelectric resistance in the topological \text{D}irac semimetal
  $\alpha$-\text{S}n},}\ }\href {https://doi.org/10.1126/sciadv.abo0052}
  {\bibfield  {journal} {\bibinfo  {journal} {\emph {Sci. Adv.}},\ }\textbf
  {\bibinfo {volume} {8}},\ \bibinfo {pages} {eabo0052}\  (\bibinfo {year}
  {2022})}\BibitemShut {NoStop}%
\bibitem [{\citenamefont {Mehraeen}\ \emph {et~al.}(2023)\citenamefont
  {Mehraeen}, \citenamefont {Shen},\ and\ \citenamefont
  {Zhang}}]{mehraeen2023quantum}%
  \BibitemOpen
  \bibfield  {author} {\bibinfo {author} {\bibfnamefont {M.}~\bibnamefont
  {Mehraeen}}, \bibinfo {author} {\bibfnamefont {P.}~\bibnamefont {Shen}}, \
  and\ \bibinfo {author} {\bibfnamefont {S.~S.-L.}\ \bibnamefont {Zhang}},\
  }\bibfield  {title} {\enquote {\bibinfo {title} {Quantum unidirectional
  magnetoresistance},}\ }\href
  {https://link.aps.org/doi/10.1103/PhysRevB.108.014411} {\bibfield  {journal}
  {\bibinfo  {journal} {\emph {Phys. Rev. B}},\ }\textbf {\bibinfo {volume}
  {108}},\ \bibinfo {pages} {014411}\  (\bibinfo {year} {2023})}\BibitemShut
  {NoStop}%
\bibitem [{\citenamefont {Mehraeen}\ and\ \citenamefont
  {Zhang}(2024)}]{mehraeen2024proximity}%
  \BibitemOpen
  \bibfield  {author} {\bibinfo {author} {\bibfnamefont {M.}~\bibnamefont
  {Mehraeen}}\ and\ \bibinfo {author} {\bibfnamefont {S.~S.-L.}\ \bibnamefont
  {Zhang}},\ }\bibfield  {title} {\enquote {\bibinfo {title} {Proximity-induced
  nonlinear magnetoresistances on topological insulators},}\ }\href
  {https://link.aps.org/doi/10.1103/PhysRevB.109.024421} {\bibfield  {journal}
  {\bibinfo  {journal} {\emph {Phys. Rev. B}},\ }\textbf {\bibinfo {volume}
  {109}},\ \bibinfo {pages} {024421}\  (\bibinfo {year} {2024})}\BibitemShut
  {NoStop}%
\bibitem [{\citenamefont {Damerio}\ \emph {et~al.}(2024)\citenamefont
  {Damerio}, \citenamefont {Sunil}, \citenamefont {Janus}, \citenamefont
  {Mehraeen}, \citenamefont {Zhang},\ and\ \citenamefont
  {Avci}}]{damerio2024magnetoresistive}%
  \BibitemOpen
  \bibfield  {author} {\bibinfo {author} {\bibfnamefont {S.}~\bibnamefont
  {Damerio}}, \bibinfo {author} {\bibfnamefont {A.}~\bibnamefont {Sunil}},
  \bibinfo {author} {\bibfnamefont {W.}~\bibnamefont {Janus}}, \bibinfo
  {author} {\bibfnamefont {M.}~\bibnamefont {Mehraeen}}, \bibinfo {author}
  {\bibfnamefont {S.~S.-L.}\ \bibnamefont {Zhang}}, \ and\ \bibinfo {author}
  {\bibfnamefont {C.~O.}\ \bibnamefont {Avci}},\ }\bibfield  {title} {\enquote
  {\bibinfo {title} {{Magnetoresistive detection of perpendicular switching in
  a magnetic insulator}},}\ }\href {https://doi.org/10.1038/s42005-024-01604-x}
  {\bibfield  {journal} {\bibinfo  {journal} {\emph {Commun. Phys.}},\ }\textbf
  {\bibinfo {volume} {7}},\ \bibinfo {pages} {114}\  (\bibinfo {year}
  {2024})}\BibitemShut {NoStop}%
\bibitem [{\citenamefont {Mehraeen}(2024)}]{mehraeen2024quantum}%
  \BibitemOpen
  \bibfield  {author} {\bibinfo {author} {\bibfnamefont {M.}~\bibnamefont
  {Mehraeen}},\ }\bibfield  {title} {\enquote {\bibinfo {title} {Quantum
  kinetic theory of quadratic responses},}\ }\href
  {https://link.aps.org/doi/10.1103/PhysRevB.110.174423} {\bibfield  {journal}
  {\bibinfo  {journal} {\emph {Phys. Rev. B}},\ }\textbf {\bibinfo {volume}
  {110}},\ \bibinfo {pages} {174423}\  (\bibinfo {year} {2024})}\BibitemShut
  {NoStop}%
\bibitem [{\citenamefont {Shim}\ \emph {et~al.}(2025)\citenamefont {Shim},
  \citenamefont {Mehraeen}, \citenamefont {Sklenar}, \citenamefont {Zhang},
  \citenamefont {Hoffmann},\ and\ \citenamefont {Mason}}]{shim2025spin}%
  \BibitemOpen
  \bibfield  {author} {\bibinfo {author} {\bibfnamefont {S.}~\bibnamefont
  {Shim}}, \bibinfo {author} {\bibfnamefont {M.}~\bibnamefont {Mehraeen}},
  \bibinfo {author} {\bibfnamefont {J.}~\bibnamefont {Sklenar}}, \bibinfo
  {author} {\bibfnamefont {S.~S.-L.}\ \bibnamefont {Zhang}}, \bibinfo {author}
  {\bibfnamefont {A.}~\bibnamefont {Hoffmann}}, \ and\ \bibinfo {author}
  {\bibfnamefont {N.}~\bibnamefont {Mason}},\ }\bibfield  {title} {\enquote
  {\bibinfo {title} {{Spin-polarized antiferromagnetic metals}},}\ }\href
  {https://doi.org/10.1146/annurev-conmatphys-042924-123620} {\bibfield
  {journal} {\bibinfo  {journal} {\emph {Annu. Rev. Condens. Matter Phys.}},\
  }\textbf {\bibinfo {volume} {16}}\  (\bibinfo {year} {2025})}\BibitemShut
  {NoStop}%
\bibitem [{\citenamefont {Mehraeen}(2025)}]{mehraeen2025quantum}%
  \BibitemOpen
  \bibfield  {author} {\bibinfo {author} {\bibfnamefont {M.}~\bibnamefont
  {Mehraeen}},\ }\bibfield  {title} {\enquote {\bibinfo {title} {{Quantum
  response theory and momentum-space gravity}},}\ }\href
  {https://arxiv.org/abs/2503.06160} {\bibfield  {journal} {\bibinfo  {journal}
  {\emph {arXiv:2503.06160}}}\  (\bibinfo {year} {2025})}\BibitemShut {NoStop}%
\bibitem [{\citenamefont {Jain}\ \emph {et~al.}(2025)\citenamefont {Jain},
  \citenamefont {Jankowski}, \citenamefont {Mehraeen},\ and\ \citenamefont
  {Slager}}]{jain2025nonlinear}%
  \BibitemOpen
  \bibfield  {author} {\bibinfo {author} {\bibfnamefont {A.}~\bibnamefont
  {Jain}}, \bibinfo {author} {\bibfnamefont {W.~J.}\ \bibnamefont {Jankowski}},
  \bibinfo {author} {\bibfnamefont {M.}~\bibnamefont {Mehraeen}}, \ and\
  \bibinfo {author} {\bibfnamefont {R.-J.}\ \bibnamefont {Slager}},\ }\bibfield
   {title} {\enquote {\bibinfo {title} {{Nonlinear Odd Viscoelastic Effect}},}\
  }\href {https://arxiv.org/abs/2511.22706} {\bibfield  {journal} {\bibinfo
  {journal} {\emph {arXiv:2511.22706}}}\  (\bibinfo {year} {2025})}\BibitemShut
  {NoStop}%
\bibitem [{\citenamefont {Jain}\ \emph {et~al.}(2026)\citenamefont {Jain},
  \citenamefont {Jankowski}, \citenamefont {Mehraeen},\ and\ \citenamefont
  {Slager}}]{jain2026topological}%
  \BibitemOpen
  \bibfield  {author} {\bibinfo {author} {\bibfnamefont {A.}~\bibnamefont
  {Jain}}, \bibinfo {author} {\bibfnamefont {W.~J.}\ \bibnamefont {Jankowski}},
  \bibinfo {author} {\bibfnamefont {M.}~\bibnamefont {Mehraeen}}, \ and\
  \bibinfo {author} {\bibfnamefont {R.-J.}\ \bibnamefont {Slager}},\ }\bibfield
   {title} {\enquote {\bibinfo {title} {{Topological Acoustic Diode}},}\ }\href
  {https://arxiv.org/abs/2601.20951} {\bibfield  {journal} {\bibinfo  {journal}
  {\emph {arXiv:2601.20951}}}\  (\bibinfo {year} {2026})}\BibitemShut {NoStop}%
\bibitem [{\citenamefont {Mihai~Miron}\ \emph {et~al.}(2010)\citenamefont
  {Mihai~Miron}, \citenamefont {Gaudin}, \citenamefont {Auffret}, \citenamefont
  {Rodmacq}, \citenamefont {Schuhl}, \citenamefont {Pizzini}, \citenamefont
  {Vogel},\ and\ \citenamefont {Gambardella}}]{mihai2010current}%
  \BibitemOpen
  \bibfield  {author} {\bibinfo {author} {\bibfnamefont {I.}~\bibnamefont
  {Mihai~Miron}}, \bibinfo {author} {\bibfnamefont {G.}~\bibnamefont {Gaudin}},
  \bibinfo {author} {\bibfnamefont {S.}~\bibnamefont {Auffret}}, \bibinfo
  {author} {\bibfnamefont {B.}~\bibnamefont {Rodmacq}}, \bibinfo {author}
  {\bibfnamefont {A.}~\bibnamefont {Schuhl}}, \bibinfo {author} {\bibfnamefont
  {S.}~\bibnamefont {Pizzini}}, \bibinfo {author} {\bibfnamefont
  {J.}~\bibnamefont {Vogel}}, \ and\ \bibinfo {author} {\bibfnamefont
  {P.}~\bibnamefont {Gambardella}},\ }\bibfield  {title} {\enquote {\bibinfo
  {title} {Current-driven spin torque induced by the \text{R}ashba effect in a
  ferromagnetic metal layer},}\ }\href {https://doi.org/10.1038/nmat2613}
  {\bibfield  {journal} {\bibinfo  {journal} {\emph {Nature Mater.}},\ }\textbf
  {\bibinfo {volume} {9}},\ \bibinfo {pages} {230--234}\  (\bibinfo {year}
  {2010})}\BibitemShut {NoStop}%
\bibitem [{\citenamefont {Park}\ \emph {et~al.}(2013)\citenamefont {Park},
  \citenamefont {Kim}, \citenamefont {Lee},\ and\ \citenamefont
  {Han}}]{park2013orbital}%
  \BibitemOpen
  \bibfield  {author} {\bibinfo {author} {\bibfnamefont {J.-H.}\ \bibnamefont
  {Park}}, \bibinfo {author} {\bibfnamefont {C.~H.}\ \bibnamefont {Kim}},
  \bibinfo {author} {\bibfnamefont {H.-W.}\ \bibnamefont {Lee}}, \ and\
  \bibinfo {author} {\bibfnamefont {J.~H.}\ \bibnamefont {Han}},\ }\bibfield
  {title} {\enquote {\bibinfo {title} {Orbital chirality and \text{R}ashba
  interaction in magnetic bands},}\ }\href
  {https://doi.org/10.1103/PhysRevB.87.041301} {\bibfield  {journal} {\bibinfo
  {journal} {\emph {Phys. Rev. B}},\ }\textbf {\bibinfo {volume} {87}},\
  \bibinfo {pages} {041301}\  (\bibinfo {year} {2013})}\BibitemShut {NoStop}%
\bibitem [{\citenamefont {Tokatly}\ \emph
  {et~al.}(2015){\natexlab{a}}\citenamefont {Tokatly}, \citenamefont
  {Krasovskii},\ and\ \citenamefont {Vignale}}]{tokatly2015current}%
  \BibitemOpen
  \bibfield  {author} {\bibinfo {author} {\bibfnamefont {I.}~\bibnamefont
  {Tokatly}}, \bibinfo {author} {\bibfnamefont {E.}~\bibnamefont {Krasovskii}},
  \ and\ \bibinfo {author} {\bibfnamefont {G.}~\bibnamefont {Vignale}},\
  }\bibfield  {title} {\enquote {\bibinfo {title} {Current-induced spin
  polarization at the surface of metallic films: A theorem and an ab initio
  calculation},}\ }\href {https://doi.org/10.1103/PhysRevB.91.035403}
  {\bibfield  {journal} {\bibinfo  {journal} {\emph {Phys. Rev. B}},\ }\textbf
  {\bibinfo {volume} {91}},\ \bibinfo {pages} {035403}\  (\bibinfo {year}
  {2015}{\natexlab{a}})}\BibitemShut {NoStop}%
\bibitem [{\citenamefont {Grytsyuk}\ \emph {et~al.}(2016)\citenamefont
  {Grytsyuk}, \citenamefont {Belabbes}, \citenamefont {Haney}, \citenamefont
  {Lee}, \citenamefont {Lee}, \citenamefont {Stiles}, \citenamefont
  {Schwingenschl{\"o}gl},\ and\ \citenamefont {Manchon}}]{grytsyuk2016k}%
  \BibitemOpen
  \bibfield  {author} {\bibinfo {author} {\bibfnamefont {S.}~\bibnamefont
  {Grytsyuk}}, \bibinfo {author} {\bibfnamefont {A.}~\bibnamefont {Belabbes}},
  \bibinfo {author} {\bibfnamefont {P.~M.}\ \bibnamefont {Haney}}, \bibinfo
  {author} {\bibfnamefont {H.-W.}\ \bibnamefont {Lee}}, \bibinfo {author}
  {\bibfnamefont {K.-J.}\ \bibnamefont {Lee}}, \bibinfo {author} {\bibfnamefont
  {M.~D.}\ \bibnamefont {Stiles}}, \bibinfo {author} {\bibfnamefont
  {U.}~\bibnamefont {Schwingenschl{\"o}gl}}, \ and\ \bibinfo {author}
  {\bibfnamefont {A.}~\bibnamefont {Manchon}},\ }\bibfield  {title} {\enquote
  {\bibinfo {title} {k-asymmetric spin splitting at the interface between
  transition metal ferromagnets and heavy metals},}\ }\href
  {https://doi.org/10.1103/PhysRevB.93.174421} {\bibfield  {journal} {\bibinfo
  {journal} {\emph {Phys. Rev. B}},\ }\textbf {\bibinfo {volume} {93}},\
  \bibinfo {pages} {174421}\  (\bibinfo {year} {2016})}\BibitemShut {NoStop}%
\bibitem [{\citenamefont {Kajiwara}\ \emph {et~al.}(2010)\citenamefont
  {Kajiwara}, \citenamefont {Harii}, \citenamefont {Takahashi}, \citenamefont
  {Ohe}, \citenamefont {Uchida}, \citenamefont {Mizuguchi}, \citenamefont
  {Umezawa}, \citenamefont {Kawai}, \citenamefont {Ando}, \citenamefont
  {Takanashi} \emph {et~al.}}]{kajiwara2010transmission}%
  \BibitemOpen
  \bibfield  {author} {\bibinfo {author} {\bibfnamefont {Y.}~\bibnamefont
  {Kajiwara}}, \bibinfo {author} {\bibfnamefont {K.}~\bibnamefont {Harii}},
  \bibinfo {author} {\bibfnamefont {S.}~\bibnamefont {Takahashi}}, \bibinfo
  {author} {\bibfnamefont {J.-i.}\ \bibnamefont {Ohe}}, \bibinfo {author}
  {\bibfnamefont {K.}~\bibnamefont {Uchida}}, \bibinfo {author} {\bibfnamefont
  {M.}~\bibnamefont {Mizuguchi}}, \bibinfo {author} {\bibfnamefont
  {H.}~\bibnamefont {Umezawa}}, \bibinfo {author} {\bibfnamefont
  {H.}~\bibnamefont {Kawai}}, \bibinfo {author} {\bibfnamefont
  {K.}~\bibnamefont {Ando}}, \bibinfo {author} {\bibfnamefont {K.}~\bibnamefont
  {Takanashi}},  \emph {et~al.},\ }\bibfield  {title} {\enquote {\bibinfo
  {title} {Transmission of electrical signals by spin-wave interconversion in a
  magnetic insulator},}\ }\href {https://doi.org/10.1038/nature08876}
  {\bibfield  {journal} {\bibinfo  {journal} {\emph {Nature}},\ }\textbf
  {\bibinfo {volume} {464}},\ \bibinfo {pages} {262--266}\  (\bibinfo {year}
  {2010})}\BibitemShut {NoStop}%
\bibitem [{\citenamefont {Qi}\ and\ \citenamefont
  {Zhang}(2011)}]{qi2011topological}%
  \BibitemOpen
  \bibfield  {author} {\bibinfo {author} {\bibfnamefont {X.-L.}\ \bibnamefont
  {Qi}}\ and\ \bibinfo {author} {\bibfnamefont {S.-C.}\ \bibnamefont {Zhang}},\
  }\bibfield  {title} {\enquote {\bibinfo {title} {Topological insulators and
  superconductors},}\ }\href
  {https://link.aps.org/doi/10.1103/RevModPhys.83.1057} {\bibfield  {journal}
  {\bibinfo  {journal} {\emph {Rev. Mod. Phys.}},\ }\textbf {\bibinfo {volume}
  {83}},\ \bibinfo {pages} {1057--1110}\  (\bibinfo {year} {2011})}\BibitemShut
  {NoStop}%
\bibitem [{\citenamefont {Kubo}(1957)}]{kubo1957statistical}%
  \BibitemOpen
  \bibfield  {author} {\bibinfo {author} {\bibfnamefont {R.}~\bibnamefont
  {Kubo}},\ }\bibfield  {title} {\enquote {\bibinfo {title}
  {Statistical-mechanical theory of irreversible processes. i. general theory
  and simple applications to magnetic and conduction problems},}\ }\href
  {https://doi.org/10.1143/JPSJ.12.570} {\bibfield  {journal} {\bibinfo
  {journal} {\emph {J. Phys. Soc. Jpn.}},\ }\textbf {\bibinfo {volume} {12}},\
  \bibinfo {pages} {570--586}\  (\bibinfo {year} {1957})}\BibitemShut {NoStop}%
\bibitem [{\citenamefont {Keldysh}(2024)}]{keldysh2024diagram}%
  \BibitemOpen
  \bibfield  {author} {\bibinfo {author} {\bibfnamefont {L.~V.}\ \bibnamefont
  {Keldysh}},\ }\bibfield  {title} {\enquote {\bibinfo {title} {{Diagram
  technique for nonequilibrium processes}},}\ }in\ \href@noop {} {\emph
  {\bibinfo {booktitle} {{Selected Papers of Leonid V Keldysh}}}}\ (\bibinfo
  {publisher} {World Scientific},\ \bibinfo {year} {2024})\ pp.\ \bibinfo
  {pages} {47--55}\BibitemShut {NoStop}%
\bibitem [{\citenamefont {Karplus}\ and\ \citenamefont
  {Luttinger}(1954)}]{karplus1954hall}%
  \BibitemOpen
  \bibfield  {author} {\bibinfo {author} {\bibfnamefont {R.}~\bibnamefont
  {Karplus}}\ and\ \bibinfo {author} {\bibfnamefont {J.~M.}\ \bibnamefont
  {Luttinger}},\ }\bibfield  {title} {\enquote {\bibinfo {title} {{Hall Effect
  in Ferromagnetics}},}\ }\href
  {https://link.aps.org/doi/10.1103/PhysRev.95.1154} {\bibfield  {journal}
  {\bibinfo  {journal} {\emph {Phys. Rev.}},\ }\textbf {\bibinfo {volume}
  {95}},\ \bibinfo {pages} {1154--1160}\  (\bibinfo {year} {1954})}\BibitemShut
  {NoStop}%
\bibitem [{\citenamefont {Kohn}\ and\ \citenamefont
  {Luttinger}(1957)}]{kohn1957quantum}%
  \BibitemOpen
  \bibfield  {author} {\bibinfo {author} {\bibfnamefont {W.}~\bibnamefont
  {Kohn}}\ and\ \bibinfo {author} {\bibfnamefont {J.~M.}\ \bibnamefont
  {Luttinger}},\ }\bibfield  {title} {\enquote {\bibinfo {title} {{Quantum
  Theory of Electrical Transport Phenomena}},}\ }\href
  {https://link.aps.org/doi/10.1103/PhysRev.108.590} {\bibfield  {journal}
  {\bibinfo  {journal} {\emph {Phys. Rev.}},\ }\textbf {\bibinfo {volume}
  {108}},\ \bibinfo {pages} {590--611}\  (\bibinfo {year} {1957})}\BibitemShut
  {NoStop}%
\bibitem [{\citenamefont {Luttinger}\ and\ \citenamefont
  {Kohn}(1958)}]{luttinger1958quantum}%
  \BibitemOpen
  \bibfield  {author} {\bibinfo {author} {\bibfnamefont {J.~M.}\ \bibnamefont
  {Luttinger}}\ and\ \bibinfo {author} {\bibfnamefont {W.}~\bibnamefont
  {Kohn}},\ }\bibfield  {title} {\enquote {\bibinfo {title} {{Quantum Theory of
  Electrical Transport Phenomena. II}},}\ }\href
  {https://link.aps.org/doi/10.1103/PhysRev.109.1892} {\bibfield  {journal}
  {\bibinfo  {journal} {\emph {Phys. Rev.}},\ }\textbf {\bibinfo {volume}
  {109}},\ \bibinfo {pages} {1892--1909}\  (\bibinfo {year}
  {1958})}\BibitemShut {NoStop}%
\bibitem [{\citenamefont {Luttinger}(1958)}]{luttinger1958theory}%
  \BibitemOpen
  \bibfield  {author} {\bibinfo {author} {\bibfnamefont {J.~M.}\ \bibnamefont
  {Luttinger}},\ }\bibfield  {title} {\enquote {\bibinfo {title} {Theory of the
  \text{H}all effect in ferromagnetic substances},}\ }\href
  {https://link.aps.org/doi/10.1103/PhysRev.112.739} {\bibfield  {journal}
  {\bibinfo  {journal} {\emph {Phys. Rev.}},\ }\textbf {\bibinfo {volume}
  {112}},\ \bibinfo {pages} {739--751}\  (\bibinfo {year} {1958})}\BibitemShut
  {NoStop}%
\bibitem [{\citenamefont {Sinitsyn}\ \emph {et~al.}(2005)\citenamefont
  {Sinitsyn}, \citenamefont {Niu}, \citenamefont {Sinova},\ and\ \citenamefont
  {Nomura}}]{sinitsyn2005disorder}%
  \BibitemOpen
  \bibfield  {author} {\bibinfo {author} {\bibfnamefont {N.~A.}\ \bibnamefont
  {Sinitsyn}}, \bibinfo {author} {\bibfnamefont {Q.}~\bibnamefont {Niu}},
  \bibinfo {author} {\bibfnamefont {J.}~\bibnamefont {Sinova}}, \ and\ \bibinfo
  {author} {\bibfnamefont {K.}~\bibnamefont {Nomura}},\ }\bibfield  {title}
  {\enquote {\bibinfo {title} {{Disorder effects in the anomalous Hall effect
  induced by Berry curvature}},}\ }\href
  {https://link.aps.org/doi/10.1103/PhysRevB.72.045346} {\bibfield  {journal}
  {\bibinfo  {journal} {\emph {Phys. Rev. B}},\ }\textbf {\bibinfo {volume}
  {72}},\ \bibinfo {pages} {045346}\  (\bibinfo {year} {2005})}\BibitemShut
  {NoStop}%
\bibitem [{\citenamefont {Sinitsyn}\ \emph {et~al.}(2007)\citenamefont
  {Sinitsyn}, \citenamefont {MacDonald}, \citenamefont {Jungwirth},
  \citenamefont {Dugaev},\ and\ \citenamefont
  {Sinova}}]{sinitsyn2007anomalous}%
  \BibitemOpen
  \bibfield  {author} {\bibinfo {author} {\bibfnamefont {N.~A.}\ \bibnamefont
  {Sinitsyn}}, \bibinfo {author} {\bibfnamefont {A.~H.}\ \bibnamefont
  {MacDonald}}, \bibinfo {author} {\bibfnamefont {T.}~\bibnamefont
  {Jungwirth}}, \bibinfo {author} {\bibfnamefont {V.~K.}\ \bibnamefont
  {Dugaev}}, \ and\ \bibinfo {author} {\bibfnamefont {J.}~\bibnamefont
  {Sinova}},\ }\bibfield  {title} {\enquote {\bibinfo {title} {{Anomalous Hall
  effect in a two-dimensional Dirac band: The link between the Kubo-Streda
  formula and the semiclassical Boltzmann equation approach}},}\ }\href
  {https://link.aps.org/doi/10.1103/PhysRevB.75.045315} {\bibfield  {journal}
  {\bibinfo  {journal} {\emph {Phys. Rev. B}},\ }\textbf {\bibinfo {volume}
  {75}},\ \bibinfo {pages} {045315}\  (\bibinfo {year} {2007})}\BibitemShut
  {NoStop}%
\bibitem [{\citenamefont {Sinitsyn}(2007)}]{sinitsyn2007semiclassical}%
  \BibitemOpen
  \bibfield  {author} {\bibinfo {author} {\bibfnamefont {N.}~\bibnamefont
  {Sinitsyn}},\ }\bibfield  {title} {\enquote {\bibinfo {title} {Semiclassical
  theories of the anomalous \text{H}all effect},}\ }\href
  {https://iopscience.iop.org/article/10.1088/0953-8984/20/02/023201/meta?casa_token=1NQD_xlIEYQAAAAA:RwHn1TnjQG64Xi4VGds8xzxrnJluOiSQDF2aMhI_9oCdxe6bUsv4VlJhSUVtHHUIjd5kfjb_Vpw}
  {\bibfield  {journal} {\bibinfo  {journal} {\emph {J. Phys.: Condens.
  Matter}},\ }\textbf {\bibinfo {volume} {20}},\ \bibinfo {pages} {023201}\
  (\bibinfo {year} {2007})}\BibitemShut {NoStop}%
\bibitem [{\citenamefont {Culcer}\ \emph {et~al.}(2017)\citenamefont {Culcer},
  \citenamefont {Sekine},\ and\ \citenamefont
  {MacDonald}}]{culcer2017interband}%
  \BibitemOpen
  \bibfield  {author} {\bibinfo {author} {\bibfnamefont {D.}~\bibnamefont
  {Culcer}}, \bibinfo {author} {\bibfnamefont {A.}~\bibnamefont {Sekine}}, \
  and\ \bibinfo {author} {\bibfnamefont {A.~H.}\ \bibnamefont {MacDonald}},\
  }\bibfield  {title} {\enquote {\bibinfo {title} {Interband coherence response
  to electric fields in crystals: \text{B}erry-phase contributions and disorder
  effects},}\ }\href {https://link.aps.org/doi/10.1103/PhysRevB.96.035106}
  {\bibfield  {journal} {\bibinfo  {journal} {\emph {Phys. Rev. B}},\ }\textbf
  {\bibinfo {volume} {96}},\ \bibinfo {pages} {035106}\  (\bibinfo {year}
  {2017})}\BibitemShut {NoStop}%
\bibitem [{\citenamefont {Sekine}\ \emph {et~al.}(2017)\citenamefont {Sekine},
  \citenamefont {Culcer},\ and\ \citenamefont {MacDonald}}]{sekine2017quantum}%
  \BibitemOpen
  \bibfield  {author} {\bibinfo {author} {\bibfnamefont {A.}~\bibnamefont
  {Sekine}}, \bibinfo {author} {\bibfnamefont {D.}~\bibnamefont {Culcer}}, \
  and\ \bibinfo {author} {\bibfnamefont {A.~H.}\ \bibnamefont {MacDonald}},\
  }\bibfield  {title} {\enquote {\bibinfo {title} {Quantum kinetic theory of
  the chiral anomaly},}\ }\href
  {https://link.aps.org/doi/10.1103/PhysRevB.96.235134} {\bibfield  {journal}
  {\bibinfo  {journal} {\emph {Phys. Rev. B}},\ }\textbf {\bibinfo {volume}
  {96}},\ \bibinfo {pages} {235134}\  (\bibinfo {year} {2017})}\BibitemShut
  {NoStop}%
\bibitem [{\citenamefont {Xiao}\ and\ \citenamefont
  {Niu}(2017)}]{xiao2017semiclassical}%
  \BibitemOpen
  \bibfield  {author} {\bibinfo {author} {\bibfnamefont {C.}~\bibnamefont
  {Xiao}}\ and\ \bibinfo {author} {\bibfnamefont {Q.}~\bibnamefont {Niu}},\
  }\bibfield  {title} {\enquote {\bibinfo {title} {Semiclassical theory of
  spin-orbit torques in disordered multiband electron systems},}\ }\href
  {https://link.aps.org/doi/10.1103/PhysRevB.96.045428} {\bibfield  {journal}
  {\bibinfo  {journal} {\emph {Phys. Rev. B}},\ }\textbf {\bibinfo {volume}
  {96}},\ \bibinfo {pages} {045428}\  (\bibinfo {year} {2017})}\BibitemShut
  {NoStop}%
\bibitem [{\citenamefont {Xiao}\ \emph
  {et~al.}(2019){\natexlab{a}}\citenamefont {Xiao}, \citenamefont {Liu},
  \citenamefont {Yuan}, \citenamefont {Yang},\ and\ \citenamefont
  {Niu}}]{xiao2019temperature}%
  \BibitemOpen
  \bibfield  {author} {\bibinfo {author} {\bibfnamefont {C.}~\bibnamefont
  {Xiao}}, \bibinfo {author} {\bibfnamefont {Y.}~\bibnamefont {Liu}}, \bibinfo
  {author} {\bibfnamefont {Z.}~\bibnamefont {Yuan}}, \bibinfo {author}
  {\bibfnamefont {S.~A.}\ \bibnamefont {Yang}}, \ and\ \bibinfo {author}
  {\bibfnamefont {Q.}~\bibnamefont {Niu}},\ }\bibfield  {title} {\enquote
  {\bibinfo {title} {{Temperature dependence of the side-jump spin Hall
  conductivity}},}\ }\href
  {https://link.aps.org/doi/10.1103/PhysRevB.100.085425} {\bibfield  {journal}
  {\bibinfo  {journal} {\emph {Phys. Rev. B}},\ }\textbf {\bibinfo {volume}
  {100}},\ \bibinfo {pages} {085425}\  (\bibinfo {year}
  {2019}{\natexlab{a}})}\BibitemShut {NoStop}%
\bibitem [{\citenamefont {Atencia}\ \emph {et~al.}(2022)\citenamefont
  {Atencia}, \citenamefont {Niu},\ and\ \citenamefont
  {Culcer}}]{atencia2022semiclassical}%
  \BibitemOpen
  \bibfield  {author} {\bibinfo {author} {\bibfnamefont {R.~B.}\ \bibnamefont
  {Atencia}}, \bibinfo {author} {\bibfnamefont {Q.}~\bibnamefont {Niu}}, \ and\
  \bibinfo {author} {\bibfnamefont {D.}~\bibnamefont {Culcer}},\ }\bibfield
  {title} {\enquote {\bibinfo {title} {Semiclassical response of disordered
  conductors: Extrinsic carrier velocity and spin and field-corrected collision
  integral},}\ }\href
  {https://link.aps.org/doi/10.1103/PhysRevResearch.4.013001} {\bibfield
  {journal} {\bibinfo  {journal} {\emph {Phys. Rev. Res.}},\ }\textbf {\bibinfo
  {volume} {4}},\ \bibinfo {pages} {013001}\  (\bibinfo {year}
  {2022})}\BibitemShut {NoStop}%
\bibitem [{\citenamefont {Mahan}(2000)}]{mahan2000many}%
  \BibitemOpen
  \bibfield  {author} {\bibinfo {author} {\bibfnamefont {G.~D.}\ \bibnamefont
  {Mahan}},\ }\href@noop {} {\emph {\bibinfo {title} {\textit{Many Particle
  Physics, Third Edition}}}}\ (\bibinfo  {publisher} {Plenum},\ \bibinfo
  {address} {New York},\ \bibinfo {year} {2000})\BibitemShut {NoStop}%
\bibitem [{\citenamefont {Bruus}\ and\ \citenamefont
  {Flensberg}(2004)}]{bruus2004many}%
  \BibitemOpen
  \bibfield  {author} {\bibinfo {author} {\bibfnamefont {H.}~\bibnamefont
  {Bruus}}\ and\ \bibinfo {author} {\bibfnamefont {K.}~\bibnamefont
  {Flensberg}},\ }\href@noop {} {\emph {\bibinfo {title} {{Many-body Quantum
  Theory in Condensed Matter Physics: An Introduction}}}}\ (\bibinfo
  {publisher} {OUP Oxford},\ \bibinfo {year} {2004})\BibitemShut {NoStop}%
\bibitem [{\citenamefont {Vasko}\ and\ \citenamefont
  {Raichev}(2006)}]{vasko2006quantum}%
  \BibitemOpen
  \bibfield  {author} {\bibinfo {author} {\bibfnamefont {F.~T.}\ \bibnamefont
  {Vasko}}\ and\ \bibinfo {author} {\bibfnamefont {O.~E.}\ \bibnamefont
  {Raichev}},\ }\href@noop {} {\emph {\bibinfo {title} {{Quantum kinetic theory
  and applications: Electrons, photons, phonons}}}}\ (\bibinfo  {publisher}
  {Springer Science \& Business Media},\ \bibinfo {year} {2006})\BibitemShut
  {NoStop}%
\bibitem [{\citenamefont {Xiao}\ \emph {et~al.}(2010)\citenamefont {Xiao},
  \citenamefont {Chang},\ and\ \citenamefont {Niu}}]{xiao2010berry}%
  \BibitemOpen
  \bibfield  {author} {\bibinfo {author} {\bibfnamefont {D.}~\bibnamefont
  {Xiao}}, \bibinfo {author} {\bibfnamefont {M.-C.}\ \bibnamefont {Chang}}, \
  and\ \bibinfo {author} {\bibfnamefont {Q.}~\bibnamefont {Niu}},\ }\bibfield
  {title} {\enquote {\bibinfo {title} {{Berry phase effects on electronic
  properties}},}\ }\href {https://link.aps.org/doi/10.1103/RevModPhys.82.1959}
  {\bibfield  {journal} {\bibinfo  {journal} {\emph {Rev. Mod. Phys.}},\
  }\textbf {\bibinfo {volume} {82}},\ \bibinfo {pages} {1959--2007}\  (\bibinfo
  {year} {2010})}\BibitemShut {NoStop}%
\bibitem [{\citenamefont {Bhalla}\ \emph {et~al.}(2020)\citenamefont {Bhalla},
  \citenamefont {MacDonald},\ and\ \citenamefont
  {Culcer}}]{bhalla2020resonant}%
  \BibitemOpen
  \bibfield  {author} {\bibinfo {author} {\bibfnamefont {P.}~\bibnamefont
  {Bhalla}}, \bibinfo {author} {\bibfnamefont {A.~H.}\ \bibnamefont
  {MacDonald}}, \ and\ \bibinfo {author} {\bibfnamefont {D.}~\bibnamefont
  {Culcer}},\ }\bibfield  {title} {\enquote {\bibinfo {title} {{Resonant
  Photovoltaic Effect in Doped Magnetic Semiconductors}},}\ }\href
  {https://link.aps.org/doi/10.1103/PhysRevLett.124.087402} {\bibfield
  {journal} {\bibinfo  {journal} {\emph {Phys. Rev. Lett.}},\ }\textbf
  {\bibinfo {volume} {124}},\ \bibinfo {pages} {087402}\  (\bibinfo {year}
  {2020})}\BibitemShut {NoStop}%
\bibitem [{\citenamefont {Bhalla}\ \emph {et~al.}(2022)\citenamefont {Bhalla},
  \citenamefont {Das}, \citenamefont {Culcer},\ and\ \citenamefont
  {Agarwal}}]{bhalla2022resonant}%
  \BibitemOpen
  \bibfield  {author} {\bibinfo {author} {\bibfnamefont {P.}~\bibnamefont
  {Bhalla}}, \bibinfo {author} {\bibfnamefont {K.}~\bibnamefont {Das}},
  \bibinfo {author} {\bibfnamefont {D.}~\bibnamefont {Culcer}}, \ and\ \bibinfo
  {author} {\bibfnamefont {A.}~\bibnamefont {Agarwal}},\ }\bibfield  {title}
  {\enquote {\bibinfo {title} {{Resonant Second-Harmonic Generation as a Probe
  of Quantum Geometry}},}\ }\href
  {https://link.aps.org/doi/10.1103/PhysRevLett.129.227401} {\bibfield
  {journal} {\bibinfo  {journal} {\emph {Phys. Rev. Lett.}},\ }\textbf
  {\bibinfo {volume} {129}},\ \bibinfo {pages} {227401}\  (\bibinfo {year}
  {2022})}\BibitemShut {NoStop}%
\bibitem [{\citenamefont {Cullen}\ \emph {et~al.}(2021)\citenamefont {Cullen},
  \citenamefont {Bhalla}, \citenamefont {Marcellina}, \citenamefont
  {Hamilton},\ and\ \citenamefont {Culcer}}]{cullen2021generating}%
  \BibitemOpen
  \bibfield  {author} {\bibinfo {author} {\bibfnamefont {J.~H.}\ \bibnamefont
  {Cullen}}, \bibinfo {author} {\bibfnamefont {P.}~\bibnamefont {Bhalla}},
  \bibinfo {author} {\bibfnamefont {E.}~\bibnamefont {Marcellina}}, \bibinfo
  {author} {\bibfnamefont {A.~R.}\ \bibnamefont {Hamilton}}, \ and\ \bibinfo
  {author} {\bibfnamefont {D.}~\bibnamefont {Culcer}},\ }\bibfield  {title}
  {\enquote {\bibinfo {title} {{Generating a Topological Anomalous Hall Effect
  in a Nonmagnetic Conductor: An In-Plane Magnetic Field as a Direct Probe of
  the Berry Curvature}},}\ }\href
  {https://link.aps.org/doi/10.1103/PhysRevLett.126.256601} {\bibfield
  {journal} {\bibinfo  {journal} {\emph {Phys. Rev. Lett.}},\ }\textbf
  {\bibinfo {volume} {126}},\ \bibinfo {pages} {256601}\  (\bibinfo {year}
  {2021})}\BibitemShut {NoStop}%
\bibitem [{\citenamefont {Bhalla}\ \emph {et~al.}(2023)\citenamefont {Bhalla},
  \citenamefont {Das}, \citenamefont {Agarwal},\ and\ \citenamefont
  {Culcer}}]{bhalla2023quantum}%
  \BibitemOpen
  \bibfield  {author} {\bibinfo {author} {\bibfnamefont {P.}~\bibnamefont
  {Bhalla}}, \bibinfo {author} {\bibfnamefont {K.}~\bibnamefont {Das}},
  \bibinfo {author} {\bibfnamefont {A.}~\bibnamefont {Agarwal}}, \ and\
  \bibinfo {author} {\bibfnamefont {D.}~\bibnamefont {Culcer}},\ }\bibfield
  {title} {\enquote {\bibinfo {title} {{Quantum kinetic theory of nonlinear
  optical currents: Finite Fermi surface and Fermi sea contributions}},}\
  }\href {https://link.aps.org/doi/10.1103/PhysRevB.107.165131} {\bibfield
  {journal} {\bibinfo  {journal} {\emph {Phys. Rev. B}},\ }\textbf {\bibinfo
  {volume} {107}},\ \bibinfo {pages} {165131}\  (\bibinfo {year}
  {2023})}\BibitemShut {NoStop}%
\bibitem [{\citenamefont {Atencia}\ \emph {et~al.}(2023)\citenamefont
  {Atencia}, \citenamefont {Xiao},\ and\ \citenamefont
  {Culcer}}]{atencia2023disorder}%
  \BibitemOpen
  \bibfield  {author} {\bibinfo {author} {\bibfnamefont {R.~B.}\ \bibnamefont
  {Atencia}}, \bibinfo {author} {\bibfnamefont {D.}~\bibnamefont {Xiao}}, \
  and\ \bibinfo {author} {\bibfnamefont {D.}~\bibnamefont {Culcer}},\
  }\bibfield  {title} {\enquote {\bibinfo {title} {{Disorder in the nonlinear
  anomalous Hall effect of $\mathcal{PT}$-symmetric Dirac fermions}},}\ }\href
  {https://link.aps.org/doi/10.1103/PhysRevB.108.L201115} {\bibfield  {journal}
  {\bibinfo  {journal} {\emph {Phys. Rev. B}},\ }\textbf {\bibinfo {volume}
  {108}},\ \bibinfo {pages} {L201115}\  (\bibinfo {year} {2023})}\BibitemShut
  {NoStop}%
\bibitem [{\citenamefont {Varshney}\ \emph {et~al.}(2023)\citenamefont
  {Varshney}, \citenamefont {Das}, \citenamefont {Bhalla},\ and\ \citenamefont
  {Agarwal}}]{varshney2023quantum}%
  \BibitemOpen
  \bibfield  {author} {\bibinfo {author} {\bibfnamefont {H.}~\bibnamefont
  {Varshney}}, \bibinfo {author} {\bibfnamefont {K.}~\bibnamefont {Das}},
  \bibinfo {author} {\bibfnamefont {P.}~\bibnamefont {Bhalla}}, \ and\ \bibinfo
  {author} {\bibfnamefont {A.}~\bibnamefont {Agarwal}},\ }\bibfield  {title}
  {\enquote {\bibinfo {title} {{Quantum kinetic theory of nonlinear thermal
  current}},}\ }\href {https://link.aps.org/doi/10.1103/PhysRevB.107.235419}
  {\bibfield  {journal} {\bibinfo  {journal} {\emph {Phys. Rev. B}},\ }\textbf
  {\bibinfo {volume} {107}},\ \bibinfo {pages} {235419}\  (\bibinfo {year}
  {2023})}\BibitemShut {NoStop}%
\bibitem [{\citenamefont {Ashcroft}\ and\ \citenamefont
  {Mermin}(1976)}]{ashcroft1976solid}%
  \BibitemOpen
  \bibfield  {author} {\bibinfo {author} {\bibfnamefont {N.~W.}\ \bibnamefont
  {Ashcroft}}\ and\ \bibinfo {author} {\bibfnamefont {N.~D.}\ \bibnamefont
  {Mermin}},\ }\href@noop {} {\emph {\bibinfo {title} {{Solid State
  Physics}}}}\ (\bibinfo  {publisher} {Holt-Saunders, New York},\ \bibinfo
  {year} {1976})\BibitemShut {NoStop}%
\bibitem [{\citenamefont {Adams}\ and\ \citenamefont
  {Blount}(1959)}]{adams1959energy}%
  \BibitemOpen
  \bibfield  {author} {\bibinfo {author} {\bibfnamefont {E.}~\bibnamefont
  {Adams}}\ and\ \bibinfo {author} {\bibfnamefont {E.}~\bibnamefont {Blount}},\
  }\bibfield  {title} {\enquote {\bibinfo {title} {{Energy bands in the
  presence of an external force field—II: Anomalous velocities}},}\ }\href
  {https://doi.org/10.1016/0022-3697(59)90004-6} {\bibfield  {journal}
  {\bibinfo  {journal} {\emph {J. Phys. Chem. Solids}},\ }\textbf {\bibinfo
  {volume} {10}},\ \bibinfo {pages} {286--303}\  (\bibinfo {year}
  {1959})}\BibitemShut {NoStop}%
\bibitem [{\citenamefont {Chang}\ and\ \citenamefont
  {Niu}(1995)}]{chang1995berry}%
  \BibitemOpen
  \bibfield  {author} {\bibinfo {author} {\bibfnamefont {M.-C.}\ \bibnamefont
  {Chang}}\ and\ \bibinfo {author} {\bibfnamefont {Q.}~\bibnamefont {Niu}},\
  }\bibfield  {title} {\enquote {\bibinfo {title} {{Berry Phase, Hyperorbits,
  and the Hofstadter Spectrum}},}\ }\href
  {https://link.aps.org/doi/10.1103/PhysRevLett.75.1348} {\bibfield  {journal}
  {\bibinfo  {journal} {\emph {Phys. Rev. Lett.}},\ }\textbf {\bibinfo {volume}
  {75}},\ \bibinfo {pages} {1348--1351}\  (\bibinfo {year} {1995})}\BibitemShut
  {NoStop}%
\bibitem [{\citenamefont {Chang}\ and\ \citenamefont
  {Niu}(1996)}]{chang1996berry}%
  \BibitemOpen
  \bibfield  {author} {\bibinfo {author} {\bibfnamefont {M.-C.}\ \bibnamefont
  {Chang}}\ and\ \bibinfo {author} {\bibfnamefont {Q.}~\bibnamefont {Niu}},\
  }\bibfield  {title} {\enquote {\bibinfo {title} {{Berry phase, hyperorbits,
  and the Hofstadter spectrum: Semiclassical dynamics in magnetic Bloch
  bands}},}\ }\href {https://link.aps.org/doi/10.1103/PhysRevB.53.7010}
  {\bibfield  {journal} {\bibinfo  {journal} {\emph {Phys. Rev. B}},\ }\textbf
  {\bibinfo {volume} {53}},\ \bibinfo {pages} {7010--7023}\  (\bibinfo {year}
  {1996})}\BibitemShut {NoStop}%
\bibitem [{\citenamefont {Sundaram}\ and\ \citenamefont
  {Niu}(1999)}]{sundaram1999wave}%
  \BibitemOpen
  \bibfield  {author} {\bibinfo {author} {\bibfnamefont {G.}~\bibnamefont
  {Sundaram}}\ and\ \bibinfo {author} {\bibfnamefont {Q.}~\bibnamefont {Niu}},\
  }\bibfield  {title} {\enquote {\bibinfo {title} {{Wave-packet dynamics in
  slowly perturbed crystals: Gradient corrections and Berry-phase effects}},}\
  }\href {https://link.aps.org/doi/10.1103/PhysRevB.59.14915} {\bibfield
  {journal} {\bibinfo  {journal} {\emph {Phys. Rev. B}},\ }\textbf {\bibinfo
  {volume} {59}},\ \bibinfo {pages} {14915--14925}\  (\bibinfo {year}
  {1999})}\BibitemShut {NoStop}%
\bibitem [{\citenamefont {Sinitsyn}\ \emph {et~al.}(2006)\citenamefont
  {Sinitsyn}, \citenamefont {Niu},\ and\ \citenamefont
  {MacDonald}}]{sinitsyn2006coordinate}%
  \BibitemOpen
  \bibfield  {author} {\bibinfo {author} {\bibfnamefont {N.~A.}\ \bibnamefont
  {Sinitsyn}}, \bibinfo {author} {\bibfnamefont {Q.}~\bibnamefont {Niu}}, \
  and\ \bibinfo {author} {\bibfnamefont {A.~H.}\ \bibnamefont {MacDonald}},\
  }\bibfield  {title} {\enquote {\bibinfo {title} {{Coordinate shift in the
  semiclassical Boltzmann equation and the anomalous Hall effect}},}\ }\href
  {https://link.aps.org/doi/10.1103/PhysRevB.73.075318} {\bibfield  {journal}
  {\bibinfo  {journal} {\emph {Phys. Rev. B}},\ }\textbf {\bibinfo {volume}
  {73}},\ \bibinfo {pages} {075318}\  (\bibinfo {year} {2006})}\BibitemShut
  {NoStop}%
\bibitem [{\citenamefont {Smit}(1958)}]{smit1958spontaneous}%
  \BibitemOpen
  \bibfield  {author} {\bibinfo {author} {\bibfnamefont {J.}~\bibnamefont
  {Smit}},\ }\bibfield  {title} {\enquote {\bibinfo {title} {{The spontaneous
  Hall effect in ferromagnetics II}},}\ }\href
  {https://doi.org/10.1016/S0031-8914(58)93541-9} {\bibfield  {journal}
  {\bibinfo  {journal} {\emph {Physica}},\ }\textbf {\bibinfo {volume} {24}},\
  \bibinfo {pages} {39--51}\  (\bibinfo {year} {1958})}\BibitemShut {NoStop}%
\bibitem [{\citenamefont {Berger}(1970)}]{berger1970side}%
  \BibitemOpen
  \bibfield  {author} {\bibinfo {author} {\bibfnamefont {L.}~\bibnamefont
  {Berger}},\ }\bibfield  {title} {\enquote {\bibinfo {title} {{Side-Jump
  Mechanism for the Hall Effect of Ferromagnets}},}\ }\href
  {https://link.aps.org/doi/10.1103/PhysRevB.2.4559} {\bibfield  {journal}
  {\bibinfo  {journal} {\emph {Phys. Rev. B}},\ }\textbf {\bibinfo {volume}
  {2}},\ \bibinfo {pages} {4559--4566}\  (\bibinfo {year} {1970})}\BibitemShut
  {NoStop}%
\bibitem [{\citenamefont {Nozi{\`e}res}\ and\ \citenamefont
  {Lewiner}(1973)}]{nozieres1973simple}%
  \BibitemOpen
  \bibfield  {author} {\bibinfo {author} {\bibfnamefont {P.}~\bibnamefont
  {Nozi{\`e}res}}\ and\ \bibinfo {author} {\bibfnamefont {C.}~\bibnamefont
  {Lewiner}},\ }\bibfield  {title} {\enquote {\bibinfo {title} {{A simple
  theory of the anomalous Hall effect in semiconductors}},}\ }\href
  {https://doi.org/10.1051/jphys:019730034010090100} {\bibfield  {journal}
  {\bibinfo  {journal} {\emph {J. Phys. (Paris)}},\ }\textbf {\bibinfo {volume}
  {34}},\ \bibinfo {pages} {901--915}\  (\bibinfo {year} {1973})}\BibitemShut
  {NoStop}%
\bibitem [{\citenamefont {Wang}\ \emph {et~al.}(2006)\citenamefont {Wang},
  \citenamefont {Yates}, \citenamefont {Souza},\ and\ \citenamefont
  {Vanderbilt}}]{wang2006abinitio}%
  \BibitemOpen
  \bibfield  {author} {\bibinfo {author} {\bibfnamefont {X.}~\bibnamefont
  {Wang}}, \bibinfo {author} {\bibfnamefont {J.~R.}\ \bibnamefont {Yates}},
  \bibinfo {author} {\bibfnamefont {I.}~\bibnamefont {Souza}}, \ and\ \bibinfo
  {author} {\bibfnamefont {D.}~\bibnamefont {Vanderbilt}},\ }\bibfield  {title}
  {\enquote {\bibinfo {title} {{Ab initio calculation of the anomalous Hall
  conductivity by Wannier interpolation}},}\ }\href
  {https://link.aps.org/doi/10.1103/PhysRevB.74.195118} {\bibfield  {journal}
  {\bibinfo  {journal} {\emph {Phys. Rev. B}},\ }\textbf {\bibinfo {volume}
  {74}},\ \bibinfo {pages} {195118}\  (\bibinfo {year} {2006})}\BibitemShut
  {NoStop}%
\bibitem [{\citenamefont {Wang}\ \emph {et~al.}(2007)\citenamefont {Wang},
  \citenamefont {Vanderbilt}, \citenamefont {Yates},\ and\ \citenamefont
  {Souza}}]{wang2007fermi}%
  \BibitemOpen
  \bibfield  {author} {\bibinfo {author} {\bibfnamefont {X.}~\bibnamefont
  {Wang}}, \bibinfo {author} {\bibfnamefont {D.}~\bibnamefont {Vanderbilt}},
  \bibinfo {author} {\bibfnamefont {J.~R.}\ \bibnamefont {Yates}}, \ and\
  \bibinfo {author} {\bibfnamefont {I.}~\bibnamefont {Souza}},\ }\bibfield
  {title} {\enquote {\bibinfo {title} {{Fermi-surface calculation of the
  anomalous Hall conductivity}},}\ }\href
  {https://link.aps.org/doi/10.1103/PhysRevB.76.195109} {\bibfield  {journal}
  {\bibinfo  {journal} {\emph {Phys. Rev. B}},\ }\textbf {\bibinfo {volume}
  {76}},\ \bibinfo {pages} {195109}\  (\bibinfo {year} {2007})}\BibitemShut
  {NoStop}%
\bibitem [{\citenamefont {Gradhand}\ \emph {et~al.}(2012)\citenamefont
  {Gradhand}, \citenamefont {Fedorov}, \citenamefont {Pientka}, \citenamefont
  {Zahn}, \citenamefont {Mertig},\ and\ \citenamefont
  {Gy{\"o}rffy}}]{gradhand2012first}%
  \BibitemOpen
  \bibfield  {author} {\bibinfo {author} {\bibfnamefont {M.}~\bibnamefont
  {Gradhand}}, \bibinfo {author} {\bibfnamefont {D.}~\bibnamefont {Fedorov}},
  \bibinfo {author} {\bibfnamefont {F.}~\bibnamefont {Pientka}}, \bibinfo
  {author} {\bibfnamefont {P.}~\bibnamefont {Zahn}}, \bibinfo {author}
  {\bibfnamefont {I.}~\bibnamefont {Mertig}}, \ and\ \bibinfo {author}
  {\bibfnamefont {B.}~\bibnamefont {Gy{\"o}rffy}},\ }\bibfield  {title}
  {\enquote {\bibinfo {title} {{First-principle calculations of the Berry
  curvature of Bloch states for charge and spin transport of electrons}},}\
  }\href {https://iopscience.iop.org/article/10.1088/0953-8984/24/21/213202}
  {\bibfield  {journal} {\bibinfo  {journal} {\emph {J. Phys. Condens.
  Matter}},\ }\textbf {\bibinfo {volume} {24}},\ \bibinfo {pages} {213202}\
  (\bibinfo {year} {2012})}\BibitemShut {NoStop}%
\bibitem [{\citenamefont {He}\ \emph {et~al.}(2012)\citenamefont {He},
  \citenamefont {Moore},\ and\ \citenamefont {Varma}}]{he2012berry}%
  \BibitemOpen
  \bibfield  {author} {\bibinfo {author} {\bibfnamefont {Y.}~\bibnamefont
  {He}}, \bibinfo {author} {\bibfnamefont {J.}~\bibnamefont {Moore}}, \ and\
  \bibinfo {author} {\bibfnamefont {C.~M.}\ \bibnamefont {Varma}},\ }\bibfield
  {title} {\enquote {\bibinfo {title} {{Berry phase and anomalous Hall effect
  in a three-orbital tight-binding Hamiltonian}},}\ }\href
  {https://link.aps.org/doi/10.1103/PhysRevB.85.155106} {\bibfield  {journal}
  {\bibinfo  {journal} {\emph {Phys. Rev. B}},\ }\textbf {\bibinfo {volume}
  {85}},\ \bibinfo {pages} {155106}\  (\bibinfo {year} {2012})}\BibitemShut
  {NoStop}%
\bibitem [{\citenamefont {Chen}\ \emph {et~al.}(2013)\citenamefont {Chen},
  \citenamefont {Bergman},\ and\ \citenamefont {Burkov}}]{chen2013weyl}%
  \BibitemOpen
  \bibfield  {author} {\bibinfo {author} {\bibfnamefont {Y.}~\bibnamefont
  {Chen}}, \bibinfo {author} {\bibfnamefont {D.~L.}\ \bibnamefont {Bergman}}, \
  and\ \bibinfo {author} {\bibfnamefont {A.~A.}\ \bibnamefont {Burkov}},\
  }\bibfield  {title} {\enquote {\bibinfo {title} {{Weyl fermions and the
  anomalous Hall effect in metallic ferromagnets}},}\ }\href
  {https://link.aps.org/doi/10.1103/PhysRevB.88.125110} {\bibfield  {journal}
  {\bibinfo  {journal} {\emph {Phys. Rev. B}},\ }\textbf {\bibinfo {volume}
  {88}},\ \bibinfo {pages} {125110}\  (\bibinfo {year} {2013})}\BibitemShut
  {NoStop}%
\bibitem [{\citenamefont {Bianco}\ \emph {et~al.}(2014)\citenamefont {Bianco},
  \citenamefont {Resta},\ and\ \citenamefont {Souza}}]{bianco2014how}%
  \BibitemOpen
  \bibfield  {author} {\bibinfo {author} {\bibfnamefont {R.}~\bibnamefont
  {Bianco}}, \bibinfo {author} {\bibfnamefont {R.}~\bibnamefont {Resta}}, \
  and\ \bibinfo {author} {\bibfnamefont {I.}~\bibnamefont {Souza}},\ }\bibfield
   {title} {\enquote {\bibinfo {title} {{How disorder affects the Berry-phase
  anomalous Hall conductivity: A reciprocal-space analysis}},}\ }\href
  {https://link.aps.org/doi/10.1103/PhysRevB.90.125153} {\bibfield  {journal}
  {\bibinfo  {journal} {\emph {Phys. Rev. B}},\ }\textbf {\bibinfo {volume}
  {90}},\ \bibinfo {pages} {125153}\  (\bibinfo {year} {2014})}\BibitemShut
  {NoStop}%
\bibitem [{\citenamefont {Chen}\ \emph {et~al.}(2014)\citenamefont {Chen},
  \citenamefont {Niu},\ and\ \citenamefont {MacDonald}}]{chen2014anomalous}%
  \BibitemOpen
  \bibfield  {author} {\bibinfo {author} {\bibfnamefont {H.}~\bibnamefont
  {Chen}}, \bibinfo {author} {\bibfnamefont {Q.}~\bibnamefont {Niu}}, \ and\
  \bibinfo {author} {\bibfnamefont {A.~H.}\ \bibnamefont {MacDonald}},\
  }\bibfield  {title} {\enquote {\bibinfo {title} {{Anomalous Hall Effect
  Arising from Noncollinear Antiferromagnetism}},}\ }\href
  {https://link.aps.org/doi/10.1103/PhysRevLett.112.017205} {\bibfield
  {journal} {\bibinfo  {journal} {\emph {Phys. Rev. Lett.}},\ }\textbf
  {\bibinfo {volume} {112}},\ \bibinfo {pages} {017205}\  (\bibinfo {year}
  {2014})}\BibitemShut {NoStop}%
\bibitem [{\citenamefont {Olsen}\ and\ \citenamefont
  {Souza}(2015)}]{olsen2015valley}%
  \BibitemOpen
  \bibfield  {author} {\bibinfo {author} {\bibfnamefont {T.}~\bibnamefont
  {Olsen}}\ and\ \bibinfo {author} {\bibfnamefont {I.}~\bibnamefont {Souza}},\
  }\bibfield  {title} {\enquote {\bibinfo {title} {{Valley Hall effect in
  disordered monolayer ${\mathrm{MoS}}_{2}$ from first principles}},}\ }\href
  {https://link.aps.org/doi/10.1103/PhysRevB.92.125146} {\bibfield  {journal}
  {\bibinfo  {journal} {\emph {Phys. Rev. B}},\ }\textbf {\bibinfo {volume}
  {92}},\ \bibinfo {pages} {125146}\  (\bibinfo {year} {2015})}\BibitemShut
  {NoStop}%
\bibitem [{\citenamefont {Feng}\ \emph {et~al.}(2016)\citenamefont {Feng},
  \citenamefont {Liu}, \citenamefont {Liu}, \citenamefont {Zhou},\ and\
  \citenamefont {Yao}}]{feng2016first}%
  \BibitemOpen
  \bibfield  {author} {\bibinfo {author} {\bibfnamefont {W.}~\bibnamefont
  {Feng}}, \bibinfo {author} {\bibfnamefont {C.-C.}\ \bibnamefont {Liu}},
  \bibinfo {author} {\bibfnamefont {G.-B.}\ \bibnamefont {Liu}}, \bibinfo
  {author} {\bibfnamefont {J.-J.}\ \bibnamefont {Zhou}}, \ and\ \bibinfo
  {author} {\bibfnamefont {Y.}~\bibnamefont {Yao}},\ }\bibfield  {title}
  {\enquote {\bibinfo {title} {{First-principles investigations on the Berry
  phase effect in spin--orbit coupling materials}},}\ }\href
  {https://doi.org/10.1016/j.commatsci.2015.09.020} {\bibfield  {journal}
  {\bibinfo  {journal} {\emph {Comput. Mater. Sci.}},\ }\textbf {\bibinfo
  {volume} {112}},\ \bibinfo {pages} {428--447}\  (\bibinfo {year}
  {2016})}\BibitemShut {NoStop}%
\bibitem [{\citenamefont {Dai}\ \emph {et~al.}(2017)\citenamefont {Dai},
  \citenamefont {Du},\ and\ \citenamefont {Lu}}]{dai2017negative}%
  \BibitemOpen
  \bibfield  {author} {\bibinfo {author} {\bibfnamefont {X.}~\bibnamefont
  {Dai}}, \bibinfo {author} {\bibfnamefont {Z.~Z.}\ \bibnamefont {Du}}, \ and\
  \bibinfo {author} {\bibfnamefont {H.-Z.}\ \bibnamefont {Lu}},\ }\bibfield
  {title} {\enquote {\bibinfo {title} {{Negative Magnetoresistance without
  Chiral Anomaly in Topological Insulators}},}\ }\href
  {https://link.aps.org/doi/10.1103/PhysRevLett.119.166601} {\bibfield
  {journal} {\bibinfo  {journal} {\emph {Phys. Rev. Lett.}},\ }\textbf
  {\bibinfo {volume} {119}},\ \bibinfo {pages} {166601}\  (\bibinfo {year}
  {2017})}\BibitemShut {NoStop}%
\bibitem [{\citenamefont {Martiny}\ \emph {et~al.}(2019)\citenamefont
  {Martiny}, \citenamefont {Kaasbjerg},\ and\ \citenamefont
  {Jauho}}]{martiny2019tunable}%
  \BibitemOpen
  \bibfield  {author} {\bibinfo {author} {\bibfnamefont {J.~H.~J.}\
  \bibnamefont {Martiny}}, \bibinfo {author} {\bibfnamefont {K.}~\bibnamefont
  {Kaasbjerg}}, \ and\ \bibinfo {author} {\bibfnamefont {A.-P.}\ \bibnamefont
  {Jauho}},\ }\bibfield  {title} {\enquote {\bibinfo {title} {{Tunable valley
  Hall effect in gate-defined graphene superlattices}},}\ }\href
  {https://link.aps.org/doi/10.1103/PhysRevB.100.155414} {\bibfield  {journal}
  {\bibinfo  {journal} {\emph {Phys. Rev. B}},\ }\textbf {\bibinfo {volume}
  {100}},\ \bibinfo {pages} {155414}\  (\bibinfo {year} {2019})}\BibitemShut
  {NoStop}%
\bibitem [{\citenamefont {Wuttke}\ \emph {et~al.}(2019)\citenamefont {Wuttke},
  \citenamefont {Caglieris}, \citenamefont {Sykora}, \citenamefont
  {Scaravaggi}, \citenamefont {Wolter}, \citenamefont {Manna}, \citenamefont
  {S\"uss}, \citenamefont {Shekhar}, \citenamefont {Felser}, \citenamefont
  {B\"uchner},\ and\ \citenamefont {Hess}}]{wuttke2019berry}%
  \BibitemOpen
  \bibfield  {author} {\bibinfo {author} {\bibfnamefont {C.}~\bibnamefont
  {Wuttke}}, \bibinfo {author} {\bibfnamefont {F.}~\bibnamefont {Caglieris}},
  \bibinfo {author} {\bibfnamefont {S.}~\bibnamefont {Sykora}}, \bibinfo
  {author} {\bibfnamefont {F.}~\bibnamefont {Scaravaggi}}, \bibinfo {author}
  {\bibfnamefont {A.~U.~B.}\ \bibnamefont {Wolter}}, \bibinfo {author}
  {\bibfnamefont {K.}~\bibnamefont {Manna}}, \bibinfo {author} {\bibfnamefont
  {V.}~\bibnamefont {S\"uss}}, \bibinfo {author} {\bibfnamefont
  {C.}~\bibnamefont {Shekhar}}, \bibinfo {author} {\bibfnamefont
  {C.}~\bibnamefont {Felser}}, \bibinfo {author} {\bibfnamefont
  {B.}~\bibnamefont {B\"uchner}}, \ and\ \bibinfo {author} {\bibfnamefont
  {C.}~\bibnamefont {Hess}},\ }\bibfield  {title} {\enquote {\bibinfo {title}
  {{Berry curvature unravelled by the anomalous Nernst effect in
  ${\mathrm{Mn}}_{3}\mathrm{Ge}$}},}\ }\href
  {https://link.aps.org/doi/10.1103/PhysRevB.100.085111} {\bibfield  {journal}
  {\bibinfo  {journal} {\emph {Phys. Rev. B}},\ }\textbf {\bibinfo {volume}
  {100}},\ \bibinfo {pages} {085111}\  (\bibinfo {year} {2019})}\BibitemShut
  {NoStop}%
\bibitem [{\citenamefont {Du}\ \emph {et~al.}(2020)\citenamefont {Du},
  \citenamefont {Tang}, \citenamefont {Li}, \citenamefont {Lin}, \citenamefont
  {Xu}, \citenamefont {Duan},\ and\ \citenamefont {Rubio}}]{du2020berry}%
  \BibitemOpen
  \bibfield  {author} {\bibinfo {author} {\bibfnamefont {S.}~\bibnamefont
  {Du}}, \bibinfo {author} {\bibfnamefont {P.}~\bibnamefont {Tang}}, \bibinfo
  {author} {\bibfnamefont {J.}~\bibnamefont {Li}}, \bibinfo {author}
  {\bibfnamefont {Z.}~\bibnamefont {Lin}}, \bibinfo {author} {\bibfnamefont
  {Y.}~\bibnamefont {Xu}}, \bibinfo {author} {\bibfnamefont {W.}~\bibnamefont
  {Duan}}, \ and\ \bibinfo {author} {\bibfnamefont {A.}~\bibnamefont {Rubio}},\
  }\bibfield  {title} {\enquote {\bibinfo {title} {{Berry curvature engineering
  by gating two-dimensional antiferromagnets}},}\ }\href
  {https://link.aps.org/doi/10.1103/PhysRevResearch.2.022025} {\bibfield
  {journal} {\bibinfo  {journal} {\emph {Phys. Rev. Res.}},\ }\textbf {\bibinfo
  {volume} {2}},\ \bibinfo {pages} {022025}\  (\bibinfo {year}
  {2020})}\BibitemShut {NoStop}%
\bibitem [{\citenamefont {He}\ \emph {et~al.}(2020)\citenamefont {He},
  \citenamefont {Goldhaber-Gordon},\ and\ \citenamefont {Law}}]{he2020giant}%
  \BibitemOpen
  \bibfield  {author} {\bibinfo {author} {\bibfnamefont {W.-Y.}\ \bibnamefont
  {He}}, \bibinfo {author} {\bibfnamefont {D.}~\bibnamefont
  {Goldhaber-Gordon}}, \ and\ \bibinfo {author} {\bibfnamefont {K.~T.}\
  \bibnamefont {Law}},\ }\bibfield  {title} {\enquote {\bibinfo {title} {{Giant
  orbital magnetoelectric effect and current-induced magnetization switching in
  twisted bilayer graphene}},}\ }\href
  {https://doi.org/10.1038/s41467-020-15473-9} {\bibfield  {journal} {\bibinfo
  {journal} {\emph {Nat. Commun.}},\ }\textbf {\bibinfo {volume} {11}},\
  \bibinfo {pages} {1650}\  (\bibinfo {year} {2020})}\BibitemShut {NoStop}%
\bibitem [{\citenamefont {He}\ and\ \citenamefont
  {Law}(2021)}]{he2021superconducting}%
  \BibitemOpen
  \bibfield  {author} {\bibinfo {author} {\bibfnamefont {W.-Y.}\ \bibnamefont
  {He}}\ and\ \bibinfo {author} {\bibfnamefont {K.~T.}\ \bibnamefont {Law}},\
  }\bibfield  {title} {\enquote {\bibinfo {title} {{Superconducting orbital
  magnetoelectric effect and its evolution across the superconductor-normal
  metal phase transition}},}\ }\href
  {https://link.aps.org/doi/10.1103/PhysRevResearch.3.L032012} {\bibfield
  {journal} {\bibinfo  {journal} {\emph {Phys. Rev. Res.}},\ }\textbf {\bibinfo
  {volume} {3}},\ \bibinfo {pages} {L032012}\  (\bibinfo {year}
  {2021})}\BibitemShut {NoStop}%
\bibitem [{\citenamefont {Du}\ \emph {et~al.}(2021){\natexlab{a}}\citenamefont
  {Du}, \citenamefont {Lu},\ and\ \citenamefont {Xie}}]{du2021nonlinear}%
  \BibitemOpen
  \bibfield  {author} {\bibinfo {author} {\bibfnamefont {Z.}~\bibnamefont
  {Du}}, \bibinfo {author} {\bibfnamefont {H.-Z.}\ \bibnamefont {Lu}}, \ and\
  \bibinfo {author} {\bibfnamefont {X.}~\bibnamefont {Xie}},\ }\bibfield
  {title} {\enquote {\bibinfo {title} {{Nonlinear Hall effects}},}\ }\href
  {https://doi.org/10.1038/s42254-021-00359-6} {\bibfield  {journal} {\bibinfo
  {journal} {\emph {Nat. Rev. Phys.}},\ }\textbf {\bibinfo {volume} {3}},\
  \bibinfo {pages} {744--752}\  (\bibinfo {year}
  {2021}{\natexlab{a}})}\BibitemShut {NoStop}%
\bibitem [{\citenamefont {Ideue}\ and\ \citenamefont
  {Iwasa}(2021)}]{ideue2021symmetry}%
  \BibitemOpen
  \bibfield  {author} {\bibinfo {author} {\bibfnamefont {T.}~\bibnamefont
  {Ideue}}\ and\ \bibinfo {author} {\bibfnamefont {Y.}~\bibnamefont {Iwasa}},\
  }\bibfield  {title} {\enquote {\bibinfo {title} {{Symmetry breaking and
  nonlinear electric transport in van der Waals nanostructures}},}\ }\href
  {https://doi.org/10.1146/annurev-conmatphys-060220-100347} {\bibfield
  {journal} {\bibinfo  {journal} {\emph {Annu. Rev. Condens. Matter Phys.}},\
  }\textbf {\bibinfo {volume} {12}},\ \bibinfo {pages} {201--223}\  (\bibinfo
  {year} {2021})}\BibitemShut {NoStop}%
\bibitem [{\citenamefont {Ortix}(2021)}]{ortix2021nonlinear}%
  \BibitemOpen
  \bibfield  {author} {\bibinfo {author} {\bibfnamefont {C.}~\bibnamefont
  {Ortix}},\ }\bibfield  {title} {\enquote {\bibinfo {title} {{Nonlinear Hall
  Effect with Time-Reversal Symmetry: Theory and Material Realizations}},}\
  }\href {https://doi.org/10.1002/qute.202100056} {\bibfield  {journal}
  {\bibinfo  {journal} {\emph {Adv. Quantum Technol.}},\ }\textbf {\bibinfo
  {volume} {4}},\ \bibinfo {pages} {2100056}\  (\bibinfo {year}
  {2021})}\BibitemShut {NoStop}%
\bibitem [{\citenamefont {Nagaosa}\ and\ \citenamefont
  {Yanase}(2024)}]{nagaosa2024nonreciprocal}%
  \BibitemOpen
  \bibfield  {author} {\bibinfo {author} {\bibfnamefont {N.}~\bibnamefont
  {Nagaosa}}\ and\ \bibinfo {author} {\bibfnamefont {Y.}~\bibnamefont
  {Yanase}},\ }\bibfield  {title} {\enquote {\bibinfo {title} {{Nonreciprocal
  transport and optical phenomena in quantum materials}},}\ }\href
  {https://doi.org/10.1146/annurev-conmatphys-032822-033734} {\bibfield
  {journal} {\bibinfo  {journal} {\emph {Annu. Rev. Condens. Matter Phys.}},\
  }\textbf {\bibinfo {volume} {15}},\ \bibinfo {pages} {63--83}\  (\bibinfo
  {year} {2024})}\BibitemShut {NoStop}%
\bibitem [{\citenamefont {Shim}\ \emph {et~al.}()\citenamefont {Shim},
  \citenamefont {Mehraeen}, \citenamefont {Sklenar}, \citenamefont {Zhang},
  \citenamefont {Hoffmann},\ and\ \citenamefont {Mason}}]{shim2024spin}%
  \BibitemOpen
  \bibfield  {author} {\bibinfo {author} {\bibfnamefont {S.}~\bibnamefont
  {Shim}}, \bibinfo {author} {\bibfnamefont {M.}~\bibnamefont {Mehraeen}},
  \bibinfo {author} {\bibfnamefont {J.}~\bibnamefont {Sklenar}}, \bibinfo
  {author} {\bibfnamefont {S.~S.-L.}\ \bibnamefont {Zhang}}, \bibinfo {author}
  {\bibfnamefont {A.}~\bibnamefont {Hoffmann}}, \ and\ \bibinfo {author}
  {\bibfnamefont {N.}~\bibnamefont {Mason}},\ }\bibfield  {title} {\enquote
  {\bibinfo {title} {{Spin-polarized antiferromagnetic metals}},}\ }\href
  {https://doi.org/10.1146/annurev-conmatphys-042924-123620} {\bibfield
  {journal} {\bibinfo  {journal} {\emph {Annu. Rev. Condens. Matter Phys.}},\
  }\ \textbf {\bibinfo {volume} {16}}}\BibitemShut {NoStop}%
\bibitem [{\citenamefont {Avci}\ \emph
  {et~al.}(2015){\natexlab{a}}\citenamefont {Avci}, \citenamefont {Garello},
  \citenamefont {Mendil}, \citenamefont {Ghosh}, \citenamefont {Blasakis},
  \citenamefont {Gabureac}, \citenamefont {Trassin}, \citenamefont {Fiebig},\
  and\ \citenamefont {Gambardella}}]{avci2015magnetoresistance}%
  \BibitemOpen
  \bibfield  {author} {\bibinfo {author} {\bibfnamefont {C.~O.}\ \bibnamefont
  {Avci}}, \bibinfo {author} {\bibfnamefont {K.}~\bibnamefont {Garello}},
  \bibinfo {author} {\bibfnamefont {J.}~\bibnamefont {Mendil}}, \bibinfo
  {author} {\bibfnamefont {A.}~\bibnamefont {Ghosh}}, \bibinfo {author}
  {\bibfnamefont {N.}~\bibnamefont {Blasakis}}, \bibinfo {author}
  {\bibfnamefont {M.}~\bibnamefont {Gabureac}}, \bibinfo {author}
  {\bibfnamefont {M.}~\bibnamefont {Trassin}}, \bibinfo {author} {\bibfnamefont
  {M.}~\bibnamefont {Fiebig}}, \ and\ \bibinfo {author} {\bibfnamefont
  {P.}~\bibnamefont {Gambardella}},\ }\bibfield  {title} {\enquote {\bibinfo
  {title} {{Magnetoresistance of heavy and light metal/ferromagnet
  bilayers}},}\ }\href {https://doi.org/10.1063/1.4935497} {\bibfield
  {journal} {\bibinfo  {journal} {\emph {Appl. Phys. Lett.}},\ }\textbf
  {\bibinfo {volume} {107}}\  (\bibinfo {year}
  {2015}{\natexlab{a}})}\BibitemShut {NoStop}%
\bibitem [{\citenamefont {Avci}\ \emph
  {et~al.}(2015){\natexlab{b}}\citenamefont {Avci}, \citenamefont {Garello},
  \citenamefont {Ghosh}, \citenamefont {Gabureac}, \citenamefont {Alvarado},\
  and\ \citenamefont {Gambardella}}]{avci2015unidirectional}%
  \BibitemOpen
  \bibfield  {author} {\bibinfo {author} {\bibfnamefont {C.~O.}\ \bibnamefont
  {Avci}}, \bibinfo {author} {\bibfnamefont {K.}~\bibnamefont {Garello}},
  \bibinfo {author} {\bibfnamefont {A.}~\bibnamefont {Ghosh}}, \bibinfo
  {author} {\bibfnamefont {M.}~\bibnamefont {Gabureac}}, \bibinfo {author}
  {\bibfnamefont {S.~F.}\ \bibnamefont {Alvarado}}, \ and\ \bibinfo {author}
  {\bibfnamefont {P.}~\bibnamefont {Gambardella}},\ }\bibfield  {title}
  {\enquote {\bibinfo {title} {{Unidirectional spin Hall magnetoresistance in
  ferromagnet/normal metal bilayers}},}\ }\href
  {https://doi.org/10.1038/nphys3356} {\bibfield  {journal} {\bibinfo
  {journal} {\emph {Nat. Phys.}},\ }\textbf {\bibinfo {volume} {11}},\ \bibinfo
  {pages} {570--575}\  (\bibinfo {year} {2015}{\natexlab{b}})}\BibitemShut
  {NoStop}%
\bibitem [{\citenamefont {Olejn\'{\i}k}\ \emph
  {et~al.}(2015){\natexlab{a}}\citenamefont {Olejn\'{\i}k}, \citenamefont
  {Nov\'ak}, \citenamefont {Wunderlich},\ and\ \citenamefont
  {Jungwirth}}]{olejnik2015electrical}%
  \BibitemOpen
  \bibfield  {author} {\bibinfo {author} {\bibfnamefont {K.}~\bibnamefont
  {Olejn\'{\i}k}}, \bibinfo {author} {\bibfnamefont {V.}~\bibnamefont
  {Nov\'ak}}, \bibinfo {author} {\bibfnamefont {J.}~\bibnamefont {Wunderlich}},
  \ and\ \bibinfo {author} {\bibfnamefont {T.}~\bibnamefont {Jungwirth}},\
  }\bibfield  {title} {\enquote {\bibinfo {title} {Electrical detection of
  magnetization reversal without auxiliary magnets},}\ }\href
  {https://link.aps.org/doi/10.1103/PhysRevB.91.180402} {\bibfield  {journal}
  {\bibinfo  {journal} {\emph {Phys. Rev. B}},\ }\textbf {\bibinfo {volume}
  {91}},\ \bibinfo {pages} {180402}\  (\bibinfo {year}
  {2015}{\natexlab{a}})}\BibitemShut {NoStop}%
\bibitem [{\citenamefont {Zhang}\ and\ \citenamefont
  {Vignale}(2016){\natexlab{a}}}]{zhang2016theory}%
  \BibitemOpen
  \bibfield  {author} {\bibinfo {author} {\bibfnamefont {S.~S.-L.}\
  \bibnamefont {Zhang}}\ and\ \bibinfo {author} {\bibfnamefont
  {G.}~\bibnamefont {Vignale}},\ }\bibfield  {title} {\enquote {\bibinfo
  {title} {{Theory of unidirectional spin Hall magnetoresistance in
  heavy-metal/ferromagnetic-metal bilayers}},}\ }\href
  {https://link.aps.org/doi/10.1103/PhysRevB.94.140411} {\bibfield  {journal}
  {\bibinfo  {journal} {\emph {Phys. Rev. B}},\ }\textbf {\bibinfo {volume}
  {94}},\ \bibinfo {pages} {140411}\  (\bibinfo {year}
  {2016}{\natexlab{a}})}\BibitemShut {NoStop}%
\bibitem [{\citenamefont {Yasuda}\ \emph
  {et~al.}(2016){\natexlab{a}}\citenamefont {Yasuda}, \citenamefont
  {Tsukazaki}, \citenamefont {Yoshimi}, \citenamefont {Takahashi},
  \citenamefont {Kawasaki},\ and\ \citenamefont {Tokura}}]{yasuda2016large}%
  \BibitemOpen
  \bibfield  {author} {\bibinfo {author} {\bibfnamefont {K.}~\bibnamefont
  {Yasuda}}, \bibinfo {author} {\bibfnamefont {A.}~\bibnamefont {Tsukazaki}},
  \bibinfo {author} {\bibfnamefont {R.}~\bibnamefont {Yoshimi}}, \bibinfo
  {author} {\bibfnamefont {K.~S.}\ \bibnamefont {Takahashi}}, \bibinfo {author}
  {\bibfnamefont {M.}~\bibnamefont {Kawasaki}}, \ and\ \bibinfo {author}
  {\bibfnamefont {Y.}~\bibnamefont {Tokura}},\ }\bibfield  {title} {\enquote
  {\bibinfo {title} {Large unidirectional magnetoresistance in a magnetic
  topological insulator},}\ }\href
  {https://link.aps.org/doi/10.1103/PhysRevLett.117.127202} {\bibfield
  {journal} {\bibinfo  {journal} {\emph {Phys. Rev. Lett.}},\ }\textbf
  {\bibinfo {volume} {117}},\ \bibinfo {pages} {127202}\  (\bibinfo {year}
  {2016}{\natexlab{a}})}\BibitemShut {NoStop}%
\bibitem [{\citenamefont {Avci}\ \emph
  {et~al.}(2018){\natexlab{a}}\citenamefont {Avci}, \citenamefont {Mendil},
  \citenamefont {Beach},\ and\ \citenamefont {Gambardella}}]{avci2018origins}%
  \BibitemOpen
  \bibfield  {author} {\bibinfo {author} {\bibfnamefont {C.~O.}\ \bibnamefont
  {Avci}}, \bibinfo {author} {\bibfnamefont {J.}~\bibnamefont {Mendil}},
  \bibinfo {author} {\bibfnamefont {G.~S.~D.}\ \bibnamefont {Beach}}, \ and\
  \bibinfo {author} {\bibfnamefont {P.}~\bibnamefont {Gambardella}},\
  }\bibfield  {title} {\enquote {\bibinfo {title} {{Origins of the
  Unidirectional Spin Hall Magnetoresistance in Metallic Bilayers}},}\ }\href
  {https://link.aps.org/doi/10.1103/PhysRevLett.121.087207} {\bibfield
  {journal} {\bibinfo  {journal} {\emph {Phys. Rev. Lett.}},\ }\textbf
  {\bibinfo {volume} {121}},\ \bibinfo {pages} {087207}\  (\bibinfo {year}
  {2018}{\natexlab{a}})}\BibitemShut {NoStop}%
\bibitem [{\citenamefont {Lv}\ \emph {et~al.}(2018){\natexlab{a}}\citenamefont
  {Lv}, \citenamefont {Kally}, \citenamefont {Zhang}, \citenamefont {Lee},
  \citenamefont {Jamali}, \citenamefont {Samarth},\ and\ \citenamefont
  {Wang}}]{lv2018unidirectional}%
  \BibitemOpen
  \bibfield  {author} {\bibinfo {author} {\bibfnamefont {Y.}~\bibnamefont
  {Lv}}, \bibinfo {author} {\bibfnamefont {J.}~\bibnamefont {Kally}}, \bibinfo
  {author} {\bibfnamefont {D.}~\bibnamefont {Zhang}}, \bibinfo {author}
  {\bibfnamefont {J.~S.}\ \bibnamefont {Lee}}, \bibinfo {author} {\bibfnamefont
  {M.}~\bibnamefont {Jamali}}, \bibinfo {author} {\bibfnamefont
  {N.}~\bibnamefont {Samarth}}, \ and\ \bibinfo {author} {\bibfnamefont
  {J.-P.}\ \bibnamefont {Wang}},\ }\bibfield  {title} {\enquote {\bibinfo
  {title} {{Unidirectional spin-Hall and Rashba- Edelstein magnetoresistance in
  topological insulator-ferromagnet layer heterostructures}},}\ }\href
  {https://doi.org/10.1038/s41467-017-02491-3} {\bibfield  {journal} {\bibinfo
  {journal} {\emph {Nat. Commun.}},\ }\textbf {\bibinfo {volume} {9}},\
  \bibinfo {pages} {1--7}\  (\bibinfo {year} {2018}{\natexlab{a}})}\BibitemShut
  {NoStop}%
\bibitem [{\citenamefont {Duy~Khang}\ and\ \citenamefont
  {Hai}(2019){\natexlab{a}}}]{duy2019giant}%
  \BibitemOpen
  \bibfield  {author} {\bibinfo {author} {\bibfnamefont {N.~H.}\ \bibnamefont
  {Duy~Khang}}\ and\ \bibinfo {author} {\bibfnamefont {P.~N.}\ \bibnamefont
  {Hai}},\ }\bibfield  {title} {\enquote {\bibinfo {title} {{Giant
  unidirectional spin Hall magnetoresistance in topological
  insulator--ferromagnetic semiconductor heterostructures}},}\ }\href
  {https://doi.org/10.1063/1.5134728} {\bibfield  {journal} {\bibinfo
  {journal} {\emph {J. Appl. Phys.}},\ }\textbf {\bibinfo {volume} {126}}\
  (\bibinfo {year} {2019}{\natexlab{a}})}\BibitemShut {NoStop}%
\bibitem [{\citenamefont {{Guillet}}\ \emph {et~al.}(2020)\citenamefont
  {{Guillet}}, \citenamefont {{Zucchetti}}, \citenamefont {{Barbedienne}},
  \citenamefont {{Marty}}, \citenamefont {{Isella}}, \citenamefont {{Cagnon}},
  \citenamefont {{Vergnaud}}, \citenamefont {{Jaffr{\`e}s}}, \citenamefont
  {{Reyren}}, \citenamefont {{George}} \emph
  {et~al.}}]{guillet2020observation}%
  \BibitemOpen
  \bibfield  {author} {\bibinfo {author} {\bibfnamefont {T.}~\bibnamefont
  {{Guillet}}}, \bibinfo {author} {\bibfnamefont {C.}~\bibnamefont
  {{Zucchetti}}}, \bibinfo {author} {\bibfnamefont {Q.}~\bibnamefont
  {{Barbedienne}}}, \bibinfo {author} {\bibfnamefont {A.}~\bibnamefont
  {{Marty}}}, \bibinfo {author} {\bibfnamefont {G.}~\bibnamefont {{Isella}}},
  \bibinfo {author} {\bibfnamefont {L.}~\bibnamefont {{Cagnon}}}, \bibinfo
  {author} {\bibfnamefont {C.}~\bibnamefont {{Vergnaud}}}, \bibinfo {author}
  {\bibfnamefont {H.}~\bibnamefont {{Jaffr{\`e}s}}}, \bibinfo {author}
  {\bibfnamefont {N.}~\bibnamefont {{Reyren}}}, \bibinfo {author}
  {\bibfnamefont {J.~M.}\ \bibnamefont {{George}}},  \emph {et~al.},\
  }\bibfield  {title} {\enquote {\bibinfo {title} {Observation of large
  unidirectional \text{R}ashba magnetoresistance in \text{G}e(111)},}\ }\href
  {https://link.aps.org/doi/10.1103/PhysRevLett.124.027201} {\bibfield
  {journal} {\bibinfo  {journal} {\emph {Phys. Rev. Lett.}},\ }\textbf
  {\bibinfo {volume} {124}},\ \bibinfo {pages} {027201}\  (\bibinfo {year}
  {2020})}\BibitemShut {NoStop}%
\bibitem [{\citenamefont {\ifmmode~\check{Z}\else \v{Z}\fi{}elezn\'y}\ \emph
  {et~al.}(2021)\citenamefont {\ifmmode~\check{Z}\else \v{Z}\fi{}elezn\'y},
  \citenamefont {Fang}, \citenamefont {Olejn\'{\i}k}, \citenamefont {Patchett},
  \citenamefont {Gerhard}, \citenamefont {Gould}, \citenamefont {Molenkamp},
  \citenamefont {Gomez-Olivella}, \citenamefont {Zemen}, \citenamefont
  {Tich\'y} \emph {et~al.}}]{zelezny2021unidirectional}%
  \BibitemOpen
  \bibfield  {author} {\bibinfo {author} {\bibfnamefont {J.}~\bibnamefont
  {\ifmmode~\check{Z}\else \v{Z}\fi{}elezn\'y}}, \bibinfo {author}
  {\bibfnamefont {Z.}~\bibnamefont {Fang}}, \bibinfo {author} {\bibfnamefont
  {K.}~\bibnamefont {Olejn\'{\i}k}}, \bibinfo {author} {\bibfnamefont
  {J.}~\bibnamefont {Patchett}}, \bibinfo {author} {\bibfnamefont
  {F.}~\bibnamefont {Gerhard}}, \bibinfo {author} {\bibfnamefont
  {C.}~\bibnamefont {Gould}}, \bibinfo {author} {\bibfnamefont {L.~W.}\
  \bibnamefont {Molenkamp}}, \bibinfo {author} {\bibfnamefont {C.}~\bibnamefont
  {Gomez-Olivella}}, \bibinfo {author} {\bibfnamefont {J.}~\bibnamefont
  {Zemen}}, \bibinfo {author} {\bibfnamefont {T.}~\bibnamefont {Tich\'y}},
  \emph {et~al.},\ }\bibfield  {title} {\enquote {\bibinfo {title}
  {{Unidirectional magnetoresistance and spin-orbit torque in NiMnSb}},}\
  }\href {https://link.aps.org/doi/10.1103/PhysRevB.104.054429} {\bibfield
  {journal} {\bibinfo  {journal} {\emph {Phys. Rev. B}},\ }\textbf {\bibinfo
  {volume} {104}},\ \bibinfo {pages} {054429}\  (\bibinfo {year}
  {2021})}\BibitemShut {NoStop}%
\bibitem [{\citenamefont {{Guillet}}\ \emph
  {et~al.}(2021){\natexlab{a}}\citenamefont {{Guillet}}, \citenamefont
  {{Marty}}, \citenamefont {{Vergnaud}}, \citenamefont {{Jamet}}, \citenamefont
  {{Zucchetti}}, \citenamefont {{Isella}}, \citenamefont {{Barbedienne}},
  \citenamefont {{Jaffr{\`e}s}}, \citenamefont {{Reyren}}, \citenamefont
  {{George}} \emph {et~al.}}]{guillet2021large}%
  \BibitemOpen
  \bibfield  {author} {\bibinfo {author} {\bibfnamefont {T.}~\bibnamefont
  {{Guillet}}}, \bibinfo {author} {\bibfnamefont {A.}~\bibnamefont {{Marty}}},
  \bibinfo {author} {\bibfnamefont {C.}~\bibnamefont {{Vergnaud}}}, \bibinfo
  {author} {\bibfnamefont {M.}~\bibnamefont {{Jamet}}}, \bibinfo {author}
  {\bibfnamefont {C.}~\bibnamefont {{Zucchetti}}}, \bibinfo {author}
  {\bibfnamefont {G.}~\bibnamefont {{Isella}}}, \bibinfo {author}
  {\bibfnamefont {Q.}~\bibnamefont {{Barbedienne}}}, \bibinfo {author}
  {\bibfnamefont {H.}~\bibnamefont {{Jaffr{\`e}s}}}, \bibinfo {author}
  {\bibfnamefont {N.}~\bibnamefont {{Reyren}}}, \bibinfo {author}
  {\bibfnamefont {J.~M.}\ \bibnamefont {{George}}},  \emph {et~al.},\
  }\bibfield  {title} {\enquote {\bibinfo {title} {{Large Rashba unidirectional
  magnetoresistance in the Fe/Ge(111) interface states}},}\ }\href
  {https://doi.org/10.1103/PhysRevB.103.064411} {\bibfield  {journal} {\bibinfo
   {journal} {\emph {Phys. Rev. B}},\ }\textbf {\bibinfo {volume} {103}},\
  \bibinfo {pages} {064411}\  (\bibinfo {year}
  {2021}{\natexlab{a}})}\BibitemShut {NoStop}%
\bibitem [{\citenamefont {Hasegawa}\ \emph {et~al.}(2021)\citenamefont
  {Hasegawa}, \citenamefont {Koyama},\ and\ \citenamefont
  {Chiba}}]{hasegawa2021enhanced}%
  \BibitemOpen
  \bibfield  {author} {\bibinfo {author} {\bibfnamefont {K.}~\bibnamefont
  {Hasegawa}}, \bibinfo {author} {\bibfnamefont {T.}~\bibnamefont {Koyama}}, \
  and\ \bibinfo {author} {\bibfnamefont {D.}~\bibnamefont {Chiba}},\ }\bibfield
   {title} {\enquote {\bibinfo {title} {{Enhanced unidirectional spin Hall
  magnetoresistance in a Pt/Co system with a Cu interlayer}},}\ }\href
  {https://link.aps.org/doi/10.1103/PhysRevB.103.L020411} {\bibfield  {journal}
  {\bibinfo  {journal} {\emph {Phys. Rev. B}},\ }\textbf {\bibinfo {volume}
  {103}},\ \bibinfo {pages} {L020411}\  (\bibinfo {year} {2021})}\BibitemShut
  {NoStop}%
\bibitem [{\citenamefont {Liu}\ \emph {et~al.}(2021){\natexlab{a}}\citenamefont
  {Liu}, \citenamefont {Wang}, \citenamefont {Luan}, \citenamefont {Zhou},
  \citenamefont {Xia}, \citenamefont {Yang}, \citenamefont {Tian},
  \citenamefont {Guo}, \citenamefont {Du},\ and\ \citenamefont
  {Wu}}]{liu2021magnonic}%
  \BibitemOpen
  \bibfield  {author} {\bibinfo {author} {\bibfnamefont {G.}~\bibnamefont
  {Liu}}, \bibinfo {author} {\bibfnamefont {X.-g.}\ \bibnamefont {Wang}},
  \bibinfo {author} {\bibfnamefont {Z.~Z.}\ \bibnamefont {Luan}}, \bibinfo
  {author} {\bibfnamefont {L.~F.}\ \bibnamefont {Zhou}}, \bibinfo {author}
  {\bibfnamefont {S.~Y.}\ \bibnamefont {Xia}}, \bibinfo {author} {\bibfnamefont
  {B.}~\bibnamefont {Yang}}, \bibinfo {author} {\bibfnamefont {Y.~Z.}\
  \bibnamefont {Tian}}, \bibinfo {author} {\bibfnamefont {G.-h.}\ \bibnamefont
  {Guo}}, \bibinfo {author} {\bibfnamefont {J.}~\bibnamefont {Du}}, \ and\
  \bibinfo {author} {\bibfnamefont {D.}~\bibnamefont {Wu}},\ }\bibfield
  {title} {\enquote {\bibinfo {title} {Magnonic unidirectional spin \text{H}all
  magnetoresistance in a heavy-metal--ferromagnetic-insulator bilayer},}\
  }\href {https://link.aps.org/doi/10.1103/PhysRevLett.127.207206} {\bibfield
  {journal} {\bibinfo  {journal} {\emph {Phys. Rev. Lett.}},\ }\textbf
  {\bibinfo {volume} {127}},\ \bibinfo {pages} {207206}\  (\bibinfo {year}
  {2021}{\natexlab{a}})}\BibitemShut {NoStop}%
\bibitem [{\citenamefont {Liu}\ \emph {et~al.}(2021){\natexlab{b}}\citenamefont
  {Liu}, \citenamefont {Holder},\ and\ \citenamefont {Yan}}]{liu2021chirality}%
  \BibitemOpen
  \bibfield  {author} {\bibinfo {author} {\bibfnamefont {Y.}~\bibnamefont
  {Liu}}, \bibinfo {author} {\bibfnamefont {T.}~\bibnamefont {Holder}}, \ and\
  \bibinfo {author} {\bibfnamefont {B.}~\bibnamefont {Yan}},\ }\bibfield
  {title} {\enquote {\bibinfo {title} {Chirality-induced giant unidirectional
  magnetoresistance in twisted bilayer graphene},}\ }\href
  {https://doi.org/10.1016/j.xinn.2021.100085} {\bibfield  {journal} {\bibinfo
  {journal} {\emph {The Innovation}},\ }\textbf {\bibinfo {volume} {2}},\
  \bibinfo {pages} {100085}\  (\bibinfo {year}
  {2021}{\natexlab{b}})}\BibitemShut {NoStop}%
\bibitem [{\citenamefont {Chang}\ \emph {et~al.}(2021)\citenamefont {Chang},
  \citenamefont {Cheng}, \citenamefont {Huang}, \citenamefont {Peng},
  \citenamefont {Huang}, \citenamefont {Chen}, \citenamefont {Liu},\ and\
  \citenamefont {Pai}}]{chang2021large}%
  \BibitemOpen
  \bibfield  {author} {\bibinfo {author} {\bibfnamefont {T.-Y.}\ \bibnamefont
  {Chang}}, \bibinfo {author} {\bibfnamefont {C.-L.}\ \bibnamefont {Cheng}},
  \bibinfo {author} {\bibfnamefont {C.-C.}\ \bibnamefont {Huang}}, \bibinfo
  {author} {\bibfnamefont {C.-W.}\ \bibnamefont {Peng}}, \bibinfo {author}
  {\bibfnamefont {Y.-H.}\ \bibnamefont {Huang}}, \bibinfo {author}
  {\bibfnamefont {T.-Y.}\ \bibnamefont {Chen}}, \bibinfo {author}
  {\bibfnamefont {Y.-T.}\ \bibnamefont {Liu}}, \ and\ \bibinfo {author}
  {\bibfnamefont {C.-F.}\ \bibnamefont {Pai}},\ }\bibfield  {title} {\enquote
  {\bibinfo {title} {{Large unidirectional magnetoresistance in metallic
  heterostructures in the spin transfer torque regime}},}\ }\href
  {https://link.aps.org/doi/10.1103/PhysRevB.104.024432} {\bibfield  {journal}
  {\bibinfo  {journal} {\emph {Phys. Rev. B}},\ }\textbf {\bibinfo {volume}
  {104}},\ \bibinfo {pages} {024432}\  (\bibinfo {year} {2021})}\BibitemShut
  {NoStop}%
\bibitem [{\citenamefont {Ding}\ \emph {et~al.}(2022)\citenamefont {Ding},
  \citenamefont {No\"el}, \citenamefont {Krishnaswamy},\ and\ \citenamefont
  {Gambardella}}]{ding2022unidirectional}%
  \BibitemOpen
  \bibfield  {author} {\bibinfo {author} {\bibfnamefont {S.}~\bibnamefont
  {Ding}}, \bibinfo {author} {\bibfnamefont {P.}~\bibnamefont {No\"el}},
  \bibinfo {author} {\bibfnamefont {G.~K.}\ \bibnamefont {Krishnaswamy}}, \
  and\ \bibinfo {author} {\bibfnamefont {P.}~\bibnamefont {Gambardella}},\
  }\bibfield  {title} {\enquote {\bibinfo {title} {Unidirectional orbital
  magnetoresistance in light-metal--ferromagnet bilayers},}\ }\href
  {https://link.aps.org/doi/10.1103/PhysRevResearch.4.L032041} {\bibfield
  {journal} {\bibinfo  {journal} {\emph {Phys. Rev. Res.}},\ }\textbf {\bibinfo
  {volume} {4}},\ \bibinfo {pages} {L032041}\  (\bibinfo {year}
  {2022})}\BibitemShut {NoStop}%
\bibitem [{\citenamefont {Lou}\ \emph {et~al.}(2022)\citenamefont {Lou},
  \citenamefont {Zhao}, \citenamefont {Jiang},\ and\ \citenamefont
  {Bi}}]{lou2022large}%
  \BibitemOpen
  \bibfield  {author} {\bibinfo {author} {\bibfnamefont {K.}~\bibnamefont
  {Lou}}, \bibinfo {author} {\bibfnamefont {Q.}~\bibnamefont {Zhao}}, \bibinfo
  {author} {\bibfnamefont {B.}~\bibnamefont {Jiang}}, \ and\ \bibinfo {author}
  {\bibfnamefont {C.}~\bibnamefont {Bi}},\ }\bibfield  {title} {\enquote
  {\bibinfo {title} {Large anomalous unidirectional magnetoresistance in a
  single ferromagnetic layer},}\ }\href
  {https://link.aps.org/doi/10.1103/PhysRevApplied.17.064052} {\bibfield
  {journal} {\bibinfo  {journal} {\emph {Phys. Rev. Appl.}},\ }\textbf
  {\bibinfo {volume} {17}},\ \bibinfo {pages} {064052}\  (\bibinfo {year}
  {2022})}\BibitemShut {NoStop}%
\bibitem [{\citenamefont {Cheng}\ \emph {et~al.}(2023)\citenamefont {Cheng},
  \citenamefont {Tang}, \citenamefont {Michel}, \citenamefont {Chong},
  \citenamefont {Yang}, \citenamefont {Cheng},\ and\ \citenamefont
  {Wang}}]{cheng2023unidirectional}%
  \BibitemOpen
  \bibfield  {author} {\bibinfo {author} {\bibfnamefont {Y.}~\bibnamefont
  {Cheng}}, \bibinfo {author} {\bibfnamefont {J.}~\bibnamefont {Tang}},
  \bibinfo {author} {\bibfnamefont {J.~J.}\ \bibnamefont {Michel}}, \bibinfo
  {author} {\bibfnamefont {S.~K.}\ \bibnamefont {Chong}}, \bibinfo {author}
  {\bibfnamefont {F.}~\bibnamefont {Yang}}, \bibinfo {author} {\bibfnamefont
  {R.}~\bibnamefont {Cheng}}, \ and\ \bibinfo {author} {\bibfnamefont {K.~L.}\
  \bibnamefont {Wang}},\ }\bibfield  {title} {\enquote {\bibinfo {title}
  {Unidirectional spin \text{H}all magnetoresistance in antiferromagnetic
  heterostructures},}\ }\href
  {https://link.aps.org/doi/10.1103/PhysRevLett.130.086703} {\bibfield
  {journal} {\bibinfo  {journal} {\emph {Phys. Rev. Lett.}},\ }\textbf
  {\bibinfo {volume} {130}},\ \bibinfo {pages} {086703}\  (\bibinfo {year}
  {2023})}\BibitemShut {NoStop}%
\bibitem [{\citenamefont {Fan}\ \emph {et~al.}(2023)\citenamefont {Fan},
  \citenamefont {Zhang}, \citenamefont {Han}, \citenamefont {Lv}, \citenamefont
  {Liu},\ and\ \citenamefont {Wang}}]{fan2023observation}%
  \BibitemOpen
  \bibfield  {author} {\bibinfo {author} {\bibfnamefont {Y.}~\bibnamefont
  {Fan}}, \bibinfo {author} {\bibfnamefont {P.}~\bibnamefont {Zhang}}, \bibinfo
  {author} {\bibfnamefont {J.}~\bibnamefont {Han}}, \bibinfo {author}
  {\bibfnamefont {Y.}~\bibnamefont {Lv}}, \bibinfo {author} {\bibfnamefont
  {L.}~\bibnamefont {Liu}}, \ and\ \bibinfo {author} {\bibfnamefont {J.-P.}\
  \bibnamefont {Wang}},\ }\bibfield  {title} {\enquote {\bibinfo {title}
  {{Observation of the unidirectional magnetoresistance in antiferromagnetic
  insulator Fe2O3/Pt bilayers}},}\ }\href
  {https://doi.org/10.1002/aelm.202300232} {\bibfield  {journal} {\bibinfo
  {journal} {\emph {Adv. Electron. Mater.}},\ \bibinfo {pages} {2300232}}\
  (\bibinfo {year} {2023})}\BibitemShut {NoStop}%
\bibitem [{\citenamefont {Zheng}\ \emph {et~al.}(2023)\citenamefont {Zheng},
  \citenamefont {Gu}, \citenamefont {Zhang}, \citenamefont {Zhang},
  \citenamefont {Zhao}, \citenamefont {Li}, \citenamefont {Ren}, \citenamefont
  {Jia}, \citenamefont {Xiao}, \citenamefont {Zhou} \emph
  {et~al.}}]{zheng2023coexistence}%
  \BibitemOpen
  \bibfield  {author} {\bibinfo {author} {\bibfnamefont {Z.}~\bibnamefont
  {Zheng}}, \bibinfo {author} {\bibfnamefont {Y.}~\bibnamefont {Gu}}, \bibinfo
  {author} {\bibfnamefont {Z.}~\bibnamefont {Zhang}}, \bibinfo {author}
  {\bibfnamefont {X.}~\bibnamefont {Zhang}}, \bibinfo {author} {\bibfnamefont
  {T.}~\bibnamefont {Zhao}}, \bibinfo {author} {\bibfnamefont {H.}~\bibnamefont
  {Li}}, \bibinfo {author} {\bibfnamefont {L.}~\bibnamefont {Ren}}, \bibinfo
  {author} {\bibfnamefont {L.}~\bibnamefont {Jia}}, \bibinfo {author}
  {\bibfnamefont {R.}~\bibnamefont {Xiao}}, \bibinfo {author} {\bibfnamefont
  {H.-A.}\ \bibnamefont {Zhou}},  \emph {et~al.},\ }\bibfield  {title}
  {\enquote {\bibinfo {title} {{Coexistence of Magnon-Induced and
  Rashba-Induced Unidirectional Magnetoresistance in Antiferromagnets}},}\
  }\href {https://doi.org/10.1021/acs.nanolett.3c01082} {\bibfield  {journal}
  {\bibinfo  {journal} {\emph {Nano Lett.}}}\  (\bibinfo {year}
  {2023})}\BibitemShut {NoStop}%
\bibitem [{\citenamefont {Zou}\ \emph {et~al.}(2024)\citenamefont {Zou},
  \citenamefont {Geng}, \citenamefont {Ma}, \citenamefont {Chen}, \citenamefont
  {Sheng},\ and\ \citenamefont {Xing}}]{zou2024nonreciprocal}%
  \BibitemOpen
  \bibfield  {author} {\bibinfo {author} {\bibfnamefont {M.~H.}\ \bibnamefont
  {Zou}}, \bibinfo {author} {\bibfnamefont {H.}~\bibnamefont {Geng}}, \bibinfo
  {author} {\bibfnamefont {R.}~\bibnamefont {Ma}}, \bibinfo {author}
  {\bibfnamefont {W.}~\bibnamefont {Chen}}, \bibinfo {author} {\bibfnamefont
  {L.}~\bibnamefont {Sheng}}, \ and\ \bibinfo {author} {\bibfnamefont {D.~Y.}\
  \bibnamefont {Xing}},\ }\bibfield  {title} {\enquote {\bibinfo {title}
  {Nonreciprocal ballistic transport in asymmetric bands},}\ }\href
  {https://link.aps.org/doi/10.1103/PhysRevB.109.155302} {\bibfield  {journal}
  {\bibinfo  {journal} {\emph {Phys. Rev. B}},\ }\textbf {\bibinfo {volume}
  {109}},\ \bibinfo {pages} {155302}\  (\bibinfo {year} {2024})}\BibitemShut
  {NoStop}%
\bibitem [{\citenamefont {Zhao}\ \emph {et~al.}(2024)\citenamefont {Zhao},
  \citenamefont {Li}, \citenamefont {Liu}, \citenamefont {Li}, \citenamefont
  {Wang}, \citenamefont {Liu}, \citenamefont {Yang}, \citenamefont {Jiang},\
  and\ \citenamefont {Gao}}]{zhao2024large}%
  \BibitemOpen
  \bibfield  {author} {\bibinfo {author} {\bibfnamefont {L.}~\bibnamefont
  {Zhao}}, \bibinfo {author} {\bibfnamefont {Y.}~\bibnamefont {Li}}, \bibinfo
  {author} {\bibfnamefont {F.}~\bibnamefont {Liu}}, \bibinfo {author}
  {\bibfnamefont {T.}~\bibnamefont {Li}}, \bibinfo {author} {\bibfnamefont
  {Y.}~\bibnamefont {Wang}}, \bibinfo {author} {\bibfnamefont {X.}~\bibnamefont
  {Liu}}, \bibinfo {author} {\bibfnamefont {D.}~\bibnamefont {Yang}}, \bibinfo
  {author} {\bibfnamefont {C.}~\bibnamefont {Jiang}}, \ and\ \bibinfo {author}
  {\bibfnamefont {C.}~\bibnamefont {Gao}},\ }\bibfield  {title} {\enquote
  {\bibinfo {title} {{Large unidirectional magnetoresistance from the dual
  functionality of copper oxide in naturally oxidized light-metal Al/Cu bilayer
  films}},}\ }\href {https://link.aps.org/doi/10.1103/PhysRevApplied.21.044020}
  {\bibfield  {journal} {\bibinfo  {journal} {\emph {Phys. Rev. Appl.}},\
  }\textbf {\bibinfo {volume} {21}},\ \bibinfo {pages} {044020}\  (\bibinfo
  {year} {2024})}\BibitemShut {NoStop}%
\bibitem [{\citenamefont {Aoki}\ \emph {et~al.}(2024)\citenamefont {Aoki},
  \citenamefont {Ohshima}, \citenamefont {Shinjo}, \citenamefont {Shiraishi},\
  and\ \citenamefont {Ando}}]{aoki2024evaluation}%
  \BibitemOpen
  \bibfield  {author} {\bibinfo {author} {\bibfnamefont {M.}~\bibnamefont
  {Aoki}}, \bibinfo {author} {\bibfnamefont {R.}~\bibnamefont {Ohshima}},
  \bibinfo {author} {\bibfnamefont {T.}~\bibnamefont {Shinjo}}, \bibinfo
  {author} {\bibfnamefont {M.}~\bibnamefont {Shiraishi}}, \ and\ \bibinfo
  {author} {\bibfnamefont {Y.}~\bibnamefont {Ando}},\ }\bibfield  {title}
  {\enquote {\bibinfo {title} {{Evaluation of Spin Hall Effect in Ferromagnets
  by Means of Unidirectional Spin Hall Magnetoresistance in Ta/Co Bilayers}},}\
  }\href {https://doi.org/10.3379/msjmag.2403R003} {\bibfield  {journal}
  {\bibinfo  {journal} {\emph {J. Magn. Soc. Jpn.}},\ }\textbf {\bibinfo
  {volume} {48}},\ \bibinfo {pages} {28--33}\  (\bibinfo {year}
  {2024})}\BibitemShut {NoStop}%
\bibitem [{\citenamefont {Huang}\ \emph {et~al.}(2024)\citenamefont {Huang},
  \citenamefont {Cui}, \citenamefont {Wang}, \citenamefont {Xie}, \citenamefont
  {Bai}, \citenamefont {Tian}, \citenamefont {Cao},\ and\ \citenamefont
  {Yan}}]{huang2024spin}%
  \BibitemOpen
  \bibfield  {author} {\bibinfo {author} {\bibfnamefont {Q.}~\bibnamefont
  {Huang}}, \bibinfo {author} {\bibfnamefont {X.}~\bibnamefont {Cui}}, \bibinfo
  {author} {\bibfnamefont {S.}~\bibnamefont {Wang}}, \bibinfo {author}
  {\bibfnamefont {R.}~\bibnamefont {Xie}}, \bibinfo {author} {\bibfnamefont
  {L.}~\bibnamefont {Bai}}, \bibinfo {author} {\bibfnamefont {Y.}~\bibnamefont
  {Tian}}, \bibinfo {author} {\bibfnamefont {Q.}~\bibnamefont {Cao}}, \ and\
  \bibinfo {author} {\bibfnamefont {S.}~\bibnamefont {Yan}},\ }\bibfield
  {title} {\enquote {\bibinfo {title} {{Spin-anomalous-Hall unidirectional
  magnetoresistance in light-metal/ferromagnetic-metal bilayers}},}\ }\href
  {https://doi.org/10.1063/5.0194720} {\bibfield  {journal} {\bibinfo
  {journal} {\emph {Appl. Phys. Rev.}},\ }\textbf {\bibinfo {volume} {11}}\
  (\bibinfo {year} {2024})}\BibitemShut {NoStop}%
\bibitem [{\citenamefont {Kao}\ \emph {et~al.}(2024)\citenamefont {Kao},
  \citenamefont {Tang}, \citenamefont {Ortiz}, \citenamefont {Zhu},
  \citenamefont {Yuan}, \citenamefont {Rao}, \citenamefont {Li}, \citenamefont
  {Edgar}, \citenamefont {Yan}, \citenamefont {Mandrus} \emph
  {et~al.}}]{kao2024unconventional}%
  \BibitemOpen
  \bibfield  {author} {\bibinfo {author} {\bibfnamefont {I.}~\bibnamefont
  {Kao}}, \bibinfo {author} {\bibfnamefont {J.}~\bibnamefont {Tang}}, \bibinfo
  {author} {\bibfnamefont {G.~C.}\ \bibnamefont {Ortiz}}, \bibinfo {author}
  {\bibfnamefont {M.}~\bibnamefont {Zhu}}, \bibinfo {author} {\bibfnamefont
  {S.}~\bibnamefont {Yuan}}, \bibinfo {author} {\bibfnamefont {R.}~\bibnamefont
  {Rao}}, \bibinfo {author} {\bibfnamefont {J.}~\bibnamefont {Li}}, \bibinfo
  {author} {\bibfnamefont {J.~H.}\ \bibnamefont {Edgar}}, \bibinfo {author}
  {\bibfnamefont {J.}~\bibnamefont {Yan}}, \bibinfo {author} {\bibfnamefont
  {D.~G.}\ \bibnamefont {Mandrus}},  \emph {et~al.},\ }\bibfield  {title}
  {\enquote {\bibinfo {title} {{Unconventional Unidirectional Magnetoresistance
  in vdW Heterostructures}},}\ }\href {https://arxiv.org/abs/2405.10889}
  {\bibfield  {journal} {\bibinfo  {journal} {\emph {arXiv:2405.10889}}}\
  (\bibinfo {year} {2024})}\BibitemShut {NoStop}%
\bibitem [{\citenamefont {Rikken}\ \emph {et~al.}(2001)\citenamefont {Rikken},
  \citenamefont {F\"olling},\ and\ \citenamefont
  {Wyder}}]{rikken2001electrical}%
  \BibitemOpen
  \bibfield  {author} {\bibinfo {author} {\bibfnamefont {G.~L. J.~A.}\
  \bibnamefont {Rikken}}, \bibinfo {author} {\bibfnamefont {J.}~\bibnamefont
  {F\"olling}}, \ and\ \bibinfo {author} {\bibfnamefont {P.}~\bibnamefont
  {Wyder}},\ }\bibfield  {title} {\enquote {\bibinfo {title} {{Electrical
  Magnetochiral Anisotropy}},}\ }\href
  {https://link.aps.org/doi/10.1103/PhysRevLett.87.236602} {\bibfield
  {journal} {\bibinfo  {journal} {\emph {Phys. Rev. Lett.}},\ }\textbf
  {\bibinfo {volume} {87}},\ \bibinfo {pages} {236602}\  (\bibinfo {year}
  {2001})}\BibitemShut {NoStop}%
\bibitem [{\citenamefont {He}\ \emph {et~al.}(2018)\citenamefont {He},
  \citenamefont {Zhang}, \citenamefont {Zhu}, \citenamefont {Liu},
  \citenamefont {Wang}, \citenamefont {Yu}, \citenamefont {Vignale},\ and\
  \citenamefont {Yang}}]{he2018bilinear}%
  \BibitemOpen
  \bibfield  {author} {\bibinfo {author} {\bibfnamefont {P.}~\bibnamefont
  {He}}, \bibinfo {author} {\bibfnamefont {S.~S.-L.}\ \bibnamefont {Zhang}},
  \bibinfo {author} {\bibfnamefont {D.}~\bibnamefont {Zhu}}, \bibinfo {author}
  {\bibfnamefont {Y.}~\bibnamefont {Liu}}, \bibinfo {author} {\bibfnamefont
  {Y.}~\bibnamefont {Wang}}, \bibinfo {author} {\bibfnamefont {J.}~\bibnamefont
  {Yu}}, \bibinfo {author} {\bibfnamefont {G.}~\bibnamefont {Vignale}}, \ and\
  \bibinfo {author} {\bibfnamefont {H.}~\bibnamefont {Yang}},\ }\bibfield
  {title} {\enquote {\bibinfo {title} {Bilinear magnetoelectric resistance as a
  probe of three-dimensional spin texture in topological surface states},}\
  }\href {https://doi.org/10.1038/s41567-017-0039-y} {\bibfield  {journal}
  {\bibinfo  {journal} {\emph {Nat. Phys.}},\ }\textbf {\bibinfo {volume}
  {14}},\ \bibinfo {pages} {495--499}\  (\bibinfo {year} {2018})}\BibitemShut
  {NoStop}%
\bibitem [{\citenamefont {Dyrda\l{}}\ \emph {et~al.}(2020)\citenamefont
  {Dyrda\l{}}, \citenamefont {Barna\ifmmode~\acute{s}\else \'{s}\fi{}},\ and\
  \citenamefont {Fert}}]{dyrdal2020spin}%
  \BibitemOpen
  \bibfield  {author} {\bibinfo {author} {\bibfnamefont {A.}~\bibnamefont
  {Dyrda\l{}}}, \bibinfo {author} {\bibfnamefont {J.}~\bibnamefont
  {Barna\ifmmode~\acute{s}\else \'{s}\fi{}}}, \ and\ \bibinfo {author}
  {\bibfnamefont {A.}~\bibnamefont {Fert}},\ }\bibfield  {title} {\enquote
  {\bibinfo {title} {Spin-momentum-locking inhomogeneities as a source of
  bilinear magnetoresistance in topological insulators},}\ }\href
  {https://link.aps.org/doi/10.1103/PhysRevLett.124.046802} {\bibfield
  {journal} {\bibinfo  {journal} {\emph {Phys. Rev. Lett.}},\ }\textbf
  {\bibinfo {volume} {124}},\ \bibinfo {pages} {046802}\  (\bibinfo {year}
  {2020})}\BibitemShut {NoStop}%
\bibitem [{\citenamefont {Wang}\ \emph
  {et~al.}(2022){\natexlab{a}}\citenamefont {Wang}, \citenamefont {Liu},
  \citenamefont {Huang}, \citenamefont {Mambakkam}, \citenamefont {Wang},
  \citenamefont {Yang}, \citenamefont {Sheng}, \citenamefont {Law},\ and\
  \citenamefont {Xiao}}]{wang2022large}%
  \BibitemOpen
  \bibfield  {author} {\bibinfo {author} {\bibfnamefont {Y.}~\bibnamefont
  {Wang}}, \bibinfo {author} {\bibfnamefont {B.}~\bibnamefont {Liu}}, \bibinfo
  {author} {\bibfnamefont {Y.-X.}\ \bibnamefont {Huang}}, \bibinfo {author}
  {\bibfnamefont {S.~V.}\ \bibnamefont {Mambakkam}}, \bibinfo {author}
  {\bibfnamefont {Y.}~\bibnamefont {Wang}}, \bibinfo {author} {\bibfnamefont
  {S.~A.}\ \bibnamefont {Yang}}, \bibinfo {author} {\bibfnamefont {X.-L.}\
  \bibnamefont {Sheng}}, \bibinfo {author} {\bibfnamefont {S.~A.}\ \bibnamefont
  {Law}}, \ and\ \bibinfo {author} {\bibfnamefont {J.~Q.}\ \bibnamefont
  {Xiao}},\ }\bibfield  {title} {\enquote {\bibinfo {title} {Large bilinear
  magnetoresistance from \text{R}ashba spin-splitting on the surface of a
  topological insulator},}\ }\href
  {https://link.aps.org/doi/10.1103/PhysRevB.106.L241401} {\bibfield  {journal}
  {\bibinfo  {journal} {\emph {Phys. Rev. B}},\ }\textbf {\bibinfo {volume}
  {106}},\ \bibinfo {pages} {L241401}\  (\bibinfo {year}
  {2022}{\natexlab{a}})}\BibitemShut {NoStop}%
\bibitem [{\citenamefont {Fu}\ \emph {et~al.}(2022)\citenamefont {Fu},
  \citenamefont {Li}, \citenamefont {Papin}, \citenamefont {Noel},
  \citenamefont {Teresi}, \citenamefont {Cosset-Ch{\'e}neau}, \citenamefont
  {Grezes}, \citenamefont {Guillet}, \citenamefont {Thomas}, \citenamefont
  {Niquet} \emph {et~al.}}]{fu2022bilinear}%
  \BibitemOpen
  \bibfield  {author} {\bibinfo {author} {\bibfnamefont {Y.}~\bibnamefont
  {Fu}}, \bibinfo {author} {\bibfnamefont {J.}~\bibnamefont {Li}}, \bibinfo
  {author} {\bibfnamefont {J.}~\bibnamefont {Papin}}, \bibinfo {author}
  {\bibfnamefont {P.}~\bibnamefont {Noel}}, \bibinfo {author} {\bibfnamefont
  {S.}~\bibnamefont {Teresi}}, \bibinfo {author} {\bibfnamefont
  {M.}~\bibnamefont {Cosset-Ch{\'e}neau}}, \bibinfo {author} {\bibfnamefont
  {C.}~\bibnamefont {Grezes}}, \bibinfo {author} {\bibfnamefont
  {T.}~\bibnamefont {Guillet}}, \bibinfo {author} {\bibfnamefont
  {C.}~\bibnamefont {Thomas}}, \bibinfo {author} {\bibfnamefont {Y.-M.}\
  \bibnamefont {Niquet}},  \emph {et~al.},\ }\bibfield  {title} {\enquote
  {\bibinfo {title} {{Bilinear magnetoresistance in HgTe topological insulator:
  opposite signs at opposite surfaces demonstrated by gate control}},}\ }\href
  {https://doi.org/10.1021/acs.nanolett.2c02585} {\bibfield  {journal}
  {\bibinfo  {journal} {\emph {Nano Lett.}},\ }\textbf {\bibinfo {volume}
  {22}},\ \bibinfo {pages} {7867--7873}\  (\bibinfo {year} {2022})}\BibitemShut
  {NoStop}%
\bibitem [{\citenamefont {Golub}\ \emph {et~al.}(2023)\citenamefont {Golub},
  \citenamefont {Ivchenko},\ and\ \citenamefont
  {Spivak}}]{golub2023electrical}%
  \BibitemOpen
  \bibfield  {author} {\bibinfo {author} {\bibfnamefont {L.~E.}\ \bibnamefont
  {Golub}}, \bibinfo {author} {\bibfnamefont {E.~L.}\ \bibnamefont {Ivchenko}},
  \ and\ \bibinfo {author} {\bibfnamefont {B.}~\bibnamefont {Spivak}},\
  }\bibfield  {title} {\enquote {\bibinfo {title} {{Electrical magnetochiral
  current in tellurium}},}\ }\href
  {https://link.aps.org/doi/10.1103/PhysRevB.108.245202} {\bibfield  {journal}
  {\bibinfo  {journal} {\emph {Phys. Rev. B}},\ }\textbf {\bibinfo {volume}
  {108}},\ \bibinfo {pages} {245202}\  (\bibinfo {year} {2023})}\BibitemShut
  {NoStop}%
\bibitem [{\citenamefont {Marx}\ \emph {et~al.}(2024)\citenamefont {Marx},
  \citenamefont {Jafari}, \citenamefont {Tekelenburg}, \citenamefont {Loi},
  \citenamefont {S\l{}awi\ifmmode~\acute{n}\else \'{n}\fi{}ska},\ and\
  \citenamefont {Guimar\~aes}}]{marx2024nonlinear}%
  \BibitemOpen
  \bibfield  {author} {\bibinfo {author} {\bibfnamefont {A.~C.}\ \bibnamefont
  {Marx}}, \bibinfo {author} {\bibfnamefont {H.}~\bibnamefont {Jafari}},
  \bibinfo {author} {\bibfnamefont {E.~K.}\ \bibnamefont {Tekelenburg}},
  \bibinfo {author} {\bibfnamefont {M.~A.}\ \bibnamefont {Loi}}, \bibinfo
  {author} {\bibfnamefont {J.}~\bibnamefont {S\l{}awi\ifmmode~\acute{n}\else
  \'{n}\fi{}ska}}, \ and\ \bibinfo {author} {\bibfnamefont {M.~H.~D.}\
  \bibnamefont {Guimar\~aes}},\ }\bibfield  {title} {\enquote {\bibinfo {title}
  {{Nonlinear magnetotransport in ${\mathrm{MoTe}}_{2}$}},}\ }\href
  {https://link.aps.org/doi/10.1103/PhysRevB.109.125408} {\bibfield  {journal}
  {\bibinfo  {journal} {\emph {Phys. Rev. B}},\ }\textbf {\bibinfo {volume}
  {109}},\ \bibinfo {pages} {125408}\  (\bibinfo {year} {2024})}\BibitemShut
  {NoStop}%
\bibitem [{\citenamefont {Boboshko}\ and\ \citenamefont
  {Dyrda\l{}}(2024)}]{boboshko2024bilinear}%
  \BibitemOpen
  \bibfield  {author} {\bibinfo {author} {\bibfnamefont {K.}~\bibnamefont
  {Boboshko}}\ and\ \bibinfo {author} {\bibfnamefont {A.}~\bibnamefont
  {Dyrda\l{}}},\ }\bibfield  {title} {\enquote {\bibinfo {title} {{Bilinear
  magnetoresistance and planar Hall effect in topological insulators: Interplay
  of scattering on spin-orbital impurities and nonequilibrium spin
  polarization}},}\ }\href
  {https://link.aps.org/doi/10.1103/PhysRevB.109.155420} {\bibfield  {journal}
  {\bibinfo  {journal} {\emph {Phys. Rev. B}},\ }\textbf {\bibinfo {volume}
  {109}},\ \bibinfo {pages} {155420}\  (\bibinfo {year} {2024})}\BibitemShut
  {NoStop}%
\bibitem [{\citenamefont {Kim}\ \emph {et~al.}(2024)\citenamefont {Kim},
  \citenamefont {Kim}, \citenamefont {Lee}, \citenamefont {Oh}, \citenamefont
  {Chen}, \citenamefont {Yang}, \citenamefont {Pu}, \citenamefont {Liu},
  \citenamefont {Hu}, \citenamefont {Cao~Van} \emph {et~al.}}]{kim2024spin}%
  \BibitemOpen
  \bibfield  {author} {\bibinfo {author} {\bibfnamefont {D.-J.}\ \bibnamefont
  {Kim}}, \bibinfo {author} {\bibfnamefont {K.-W.}\ \bibnamefont {Kim}},
  \bibinfo {author} {\bibfnamefont {K.}~\bibnamefont {Lee}}, \bibinfo {author}
  {\bibfnamefont {J.~H.}\ \bibnamefont {Oh}}, \bibinfo {author} {\bibfnamefont
  {X.}~\bibnamefont {Chen}}, \bibinfo {author} {\bibfnamefont {S.}~\bibnamefont
  {Yang}}, \bibinfo {author} {\bibfnamefont {Y.}~\bibnamefont {Pu}}, \bibinfo
  {author} {\bibfnamefont {Y.}~\bibnamefont {Liu}}, \bibinfo {author}
  {\bibfnamefont {F.}~\bibnamefont {Hu}}, \bibinfo {author} {\bibfnamefont
  {P.}~\bibnamefont {Cao~Van}},  \emph {et~al.},\ }\bibfield  {title} {\enquote
  {\bibinfo {title} {{Spin Hall-induced bilinear magnetoelectric
  resistance}},}\ }\href {https://doi.org/10.1038/s41563-024-02000-0}
  {\bibfield  {journal} {\bibinfo  {journal} {\emph {Nat. Mater.}},\ \bibinfo
  {pages} {1--6}}\  (\bibinfo {year} {2024})}\BibitemShut {NoStop}%
\bibitem [{\citenamefont {Sodemann}\ and\ \citenamefont
  {Fu}(2015)}]{sodemann2015quantum}%
  \BibitemOpen
  \bibfield  {author} {\bibinfo {author} {\bibfnamefont {I.}~\bibnamefont
  {Sodemann}}\ and\ \bibinfo {author} {\bibfnamefont {L.}~\bibnamefont {Fu}},\
  }\bibfield  {title} {\enquote {\bibinfo {title} {{Quantum Nonlinear Hall
  Effect Induced by Berry Curvature Dipole in Time-Reversal Invariant
  Materials}},}\ }\href
  {https://link.aps.org/doi/10.1103/PhysRevLett.115.216806} {\bibfield
  {journal} {\bibinfo  {journal} {\emph {Phys. Rev. Lett.}},\ }\textbf
  {\bibinfo {volume} {115}},\ \bibinfo {pages} {216806}\  (\bibinfo {year}
  {2015})}\BibitemShut {NoStop}%
\bibitem [{\citenamefont {Low}\ \emph {et~al.}(2015)\citenamefont {Low},
  \citenamefont {Jiang},\ and\ \citenamefont {Guinea}}]{low2015topological}%
  \BibitemOpen
  \bibfield  {author} {\bibinfo {author} {\bibfnamefont {T.}~\bibnamefont
  {Low}}, \bibinfo {author} {\bibfnamefont {Y.}~\bibnamefont {Jiang}}, \ and\
  \bibinfo {author} {\bibfnamefont {F.}~\bibnamefont {Guinea}},\ }\bibfield
  {title} {\enquote {\bibinfo {title} {{Topological currents in black
  phosphorus with broken inversion symmetry}},}\ }\href
  {https://link.aps.org/doi/10.1103/PhysRevB.92.235447} {\bibfield  {journal}
  {\bibinfo  {journal} {\emph {Phys. Rev. B}},\ }\textbf {\bibinfo {volume}
  {92}},\ \bibinfo {pages} {235447}\  (\bibinfo {year} {2015})}\BibitemShut
  {NoStop}%
\bibitem [{\citenamefont {Yasuda}\ \emph {et~al.}(2017)\citenamefont {Yasuda},
  \citenamefont {Tsukazaki}, \citenamefont {Yoshimi}, \citenamefont {Kondou},
  \citenamefont {Takahashi}, \citenamefont {Otani}, \citenamefont {Kawasaki},\
  and\ \citenamefont {Tokura}}]{yasuda2017current}%
  \BibitemOpen
  \bibfield  {author} {\bibinfo {author} {\bibfnamefont {K.}~\bibnamefont
  {Yasuda}}, \bibinfo {author} {\bibfnamefont {A.}~\bibnamefont {Tsukazaki}},
  \bibinfo {author} {\bibfnamefont {R.}~\bibnamefont {Yoshimi}}, \bibinfo
  {author} {\bibfnamefont {K.}~\bibnamefont {Kondou}}, \bibinfo {author}
  {\bibfnamefont {K.~S.}\ \bibnamefont {Takahashi}}, \bibinfo {author}
  {\bibfnamefont {Y.}~\bibnamefont {Otani}}, \bibinfo {author} {\bibfnamefont
  {M.}~\bibnamefont {Kawasaki}}, \ and\ \bibinfo {author} {\bibfnamefont
  {Y.}~\bibnamefont {Tokura}},\ }\bibfield  {title} {\enquote {\bibinfo {title}
  {{Current-Nonlinear Hall Effect and Spin-Orbit Torque Magnetization Switching
  in a Magnetic Topological Insulator}},}\ }\href
  {https://link.aps.org/doi/10.1103/PhysRevLett.119.137204} {\bibfield
  {journal} {\bibinfo  {journal} {\emph {Phys. Rev. Lett.}},\ }\textbf
  {\bibinfo {volume} {119}},\ \bibinfo {pages} {137204}\  (\bibinfo {year}
  {2017})}\BibitemShut {NoStop}%
\bibitem [{\citenamefont {Du}\ \emph {et~al.}(2018)\citenamefont {Du},
  \citenamefont {Wang}, \citenamefont {Lu},\ and\ \citenamefont
  {Xie}}]{du2018band}%
  \BibitemOpen
  \bibfield  {author} {\bibinfo {author} {\bibfnamefont {Z.~Z.}\ \bibnamefont
  {Du}}, \bibinfo {author} {\bibfnamefont {C.~M.}\ \bibnamefont {Wang}},
  \bibinfo {author} {\bibfnamefont {H.-Z.}\ \bibnamefont {Lu}}, \ and\ \bibinfo
  {author} {\bibfnamefont {X.~C.}\ \bibnamefont {Xie}},\ }\bibfield  {title}
  {\enquote {\bibinfo {title} {{Band Signatures for Strong Nonlinear Hall
  Effect in Bilayer ${\mathrm{WTe}}_{2}$}},}\ }\href
  {https://link.aps.org/doi/10.1103/PhysRevLett.121.266601} {\bibfield
  {journal} {\bibinfo  {journal} {\emph {Phys. Rev. Lett.}},\ }\textbf
  {\bibinfo {volume} {121}},\ \bibinfo {pages} {266601}\  (\bibinfo {year}
  {2018})}\BibitemShut {NoStop}%
\bibitem [{\citenamefont {Facio}\ \emph {et~al.}(2018)\citenamefont {Facio},
  \citenamefont {Efremov}, \citenamefont {Koepernik}, \citenamefont {You},
  \citenamefont {Sodemann},\ and\ \citenamefont {van~den
  Brink}}]{facio2018strongly}%
  \BibitemOpen
  \bibfield  {author} {\bibinfo {author} {\bibfnamefont {J.~I.}\ \bibnamefont
  {Facio}}, \bibinfo {author} {\bibfnamefont {D.}~\bibnamefont {Efremov}},
  \bibinfo {author} {\bibfnamefont {K.}~\bibnamefont {Koepernik}}, \bibinfo
  {author} {\bibfnamefont {J.-S.}\ \bibnamefont {You}}, \bibinfo {author}
  {\bibfnamefont {I.}~\bibnamefont {Sodemann}}, \ and\ \bibinfo {author}
  {\bibfnamefont {J.}~\bibnamefont {van~den Brink}},\ }\bibfield  {title}
  {\enquote {\bibinfo {title} {{Strongly Enhanced Berry Dipole at Topological
  Phase Transitions in BiTeI}},}\ }\href
  {https://link.aps.org/doi/10.1103/PhysRevLett.121.246403} {\bibfield
  {journal} {\bibinfo  {journal} {\emph {Phys. Rev. Lett.}},\ }\textbf
  {\bibinfo {volume} {121}},\ \bibinfo {pages} {246403}\  (\bibinfo {year}
  {2018})}\BibitemShut {NoStop}%
\bibitem [{\citenamefont {You}\ \emph {et~al.}(2018)\citenamefont {You},
  \citenamefont {Fang}, \citenamefont {Xu}, \citenamefont {Kaxiras},\ and\
  \citenamefont {Low}}]{you2018berry}%
  \BibitemOpen
  \bibfield  {author} {\bibinfo {author} {\bibfnamefont {J.-S.}\ \bibnamefont
  {You}}, \bibinfo {author} {\bibfnamefont {S.}~\bibnamefont {Fang}}, \bibinfo
  {author} {\bibfnamefont {S.-Y.}\ \bibnamefont {Xu}}, \bibinfo {author}
  {\bibfnamefont {E.}~\bibnamefont {Kaxiras}}, \ and\ \bibinfo {author}
  {\bibfnamefont {T.}~\bibnamefont {Low}},\ }\bibfield  {title} {\enquote
  {\bibinfo {title} {{Berry curvature dipole current in the transition metal
  dichalcogenides family}},}\ }\href
  {https://link.aps.org/doi/10.1103/PhysRevB.98.121109} {\bibfield  {journal}
  {\bibinfo  {journal} {\emph {Phys. Rev. B}},\ }\textbf {\bibinfo {volume}
  {98}},\ \bibinfo {pages} {121109}\  (\bibinfo {year} {2018})}\BibitemShut
  {NoStop}%
\bibitem [{\citenamefont {Zhang}\ \emph
  {et~al.}(2018){\natexlab{a}}\citenamefont {Zhang}, \citenamefont {Van
  Den~Brink}, \citenamefont {Felser},\ and\ \citenamefont
  {Yan}}]{zhang2018electrically}%
  \BibitemOpen
  \bibfield  {author} {\bibinfo {author} {\bibfnamefont {Y.}~\bibnamefont
  {Zhang}}, \bibinfo {author} {\bibfnamefont {J.}~\bibnamefont {Van
  Den~Brink}}, \bibinfo {author} {\bibfnamefont {C.}~\bibnamefont {Felser}}, \
  and\ \bibinfo {author} {\bibfnamefont {B.}~\bibnamefont {Yan}},\ }\bibfield
  {title} {\enquote {\bibinfo {title} {{Electrically tuneable nonlinear
  anomalous Hall effect in two-dimensional transition-metal dichalcogenides
  WTe2 and MoTe2}},}\ }\href
  {https://iopscience.iop.org/article/10.1088/2053-1583/aad1ae/meta?casa_token=6lPJBbfismEAAAAA:61V2Scxg7xVW8nSGLjqtf3Goo7fEs_54K-CQ_yEIRBwRJhUsCpqNetZacfTpQJ6OFqSx6bQu5q-cOr5KO0GE-Q7FOVh3}
  {\bibfield  {journal} {\bibinfo  {journal} {\emph {2D Mater.}},\ }\textbf
  {\bibinfo {volume} {5}},\ \bibinfo {pages} {044001}\  (\bibinfo {year}
  {2018}{\natexlab{a}})}\BibitemShut {NoStop}%
\bibitem [{\citenamefont {Zhang}\ \emph
  {et~al.}(2018){\natexlab{b}}\citenamefont {Zhang}, \citenamefont {Sun},\ and\
  \citenamefont {Yan}}]{zhang2018berry}%
  \BibitemOpen
  \bibfield  {author} {\bibinfo {author} {\bibfnamefont {Y.}~\bibnamefont
  {Zhang}}, \bibinfo {author} {\bibfnamefont {Y.}~\bibnamefont {Sun}}, \ and\
  \bibinfo {author} {\bibfnamefont {B.}~\bibnamefont {Yan}},\ }\bibfield
  {title} {\enquote {\bibinfo {title} {{Berry curvature dipole in Weyl
  semimetal materials: An ab initio study}},}\ }\href
  {https://link.aps.org/doi/10.1103/PhysRevB.97.041101} {\bibfield  {journal}
  {\bibinfo  {journal} {\emph {Phys. Rev. B}},\ }\textbf {\bibinfo {volume}
  {97}},\ \bibinfo {pages} {041101}\  (\bibinfo {year}
  {2018}{\natexlab{b}})}\BibitemShut {NoStop}%
\bibitem [{\citenamefont {He}\ \emph {et~al.}(2019)\citenamefont {He},
  \citenamefont {Zhang}, \citenamefont {Zhu}, \citenamefont {Shi},
  \citenamefont {Heinonen}, \citenamefont {Vignale},\ and\ \citenamefont
  {Yang}}]{he2019nonlinear}%
  \BibitemOpen
  \bibfield  {author} {\bibinfo {author} {\bibfnamefont {P.}~\bibnamefont
  {He}}, \bibinfo {author} {\bibfnamefont {S.~S.-L.}\ \bibnamefont {Zhang}},
  \bibinfo {author} {\bibfnamefont {D.}~\bibnamefont {Zhu}}, \bibinfo {author}
  {\bibfnamefont {S.}~\bibnamefont {Shi}}, \bibinfo {author} {\bibfnamefont
  {O.~G.}\ \bibnamefont {Heinonen}}, \bibinfo {author} {\bibfnamefont
  {G.}~\bibnamefont {Vignale}}, \ and\ \bibinfo {author} {\bibfnamefont
  {H.}~\bibnamefont {Yang}},\ }\bibfield  {title} {\enquote {\bibinfo {title}
  {Nonlinear planar \text{H}all effect},}\ }\href
  {https://link.aps.org/doi/10.1103/PhysRevLett.123.016801} {\bibfield
  {journal} {\bibinfo  {journal} {\emph {Phys. Rev. Lett.}},\ }\textbf
  {\bibinfo {volume} {123}},\ \bibinfo {pages} {016801}\  (\bibinfo {year}
  {2019})}\BibitemShut {NoStop}%
\bibitem [{\citenamefont {Ma}\ \emph {et~al.}(2019)\citenamefont {Ma},
  \citenamefont {Xu}, \citenamefont {Shen}, \citenamefont {MacNeill},
  \citenamefont {Fatemi}, \citenamefont {Chang}, \citenamefont {Mier~Valdivia},
  \citenamefont {Wu}, \citenamefont {Du}, \citenamefont {Hsu} \emph
  {et~al.}}]{ma2019observation}%
  \BibitemOpen
  \bibfield  {author} {\bibinfo {author} {\bibfnamefont {Q.}~\bibnamefont
  {Ma}}, \bibinfo {author} {\bibfnamefont {S.-Y.}\ \bibnamefont {Xu}}, \bibinfo
  {author} {\bibfnamefont {H.}~\bibnamefont {Shen}}, \bibinfo {author}
  {\bibfnamefont {D.}~\bibnamefont {MacNeill}}, \bibinfo {author}
  {\bibfnamefont {V.}~\bibnamefont {Fatemi}}, \bibinfo {author} {\bibfnamefont
  {T.-R.}\ \bibnamefont {Chang}}, \bibinfo {author} {\bibfnamefont {A.~M.}\
  \bibnamefont {Mier~Valdivia}}, \bibinfo {author} {\bibfnamefont
  {S.}~\bibnamefont {Wu}}, \bibinfo {author} {\bibfnamefont {Z.}~\bibnamefont
  {Du}}, \bibinfo {author} {\bibfnamefont {C.-H.}\ \bibnamefont {Hsu}},  \emph
  {et~al.},\ }\bibfield  {title} {\enquote {\bibinfo {title} {{Observation of
  the nonlinear Hall effect under time-reversal-symmetric conditions}},}\
  }\href {https://doi.org/10.1038/s41586-018-0807-6} {\bibfield  {journal}
  {\bibinfo  {journal} {\emph {Nature}},\ }\textbf {\bibinfo {volume} {565}},\
  \bibinfo {pages} {337--342}\  (\bibinfo {year} {2019})}\BibitemShut {NoStop}%
\bibitem [{\citenamefont {Kang}\ \emph {et~al.}(2019)\citenamefont {Kang},
  \citenamefont {Li}, \citenamefont {Sohn}, \citenamefont {Shan},\ and\
  \citenamefont {Mak}}]{kang2019nonlinear}%
  \BibitemOpen
  \bibfield  {author} {\bibinfo {author} {\bibfnamefont {K.}~\bibnamefont
  {Kang}}, \bibinfo {author} {\bibfnamefont {T.}~\bibnamefont {Li}}, \bibinfo
  {author} {\bibfnamefont {E.}~\bibnamefont {Sohn}}, \bibinfo {author}
  {\bibfnamefont {J.}~\bibnamefont {Shan}}, \ and\ \bibinfo {author}
  {\bibfnamefont {K.~F.}\ \bibnamefont {Mak}},\ }\bibfield  {title} {\enquote
  {\bibinfo {title} {{Nonlinear anomalous Hall effect in few-layer WTe2}},}\
  }\href {https://doi.org/10.1038/s41563-019-0294-7} {\bibfield  {journal}
  {\bibinfo  {journal} {\emph {Nat. Mater.}},\ }\textbf {\bibinfo {volume}
  {18}},\ \bibinfo {pages} {324--328}\  (\bibinfo {year} {2019})}\BibitemShut
  {NoStop}%
\bibitem [{\citenamefont {Du}\ \emph {et~al.}(2019)\citenamefont {Du},
  \citenamefont {Wang}, \citenamefont {Li}, \citenamefont {Lu},\ and\
  \citenamefont {Xie}}]{du2019disorder}%
  \BibitemOpen
  \bibfield  {author} {\bibinfo {author} {\bibfnamefont {Z.}~\bibnamefont
  {Du}}, \bibinfo {author} {\bibfnamefont {C.}~\bibnamefont {Wang}}, \bibinfo
  {author} {\bibfnamefont {S.}~\bibnamefont {Li}}, \bibinfo {author}
  {\bibfnamefont {H.-Z.}\ \bibnamefont {Lu}}, \ and\ \bibinfo {author}
  {\bibfnamefont {X.}~\bibnamefont {Xie}},\ }\bibfield  {title} {\enquote
  {\bibinfo {title} {{Disorder-induced nonlinear Hall effect with time-reversal
  symmetry}},}\ }\href {https://doi.org/10.1038/s41467-019-10941-3} {\bibfield
  {journal} {\bibinfo  {journal} {\emph {Nat. Commun.}},\ }\textbf {\bibinfo
  {volume} {10}},\ \bibinfo {pages} {3047}\  (\bibinfo {year}
  {2019})}\BibitemShut {NoStop}%
\bibitem [{\citenamefont {Wang}\ \emph {et~al.}(2021)\citenamefont {Wang},
  \citenamefont {Gao},\ and\ \citenamefont {Xiao}}]{wang2021intrinsic}%
  \BibitemOpen
  \bibfield  {author} {\bibinfo {author} {\bibfnamefont {C.}~\bibnamefont
  {Wang}}, \bibinfo {author} {\bibfnamefont {Y.}~\bibnamefont {Gao}}, \ and\
  \bibinfo {author} {\bibfnamefont {D.}~\bibnamefont {Xiao}},\ }\bibfield
  {title} {\enquote {\bibinfo {title} {{Intrinsic Nonlinear \text{H}all Effect
  in Antiferromagnetic Tetragonal CuMnAs}},}\ }\href
  {https://link.aps.org/doi/10.1103/PhysRevLett.127.277201} {\bibfield
  {journal} {\bibinfo  {journal} {\emph {Phys. Rev. Lett.}},\ }\textbf
  {\bibinfo {volume} {127}},\ \bibinfo {pages} {277201}\  (\bibinfo {year}
  {2021})}\BibitemShut {NoStop}%
\bibitem [{\citenamefont {Li}\ \emph {et~al.}(2021){\natexlab{a}}\citenamefont
  {Li}, \citenamefont {Heinonen}, \citenamefont {Burkov},\ and\ \citenamefont
  {Zhang}}]{li2021nonlinear}%
  \BibitemOpen
  \bibfield  {author} {\bibinfo {author} {\bibfnamefont {R.-H.}\ \bibnamefont
  {Li}}, \bibinfo {author} {\bibfnamefont {O.~G.}\ \bibnamefont {Heinonen}},
  \bibinfo {author} {\bibfnamefont {A.~A.}\ \bibnamefont {Burkov}}, \ and\
  \bibinfo {author} {\bibfnamefont {S.~S.-L.}\ \bibnamefont {Zhang}},\
  }\bibfield  {title} {\enquote {\bibinfo {title} {{Nonlinear Hall effect in
  Weyl semimetals induced by chiral anomaly}},}\ }\href
  {https://link.aps.org/doi/10.1103/PhysRevB.103.045105} {\bibfield  {journal}
  {\bibinfo  {journal} {\emph {Phys. Rev. B}},\ }\textbf {\bibinfo {volume}
  {103}},\ \bibinfo {pages} {045105}\  (\bibinfo {year}
  {2021}{\natexlab{a}})}\BibitemShut {NoStop}%
\bibitem [{\citenamefont {Zeng}\ \emph {et~al.}(2021)\citenamefont {Zeng},
  \citenamefont {Nandy},\ and\ \citenamefont {Tewari}}]{zeng2021nonlinear}%
  \BibitemOpen
  \bibfield  {author} {\bibinfo {author} {\bibfnamefont {C.}~\bibnamefont
  {Zeng}}, \bibinfo {author} {\bibfnamefont {S.}~\bibnamefont {Nandy}}, \ and\
  \bibinfo {author} {\bibfnamefont {S.}~\bibnamefont {Tewari}},\ }\bibfield
  {title} {\enquote {\bibinfo {title} {{Nonlinear transport in Weyl semimetals
  induced by Berry curvature dipole}},}\ }\href
  {https://link.aps.org/doi/10.1103/PhysRevB.103.245119} {\bibfield  {journal}
  {\bibinfo  {journal} {\emph {Phys. Rev. B}},\ }\textbf {\bibinfo {volume}
  {103}},\ \bibinfo {pages} {245119}\  (\bibinfo {year} {2021})}\BibitemShut
  {NoStop}%
\bibitem [{\citenamefont {Wang}\ \emph
  {et~al.}(2022){\natexlab{b}}\citenamefont {Wang}, \citenamefont {Mambakkam},
  \citenamefont {Huang}, \citenamefont {Wang}, \citenamefont {Ji},
  \citenamefont {Xiao}, \citenamefont {Yang}, \citenamefont {Law},\ and\
  \citenamefont {Xiao}}]{wang2022observation}%
  \BibitemOpen
  \bibfield  {author} {\bibinfo {author} {\bibfnamefont {Y.}~\bibnamefont
  {Wang}}, \bibinfo {author} {\bibfnamefont {S.~V.}\ \bibnamefont {Mambakkam}},
  \bibinfo {author} {\bibfnamefont {Y.-X.}\ \bibnamefont {Huang}}, \bibinfo
  {author} {\bibfnamefont {Y.}~\bibnamefont {Wang}}, \bibinfo {author}
  {\bibfnamefont {Y.}~\bibnamefont {Ji}}, \bibinfo {author} {\bibfnamefont
  {C.}~\bibnamefont {Xiao}}, \bibinfo {author} {\bibfnamefont {S.~A.}\
  \bibnamefont {Yang}}, \bibinfo {author} {\bibfnamefont {S.~A.}\ \bibnamefont
  {Law}}, \ and\ \bibinfo {author} {\bibfnamefont {J.~Q.}\ \bibnamefont
  {Xiao}},\ }\bibfield  {title} {\enquote {\bibinfo {title} {Observation of
  nonlinear planar \text{H}all effect in
  magnetic-insulator--topological-insulator heterostructures},}\ }\href
  {https://link.aps.org/doi/10.1103/PhysRevB.106.155408} {\bibfield  {journal}
  {\bibinfo  {journal} {\emph {Phys. Rev. B}},\ }\textbf {\bibinfo {volume}
  {106}},\ \bibinfo {pages} {155408}\  (\bibinfo {year}
  {2022}{\natexlab{b}})}\BibitemShut {NoStop}%
\bibitem [{\citenamefont {Gao}\ \emph {et~al.}(2023)\citenamefont {Gao},
  \citenamefont {Liu}, \citenamefont {Qiu}, \citenamefont {Ghosh},
  \citenamefont {V.~Trevisan}, \citenamefont {Onishi}, \citenamefont {Hu},
  \citenamefont {Qian}, \citenamefont {Tien}, \citenamefont {Chen} \emph
  {et~al.}}]{gao2023quantum}%
  \BibitemOpen
  \bibfield  {author} {\bibinfo {author} {\bibfnamefont {A.}~\bibnamefont
  {Gao}}, \bibinfo {author} {\bibfnamefont {Y.-F.}\ \bibnamefont {Liu}},
  \bibinfo {author} {\bibfnamefont {J.-X.}\ \bibnamefont {Qiu}}, \bibinfo
  {author} {\bibfnamefont {B.}~\bibnamefont {Ghosh}}, \bibinfo {author}
  {\bibfnamefont {T.}~\bibnamefont {V.~Trevisan}}, \bibinfo {author}
  {\bibfnamefont {Y.}~\bibnamefont {Onishi}}, \bibinfo {author} {\bibfnamefont
  {C.}~\bibnamefont {Hu}}, \bibinfo {author} {\bibfnamefont {T.}~\bibnamefont
  {Qian}}, \bibinfo {author} {\bibfnamefont {H.-J.}\ \bibnamefont {Tien}},
  \bibinfo {author} {\bibfnamefont {S.-W.}\ \bibnamefont {Chen}},  \emph
  {et~al.},\ }\bibfield  {title} {\enquote {\bibinfo {title} {{Quantum metric
  nonlinear Hall effect in a topological antiferromagnetic heterostructure}},}\
  }\href {https://doi.org/10.1126/science.adf1506} {\bibfield  {journal}
  {\bibinfo  {journal} {\emph {Science}},\ \bibinfo {pages} {eadf1506}}\
  (\bibinfo {year} {2023})}\BibitemShut {NoStop}%
\bibitem [{\citenamefont {Wang}\ \emph
  {et~al.}(2023){\natexlab{a}}\citenamefont {Wang}, \citenamefont {Kaplan},
  \citenamefont {Zhang}, \citenamefont {Holder}, \citenamefont {Cao},
  \citenamefont {Wang}, \citenamefont {Zhou}, \citenamefont {Zhou},
  \citenamefont {Jiang}, \citenamefont {Zhang} \emph
  {et~al.}}]{wang2023quantum}%
  \BibitemOpen
  \bibfield  {author} {\bibinfo {author} {\bibfnamefont {N.}~\bibnamefont
  {Wang}}, \bibinfo {author} {\bibfnamefont {D.}~\bibnamefont {Kaplan}},
  \bibinfo {author} {\bibfnamefont {Z.}~\bibnamefont {Zhang}}, \bibinfo
  {author} {\bibfnamefont {T.}~\bibnamefont {Holder}}, \bibinfo {author}
  {\bibfnamefont {N.}~\bibnamefont {Cao}}, \bibinfo {author} {\bibfnamefont
  {A.}~\bibnamefont {Wang}}, \bibinfo {author} {\bibfnamefont {X.}~\bibnamefont
  {Zhou}}, \bibinfo {author} {\bibfnamefont {F.}~\bibnamefont {Zhou}}, \bibinfo
  {author} {\bibfnamefont {Z.}~\bibnamefont {Jiang}}, \bibinfo {author}
  {\bibfnamefont {C.}~\bibnamefont {Zhang}},  \emph {et~al.},\ }\bibfield
  {title} {\enquote {\bibinfo {title} {{Quantum-metric-induced nonlinear
  transport in a topological antiferromagnet}},}\ }\href
  {https://doi.org/10.1038/s41586-023-06363-3} {\bibfield  {journal} {\bibinfo
  {journal} {\emph {Nature}},\ }\textbf {\bibinfo {volume} {621}},\ \bibinfo
  {pages} {487--492}\  (\bibinfo {year} {2023}{\natexlab{a}})}\BibitemShut
  {NoStop}%
\bibitem [{\citenamefont {Kaplan}\ \emph
  {et~al.}(2023){\natexlab{a}}\citenamefont {Kaplan}, \citenamefont {Holder},\
  and\ \citenamefont {Yan}}]{kaplan2023general}%
  \BibitemOpen
  \bibfield  {author} {\bibinfo {author} {\bibfnamefont {D.}~\bibnamefont
  {Kaplan}}, \bibinfo {author} {\bibfnamefont {T.}~\bibnamefont {Holder}}, \
  and\ \bibinfo {author} {\bibfnamefont {B.}~\bibnamefont {Yan}},\ }\bibfield
  {title} {\enquote {\bibinfo {title} {{General nonlinear Hall current in
  magnetic insulators beyond the quantum anomalous Hall effect}},}\ }\href
  {https://doi.org/10.1038/s41467-023-38734-9} {\bibfield  {journal} {\bibinfo
  {journal} {\emph {Nat. Commun.}},\ }\textbf {\bibinfo {volume} {14}},\
  \bibinfo {pages} {3053}\  (\bibinfo {year} {2023}{\natexlab{a}})}\BibitemShut
  {NoStop}%
\bibitem [{\citenamefont {Das}\ \emph {et~al.}(2023)\citenamefont {Das},
  \citenamefont {Lahiri}, \citenamefont {Atencia}, \citenamefont {Culcer},\
  and\ \citenamefont {Agarwal}}]{das2023intrinsic}%
  \BibitemOpen
  \bibfield  {author} {\bibinfo {author} {\bibfnamefont {K.}~\bibnamefont
  {Das}}, \bibinfo {author} {\bibfnamefont {S.}~\bibnamefont {Lahiri}},
  \bibinfo {author} {\bibfnamefont {R.~B.}\ \bibnamefont {Atencia}}, \bibinfo
  {author} {\bibfnamefont {D.}~\bibnamefont {Culcer}}, \ and\ \bibinfo {author}
  {\bibfnamefont {A.}~\bibnamefont {Agarwal}},\ }\bibfield  {title} {\enquote
  {\bibinfo {title} {{Intrinsic nonlinear conductivities induced by the quantum
  metric}},}\ }\href {https://link.aps.org/doi/10.1103/PhysRevB.108.L201405}
  {\bibfield  {journal} {\bibinfo  {journal} {\emph {Phys. Rev. B}},\ }\textbf
  {\bibinfo {volume} {108}},\ \bibinfo {pages} {L201405}\  (\bibinfo {year}
  {2023})}\BibitemShut {NoStop}%
\bibitem [{\citenamefont {Ma}\ \emph {et~al.}(2023)\citenamefont {Ma},
  \citenamefont {Arora}, \citenamefont {Vignale},\ and\ \citenamefont
  {Song}}]{ma2023anomalous}%
  \BibitemOpen
  \bibfield  {author} {\bibinfo {author} {\bibfnamefont {D.}~\bibnamefont
  {Ma}}, \bibinfo {author} {\bibfnamefont {A.}~\bibnamefont {Arora}}, \bibinfo
  {author} {\bibfnamefont {G.}~\bibnamefont {Vignale}}, \ and\ \bibinfo
  {author} {\bibfnamefont {J.~C.~W.}\ \bibnamefont {Song}},\ }\bibfield
  {title} {\enquote {\bibinfo {title} {{Anomalous Skew-Scattering Nonlinear
  Hall Effect and Chiral Photocurrents in $\mathcal{PT}$-Symmetric
  Antiferromagnets}},}\ }\href
  {https://link.aps.org/doi/10.1103/PhysRevLett.131.076601} {\bibfield
  {journal} {\bibinfo  {journal} {\emph {Phys. Rev. Lett.}},\ }\textbf
  {\bibinfo {volume} {131}},\ \bibinfo {pages} {076601}\  (\bibinfo {year}
  {2023})}\BibitemShut {NoStop}%
\bibitem [{\citenamefont {Zhuang}\ and\ \citenamefont
  {Yan}(2024)}]{zhuang2024intrinsic}%
  \BibitemOpen
  \bibfield  {author} {\bibinfo {author} {\bibfnamefont {Z.-Y.}\ \bibnamefont
  {Zhuang}}\ and\ \bibinfo {author} {\bibfnamefont {Z.}~\bibnamefont {Yan}},\
  }\bibfield  {title} {\enquote {\bibinfo {title} {{Intrinsic nonlinear Hall
  effect in two-dimensional honeycomb topological antiferromagnets}},}\ }\href
  {https://link.aps.org/doi/10.1103/PhysRevB.109.174443} {\bibfield  {journal}
  {\bibinfo  {journal} {\emph {Phys. Rev. B}},\ }\textbf {\bibinfo {volume}
  {109}},\ \bibinfo {pages} {174443}\  (\bibinfo {year} {2024})}\BibitemShut
  {NoStop}%
\bibitem [{\citenamefont {Wang}\ \emph {et~al.}(2024)\citenamefont {Wang},
  \citenamefont {Zhang}, \citenamefont {Zhu},\ and\ \citenamefont
  {Su}}]{wang2024intrinsic}%
  \BibitemOpen
  \bibfield  {author} {\bibinfo {author} {\bibfnamefont {Y.}~\bibnamefont
  {Wang}}, \bibinfo {author} {\bibfnamefont {Z.}~\bibnamefont {Zhang}},
  \bibinfo {author} {\bibfnamefont {Z.-G.}\ \bibnamefont {Zhu}}, \ and\
  \bibinfo {author} {\bibfnamefont {G.}~\bibnamefont {Su}},\ }\bibfield
  {title} {\enquote {\bibinfo {title} {{Intrinsic nonlinear Ohmic current}},}\
  }\href {https://link.aps.org/doi/10.1103/PhysRevB.109.085419} {\bibfield
  {journal} {\bibinfo  {journal} {\emph {Phys. Rev. B}},\ }\textbf {\bibinfo
  {volume} {109}},\ \bibinfo {pages} {085419}\  (\bibinfo {year}
  {2024})}\BibitemShut {NoStop}%
\bibitem [{\citenamefont {Liu}\ \emph {et~al.}(2024){\natexlab{a}}\citenamefont
  {Liu}, \citenamefont {Qiang}, \citenamefont {Lu},\ and\ \citenamefont
  {Xie}}]{liu2024quantum}%
  \BibitemOpen
  \bibfield  {author} {\bibinfo {author} {\bibfnamefont {T.}~\bibnamefont
  {Liu}}, \bibinfo {author} {\bibfnamefont {X.-B.}\ \bibnamefont {Qiang}},
  \bibinfo {author} {\bibfnamefont {H.-Z.}\ \bibnamefont {Lu}}, \ and\ \bibinfo
  {author} {\bibfnamefont {X.}~\bibnamefont {Xie}},\ }\bibfield  {title}
  {\enquote {\bibinfo {title} {{Quantum geometry in condensed matter}},}\
  }\href {https://doi.org/10.1093/nsr/nwae334} {\bibfield  {journal} {\bibinfo
  {journal} {\emph {Natl. Sci. Rev.}},\ \bibinfo {pages} {nwae334}}\  (\bibinfo
  {year} {2024}{\natexlab{a}})}\BibitemShut {NoStop}%
\bibitem [{\citenamefont {Liu}\ \emph {et~al.}(2021){\natexlab{c}}\citenamefont
  {Liu}, \citenamefont {Zhao}, \citenamefont {Huang}, \citenamefont {Wu},
  \citenamefont {Sheng}, \citenamefont {Xiao},\ and\ \citenamefont
  {Yang}}]{liu2021intrinsic}%
  \BibitemOpen
  \bibfield  {author} {\bibinfo {author} {\bibfnamefont {H.}~\bibnamefont
  {Liu}}, \bibinfo {author} {\bibfnamefont {J.}~\bibnamefont {Zhao}}, \bibinfo
  {author} {\bibfnamefont {Y.-X.}\ \bibnamefont {Huang}}, \bibinfo {author}
  {\bibfnamefont {W.}~\bibnamefont {Wu}}, \bibinfo {author} {\bibfnamefont
  {X.-L.}\ \bibnamefont {Sheng}}, \bibinfo {author} {\bibfnamefont
  {C.}~\bibnamefont {Xiao}}, \ and\ \bibinfo {author} {\bibfnamefont {S.~A.}\
  \bibnamefont {Yang}},\ }\bibfield  {title} {\enquote {\bibinfo {title}
  {{Intrinsic Second-Order Anomalous Hall Effect and Its Application in
  Compensated Antiferromagnets}},}\ }\href
  {https://link.aps.org/doi/10.1103/PhysRevLett.127.277202} {\bibfield
  {journal} {\bibinfo  {journal} {\emph {Phys. Rev. Lett.}},\ }\textbf
  {\bibinfo {volume} {127}},\ \bibinfo {pages} {277202}\  (\bibinfo {year}
  {2021}{\natexlab{c}})}\BibitemShut {NoStop}%
\bibitem [{\citenamefont {Parker}\ \emph {et~al.}(2019)\citenamefont {Parker},
  \citenamefont {Morimoto}, \citenamefont {Orenstein},\ and\ \citenamefont
  {Moore}}]{parker2019diagrammatic}%
  \BibitemOpen
  \bibfield  {author} {\bibinfo {author} {\bibfnamefont {D.~E.}\ \bibnamefont
  {Parker}}, \bibinfo {author} {\bibfnamefont {T.}~\bibnamefont {Morimoto}},
  \bibinfo {author} {\bibfnamefont {J.}~\bibnamefont {Orenstein}}, \ and\
  \bibinfo {author} {\bibfnamefont {J.~E.}\ \bibnamefont {Moore}},\ }\bibfield
  {title} {\enquote {\bibinfo {title} {{Diagrammatic approach to nonlinear
  optical response with application to Weyl semimetals}},}\ }\href
  {https://link.aps.org/doi/10.1103/PhysRevB.99.045121} {\bibfield  {journal}
  {\bibinfo  {journal} {\emph {Phys. Rev. B}},\ }\textbf {\bibinfo {volume}
  {99}},\ \bibinfo {pages} {045121}\  (\bibinfo {year} {2019})}\BibitemShut
  {NoStop}%
\bibitem [{\citenamefont {Du}\ \emph {et~al.}(2021){\natexlab{b}}\citenamefont
  {Du}, \citenamefont {Wang}, \citenamefont {Sun}, \citenamefont {Lu},\ and\
  \citenamefont {Xie}}]{du2021quantum}%
  \BibitemOpen
  \bibfield  {author} {\bibinfo {author} {\bibfnamefont {Z.}~\bibnamefont
  {Du}}, \bibinfo {author} {\bibfnamefont {C.}~\bibnamefont {Wang}}, \bibinfo
  {author} {\bibfnamefont {H.-P.}\ \bibnamefont {Sun}}, \bibinfo {author}
  {\bibfnamefont {H.-Z.}\ \bibnamefont {Lu}}, \ and\ \bibinfo {author}
  {\bibfnamefont {X.}~\bibnamefont {Xie}},\ }\bibfield  {title} {\enquote
  {\bibinfo {title} {Quantum theory of the nonlinear \text{H}all effect},}\
  }\href {https://doi.org/10.1038/s41467-021-25273-4} {\bibfield  {journal}
  {\bibinfo  {journal} {\emph {Nat. Commun.}},\ }\textbf {\bibinfo {volume}
  {12}},\ \bibinfo {pages} {1--7}\  (\bibinfo {year}
  {2021}{\natexlab{b}})}\BibitemShut {NoStop}%
\bibitem [{\citenamefont {Rostami}\ \emph {et~al.}(2021)\citenamefont
  {Rostami}, \citenamefont {Katsnelson}, \citenamefont {Vignale},\ and\
  \citenamefont {Polini}}]{rostami2021gauge}%
  \BibitemOpen
  \bibfield  {author} {\bibinfo {author} {\bibfnamefont {H.}~\bibnamefont
  {Rostami}}, \bibinfo {author} {\bibfnamefont {M.~I.}\ \bibnamefont
  {Katsnelson}}, \bibinfo {author} {\bibfnamefont {G.}~\bibnamefont {Vignale}},
  \ and\ \bibinfo {author} {\bibfnamefont {M.}~\bibnamefont {Polini}},\
  }\bibfield  {title} {\enquote {\bibinfo {title} {{Gauge invariance and Ward
  identities in nonlinear response theory}},}\ }\href
  {https://doi.org/10.1016/j.aop.2021.168523} {\bibfield  {journal} {\bibinfo
  {journal} {\emph {Ann. Phys.}},\ }\textbf {\bibinfo {volume} {431}},\
  \bibinfo {pages} {168523}\  (\bibinfo {year} {2021})}\BibitemShut {NoStop}%
\bibitem [{\citenamefont {Kaplan}\ \emph
  {et~al.}(2023){\natexlab{b}}\citenamefont {Kaplan}, \citenamefont {Holder},\
  and\ \citenamefont {Yan}}]{kaplan2023unifying}%
  \BibitemOpen
  \bibfield  {author} {\bibinfo {author} {\bibfnamefont {D.}~\bibnamefont
  {Kaplan}}, \bibinfo {author} {\bibfnamefont {T.}~\bibnamefont {Holder}}, \
  and\ \bibinfo {author} {\bibfnamefont {B.}~\bibnamefont {Yan}},\ }\bibfield
  {title} {\enquote {\bibinfo {title} {{Unifying semiclassics and quantum
  perturbation theory at nonlinear order}},}\ }\href
  {https://scipost.org/10.21468/SciPostPhys.14.4.082} {\bibfield  {journal}
  {\bibinfo  {journal} {\emph {SciPost Phys.}},\ }\textbf {\bibinfo {volume}
  {14}},\ \bibinfo {pages} {082}\  (\bibinfo {year}
  {2023}{\natexlab{b}})}\BibitemShut {NoStop}%
\bibitem [{\citenamefont {McKay}\ \emph {et~al.}(2024)\citenamefont {McKay},
  \citenamefont {Mahmood},\ and\ \citenamefont {Bradlyn}}]{mckay2024charge}%
  \BibitemOpen
  \bibfield  {author} {\bibinfo {author} {\bibfnamefont {R.~C.}\ \bibnamefont
  {McKay}}, \bibinfo {author} {\bibfnamefont {F.}~\bibnamefont {Mahmood}}, \
  and\ \bibinfo {author} {\bibfnamefont {B.}~\bibnamefont {Bradlyn}},\
  }\bibfield  {title} {\enquote {\bibinfo {title} {{Charge Conservation beyond
  Uniformity: Spatially Inhomogeneous Electromagnetic Response in Periodic
  Solids}},}\ }\href {https://link.aps.org/doi/10.1103/PhysRevX.14.011058}
  {\bibfield  {journal} {\bibinfo  {journal} {\emph {Phys. Rev. X}},\ }\textbf
  {\bibinfo {volume} {14}},\ \bibinfo {pages} {011058}\  (\bibinfo {year}
  {2024})}\BibitemShut {NoStop}%
\bibitem [{\citenamefont {Xiao}\ \emph
  {et~al.}(2019){\natexlab{b}}\citenamefont {Xiao}, \citenamefont {Du},\ and\
  \citenamefont {Niu}}]{xiao2019theory}%
  \BibitemOpen
  \bibfield  {author} {\bibinfo {author} {\bibfnamefont {C.}~\bibnamefont
  {Xiao}}, \bibinfo {author} {\bibfnamefont {Z.~Z.}\ \bibnamefont {Du}}, \ and\
  \bibinfo {author} {\bibfnamefont {Q.}~\bibnamefont {Niu}},\ }\bibfield
  {title} {\enquote {\bibinfo {title} {{Theory of nonlinear Hall effects:
  Modified semiclassics from quantum kinetics}},}\ }\href
  {https://link.aps.org/doi/10.1103/PhysRevB.100.165422} {\bibfield  {journal}
  {\bibinfo  {journal} {\emph {Phys. Rev. B}},\ }\textbf {\bibinfo {volume}
  {100}},\ \bibinfo {pages} {165422}\  (\bibinfo {year}
  {2019}{\natexlab{b}})}\BibitemShut {NoStop}%
\bibitem [{\citenamefont {Nandy}\ and\ \citenamefont
  {Sodemann}(2019)}]{nandy2019symmetry}%
  \BibitemOpen
  \bibfield  {author} {\bibinfo {author} {\bibfnamefont {S.}~\bibnamefont
  {Nandy}}\ and\ \bibinfo {author} {\bibfnamefont {I.}~\bibnamefont
  {Sodemann}},\ }\bibfield  {title} {\enquote {\bibinfo {title} {{Symmetry and
  quantum kinetics of the nonlinear Hall effect}},}\ }\href
  {https://link.aps.org/doi/10.1103/PhysRevB.100.195117} {\bibfield  {journal}
  {\bibinfo  {journal} {\emph {Phys. Rev. B}},\ }\textbf {\bibinfo {volume}
  {100}},\ \bibinfo {pages} {195117}\  (\bibinfo {year} {2019})}\BibitemShut
  {NoStop}%
\bibitem [{\citenamefont {Ba}\ \emph {et~al.}(2023)\citenamefont {Ba},
  \citenamefont {Wang}, \citenamefont {Duan}, \citenamefont {Deng},\ and\
  \citenamefont {Wang}}]{ba2023nonlinear}%
  \BibitemOpen
  \bibfield  {author} {\bibinfo {author} {\bibfnamefont {J.-Y.}\ \bibnamefont
  {Ba}}, \bibinfo {author} {\bibfnamefont {Y.-M.}\ \bibnamefont {Wang}},
  \bibinfo {author} {\bibfnamefont {H.-J.}\ \bibnamefont {Duan}}, \bibinfo
  {author} {\bibfnamefont {M.-X.}\ \bibnamefont {Deng}}, \ and\ \bibinfo
  {author} {\bibfnamefont {R.-Q.}\ \bibnamefont {Wang}},\ }\bibfield  {title}
  {\enquote {\bibinfo {title} {{Nonlinear planar Hall effect induced by
  interband transitions: Application to surface states of topological
  insulators}},}\ }\href
  {https://link.aps.org/doi/10.1103/PhysRevB.108.L241104} {\bibfield  {journal}
  {\bibinfo  {journal} {\emph {Phys. Rev. B}},\ }\textbf {\bibinfo {volume}
  {108}},\ \bibinfo {pages} {L241104}\  (\bibinfo {year} {2023})}\BibitemShut
  {NoStop}%
\bibitem [{\citenamefont {Huang}\ \emph {et~al.}(2023)\citenamefont {Huang},
  \citenamefont {Xiao}, \citenamefont {Yang},\ and\ \citenamefont
  {Li}}]{huang2023scaling}%
  \BibitemOpen
  \bibfield  {author} {\bibinfo {author} {\bibfnamefont {Y.-X.}\ \bibnamefont
  {Huang}}, \bibinfo {author} {\bibfnamefont {C.}~\bibnamefont {Xiao}},
  \bibinfo {author} {\bibfnamefont {S.~A.}\ \bibnamefont {Yang}}, \ and\
  \bibinfo {author} {\bibfnamefont {X.}~\bibnamefont {Li}},\ }\bibfield
  {title} {\enquote {\bibinfo {title} {{Scaling law for time-reversal-odd
  nonlinear transport}},}\ }\href {https://arxiv.org/abs/2311.01219} {\bibfield
   {journal} {\bibinfo  {journal} {\emph {arXiv:2311.01219}}}\  (\bibinfo
  {year} {2023})}\BibitemShut {NoStop}%
\bibitem [{\citenamefont {Freimuth}\ \emph {et~al.}(2021)\citenamefont
  {Freimuth}, \citenamefont {Bl{\"u}gel},\ and\ \citenamefont
  {Mokrousov}}]{freimuth2021theory}%
  \BibitemOpen
  \bibfield  {author} {\bibinfo {author} {\bibfnamefont {F.}~\bibnamefont
  {Freimuth}}, \bibinfo {author} {\bibfnamefont {S.}~\bibnamefont
  {Bl{\"u}gel}}, \ and\ \bibinfo {author} {\bibfnamefont {Y.}~\bibnamefont
  {Mokrousov}},\ }\bibfield  {title} {\enquote {\bibinfo {title} {{Theory of
  unidirectional magnetoresistance and nonlinear Hall effect}},}\ }\href
  {https://iopscience.iop.org/article/10.1088/1361-648X/ac327f/meta?casa_token=kQEn_-yG0EMAAAAA:01Alo9_HDofnEUbQF5aftDg3cEvoSsGHJEBnt6WDz2-lrGmQGaB2oTTUOq7RbGn8q3TPVput5WjINphCZZ3giKjXRAX_}
  {\bibfield  {journal} {\bibinfo  {journal} {\emph {J. Phys. Condens.
  Matter}},\ }\textbf {\bibinfo {volume} {34}},\ \bibinfo {pages} {055301}\
  (\bibinfo {year} {2021})}\BibitemShut {NoStop}%
\bibitem [{\citenamefont {Nagaosa}\ \emph
  {et~al.}(2010){\natexlab{a}}\citenamefont {Nagaosa}, \citenamefont {Sinova},
  \citenamefont {Onoda}, \citenamefont {MacDonald},\ and\ \citenamefont
  {Ong}}]{nagaosa2010anomalous}%
  \BibitemOpen
  \bibfield  {author} {\bibinfo {author} {\bibfnamefont {N.}~\bibnamefont
  {Nagaosa}}, \bibinfo {author} {\bibfnamefont {J.}~\bibnamefont {Sinova}},
  \bibinfo {author} {\bibfnamefont {S.}~\bibnamefont {Onoda}}, \bibinfo
  {author} {\bibfnamefont {A.~H.}\ \bibnamefont {MacDonald}}, \ and\ \bibinfo
  {author} {\bibfnamefont {N.~P.}\ \bibnamefont {Ong}},\ }\bibfield  {title}
  {\enquote {\bibinfo {title} {{Anomalous Hall effect}},}\ }\href
  {https://link.aps.org/doi/10.1103/RevModPhys.82.1539} {\bibfield  {journal}
  {\bibinfo  {journal} {\emph {Rev. Mod. Phys.}},\ }\textbf {\bibinfo {volume}
  {82}},\ \bibinfo {pages} {1539--1592}\  (\bibinfo {year}
  {2010}{\natexlab{a}})}\BibitemShut {NoStop}%
\bibitem [{\citenamefont {Provost}\ and\ \citenamefont
  {Vallee}(1980)}]{provost1980riemannian}%
  \BibitemOpen
  \bibfield  {author} {\bibinfo {author} {\bibfnamefont {J.}~\bibnamefont
  {Provost}}\ and\ \bibinfo {author} {\bibfnamefont {G.}~\bibnamefont
  {Vallee}},\ }\bibfield  {title} {\enquote {\bibinfo {title} {{Riemannian
  structure on manifolds of quantum states}},}\ }\href
  {https://doi.org/10.1007/BF02193559} {\bibfield  {journal} {\bibinfo
  {journal} {\emph {Commun. Math. Phys.}},\ }\textbf {\bibinfo {volume} {76}},\
  \bibinfo {pages} {289--301}\  (\bibinfo {year} {1980})}\BibitemShut {NoStop}%
\bibitem [{\citenamefont {Cheng}(2010)}]{cheng2010quantum}%
  \BibitemOpen
  \bibfield  {author} {\bibinfo {author} {\bibfnamefont {R.}~\bibnamefont
  {Cheng}},\ }\bibfield  {title} {\enquote {\bibinfo {title} {{Quantum
  geometric tensor (Fubini-Study metric) in simple quantum system: A
  pedagogical introduction}},}\ }\href {https://arxiv.org/abs/1012.1337}
  {\bibfield  {journal} {\bibinfo  {journal} {\emph {arXiv preprint
  arXiv:1012.1337}}}\  (\bibinfo {year} {2010})}\BibitemShut {NoStop}%
\bibitem [{\citenamefont {Blount}(1962)}]{blount1962formalisms}%
  \BibitemOpen
  \bibfield  {author} {\bibinfo {author} {\bibfnamefont {E.}~\bibnamefont
  {Blount}},\ }\bibfield  {title} {\enquote {\bibinfo {title} {Formalisms of
  band theory},}\ }in\ \href@noop {} {\emph {\bibinfo {booktitle} {Solid state
  physics}}},\ Vol.~\bibinfo {volume} {13}\ (\bibinfo  {publisher} {Elsevier},\
  \bibinfo {year} {1962})\ pp.\ \bibinfo {pages} {305--373}\BibitemShut
  {NoStop}%
\bibitem [{\citenamefont {Gao}\ \emph {et~al.}(2014)\citenamefont {Gao},
  \citenamefont {Yang},\ and\ \citenamefont {Niu}}]{gao2014field}%
  \BibitemOpen
  \bibfield  {author} {\bibinfo {author} {\bibfnamefont {Y.}~\bibnamefont
  {Gao}}, \bibinfo {author} {\bibfnamefont {S.~A.}\ \bibnamefont {Yang}}, \
  and\ \bibinfo {author} {\bibfnamefont {Q.}~\bibnamefont {Niu}},\ }\bibfield
  {title} {\enquote {\bibinfo {title} {{Field Induced Positional Shift of Bloch
  Electrons and Its Dynamical Implications}},}\ }\href
  {https://link.aps.org/doi/10.1103/PhysRevLett.112.166601} {\bibfield
  {journal} {\bibinfo  {journal} {\emph {Phys. Rev. Lett.}},\ }\textbf
  {\bibinfo {volume} {112}},\ \bibinfo {pages} {166601}\  (\bibinfo {year}
  {2014})}\BibitemShut {NoStop}%
\bibitem [{\citenamefont {Het\'enyi}\ and\ \citenamefont
  {L\'evay}(2023)}]{hetenyi2023fluctuations}%
  \BibitemOpen
  \bibfield  {author} {\bibinfo {author} {\bibfnamefont {B.}~\bibnamefont
  {Het\'enyi}}\ and\ \bibinfo {author} {\bibfnamefont {P.}~\bibnamefont
  {L\'evay}},\ }\bibfield  {title} {\enquote {\bibinfo {title} {{Fluctuations,
  uncertainty relations, and the geometry of quantum state manifolds}},}\
  }\href {https://link.aps.org/doi/10.1103/PhysRevA.108.032218} {\bibfield
  {journal} {\bibinfo  {journal} {\emph {Phys. Rev. A}},\ }\textbf {\bibinfo
  {volume} {108}},\ \bibinfo {pages} {032218}\  (\bibinfo {year}
  {2023})}\BibitemShut {NoStop}%
\bibitem [{\citenamefont {Goerbig}\ \emph {et~al.}(2008)\citenamefont
  {Goerbig}, \citenamefont {Fuchs}, \citenamefont {Montambaux},\ and\
  \citenamefont {Pi\'echon}}]{goerbig2008tilted}%
  \BibitemOpen
  \bibfield  {author} {\bibinfo {author} {\bibfnamefont {M.~O.}\ \bibnamefont
  {Goerbig}}, \bibinfo {author} {\bibfnamefont {J.-N.}\ \bibnamefont {Fuchs}},
  \bibinfo {author} {\bibfnamefont {G.}~\bibnamefont {Montambaux}}, \ and\
  \bibinfo {author} {\bibfnamefont {F.}~\bibnamefont {Pi\'echon}},\ }\bibfield
  {title} {\enquote {\bibinfo {title} {{Tilted anisotropic Dirac cones in
  quinoid-type graphene and
  $\ensuremath{\alpha}\text{\ensuremath{-}}{(\text{BEDT-TTF})}_{2}{\text{I}}_{3}$}},}\
  }\href {https://link.aps.org/doi/10.1103/PhysRevB.78.045415} {\bibfield
  {journal} {\bibinfo  {journal} {\emph {Phys. Rev. B}},\ }\textbf {\bibinfo
  {volume} {78}},\ \bibinfo {pages} {045415}\  (\bibinfo {year}
  {2008})}\BibitemShut {NoStop}%
\bibitem [{\citenamefont {Marinescu}\ and\ \citenamefont
  {Tewari}(2023)}]{marinescu2023magnetochiral}%
  \BibitemOpen
  \bibfield  {author} {\bibinfo {author} {\bibfnamefont {D.~C.}\ \bibnamefont
  {Marinescu}}\ and\ \bibinfo {author} {\bibfnamefont {S.}~\bibnamefont
  {Tewari}},\ }\bibfield  {title} {\enquote {\bibinfo {title} {{Magnetochiral
  anisotropy-induced nonlinear Hall effect in spin-orbit coupled Rashba
  conductors}},}\ }\href {https://link.aps.org/doi/10.1103/PhysRevB.108.195303}
  {\bibfield  {journal} {\bibinfo  {journal} {\emph {Phys. Rev. B}},\ }\textbf
  {\bibinfo {volume} {108}},\ \bibinfo {pages} {195303}\  (\bibinfo {year}
  {2023})}\BibitemShut {NoStop}%
\bibitem [{\citenamefont {Gong}\ \emph {et~al.}(2024)\citenamefont {Gong},
  \citenamefont {Du}, \citenamefont {Sun}, \citenamefont {Lu},\ and\
  \citenamefont {Xie}}]{gong2024nonlinear}%
  \BibitemOpen
  \bibfield  {author} {\bibinfo {author} {\bibfnamefont {Z.-H.}\ \bibnamefont
  {Gong}}, \bibinfo {author} {\bibfnamefont {Z.}~\bibnamefont {Du}}, \bibinfo
  {author} {\bibfnamefont {H.-P.}\ \bibnamefont {Sun}}, \bibinfo {author}
  {\bibfnamefont {H.-Z.}\ \bibnamefont {Lu}}, \ and\ \bibinfo {author}
  {\bibfnamefont {X.}~\bibnamefont {Xie}},\ }\bibfield  {title} {\enquote
  {\bibinfo {title} {{Nonlinear transport theory at the order of quantum
  metric}},}\ }\href {https://arxiv.org/abs/2410.04995} {\bibfield  {journal}
  {\bibinfo  {journal} {\emph {arXiv:2410.04995}}}\  (\bibinfo {year}
  {2024})}\BibitemShut {NoStop}%
\bibitem [{\citenamefont {Nakahara}(2018)}]{nakahara2018geometry}%
  \BibitemOpen
  \bibfield  {author} {\bibinfo {author} {\bibfnamefont {M.}~\bibnamefont
  {Nakahara}},\ }\href@noop {} {\emph {\bibinfo {title} {Geometry, topology and
  physics}}}\ (\bibinfo  {publisher} {CRC press},\ \bibinfo {year}
  {2018})\BibitemShut {NoStop}%
\bibitem [{\citenamefont {Ashtekar}\ and\ \citenamefont
  {Schilling}(1999)}]{ashtekar1999geometrical}%
  \BibitemOpen
  \bibfield  {author} {\bibinfo {author} {\bibfnamefont {A.}~\bibnamefont
  {Ashtekar}}\ and\ \bibinfo {author} {\bibfnamefont {T.~A.}\ \bibnamefont
  {Schilling}},\ }\bibfield  {title} {\enquote {\bibinfo {title} {Geometrical
  formulation of quantum mechanics},}\ }in\ \href@noop {} {\emph {\bibinfo
  {booktitle} {On Einstein’s Path: Essays in Honor of Engelbert Schucking}}}\
  (\bibinfo  {publisher} {Springer},\ \bibinfo {year} {1999})\ pp.\ \bibinfo
  {pages} {23--65}\BibitemShut {NoStop}%
\bibitem [{\citenamefont {T\"orm\"a}(2023)}]{torma2023essay}%
  \BibitemOpen
  \bibfield  {author} {\bibinfo {author} {\bibfnamefont {P.}~\bibnamefont
  {T\"orm\"a}},\ }\bibfield  {title} {\enquote {\bibinfo {title} {{Essay: Where
  Can Quantum Geometry Lead Us?}}}\ }\href
  {https://link.aps.org/doi/10.1103/PhysRevLett.131.240001} {\bibfield
  {journal} {\bibinfo  {journal} {\emph {Phys. Rev. Lett.}},\ }\textbf
  {\bibinfo {volume} {131}},\ \bibinfo {pages} {240001}\  (\bibinfo {year}
  {2023})}\BibitemShut {NoStop}%
\bibitem [{\citenamefont {Yu}\ \emph {et~al.}(2024)\citenamefont {Yu},
  \citenamefont {Bernevig}, \citenamefont {Queiroz}, \citenamefont {Rossi},
  \citenamefont {T{\"o}rm{\"a}},\ and\ \citenamefont {Yang}}]{yu2024quantum}%
  \BibitemOpen
  \bibfield  {author} {\bibinfo {author} {\bibfnamefont {J.}~\bibnamefont
  {Yu}}, \bibinfo {author} {\bibfnamefont {B.~A.}\ \bibnamefont {Bernevig}},
  \bibinfo {author} {\bibfnamefont {R.}~\bibnamefont {Queiroz}}, \bibinfo
  {author} {\bibfnamefont {E.}~\bibnamefont {Rossi}}, \bibinfo {author}
  {\bibfnamefont {P.}~\bibnamefont {T{\"o}rm{\"a}}}, \ and\ \bibinfo {author}
  {\bibfnamefont {B.-J.}\ \bibnamefont {Yang}},\ }\bibfield  {title} {\enquote
  {\bibinfo {title} {{Quantum Geometry in Quantum Materials}},}\ }\href
  {https://arxiv.org/abs/2501.00098} {\bibfield  {journal} {\bibinfo  {journal}
  {\emph {arXiv:2501.00098}}}\  (\bibinfo {year} {2024})}\BibitemShut {NoStop}%
\bibitem [{\citenamefont {Chen}(2024)}]{chen2024quantum}%
  \BibitemOpen
  \bibfield  {author} {\bibinfo {author} {\bibfnamefont {W.}~\bibnamefont
  {Chen}},\ }\bibfield  {title} {\enquote {\bibinfo {title} {Quantum
  geometrical properties of topological materials},}\ }\href
  {http://dx.doi.org/10.1088/1361-648X/ad8619} {\bibfield  {journal} {\bibinfo
  {journal} {\emph {J. Phys.: Condens. Matter}},\ }\textbf {\bibinfo {volume}
  {37}},\ \bibinfo {pages} {025605}\  (\bibinfo {year} {2024})}\BibitemShut
  {NoStop}%
\bibitem [{\citenamefont {Jiang}\ \emph {et~al.}(2025)\citenamefont {Jiang},
  \citenamefont {Holder},\ and\ \citenamefont {Yan}}]{jiang2025revealing}%
  \BibitemOpen
  \bibfield  {author} {\bibinfo {author} {\bibfnamefont {Y.}~\bibnamefont
  {Jiang}}, \bibinfo {author} {\bibfnamefont {T.}~\bibnamefont {Holder}}, \
  and\ \bibinfo {author} {\bibfnamefont {B.}~\bibnamefont {Yan}},\ }\bibfield
  {title} {\enquote {\bibinfo {title} {Revealing quantum geometry in nonlinear
  quantum materials},}\ }\href {https://arxiv.org/abs/2503.04943} {\bibfield
  {journal} {\bibinfo  {journal} {\emph {arXiv:2503.04943}}}\  (\bibinfo {year}
  {2025})}\BibitemShut {NoStop}%
\bibitem [{\citenamefont {Verma}\ \emph {et~al.}(2025)\citenamefont {Verma},
  \citenamefont {Moll}, \citenamefont {Holder},\ and\ \citenamefont
  {Queiroz}}]{verma2025quantum}%
  \BibitemOpen
  \bibfield  {author} {\bibinfo {author} {\bibfnamefont {N.}~\bibnamefont
  {Verma}}, \bibinfo {author} {\bibfnamefont {P.~J.}\ \bibnamefont {Moll}},
  \bibinfo {author} {\bibfnamefont {T.}~\bibnamefont {Holder}}, \ and\ \bibinfo
  {author} {\bibfnamefont {R.}~\bibnamefont {Queiroz}},\ }\bibfield  {title}
  {\enquote {\bibinfo {title} {{Quantum Geometry: Revisiting electronic scales
  in quantum matter}},}\ }\href {https://arxiv.org/abs/2504.07173} {\bibfield
  {journal} {\bibinfo  {journal} {\emph {arXiv:2504.07173}}}\  (\bibinfo {year}
  {2025})}\BibitemShut {NoStop}%
\bibitem [{\citenamefont {Neupert}\ \emph {et~al.}(2013)\citenamefont
  {Neupert}, \citenamefont {Chamon},\ and\ \citenamefont
  {Mudry}}]{neupert2013measuring}%
  \BibitemOpen
  \bibfield  {author} {\bibinfo {author} {\bibfnamefont {T.}~\bibnamefont
  {Neupert}}, \bibinfo {author} {\bibfnamefont {C.}~\bibnamefont {Chamon}}, \
  and\ \bibinfo {author} {\bibfnamefont {C.}~\bibnamefont {Mudry}},\ }\bibfield
   {title} {\enquote {\bibinfo {title} {{Measuring the quantum geometry of
  Bloch bands with current noise}},}\ }\href
  {https://link.aps.org/doi/10.1103/PhysRevB.87.245103} {\bibfield  {journal}
  {\bibinfo  {journal} {\emph {Phys. Rev. B}},\ }\textbf {\bibinfo {volume}
  {87}},\ \bibinfo {pages} {245103}\  (\bibinfo {year} {2013})}\BibitemShut
  {NoStop}%
\bibitem [{\citenamefont {Claassen}\ \emph {et~al.}(2015)\citenamefont
  {Claassen}, \citenamefont {Lee}, \citenamefont {Thomale}, \citenamefont
  {Qi},\ and\ \citenamefont {Devereaux}}]{claassen2015position}%
  \BibitemOpen
  \bibfield  {author} {\bibinfo {author} {\bibfnamefont {M.}~\bibnamefont
  {Claassen}}, \bibinfo {author} {\bibfnamefont {C.~H.}\ \bibnamefont {Lee}},
  \bibinfo {author} {\bibfnamefont {R.}~\bibnamefont {Thomale}}, \bibinfo
  {author} {\bibfnamefont {X.-L.}\ \bibnamefont {Qi}}, \ and\ \bibinfo {author}
  {\bibfnamefont {T.~P.}\ \bibnamefont {Devereaux}},\ }\bibfield  {title}
  {\enquote {\bibinfo {title} {{Position-Momentum Duality and Fractional
  Quantum Hall Effect in Chern Insulators}},}\ }\href
  {https://link.aps.org/doi/10.1103/PhysRevLett.114.236802} {\bibfield
  {journal} {\bibinfo  {journal} {\emph {Phys. Rev. Lett.}},\ }\textbf
  {\bibinfo {volume} {114}},\ \bibinfo {pages} {236802}\  (\bibinfo {year}
  {2015})}\BibitemShut {NoStop}%
\bibitem [{\citenamefont {Ahn}\ \emph {et~al.}(2020)\citenamefont {Ahn},
  \citenamefont {Guo},\ and\ \citenamefont {Nagaosa}}]{ahn2020low}%
  \BibitemOpen
  \bibfield  {author} {\bibinfo {author} {\bibfnamefont {J.}~\bibnamefont
  {Ahn}}, \bibinfo {author} {\bibfnamefont {G.-Y.}\ \bibnamefont {Guo}}, \ and\
  \bibinfo {author} {\bibfnamefont {N.}~\bibnamefont {Nagaosa}},\ }\bibfield
  {title} {\enquote {\bibinfo {title} {{Low-Frequency Divergence and Quantum
  Geometry of the Bulk Photovoltaic Effect in Topological Semimetals}},}\
  }\href {https://link.aps.org/doi/10.1103/PhysRevX.10.041041} {\bibfield
  {journal} {\bibinfo  {journal} {\emph {Phys. Rev. X}},\ }\textbf {\bibinfo
  {volume} {10}},\ \bibinfo {pages} {041041}\  (\bibinfo {year}
  {2020})}\BibitemShut {NoStop}%
\bibitem [{\citenamefont {Watanabe}\ and\ \citenamefont
  {Yanase}(2021)}]{watanabe2021chiral}%
  \BibitemOpen
  \bibfield  {author} {\bibinfo {author} {\bibfnamefont {H.}~\bibnamefont
  {Watanabe}}\ and\ \bibinfo {author} {\bibfnamefont {Y.}~\bibnamefont
  {Yanase}},\ }\bibfield  {title} {\enquote {\bibinfo {title} {{Chiral
  Photocurrent in Parity-Violating Magnet and Enhanced Response in Topological
  Antiferromagnet}},}\ }\href
  {https://link.aps.org/doi/10.1103/PhysRevX.11.011001} {\bibfield  {journal}
  {\bibinfo  {journal} {\emph {Phys. Rev. X}},\ }\textbf {\bibinfo {volume}
  {11}},\ \bibinfo {pages} {011001}\  (\bibinfo {year} {2021})}\BibitemShut
  {NoStop}%
\bibitem [{\citenamefont {Komissarov}\ \emph {et~al.}(2024)\citenamefont
  {Komissarov}, \citenamefont {Holder},\ and\ \citenamefont
  {Queiroz}}]{komissarov2024quantum}%
  \BibitemOpen
  \bibfield  {author} {\bibinfo {author} {\bibfnamefont {I.}~\bibnamefont
  {Komissarov}}, \bibinfo {author} {\bibfnamefont {T.}~\bibnamefont {Holder}},
  \ and\ \bibinfo {author} {\bibfnamefont {R.}~\bibnamefont {Queiroz}},\
  }\bibfield  {title} {\enquote {\bibinfo {title} {The quantum geometric origin
  of capacitance in insulators},}\ }\href
  {https://doi.org/10.1038/s41467-024-48808-x} {\bibfield  {journal} {\bibinfo
  {journal} {\emph {Nat. Commun.}},\ }\textbf {\bibinfo {volume} {15}},\
  \bibinfo {pages} {4621}\  (\bibinfo {year} {2024})}\BibitemShut {NoStop}%
\bibitem [{\citenamefont {Onishi}\ and\ \citenamefont
  {Fu}(2024)}]{onishi2024fundamental}%
  \BibitemOpen
  \bibfield  {author} {\bibinfo {author} {\bibfnamefont {Y.}~\bibnamefont
  {Onishi}}\ and\ \bibinfo {author} {\bibfnamefont {L.}~\bibnamefont {Fu}},\
  }\bibfield  {title} {\enquote {\bibinfo {title} {Fundamental bound on
  topological gap},}\ }\href
  {https://link.aps.org/doi/10.1103/PhysRevX.14.011052} {\bibfield  {journal}
  {\bibinfo  {journal} {\emph {Phys. Rev. X}},\ }\textbf {\bibinfo {volume}
  {14}},\ \bibinfo {pages} {011052}\  (\bibinfo {year} {2024})}\BibitemShut
  {NoStop}%
\bibitem [{\citenamefont {Fang}\ \emph {et~al.}(2024)\citenamefont {Fang},
  \citenamefont {Cano},\ and\ \citenamefont {Ghorashi}}]{fang2024quantum}%
  \BibitemOpen
  \bibfield  {author} {\bibinfo {author} {\bibfnamefont {Y.}~\bibnamefont
  {Fang}}, \bibinfo {author} {\bibfnamefont {J.}~\bibnamefont {Cano}}, \ and\
  \bibinfo {author} {\bibfnamefont {S.~A.~A.}\ \bibnamefont {Ghorashi}},\
  }\bibfield  {title} {\enquote {\bibinfo {title} {{Quantum Geometry Induced
  Nonlinear Transport in Altermagnets}},}\ }\href
  {https://link.aps.org/doi/10.1103/PhysRevLett.133.106701} {\bibfield
  {journal} {\bibinfo  {journal} {\emph {Phys. Rev. Lett.}},\ }\textbf
  {\bibinfo {volume} {133}},\ \bibinfo {pages} {106701}\  (\bibinfo {year}
  {2024})}\BibitemShut {NoStop}%
\bibitem [{\citenamefont {Kang}\ \emph {et~al.}(2024)\citenamefont {Kang},
  \citenamefont {Kim}, \citenamefont {Qian}, \citenamefont {Neves},
  \citenamefont {Ye}, \citenamefont {Jung}, \citenamefont {Puntel},
  \citenamefont {Mazzola}, \citenamefont {Fang}, \citenamefont {Jozwiak} \emph
  {et~al.}}]{kang2024measurements}%
  \BibitemOpen
  \bibfield  {author} {\bibinfo {author} {\bibfnamefont {M.}~\bibnamefont
  {Kang}}, \bibinfo {author} {\bibfnamefont {S.}~\bibnamefont {Kim}}, \bibinfo
  {author} {\bibfnamefont {Y.}~\bibnamefont {Qian}}, \bibinfo {author}
  {\bibfnamefont {P.~M.}\ \bibnamefont {Neves}}, \bibinfo {author}
  {\bibfnamefont {L.}~\bibnamefont {Ye}}, \bibinfo {author} {\bibfnamefont
  {J.}~\bibnamefont {Jung}}, \bibinfo {author} {\bibfnamefont {D.}~\bibnamefont
  {Puntel}}, \bibinfo {author} {\bibfnamefont {F.}~\bibnamefont {Mazzola}},
  \bibinfo {author} {\bibfnamefont {S.}~\bibnamefont {Fang}}, \bibinfo {author}
  {\bibfnamefont {C.}~\bibnamefont {Jozwiak}},  \emph {et~al.},\ }\bibfield
  {title} {\enquote {\bibinfo {title} {Measurements of the quantum geometric
  tensor in solids},}\ }\href {https://doi.org/10.1038/s41567-024-02678-8}
  {\bibfield  {journal} {\bibinfo  {journal} {\emph {Nat. Phys.}},\ \bibinfo
  {pages} {1--8}}\  (\bibinfo {year} {2024})}\BibitemShut {NoStop}%
\bibitem [{\citenamefont {Jankowski}\ \emph
  {et~al.}(2025){\natexlab{a}}\citenamefont {Jankowski}, \citenamefont
  {Morris}, \citenamefont {Bouhon}, \citenamefont {\"Unal},\ and\ \citenamefont
  {Slager}}]{jankowski2025optical}%
  \BibitemOpen
  \bibfield  {author} {\bibinfo {author} {\bibfnamefont {W.~J.}\ \bibnamefont
  {Jankowski}}, \bibinfo {author} {\bibfnamefont {A.~S.}\ \bibnamefont
  {Morris}}, \bibinfo {author} {\bibfnamefont {A.}~\bibnamefont {Bouhon}},
  \bibinfo {author} {\bibfnamefont {F.~N.}\ \bibnamefont {\"Unal}}, \ and\
  \bibinfo {author} {\bibfnamefont {R.-J.}\ \bibnamefont {Slager}},\ }\bibfield
   {title} {\enquote {\bibinfo {title} {{Optical manifestations and bounds of
  topological Euler class}},}\ }\href
  {https://link.aps.org/doi/10.1103/PhysRevB.111.L081103} {\bibfield  {journal}
  {\bibinfo  {journal} {\emph {Phys. Rev. B}},\ }\textbf {\bibinfo {volume}
  {111}},\ \bibinfo {pages} {L081103}\  (\bibinfo {year}
  {2025}{\natexlab{a}})}\BibitemShut {NoStop}%
\bibitem [{\citenamefont {Smith}\ \emph {et~al.}(2022)\citenamefont {Smith},
  \citenamefont {Pullasseri},\ and\ \citenamefont
  {Srivastava}}]{smith2022momentum}%
  \BibitemOpen
  \bibfield  {author} {\bibinfo {author} {\bibfnamefont {T.~B.}\ \bibnamefont
  {Smith}}, \bibinfo {author} {\bibfnamefont {L.}~\bibnamefont {Pullasseri}}, \
  and\ \bibinfo {author} {\bibfnamefont {A.}~\bibnamefont {Srivastava}},\
  }\bibfield  {title} {\enquote {\bibinfo {title} {{Momentum-space gravity from
  the quantum geometry and entropy of Bloch electrons}},}\ }\href
  {https://link.aps.org/doi/10.1103/PhysRevResearch.4.013217} {\bibfield
  {journal} {\bibinfo  {journal} {\emph {Phys. Rev. Res.}},\ }\textbf {\bibinfo
  {volume} {4}},\ \bibinfo {pages} {013217}\  (\bibinfo {year}
  {2022})}\BibitemShut {NoStop}%
\bibitem [{\citenamefont {Bekenstein}(1973)}]{bekenstein1973black}%
  \BibitemOpen
  \bibfield  {author} {\bibinfo {author} {\bibfnamefont {J.~D.}\ \bibnamefont
  {Bekenstein}},\ }\bibfield  {title} {\enquote {\bibinfo {title} {{Black Holes
  and Entropy}},}\ }\href {https://link.aps.org/doi/10.1103/PhysRevD.7.2333}
  {\bibfield  {journal} {\bibinfo  {journal} {\emph {Phys. Rev. D}},\ }\textbf
  {\bibinfo {volume} {7}},\ \bibinfo {pages} {2333--2346}\  (\bibinfo {year}
  {1973})}\BibitemShut {NoStop}%
\bibitem [{\citenamefont {Hawking}(1975)}]{hawking1975particle}%
  \BibitemOpen
  \bibfield  {author} {\bibinfo {author} {\bibfnamefont {S.~W.}\ \bibnamefont
  {Hawking}},\ }\bibfield  {title} {\enquote {\bibinfo {title} {Particle
  creation by black holes},}\ }\href {https://doi.org/10.1007/BF02345020}
  {\bibfield  {journal} {\bibinfo  {journal} {\emph {Commun. Math. Phys.}},\
  }\textbf {\bibinfo {volume} {43}},\ \bibinfo {pages} {199--220}\  (\bibinfo
  {year} {1975})}\BibitemShut {NoStop}%
\bibitem [{\citenamefont {Ruppeiner}(1979)}]{ruppeiner1979thermodynamics}%
  \BibitemOpen
  \bibfield  {author} {\bibinfo {author} {\bibfnamefont {G.}~\bibnamefont
  {Ruppeiner}},\ }\bibfield  {title} {\enquote {\bibinfo {title}
  {{Thermodynamics: A Riemannian geometric model}},}\ }\href
  {https://link.aps.org/doi/10.1103/PhysRevA.20.1608} {\bibfield  {journal}
  {\bibinfo  {journal} {\emph {Phys. Rev. A}},\ }\textbf {\bibinfo {volume}
  {20}},\ \bibinfo {pages} {1608--1613}\  (\bibinfo {year} {1979})}\BibitemShut
  {NoStop}%
\bibitem [{\citenamefont {Jacobson}(1995)}]{jacobson1995thermodynamics}%
  \BibitemOpen
  \bibfield  {author} {\bibinfo {author} {\bibfnamefont {T.}~\bibnamefont
  {Jacobson}},\ }\bibfield  {title} {\enquote {\bibinfo {title}
  {{Thermodynamics of Spacetime: The Einstein Equation of State}},}\ }\href
  {https://link.aps.org/doi/10.1103/PhysRevLett.75.1260} {\bibfield  {journal}
  {\bibinfo  {journal} {\emph {Phys. Rev. Lett.}},\ }\textbf {\bibinfo {volume}
  {75}},\ \bibinfo {pages} {1260--1263}\  (\bibinfo {year} {1995})}\BibitemShut
  {NoStop}%
\bibitem [{\citenamefont {Padmanabhan}(2010)}]{padmanabhan2010thermodynamical}%
  \BibitemOpen
  \bibfield  {author} {\bibinfo {author} {\bibfnamefont {T.}~\bibnamefont
  {Padmanabhan}},\ }\bibfield  {title} {\enquote {\bibinfo {title}
  {Thermodynamical aspects of gravity: new insights},}\ }\href
  {https://iopscience.iop.org/article/10.1088/0034-4885/73/4/046901} {\bibfield
   {journal} {\bibinfo  {journal} {\emph {Rep. Prog. Phys.}},\ }\textbf
  {\bibinfo {volume} {73}},\ \bibinfo {pages} {046901}\  (\bibinfo {year}
  {2010})}\BibitemShut {NoStop}%
\bibitem [{\citenamefont {Verlinde}(2011)}]{verlinde2011origin}%
  \BibitemOpen
  \bibfield  {author} {\bibinfo {author} {\bibfnamefont {E.}~\bibnamefont
  {Verlinde}},\ }\bibfield  {title} {\enquote {\bibinfo {title} {{On the origin
  of gravity and the laws of Newton}},}\ }\href
  {https://doi.org/10.1007/JHEP04(2011)029} {\bibfield  {journal} {\bibinfo
  {journal} {\emph {J. High Energy Phys.}},\ }\textbf {\bibinfo {volume}
  {2011}},\ \bibinfo {pages} {1--27}\  (\bibinfo {year} {2011})}\BibitemShut
  {NoStop}%
\bibitem [{\citenamefont {Carroll}\ and\ \citenamefont
  {Remmen}(2016)}]{carroll2016what}%
  \BibitemOpen
  \bibfield  {author} {\bibinfo {author} {\bibfnamefont {S.~M.}\ \bibnamefont
  {Carroll}}\ and\ \bibinfo {author} {\bibfnamefont {G.~N.}\ \bibnamefont
  {Remmen}},\ }\bibfield  {title} {\enquote {\bibinfo {title} {What is the
  entropy in entropic gravity?}}\ }\href
  {https://link.aps.org/doi/10.1103/PhysRevD.93.124052} {\bibfield  {journal}
  {\bibinfo  {journal} {\emph {Phys. Rev. D}},\ }\textbf {\bibinfo {volume}
  {93}},\ \bibinfo {pages} {124052}\  (\bibinfo {year} {2016})}\BibitemShut
  {NoStop}%
\bibitem [{\citenamefont {Bianconi}(2025)}]{bianconi2025gravity}%
  \BibitemOpen
  \bibfield  {author} {\bibinfo {author} {\bibfnamefont {G.}~\bibnamefont
  {Bianconi}},\ }\bibfield  {title} {\enquote {\bibinfo {title} {Gravity from
  entropy},}\ }\href {https://link.aps.org/doi/10.1103/PhysRevD.111.066001}
  {\bibfield  {journal} {\bibinfo  {journal} {\emph {Phys. Rev. D}},\ }\textbf
  {\bibinfo {volume} {111}},\ \bibinfo {pages} {066001}\  (\bibinfo {year}
  {2025})}\BibitemShut {NoStop}%
\bibitem [{\citenamefont {Ahn}\ \emph {et~al.}(2022)\citenamefont {Ahn},
  \citenamefont {Guo}, \citenamefont {Nagaosa},\ and\ \citenamefont
  {Vishwanath}}]{ahn2022riemannian}%
  \BibitemOpen
  \bibfield  {author} {\bibinfo {author} {\bibfnamefont {J.}~\bibnamefont
  {Ahn}}, \bibinfo {author} {\bibfnamefont {G.-Y.}\ \bibnamefont {Guo}},
  \bibinfo {author} {\bibfnamefont {N.}~\bibnamefont {Nagaosa}}, \ and\
  \bibinfo {author} {\bibfnamefont {A.}~\bibnamefont {Vishwanath}},\ }\bibfield
   {title} {\enquote {\bibinfo {title} {Riemannian geometry of resonant optical
  responses},}\ }\href {https://doi.org/10.1038/s41567-021-01465-z} {\bibfield
  {journal} {\bibinfo  {journal} {\emph {Nat. Phys.}},\ }\textbf {\bibinfo
  {volume} {18}},\ \bibinfo {pages} {290--295}\  (\bibinfo {year}
  {2022})}\BibitemShut {NoStop}%
\bibitem [{\citenamefont {Bouhon}\ \emph {et~al.}(2023)\citenamefont {Bouhon},
  \citenamefont {Timmel},\ and\ \citenamefont {Slager}}]{bouhon2023quantum}%
  \BibitemOpen
  \bibfield  {author} {\bibinfo {author} {\bibfnamefont {A.}~\bibnamefont
  {Bouhon}}, \bibinfo {author} {\bibfnamefont {A.}~\bibnamefont {Timmel}}, \
  and\ \bibinfo {author} {\bibfnamefont {R.-J.}\ \bibnamefont {Slager}},\
  }\bibfield  {title} {\enquote {\bibinfo {title} {Quantum geometry beyond
  projective single bands},}\ }\href {https://arxiv.org/abs/2303.02180}
  {\bibfield  {journal} {\bibinfo  {journal} {\emph {arXiv:2303.02180}}}\
  (\bibinfo {year} {2023})}\BibitemShut {NoStop}%
\bibitem [{\citenamefont {Mitscherling}\ \emph {et~al.}(2024)\citenamefont
  {Mitscherling}, \citenamefont {Avdoshkin},\ and\ \citenamefont
  {Moore}}]{mitscherling2024gauge}%
  \BibitemOpen
  \bibfield  {author} {\bibinfo {author} {\bibfnamefont {J.}~\bibnamefont
  {Mitscherling}}, \bibinfo {author} {\bibfnamefont {A.}~\bibnamefont
  {Avdoshkin}}, \ and\ \bibinfo {author} {\bibfnamefont {J.~E.}\ \bibnamefont
  {Moore}},\ }\bibfield  {title} {\enquote {\bibinfo {title} {Gauge-invariant
  projector calculus for quantum state geometry and applications to observables
  in crystals},}\ }\href {https://arxiv.org/abs/2412.03637} {\bibfield
  {journal} {\bibinfo  {journal} {\emph {arXiv:2412.03637}}}\  (\bibinfo {year}
  {2024})}\BibitemShut {NoStop}%
\bibitem [{\citenamefont {Avdoshkin}\ \emph {et~al.}(2024)\citenamefont
  {Avdoshkin}, \citenamefont {Mitscherling},\ and\ \citenamefont
  {Moore}}]{avdoshkin2024multi}%
  \BibitemOpen
  \bibfield  {author} {\bibinfo {author} {\bibfnamefont {A.}~\bibnamefont
  {Avdoshkin}}, \bibinfo {author} {\bibfnamefont {J.}~\bibnamefont
  {Mitscherling}}, \ and\ \bibinfo {author} {\bibfnamefont {J.~E.}\
  \bibnamefont {Moore}},\ }\bibfield  {title} {\enquote {\bibinfo {title} {The
  multi-state geometry of shift current and polarization},}\ }\href
  {https://arxiv.org/abs/2409.16358} {\bibfield  {journal} {\bibinfo  {journal}
  {\emph {arXiv:2409.16358}}}\  (\bibinfo {year} {2024})}\BibitemShut {NoStop}%
\bibitem [{\citenamefont {Jankowski}\ and\ \citenamefont
  {Slager}(2024)}]{jankowski2024quantized}%
  \BibitemOpen
  \bibfield  {author} {\bibinfo {author} {\bibfnamefont {W.~J.}\ \bibnamefont
  {Jankowski}}\ and\ \bibinfo {author} {\bibfnamefont {R.-J.}\ \bibnamefont
  {Slager}},\ }\bibfield  {title} {\enquote {\bibinfo {title} {{Quantized
  Integrated Shift Effect in Multigap Topological Phases}},}\ }\href
  {https://link.aps.org/doi/10.1103/PhysRevLett.133.186601} {\bibfield
  {journal} {\bibinfo  {journal} {\emph {Phys. Rev. Lett.}},\ }\textbf
  {\bibinfo {volume} {133}},\ \bibinfo {pages} {186601}\  (\bibinfo {year}
  {2024})}\BibitemShut {NoStop}%
\bibitem [{\citenamefont {Jankowski}\ \emph
  {et~al.}(2025){\natexlab{b}}\citenamefont {Jankowski}, \citenamefont
  {Slager},\ and\ \citenamefont {Pizzochero}}]{jankowski2025enhancing}%
  \BibitemOpen
  \bibfield  {author} {\bibinfo {author} {\bibfnamefont {W.~J.}\ \bibnamefont
  {Jankowski}}, \bibinfo {author} {\bibfnamefont {R.-J.}\ \bibnamefont
  {Slager}}, \ and\ \bibinfo {author} {\bibfnamefont {M.}~\bibnamefont
  {Pizzochero}},\ }\bibfield  {title} {\enquote {\bibinfo {title} {Enhancing
  the hyperpolarizability of crystals with quantum geometry},}\ }\href
  {https://arxiv.org/abs/2502.02660} {\bibfield  {journal} {\bibinfo  {journal}
  {\emph {arXiv:2502.02660}}}\  (\bibinfo {year}
  {2025}{\natexlab{b}})}\BibitemShut {NoStop}%
\bibitem [{\citenamefont {Misner}\ \emph {et~al.}(1973)\citenamefont {Misner},
  \citenamefont {Thorne},\ and\ \citenamefont
  {Wheeler}}]{misner1973gravitation}%
  \BibitemOpen
  \bibfield  {author} {\bibinfo {author} {\bibfnamefont {C.~W.}\ \bibnamefont
  {Misner}}, \bibinfo {author} {\bibfnamefont {K.~S.}\ \bibnamefont {Thorne}},
  \ and\ \bibinfo {author} {\bibfnamefont {J.~A.}\ \bibnamefont {Wheeler}},\
  }\href@noop {} {\emph {\bibinfo {title} {Gravitation}}}\ (\bibinfo
  {publisher} {Macmillan},\ \bibinfo {year} {1973})\BibitemShut {NoStop}%
\bibitem [{\citenamefont {Wald}(2010)}]{wald2010general}%
  \BibitemOpen
  \bibfield  {author} {\bibinfo {author} {\bibfnamefont {R.~M.}\ \bibnamefont
  {Wald}},\ }\href@noop {} {\emph {\bibinfo {title} {General relativity}}}\
  (\bibinfo  {publisher} {University of Chicago press},\ \bibinfo {year}
  {2010})\BibitemShut {NoStop}%
\bibitem [{\citenamefont {Souza}\ \emph {et~al.}(2000)\citenamefont {Souza},
  \citenamefont {Wilkens},\ and\ \citenamefont
  {Martin}}]{souza2000polarization}%
  \BibitemOpen
  \bibfield  {author} {\bibinfo {author} {\bibfnamefont {I.}~\bibnamefont
  {Souza}}, \bibinfo {author} {\bibfnamefont {T.}~\bibnamefont {Wilkens}}, \
  and\ \bibinfo {author} {\bibfnamefont {R.~M.}\ \bibnamefont {Martin}},\
  }\bibfield  {title} {\enquote {\bibinfo {title} {{Polarization and
  localization in insulators: Generating function approach}},}\ }\href
  {https://link.aps.org/doi/10.1103/PhysRevB.62.1666} {\bibfield  {journal}
  {\bibinfo  {journal} {\emph {Phys. Rev. B}},\ }\textbf {\bibinfo {volume}
  {62}},\ \bibinfo {pages} {1666--1683}\  (\bibinfo {year} {2000})}\BibitemShut
  {NoStop}%
\bibitem [{\citenamefont {Michishita}\ and\ \citenamefont
  {Nagaosa}(2022)}]{Michishita2022dissipation}%
  \BibitemOpen
  \bibfield  {author} {\bibinfo {author} {\bibfnamefont {Y.}~\bibnamefont
  {Michishita}}\ and\ \bibinfo {author} {\bibfnamefont {N.}~\bibnamefont
  {Nagaosa}},\ }\bibfield  {title} {\enquote {\bibinfo {title} {Dissipation and
  geometry in nonlinear quantum transports of multiband electronic systems},}\
  }\href {https://link.aps.org/doi/10.1103/PhysRevB.106.125114} {\bibfield
  {journal} {\bibinfo  {journal} {\emph {Phys. Rev. B}},\ }\textbf {\bibinfo
  {volume} {106}},\ \bibinfo {pages} {125114}\  (\bibinfo {year}
  {2022})}\BibitemShut {NoStop}%
\bibitem [{\citenamefont {Chen}\ and\ \citenamefont {von
  Gersdorff}(2022)}]{chen2022measurement}%
  \BibitemOpen
  \bibfield  {author} {\bibinfo {author} {\bibfnamefont {W.}~\bibnamefont
  {Chen}}\ and\ \bibinfo {author} {\bibfnamefont {G.}~\bibnamefont {von
  Gersdorff}},\ }\bibfield  {title} {\enquote {\bibinfo {title} {{Measurement
  of interaction-dressed Berry curvature and quantum metric in solids by
  optical absorption}},}\ }\href
  {https://scipost.org/10.21468/SciPostPhysCore.5.3.040} {\bibfield  {journal}
  {\bibinfo  {journal} {\emph {SciPost Phys. Core}},\ }\textbf {\bibinfo
  {volume} {5}},\ \bibinfo {pages} {040}\  (\bibinfo {year}
  {2022})}\BibitemShut {NoStop}%
\bibitem [{\citenamefont {Kashihara}\ \emph {et~al.}(2023)\citenamefont
  {Kashihara}, \citenamefont {Michishita},\ and\ \citenamefont
  {Peters}}]{kashihara2023quantum}%
  \BibitemOpen
  \bibfield  {author} {\bibinfo {author} {\bibfnamefont {T.}~\bibnamefont
  {Kashihara}}, \bibinfo {author} {\bibfnamefont {Y.}~\bibnamefont
  {Michishita}}, \ and\ \bibinfo {author} {\bibfnamefont {R.}~\bibnamefont
  {Peters}},\ }\bibfield  {title} {\enquote {\bibinfo {title} {{Quantum metric
  on the Brillouin zone in correlated electron systems and its relation to
  topology for Chern insulators}},}\ }\href
  {https://link.aps.org/doi/10.1103/PhysRevB.107.125116} {\bibfield  {journal}
  {\bibinfo  {journal} {\emph {Phys. Rev. B}},\ }\textbf {\bibinfo {volume}
  {107}},\ \bibinfo {pages} {125116}\  (\bibinfo {year} {2023})}\BibitemShut
  {NoStop}%
\bibitem [{\citenamefont {Zhou}\ \emph {et~al.}(2024)\citenamefont {Zhou},
  \citenamefont {Hou}, \citenamefont {Wang}, \citenamefont {Tang},
  \citenamefont {Guo},\ and\ \citenamefont {Chien}}]{zhou2024sloqvist}%
  \BibitemOpen
  \bibfield  {author} {\bibinfo {author} {\bibfnamefont {Z.}~\bibnamefont
  {Zhou}}, \bibinfo {author} {\bibfnamefont {X.-Y.}\ \bibnamefont {Hou}},
  \bibinfo {author} {\bibfnamefont {X.}~\bibnamefont {Wang}}, \bibinfo {author}
  {\bibfnamefont {J.-C.}\ \bibnamefont {Tang}}, \bibinfo {author}
  {\bibfnamefont {H.}~\bibnamefont {Guo}}, \ and\ \bibinfo {author}
  {\bibfnamefont {C.-C.}\ \bibnamefont {Chien}},\ }\bibfield  {title} {\enquote
  {\bibinfo {title} {Sj\"oqvist quantum geometric tensor of finite-temperature
  mixed states},}\ }\href
  {https://link.aps.org/doi/10.1103/PhysRevB.110.035404} {\bibfield  {journal}
  {\bibinfo  {journal} {\emph {Phys. Rev. B}},\ }\textbf {\bibinfo {volume}
  {110}},\ \bibinfo {pages} {035404}\  (\bibinfo {year} {2024})}\BibitemShut
  {NoStop}%
\bibitem [{\citenamefont {Romeral}\ \emph {et~al.}(2025)\citenamefont
  {Romeral}, \citenamefont {Cummings},\ and\ \citenamefont
  {Roche}}]{romeral2025scaling}%
  \BibitemOpen
  \bibfield  {author} {\bibinfo {author} {\bibfnamefont {J.~M.}\ \bibnamefont
  {Romeral}}, \bibinfo {author} {\bibfnamefont {A.~W.}\ \bibnamefont
  {Cummings}}, \ and\ \bibinfo {author} {\bibfnamefont {S.}~\bibnamefont
  {Roche}},\ }\bibfield  {title} {\enquote {\bibinfo {title} {{Scaling of the
  integrated quantum metric in disordered topological phases}},}\ }\href
  {https://link.aps.org/doi/10.1103/PhysRevB.111.134201} {\bibfield  {journal}
  {\bibinfo  {journal} {\emph {Phys. Rev. B}},\ }\textbf {\bibinfo {volume}
  {111}},\ \bibinfo {pages} {134201}\  (\bibinfo {year} {2025})}\BibitemShut
  {NoStop}%
\bibitem [{\citenamefont {Sukhachov}\ \emph {et~al.}(2025)\citenamefont
  {Sukhachov}, \citenamefont {Aase}, \citenamefont {M\ae{}land},\ and\
  \citenamefont {Sudb\o{}}}]{sukhachov2025effect}%
  \BibitemOpen
  \bibfield  {author} {\bibinfo {author} {\bibfnamefont {P.}~\bibnamefont
  {Sukhachov}}, \bibinfo {author} {\bibfnamefont {N.~H.}\ \bibnamefont {Aase}},
  \bibinfo {author} {\bibfnamefont {K.}~\bibnamefont {M\ae{}land}}, \ and\
  \bibinfo {author} {\bibfnamefont {A.}~\bibnamefont {Sudb\o{}}},\ }\bibfield
  {title} {\enquote {\bibinfo {title} {{Effect of the Hubbard interaction on
  the quantum metric}},}\ }\href
  {https://link.aps.org/doi/10.1103/PhysRevB.111.085143} {\bibfield  {journal}
  {\bibinfo  {journal} {\emph {Phys. Rev. B}},\ }\textbf {\bibinfo {volume}
  {111}},\ \bibinfo {pages} {085143}\  (\bibinfo {year} {2025})}\BibitemShut
  {NoStop}%
\bibitem [{\citenamefont {Ma}\ \emph {et~al.}(2010)\citenamefont {Ma},
  \citenamefont {Chen}, \citenamefont {Fan},\ and\ \citenamefont
  {Liu}}]{ma2010abelian}%
  \BibitemOpen
  \bibfield  {author} {\bibinfo {author} {\bibfnamefont {Y.-Q.}\ \bibnamefont
  {Ma}}, \bibinfo {author} {\bibfnamefont {S.}~\bibnamefont {Chen}}, \bibinfo
  {author} {\bibfnamefont {H.}~\bibnamefont {Fan}}, \ and\ \bibinfo {author}
  {\bibfnamefont {W.-M.}\ \bibnamefont {Liu}},\ }\bibfield  {title} {\enquote
  {\bibinfo {title} {{Abelian and non-Abelian quantum geometric tensor}},}\
  }\href {https://link.aps.org/doi/10.1103/PhysRevB.81.245129} {\bibfield
  {journal} {\bibinfo  {journal} {\emph {Phys. Rev. B}},\ }\textbf {\bibinfo
  {volume} {81}},\ \bibinfo {pages} {245129}\  (\bibinfo {year}
  {2010})}\BibitemShut {NoStop}%
\bibitem [{\citenamefont {Michishita}\ and\ \citenamefont
  {Peters}(2021)}]{Michishita2021effects}%
  \BibitemOpen
  \bibfield  {author} {\bibinfo {author} {\bibfnamefont {Y.}~\bibnamefont
  {Michishita}}\ and\ \bibinfo {author} {\bibfnamefont {R.}~\bibnamefont
  {Peters}},\ }\bibfield  {title} {\enquote {\bibinfo {title} {{Effects of
  renormalization and non-Hermiticity on nonlinear responses in strongly
  correlated electron systems}},}\ }\href
  {https://link.aps.org/doi/10.1103/PhysRevB.103.195133} {\bibfield  {journal}
  {\bibinfo  {journal} {\emph {Phys. Rev. B}},\ }\textbf {\bibinfo {volume}
  {103}},\ \bibinfo {pages} {195133}\  (\bibinfo {year} {2021})}\BibitemShut
  {NoStop}%
\bibitem [{\citenamefont {Ziman}(2001)}]{ziman2001electrons}%
  \BibitemOpen
  \bibfield  {author} {\bibinfo {author} {\bibfnamefont {J.~M.}\ \bibnamefont
  {Ziman}},\ }\href@noop {} {\emph {\bibinfo {title} {Electrons and phonons:
  the theory of transport phenomena in solids}}}\ (\bibinfo  {publisher}
  {Oxford university press},\ \bibinfo {year} {2001})\BibitemShut {NoStop}%
\bibitem [{\citenamefont {Carroll}(2019)}]{carroll2019spacetime}%
  \BibitemOpen
  \bibfield  {author} {\bibinfo {author} {\bibfnamefont {S.~M.}\ \bibnamefont
  {Carroll}},\ }\href@noop {} {\emph {\bibinfo {title} {Spacetime and
  geometry}}}\ (\bibinfo  {publisher} {Cambridge University Press},\ \bibinfo
  {year} {2019})\BibitemShut {NoStop}%
\bibitem [{\citenamefont {Huang}\ \emph {et~al.}(2025)\citenamefont {Huang},
  \citenamefont {Xiao}, \citenamefont {Yang},\ and\ \citenamefont
  {Li}}]{huang2025scaling}%
  \BibitemOpen
  \bibfield  {author} {\bibinfo {author} {\bibfnamefont {Y.-X.}\ \bibnamefont
  {Huang}}, \bibinfo {author} {\bibfnamefont {C.}~\bibnamefont {Xiao}},
  \bibinfo {author} {\bibfnamefont {S.~A.}\ \bibnamefont {Yang}}, \ and\
  \bibinfo {author} {\bibfnamefont {X.}~\bibnamefont {Li}},\ }\bibfield
  {title} {\enquote {\bibinfo {title} {Scaling law and extrinsic mechanisms for
  time-reversal-odd second-order nonlinear transport},}\ }\href
  {https://link.aps.org/doi/10.1103/PhysRevB.111.155127} {\bibfield  {journal}
  {\bibinfo  {journal} {\emph {Phys. Rev. B}},\ }\textbf {\bibinfo {volume}
  {111}},\ \bibinfo {pages} {155127}\  (\bibinfo {year} {2025})}\BibitemShut
  {NoStop}%
\bibitem [{\citenamefont {Liu}\ \emph {et~al.}(2024){\natexlab{b}}\citenamefont
  {Liu}, \citenamefont {Zhang}, \citenamefont {Zhu},\ and\ \citenamefont
  {Su}}]{liu2024effect}%
  \BibitemOpen
  \bibfield  {author} {\bibinfo {author} {\bibfnamefont {Z.}~\bibnamefont
  {Liu}}, \bibinfo {author} {\bibfnamefont {Z.-F.}\ \bibnamefont {Zhang}},
  \bibinfo {author} {\bibfnamefont {Z.-G.}\ \bibnamefont {Zhu}}, \ and\
  \bibinfo {author} {\bibfnamefont {G.}~\bibnamefont {Su}},\ }\bibfield
  {title} {\enquote {\bibinfo {title} {{Effect of disorder on Berry curvature
  and quantum metric in two-band gapped graphene}},}\ }\href
  {https://link.aps.org/doi/10.1103/PhysRevB.110.245419} {\bibfield  {journal}
  {\bibinfo  {journal} {\emph {Phys. Rev. B}},\ }\textbf {\bibinfo {volume}
  {110}},\ \bibinfo {pages} {245419}\  (\bibinfo {year}
  {2024}{\natexlab{b}})}\BibitemShut {NoStop}%
\bibitem [{\citenamefont {Konopik}\ and\ \citenamefont
  {Lutz}(2019)}]{konopik2019quantum}%
  \BibitemOpen
  \bibfield  {author} {\bibinfo {author} {\bibfnamefont {M.}~\bibnamefont
  {Konopik}}\ and\ \bibinfo {author} {\bibfnamefont {E.}~\bibnamefont {Lutz}},\
  }\bibfield  {title} {\enquote {\bibinfo {title} {Quantum response theory for
  nonequilibrium steady states},}\ }\href
  {https://link.aps.org/doi/10.1103/PhysRevResearch.1.033156} {\bibfield
  {journal} {\bibinfo  {journal} {\emph {Phys. Rev. Res.}},\ }\textbf {\bibinfo
  {volume} {1}},\ \bibinfo {pages} {033156}\  (\bibinfo {year}
  {2019})}\BibitemShut {NoStop}%
\bibitem [{\citenamefont {Facchi}\ \emph {et~al.}(2010)\citenamefont {Facchi},
  \citenamefont {Kulkarni}, \citenamefont {Man'Ko}, \citenamefont {Marmo},
  \citenamefont {Sudarshan},\ and\ \citenamefont
  {Ventriglia}}]{facchi2010classical}%
  \BibitemOpen
  \bibfield  {author} {\bibinfo {author} {\bibfnamefont {P.}~\bibnamefont
  {Facchi}}, \bibinfo {author} {\bibfnamefont {R.}~\bibnamefont {Kulkarni}},
  \bibinfo {author} {\bibfnamefont {V.}~\bibnamefont {Man'Ko}}, \bibinfo
  {author} {\bibfnamefont {G.}~\bibnamefont {Marmo}}, \bibinfo {author}
  {\bibfnamefont {E.}~\bibnamefont {Sudarshan}}, \ and\ \bibinfo {author}
  {\bibfnamefont {F.}~\bibnamefont {Ventriglia}},\ }\bibfield  {title}
  {\enquote {\bibinfo {title} {{Classical and quantum Fisher information in the
  geometrical formulation of quantum mechanics}},}\ }\href
  {https://doi.org/10.1016/j.physleta.2010.10.005} {\bibfield  {journal}
  {\bibinfo  {journal} {\emph {Phys. Lett. A}},\ }\textbf {\bibinfo {volume}
  {374}},\ \bibinfo {pages} {4801--4803}\  (\bibinfo {year}
  {2010})}\BibitemShut {NoStop}%
\bibitem [{\citenamefont {Loaiza}\ \emph {et~al.}(2024)\citenamefont {Loaiza},
  \citenamefont {Motlagh}, \citenamefont {Hejazi}, \citenamefont {Zini},
  \citenamefont {Delgado},\ and\ \citenamefont
  {Arrazola}}]{loaiza2024nonlinear}%
  \BibitemOpen
  \bibfield  {author} {\bibinfo {author} {\bibfnamefont {I.}~\bibnamefont
  {Loaiza}}, \bibinfo {author} {\bibfnamefont {D.}~\bibnamefont {Motlagh}},
  \bibinfo {author} {\bibfnamefont {K.}~\bibnamefont {Hejazi}}, \bibinfo
  {author} {\bibfnamefont {M.~S.}\ \bibnamefont {Zini}}, \bibinfo {author}
  {\bibfnamefont {A.}~\bibnamefont {Delgado}}, \ and\ \bibinfo {author}
  {\bibfnamefont {J.~M.}\ \bibnamefont {Arrazola}},\ }\bibfield  {title}
  {\enquote {\bibinfo {title} {Nonlinear spectroscopy via generalized quantum
  phase estimation},}\ }\href {https://arxiv.org/abs/2405.13885} {\bibfield
  {journal} {\bibinfo  {journal} {\emph {arXiv:2405.13885}}}\  (\bibinfo {year}
  {2024})}\BibitemShut {NoStop}%
\bibitem [{\citenamefont {Hall}(1881)}]{Hall1881AHE}%
  \BibitemOpen
  \bibfield  {author} {\bibinfo {author} {\bibfnamefont {E.}~\bibnamefont
  {Hall}},\ }\bibfield  {title} {\enquote {\bibinfo {title} {On the
  “rotational coefficient” in nickel and cobalt},}\ }\href
  {https://doi.org/10.1080/14786448108627086} {\bibfield  {journal} {\bibinfo
  {journal} {\emph {Philos. Mag.}},\ }\textbf {\bibinfo {volume} {12}},\
  \bibinfo {pages} {157--172}\  (\bibinfo {year} {1881})}\BibitemShut {NoStop}%
\bibitem [{\citenamefont {Nagaosa}\ \emph
  {et~al.}(2010){\natexlab{b}}\citenamefont {Nagaosa}, \citenamefont {Sinova},
  \citenamefont {Onoda}, \citenamefont {MacDonald},\ and\ \citenamefont
  {Ong}}]{Nagaosa10RMP}%
  \BibitemOpen
  \bibfield  {author} {\bibinfo {author} {\bibfnamefont {N.}~\bibnamefont
  {Nagaosa}}, \bibinfo {author} {\bibfnamefont {J.}~\bibnamefont {Sinova}},
  \bibinfo {author} {\bibfnamefont {S.}~\bibnamefont {Onoda}}, \bibinfo
  {author} {\bibfnamefont {A.~H.}\ \bibnamefont {MacDonald}}, \ and\ \bibinfo
  {author} {\bibfnamefont {N.~P.}\ \bibnamefont {Ong}},\ }\bibfield  {title}
  {\enquote {\bibinfo {title} {Anomalous \text{H}all effect},}\ }\href
  {http://link.aps.org/doi/10.1103/RevModPhys.82.1539} {\bibfield  {journal}
  {\bibinfo  {journal} {\emph {Rev. Mod. Phys.}},\ }\textbf {\bibinfo {volume}
  {82}},\ \bibinfo {pages} {1539--1592}\  (\bibinfo {year}
  {2010}{\natexlab{b}})}\BibitemShut {NoStop}%
\bibitem [{\citenamefont {Zhang}\ and\ \citenamefont
  {Zhang}(2014)}]{sZhang14JAP_AMR}%
  \BibitemOpen
  \bibfield  {author} {\bibinfo {author} {\bibfnamefont {S.~S.-L.}\
  \bibnamefont {Zhang}}\ and\ \bibinfo {author} {\bibfnamefont
  {S.}~\bibnamefont {Zhang}},\ }\bibfield  {title} {\enquote {\bibinfo {title}
  {Angular dependence of anisotropic magnetoresistance in magnetic systems},}\
  }\href {https://doi.org/10.1063/1.4855935} {\bibfield  {journal} {\bibinfo
  {journal} {\emph {J. Appl. Phys.}},\ }\textbf {\bibinfo {volume} {115}},\
  \bibinfo {pages} {17C703}\  (\bibinfo {year} {2014})}\BibitemShut {NoStop}%
\bibitem [{\citenamefont {Zhang}(2014)}]{zhang14Thesis}%
  \BibitemOpen
  \bibfield  {author} {\bibinfo {author} {\bibfnamefont {S.~S.-L.}\
  \bibnamefont {Zhang}},\ }\emph {\bibinfo {title} {Spin Transport and
  Magnetization Dynamics in Various Magnetic Systems}},\ \href
  {http://hdl.handle.net/10150/333352} {Ph.D. thesis},\ \bibinfo  {school}
  {University of Arizona} (\bibinfo {year} {2014})\BibitemShut {NoStop}%
\bibitem [{\citenamefont {Taniguchi}\ \emph {et~al.}(2015)\citenamefont
  {Taniguchi}, \citenamefont {Grollier},\ and\ \citenamefont
  {Stiles}}]{Tomo15PRAppl_AHE-SOT}%
  \BibitemOpen
  \bibfield  {author} {\bibinfo {author} {\bibfnamefont {T.}~\bibnamefont
  {Taniguchi}}, \bibinfo {author} {\bibfnamefont {J.}~\bibnamefont {Grollier}},
  \ and\ \bibinfo {author} {\bibfnamefont {M.~D.}\ \bibnamefont {Stiles}},\
  }\bibfield  {title} {\enquote {\bibinfo {title} {Spin-transfer torques
  generated by the anomalous \text{H}all effect and anisotropic
  magnetoresistance},}\ }\href
  {https://link.aps.org/doi/10.1103/PhysRevApplied.3.044001} {\bibfield
  {journal} {\bibinfo  {journal} {\emph {Phys. Rev. Applied}},\ }\textbf
  {\bibinfo {volume} {3}},\ \bibinfo {pages} {044001}\  (\bibinfo {year}
  {2015})}\BibitemShut {NoStop}%
\bibitem [{\citenamefont {Zhang}\ \emph {et~al.}(2016)\citenamefont {Zhang},
  \citenamefont {Wang},\ and\ \citenamefont {Zhang}}]{xrWang16EPL_AMR}%
  \BibitemOpen
  \bibfield  {author} {\bibinfo {author} {\bibfnamefont {Y.}~\bibnamefont
  {Zhang}}, \bibinfo {author} {\bibfnamefont {X.~R.}\ \bibnamefont {Wang}}, \
  and\ \bibinfo {author} {\bibfnamefont {H.~W.}\ \bibnamefont {Zhang}},\
  }\bibfield  {title} {\enquote {\bibinfo {title} {Extraordinary
  galvanomagnetic effects in polycrystalline magnetic films},}\ }\href
  {https://doi.org/10.1209/0295-5075/113/47003} {\bibfield  {journal} {\bibinfo
   {journal} {\emph {Europhys. Lett.}},\ }\textbf {\bibinfo {volume} {113}},\
  \bibinfo {pages} {47003}\  (\bibinfo {year} {2016})}\BibitemShut {NoStop}%
\bibitem [{\citenamefont {Taniguchi}(2016)}]{Tomo16PRB_AH-AMR}%
  \BibitemOpen
  \bibfield  {author} {\bibinfo {author} {\bibfnamefont {T.}~\bibnamefont
  {Taniguchi}},\ }\bibfield  {title} {\enquote {\bibinfo {title}
  {Magnetoresistance generated from charge-spin conversion by anomalous
  \text{H}all effect in metallic ferromagnetic/nonmagnetic bilayers},}\ }\href
  {https://link.aps.org/doi/10.1103/PhysRevB.94.174440} {\bibfield  {journal}
  {\bibinfo  {journal} {\emph {Phys. Rev. B}},\ }\textbf {\bibinfo {volume}
  {94}},\ \bibinfo {pages} {174440}\  (\bibinfo {year} {2016})}\BibitemShut
  {NoStop}%
\bibitem [{\citenamefont {Yang}\ \emph {et~al.}(2018)\citenamefont {Yang},
  \citenamefont {Luo}, \citenamefont {Wu}, \citenamefont {Xu}, \citenamefont
  {Li}, \citenamefont {Pennycook}, \citenamefont {Zhang},\ and\ \citenamefont
  {Wu}}]{yhWu18NC_AHMR}%
  \BibitemOpen
  \bibfield  {author} {\bibinfo {author} {\bibfnamefont {Y.}~\bibnamefont
  {Yang}}, \bibinfo {author} {\bibfnamefont {Z.}~\bibnamefont {Luo}}, \bibinfo
  {author} {\bibfnamefont {H.}~\bibnamefont {Wu}}, \bibinfo {author}
  {\bibfnamefont {Y.}~\bibnamefont {Xu}}, \bibinfo {author} {\bibfnamefont
  {R.-W.}\ \bibnamefont {Li}}, \bibinfo {author} {\bibfnamefont {S.~J.}\
  \bibnamefont {Pennycook}}, \bibinfo {author} {\bibfnamefont {S.}~\bibnamefont
  {Zhang}}, \ and\ \bibinfo {author} {\bibfnamefont {Y.}~\bibnamefont {Wu}},\
  }\bibfield  {title} {\enquote {\bibinfo {title} {Anomalous \text{H}all
  magnetoresistance in a ferromagnet},}\ }\href
  {https://doi.org/10.1038/s41467-018-04712-9} {\bibfield  {journal} {\bibinfo
  {journal} {\emph {Nat. Commun.}},\ }\textbf {\bibinfo {volume} {9}},\
  \bibinfo {pages} {2255}\  (\bibinfo {year} {2018})}\BibitemShut {NoStop}%
\bibitem [{\citenamefont {Jia}\ \emph {et~al.}(2020)\citenamefont {Jia},
  \citenamefont {Li}, \citenamefont {Chen}, \citenamefont {Zeng}, \citenamefont
  {Xiao},\ and\ \citenamefont {Wu}}]{yzWu20NJP_AHMR}%
  \BibitemOpen
  \bibfield  {author} {\bibinfo {author} {\bibfnamefont {M.~W.}\ \bibnamefont
  {Jia}}, \bibinfo {author} {\bibfnamefont {J.~X.}\ \bibnamefont {Li}},
  \bibinfo {author} {\bibfnamefont {H.~R.}\ \bibnamefont {Chen}}, \bibinfo
  {author} {\bibfnamefont {F.~L.}\ \bibnamefont {Zeng}}, \bibinfo {author}
  {\bibfnamefont {X.}~\bibnamefont {Xiao}}, \ and\ \bibinfo {author}
  {\bibfnamefont {Y.~Z.}\ \bibnamefont {Wu}},\ }\bibfield  {title} {\enquote
  {\bibinfo {title} {Anomalous \text{H}all magnetoresistance in single-crystal
  \textnormal{Fe}(001) films},}\ }\href
  {https://doi.org/10.1088/1367-2630/ab7d7b} {\bibfield  {journal} {\bibinfo
  {journal} {\emph {New J. Phys.}},\ }\textbf {\bibinfo {volume} {22}},\
  \bibinfo {pages} {043014}\  (\bibinfo {year} {2020})}\BibitemShut {NoStop}%
\bibitem [{\citenamefont {Avci}\ \emph
  {et~al.}(2015){\natexlab{c}}\citenamefont {Avci}, \citenamefont {Garello},
  \citenamefont {Ghosh}, \citenamefont {Gabureac}, \citenamefont {Alvarado},\
  and\ \citenamefont {Gambardella}}]{avci2015natphys}%
  \BibitemOpen
  \bibfield  {author} {\bibinfo {author} {\bibfnamefont {C.~O.}\ \bibnamefont
  {Avci}}, \bibinfo {author} {\bibfnamefont {K.}~\bibnamefont {Garello}},
  \bibinfo {author} {\bibfnamefont {A.}~\bibnamefont {Ghosh}}, \bibinfo
  {author} {\bibfnamefont {M.}~\bibnamefont {Gabureac}}, \bibinfo {author}
  {\bibfnamefont {S.~F.}\ \bibnamefont {Alvarado}}, \ and\ \bibinfo {author}
  {\bibfnamefont {P.}~\bibnamefont {Gambardella}},\ }\bibfield  {title}
  {\enquote {\bibinfo {title} {Unidirectional spin hall magnetoresistance in
  ferromagnet/normal metal bilayers},}\ }\href
  {https://doi.org/10.1038/nphys3356} {\bibfield  {journal} {\bibinfo
  {journal} {\emph {Nat. Phys.}},\ }\textbf {\bibinfo {volume} {11}},\ \bibinfo
  {pages} {570--575}\  (\bibinfo {year} {2015}{\natexlab{c}})}\BibitemShut
  {NoStop}%
\bibitem [{\citenamefont {Olejn\'{\i}k}\ \emph
  {et~al.}(2015){\natexlab{b}}\citenamefont {Olejn\'{\i}k}, \citenamefont
  {Nov\'ak}, \citenamefont {Wunderlich},\ and\ \citenamefont
  {Jungwirth}}]{Jungwirth15PRB_UMR-DMS}%
  \BibitemOpen
  \bibfield  {author} {\bibinfo {author} {\bibfnamefont {K.}~\bibnamefont
  {Olejn\'{\i}k}}, \bibinfo {author} {\bibfnamefont {V.}~\bibnamefont
  {Nov\'ak}}, \bibinfo {author} {\bibfnamefont {J.}~\bibnamefont {Wunderlich}},
  \ and\ \bibinfo {author} {\bibfnamefont {T.}~\bibnamefont {Jungwirth}},\
  }\bibfield  {title} {\enquote {\bibinfo {title} {Electrical detection of
  magnetization reversal without auxiliary magnets},}\ }\href
  {https://link.aps.org/doi/10.1103/PhysRevB.91.180402} {\bibfield  {journal}
  {\bibinfo  {journal} {\emph {Phys. Rev. B}},\ }\textbf {\bibinfo {volume}
  {91}},\ \bibinfo {pages} {180402}\  (\bibinfo {year}
  {2015}{\natexlab{b}})}\BibitemShut {NoStop}%
\bibitem [{\citenamefont {Zhang}\ and\ \citenamefont
  {Vignale}(2016){\natexlab{b}}}]{shulei2016prb}%
  \BibitemOpen
  \bibfield  {author} {\bibinfo {author} {\bibfnamefont {S.~S.-L.}\
  \bibnamefont {Zhang}}\ and\ \bibinfo {author} {\bibfnamefont
  {G.}~\bibnamefont {Vignale}},\ }\bibfield  {title} {\enquote {\bibinfo
  {title} {Theory of unidirectional spin hall magnetoresistance in
  heavy-metal/ferromagnetic-metal bilayers},}\ }\href
  {https://link.aps.org/doi/10.1103/PhysRevB.94.140411} {\bibfield  {journal}
  {\bibinfo  {journal} {\emph {Phys. Rev. B}},\ }\textbf {\bibinfo {volume}
  {94}},\ \bibinfo {pages} {140411}\  (\bibinfo {year}
  {2016}{\natexlab{b}})}\BibitemShut {NoStop}%
\bibitem [{\citenamefont {Langenfeld}\ \emph
  {et~al.}(2016){\natexlab{a}}\citenamefont {Langenfeld}, \citenamefont
  {Tshitoyan}, \citenamefont {Fang}, \citenamefont {Wells}, \citenamefont
  {Moore},\ and\ \citenamefont {Ferguson}}]{Ferguson16APL_UMR-STO}%
  \BibitemOpen
  \bibfield  {author} {\bibinfo {author} {\bibfnamefont {S.}~\bibnamefont
  {Langenfeld}}, \bibinfo {author} {\bibfnamefont {V.}~\bibnamefont
  {Tshitoyan}}, \bibinfo {author} {\bibfnamefont {Z.}~\bibnamefont {Fang}},
  \bibinfo {author} {\bibfnamefont {A.}~\bibnamefont {Wells}}, \bibinfo
  {author} {\bibfnamefont {T.~A.}\ \bibnamefont {Moore}}, \ and\ \bibinfo
  {author} {\bibfnamefont {A.~J.}\ \bibnamefont {Ferguson}},\ }\bibfield
  {title} {\enquote {\bibinfo {title} {Exchange magnon induced resistance
  asymmetry in permalloy spin-\text{H}all oscillators},}\ }\href
  {https://doi.org/10.1063/1.4948921} {\bibfield  {journal} {\bibinfo
  {journal} {\emph {Appl. Phys. Lett.}},\ }\textbf {\bibinfo {volume} {108}},\
  \bibinfo {pages} {192402}\  (\bibinfo {year}
  {2016}{\natexlab{a}})}\BibitemShut {NoStop}%
\bibitem [{\citenamefont {Yasuda}\ \emph
  {et~al.}(2016){\natexlab{b}}\citenamefont {Yasuda}, \citenamefont
  {Tsukazaki}, \citenamefont {Yoshimi}, \citenamefont {Takahashi},
  \citenamefont {Kawasaki},\ and\ \citenamefont {Tokura}}]{yasuda2016prl}%
  \BibitemOpen
  \bibfield  {author} {\bibinfo {author} {\bibfnamefont {K.}~\bibnamefont
  {Yasuda}}, \bibinfo {author} {\bibfnamefont {A.}~\bibnamefont {Tsukazaki}},
  \bibinfo {author} {\bibfnamefont {R.}~\bibnamefont {Yoshimi}}, \bibinfo
  {author} {\bibfnamefont {K.~S.}\ \bibnamefont {Takahashi}}, \bibinfo {author}
  {\bibfnamefont {M.}~\bibnamefont {Kawasaki}}, \ and\ \bibinfo {author}
  {\bibfnamefont {Y.}~\bibnamefont {Tokura}},\ }\bibfield  {title} {\enquote
  {\bibinfo {title} {Large unidirectional magnetoresistance in a magnetic
  topological insulator},}\ }\href
  {https://link.aps.org/doi/10.1103/PhysRevLett.117.127202} {\bibfield
  {journal} {\bibinfo  {journal} {\emph {Phys. Rev. Lett.}},\ }\textbf
  {\bibinfo {volume} {117}},\ \bibinfo {pages} {127202}\  (\bibinfo {year}
  {2016}{\natexlab{b}})}\BibitemShut {NoStop}%
\bibitem [{\citenamefont {Avci}\ \emph
  {et~al.}(2018){\natexlab{b}}\citenamefont {Avci}, \citenamefont {Mendil},
  \citenamefont {Beach},\ and\ \citenamefont {Gambardella}}]{avci2018prl}%
  \BibitemOpen
  \bibfield  {author} {\bibinfo {author} {\bibfnamefont {C.~O.}\ \bibnamefont
  {Avci}}, \bibinfo {author} {\bibfnamefont {J.}~\bibnamefont {Mendil}},
  \bibinfo {author} {\bibfnamefont {G.~S.~D.}\ \bibnamefont {Beach}}, \ and\
  \bibinfo {author} {\bibfnamefont {P.}~\bibnamefont {Gambardella}},\
  }\bibfield  {title} {\enquote {\bibinfo {title} {Origins of the
  unidirectional spin hall magnetoresistance in metallic bilayers},}\ }\href
  {https://link.aps.org/doi/10.1103/PhysRevLett.121.087207} {\bibfield
  {journal} {\bibinfo  {journal} {\emph {Phys. Rev. Lett.}},\ }\textbf
  {\bibinfo {volume} {121}},\ \bibinfo {pages} {087207}\  (\bibinfo {year}
  {2018}{\natexlab{b}})}\BibitemShut {NoStop}%
\bibitem [{\citenamefont {Lv}\ \emph {et~al.}(2018){\natexlab{b}}\citenamefont
  {Lv}, \citenamefont {Kally}, \citenamefont {Zhang}, \citenamefont {Lee},
  \citenamefont {Jamali}, \citenamefont {Samarth},\ and\ \citenamefont
  {Wang}}]{lv2018natcomm}%
  \BibitemOpen
  \bibfield  {author} {\bibinfo {author} {\bibfnamefont {Y.}~\bibnamefont
  {Lv}}, \bibinfo {author} {\bibfnamefont {J.}~\bibnamefont {Kally}}, \bibinfo
  {author} {\bibfnamefont {D.}~\bibnamefont {Zhang}}, \bibinfo {author}
  {\bibfnamefont {J.~S.}\ \bibnamefont {Lee}}, \bibinfo {author} {\bibfnamefont
  {M.}~\bibnamefont {Jamali}}, \bibinfo {author} {\bibfnamefont
  {N.}~\bibnamefont {Samarth}}, \ and\ \bibinfo {author} {\bibfnamefont
  {J.-P.}\ \bibnamefont {Wang}},\ }\bibfield  {title} {\enquote {\bibinfo
  {title} {Unidirectional spin-\text{H}all and \text{R}ashba-\text{E}delstein
  magnetoresistance in topological insulator-ferromagnet layer
  heterostructures},}\ }\href {https://doi.org/10.1038/s41467-017-02491-3}
  {\bibfield  {journal} {\bibinfo  {journal} {\emph {Nat. Commun.}},\ }\textbf
  {\bibinfo {volume} {9}},\ \bibinfo {pages} {1--7}\  (\bibinfo {year}
  {2018}{\natexlab{b}})}\BibitemShut {NoStop}%
\bibitem [{\citenamefont {Duy~Khang}\ and\ \citenamefont
  {Hai}(2019){\natexlab{b}}}]{pnHai19JAP_UMR_TI-DMS}%
  \BibitemOpen
  \bibfield  {author} {\bibinfo {author} {\bibfnamefont {N.~H.}\ \bibnamefont
  {Duy~Khang}}\ and\ \bibinfo {author} {\bibfnamefont {P.~N.}\ \bibnamefont
  {Hai}},\ }\bibfield  {title} {\enquote {\bibinfo {title} {Giant
  unidirectional spin \text{H}all magnetoresistance in topological insulator
  – ferromagnetic semiconductor heterostructures},}\ }\href
  {https://doi.org/10.1063/1.5134728} {\bibfield  {journal} {\bibinfo
  {journal} {\emph {J. Appl. Phys.}},\ }\textbf {\bibinfo {volume} {126}},\
  \bibinfo {pages} {233903}\  (\bibinfo {year}
  {2019}{\natexlab{b}})}\BibitemShut {NoStop}%
\bibitem [{\citenamefont {{Guillet}}\ \emph
  {et~al.}(2021){\natexlab{b}}\citenamefont {{Guillet}}, \citenamefont
  {{Marty}}, \citenamefont {{Vergnaud}}, \citenamefont {{Jamet}}, \citenamefont
  {{Zucchetti}}, \citenamefont {{Isella}}, \citenamefont {{Barbedienne}},
  \citenamefont {{Jaffr{\`e}s}}, \citenamefont {{Reyren}}, \citenamefont
  {{George}} \emph {et~al.}}]{guillet2021prb}%
  \BibitemOpen
  \bibfield  {author} {\bibinfo {author} {\bibfnamefont {T.}~\bibnamefont
  {{Guillet}}}, \bibinfo {author} {\bibfnamefont {A.}~\bibnamefont {{Marty}}},
  \bibinfo {author} {\bibfnamefont {C.}~\bibnamefont {{Vergnaud}}}, \bibinfo
  {author} {\bibfnamefont {M.}~\bibnamefont {{Jamet}}}, \bibinfo {author}
  {\bibfnamefont {C.}~\bibnamefont {{Zucchetti}}}, \bibinfo {author}
  {\bibfnamefont {G.}~\bibnamefont {{Isella}}}, \bibinfo {author}
  {\bibfnamefont {Q.}~\bibnamefont {{Barbedienne}}}, \bibinfo {author}
  {\bibfnamefont {H.}~\bibnamefont {{Jaffr{\`e}s}}}, \bibinfo {author}
  {\bibfnamefont {N.}~\bibnamefont {{Reyren}}}, \bibinfo {author}
  {\bibfnamefont {J.~M.}\ \bibnamefont {{George}}},  \emph {et~al.},\
  }\bibfield  {title} {\enquote {\bibinfo {title} {{Large Rashba unidirectional
  magnetoresistance in the Fe/Ge(111) interface states}},}\ }\href
  {https://doi.org/10.1103/PhysRevB.103.064411} {\bibfield  {journal} {\bibinfo
   {journal} {\emph {Phys. Rev. B}},\ }\textbf {\bibinfo {volume} {103}},\
  \bibinfo {pages} {064411}\  (\bibinfo {year}
  {2021}{\natexlab{b}})}\BibitemShut {NoStop}%
\bibitem [{\citenamefont {Liu}\ \emph {et~al.}(2021){\natexlab{d}}\citenamefont
  {Liu}, \citenamefont {Wang}, \citenamefont {Luan}, \citenamefont {Zhou},
  \citenamefont {Xia}, \citenamefont {Yang}, \citenamefont {Tian},
  \citenamefont {Guo}, \citenamefont {Du},\ and\ \citenamefont
  {Wu}}]{dWu21PRL_magnon-USMR}%
  \BibitemOpen
  \bibfield  {author} {\bibinfo {author} {\bibfnamefont {G.}~\bibnamefont
  {Liu}}, \bibinfo {author} {\bibfnamefont {X.-g.}\ \bibnamefont {Wang}},
  \bibinfo {author} {\bibfnamefont {Z.~Z.}\ \bibnamefont {Luan}}, \bibinfo
  {author} {\bibfnamefont {L.~F.}\ \bibnamefont {Zhou}}, \bibinfo {author}
  {\bibfnamefont {S.~Y.}\ \bibnamefont {Xia}}, \bibinfo {author} {\bibfnamefont
  {B.}~\bibnamefont {Yang}}, \bibinfo {author} {\bibfnamefont {Y.~Z.}\
  \bibnamefont {Tian}}, \bibinfo {author} {\bibfnamefont {G.-h.}\ \bibnamefont
  {Guo}}, \bibinfo {author} {\bibfnamefont {J.}~\bibnamefont {Du}}, \ and\
  \bibinfo {author} {\bibfnamefont {D.}~\bibnamefont {Wu}},\ }\bibfield
  {title} {\enquote {\bibinfo {title} {Magnonic unidirectional spin \text{H}all
  magnetoresistance in a heavy-metal--ferromagnetic-insulator bilayer},}\
  }\href {https://link.aps.org/doi/10.1103/PhysRevLett.127.207206} {\bibfield
  {journal} {\bibinfo  {journal} {\emph {Phys. Rev. Lett.}},\ }\textbf
  {\bibinfo {volume} {127}},\ \bibinfo {pages} {207206}\  (\bibinfo {year}
  {2021}{\natexlab{d}})}\BibitemShut {NoStop}%
\bibitem [{\citenamefont {{Mehraeen}}\ \emph {et~al.}(2021)\citenamefont
  {{Mehraeen}}, \citenamefont {{Shen}},\ and\ \citenamefont
  {{Zhang}}}]{mandela21_QUMR}%
  \BibitemOpen
  \bibfield  {author} {\bibinfo {author} {\bibfnamefont {M.}~\bibnamefont
  {{Mehraeen}}}, \bibinfo {author} {\bibfnamefont {P.}~\bibnamefont {{Shen}}},
  \ and\ \bibinfo {author} {\bibfnamefont {S.~S.-L.}\ \bibnamefont {{Zhang}}},\
  }\bibfield  {title} {\enquote {\bibinfo {title} {{Quantum unidirectional
  magnetoresistance}},}\ }\href {https://arxiv.org/abs/2108.13711} {\bibfield
  {journal} {\bibinfo  {journal} {\emph {arXiv e-prints}},\ \bibinfo {eid}
  {arXiv:2108.13711}}\  (\bibinfo {year} {2021})}\BibitemShut {NoStop}%
\bibitem [{\citenamefont {Amin}\ \emph {et~al.}(2019)\citenamefont {Amin},
  \citenamefont {Li}, \citenamefont {Stiles},\ and\ \citenamefont
  {Haney}}]{amin2019intrinsic}%
  \BibitemOpen
  \bibfield  {author} {\bibinfo {author} {\bibfnamefont {V.~P.}\ \bibnamefont
  {Amin}}, \bibinfo {author} {\bibfnamefont {J.}~\bibnamefont {Li}}, \bibinfo
  {author} {\bibfnamefont {M.~D.}\ \bibnamefont {Stiles}}, \ and\ \bibinfo
  {author} {\bibfnamefont {P.~M.}\ \bibnamefont {Haney}},\ }\bibfield  {title}
  {\enquote {\bibinfo {title} {Intrinsic spin currents in ferromagnets},}\
  }\href {https://link.aps.org/doi/10.1103/PhysRevB.99.220405} {\bibfield
  {journal} {\bibinfo  {journal} {\emph {Phys. Rev. B}},\ }\textbf {\bibinfo
  {volume} {99}},\ \bibinfo {pages} {220405}\  (\bibinfo {year}
  {2019})}\BibitemShut {NoStop}%
\bibitem [{\citenamefont {McGuire}\ and\ \citenamefont
  {Potter}(1975)}]{McGuire&Potter75IEEE_AMR}%
  \BibitemOpen
  \bibfield  {author} {\bibinfo {author} {\bibfnamefont {T.}~\bibnamefont
  {McGuire}}\ and\ \bibinfo {author} {\bibfnamefont {R.}~\bibnamefont
  {Potter}},\ }\bibfield  {title} {\enquote {\bibinfo {title} {Anisotropic
  magnetoresistance in ferromagnetic 3d alloys},}\ }\href
  {https://doi.org/10.1109/TMAG.1975.1058782} {\bibfield  {journal} {\bibinfo
  {journal} {\emph {IEEE Trans. Magn.}},\ }\textbf {\bibinfo {volume} {11}},\
  \bibinfo {pages} {1018--1038}\  (\bibinfo {year} {1975})}\BibitemShut
  {NoStop}%
\bibitem [{\citenamefont {Kim}\ and\ \citenamefont
  {Lee}(2020)}]{kim2020generalized}%
  \BibitemOpen
  \bibfield  {author} {\bibinfo {author} {\bibfnamefont {K.-W.}\ \bibnamefont
  {Kim}}\ and\ \bibinfo {author} {\bibfnamefont {K.-J.}\ \bibnamefont {Lee}},\
  }\bibfield  {title} {\enquote {\bibinfo {title} {Generalized spin
  drift-diffusion formalism in the presence of spin-orbit interaction of
  ferromagnets},}\ }\href
  {https://link.aps.org/doi/10.1103/PhysRevLett.125.207205} {\bibfield
  {journal} {\bibinfo  {journal} {\emph {Phys. Rev. Lett.}},\ }\textbf
  {\bibinfo {volume} {125}},\ \bibinfo {pages} {207205}\  (\bibinfo {year}
  {2020})}\BibitemShut {NoStop}%
\bibitem [{\citenamefont {Valet}\ and\ \citenamefont
  {Fert}(1993)}]{valet1993prb}%
  \BibitemOpen
  \bibfield  {author} {\bibinfo {author} {\bibfnamefont {T.}~\bibnamefont
  {Valet}}\ and\ \bibinfo {author} {\bibfnamefont {A.}~\bibnamefont {Fert}},\
  }\bibfield  {title} {\enquote {\bibinfo {title} {Theory of the perpendicular
  magnetoresistance in magnetic multilayers},}\ }\href
  {https://doi.org/10.1103/PhysRevB.48.7099} {\bibfield  {journal} {\bibinfo
  {journal} {\emph {Phys. Rev. B}},\ }\textbf {\bibinfo {volume} {48}},\
  \bibinfo {pages} {7099}\  (\bibinfo {year} {1993})}\BibitemShut {NoStop}%
\bibitem [{\citenamefont {Bass}\ and\ \citenamefont
  {Pratt}(2007)}]{Bass07JP-CM_spin-diffusion}%
  \BibitemOpen
  \bibfield  {author} {\bibinfo {author} {\bibfnamefont {J.}~\bibnamefont
  {Bass}}\ and\ \bibinfo {author} {\bibfnamefont {W.~P.}\ \bibnamefont
  {Pratt}},\ }\bibfield  {title} {\enquote {\bibinfo {title} {Spin-diffusion
  lengths in metals and alloys, and spin-flipping at metal/metal interfaces: an
  experimentalist's critical review},}\ }\href
  {https://doi.org/10.1088/0953-8984/19/18/183201} {\bibfield  {journal}
  {\bibinfo  {journal} {\emph {J. Phys.: Condens. Matter}},\ }\textbf {\bibinfo
  {volume} {19}},\ \bibinfo {pages} {183201}\  (\bibinfo {year}
  {2007})}\BibitemShut {NoStop}%
\bibitem [{\citenamefont {Avci}\ \emph
  {et~al.}(2015){\natexlab{d}}\citenamefont {Avci}, \citenamefont {Garello},
  \citenamefont {Mendil}, \citenamefont {Ghosh}, \citenamefont {Blasakis},
  \citenamefont {Gabureac}, \citenamefont {Trassin}, \citenamefont {Fiebig},\
  and\ \citenamefont {Gambardella}}]{Gambardella15APL_MR-HL}%
  \BibitemOpen
  \bibfield  {author} {\bibinfo {author} {\bibfnamefont {C.~O.}\ \bibnamefont
  {Avci}}, \bibinfo {author} {\bibfnamefont {K.}~\bibnamefont {Garello}},
  \bibinfo {author} {\bibfnamefont {J.}~\bibnamefont {Mendil}}, \bibinfo
  {author} {\bibfnamefont {A.}~\bibnamefont {Ghosh}}, \bibinfo {author}
  {\bibfnamefont {N.}~\bibnamefont {Blasakis}}, \bibinfo {author}
  {\bibfnamefont {M.}~\bibnamefont {Gabureac}}, \bibinfo {author}
  {\bibfnamefont {M.}~\bibnamefont {Trassin}}, \bibinfo {author} {\bibfnamefont
  {M.}~\bibnamefont {Fiebig}}, \ and\ \bibinfo {author} {\bibfnamefont
  {P.}~\bibnamefont {Gambardella}},\ }\bibfield  {title} {\enquote {\bibinfo
  {title} {Magnetoresistance of heavy and light metal/ferromagnet bilayers},}\
  }\href {https://doi.org/10.1063/1.4935497} {\bibfield  {journal} {\bibinfo
  {journal} {\emph {Appl. Phys. Lett.}},\ }\textbf {\bibinfo {volume} {107}},\
  \bibinfo {pages} {192405}\  (\bibinfo {year}
  {2015}{\natexlab{d}})}\BibitemShut {NoStop}%
\bibitem [{\citenamefont {Liu}\ \emph {et~al.}(2012)\citenamefont {Liu},
  \citenamefont {Pai}, \citenamefont {Li}, \citenamefont {Tseng}, \citenamefont
  {Ralph},\ and\ \citenamefont {Buhrman}}]{liu2012spin}%
  \BibitemOpen
  \bibfield  {author} {\bibinfo {author} {\bibfnamefont {L.}~\bibnamefont
  {Liu}}, \bibinfo {author} {\bibfnamefont {C.-F.}\ \bibnamefont {Pai}},
  \bibinfo {author} {\bibfnamefont {Y.}~\bibnamefont {Li}}, \bibinfo {author}
  {\bibfnamefont {H.~W.}\ \bibnamefont {Tseng}}, \bibinfo {author}
  {\bibfnamefont {D.~C.}\ \bibnamefont {Ralph}}, \ and\ \bibinfo {author}
  {\bibfnamefont {R.~A.}\ \bibnamefont {Buhrman}},\ }\bibfield  {title}
  {\enquote {\bibinfo {title} {Spin-torque switching with the giant spin
  \text{H}all effect of tantalum},}\ }\href
  {https://doi.org/10.1126/science.1218197} {\bibfield  {journal} {\bibinfo
  {journal} {\emph {Science}},\ }\textbf {\bibinfo {volume} {336}},\ \bibinfo
  {pages} {555--558}\  (\bibinfo {year} {2012})}\BibitemShut {NoStop}%
\bibitem [{\citenamefont {Xu}\ \emph {et~al.}(2010)\citenamefont {Xu},
  \citenamefont {Zhang}, \citenamefont {Liu}, \citenamefont {Wang},
  \citenamefont {Li}, \citenamefont {Wu}, \citenamefont {Yu},\ and\
  \citenamefont {Zhang}}]{xxZhang10EPL_AH_Fe-Gd}%
  \BibitemOpen
  \bibfield  {author} {\bibinfo {author} {\bibfnamefont {W.~J.}\ \bibnamefont
  {Xu}}, \bibinfo {author} {\bibfnamefont {B.}~\bibnamefont {Zhang}}, \bibinfo
  {author} {\bibfnamefont {Z.~X.}\ \bibnamefont {Liu}}, \bibinfo {author}
  {\bibfnamefont {Z.}~\bibnamefont {Wang}}, \bibinfo {author} {\bibfnamefont
  {W.}~\bibnamefont {Li}}, \bibinfo {author} {\bibfnamefont {Z.~B.}\
  \bibnamefont {Wu}}, \bibinfo {author} {\bibfnamefont {R.~H.}\ \bibnamefont
  {Yu}}, \ and\ \bibinfo {author} {\bibfnamefont {X.~X.}\ \bibnamefont
  {Zhang}},\ }\bibfield  {title} {\enquote {\bibinfo {title} {Anomalous
  \text{H}all effect in \textnormal{Fe/Gd} bilayers},}\ }\href
  {https://doi.org/10.1209/0295-5075/90/27004} {\bibfield  {journal} {\bibinfo
  {journal} {\emph {Europhys. Lett.}},\ }\textbf {\bibinfo {volume} {90}},\
  \bibinfo {pages} {27004}\  (\bibinfo {year} {2010})}\BibitemShut {NoStop}%
\bibitem [{\citenamefont {Zhou}\ \emph
  {et~al.}(2021){\natexlab{a}}\citenamefont {Zhou}, \citenamefont {Zeng},
  \citenamefont {Jia}, \citenamefont {Chen},\ and\ \citenamefont
  {Wu}}]{yWu21PRB_UMR-sign-revs}%
  \BibitemOpen
  \bibfield  {author} {\bibinfo {author} {\bibfnamefont {X.}~\bibnamefont
  {Zhou}}, \bibinfo {author} {\bibfnamefont {F.}~\bibnamefont {Zeng}}, \bibinfo
  {author} {\bibfnamefont {M.}~\bibnamefont {Jia}}, \bibinfo {author}
  {\bibfnamefont {H.}~\bibnamefont {Chen}}, \ and\ \bibinfo {author}
  {\bibfnamefont {Y.}~\bibnamefont {Wu}},\ }\bibfield  {title} {\enquote
  {\bibinfo {title} {Sign reversal of unidirectional magnetoresistance in
  monocrystalline \textnormal{Fe/Pt} bilayers},}\ }\href
  {https://link.aps.org/doi/10.1103/PhysRevB.104.184413} {\bibfield  {journal}
  {\bibinfo  {journal} {\emph {Phys. Rev. B}},\ }\textbf {\bibinfo {volume}
  {104}},\ \bibinfo {pages} {184413}\  (\bibinfo {year}
  {2021}{\natexlab{a}})}\BibitemShut {NoStop}%
\bibitem [{\citenamefont {Weiler}\ \emph {et~al.}(2012)\citenamefont {Weiler},
  \citenamefont {Althammer}, \citenamefont {Czeschka}, \citenamefont {Huebl},
  \citenamefont {Wagner}, \citenamefont {Opel}, \citenamefont {Imort},
  \citenamefont {Reiss}, \citenamefont {Thomas}, \citenamefont {Gross} \emph
  {et~al.}}]{weiler2012local}%
  \BibitemOpen
  \bibfield  {author} {\bibinfo {author} {\bibfnamefont {M.}~\bibnamefont
  {Weiler}}, \bibinfo {author} {\bibfnamefont {M.}~\bibnamefont {Althammer}},
  \bibinfo {author} {\bibfnamefont {F.~D.}\ \bibnamefont {Czeschka}}, \bibinfo
  {author} {\bibfnamefont {H.}~\bibnamefont {Huebl}}, \bibinfo {author}
  {\bibfnamefont {M.~S.}\ \bibnamefont {Wagner}}, \bibinfo {author}
  {\bibfnamefont {M.}~\bibnamefont {Opel}}, \bibinfo {author} {\bibfnamefont
  {I.-M.}\ \bibnamefont {Imort}}, \bibinfo {author} {\bibfnamefont
  {G.}~\bibnamefont {Reiss}}, \bibinfo {author} {\bibfnamefont
  {A.}~\bibnamefont {Thomas}}, \bibinfo {author} {\bibfnamefont
  {R.}~\bibnamefont {Gross}},  \emph {et~al.},\ }\bibfield  {title} {\enquote
  {\bibinfo {title} {Local charge and spin currents in magnetothermal
  landscapes},}\ }\href
  {https://link.aps.org/doi/10.1103/PhysRevLett.108.106602} {\bibfield
  {journal} {\bibinfo  {journal} {\emph {Phys. Rev. Lett.}},\ }\textbf
  {\bibinfo {volume} {108}},\ \bibinfo {pages} {106602}\  (\bibinfo {year}
  {2012})}\BibitemShut {NoStop}%
\bibitem [{\citenamefont {Kikkawa}\ \emph {et~al.}(2013)\citenamefont
  {Kikkawa}, \citenamefont {Uchida}, \citenamefont {Shiomi}, \citenamefont
  {Qiu}, \citenamefont {Hou}, \citenamefont {Tian}, \citenamefont {Nakayama},
  \citenamefont {Jin},\ and\ \citenamefont {Saitoh}}]{kikkawa2013longitudinal}%
  \BibitemOpen
  \bibfield  {author} {\bibinfo {author} {\bibfnamefont {T.}~\bibnamefont
  {Kikkawa}}, \bibinfo {author} {\bibfnamefont {K.}~\bibnamefont {Uchida}},
  \bibinfo {author} {\bibfnamefont {Y.}~\bibnamefont {Shiomi}}, \bibinfo
  {author} {\bibfnamefont {Z.}~\bibnamefont {Qiu}}, \bibinfo {author}
  {\bibfnamefont {D.}~\bibnamefont {Hou}}, \bibinfo {author} {\bibfnamefont
  {D.}~\bibnamefont {Tian}}, \bibinfo {author} {\bibfnamefont {H.}~\bibnamefont
  {Nakayama}}, \bibinfo {author} {\bibfnamefont {X.-F.}\ \bibnamefont {Jin}}, \
  and\ \bibinfo {author} {\bibfnamefont {E.}~\bibnamefont {Saitoh}},\
  }\bibfield  {title} {\enquote {\bibinfo {title} {Longitudinal spin
  \text{S}eebeck effect free from the proximity \text{N}ernst effect},}\ }\href
  {https://link.aps.org/doi/10.1103/PhysRevLett.110.067207} {\bibfield
  {journal} {\bibinfo  {journal} {\emph {Phys. Rev. Lett.}},\ }\textbf
  {\bibinfo {volume} {110}},\ \bibinfo {pages} {067207}\  (\bibinfo {year}
  {2013})}\BibitemShut {NoStop}%
\bibitem [{\citenamefont {Anderson}(1958)}]{Anderson58pr_localization}%
  \BibitemOpen
  \bibfield  {author} {\bibinfo {author} {\bibfnamefont {P.~W.}\ \bibnamefont
  {Anderson}},\ }\bibfield  {title} {\enquote {\bibinfo {title} {Absence of
  diffusion in certain random lattices},}\ }\href
  {https://link.aps.org/doi/10.1103/PhysRev.109.1492} {\bibfield  {journal}
  {\bibinfo  {journal} {\emph {Phys. Rev.}},\ }\textbf {\bibinfo {volume}
  {109}},\ \bibinfo {pages} {1492--1505}\  (\bibinfo {year}
  {1958})}\BibitemShut {NoStop}%
\bibitem [{\citenamefont {Abrahams}\ \emph {et~al.}(1979)\citenamefont
  {Abrahams}, \citenamefont {Anderson}, \citenamefont {Licciardello},\ and\
  \citenamefont {Ramakrishnan}}]{AbrahamsPRL79_WL-2D}%
  \BibitemOpen
  \bibfield  {author} {\bibinfo {author} {\bibfnamefont {E.}~\bibnamefont
  {Abrahams}}, \bibinfo {author} {\bibfnamefont {P.~W.}\ \bibnamefont
  {Anderson}}, \bibinfo {author} {\bibfnamefont {D.~C.}\ \bibnamefont
  {Licciardello}}, \ and\ \bibinfo {author} {\bibfnamefont {T.~V.}\
  \bibnamefont {Ramakrishnan}},\ }\bibfield  {title} {\enquote {\bibinfo
  {title} {Scaling theory of localization: Absence of quantum diffusion in two
  dimensions},}\ }\href {https://link.aps.org/doi/10.1103/PhysRevLett.42.673}
  {\bibfield  {journal} {\bibinfo  {journal} {\emph {Phys. Rev. Lett.}},\
  }\textbf {\bibinfo {volume} {42}},\ \bibinfo {pages} {673--676}\  (\bibinfo
  {year} {1979})}\BibitemShut {NoStop}%
\bibitem [{\citenamefont {Bergmann}(1984)}]{BERGMANN84Phys-rep_WL-thin-film}%
  \BibitemOpen
  \bibfield  {author} {\bibinfo {author} {\bibfnamefont {G.}~\bibnamefont
  {Bergmann}},\ }\bibfield  {title} {\enquote {\bibinfo {title} {Weak
  localization in thin films: a time-of-flight experiment with conduction
  electrons},}\ }\href
  {https://www.sciencedirect.com/science/article/pii/0370157384901030}
  {\bibfield  {journal} {\bibinfo  {journal} {\emph {Phys. Rep.}},\ }\textbf
  {\bibinfo {volume} {107}},\ \bibinfo {pages} {1--58}\  (\bibinfo {year}
  {1984})}\BibitemShut {NoStop}%
\bibitem [{\citenamefont {Webb}\ \emph {et~al.}(1985)\citenamefont {Webb},
  \citenamefont {Washburn}, \citenamefont {Umbach},\ and\ \citenamefont
  {Laibowitz}}]{Webb85PRL_AB-oscillation_NM-ring}%
  \BibitemOpen
  \bibfield  {author} {\bibinfo {author} {\bibfnamefont {R.~A.}\ \bibnamefont
  {Webb}}, \bibinfo {author} {\bibfnamefont {S.}~\bibnamefont {Washburn}},
  \bibinfo {author} {\bibfnamefont {C.~P.}\ \bibnamefont {Umbach}}, \ and\
  \bibinfo {author} {\bibfnamefont {R.~B.}\ \bibnamefont {Laibowitz}},\
  }\bibfield  {title} {\enquote {\bibinfo {title} {Observation of $\frac{h}{e}$
  aharonov-bohm oscillations in normal-metal rings},}\ }\href
  {https://link.aps.org/doi/10.1103/PhysRevLett.54.2696} {\bibfield  {journal}
  {\bibinfo  {journal} {\emph {Phys. Rev. Lett.}},\ }\textbf {\bibinfo {volume}
  {54}},\ \bibinfo {pages} {2696--2699}\  (\bibinfo {year} {1985})}\BibitemShut
  {NoStop}%
\bibitem [{\citenamefont {Bachtold}\ \emph {et~al.}(1999)\citenamefont
  {Bachtold}, \citenamefont {Strunk}, \citenamefont {Salvetat}, \citenamefont
  {Bonard}, \citenamefont {Forr{\'o}}, \citenamefont {Nussbaumer},\ and\
  \citenamefont {Sch{\"o}nenberger}}]{Bachtold99_AB-osc_Carbon-NT}%
  \BibitemOpen
  \bibfield  {author} {\bibinfo {author} {\bibfnamefont {A.}~\bibnamefont
  {Bachtold}}, \bibinfo {author} {\bibfnamefont {C.}~\bibnamefont {Strunk}},
  \bibinfo {author} {\bibfnamefont {J.-P.}\ \bibnamefont {Salvetat}}, \bibinfo
  {author} {\bibfnamefont {J.-M.}\ \bibnamefont {Bonard}}, \bibinfo {author}
  {\bibfnamefont {L.}~\bibnamefont {Forr{\'o}}}, \bibinfo {author}
  {\bibfnamefont {T.}~\bibnamefont {Nussbaumer}}, \ and\ \bibinfo {author}
  {\bibfnamefont {C.}~\bibnamefont {Sch{\"o}nenberger}},\ }\bibfield  {title}
  {\enquote {\bibinfo {title} {Aharonov--bohm oscillations in carbon
  nanotubes},}\ }\href {https://doi.org/10.1038/17755} {\bibfield  {journal}
  {\bibinfo  {journal} {\emph {Nature}},\ }\textbf {\bibinfo {volume} {397}},\
  \bibinfo {pages} {673--675}\  (\bibinfo {year} {1999})}\BibitemShut {NoStop}%
\bibitem [{\citenamefont {Peng}\ \emph {et~al.}(2010)\citenamefont {Peng},
  \citenamefont {Lai}, \citenamefont {Kong}, \citenamefont {Meister},
  \citenamefont {Chen}, \citenamefont {Qi}, \citenamefont {Zhang},
  \citenamefont {Shen},\ and\ \citenamefont {Cui}}]{Peng10NM_AB-TI}%
  \BibitemOpen
  \bibfield  {author} {\bibinfo {author} {\bibfnamefont {H.}~\bibnamefont
  {Peng}}, \bibinfo {author} {\bibfnamefont {K.}~\bibnamefont {Lai}}, \bibinfo
  {author} {\bibfnamefont {D.}~\bibnamefont {Kong}}, \bibinfo {author}
  {\bibfnamefont {S.}~\bibnamefont {Meister}}, \bibinfo {author} {\bibfnamefont
  {Y.}~\bibnamefont {Chen}}, \bibinfo {author} {\bibfnamefont {X.-L.}\
  \bibnamefont {Qi}}, \bibinfo {author} {\bibfnamefont {S.-C.}\ \bibnamefont
  {Zhang}}, \bibinfo {author} {\bibfnamefont {Z.-X.}\ \bibnamefont {Shen}}, \
  and\ \bibinfo {author} {\bibfnamefont {Y.}~\bibnamefont {Cui}},\ }\bibfield
  {title} {\enquote {\bibinfo {title} {Aharonov--bohm interference in
  topological insulator nanoribbons},}\ }\href
  {https://doi.org/10.1038/nmat2609} {\bibfield  {journal} {\bibinfo  {journal}
  {\emph {Nat. Mater.}},\ }\textbf {\bibinfo {volume} {9}},\ \bibinfo {pages}
  {225--229}\  (\bibinfo {year} {2010})}\BibitemShut {NoStop}%
\bibitem [{\citenamefont {Spivak}\ \emph {et~al.}(2010)\citenamefont {Spivak},
  \citenamefont {Kravchenko}, \citenamefont {Kivelson},\ and\ \citenamefont
  {Gao}}]{Spivak10RMP_QM-transport_2DEF}%
  \BibitemOpen
  \bibfield  {author} {\bibinfo {author} {\bibfnamefont {B.}~\bibnamefont
  {Spivak}}, \bibinfo {author} {\bibfnamefont {S.~V.}\ \bibnamefont
  {Kravchenko}}, \bibinfo {author} {\bibfnamefont {S.~A.}\ \bibnamefont
  {Kivelson}}, \ and\ \bibinfo {author} {\bibfnamefont {X.~P.~A.}\ \bibnamefont
  {Gao}},\ }\bibfield  {title} {\enquote {\bibinfo {title} {Colloquium:
  Transport in strongly correlated two dimensional electron fluids},}\ }\href
  {https://link.aps.org/doi/10.1103/RevModPhys.82.1743} {\bibfield  {journal}
  {\bibinfo  {journal} {\emph {Rev. Mod. Phys.}},\ }\textbf {\bibinfo {volume}
  {82}},\ \bibinfo {pages} {1743--1766}\  (\bibinfo {year} {2010})}\BibitemShut
  {NoStop}%
\bibitem [{\citenamefont {Whiticar}\ \emph {et~al.}(2020)\citenamefont
  {Whiticar}, \citenamefont {Fornieri}, \citenamefont {O'Farrell},
  \citenamefont {Drachmann}, \citenamefont {Wang}, \citenamefont {Thomas},
  \citenamefont {Gronin}, \citenamefont {Kallaher}, \citenamefont {Gardner},
  \citenamefont {Manfra} \emph {et~al.}}]{Whiticar20NC_Majorana-AB}%
  \BibitemOpen
  \bibfield  {author} {\bibinfo {author} {\bibfnamefont {A.~M.}\ \bibnamefont
  {Whiticar}}, \bibinfo {author} {\bibfnamefont {A.}~\bibnamefont {Fornieri}},
  \bibinfo {author} {\bibfnamefont {E.~C.~T.}\ \bibnamefont {O'Farrell}},
  \bibinfo {author} {\bibfnamefont {A.~C.~C.}\ \bibnamefont {Drachmann}},
  \bibinfo {author} {\bibfnamefont {T.}~\bibnamefont {Wang}}, \bibinfo {author}
  {\bibfnamefont {C.}~\bibnamefont {Thomas}}, \bibinfo {author} {\bibfnamefont
  {S.}~\bibnamefont {Gronin}}, \bibinfo {author} {\bibfnamefont
  {R.}~\bibnamefont {Kallaher}}, \bibinfo {author} {\bibfnamefont {G.~C.}\
  \bibnamefont {Gardner}}, \bibinfo {author} {\bibfnamefont {M.~J.}\
  \bibnamefont {Manfra}},  \emph {et~al.},\ }\bibfield  {title} {\enquote
  {\bibinfo {title} {Coherent transport through a majorana island in an
  aharonov--bohm interferometer},}\ }\href
  {https://doi.org/10.1038/s41467-020-16988-x} {\bibfield  {journal} {\bibinfo
  {journal} {\emph {Nat. Commun.}},\ }\textbf {\bibinfo {volume} {11}},\
  \bibinfo {pages} {3212}\  (\bibinfo {year} {2020})}\BibitemShut {NoStop}%
\bibitem [{\citenamefont {Altshuler}\ \emph {et~al.}(1980)\citenamefont
  {Altshuler}, \citenamefont {Khmel'nitzkii}, \citenamefont {Larkin},\ and\
  \citenamefont {Lee}}]{altshuler1980prb}%
  \BibitemOpen
  \bibfield  {author} {\bibinfo {author} {\bibfnamefont {B.~L.}\ \bibnamefont
  {Altshuler}}, \bibinfo {author} {\bibfnamefont {D.}~\bibnamefont
  {Khmel'nitzkii}}, \bibinfo {author} {\bibfnamefont {A.~I.}\ \bibnamefont
  {Larkin}}, \ and\ \bibinfo {author} {\bibfnamefont {P.~A.}\ \bibnamefont
  {Lee}},\ }\bibfield  {title} {\enquote {\bibinfo {title} {Magnetoresistance
  and hall effect in a disordered two-dimensional electron gas},}\ }\href
  {https://link.aps.org/doi/10.1103/PhysRevB.22.5142} {\bibfield  {journal}
  {\bibinfo  {journal} {\emph {Phys. Rev. B}},\ }\textbf {\bibinfo {volume}
  {22}},\ \bibinfo {pages} {5142--5153}\  (\bibinfo {year} {1980})}\BibitemShut
  {NoStop}%
\bibitem [{\citenamefont {{Hikami}}\ \emph {et~al.}(1980)\citenamefont
  {{Hikami}}, \citenamefont {{Larkin}},\ and\ \citenamefont
  {{Nagaoka}}}]{hikami1980}%
  \BibitemOpen
  \bibfield  {author} {\bibinfo {author} {\bibfnamefont {S.}~\bibnamefont
  {{Hikami}}}, \bibinfo {author} {\bibfnamefont {A.~I.}\ \bibnamefont
  {{Larkin}}}, \ and\ \bibinfo {author} {\bibfnamefont {Y.}~\bibnamefont
  {{Nagaoka}}},\ }\bibfield  {title} {\enquote {\bibinfo {title} {{Spin-Orbit
  Interaction and Magnetoresistance in the Two Dimensional Random System}},}\
  }\href {https://doi.org/10.1143/PTP.63.707} {\bibfield  {journal} {\bibinfo
  {journal} {\emph {Prog. Theor. Phys.}},\ }\textbf {\bibinfo {volume} {63}},\
  \bibinfo {pages} {707--710}\  (\bibinfo {year} {1980})}\BibitemShut {NoStop}%
\bibitem [{\citenamefont {Olejn\'{\i}k}\ \emph
  {et~al.}(2015){\natexlab{c}}\citenamefont {Olejn\'{\i}k}, \citenamefont
  {Nov\'ak}, \citenamefont {Wunderlich},\ and\ \citenamefont
  {Jungwirth}}]{Olejnik15PRB_UMR-semicond}%
  \BibitemOpen
  \bibfield  {author} {\bibinfo {author} {\bibfnamefont {K.}~\bibnamefont
  {Olejn\'{\i}k}}, \bibinfo {author} {\bibfnamefont {V.}~\bibnamefont
  {Nov\'ak}}, \bibinfo {author} {\bibfnamefont {J.}~\bibnamefont {Wunderlich}},
  \ and\ \bibinfo {author} {\bibfnamefont {T.}~\bibnamefont {Jungwirth}},\
  }\bibfield  {title} {\enquote {\bibinfo {title} {Electrical detection of
  magnetization reversal without auxiliary magnets},}\ }\href
  {https://link.aps.org/doi/10.1103/PhysRevB.91.180402} {\bibfield  {journal}
  {\bibinfo  {journal} {\emph {Phys. Rev. B}},\ }\textbf {\bibinfo {volume}
  {91}},\ \bibinfo {pages} {180402}\  (\bibinfo {year}
  {2015}{\natexlab{c}})}\BibitemShut {NoStop}%
\bibitem [{\citenamefont {Langenfeld}\ \emph
  {et~al.}(2016){\natexlab{b}}\citenamefont {Langenfeld}, \citenamefont
  {Tshitoyan}, \citenamefont {Fang}, \citenamefont {Wells}, \citenamefont
  {Moore},\ and\ \citenamefont {Ferguson}}]{Langenfeld16APL_UMR-FMR}%
  \BibitemOpen
  \bibfield  {author} {\bibinfo {author} {\bibfnamefont {S.}~\bibnamefont
  {Langenfeld}}, \bibinfo {author} {\bibfnamefont {V.}~\bibnamefont
  {Tshitoyan}}, \bibinfo {author} {\bibfnamefont {Z.}~\bibnamefont {Fang}},
  \bibinfo {author} {\bibfnamefont {A.}~\bibnamefont {Wells}}, \bibinfo
  {author} {\bibfnamefont {T.~A.}\ \bibnamefont {Moore}}, \ and\ \bibinfo
  {author} {\bibfnamefont {A.~J.}\ \bibnamefont {Ferguson}},\ }\bibfield
  {title} {\enquote {\bibinfo {title} {Exchange magnon induced resistance
  asymmetry in permalloy spin-hall oscillators},}\ }\href
  {https://doi.org/10.1063/1.4948921} {\bibfield  {journal} {\bibinfo
  {journal} {\emph {Appl. Phys. Lett.}},\ }\textbf {\bibinfo {volume} {108}},\
  \bibinfo {pages} {192402}\  (\bibinfo {year}
  {2016}{\natexlab{b}})}\BibitemShut {NoStop}%
\bibitem [{\citenamefont {Haney}\ \emph {et~al.}(2013)\citenamefont {Haney},
  \citenamefont {Lee}, \citenamefont {Lee}, \citenamefont {Manchon},\ and\
  \citenamefont {Stiles}}]{Stiles13PRB_SOT}%
  \BibitemOpen
  \bibfield  {author} {\bibinfo {author} {\bibfnamefont {P.~M.}\ \bibnamefont
  {Haney}}, \bibinfo {author} {\bibfnamefont {H.-W.}\ \bibnamefont {Lee}},
  \bibinfo {author} {\bibfnamefont {K.-J.}\ \bibnamefont {Lee}}, \bibinfo
  {author} {\bibfnamefont {A.}~\bibnamefont {Manchon}}, \ and\ \bibinfo
  {author} {\bibfnamefont {M.~D.}\ \bibnamefont {Stiles}},\ }\bibfield  {title}
  {\enquote {\bibinfo {title} {Current induced torques and interfacial
  spin-orbit coupling: Semiclassical modeling},}\ }\href
  {https://link.aps.org/doi/10.1103/PhysRevB.87.174411} {\bibfield  {journal}
  {\bibinfo  {journal} {\emph {Phys. Rev. B}},\ }\textbf {\bibinfo {volume}
  {87}},\ \bibinfo {pages} {174411}\  (\bibinfo {year} {2013})}\BibitemShut
  {NoStop}%
\bibitem [{\citenamefont {Tokatly}\ \emph
  {et~al.}(2015){\natexlab{b}}\citenamefont {Tokatly}, \citenamefont
  {Krasovskii},\ and\ \citenamefont {Vignale}}]{tokatly2015prb}%
  \BibitemOpen
  \bibfield  {author} {\bibinfo {author} {\bibfnamefont {I.~V.}\ \bibnamefont
  {Tokatly}}, \bibinfo {author} {\bibfnamefont {E.~E.}\ \bibnamefont
  {Krasovskii}}, \ and\ \bibinfo {author} {\bibfnamefont {G.}~\bibnamefont
  {Vignale}},\ }\bibfield  {title} {\enquote {\bibinfo {title} {Current-induced
  spin polarization at the surface of metallic films: A theorem and an ab
  initio calculation},}\ }\href
  {https://link.aps.org/doi/10.1103/PhysRevB.91.035403} {\bibfield  {journal}
  {\bibinfo  {journal} {\emph {Phys. Rev. B}},\ }\textbf {\bibinfo {volume}
  {91}},\ \bibinfo {pages} {035403}\  (\bibinfo {year}
  {2015}{\natexlab{b}})}\BibitemShut {NoStop}%
\bibitem [{\citenamefont {Nakayama}\ \emph {et~al.}(2013)\citenamefont
  {Nakayama} \emph {et~al.}}]{Saitoh13PRL_SH-MR}%
  \BibitemOpen
  \bibfield  {author} {\bibinfo {author} {\bibfnamefont {H.}~\bibnamefont
  {Nakayama}} \emph {et~al.},\ }\bibfield  {title} {\enquote {\bibinfo {title}
  {Spin hall magnetoresistance induced by a nonequilibrium proximity effect},}\
  }\href {http://link.aps.org/doi/10.1103/PhysRevLett.110.206601} {\bibfield
  {journal} {\bibinfo  {journal} {\emph {Phys. Rev. Lett.}},\ }\textbf
  {\bibinfo {volume} {110}},\ \bibinfo {pages} {206601}\  (\bibinfo {year}
  {2013})}\BibitemShut {NoStop}%
\bibitem [{\citenamefont {Grigoryan}\ \emph {et~al.}(2014)\citenamefont
  {Grigoryan}, \citenamefont {Guo}, \citenamefont {Bauer},\ and\ \citenamefont
  {Xiao}}]{xJiang14prb_Rashba-AMR}%
  \BibitemOpen
  \bibfield  {author} {\bibinfo {author} {\bibfnamefont {V.~L.}\ \bibnamefont
  {Grigoryan}}, \bibinfo {author} {\bibfnamefont {W.}~\bibnamefont {Guo}},
  \bibinfo {author} {\bibfnamefont {G.~E.~W.}\ \bibnamefont {Bauer}}, \ and\
  \bibinfo {author} {\bibfnamefont {J.}~\bibnamefont {Xiao}},\ }\bibfield
  {title} {\enquote {\bibinfo {title} {Intrinsic magnetoresistance in metal
  films on ferromagnetic insulators},}\ }\href
  {https://link.aps.org/doi/10.1103/PhysRevB.90.161412} {\bibfield  {journal}
  {\bibinfo  {journal} {\emph {Phys. Rev. B}},\ }\textbf {\bibinfo {volume}
  {90}},\ \bibinfo {pages} {161412}\  (\bibinfo {year} {2014})}\BibitemShut
  {NoStop}%
\bibitem [{\citenamefont {Zhang}\ \emph {et~al.}(2015)\citenamefont {Zhang},
  \citenamefont {Vignale},\ and\ \citenamefont {Zhang}}]{slzhang15PRB_AMR}%
  \BibitemOpen
  \bibfield  {author} {\bibinfo {author} {\bibfnamefont {S.~S.-L.}\
  \bibnamefont {Zhang}}, \bibinfo {author} {\bibfnamefont {G.}~\bibnamefont
  {Vignale}}, \ and\ \bibinfo {author} {\bibfnamefont {S.}~\bibnamefont
  {Zhang}},\ }\bibfield  {title} {\enquote {\bibinfo {title} {Anisotropic
  magnetoresistance driven by surface spin-orbit scattering},}\ }\href
  {http://link.aps.org/doi/10.1103/PhysRevB.92.024412} {\bibfield  {journal}
  {\bibinfo  {journal} {\emph {Phys. Rev. B}},\ }\textbf {\bibinfo {volume}
  {92}},\ \bibinfo {pages} {024412}\  (\bibinfo {year} {2015})}\BibitemShut
  {NoStop}%
\bibitem [{\citenamefont {Camblong}\ \emph {et~al.}(1994)\citenamefont
  {Camblong}, \citenamefont {Zhang},\ and\ \citenamefont
  {Levy}}]{camblong1994theory}%
  \BibitemOpen
  \bibfield  {author} {\bibinfo {author} {\bibfnamefont {H.~E.}\ \bibnamefont
  {Camblong}}, \bibinfo {author} {\bibfnamefont {S.}~\bibnamefont {Zhang}}, \
  and\ \bibinfo {author} {\bibfnamefont {P.~M.}\ \bibnamefont {Levy}},\
  }\bibfield  {title} {\enquote {\bibinfo {title} {Theory of magnetotransport
  in inhomogeneous magnetic structures},}\ }\href
  {https://doi.org/10.1063/1.356776} {\bibfield  {journal} {\bibinfo  {journal}
  {\emph {J. Appl. Phys.}},\ }\textbf {\bibinfo {volume} {75}},\ \bibinfo
  {pages} {6906--6908}\  (\bibinfo {year} {1994})}\BibitemShut {NoStop}%
\bibitem [{\citenamefont {Dyakonov}\ and\ \citenamefont
  {Perel}(1971)}]{dyakonov1971}%
  \BibitemOpen
  \bibfield  {author} {\bibinfo {author} {\bibfnamefont {M.~I.}\ \bibnamefont
  {Dyakonov}}\ and\ \bibinfo {author} {\bibfnamefont {V.}~\bibnamefont
  {Perel}},\ }\bibfield  {title} {\enquote {\bibinfo {title} {Current-induced
  spin orientation of electrons in semiconductors},}\ }\href
  {https://doi.org/10.1016/0375-9601(71)90196-4} {\bibfield  {journal}
  {\bibinfo  {journal} {\emph {Phys. Lett. A}},\ }\textbf {\bibinfo {volume}
  {35}},\ \bibinfo {pages} {459--460}\  (\bibinfo {year} {1971})}\BibitemShut
  {NoStop}%
\bibitem [{\citenamefont {Hirsch}(1999)}]{hirsch1999prl}%
  \BibitemOpen
  \bibfield  {author} {\bibinfo {author} {\bibfnamefont {J.~E.}\ \bibnamefont
  {Hirsch}},\ }\bibfield  {title} {\enquote {\bibinfo {title} {Spin hall
  effect},}\ }\href {https://link.aps.org/doi/10.1103/PhysRevLett.83.1834}
  {\bibfield  {journal} {\bibinfo  {journal} {\emph {Phys. Rev. Lett.}},\
  }\textbf {\bibinfo {volume} {83}},\ \bibinfo {pages} {1834--1837}\  (\bibinfo
  {year} {1999})}\BibitemShut {NoStop}%
\bibitem [{\citenamefont {Zhang}(2000)}]{shufeng2000prl}%
  \BibitemOpen
  \bibfield  {author} {\bibinfo {author} {\bibfnamefont {S.}~\bibnamefont
  {Zhang}},\ }\bibfield  {title} {\enquote {\bibinfo {title} {Spin hall effect
  in the presence of spin diffusion},}\ }\href
  {https://link.aps.org/doi/10.1103/PhysRevLett.85.393} {\bibfield  {journal}
  {\bibinfo  {journal} {\emph {Phys. Rev. Lett.}},\ }\textbf {\bibinfo {volume}
  {85}},\ \bibinfo {pages} {393--396}\  (\bibinfo {year} {2000})}\BibitemShut
  {NoStop}%
\bibitem [{\citenamefont {Huang}\ \emph {et~al.}(2011)\citenamefont {Huang},
  \citenamefont {Wang}, \citenamefont {Lee}, \citenamefont {Kwo},\ and\
  \citenamefont {Chien}}]{syHuang11PRL_ANE}%
  \BibitemOpen
  \bibfield  {author} {\bibinfo {author} {\bibfnamefont {S.~Y.}\ \bibnamefont
  {Huang}}, \bibinfo {author} {\bibfnamefont {W.~G.}\ \bibnamefont {Wang}},
  \bibinfo {author} {\bibfnamefont {S.~F.}\ \bibnamefont {Lee}}, \bibinfo
  {author} {\bibfnamefont {J.}~\bibnamefont {Kwo}}, \ and\ \bibinfo {author}
  {\bibfnamefont {C.~L.}\ \bibnamefont {Chien}},\ }\bibfield  {title} {\enquote
  {\bibinfo {title} {Intrinsic spin-dependent thermal transport},}\ }\href
  {https://link.aps.org/doi/10.1103/PhysRevLett.107.216604} {\bibfield
  {journal} {\bibinfo  {journal} {\emph {Phys. Rev. Lett.}},\ }\textbf
  {\bibinfo {volume} {107}},\ \bibinfo {pages} {216604}\  (\bibinfo {year}
  {2011})}\BibitemShut {NoStop}%
\bibitem [{\citenamefont {Uchida}\ \emph {et~al.}(2008)\citenamefont {Uchida},
  \citenamefont {Takahashi}, \citenamefont {Harii}, \citenamefont {Ieda},
  \citenamefont {Koshibae}, \citenamefont {Ando}, \citenamefont {Maekawa},\
  and\ \citenamefont {Saitoh}}]{uchida2008observation}%
  \BibitemOpen
  \bibfield  {author} {\bibinfo {author} {\bibfnamefont {K.}~\bibnamefont
  {Uchida}}, \bibinfo {author} {\bibfnamefont {S.}~\bibnamefont {Takahashi}},
  \bibinfo {author} {\bibfnamefont {K.}~\bibnamefont {Harii}}, \bibinfo
  {author} {\bibfnamefont {J.}~\bibnamefont {Ieda}}, \bibinfo {author}
  {\bibfnamefont {W.}~\bibnamefont {Koshibae}}, \bibinfo {author}
  {\bibfnamefont {K.}~\bibnamefont {Ando}}, \bibinfo {author} {\bibfnamefont
  {S.}~\bibnamefont {Maekawa}}, \ and\ \bibinfo {author} {\bibfnamefont
  {E.}~\bibnamefont {Saitoh}},\ }\bibfield  {title} {\enquote {\bibinfo {title}
  {Observation of the spin \text{S}eebeck effect},}\ }\href
  {https://doi.org/10.1038/nature07321} {\bibfield  {journal} {\bibinfo
  {journal} {\emph {Nature}},\ }\textbf {\bibinfo {volume} {455}},\ \bibinfo
  {pages} {778--781}\  (\bibinfo {year} {2008})}\BibitemShut {NoStop}%
\bibitem [{\citenamefont {Uchida}\ \emph {et~al.}(2010)\citenamefont {Uchida},
  \citenamefont {Adachi}, \citenamefont {Ota}, \citenamefont {Nakayama},
  \citenamefont {Maekawa},\ and\ \citenamefont
  {Saitoh}}]{uchida2010observation}%
  \BibitemOpen
  \bibfield  {author} {\bibinfo {author} {\bibfnamefont {K.-i.}\ \bibnamefont
  {Uchida}}, \bibinfo {author} {\bibfnamefont {H.}~\bibnamefont {Adachi}},
  \bibinfo {author} {\bibfnamefont {T.}~\bibnamefont {Ota}}, \bibinfo {author}
  {\bibfnamefont {H.}~\bibnamefont {Nakayama}}, \bibinfo {author}
  {\bibfnamefont {S.}~\bibnamefont {Maekawa}}, \ and\ \bibinfo {author}
  {\bibfnamefont {E.}~\bibnamefont {Saitoh}},\ }\bibfield  {title} {\enquote
  {\bibinfo {title} {Observation of longitudinal spin-\text{S}eebeck effect in
  magnetic insulators},}\ }\href {https://doi.org/10.1063/1.3507386} {\bibfield
   {journal} {\bibinfo  {journal} {\emph {Appl. Phys. Lett.}},\ }\textbf
  {\bibinfo {volume} {97}},\ \bibinfo {pages} {172505}\  (\bibinfo {year}
  {2010})}\BibitemShut {NoStop}%
\bibitem [{\citenamefont {Qu}\ \emph {et~al.}(2013)\citenamefont {Qu},
  \citenamefont {Huang}, \citenamefont {Hu}, \citenamefont {Wu},\ and\
  \citenamefont {Chien}}]{qu2013prl}%
  \BibitemOpen
  \bibfield  {author} {\bibinfo {author} {\bibfnamefont {D.}~\bibnamefont
  {Qu}}, \bibinfo {author} {\bibfnamefont {S.~Y.}\ \bibnamefont {Huang}},
  \bibinfo {author} {\bibfnamefont {J.}~\bibnamefont {Hu}}, \bibinfo {author}
  {\bibfnamefont {R.}~\bibnamefont {Wu}}, \ and\ \bibinfo {author}
  {\bibfnamefont {C.~L.}\ \bibnamefont {Chien}},\ }\bibfield  {title} {\enquote
  {\bibinfo {title} {Intrinsic spin seebeck effect in
  $\mathrm{Au}/\mathrm{YIG}$},}\ }\href
  {https://link.aps.org/doi/10.1103/PhysRevLett.110.067206} {\bibfield
  {journal} {\bibinfo  {journal} {\emph {Phys. Rev. Lett.}},\ }\textbf
  {\bibinfo {volume} {110}},\ \bibinfo {pages} {067206}\  (\bibinfo {year}
  {2013})}\BibitemShut {NoStop}%
\bibitem [{\citenamefont {Adachi}\ \emph {et~al.}(2013)\citenamefont {Adachi},
  \citenamefont {Uchida}, \citenamefont {Saitoh},\ and\ \citenamefont
  {Maekawa}}]{adachi2013theory}%
  \BibitemOpen
  \bibfield  {author} {\bibinfo {author} {\bibfnamefont {H.}~\bibnamefont
  {Adachi}}, \bibinfo {author} {\bibfnamefont {K.-i.}\ \bibnamefont {Uchida}},
  \bibinfo {author} {\bibfnamefont {E.}~\bibnamefont {Saitoh}}, \ and\ \bibinfo
  {author} {\bibfnamefont {S.}~\bibnamefont {Maekawa}},\ }\bibfield  {title}
  {\enquote {\bibinfo {title} {Theory of the spin \text{S}eebeck effect},}\
  }\href {https://doi.org/10.1088/0034-4885/76/3/036501} {\bibfield  {journal}
  {\bibinfo  {journal} {\emph {Rep. Prog. Phys.}},\ }\textbf {\bibinfo {volume}
  {76}},\ \bibinfo {pages} {036501}\  (\bibinfo {year} {2013})}\BibitemShut
  {NoStop}%
\bibitem [{\citenamefont {Avci}\ \emph {et~al.}(2014)\citenamefont {Avci},
  \citenamefont {Garello}, \citenamefont {Gabureac}, \citenamefont {Ghosh},
  \citenamefont {Fuhrer}, \citenamefont {Alvarado},\ and\ \citenamefont
  {Gambardella}}]{avci2014interplay}%
  \BibitemOpen
  \bibfield  {author} {\bibinfo {author} {\bibfnamefont {C.~O.}\ \bibnamefont
  {Avci}}, \bibinfo {author} {\bibfnamefont {K.}~\bibnamefont {Garello}},
  \bibinfo {author} {\bibfnamefont {M.}~\bibnamefont {Gabureac}}, \bibinfo
  {author} {\bibfnamefont {A.}~\bibnamefont {Ghosh}}, \bibinfo {author}
  {\bibfnamefont {A.}~\bibnamefont {Fuhrer}}, \bibinfo {author} {\bibfnamefont
  {S.~F.}\ \bibnamefont {Alvarado}}, \ and\ \bibinfo {author} {\bibfnamefont
  {P.}~\bibnamefont {Gambardella}},\ }\bibfield  {title} {\enquote {\bibinfo
  {title} {Interplay of spin-orbit torque and thermoelectric effects in
  ferromagnet/normal-metal bilayers},}\ }\href
  {https://link.aps.org/doi/10.1103/PhysRevB.90.224427} {\bibfield  {journal}
  {\bibinfo  {journal} {\emph {Phys. Rev. B}},\ }\textbf {\bibinfo {volume}
  {90}},\ \bibinfo {pages} {224427}\  (\bibinfo {year} {2014})}\BibitemShut
  {NoStop}%
\bibitem [{\citenamefont {Stewart}\ \emph {et~al.}(2003)\citenamefont
  {Stewart}, \citenamefont {Butler}, \citenamefont {Zhang},\ and\ \citenamefont
  {Los}}]{stewart2003interfacial}%
  \BibitemOpen
  \bibfield  {author} {\bibinfo {author} {\bibfnamefont {D.~A.}\ \bibnamefont
  {Stewart}}, \bibinfo {author} {\bibfnamefont {W.~H.}\ \bibnamefont {Butler}},
  \bibinfo {author} {\bibfnamefont {X.-G.}\ \bibnamefont {Zhang}}, \ and\
  \bibinfo {author} {\bibfnamefont {V.~F.}\ \bibnamefont {Los}},\ }\bibfield
  {title} {\enquote {\bibinfo {title} {Interfacial scattering in magnetic
  multilayers and spin valves},}\ }\href
  {https://link.aps.org/doi/10.1103/PhysRevB.68.014433} {\bibfield  {journal}
  {\bibinfo  {journal} {\emph {Phys. Rev. B}},\ }\textbf {\bibinfo {volume}
  {68}},\ \bibinfo {pages} {014433}\  (\bibinfo {year} {2003})}\BibitemShut
  {NoStop}%
\bibitem [{\citenamefont {Brataas}\ and\ \citenamefont
  {Bauer}(1994)}]{brataas1994semiclassical}%
  \BibitemOpen
  \bibfield  {author} {\bibinfo {author} {\bibfnamefont {A.}~\bibnamefont
  {Brataas}}\ and\ \bibinfo {author} {\bibfnamefont {G.~E.~W.}\ \bibnamefont
  {Bauer}},\ }\bibfield  {title} {\enquote {\bibinfo {title} {Semiclassical
  theory of perpendicular transport and giant magnetoresistance in disordered
  metallic multilayers},}\ }\href
  {https://link.aps.org/doi/10.1103/PhysRevB.49.14684} {\bibfield  {journal}
  {\bibinfo  {journal} {\emph {Phys. Rev. B}},\ }\textbf {\bibinfo {volume}
  {49}},\ \bibinfo {pages} {14684--14699}\  (\bibinfo {year}
  {1994})}\BibitemShut {NoStop}%
\bibitem [{\citenamefont {Okada}\ \emph {et~al.}(2016)\citenamefont {Okada},
  \citenamefont {Ogawa}, \citenamefont {Yoshimi}, \citenamefont {Tsukazaki},
  \citenamefont {Takahashi}, \citenamefont {Kawasaki},\ and\ \citenamefont
  {Tokura}}]{Okada16PRB_Fermi-level_TI}%
  \BibitemOpen
  \bibfield  {author} {\bibinfo {author} {\bibfnamefont {K.~N.}\ \bibnamefont
  {Okada}}, \bibinfo {author} {\bibfnamefont {N.}~\bibnamefont {Ogawa}},
  \bibinfo {author} {\bibfnamefont {R.}~\bibnamefont {Yoshimi}}, \bibinfo
  {author} {\bibfnamefont {A.}~\bibnamefont {Tsukazaki}}, \bibinfo {author}
  {\bibfnamefont {K.~S.}\ \bibnamefont {Takahashi}}, \bibinfo {author}
  {\bibfnamefont {M.}~\bibnamefont {Kawasaki}}, \ and\ \bibinfo {author}
  {\bibfnamefont {Y.}~\bibnamefont {Tokura}},\ }\bibfield  {title} {\enquote
  {\bibinfo {title} {Enhanced photogalvanic current in topological insulators
  via fermi energy tuning},}\ }\href
  {https://link.aps.org/doi/10.1103/PhysRevB.93.081403} {\bibfield  {journal}
  {\bibinfo  {journal} {\emph {Phys. Rev. B}},\ }\textbf {\bibinfo {volume}
  {93}},\ \bibinfo {pages} {081403}\  (\bibinfo {year} {2016})}\BibitemShut
  {NoStop}%
\bibitem [{\citenamefont {Kondou}\ \emph {et~al.}(2016)\citenamefont {Kondou},
  \citenamefont {Yoshimi}, \citenamefont {Tsukazaki}, \citenamefont {Fukuma},
  \citenamefont {Matsuno}, \citenamefont {Takahashi}, \citenamefont {Kawasaki},
  \citenamefont {Tokura},\ and\ \citenamefont
  {Otani}}]{Kondou16NP_Fermi-level-sc_TI}%
  \BibitemOpen
  \bibfield  {author} {\bibinfo {author} {\bibfnamefont {K.}~\bibnamefont
  {Kondou}}, \bibinfo {author} {\bibfnamefont {R.}~\bibnamefont {Yoshimi}},
  \bibinfo {author} {\bibfnamefont {A.}~\bibnamefont {Tsukazaki}}, \bibinfo
  {author} {\bibfnamefont {Y.}~\bibnamefont {Fukuma}}, \bibinfo {author}
  {\bibfnamefont {J.}~\bibnamefont {Matsuno}}, \bibinfo {author} {\bibfnamefont
  {K.~S.}\ \bibnamefont {Takahashi}}, \bibinfo {author} {\bibfnamefont
  {M.}~\bibnamefont {Kawasaki}}, \bibinfo {author} {\bibfnamefont
  {Y.}~\bibnamefont {Tokura}}, \ and\ \bibinfo {author} {\bibfnamefont
  {Y.}~\bibnamefont {Otani}},\ }\bibfield  {title} {\enquote {\bibinfo {title}
  {Fermi-level-dependent charge-to-spin current conversion by dirac surface
  states of topological insulators},}\ }\href
  {https://doi.org/10.1038/nphys3833} {\bibfield  {journal} {\bibinfo
  {journal} {\emph {Nat. Phys.}},\ }\textbf {\bibinfo {volume} {12}},\ \bibinfo
  {pages} {1027--1031}\  (\bibinfo {year} {2016})}\BibitemShut {NoStop}%
\bibitem [{\citenamefont {Sun}\ \emph {et~al.}(2019)\citenamefont {Sun},
  \citenamefont {Yang}, \citenamefont {Yang}, \citenamefont {Vetter},
  \citenamefont {Sun}, \citenamefont {Li}, \citenamefont {Su}, \citenamefont
  {Li}, \citenamefont {Li}, \citenamefont {Gong} \emph
  {et~al.}}]{sun2019large}%
  \BibitemOpen
  \bibfield  {author} {\bibinfo {author} {\bibfnamefont {R.}~\bibnamefont
  {Sun}}, \bibinfo {author} {\bibfnamefont {S.}~\bibnamefont {Yang}}, \bibinfo
  {author} {\bibfnamefont {X.}~\bibnamefont {Yang}}, \bibinfo {author}
  {\bibfnamefont {E.}~\bibnamefont {Vetter}}, \bibinfo {author} {\bibfnamefont
  {D.}~\bibnamefont {Sun}}, \bibinfo {author} {\bibfnamefont {N.}~\bibnamefont
  {Li}}, \bibinfo {author} {\bibfnamefont {L.}~\bibnamefont {Su}}, \bibinfo
  {author} {\bibfnamefont {Y.}~\bibnamefont {Li}}, \bibinfo {author}
  {\bibfnamefont {Y.}~\bibnamefont {Li}}, \bibinfo {author} {\bibfnamefont
  {Z.-z.}\ \bibnamefont {Gong}},  \emph {et~al.},\ }\bibfield  {title}
  {\enquote {\bibinfo {title} {Large tunable spin-to-charge conversion induced
  by hybrid rashba and dirac surface states in topological insulator
  heterostructures},}\ }\href {https://doi.org/10.1021/acs.nanolett.9b01151}
  {\bibfield  {journal} {\bibinfo  {journal} {\emph {Nano Lett.}},\ }\textbf
  {\bibinfo {volume} {19}},\ \bibinfo {pages} {4420--4426}\  (\bibinfo {year}
  {2019})}\BibitemShut {NoStop}%
\bibitem [{\citenamefont {Su}\ \emph {et~al.}(2021)\citenamefont {Su},
  \citenamefont {Chuang}, \citenamefont {Lee}, \citenamefont {Chong},
  \citenamefont {Li}, \citenamefont {Lin}, \citenamefont {Chen}, \citenamefont
  {Cheng},\ and\ \citenamefont {Huang}}]{Su21ACS_Fermi-level_TI}%
  \BibitemOpen
  \bibfield  {author} {\bibinfo {author} {\bibfnamefont {S.~H.}\ \bibnamefont
  {Su}}, \bibinfo {author} {\bibfnamefont {P.-Y.}\ \bibnamefont {Chuang}},
  \bibinfo {author} {\bibfnamefont {J.-C.}\ \bibnamefont {Lee}}, \bibinfo
  {author} {\bibfnamefont {C.-W.}\ \bibnamefont {Chong}}, \bibinfo {author}
  {\bibfnamefont {Y.~W.}\ \bibnamefont {Li}}, \bibinfo {author} {\bibfnamefont
  {Z.~M.}\ \bibnamefont {Lin}}, \bibinfo {author} {\bibfnamefont {Y.-C.}\
  \bibnamefont {Chen}}, \bibinfo {author} {\bibfnamefont {C.-M.}\ \bibnamefont
  {Cheng}}, \ and\ \bibinfo {author} {\bibfnamefont {J.-C.-A.}\ \bibnamefont
  {Huang}},\ }\bibfield  {title} {\enquote {\bibinfo {title} {Spin-to-charge
  conversion manipulated by fine-tuning the fermi level of topological
  insulator (bi1-xsbx)2te3},}\ }\href {https://doi.org/10.1021/acsaelm.1c00182}
  {\bibfield  {journal} {\bibinfo  {journal} {\emph {ACS Appl. Electron.
  Mater.}},\ }\textbf {\bibinfo {volume} {3}},\ \bibinfo {pages} {2988--2994}\
  (\bibinfo {year} {2021})}\BibitemShut {NoStop}%
\bibitem [{\citenamefont {Qi}\ \emph {et~al.}(2006)\citenamefont {Qi},
  \citenamefont {Wu},\ and\ \citenamefont {Zhang}}]{qi2006topological}%
  \BibitemOpen
  \bibfield  {author} {\bibinfo {author} {\bibfnamefont {X.-L.}\ \bibnamefont
  {Qi}}, \bibinfo {author} {\bibfnamefont {Y.-S.}\ \bibnamefont {Wu}}, \ and\
  \bibinfo {author} {\bibfnamefont {S.-C.}\ \bibnamefont {Zhang}},\ }\bibfield
  {title} {\enquote {\bibinfo {title} {Topological quantization of the spin
  hall effect in two-dimensional paramagnetic semiconductors},}\ }\href
  {https://link.aps.org/doi/10.1103/PhysRevB.74.085308} {\bibfield  {journal}
  {\bibinfo  {journal} {\emph {Phys. Rev. B}},\ }\textbf {\bibinfo {volume}
  {74}},\ \bibinfo {pages} {085308}\  (\bibinfo {year} {2006})}\BibitemShut
  {NoStop}%
\bibitem [{\citenamefont {Yu}\ \emph {et~al.}(2010)\citenamefont {Yu},
  \citenamefont {Zhang}, \citenamefont {Zhang}, \citenamefont {Zhang},
  \citenamefont {Dai},\ and\ \citenamefont {Fang}}]{yu2010quantized}%
  \BibitemOpen
  \bibfield  {author} {\bibinfo {author} {\bibfnamefont {R.}~\bibnamefont
  {Yu}}, \bibinfo {author} {\bibfnamefont {W.}~\bibnamefont {Zhang}}, \bibinfo
  {author} {\bibfnamefont {H.-J.}\ \bibnamefont {Zhang}}, \bibinfo {author}
  {\bibfnamefont {S.-C.}\ \bibnamefont {Zhang}}, \bibinfo {author}
  {\bibfnamefont {X.}~\bibnamefont {Dai}}, \ and\ \bibinfo {author}
  {\bibfnamefont {Z.}~\bibnamefont {Fang}},\ }\bibfield  {title} {\enquote
  {\bibinfo {title} {Quantized anomalous \text{H}all effect in magnetic
  topological insulators},}\ }\href {https://doi.org/10.1126/science.1187485}
  {\bibfield  {journal} {\bibinfo  {journal} {\emph {Science}},\ }\textbf
  {\bibinfo {volume} {329}},\ \bibinfo {pages} {61--64}\  (\bibinfo {year}
  {2010})}\BibitemShut {NoStop}%
\bibitem [{\citenamefont {Chang}\ \emph {et~al.}(2013)\citenamefont {Chang},
  \citenamefont {Zhang}, \citenamefont {Feng}, \citenamefont {Shen},
  \citenamefont {Zhang}, \citenamefont {Guo}, \citenamefont {Li}, \citenamefont
  {Ou}, \citenamefont {Wei}, \citenamefont {Wang} \emph
  {et~al.}}]{chang2013experimental}%
  \BibitemOpen
  \bibfield  {author} {\bibinfo {author} {\bibfnamefont {C.-Z.}\ \bibnamefont
  {Chang}}, \bibinfo {author} {\bibfnamefont {J.}~\bibnamefont {Zhang}},
  \bibinfo {author} {\bibfnamefont {X.}~\bibnamefont {Feng}}, \bibinfo {author}
  {\bibfnamefont {J.}~\bibnamefont {Shen}}, \bibinfo {author} {\bibfnamefont
  {Z.}~\bibnamefont {Zhang}}, \bibinfo {author} {\bibfnamefont
  {M.}~\bibnamefont {Guo}}, \bibinfo {author} {\bibfnamefont {K.}~\bibnamefont
  {Li}}, \bibinfo {author} {\bibfnamefont {Y.}~\bibnamefont {Ou}}, \bibinfo
  {author} {\bibfnamefont {P.}~\bibnamefont {Wei}}, \bibinfo {author}
  {\bibfnamefont {L.-L.}\ \bibnamefont {Wang}},  \emph {et~al.},\ }\bibfield
  {title} {\enquote {\bibinfo {title} {Experimental observation of the quantum
  anomalous \text{H}all effect in a magnetic topological insulator},}\ }\href
  {https://doi.org/10.1126/science.1234414} {\bibfield  {journal} {\bibinfo
  {journal} {\emph {Science}},\ }\textbf {\bibinfo {volume} {340}},\ \bibinfo
  {pages} {167--170}\  (\bibinfo {year} {2013})}\BibitemShut {NoStop}%
\bibitem [{\citenamefont {Checkelsky}\ \emph {et~al.}(2014)\citenamefont
  {Checkelsky}, \citenamefont {Yoshimi}, \citenamefont {Tsukazaki},
  \citenamefont {Takahashi}, \citenamefont {Kozuka}, \citenamefont {Falson},
  \citenamefont {Kawasaki},\ and\ \citenamefont
  {Tokura}}]{checkelsky2014trajectory}%
  \BibitemOpen
  \bibfield  {author} {\bibinfo {author} {\bibfnamefont {J.}~\bibnamefont
  {Checkelsky}}, \bibinfo {author} {\bibfnamefont {R.}~\bibnamefont {Yoshimi}},
  \bibinfo {author} {\bibfnamefont {A.}~\bibnamefont {Tsukazaki}}, \bibinfo
  {author} {\bibfnamefont {K.}~\bibnamefont {Takahashi}}, \bibinfo {author}
  {\bibfnamefont {Y.}~\bibnamefont {Kozuka}}, \bibinfo {author} {\bibfnamefont
  {J.}~\bibnamefont {Falson}}, \bibinfo {author} {\bibfnamefont
  {M.}~\bibnamefont {Kawasaki}}, \ and\ \bibinfo {author} {\bibfnamefont
  {Y.}~\bibnamefont {Tokura}},\ }\bibfield  {title} {\enquote {\bibinfo {title}
  {Trajectory of the anomalous \text{H}all effect towards the quantized state
  in a ferromagnetic topological insulator},}\ }\href
  {https://doi.org/10.1038/nphys3053} {\bibfield  {journal} {\bibinfo
  {journal} {\emph {Nat. Phys.}},\ }\textbf {\bibinfo {volume} {10}},\ \bibinfo
  {pages} {731--736}\  (\bibinfo {year} {2014})}\BibitemShut {NoStop}%
\bibitem [{\citenamefont {Liu}\ \emph {et~al.}(2016)\citenamefont {Liu},
  \citenamefont {Zhang},\ and\ \citenamefont {Qi}}]{liu2016quantum}%
  \BibitemOpen
  \bibfield  {author} {\bibinfo {author} {\bibfnamefont {C.-X.}\ \bibnamefont
  {Liu}}, \bibinfo {author} {\bibfnamefont {S.-C.}\ \bibnamefont {Zhang}}, \
  and\ \bibinfo {author} {\bibfnamefont {X.-L.}\ \bibnamefont {Qi}},\
  }\bibfield  {title} {\enquote {\bibinfo {title} {The quantum anomalous
  \text{H}all effect: theory and experiment},}\ }\href
  {https://doi.org/10.1146/annurev-conmatphys-031115-011417} {\bibfield
  {journal} {\bibinfo  {journal} {\emph {Annu. Rev. Condens. Matter Phys.}},\
  }\textbf {\bibinfo {volume} {7}},\ \bibinfo {pages} {301--321}\  (\bibinfo
  {year} {2016})}\BibitemShut {NoStop}%
\bibitem [{\citenamefont {Chang}\ \emph {et~al.}(2023)\citenamefont {Chang},
  \citenamefont {Liu},\ and\ \citenamefont {MacDonald}}]{chang2023colloquium}%
  \BibitemOpen
  \bibfield  {author} {\bibinfo {author} {\bibfnamefont {C.-Z.}\ \bibnamefont
  {Chang}}, \bibinfo {author} {\bibfnamefont {C.-X.}\ \bibnamefont {Liu}}, \
  and\ \bibinfo {author} {\bibfnamefont {A.~H.}\ \bibnamefont {MacDonald}},\
  }\bibfield  {title} {\enquote {\bibinfo {title} {Colloquium: Quantum
  anomalous \text{H}all effect},}\ }\href
  {https://link.aps.org/doi/10.1103/RevModPhys.95.011002} {\bibfield  {journal}
  {\bibinfo  {journal} {\emph {Rev. Mod. Phys.}},\ }\textbf {\bibinfo {volume}
  {95}},\ \bibinfo {pages} {011002}\  (\bibinfo {year} {2023})}\BibitemShut
  {NoStop}%
\bibitem [{\citenamefont {Wu}\ \emph {et~al.}(2020)\citenamefont {Wu},
  \citenamefont {Gro{\ss}}, \citenamefont {Dai}, \citenamefont {Lujan},
  \citenamefont {Razavi}, \citenamefont {Zhang}, \citenamefont {Liu},
  \citenamefont {Sobotkiewich}, \citenamefont {F{\"o}rster}, \citenamefont
  {Weigand} \emph {et~al.}}]{wu2020ferrimagnetic}%
  \BibitemOpen
  \bibfield  {author} {\bibinfo {author} {\bibfnamefont {H.}~\bibnamefont
  {Wu}}, \bibinfo {author} {\bibfnamefont {F.}~\bibnamefont {Gro{\ss}}},
  \bibinfo {author} {\bibfnamefont {B.}~\bibnamefont {Dai}}, \bibinfo {author}
  {\bibfnamefont {D.}~\bibnamefont {Lujan}}, \bibinfo {author} {\bibfnamefont
  {S.~A.}\ \bibnamefont {Razavi}}, \bibinfo {author} {\bibfnamefont
  {P.}~\bibnamefont {Zhang}}, \bibinfo {author} {\bibfnamefont
  {Y.}~\bibnamefont {Liu}}, \bibinfo {author} {\bibfnamefont {K.}~\bibnamefont
  {Sobotkiewich}}, \bibinfo {author} {\bibfnamefont {J.}~\bibnamefont
  {F{\"o}rster}}, \bibinfo {author} {\bibfnamefont {M.}~\bibnamefont
  {Weigand}},  \emph {et~al.},\ }\bibfield  {title} {\enquote {\bibinfo {title}
  {Ferrimagnetic {S}kyrmions in {T}opological {I}nsulator/{F}errimagnet
  {H}eterostructures},}\ }\href
  {https://onlinelibrary.wiley.com/doi/abs/10.1002/adma.202003380} {\bibfield
  {journal} {\bibinfo  {journal} {\emph {Adv. Mater.}},\ }\textbf {\bibinfo
  {volume} {32}},\ \bibinfo {pages} {2003380}\  (\bibinfo {year}
  {2020})}\BibitemShut {NoStop}%
\bibitem [{\citenamefont {Li}\ \emph {et~al.}(2021){\natexlab{b}}\citenamefont
  {Li}, \citenamefont {Ding}, \citenamefont {Zhang}, \citenamefont {Kally},
  \citenamefont {Pillsbury}, \citenamefont {Heinonen}, \citenamefont {Rimal},
  \citenamefont {Bi}, \citenamefont {DeMann}, \citenamefont {Field} \emph
  {et~al.}}]{li2021topological}%
  \BibitemOpen
  \bibfield  {author} {\bibinfo {author} {\bibfnamefont {P.}~\bibnamefont
  {Li}}, \bibinfo {author} {\bibfnamefont {J.}~\bibnamefont {Ding}}, \bibinfo
  {author} {\bibfnamefont {S.~S.-L.}\ \bibnamefont {Zhang}}, \bibinfo {author}
  {\bibfnamefont {J.}~\bibnamefont {Kally}}, \bibinfo {author} {\bibfnamefont
  {T.}~\bibnamefont {Pillsbury}}, \bibinfo {author} {\bibfnamefont {O.~G.}\
  \bibnamefont {Heinonen}}, \bibinfo {author} {\bibfnamefont {G.}~\bibnamefont
  {Rimal}}, \bibinfo {author} {\bibfnamefont {C.}~\bibnamefont {Bi}}, \bibinfo
  {author} {\bibfnamefont {A.}~\bibnamefont {DeMann}}, \bibinfo {author}
  {\bibfnamefont {S.~B.}\ \bibnamefont {Field}},  \emph {et~al.},\ }\bibfield
  {title} {\enquote {\bibinfo {title} {Topological {H}all {E}ffect in a
  {T}opological {I}nsulator {I}nterfaced with a {M}agnetic {I}nsulator},}\
  }\href {https://doi.org/10.1021/acs.nanolett.0c03195} {\bibfield  {journal}
  {\bibinfo  {journal} {\emph {Nano Lett.}},\ }\textbf {\bibinfo {volume}
  {21}},\ \bibinfo {pages} {84--90}\  (\bibinfo {year}
  {2021}{\natexlab{b}})}\BibitemShut {NoStop}%
\bibitem [{\citenamefont {Zhang}\ \emph {et~al.}(2021)\citenamefont {Zhang},
  \citenamefont {Ambhire}, \citenamefont {Lu}, \citenamefont {Niu},
  \citenamefont {Cook}, \citenamefont {Jiang}, \citenamefont {Hong},
  \citenamefont {Alahmed}, \citenamefont {He}, \citenamefont {Zhang} \emph
  {et~al.}}]{zhang2021giant}%
  \BibitemOpen
  \bibfield  {author} {\bibinfo {author} {\bibfnamefont {X.}~\bibnamefont
  {Zhang}}, \bibinfo {author} {\bibfnamefont {S.~C.}\ \bibnamefont {Ambhire}},
  \bibinfo {author} {\bibfnamefont {Q.}~\bibnamefont {Lu}}, \bibinfo {author}
  {\bibfnamefont {W.}~\bibnamefont {Niu}}, \bibinfo {author} {\bibfnamefont
  {J.}~\bibnamefont {Cook}}, \bibinfo {author} {\bibfnamefont {J.~S.}\
  \bibnamefont {Jiang}}, \bibinfo {author} {\bibfnamefont {D.}~\bibnamefont
  {Hong}}, \bibinfo {author} {\bibfnamefont {L.}~\bibnamefont {Alahmed}},
  \bibinfo {author} {\bibfnamefont {L.}~\bibnamefont {He}}, \bibinfo {author}
  {\bibfnamefont {R.}~\bibnamefont {Zhang}},  \emph {et~al.},\ }\bibfield
  {title} {\enquote {\bibinfo {title} {Giant topological {Hall} effect in van
  der {Waals} heterostructures of {CrTe$_2$}/{Bi$_2$Te$_3$}},}\ }\href
  {https://doi.org/10.1021/acsnano.1c05519} {\bibfield  {journal} {\bibinfo
  {journal} {\emph {ACS Nano}},\ }\textbf {\bibinfo {volume} {15}},\ \bibinfo
  {pages} {15710--15719}\  (\bibinfo {year} {2021})}\BibitemShut {NoStop}%
\bibitem [{\citenamefont {Mellnik}\ \emph {et~al.}(2014)\citenamefont
  {Mellnik}, \citenamefont {Lee}, \citenamefont {Richardella}, \citenamefont
  {Grab}, \citenamefont {Mintun}, \citenamefont {Fischer}, \citenamefont
  {Vaezi}, \citenamefont {Manchon}, \citenamefont {Kim}, \citenamefont
  {Samarth} \emph {et~al.}}]{Mellnik2014}%
  \BibitemOpen
  \bibfield  {author} {\bibinfo {author} {\bibfnamefont {A.~R.}\ \bibnamefont
  {Mellnik}}, \bibinfo {author} {\bibfnamefont {J.~S.}\ \bibnamefont {Lee}},
  \bibinfo {author} {\bibfnamefont {A.}~\bibnamefont {Richardella}}, \bibinfo
  {author} {\bibfnamefont {J.~L.}\ \bibnamefont {Grab}}, \bibinfo {author}
  {\bibfnamefont {P.~J.}\ \bibnamefont {Mintun}}, \bibinfo {author}
  {\bibfnamefont {M.~H.}\ \bibnamefont {Fischer}}, \bibinfo {author}
  {\bibfnamefont {A.}~\bibnamefont {Vaezi}}, \bibinfo {author} {\bibfnamefont
  {A.}~\bibnamefont {Manchon}}, \bibinfo {author} {\bibfnamefont {E.-A.}\
  \bibnamefont {Kim}}, \bibinfo {author} {\bibfnamefont {N.}~\bibnamefont
  {Samarth}},  \emph {et~al.},\ }\bibfield  {title} {\enquote {\bibinfo {title}
  {{Spin-transfer torque generated by a topological insulator}},}\ }\href
  {https://doi.org/10.1038/nature13534} {\bibfield  {journal} {\bibinfo
  {journal} {\emph {Nature}},\ }\textbf {\bibinfo {volume} {511}},\ \bibinfo
  {pages} {449--451}\  (\bibinfo {year} {2014})}\BibitemShut {NoStop}%
\bibitem [{\citenamefont {Han}\ \emph {et~al.}(2017)\citenamefont {Han},
  \citenamefont {Richardella}, \citenamefont {Siddiqui}, \citenamefont
  {Finley}, \citenamefont {Samarth},\ and\ \citenamefont
  {Liu}}]{Han17PRL_SOT-TI}%
  \BibitemOpen
  \bibfield  {author} {\bibinfo {author} {\bibfnamefont {J.}~\bibnamefont
  {Han}}, \bibinfo {author} {\bibfnamefont {A.}~\bibnamefont {Richardella}},
  \bibinfo {author} {\bibfnamefont {S.~A.}\ \bibnamefont {Siddiqui}}, \bibinfo
  {author} {\bibfnamefont {J.}~\bibnamefont {Finley}}, \bibinfo {author}
  {\bibfnamefont {N.}~\bibnamefont {Samarth}}, \ and\ \bibinfo {author}
  {\bibfnamefont {L.}~\bibnamefont {Liu}},\ }\bibfield  {title} {\enquote
  {\bibinfo {title} {Room-temperature spin-orbit torque switching induced by a
  topological insulator},}\ }\href
  {https://link.aps.org/doi/10.1103/PhysRevLett.119.077702} {\bibfield
  {journal} {\bibinfo  {journal} {\emph {Phys. Rev. Lett.}},\ }\textbf
  {\bibinfo {volume} {119}},\ \bibinfo {pages} {077702}\  (\bibinfo {year}
  {2017})}\BibitemShut {NoStop}%
\bibitem [{\citenamefont {Li}\ \emph {et~al.}(2019)\citenamefont {Li},
  \citenamefont {Kally}, \citenamefont {Zhang}, \citenamefont {Pillsbury},
  \citenamefont {Ding}, \citenamefont {Csaba}, \citenamefont {Ding},
  \citenamefont {Jiang}, \citenamefont {Liu}, \citenamefont {Sinclair} \emph
  {et~al.}}]{li2019magnetization}%
  \BibitemOpen
  \bibfield  {author} {\bibinfo {author} {\bibfnamefont {P.}~\bibnamefont
  {Li}}, \bibinfo {author} {\bibfnamefont {J.}~\bibnamefont {Kally}}, \bibinfo
  {author} {\bibfnamefont {S.~S.-L.}\ \bibnamefont {Zhang}}, \bibinfo {author}
  {\bibfnamefont {T.}~\bibnamefont {Pillsbury}}, \bibinfo {author}
  {\bibfnamefont {J.}~\bibnamefont {Ding}}, \bibinfo {author} {\bibfnamefont
  {G.}~\bibnamefont {Csaba}}, \bibinfo {author} {\bibfnamefont
  {J.}~\bibnamefont {Ding}}, \bibinfo {author} {\bibfnamefont {J.}~\bibnamefont
  {Jiang}}, \bibinfo {author} {\bibfnamefont {Y.}~\bibnamefont {Liu}}, \bibinfo
  {author} {\bibfnamefont {R.}~\bibnamefont {Sinclair}},  \emph {et~al.},\
  }\bibfield  {title} {\enquote {\bibinfo {title} {Magnetization switching
  using topological surface states},}\ }\href
  {https://doi.org/10.1126/sciadv.aaw3415} {\bibfield  {journal} {\bibinfo
  {journal} {\emph {Sci. Adv.}},\ }\textbf {\bibinfo {volume} {5}},\ \bibinfo
  {pages} {eaaw3415}\  (\bibinfo {year} {2019})}\BibitemShut {NoStop}%
\bibitem [{\citenamefont {Moghaddam}\ \emph {et~al.}(2020)\citenamefont
  {Moghaddam}, \citenamefont {Qaiumzadeh}, \citenamefont {Dyrda\l{}},\ and\
  \citenamefont {Berakdar}}]{Moghaddam20PRL_SOT-AMR-proximity-TI}%
  \BibitemOpen
  \bibfield  {author} {\bibinfo {author} {\bibfnamefont {A.~G.}\ \bibnamefont
  {Moghaddam}}, \bibinfo {author} {\bibfnamefont {A.}~\bibnamefont
  {Qaiumzadeh}}, \bibinfo {author} {\bibfnamefont {A.}~\bibnamefont
  {Dyrda\l{}}}, \ and\ \bibinfo {author} {\bibfnamefont {J.}~\bibnamefont
  {Berakdar}},\ }\bibfield  {title} {\enquote {\bibinfo {title} {Highly tunable
  spin-orbit torque and anisotropic magnetoresistance in a topological
  insulator thin film attached to ferromagnetic layer},}\ }\href
  {https://link.aps.org/doi/10.1103/PhysRevLett.125.196801} {\bibfield
  {journal} {\bibinfo  {journal} {\emph {Phys. Rev. Lett.}},\ }\textbf
  {\bibinfo {volume} {125}},\ \bibinfo {pages} {196801}\  (\bibinfo {year}
  {2020})}\BibitemShut {NoStop}%
\bibitem [{\citenamefont {Chiba}\ \emph {et~al.}(2017)\citenamefont {Chiba},
  \citenamefont {Takahashi},\ and\ \citenamefont {Bauer}}]{chiba2017magnetic}%
  \BibitemOpen
  \bibfield  {author} {\bibinfo {author} {\bibfnamefont {T.}~\bibnamefont
  {Chiba}}, \bibinfo {author} {\bibfnamefont {S.}~\bibnamefont {Takahashi}}, \
  and\ \bibinfo {author} {\bibfnamefont {G.~E.~W.}\ \bibnamefont {Bauer}},\
  }\bibfield  {title} {\enquote {\bibinfo {title} {Magnetic-proximity-induced
  magnetoresistance on topological insulators},}\ }\href
  {https://link.aps.org/doi/10.1103/PhysRevB.95.094428} {\bibfield  {journal}
  {\bibinfo  {journal} {\emph {Phys. Rev. B}},\ }\textbf {\bibinfo {volume}
  {95}},\ \bibinfo {pages} {094428}\  (\bibinfo {year} {2017})}\BibitemShut
  {NoStop}%
\bibitem [{\citenamefont {Sklenar}\ \emph {et~al.}(2021)\citenamefont
  {Sklenar}, \citenamefont {Zhang}, \citenamefont {Jungfleisch}, \citenamefont
  {Kim}, \citenamefont {Xiao}, \citenamefont {MacDougall}, \citenamefont
  {Gilbert}, \citenamefont {Hoffmann}, \citenamefont {Schiffer},\ and\
  \citenamefont {Mason}}]{sklenar2021proximity}%
  \BibitemOpen
  \bibfield  {author} {\bibinfo {author} {\bibfnamefont {J.}~\bibnamefont
  {Sklenar}}, \bibinfo {author} {\bibfnamefont {Y.}~\bibnamefont {Zhang}},
  \bibinfo {author} {\bibfnamefont {M.~B.}\ \bibnamefont {Jungfleisch}},
  \bibinfo {author} {\bibfnamefont {Y.}~\bibnamefont {Kim}}, \bibinfo {author}
  {\bibfnamefont {Y.}~\bibnamefont {Xiao}}, \bibinfo {author} {\bibfnamefont
  {G.~J.}\ \bibnamefont {MacDougall}}, \bibinfo {author} {\bibfnamefont
  {M.~J.}\ \bibnamefont {Gilbert}}, \bibinfo {author} {\bibfnamefont
  {A.}~\bibnamefont {Hoffmann}}, \bibinfo {author} {\bibfnamefont
  {P.}~\bibnamefont {Schiffer}}, \ and\ \bibinfo {author} {\bibfnamefont
  {N.}~\bibnamefont {Mason}},\ }\bibfield  {title} {\enquote {\bibinfo {title}
  {Proximity-induced anisotropic magnetoresistance in magnetized topological
  insulators},}\ }\href {https://doi.org/10.1063/5.0052301} {\bibfield
  {journal} {\bibinfo  {journal} {\emph {Appl. Phys. Lett.}},\ }\textbf
  {\bibinfo {volume} {118}},\ \bibinfo {pages} {232402}\  (\bibinfo {year}
  {2021})}\BibitemShut {NoStop}%
\bibitem [{\citenamefont {Wu}\ \emph {et~al.}(2021)\citenamefont {Wu},
  \citenamefont {Chen}, \citenamefont {Zhang}, \citenamefont {He},
  \citenamefont {Nance}, \citenamefont {Guo}, \citenamefont {Sasaki},
  \citenamefont {Shirokura}, \citenamefont {Hai}, \citenamefont {Fang} \emph
  {et~al.}}]{Wu21NC_MRAM-TI}%
  \BibitemOpen
  \bibfield  {author} {\bibinfo {author} {\bibfnamefont {H.}~\bibnamefont
  {Wu}}, \bibinfo {author} {\bibfnamefont {A.}~\bibnamefont {Chen}}, \bibinfo
  {author} {\bibfnamefont {P.}~\bibnamefont {Zhang}}, \bibinfo {author}
  {\bibfnamefont {H.}~\bibnamefont {He}}, \bibinfo {author} {\bibfnamefont
  {J.}~\bibnamefont {Nance}}, \bibinfo {author} {\bibfnamefont
  {C.}~\bibnamefont {Guo}}, \bibinfo {author} {\bibfnamefont {J.}~\bibnamefont
  {Sasaki}}, \bibinfo {author} {\bibfnamefont {T.}~\bibnamefont {Shirokura}},
  \bibinfo {author} {\bibfnamefont {P.~N.}\ \bibnamefont {Hai}}, \bibinfo
  {author} {\bibfnamefont {B.}~\bibnamefont {Fang}},  \emph {et~al.},\
  }\bibfield  {title} {\enquote {\bibinfo {title} {Magnetic memory driven by
  topological insulators},}\ }\href
  {https://doi.org/10.1038/s41467-021-26478-3} {\bibfield  {journal} {\bibinfo
  {journal} {\emph {Nat. Commun.}},\ }\textbf {\bibinfo {volume} {12}},\
  \bibinfo {pages} {6251}\  (\bibinfo {year} {2021})}\BibitemShut {NoStop}%
\bibitem [{\citenamefont {Fu}\ and\ \citenamefont
  {Kane}(2008)}]{Fu08PRL_proximity-SC_TI_majorana}%
  \BibitemOpen
  \bibfield  {author} {\bibinfo {author} {\bibfnamefont {L.}~\bibnamefont
  {Fu}}\ and\ \bibinfo {author} {\bibfnamefont {C.~L.}\ \bibnamefont {Kane}},\
  }\bibfield  {title} {\enquote {\bibinfo {title} {Superconducting proximity
  effect and majorana fermions at the surface of a topological insulator},}\
  }\href {https://link.aps.org/doi/10.1103/PhysRevLett.100.096407} {\bibfield
  {journal} {\bibinfo  {journal} {\emph {Phys. Rev. Lett.}},\ }\textbf
  {\bibinfo {volume} {100}},\ \bibinfo {pages} {096407}\  (\bibinfo {year}
  {2008})}\BibitemShut {NoStop}%
\bibitem [{\citenamefont {Lian}\ \emph {et~al.}(2018)\citenamefont {Lian},
  \citenamefont {Sun}, \citenamefont {Vaezi}, \citenamefont {Qi},\ and\
  \citenamefont {Zhang}}]{Lian18PNAS_Topo-quant-comput_MZM}%
  \BibitemOpen
  \bibfield  {author} {\bibinfo {author} {\bibfnamefont {B.}~\bibnamefont
  {Lian}}, \bibinfo {author} {\bibfnamefont {X.-Q.}\ \bibnamefont {Sun}},
  \bibinfo {author} {\bibfnamefont {A.}~\bibnamefont {Vaezi}}, \bibinfo
  {author} {\bibfnamefont {X.-L.}\ \bibnamefont {Qi}}, \ and\ \bibinfo {author}
  {\bibfnamefont {S.-C.}\ \bibnamefont {Zhang}},\ }\bibfield  {title} {\enquote
  {\bibinfo {title} {Topological quantum computation based on chiral majorana
  fermions},}\ }\href {https://www.pnas.org/doi/abs/10.1073/pnas.1810003115}
  {\bibfield  {journal} {\bibinfo  {journal} {\emph {Proc. Natl. Acad. Sci.}},\
  }\textbf {\bibinfo {volume} {115}},\ \bibinfo {pages} {10938--10942}\
  (\bibinfo {year} {2018})}\BibitemShut {NoStop}%
\bibitem [{\citenamefont {Lv}\ \emph {et~al.}(2022)\citenamefont {Lv},
  \citenamefont {Kally}, \citenamefont {Liu}, \citenamefont {Quarterman},
  \citenamefont {Pillsbury}, \citenamefont {Kirby}, \citenamefont {Grutter},
  \citenamefont {Sahu}, \citenamefont {Borchers}, \citenamefont {Wu} \emph
  {et~al.}}]{lv2022large}%
  \BibitemOpen
  \bibfield  {author} {\bibinfo {author} {\bibfnamefont {Y.}~\bibnamefont
  {Lv}}, \bibinfo {author} {\bibfnamefont {J.}~\bibnamefont {Kally}}, \bibinfo
  {author} {\bibfnamefont {T.}~\bibnamefont {Liu}}, \bibinfo {author}
  {\bibfnamefont {P.}~\bibnamefont {Quarterman}}, \bibinfo {author}
  {\bibfnamefont {T.}~\bibnamefont {Pillsbury}}, \bibinfo {author}
  {\bibfnamefont {B.~J.}\ \bibnamefont {Kirby}}, \bibinfo {author}
  {\bibfnamefont {A.~J.}\ \bibnamefont {Grutter}}, \bibinfo {author}
  {\bibfnamefont {P.}~\bibnamefont {Sahu}}, \bibinfo {author} {\bibfnamefont
  {J.~A.}\ \bibnamefont {Borchers}}, \bibinfo {author} {\bibfnamefont
  {M.}~\bibnamefont {Wu}},  \emph {et~al.},\ }\bibfield  {title} {\enquote
  {\bibinfo {title} {Large unidirectional spin \text{H}all and \text{R}ashba-
  \text{E}delstein magnetoresistance in topological insulator/magnetic
  insulator heterostructures},}\ }\href {https://doi.org/10.1063/5.0073976}
  {\bibfield  {journal} {\bibinfo  {journal} {\emph {Appl. Phys. Rev.}},\
  }\textbf {\bibinfo {volume} {9}},\ \bibinfo {pages} {011406}\  (\bibinfo
  {year} {2022})}\BibitemShut {NoStop}%
\bibitem [{\citenamefont {Zhou}\ \emph
  {et~al.}(2021){\natexlab{b}}\citenamefont {Zhou}, \citenamefont {Zeng},
  \citenamefont {Jia}, \citenamefont {Chen},\ and\ \citenamefont
  {Wu}}]{zhou2021sign}%
  \BibitemOpen
  \bibfield  {author} {\bibinfo {author} {\bibfnamefont {X.}~\bibnamefont
  {Zhou}}, \bibinfo {author} {\bibfnamefont {F.}~\bibnamefont {Zeng}}, \bibinfo
  {author} {\bibfnamefont {M.}~\bibnamefont {Jia}}, \bibinfo {author}
  {\bibfnamefont {H.}~\bibnamefont {Chen}}, \ and\ \bibinfo {author}
  {\bibfnamefont {Y.}~\bibnamefont {Wu}},\ }\bibfield  {title} {\enquote
  {\bibinfo {title} {{Sign reversal of unidirectional magnetoresistance in
  monocrystalline Fe/Pt bilayers}},}\ }\href
  {https://link.aps.org/doi/10.1103/PhysRevB.104.184413} {\bibfield  {journal}
  {\bibinfo  {journal} {\emph {Phys. Rev. B}},\ }\textbf {\bibinfo {volume}
  {104}},\ \bibinfo {pages} {184413}\  (\bibinfo {year}
  {2021}{\natexlab{b}})}\BibitemShut {NoStop}%
\bibitem [{\citenamefont {Nguyen}\ \emph {et~al.}(2021)\citenamefont {Nguyen},
  \citenamefont {Nguyen}, \citenamefont {Jeong}, \citenamefont {Park},
  \citenamefont {Jang}, \citenamefont {Lee}, \citenamefont {Lee}, \citenamefont
  {Park}, \citenamefont {Cho}, \citenamefont {Lee} \emph
  {et~al.}}]{nguyen2021unidirectional}%
  \BibitemOpen
  \bibfield  {author} {\bibinfo {author} {\bibfnamefont {T.~H.~T.}\
  \bibnamefont {Nguyen}}, \bibinfo {author} {\bibfnamefont {V.~Q.}\
  \bibnamefont {Nguyen}}, \bibinfo {author} {\bibfnamefont {S.}~\bibnamefont
  {Jeong}}, \bibinfo {author} {\bibfnamefont {E.}~\bibnamefont {Park}},
  \bibinfo {author} {\bibfnamefont {H.}~\bibnamefont {Jang}}, \bibinfo {author}
  {\bibfnamefont {N.~J.}\ \bibnamefont {Lee}}, \bibinfo {author} {\bibfnamefont
  {S.}~\bibnamefont {Lee}}, \bibinfo {author} {\bibfnamefont {B.-G.}\
  \bibnamefont {Park}}, \bibinfo {author} {\bibfnamefont {S.}~\bibnamefont
  {Cho}}, \bibinfo {author} {\bibfnamefont {H.-W.}\ \bibnamefont {Lee}},  \emph
  {et~al.},\ }\bibfield  {title} {\enquote {\bibinfo {title} {{Unidirectional
  spin Hall magnetoresistance in epitaxial Cr/Fe bilayer from electron-magnon
  scattering}},}\ }\href {https://doi.org/10.1038/s42005-021-00743-9}
  {\bibfield  {journal} {\bibinfo  {journal} {\emph {Commun. Phys.}},\ }\textbf
  {\bibinfo {volume} {4}},\ \bibinfo {pages} {247}\  (\bibinfo {year}
  {2021})}\BibitemShut {NoStop}%
\bibitem [{\citenamefont {Fan}\ \emph {et~al.}(2022)\citenamefont {Fan},
  \citenamefont {Zhang}, \citenamefont {Han}, \citenamefont {Lv}, \citenamefont
  {Liu},\ and\ \citenamefont {Wang}}]{fan2022observation}%
  \BibitemOpen
  \bibfield  {author} {\bibinfo {author} {\bibfnamefont {Y.}~\bibnamefont
  {Fan}}, \bibinfo {author} {\bibfnamefont {P.}~\bibnamefont {Zhang}}, \bibinfo
  {author} {\bibfnamefont {J.}~\bibnamefont {Han}}, \bibinfo {author}
  {\bibfnamefont {Y.}~\bibnamefont {Lv}}, \bibinfo {author} {\bibfnamefont
  {L.}~\bibnamefont {Liu}}, \ and\ \bibinfo {author} {\bibfnamefont {J.-P.}\
  \bibnamefont {Wang}},\ }\bibfield  {title} {\enquote {\bibinfo {title}
  {{Observation of the unidirectional magnetoresistance in antiferromagnetic
  insulator Fe2O3/Pt bilayers}},}\ }\href
  {https://doi.org/10.1002/aelm.202300232} {\bibfield  {journal} {\bibinfo
  {journal} {\emph {Adv. Electron. Mater.}},\ \bibinfo {pages} {2300232}}\
  (\bibinfo {year} {2022})}\BibitemShut {NoStop}%
\bibitem [{\citenamefont {Wang}\ \emph
  {et~al.}(2023){\natexlab{b}}\citenamefont {Wang}, \citenamefont {Cui},
  \citenamefont {Xie}, \citenamefont {Zhang}, \citenamefont {Tian},
  \citenamefont {Bai}, \citenamefont {Huang}, \citenamefont {Cao},\ and\
  \citenamefont {Yan}}]{wang2023controllable}%
  \BibitemOpen
  \bibfield  {author} {\bibinfo {author} {\bibfnamefont {S.}~\bibnamefont
  {Wang}}, \bibinfo {author} {\bibfnamefont {X.}~\bibnamefont {Cui}}, \bibinfo
  {author} {\bibfnamefont {R.}~\bibnamefont {Xie}}, \bibinfo {author}
  {\bibfnamefont {C.}~\bibnamefont {Zhang}}, \bibinfo {author} {\bibfnamefont
  {Y.}~\bibnamefont {Tian}}, \bibinfo {author} {\bibfnamefont {L.}~\bibnamefont
  {Bai}}, \bibinfo {author} {\bibfnamefont {Q.}~\bibnamefont {Huang}}, \bibinfo
  {author} {\bibfnamefont {Q.}~\bibnamefont {Cao}}, \ and\ \bibinfo {author}
  {\bibfnamefont {S.}~\bibnamefont {Yan}},\ }\bibfield  {title} {\enquote
  {\bibinfo {title} {Controllable unidirectional magnetoresistance in
  ferromagnetic films with broken symmetry},}\ }\href
  {https://link.aps.org/doi/10.1103/PhysRevB.107.094410} {\bibfield  {journal}
  {\bibinfo  {journal} {\emph {Phys. Rev. B}},\ }\textbf {\bibinfo {volume}
  {107}},\ \bibinfo {pages} {094410}\  (\bibinfo {year}
  {2023}{\natexlab{b}})}\BibitemShut {NoStop}%
\bibitem [{\citenamefont {Sherman}(2003)}]{sherman2003minimum}%
  \BibitemOpen
  \bibfield  {author} {\bibinfo {author} {\bibfnamefont {E.~Y.}\ \bibnamefont
  {Sherman}},\ }\bibfield  {title} {\enquote {\bibinfo {title} {Minimum of
  spin-orbit coupling in two-dimensional structures},}\ }\href
  {https://link.aps.org/doi/10.1103/PhysRevB.67.161303} {\bibfield  {journal}
  {\bibinfo  {journal} {\emph {Phys. Rev. B}},\ }\textbf {\bibinfo {volume}
  {67}},\ \bibinfo {pages} {161303}\  (\bibinfo {year} {2003})}\BibitemShut
  {NoStop}%
\bibitem [{\citenamefont {Golub}\ and\ \citenamefont
  {Ivchenko}(2004)}]{golub2004spin}%
  \BibitemOpen
  \bibfield  {author} {\bibinfo {author} {\bibfnamefont {L.~E.}\ \bibnamefont
  {Golub}}\ and\ \bibinfo {author} {\bibfnamefont {E.~L.}\ \bibnamefont
  {Ivchenko}},\ }\bibfield  {title} {\enquote {\bibinfo {title} {Spin splitting
  in symmetrical \text{SiGe} quantum wells},}\ }\href
  {https://link.aps.org/doi/10.1103/PhysRevB.69.115333} {\bibfield  {journal}
  {\bibinfo  {journal} {\emph {Phys. Rev. B}},\ }\textbf {\bibinfo {volume}
  {69}},\ \bibinfo {pages} {115333}\  (\bibinfo {year} {2004})}\BibitemShut
  {NoStop}%
\bibitem [{\citenamefont {Str\"om}\ \emph {et~al.}(2010)\citenamefont
  {Str\"om}, \citenamefont {Johannesson},\ and\ \citenamefont
  {Japaridze}}]{strom2010edge}%
  \BibitemOpen
  \bibfield  {author} {\bibinfo {author} {\bibfnamefont {A.}~\bibnamefont
  {Str\"om}}, \bibinfo {author} {\bibfnamefont {H.}~\bibnamefont
  {Johannesson}}, \ and\ \bibinfo {author} {\bibfnamefont {G.~I.}\ \bibnamefont
  {Japaridze}},\ }\bibfield  {title} {\enquote {\bibinfo {title} {Edge dynamics
  in a quantum spin \text{H}all state: Effects from \text{R}ashba spin-orbit
  interaction},}\ }\href
  {https://link.aps.org/doi/10.1103/PhysRevLett.104.256804} {\bibfield
  {journal} {\bibinfo  {journal} {\emph {Phys. Rev. Lett.}},\ }\textbf
  {\bibinfo {volume} {104}},\ \bibinfo {pages} {256804}\  (\bibinfo {year}
  {2010})}\BibitemShut {NoStop}%
\bibitem [{\citenamefont {Kimme}\ \emph {et~al.}(2016)\citenamefont {Kimme},
  \citenamefont {Rosenow},\ and\ \citenamefont
  {Brataas}}]{kimme2016backscattering}%
  \BibitemOpen
  \bibfield  {author} {\bibinfo {author} {\bibfnamefont {L.}~\bibnamefont
  {Kimme}}, \bibinfo {author} {\bibfnamefont {B.}~\bibnamefont {Rosenow}}, \
  and\ \bibinfo {author} {\bibfnamefont {A.}~\bibnamefont {Brataas}},\
  }\bibfield  {title} {\enquote {\bibinfo {title} {Backscattering in helical
  edge states from a magnetic impurity and \text{R}ashba disorder},}\ }\href
  {https://link.aps.org/doi/10.1103/PhysRevB.93.081301} {\bibfield  {journal}
  {\bibinfo  {journal} {\emph {Phys. Rev. B}},\ }\textbf {\bibinfo {volume}
  {93}},\ \bibinfo {pages} {081301}\  (\bibinfo {year} {2016})}\BibitemShut
  {NoStop}%
\bibitem [{\citenamefont {Langenfeld}\ \emph
  {et~al.}(2016){\natexlab{c}}\citenamefont {Langenfeld}, \citenamefont
  {Tshitoyan}, \citenamefont {Fang}, \citenamefont {Wells}, \citenamefont
  {Moore},\ and\ \citenamefont {Ferguson}}]{langenfeld2016exchange}%
  \BibitemOpen
  \bibfield  {author} {\bibinfo {author} {\bibfnamefont {S.}~\bibnamefont
  {Langenfeld}}, \bibinfo {author} {\bibfnamefont {V.}~\bibnamefont
  {Tshitoyan}}, \bibinfo {author} {\bibfnamefont {Z.}~\bibnamefont {Fang}},
  \bibinfo {author} {\bibfnamefont {A.}~\bibnamefont {Wells}}, \bibinfo
  {author} {\bibfnamefont {T.}~\bibnamefont {Moore}}, \ and\ \bibinfo {author}
  {\bibfnamefont {A.}~\bibnamefont {Ferguson}},\ }\bibfield  {title} {\enquote
  {\bibinfo {title} {{Exchange magnon induced resistance asymmetry in permalloy
  spin-Hall oscillators}},}\ }\href {https://doi.org/10.1063/1.4948921}
  {\bibfield  {journal} {\bibinfo  {journal} {\emph {Appl. Phys. Lett.}},\
  }\textbf {\bibinfo {volume} {108}}\  (\bibinfo {year}
  {2016}{\natexlab{c}})}\BibitemShut {NoStop}%
\bibitem [{\citenamefont {Sterk}\ \emph {et~al.}(2019)\citenamefont {Sterk},
  \citenamefont {Peerlings},\ and\ \citenamefont {Duine}}]{sterk2019magnon}%
  \BibitemOpen
  \bibfield  {author} {\bibinfo {author} {\bibfnamefont {W.~P.}\ \bibnamefont
  {Sterk}}, \bibinfo {author} {\bibfnamefont {D.}~\bibnamefont {Peerlings}}, \
  and\ \bibinfo {author} {\bibfnamefont {R.~A.}\ \bibnamefont {Duine}},\
  }\bibfield  {title} {\enquote {\bibinfo {title} {{Magnon contribution to
  unidirectional spin Hall magnetoresistance in
  ferromagnetic-insulator/heavy-metal bilayers}},}\ }\href
  {https://link.aps.org/doi/10.1103/PhysRevB.99.064438} {\bibfield  {journal}
  {\bibinfo  {journal} {\emph {Phys. Rev. B}},\ }\textbf {\bibinfo {volume}
  {99}},\ \bibinfo {pages} {064438}\  (\bibinfo {year} {2019})}\BibitemShut
  {NoStop}%
\bibitem [{\citenamefont {Zhang}\ \emph {et~al.}(2019)\citenamefont {Zhang},
  \citenamefont {Shi}, \citenamefont {Zhu}, \citenamefont {Xing}, \citenamefont
  {Zhang},\ and\ \citenamefont {Wang}}]{zhang2019topological}%
  \BibitemOpen
  \bibfield  {author} {\bibinfo {author} {\bibfnamefont {D.}~\bibnamefont
  {Zhang}}, \bibinfo {author} {\bibfnamefont {M.}~\bibnamefont {Shi}}, \bibinfo
  {author} {\bibfnamefont {T.}~\bibnamefont {Zhu}}, \bibinfo {author}
  {\bibfnamefont {D.}~\bibnamefont {Xing}}, \bibinfo {author} {\bibfnamefont
  {H.}~\bibnamefont {Zhang}}, \ and\ \bibinfo {author} {\bibfnamefont
  {J.}~\bibnamefont {Wang}},\ }\bibfield  {title} {\enquote {\bibinfo {title}
  {{Topological Axion States in the Magnetic Insulator
  ${\mathrm{MnBi}}_{2}{\mathrm{Te}}_{4}$ with the Quantized Magnetoelectric
  Effect}},}\ }\href {https://link.aps.org/doi/10.1103/PhysRevLett.122.206401}
  {\bibfield  {journal} {\bibinfo  {journal} {\emph {Phys. Rev. Lett.}},\
  }\textbf {\bibinfo {volume} {122}},\ \bibinfo {pages} {206401}\  (\bibinfo
  {year} {2019})}\BibitemShut {NoStop}%
\bibitem [{\citenamefont {Wang}\ \emph {et~al.}(2020)\citenamefont {Wang},
  \citenamefont {Wang}, \citenamefont {Yang}, \citenamefont {Shi},
  \citenamefont {Ruan}, \citenamefont {Xing}, \citenamefont {Wang},\ and\
  \citenamefont {Zhang}}]{wang2020dynamical}%
  \BibitemOpen
  \bibfield  {author} {\bibinfo {author} {\bibfnamefont {H.}~\bibnamefont
  {Wang}}, \bibinfo {author} {\bibfnamefont {D.}~\bibnamefont {Wang}}, \bibinfo
  {author} {\bibfnamefont {Z.}~\bibnamefont {Yang}}, \bibinfo {author}
  {\bibfnamefont {M.}~\bibnamefont {Shi}}, \bibinfo {author} {\bibfnamefont
  {J.}~\bibnamefont {Ruan}}, \bibinfo {author} {\bibfnamefont {D.}~\bibnamefont
  {Xing}}, \bibinfo {author} {\bibfnamefont {J.}~\bibnamefont {Wang}}, \ and\
  \bibinfo {author} {\bibfnamefont {H.}~\bibnamefont {Zhang}},\ }\bibfield
  {title} {\enquote {\bibinfo {title} {{Dynamical axion state with hidden
  pseudospin Chern numbers in ${\mathrm{MnBi}}_{2}{\mathrm{Te}}_{4}$-based
  heterostructures}},}\ }\href
  {https://link.aps.org/doi/10.1103/PhysRevB.101.081109} {\bibfield  {journal}
  {\bibinfo  {journal} {\emph {Phys. Rev. B}},\ }\textbf {\bibinfo {volume}
  {101}},\ \bibinfo {pages} {081109}\  (\bibinfo {year} {2020})}\BibitemShut
  {NoStop}%
\bibitem [{\citenamefont {Liu}\ \emph {et~al.}(2009)\citenamefont {Liu},
  \citenamefont {Liu}, \citenamefont {Xu}, \citenamefont {Qi},\ and\
  \citenamefont {Zhang}}]{liu2009magnetic}%
  \BibitemOpen
  \bibfield  {author} {\bibinfo {author} {\bibfnamefont {Q.}~\bibnamefont
  {Liu}}, \bibinfo {author} {\bibfnamefont {C.-X.}\ \bibnamefont {Liu}},
  \bibinfo {author} {\bibfnamefont {C.}~\bibnamefont {Xu}}, \bibinfo {author}
  {\bibfnamefont {X.-L.}\ \bibnamefont {Qi}}, \ and\ \bibinfo {author}
  {\bibfnamefont {S.-C.}\ \bibnamefont {Zhang}},\ }\bibfield  {title} {\enquote
  {\bibinfo {title} {{Magnetic Impurities on the Surface of a Topological
  Insulator}},}\ }\href
  {https://link.aps.org/doi/10.1103/PhysRevLett.102.156603} {\bibfield
  {journal} {\bibinfo  {journal} {\emph {Phys. Rev. Lett.}},\ }\textbf
  {\bibinfo {volume} {102}},\ \bibinfo {pages} {156603}\  (\bibinfo {year}
  {2009})}\BibitemShut {NoStop}%
\bibitem [{\citenamefont {Abanin}\ and\ \citenamefont
  {Pesin}(2011)}]{abanin2011ordering}%
  \BibitemOpen
  \bibfield  {author} {\bibinfo {author} {\bibfnamefont {D.~A.}\ \bibnamefont
  {Abanin}}\ and\ \bibinfo {author} {\bibfnamefont {D.~A.}\ \bibnamefont
  {Pesin}},\ }\bibfield  {title} {\enquote {\bibinfo {title} {{Ordering of
  Magnetic Impurities and Tunable Electronic Properties of Topological
  Insulators}},}\ }\href
  {https://link.aps.org/doi/10.1103/PhysRevLett.106.136802} {\bibfield
  {journal} {\bibinfo  {journal} {\emph {Phys. Rev. Lett.}},\ }\textbf
  {\bibinfo {volume} {106}},\ \bibinfo {pages} {136802}\  (\bibinfo {year}
  {2011})}\BibitemShut {NoStop}%
\bibitem [{\citenamefont {Nomura}\ and\ \citenamefont
  {Nagaosa}(2011)}]{nomura2011surface}%
  \BibitemOpen
  \bibfield  {author} {\bibinfo {author} {\bibfnamefont {K.}~\bibnamefont
  {Nomura}}\ and\ \bibinfo {author} {\bibfnamefont {N.}~\bibnamefont
  {Nagaosa}},\ }\bibfield  {title} {\enquote {\bibinfo {title}
  {{Surface-Quantized Anomalous Hall Current and the Magnetoelectric Effect in
  Magnetically Disordered Topological Insulators}},}\ }\href
  {https://link.aps.org/doi/10.1103/PhysRevLett.106.166802} {\bibfield
  {journal} {\bibinfo  {journal} {\emph {Phys. Rev. Lett.}},\ }\textbf
  {\bibinfo {volume} {106}},\ \bibinfo {pages} {166802}\  (\bibinfo {year}
  {2011})}\BibitemShut {NoStop}%
\end{thebibliography}%

%mypapers.bib, qmsg.bib, sahumr.bib, qumr.bib, proximity.bib, qkt.bib

\end{document}